\documentclass[11pt]{article}
\usepackage{epubtk}    
\usepackage{booktabs}  
\usepackage{graphicx}  
\usepackage{subfigure}
\usepackage{dsfont}
\usepackage{cancel}
\usepackage{placeins}
\usepackage{color}
\usepackage{enumerate}

\usepackage{amssymb,amsmath,amsthm}
\usepackage{aas_macros}
\usepackage{epsf}
\usepackage{longtable}
\numberwithin{equation}{section}

\def\be{\begin{equation}}
\def\ee{\end{equation}}
\def\bi{\begin{itemize}}
\def\ei{\end{itemize}}
\def\ben{\begin{enumerate}}
\def\een{\end{enumerate}}
\def\i{\item{}}
\def\edth{\check\partial}

\newcommand{\mb}[1]{\mathbf{#1}}
\newcommand{\mc}[1]{\mathcal{#1}}
\newcommand{\mr}[1]{\mathrm{#1}}

\newcommand{\unit}{\mathds{1}}
\newcommand{\zero}{\mb{0}}

\begin{document}

\title{Detection methods for stochastic gravitational-wave
backgrounds: a unified treatment}

\author{
\epubtkAuthorData{Joseph D.\ Romano}{%
Department of Physics and Astronomy\\
University of Texas Rio Grande Valley, Brownsville, TX 78520}{%
joseph.romano@utrgv.edu}{
}
\and
\epubtkAuthorData{Neil J.\ Cornish}{%
Department of Physics\\
Montana State University, Bozeman, MT 59717}{%
cornish@physics.montana.edu}{
}
}

\date{}
\maketitle

\begin{abstract}
We review detection methods that are currently in use 
or have been proposed to search for a stochastic background of 
gravitational radiation.
We consider both Bayesian and frequentist searches using 
ground-based and space-based laser interferometers, 
spacecraft Doppler tracking, and pulsar timing arrays; and we 
allow for anisotropy, non-Gaussianity, and non-standard 
polarization states.
Our focus is on relevant data analysis issues, and not 
on the particular astrophysical or early Universe sources 
that might give rise to such backgrounds.
We provide a unified treatment of these searches at the level of
detector response functions, detection sensitivity curves, and,
more generally, at the level of the likelihood function, since 
the choice of signal and noise models and prior probability 
distributions are actually what define the search.
Pedagogical examples are given whenever possible to compare and 
contrast different approaches.
We have tried to make the article as self-contained and 
comprehensive as possible, targeting graduate students and
new researchers looking to enter this field.
\end{abstract}

\epubtkKeywords{Gravitational waves, Data analysis, Stochastic backgrounds}

\pagebreak
\tableofcontents


\newpage
\section{Introduction}
\label{s:intro}

\begin{quotation}
The real voyage of discovery consists not in seeking new landscapes, 
but in having new eyes.
{\em Marcel Proust}
\end{quotation}

\noindent
It is an exciting time for the field of gravitational-wave 
astronomy.
The observation, on September 14th, 2015, of gravitational waves 
from the inspiral and merger of a pair of black holes~\cite{TheLIGOScientific:2016-GW150914}
has opened a radically new way of observing the Universe.
The event, denoted GW150914, was observed simultaneously by 
the two detectors of the Laser Interferometer 
Gravitational-wave Observatory (LIGO)~\cite{TheLIGOScientific:2014-advancedLIGO}.
[LIGO consists of two 4~km-long laser interferometers, 
one located in Hanford, Washington, the other in Livingston, LA.]
The merger event that produced the gravitational waves
occured in a distant galaxy roughly $1.3$~billion light~years from Earth.
The initial masses of the two black holes were estimated to be 
$36^{+5}_{-4}\ {\rm M}_\odot$ and $29^{+4}_{-4}\ {\rm M}_\odot$, 
and that of the post-merger black hole as
$62^{+4}_{-4}\ {\rm M}_\odot$ \cite{TheLIGOScientific:2016-parameters}.
The difference between the initial and final masses 
corresponds to 
$3.0^{+0.5}_{-0.5}\ {\rm M}_\odot c^2$ of energy radiated in
gravitational waves, with a peak luminosity of 
{\em more than ten times the combined luminosity of all 
the stars in all the galaxies in the visible universe}!
The fact that this event was observed {\em only} in 
gravitational waves---and not in electromagnetic 
waves---illustrates the complementarity and potential 
for new discoveries that comes with the opening of the 
gravitational-wave window onto the universe.

GW150914 is just the first of many gravitational-wave 
signals that we expect to observe over the next several years.
Indeed, roughly three months after the detection of
GW150914, a second event, GW151226, was observed
by the two LIGO detectors~\cite{TheLIGOScientific:2016-GW151226}.
This event also involved the inspiral and merger of a 
pair of stellar mass black holes, with initial component masses 
$14.2^{+8.3}_{-3.7}\ {\rm M}_\odot$ and $7.5^{+2.3}_{-2.3}\ {\rm M}_\odot$, 
and a final black hole mass of 
$20.8^{+6.1}_{-1.7}\ {\rm M}_\odot$.
The source was at a distance of roughly 1.4~billion light-years 
from Earth, comparable to that of GW150914.
Advanced LIGO will continue interleaving observation
runs and commissioning activities to reach design 
sensivity around 2020~\cite{TheLIGOScientific:2014-advancedLIGO}, 
which will allow detections of signals like GW150914 and GW151226 
with more than three times the signal-to-noise ratio than was 
observed for GW150914 (which was 24).
In addition, the Advanced Virgo detector~\cite{TheVirgo:2014-advancedVirgo} 
(a 3~km-long laser interferometer in Cascina, Italy) 
and KAGRA~\cite{KAGRA:2013} 
(a 3~km-long cryogenic laser interferometer in Kamioka mine in Japan)
should both be taking data by the end of 2016.
There are also plans for a third LIGO detector in India~\cite{Indigo}.
A global network of detectors such as this will allow
for much improved position reconstruction and parameter 
estimation of the sources~\cite{TheLIGOScientific:2016-network}.

\subsection{Motivation and context}
\label{s:motivation}

GW150914 and GW151226 were single events---binary black hole 
mergers that were observed with both template-based searches 
for compact binary inspirals and searches for 
generic gravitational-wave transients in the two LIGO 
detectors~\cite{TheLIGOScientific:2016-GW150914, TheLIGOScientific:2016-GW151226}.
The network matched-filter signal-to-noise ratio~\cite{Owen-Sathyaprakash:1999}
for these two events, using relativitistic waveform models
for binary black holes, was 24 and 13, respectively.
The probability that these detections were due to noise alone 
is $< 2\times 10^{-7}$, corresponding to a significance greater
than $5\sigma$---the standard for so-called ``gold-plated" detections.
But for every loud event like GW150914 or GW151226, 
we expect many more quiet events that are too distant to 
be individually detected, since the associated signal-to-noise 
ratios are too low.

The total rate of merger events from the population of 
stellar-mass binary black holes of which GW150914 and GW151226 
are members can be estimated%
\footnote{The coalescence rate is expected to vary significantly
with redshift $z$, so this simple calculation, which assumes a 
constant coalescence rate, provides only a rough estimate.}
by multiplying the local rate estimate of
9--240~${\rm Gpc}^{-3}\, {\rm yr}^{-1}$~\cite{TheLIGOScientific:2016-O1-rates}
by the comoving volume out to some large redshift, 
e.g., $z\sim 6$.
This yields a total rate of binary black hole mergers 
between $\sim\!1$~per minute and a few per hour.
Since the duration of each merger signal in the 
sensitive band of a LIGO-like detector is of 
order a few tenths of a second to $\sim\! 1$~second,
the {\em duty cycle} (the fraction of time that the signal 
is ``on" in the data) is $\ll 1$.
This means that the combined signal from such a population of binary 
black holes will be ``popcorn-like", with the majority of the individual 
signals being too weak to individually detect.
Since the arrival times of the merger signals are 
randomly-distributed, the combined signal from the population of 
binary black holes is itself random---it is an example of a
{\em stochastic background} of gravitational radiation.

More generally, a stochastic background of gravitational radiation
is {\em any} random gravitational-wave signal produced by a 
large number of weak, independent, and unresolved sources.
The background doesn't have to be popcorn-like, like the
expected signal from the population of binary black holes
which gave rise to GW150914 and GW151226.
It can be composed of individual deterministic signals 
that overlap in time (or in frequency) 
producing a ``confusion" noise analogous to conversations
at a cocktail party.
Such a confusion noise is produced by the galactic population 
of compact white dwarf binaries. 
(For this case, the stochastic signal is so strong 
that it becomes a {\em foreground}, acting as an additional 
source of noise when trying to detect {\em other} weak 
gravitational-wave signals in the same frequency band.)
Alternatively, the signal can be {\em intrinsically} random, 
associated with stochastic processes in the early Universe 
or with unmodeled sources, like supernovae, which produce 
signals that are not described by deterministic waveforms.

The focus of this review article is on data analysis 
strategies (i.e., detection methods) that can be used to 
detect and ultimately characterize a stochastic gravitational-wave
background.
To introduce this topic and to set the stage for the 
more detailed discussions to follow in later sections, we ask 
(and start to answer) the following questions:

\subsubsection{Why do we care about detecting a stochastic
background?}

Detecting a stochastic background of gravitational radiation
can provide information about astrophysical source populations
and processes in the very early Universe, which are 
inaccessible by any other means.
For example, electromagnetic radiation cannot provide a 
picture of the 
Universe any earlier than the time of last of scattering
(roughly 400,000 years after the Big Bang).
Gravitational waves, on the other hand, can give us 
information all the way back to the
onset of inflation, a mere 
$\sim\! 10^{-32}~{\rm s}$ after the Big Bang.
(See~\cite{Maggiore:2000} for a detailed discussion 
of both cosmological and astrophysical sources of a
stochastic gravitational-wave background.)

\subsubsection{Why is detection challenging?}

Stochastic signals are effectively another source of noise in 
a single detector.
So the fundamental problem is how to distinguish between 
gravitational-wave ``noise" and instrumental noise.
It turns out that there are several ways to do this, as we 
will discuss in the later sections of this article.

\subsubsection{What detection methods can one use?}

Cross-correlation methods can be used whenever one has 
multiple detectors that respond to the common 
gravitational-wave background.
For single detector analyses, 
e.g., for the Laser Interferometer Space Antenna (LISA), 
one needs 
to take advantage of null combinations of the data 
(which act as instrument noise monitors) or use instrument 
noise modeling to try to distinguish the gravitational-wave
signal from instrumental noise.
Over the past 15 years or so, the number of detection methods 
for stochastic backgrounds has increased considerably.
So now, in addition to the standard cross-correlation
search for a ``vanilla" (Gaussian-stationary, unpolarized,
isotropic) background,
one can search for 
non-Gaussian backgrounds,
anisotropic backgrounds,
circularly-polarized backgrounds,
and backgrounds with polarization components predicted by 
alternative (non-general-relativity) theories of gravity.
These searches are discussed in Sections~\ref{s:anisotropic} and \ref{s:extensions}.

Table~\ref{t:analysis-methods} summarizes the basic properties 
of various analysis methods that have been used 
(or proposed) for stochastic background searches.
\begin{table}[htbp]
\addtocounter{table}{-1} 
\centering
\begin{longtable}{p{2.5in} | p{3in}}
\toprule
EARLY ANALYSES (before 2000) & MORE RECENT ANALYSES 
\\
\midrule
used frequentist statistics & use both frequentist and Bayesian inference 
\\
\midrule
used cross-correlation methods & use cross-correlation methods and
stochastic templates; 
use null channels or knowledge about instrumental noise 
when cross-correlation is not available 
\\
\midrule
assumed Gaussian noise & have allowed non-Gaussian noise 
\\
\midrule
assumed stationary, Gaussian, unpolarized, and isotropic 
gravitational-wave backgrounds &
have allowed non-Gaussian, polarized, and anisotropic 
gravitational-wave backgrounds 
\\
\midrule
were done primarily in the context of ground-based detectors
(e.g., resonant bars and LIGO-like interferometers)
where the small-antenna (i.e., long-wavelength) approximation 
was valid &
have been done in the context of space-based detectors
(e.g., spacecraft tracking, LISA) and pulsar timing
arrays for which the small-antenna approximation is not valid
\\
\bottomrule
\end{longtable}
\caption{Overview of analysis methods for stochastic
gravitational-wave backgrounds.
The number and flexibility of the methods have increased 
considerably since the year 2000.}
\label{t:analysis-methods}
\end{table}
Despite apparent differences, {\em all} analyses
use a likelihood function, e.g., for defining 
frequentist statistics or for calculating posterior
distributions for Bayesian inference (as will be 
described in more detail in Section~\ref{s:inference}),
and take advantage of cross-correlations if multiple 
detectors are available (as will be described in more 
detail in Section~\ref{s:corr}).

\subsubsection{What are the prospects for detection?}

The prospects for detection 
depend on the source of the background (i.e., astrophysical
or cosmological) and the type of detector being used.
For example, 
a space-based interferometer like LISA 
is {\em guaranteed} to detect the gravitational-wave 
confusion noise produced by the galactic population of 
compact white dwarf binaries.
Pulsar timing arrays, on the other hand, should be
able to detect the confusion noise from supermassive 
black hole binaries 
(SMBHBs) at the centers of merging galaxies, provided
the binaries are not affected by their environments in
a way that severely diminishes the strength of the 
background~\cite{2015Sci...349.1522S}.
Detection sensitivity curves are a very convenient way 
of comparing theoretical predictions of source strengths 
to the sensivity levels of the various
detectors (as we will discuss in Section~\ref{s:obs}).

\subsection{Searches across the gravitational-wave spectrum}

\begin{figure}[h!tbp]
\begin{center}
\includegraphics[angle=0,width=\columnwidth]{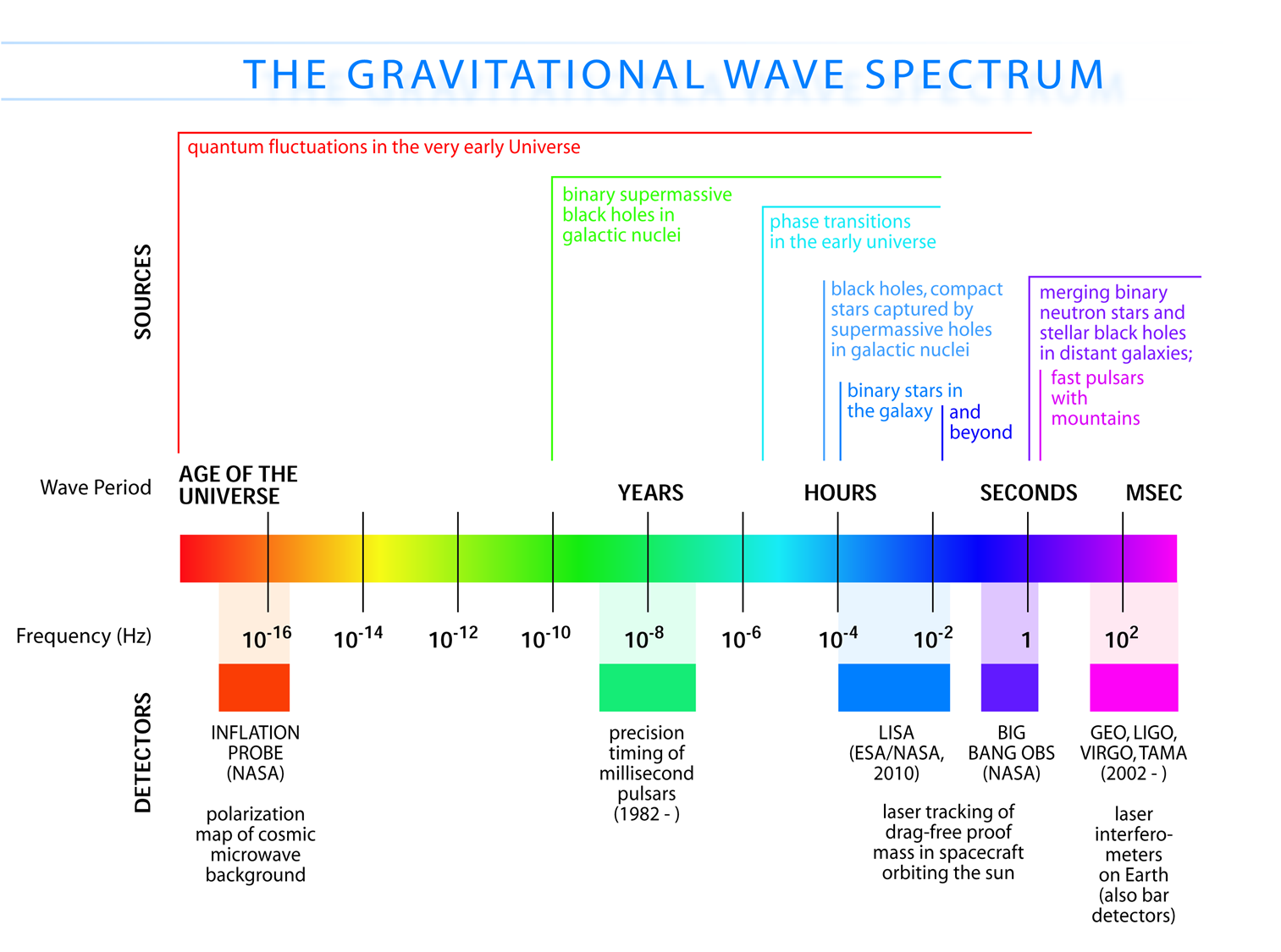}
\caption{Gravitational-wave spectrum, together with
potential sources and relevant detectors.
Image credit: Institute of Gravitational Research/ University 
of Glasgow.}
\label{f:GWspectrum}
\end{center}
\end{figure}

The frequency band of ground-based laser interferometers 
like LIGO, Virgo, and KAGRA is 
between $\sim\!10~{\rm Hz}$ and a few kHz (gravity gradient 
and seismic noise are the limiting%
\footnote{Actually, even if the gravity-gradient and seismic
noise were zero, one couldn't go below $\sim\!1~\mr{Hz}$ 
with the current generation of ground-based laser interferometers,
since the suspended mirrors (i.e., the test masses) are no 
longer freely floating when you go below their resonant 
frequencies: $\sim\!1~\mr{Hz}$.}
noise sources below 10~Hz,
and photon shot noise above a couple of kHz).
Outside this band there are several other 
experiments---both currently operating and planned---that 
should also be able to detect gravitational waves.
An illustration of the gravitational-wave spectrum, 
together with potential sources and 
relevant detectors, is shown in Figure~\ref{f:GWspectrum}.
We highlight a few of these experiments below.

\subsubsection{Cosmic microwave background experiments}

At the extreme low-frequency end of the spectrum, 
corresponding to gravitational-wave periods of 
order the age of the Universe, the Planck satellite~\cite{Planck:web} 
and other cosmic microwave background (CMB) experiments,
e.g., BICEP and Keck~\cite{BICEP/Keck:web} are looking 
for evidence of relic gravitational waves from the Big Bang 
in the $B$-mode component of CMB polarization 
maps~\cite{KKS:1997, Hu-White:1997, BICEP2/Keck-Planck:2015}.
In 2014, BICEP2 announced the detection of relic 
gravitational waves~\cite{BICEP2:2014}, but it was later 
shown that the observed $B$-mode signal was due to 
contamination by intervening dust in the 
galaxy~\cite{Flauger-et-al:2014, Mortonson-Seljack:2014}.
So at present, these experiments have been able
to only {\em constrain} (i.e., set upper limits on)
the amount of gravitational waves in  the very
early Universe~\cite{BICEP2/Keck-Planck:2015}.
But these constraints severely limit the possibility
of detecting the relic gravitational-wave background
with any of the higher-frequency detection methods,
unless its spectrum increases with frequency.
[Note that standard models of inflation predict a relic 
background whose energy density is almost constant in frequency,
leading to a strain spectral density that decreases with frequency.]
Needless to say, the detection of a primordial 
gravitational-wave background is a ``holy grail" of  
gravitational-wave astronomy.

\subsubsection{Pulsar timing arrays}

At frequencies between $\sim\!10^{-9}~{\rm Hz}$ 
and $10^{-7}~{\rm Hz}$, corresponding to 
gravitational-wave periods of order decades to years, 
pulsar timing arrays (PTAs) can be used to 
search for gravitational waves.
This is done by carefully monitoring the arrival
times of radio pulses from an array of 
galactic millisecond pulsars, looking for 
{\em correlated} modulations in the arrival times
induced by a passing gravitational 
wave~\cite{Detweiler:1979, Hellings-Downs:1983}.
The most-likely gravitational-wave source for PTAs is a 
gravitational-wave background formed from the incoherent 
superposition of signals produced by the inspirals and 
mergers of SMBHBs in the centers of distant galaxies~\cite{Jaffe-Backer:2003}.
These searches continue to improve their 
sensitivity by upgrading instrument back-ends and 
discovering more millisecond pulsars that can be 
added to the array.
These improvements have led to more
constraining upper limits on the amplitude of the 
gravitational-wave background~\cite{PPTA:2015-Science, NANOGrav:2016-9year}, 
with a detection being likely before the end  
of this decade~\cite{Siemens-et-al:2013, Taylor-et-al:2016a}.

\subsubsection{Space-based interferometers}

At frequencies between $\sim\!10^{-4}~{\rm Hz}$ 
and $10^{-1}~{\rm Hz}$, corresponding to 
gravitational-wave periods of order hours to 
minutes, proposed space-based interferometers like 
LISA can search for gravitational waves from a wide 
variety of sources~\cite{Gair-et-al:2013-LRR}.
These include:
(i) inspirals and mergers of SMBHBs with masses 
of order $10^6~{\rm M}_\odot$,
(ii) captures of compact stellar-mass objects
around supermassive black holes, and
(iii) the stochastic confusion noise produced 
by compact white-dwarf binaries in our galaxy.
In fact, hundreds of binary black holes that are 
individually resolvable by LISA will coalesce in 
the aLIGO band within a 10~year period, opening 
up the possibility of doing {\em multi-band} 
gravitational-wave astronomy~\cite{Sesana:2016}.

The basic space-based interferometer configuration 
consists of three satellites 
(each housing two lasers, two telescopes, and two 
test masses) that fly in an 
equilateral-triangle formation, with arm lengths 
of order several million km.
A variant of the original LISA design was selected 
in February 2017 by the European Space Agency (ESA) 
as the 3rd large mission in its Cosmic Vision 
Program~\cite{ESA-CosmicVision:web}.
The earliest launch date for LISA would be around 2030.
A technology-demonstration mission, called 
LISA Pathfinder~\cite{Pathfinder:web}, was 
launched in December 2015, meeting or exceeding all 
of the requirements for an important subset of the 
LISA technologies~\cite{Pathfinder:2016}.

\subsubsection{Other detectors}

Finally, in the frequency band between 
$\sim\!0.1~\mr{Hz}$ and $10~\mr{Hz}$, there are 
proposals for both 
Earth-based detectors~\cite{Harms-et-al:2013} 
and also second-generation space-based interferometers---the 
Big-Bang Observer (BBO)~\cite{Phinney-et-al:2004} and 
the DECI-hertz interferometer Gravitational-wave Observatory
(DECIGO)~\cite{Ando-et-al:DECIGO2010}.
Such detectors would be sensitive to gravitational waves
with periods between $\sim\!10~\mr{s}$ and $0.1~\mr{s}$.
The primary sources in this band are intermediate-mass
($10^3$--$10^4~M_\odot$) binary black holes,
galactic and extra-galactic neutron star binaries, 
and a cosmologically-generated stochastic background.

\subsection{Goal of this article}

Starting with the pioneering work of 
Grishchuk~\cite{Grishchuk:1976},
Detweiler~\cite{Detweiler:1979},
Hellings and Downs~\cite{Hellings-Downs:1983}, and
Michelson~\cite{Michelson:1987},
detection methods for gravitational-wave backgrounds
have increased in scope and sophistication over the
years, with several new developments occuring rather recently.
As mentioned above, we have search methods now that 
target different properties of the background
(e.g., isotropic or anisotropic, Gaussian or non-Gaussian, 
polarized or unpolarized, etc.).
These searches are necessarily implemented differently for different 
detectors, since, for example, ground-based detectors
like LIGO and Virgo operate in the {\em small-antenna}
(or {\em long-wavelength}) limit, while pulsar timing
arrays operate in the {\em short-wavelength} limit.
Moreover, each of these searches can be formulated in 
terms of either Bayesian or frequentist statistics.
{\em The goal of this review article is to discuss these 
different detection methods
from a perspective that attempts to {\em unify} the 
different treatments, emphasizing the similarities that 
exist when viewed from this broader perspective.}


\subsection{Unification}

The extensive literature describing stochastic background analyses
leaves the reader with the impression that highly specialized
techniques are needed for ground-based, space-based, and pulsar timing
observations. Moreover, reviews of gravitational-wave data analysis
leave the impression that the analysis of stochastic signals is
somehow fundamentally different from that of any other signal
type. Both of these impressions are misleading. The apparent
differences are due to differences in terminology and perspective. By
adopting a common analysis framework and notation, we are able to
present a {\em unified} treatment of gravitational-wave data analysis across
source classes and observation techniques.

We will provide a unified treatment of the various methods at the
level of detector response functions, detection sensitivity curves,
and, more generally, at the level of the likelihood function, since
the choice of signal and noise models and prior probability
distributions are actually what define the search.  The same
photon time-of-flight calculation underpins the detector response
functions, and the choice of prior for the gravitational-wave template
defines the search.  A {\em matched-filter} search for binary mergers
and a {\em cross-correlation} search for stochastic signals are both
derived from the same likelihood function, the difference being that
the former uses a parameterized, deterministic template, while the
latter uses a stochastic template. Hopefully, by the end of this
article, the reader will see that the plethora of searches for
different types of backgrounds, using different types of detectors,
and using different statistical inference frameworks are not all that
different after all.

\subsection{Outline}

The rest of the article is organized as follows:
We begin in Section~\ref{s:definition} by 
specifying the quantities that one uses to characterize 
a stochastic gravitational-wave background.
In Section~\ref{s:inference}, we give an overview of
statistical inference by comparing and contrasting how 
the Bayesian and frequentist formalisms address issues
related to hypothesis testing, model selection, setting 
upper limits, parameter estimation, etc.
We then illustrate these concepts in the context of a 
very simple toy problem.
In Section~\ref{s:corr}, we introduce the key
concept of correlation, which forms the basis for the 
majority of detection methods used 
for gravitational-wave backgrounds, and show how these techniques
arise naturally from the standard template-based approach.
We derive the frequentist cross-correlation statistic 
for a simple example.
We also describe how a null channel is useful when 
correlation methods are not possible.

In Section~\ref{s:geom}, we go into more detail 
regarding the different types of detectors.
In particular, we calculate single-detector response 
functions and the associated antenna patterns for 
ground-based and space-based laser interferometers,
spacecraft Doppler tracking, and 
pulsar timing measurements.
(We do not discuss resonant bar detectors or 
CMB-based detection methods in this review article.
However, current bounds from CMB observations will be 
reviewed in Section~\ref{s:obs}.)
By correlating the outputs of two such detectors, 
we obtain expressions for the correlation coefficient
(or {\em overlap reduction function}) for a Gaussian-stationary,
unpolarized, isotropic background as a function of 
the separation and orientation of the two detectors.
In Section~\ref{s:optimal_filtering}, we discuss 
optimal filtering.
Section~\ref{s:anisotropic} extends the analysis
of the previous sections to {\em anisotropic} backgrounds.
Here we describe several different analyses that 
produce maps of the gravitational-wave sky: 
(i) a frequentist gravitational-wave radiometer search, which is 
optimal for point sources, (ii) searches that decompose 
the gravitational-wave power on the sky in terms of
spherical harmonics, and (iii) a phase-coherent search
that can map both the amplitude and phase of a 
gravitational-wave background at each location on the sky.
In Section~\ref{s:extensions}, we discuss searches for:
(i) non-Gaussian backgrounds,
(ii) circularly-polarized backgrounds, and 
(iii) backgrounds having non-standard (i.e., non-general-relativity)
polarization modes.
We also briefly describe extensions of the 
cross-correlation search method to look for 
{\em non-stochastic-background-type} signals---in
particular, 
long-duration unmodelled transients and continuous
(nearly-monochromatic) gravitational-wave signals from 
sources like Sco X-1.

In Section~\ref{s:complications}, we discuss real-world 
complications introduced by irregular sampling, 
non-stationary and non-Gaussian detector noise,
and correlated environmental noise (e.g., Schumann resonances).
We also describe what one can do if one has only a single
detector, as is the case for LISA.
Finally, we conclude in Section~\ref{s:obs} by
discussing prospects for detection, including detection
sensitivity curves and current observational results.

We also include several appendices:
In Appendix~\ref{s:polarization_tensors} we discuss 
different polarization basis tensors, and a Stokes' 
parameter characterization of gravitational-waves.
In Appendices~\ref{s:basics} and \ref{s:real}, we 
summarize some 
standard statistical results for a Gaussian random 
variable, and then discuss how to define and test for 
non-stationarity and non-Gaussianity.
In Appendix~\ref{s:discretely-sampled-data} we
describe the relationship between continuous functions 
of time and frequency and their discretely-sampled counterparts.
Appendices~\ref{s:spinweightedY}, \ref{s:grad-curl-vector}, 
\ref{s:grad-curl-tensor}
are adapted from~\cite{Gair-et-al:2015}, with details 
regarding spin-weighted scalar, vector, and tensor 
spherical harmonics.
Finally, Appendix~\ref{s:translation} gives a 
``Rosetta stone" for translating back and forth between
different response function conventions for 
gravitational-wave backgrounds.

\section{Characterizing a stochastic gravitational-wave background}
\label{s:definition}

\begin{quotation}
When you can measure what you are speaking about, and express it in numbers, you know something about it,
when you cannot express it in numbers, your knowledge is of a meager and unsatisfactory kind; it may be the
beginning of knowledge, but you have scarely, in your thoughts, advanced to the stage of science.
{\em William Thomson, Baron Kelvin of Largs}
\end{quotation}

\noindent
In this section, we define several key quantities 
(e.g., fractional energy density spectrum, characteristic 
strain, distribution of gravitational-wave power on the sky), 
which are used to characterize 
a stochastic background of gravitational radiation.
The definitions are appropriate for both isotropic 
and anisotropic backgrounds.
Our approach is similar to that found in \cite{Allen-Romano:1999}
for isotropic backgrounds and for the standard polarization basis.
For the plane-wave decomposition in terms of tensor 
spherical harmonics, we follow \cite{Gair-et-al:2014, Gair-et-al:2015}.
Detailed derivations can be found in those papers.

\subsection{When is a gravitational-wave signal stochastic?}
\label{s:when-stochastic}

The standard ``textbook" definition of a stochastic 
background of gravitational radiation is 
{\em a random gravitational-wave signal produced by a 
large number of weak, independent, and unresolved sources}.
To say that it is random means that it can be 
characterized only statistically, in terms of expectation 
values of the field variables or, equivalently, in terms 
of the Fourier components of a plane-wave expansion
of the metric perturbations (Section~\ref{s:expectationvalues}).  
If the number of independent sources is sufficiently large,
the background will be Gaussian by the central limit theorem.  
Knowledge of the first two moments of the distribution
will then suffice to determine all higher-order moments (Appendix~\ref{s:basics}).
For non-Gaussian backgrounds, third and/or higher-order 
moments will also be needed.

Although there is general agreement with the above definition, 
there has been some confusion and disagreement in the
literature~\cite{Rosado:2011, Regimbau-Mandic:2008, Regimbau-Hughes:2009, Regimbau:2011}
regarding some of the defining properties of a stochastic background.
This is because terms like {\em weak} and {\em unresolved} depend
on details of the observation (e.g., the sensitivity of the 
detector, the total observation time, etc.), which are not 
intrinsic properties of the background.
So the answer to the question 
``When is a gravitational-wave signal stochastic?" is not as
simple or obvious as it might initially seem.

In \cite{Cornish-Romano:2015}, we addressed this question in the 
context of searches for gravitational-wave backgrounds 
produced by a population of astrophysical sources.
We found that it is best to give {\em operational} definitions 
for these properties, framed in the context of Bayesian inference.  
We will discuss Bayesian inference in more detail in 
Section~\ref{s:inference}, but for now the most important
thing to know is that by using Bayesian inference we can 
calculate the probabilities of different 
signal-plus-noise models, given the observed data.
The signal-plus-noise model with the largest probability 
is the preferred model, i.e., the one that is most 
consistent with the data.
This is the essence of Bayesian model selection.

So we define a signal to be {\em stochastic} if a Bayesian
model selection calculation prefers a stochastic signal 
model over any deterministic signal model.  
We also define a signal to be {\em resolvable} if it can 
be decomposed into {\em separate} 
(e.g., non-overlapping in either time or frequency) 
and {\em individually  detectable} signals, 
again in a Bayesian model selection sense.%
\footnote{Signals may be separable even when overlapping in time and frequency if the
detector has good sky resolution, or if the signals have additional 
complexities due to effects such as orbital evolution and precession.}
If the background is associated with the superposition
of signals from many astrophysical sources---as we 
expect for the population of binary black holes which
gave rise to GW150914 and GW151226---then we should 
{\em subtract out} any bright deterministic signals that standout 
above the  lower-amplitude background, 
leaving behind a residual non-deterministic signal
whose statistical properties we would like to determine. 
In the context of Bayesian inference, this `subtraction' is 
done by allowing {\em hybrid} signal models, which consist of both 
parametrized deterministic signals and non-deterministic backgrounds. 
By using such hybrid models we can investigate the statistical 
properties of the 
residual background without the influence of the resolvable signals. 

We will return to these ideas in Section~\ref{s:nongaussian}, when
we discuss searches for non-Gaussian backgrounds in more detail.

\subsection{Plane-wave expansions}
\label{s:planewave}

Gravitational waves are time-varying perturbations to the 
spacetime metric, which propagate at the speed of light.
In transverse-traceless coordinates, the metric perturbations
$h_{ab}(t,\vec x)$ corresponding to a gravitational-wave 
background can be written as a superposition of sinusoidal 
plane waves having frequency $f$, and coming from different
directions $\hat n$ on the sky:%
\footnote{The gravitational-wave propagation direction, which
we will denote by $\hat k$, is given by $\hat k=-\hat n$.}
\be
h_{ab}(t,\vec x) =
\int_{-\infty}^\infty df\>
\int d^2\Omega_{\hat n}\>
h_{ab}(f,\hat n) e^{i2\pi f(t+\hat n\cdot \vec x/c)}\,.
\label{e:planewave}
\ee
For a stochastic background, the metric perturbations
$h_{ab}(t,\vec x)$ and hence the Fourier coefficients
$h_{ab}(f,\hat n)$ are random variables, whose probability 
distributions define the statistical properties of the background.

\subsubsection{Polarization basis}
\label{s:polarizationbasis}

\begin{figure}[h!tbp]
\begin{center}
\includegraphics[angle=0,width=0.7\textwidth]{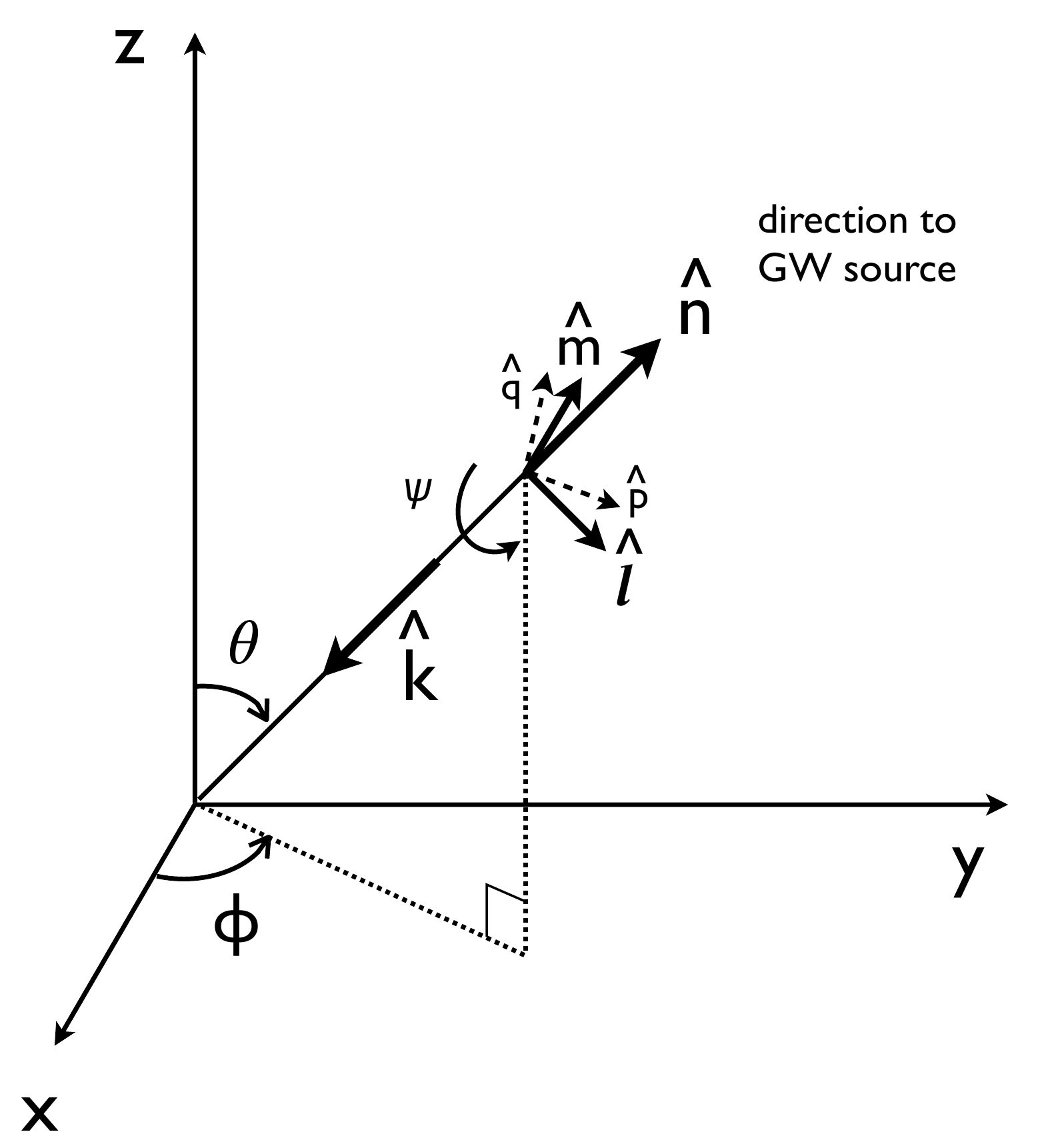}
\caption{Our convention for the unit vectors 
$\{\hat n, \hat l, \hat m\}$ in terms of which the
polarization basis tensors
$e^+_{ab}(\hat n)$ and $e^\times_{ab}(\hat n)$ are
defined.
The unit vector
$\hat n$ points in the direction of the gravitational-wave
source (the gravitational wave propagates in 
direction $\hat k=-\hat n$);
$\hat l=\hat \theta$ and 
$\hat m=\hat\phi$ are two unit vectors that lie
in the plane perpendicular to $\hat n$.
Another choice for the polarization basis tensors,
defined in terms of the `rotated' unit vectors 
$\hat p$ and $\hat q$, is given
in Appendix~\ref{s:polarization_tensors}.}
\label{f:nlm_convention}
\end{center}
\end{figure}

Typically, one expands the Fourier coefficients $h_{ab}(f,\hat n)$ 
in terms of the standard $+$ and $\times$ polarization tensors:
\be
h_{ab}(f,\hat n)
=h_+(f,\hat n) e^+_{ab}(\hat n) + h_\times (f,\hat n) e^\times_{ab}(\hat n)\,,
\ee
where
\be
\begin{aligned}
e_{ab}^+(\hat n)
&=\hat l_a\hat l_b-\hat m_a\hat m_b\,,
\\
e_{ab}^\times(\hat n)
&=\hat l_a\hat m_b+\hat m_a\hat l_b\,,
\end{aligned}
\ee
and $\hat l$, $\hat m$ are the standard
angular unit vectors tangent to the sphere:
\be
\begin{aligned}
\hat n
&=\sin\theta\cos\phi\,\hat x
+\sin\theta\sin\phi\,\hat y
+\cos\theta\,\hat z
\equiv \hat r\,,
\\
\hat l
&=\cos\theta\cos\phi\,\hat x
+\cos\theta\sin\phi\,\hat y
-\sin\theta\,\hat z 
\equiv \hat\theta\,,
\\
\hat m
&=-\sin\phi\,\hat x
+\cos\phi\,\hat y
\equiv \hat\phi\,.
\end{aligned}
\label{e:nlm_def}
\ee
(See Figure~\ref{f:nlm_convention}.)
Searches for stochastic backgrounds having alternative 
polarization modes, as predicted by modified (metric) theories of 
gravity, will be discussed in Section~\ref{s:altpol}.
%
\subsubsection{Tensor spherical harmonic basis}
\label{s:sphericalharmonicbasis}

It is also possible to expand the Fourier coefficients
$h_{ab}(f,\hat n)$ in terms of the {\em gradient} and {\em curl} 
tensor spherical harmonics~\cite{Gair-et-al:2014}:
\be
h_{ab}(f,\hat n)
=\sum_{l=2}^\infty \sum_{m=-l}^l
\left[a^G_{(lm)}(f)Y^G_{(lm)ab}(\hat n)
+a^C_{(lm)}(f)Y^C_{(lm)ab}(\hat n)\right]\,,
\ee
where
\be
\begin{aligned}
Y^G_{(lm)ab} &= {}^{(2)}\!N_l 
\left(Y_{(lm);ab} - \frac{1}{2} g_{ab}  Y_{(lm);c}{}^{c} \right)\,,
\\
Y^C_{(lm)ab} &= \frac{{}^{(2)}\!N_l}{2} 
\left(Y_{(lm);ac}\epsilon^c{}_b +  Y_{(lm);bc} \epsilon^c{}_a \right)\,.
\end{aligned}
\ee
In the above expressions,
a semi-colon denotes covariant differentiation, $g_{ab}$ is the
metric tensor on the sphere, and $\epsilon_{ab}$ is the Levi-Civita
anti-symmetric tensor.
In standard spherical coordinates $(\theta,\phi)$,
\be
g_{ab}=\left(
\begin{array}{cc}
1&0\\
0&\sin^2\theta\\
\end{array}
\right)\,,
\qquad
\epsilon_{ab} 
= \sqrt{g} \left( \begin{array}{cc}0&1\\-1&0\end{array}\right)\,,
\qquad
\sqrt{g}=\sin\theta\,.
\ee
The normalization constant
\be
{}^{(2)}\!N_l = \sqrt{\frac{2 (l-2)!}{(l+2)!}}\,,
\label{e:N_l}
\ee
was chosen so that 
$\{Y^G_{(lm)ab}(\hat n), Y^C_{(lm)ab}(\hat n)\}$ 
is a set of orthonormal functions (with respect to the 
multipole indices $l$ and $m$) on the 2-sphere.
Appendix~\ref{s:grad-curl-tensor} contains additional details
regarding gradient and curl spherical harmonics.

NOTE: we have adopted the notational convention used in the
CMB literature, e.g., \cite{KKS:1997},
by putting parentheses around
the $lm$ indices to distinguish them from the spatial tensor 
indices $a$, $b$, etc.
In addition, summations over $l$ and $m$ start at $l=2$,
and not $l=0$ as would be the case for the expansion of
a scalar field on the 2-sphere in terms of ordinary 
(i.e., undifferentiated) spherical harmonics. 
In what follows, we will use $\sum_{(lm)}$ as shorthand 
notation for $\sum_{l=2}^\infty \sum_{m=-l}^l$ unless indicated
otherwise.

\subsubsection{Relating the two expansions}

The gradient and curl spherical harmonics have been used
extensively in the CMB community for decomposing
CMB-polarization maps in terms of $E$-modes and $B$-modes 
(corresponding to the gradient and curl spherical harmonics).
The most relevant property of the gradient and curl spherical 
harmonics is that they transform like combinations of
spin-weight~$\pm 2$ fields with respect to rotations of an orthonomal 
basis at points on the 2-sphere.
Explicitly,
\begin{align}
Y^G_{(lm)ab}(\hat n) \pm i Y^C_{(lm)ab}(\hat n)
&=\frac{1}{\sqrt{2}}
\left(e_{ab}^+(\hat n) \pm i e_{ab}^\times(\hat n)\right)
\,{}_{\mp 2}Y_{lm}(\hat n)\,,
\end{align}
where ${}_{\pm 2}Y_{lm}(\hat n)$ are the spin-weight~$\pm2$ 
spherical harmonics (Appendix~\ref{s:spinweightedY}).
Using this relationship between the
tensor spherical harmonic and  
$(+,\times)$ polarization bases, one 
can show~\cite{Gair-et-al:2014}:
\be
h_+(f,\hat n)\pm i h_\times(f,\hat n)
=\frac{1}{\sqrt{2}} \sum_{(lm)}
\left(a^G_{(lm)}(f)\pm i a^C_{(lm)}(f)\right)\,
{}_{\pm 2}Y_{lm}(\hat n)\,,
\label{e:h++ihx}
\ee
or, equivalently,
\be
a^G_{(lm)}(f)\pm i a^C_{(lm)}(f)
= \sqrt{2} \int {\rm d}^2\Omega_{\hat n}\>
\left(h_+(f,\hat n) \pm i h_\times(f,\hat n)\right)\,
{}_{\pm 2}Y_{lm}^*(\hat n)\,.
\label{e:aG+iaC}
\ee
These two expressions allow us to go back and forth 
between the expansion coefficients for the two different bases.

\subsection{Statistical properties}
\label{s:statisticalproperties}

The statistical properties of a stochastic 
gravitational-wave background are specified 
in terms of the probability distribution or
{\em moments} (Appendix~\ref{s:basics}) 
of the metric perturbations:
\be
\langle h_{ab}(t,\vec x)\rangle\,,
\quad
\langle h_{ab}(t,\vec x)h_{cd}(t', \vec x')\rangle\,,
\quad
\langle h_{ab}(t,\vec x)h_{cd}(t', \vec x')h_{ef}(t'',\vec x'')\rangle\,,
\quad
\cdots
\ee
or similar expressions in terms of the Fourier coefficients 
$h_A(f,\hat n)$, where $A\equiv\{+,\times\}$ labels the standard 
polarization modes of general relativity, 
or $a^P_{(lm)}(f)$, where $P\equiv\{G,C\}$ and $(lm)$ label 
the multipole components for 
the gradient and curl tensor spherical harmonic decomposition.
Without loss of generality we can assume that the
background has zero mean:
\be
\langle h_{ab}(t,\vec x)\rangle=0
\quad{\Leftrightarrow}\quad
\langle h_A(f,\hat n)\rangle=0
\quad{\Leftrightarrow}\quad
\langle a_{(lm)}^P(f)\rangle=0\,.
\ee
We will also assume that the background is {\em stationary}
(Appendix~\ref{s:real}).
This means that all statistical 
quantities constructed from the metric perturbations 
at times $t$, $t'$, etc., depend only on the 
difference between times, e.g., $t-t'$, and not on 
the choice of time origin. 
We expect this to be true given that the age of 
the universe is roughly 9~orders of magnitude larger
than realistic observation times, $\sim\!10~\mr{yr}$.
It is thus unlikely that a stochastic gravitational-wave
background has statistical properties that
vary over the time scale of the observation.

For Gaussian backgrounds we need only consider
quadratic expectation values, since all higher-order 
moments are either zero or can be written 
in terms of the quadractic moments (Appendix~\ref{s:basics}).
For non-Gaussian backgrounds (Section~\ref{s:nongaussian}), 
third and/or higher order moments will also be needed.

Beyond our assumption of stationarity, the specific 
form of the expectation values will depend, 
in general, on the source of the background. 
For example, a cosmological background produced by the
superposition of a large number of independent 
gravitational-wave signals from the early Universe 
is expected to be Gaussian (via the central limit theorem),
as well as isotropically-distributed on the sky. 
Contrast this with the superposition of gravitational
waves produced by unresolved Galactic white-dwarf binaries 
radiating in the LISA band ($10^{-4}~{\rm Hz}-10^{-1}~{\rm Hz}$). 
Although this confusion-limited astrophysical foreground 
is also expected to be Gaussian and stationary, 
it will have an {\em anisotropic distribution}, following 
the spatial distribution of the Milky Way. 
The anistropy will be encoded as a modulation in the 
LISA output, due to the changing antenna pattern of the 
LISA constellation in its yearly orbit around the Sun. 
Hence, different sources will give rise to different 
statistical distributions, which we will need to consider 
when formulating our data analysis strategies.

\subsubsection{Quadratic expectation values for Gaussian-stationary backgrounds}
\label{s:expectationvalues}

The simplest type of stochastic background will be
Gaussian-stationary, unpolarized, and spatially
homogenous and isotropic.
The quadratic expectation values for such a background
are then
\be
\langle h_A(f,\hat n) h_{A'}^*(f',\hat n')\rangle
=\frac{1}{16\pi}S_h(f)
\delta(f-f')
\delta_{AA'}
\delta^2(\hat n,\hat n')\,,
\label{e:iso_hh}
\ee
or, equivalently,
\be
\langle a^P_{(lm)}(f)a^{P'*}_{(l'm')}(f')\rangle
=\frac{1}{8\pi} S_h(f)\delta(f-f')\delta^{PP'}\delta_{ll'}\delta_{mm'}\,. 
\label{e:iso_aa}
\ee 
The numerical factors out front have been included so that
$S_h(f)$ has the interpretation of being the one-sided 
gravitational-wave {\em strain power spectral density} 
function (units of ${\rm strain}^2/{\rm Hz}$), 
summed over both polarizations and integrated over the sky.
The factor of $\delta(f-f')$ arises due to our
assumption of stationarity;
the factor of $\delta_{AA'}$ (or $\delta^{PP'}$) is due to 
our assumption that the polarization modes are 
statistically independent of one another and have
no preferred component; and 
the factor of  
$\delta^2(\hat n,\hat n')$ (or $\delta_{ll'}\delta_{mm'}$) 
is due to our assumption of spatial homogeneity and isotropy.

Anisotropic, unpolarized, Gaussian-stationary backgrounds,
whose radiation from different directions on the sky 
are uncorrelated with one another, are also simply represented
in terms of the quadratic expectation values:
\be
\langle h_A(f,\hat n) h_{A'}^*(f',\hat n')\rangle
=\frac{1}{4}{\cal P}(f,\hat n)
\delta(f-f')
\delta_{AA'}
\delta^2(\hat n,\hat n')\,.
\label{e:aniso_hh}
\ee
The function ${\cal P}(f,\hat n)$ describes the spatial 
distribution of gravitational-wave power on the sky at
frequecy $f$.
It is related to $S_h(f)$ via
\be
S_h(f) = \int d^2\Omega_{\hat n}\> {\cal P}(f,\hat n)\,.
\ee
The corresponding expectation values in terms 
of the tensor spherical harmonic expansion 
coefficients $a^P_{(lm)}(f)$ are more complicated, 
since an individual mode in this basis corresponds
to a gravitational-wave background whose radiation
is correlated between different angular directions
on the sky.
(See \cite{Gair-et-al:2014} for a discussion of
backgrounds that have such correlations.)
We will discuss searches for anisotropic backgrounds 
in more detail in Section~\ref{s:anisotropic}.

More general Gaussian-stationary backgrounds
(e.g., polarized, statistically isotropic but with 
correlated radiation, etc.)~can be represented by 
appropriately changing the right-hand-side of the 
quadratic expectation values.
However, for the remainder of this section and for 
most of the article, we will consider ``vanilla" 
isotropic backgrounds, 
whose quadratic expectation values 
(\ref{e:iso_hh}) or (\ref{e:iso_aa}) are completely
specified by the power spectral density $S_h(f)$.

\subsection{Fractional energy density spectrum}

The gravitational-wave strain power spectral density
$S_h(f)$ is simply related to the fractional energy density spectrum
in gravitational waves $\Omega_{\rm gw}(f)$, 
see e.g.,~\cite{Allen-Romano:1999}:
\be
S_h(f) = \frac{3 H_0^2}{2\pi^2}\frac{\Omega_{\rm gw}(f)}{f^3}\,,
\label{e:Sh-Omega_gw}
\ee
where
\be
\Omega_{\rm gw}(f)
=\frac{1}{\rho_c}\frac{d\rho_{\rm gw}}{d\ln f}\,.
\ee
Here $d\rho_{\rm gw}$ is the energy density in 
gravitational waves contained in the frequency interval 
$f$ to $f+df$, and $\rho_c\equiv 3c^2 H_0^2/8\pi G$ is the 
critical energy density need to close the universe.
The {\em total} energy density in gravitational waves 
normalized by the critical energy density is thus
\begin{equation}
\Omega_\mr{gw} = 
\int_{f=0}^{f_\mr{max}} d(\ln f)\>\Omega_\mr{gw}(f)
\,,
\end{equation}
where $f_\mr{max}$ is some maximum cutoff frequency (e.g., 
associated with the Planck scale), beyond which our current
understanding of gravity breaks down.
$\Omega_\mr{gw}$ can be compared, for example, to the total fractional energy density
$\Omega_\mr{b}$, $\Omega_{\Lambda}$, in baryons, dark energy, etc.
Since $\rho_\mr{c}$ involves the Hubble constant, one sometimes
writes $H_0=h_0\,100$~km~s${}^{-1}$~Mpc${}^{-1}$, and then absorbs a factor of 
$h_0^2$ in $\Omega_\mr{gw}(f)$.
The quantity $h_0^2\,\Omega_\mr{gw}(f)$ is then {\em independent} of 
the value of the Hubble constant.
However, since recent measurements by Planck~\cite{Ade-et-al:2015-Planck, Planck:web}
have shown that
$h_0=0.68$ to a high degree of precision, we have assumed
this value in this review article and quote limits directly on
$\Omega_\mr{gw}(f)$ (Section~\ref{s:obs}).
The specific functional form for $\Omega_\mr{gw}(f)$ depends 
on the source of the background, as we shall see explicitly below.

\subsection{Characteristic strain}
\label{s:characteristic-strain}

Although the fractional energy density spectrum 
$\Omega_{\rm gw}(f)$ completely characterizes 
the statistical properties of a Gaussian-stationary isotropic
background, it is often convenient to work with the 
(dimensionless) characteristic strain amplitude $h_c(f)$ 
defined by
\be
h_c(f) \equiv \sqrt{f S_h(f)}\,.
\ee
It is related to $\Omega_{\rm gw}(f)$ via:
\be
\Omega_{\rm gw}(f) = \frac{2\pi^2}{3 H_0^2}f^2 h^2_c(f)\,.
\label{e:Omega-hc}
\ee
Several theoretical models of gravitational-wave
backgrounds predict characteristic strains that have 
a power-law form
\be
h_c(f) 
=A_\alpha\left(\frac{f}{f_{\rm ref}}\right)^\alpha\,,
\label{e:hc-A}
\ee
where $\alpha$ is spectral index and $f_{\rm ref}$ is
typically set to $1/{\rm yr}$.
(There is no sum over $\alpha$ in the above expression,
and no sum over $\beta$ in the following expression.)
Using equations (\ref{e:Omega-hc}) and (\ref{e:hc-A})
it follows that 
\be
\Omega_{\rm gw}(f) 
= \Omega_\beta \left(\frac{f}{f_{\rm ref}}\right)^\beta\,,
\ee
where
\be
\Omega_\beta = \frac{2\pi^2}{3 H_0^2}
f_{\rm ref}^2\,A_\alpha^2\,,
\quad
\beta = 2\alpha+2\,.
\label{e:alpha-beta}
\ee
For inflationary backgrounds relevant for cosmology, it is often assumed that
$\Omega_{\rm gw}(f)={\rm const}$,
for which $\beta=0$ and $\alpha=-1$.
For a background arising from binary coalescence,
$\Omega_{\rm gw}(f) \propto f^{2/3}$,
for which $\beta = 2/3$ and $\alpha=-2/3$.
This power-law dependence is applicable to 
super-massive black-hole binary (SMBHB) 
coalescences targeted by pulsar timing observations as well 
as to compact binary coalescences 
relevant for ground-based and space-based detectors.

\section{Statistical inference}
\label{s:inference}

\begin{quotation}
If your experiment needs statistics, you ought to have done a better experiment.
{\em Ernest Rutherford}
\end{quotation}

\noindent
In this section, we review statistical inference from both the
Bayesian and frequentist perspectives.
Our discussion of frequentist and Bayesian upper limits, and 
the example given in Section~\ref{s:example-bayes-freq} comparing
Bayesian and frequentist analyses is modelled in part
after~\cite{Rover-et-al:2011}.
Readers interested in more details about Bayesian statistical 
inference should see e.g., \cite{Howson-Urbach:1991, Howson-Urbach:2006,
Jaynes:2003, Gregory:2005, Sivia-Skilling:2006}.
For a description of frequentist statistics, we recommend
\cite{Helstrom:1968, Wainstein-Zubakov:1971, Feldman-Cousins:1998}.
 
\subsection{Introduction to Bayesian and frequentist inference}
\label{s:intro-bayes-freq}

Statistical inference can be used to answer questions such as ``Is a
gravitational-wave signal present in the data?''~and, if so,
``What are the physical characteristics of the source?"
These questions are addressed using the techniques of 
classical (also known as {\em frequentist})
inference and {\em Bayesian} inference. Many of the early theoretical
studies and observational papers in gravitational-wave astronomy
followed the frequentist approach, but the use of Bayesian inference is
growing in popularity. Moreover, many contemporary analyses cannot be
classified as purely frequentist or Bayesian. 

The textbook definition
states that the difference between the two approaches comes down to
their different interpretations of probability: for frequentists,
probabilities are fundamentally related to frequencies of events,
while for Bayesians, probabilities are fundamentally related to our own
knowledge about an event. For example, when inferring the mass of a
star, the frequentist interpretation is that the star has a true,
fixed (albeit unknown) mass, so it is meaningless to talk about a probability
distribution for it. Rather, the uncertainty is in the data, and
the relevant probability is that of observing the data $d$, given 
that the star has mass $m$. This probability distribution is the
{\em likelihood}, denoted $p(d\vert m)$. 
In contrast, in the Bayesian interpretation the data are known 
(after all, it is what is measured!), 
and the mass of the star is what we are uncertain
about\footnote{In some treatments, the Bayesian interpretation is
  equated to philosophical schools such as Berkeley's empiricist
  idealism, or subjectivism, which holds that things only exist to the
  extent that they are perceived, while the frequentist interpretation
  is equated to Platonic realism, or metaphysical objectivism, holding
  that things exist objectively and independently of
  observation. These equivalences are false. A physical object can
  have a definite, Platonic existence, and Bayesians can still assign
  probabilities to its attributes since our ability to measure is
  limited by imperfect equipment.}, so the relevant probability is
that the mass has a certain value, given the data. This probability
distribution is the {\em posterior}, $p(m \vert d)$.  The likelihood and
posterior are related via  Bayes' theorem:
\begin{equation}\label{bayes}
p( m \vert d) = \frac{ p( d \vert m) p(m)}{p(d)} \, ,
\end{equation}
where $p(m)$ is the prior probability distribution for $m$, 
and the normalization constant,
\begin{equation}
p(d) = \int p( d\vert m) p(m) \, dm \, ,
\end{equation}
is the {\em marginalized likelihood}, or {\em evidence}.  
For uniform (flat) priors
the frequentist confidence intervals for the parameters will coincide
with the Bayesian credible intervals, but the interpretation remains
quiet distinct.  

The choice of prior probability distributions is a
source of much consternation and debate, and is often cited as a
weakness of the Bayesian approach. But the choice of
probability distribution for the likelihood (which is also 
important for the frequentist approach) is often no less
fraught. The prior quantifies what we know about the range and
distribution of the parameters in our model, while the likelihood
quantifies what we know about our measurement apparatus, and, in 
particular, the nature of the measurement noise. The choice of prior
is especially problematic in a new field where there is little to
guide the choice.  For example, electromagnetic observations and
population synthesis models give some guidance about black hole
masses, but the mass range and distribution is currently not well
constrained.  The choice of likelihood can also be challenging when
the measurement noise deviates from the stationary, Gaussian ideal.
More details related to the choice of likelihood and choice of 
prior will be given in Section~\ref{s:likelihood+prior}.

In addition to parameter estimation, statistical inference is used to
select between competing models, or hypotheses, such as, ``is there a
gravitational-wave signal in the data or not?" 
Thanks to GW150914 and GW151226, we know that gravitational-wave
signals {\em are} already present in existing data sets, but most are
at levels where we are unable to distinguish them from noise processes. For
detection we demand that a model for the data that includes a
gravitational-wave signal be favored over a model having no
gravitational-wave signal. In Bayesian inference a detection might be
announced when the odds ratio between models with and without
gravitational-wave signals gets sufficiently large, while in 
frequentist inference a detection might be announced when the 
$p$-value for some test statistic is less than some prescribed threshold.
These different approaches to deciding whether or not to
claim a detection (e.g., Bayesian model selection or frequentist
hypothesis testing), as well as differences in regard to 
parameter estimation, are described in the following subsections.
Table~\ref{t:bayesfreq} provides an overview of the key 
similarities and differences between frequentist and 
Bayesian inference, to be described in detail below.
\begin{table}[htbp]
\addtocounter{table}{-1} 
\centering
\begin{longtable}{p{2.75in} | p{2.75in}}
\toprule
FREQUENTIST & BAYESIAN
\\
\midrule
probabilities assigned only to propositions about
outcomes of repeatable experiments (i.e., random variables),
not to hypotheses or parameters which have fixed but
unknown values
&
probabilities can be assigned to hypotheses and parameters
since probability is degree of belief (or confidence,
plausibility) in any proposition
\\
\midrule
assumes measured data are drawn from an underlying 
probability distribution, which assumes the truth of a 
particular hypothesis or model (likelihood function)
&
same
\\
\midrule
constructs a statistic to estimate a parameter or
to decide whether or not to claim a detection
&
needs to specify prior degree of belief in a particular
hypothesis or parameter
\\
\midrule
calculates the probability distribution of the statistic
(sampling distribution)
&
uses Bayes' theorem to update the prior degree of belief in
light of new data (i.e., likelihood ``plus" prior yields posterior)
\\
\midrule
constructs confidence intervals and $p$-values 
for parameter estimation and hypothesis testing
&
constructs posteriors and odds ratios 
for parameter estimation and hypothesis testing / 
model comparison
\\
\bottomrule
\end{longtable}
\caption{Comparison of frequentist and Bayesian approaches
to statistical inference.
See Sections~\ref{s:frequentist} and \ref{s:bayesian} for details.}
\label{t:bayesfreq}
\end{table}
%
\subsection{Frequentist statistics}
\label{s:frequentist}

As mentioned above, 
classical or {\em frequentist} statistics is a branch of 
statistical inference that interprets probability as the 
``long-run relative occurrence of 
an event in a set of identical experiments."
Thus, for a frequentist, 
probabilities can only be assigned to propositions
about outcomes of (in principle) repeated 
experiments (i.e., {\em random variables})
and not to hypotheses or parameters describing 
the state of nature, which have fixed but unknown values.
In this interpretation, the measured data are drawn from 
an underlying probability distribution, which assumes the 
truth of a particular hypothesis or model.
The probability distribution for the data is just the 
likelihood function, which we can write as
$p(d|H)$, where $d$ denotes the data and $H$ denotes an
hypothesis.

Statistics play an important role in the frequentist framework.
These are random variables constructed from the data,
which typically estimate a signal parameter or 
indicate how well the data fit a particular hypothesis.
Although it is common to construct statistics from the 
likelihood function (e.g., the maximum-likelihood 
statistic for a particular parameter, or the maximum-likelihood 
ratio to compare a signal-plus-noise model to a 
noise-only model), there is no a~priori 
restriction on the form of a statistic other than 
it be {\em some} function of the data.
Ultimately, it is the goal of the analysis and the 
cleverness of the data analyst that dictate which 
statistic (or statistics) to use.

To make statistical inferences in the frequentist 
framework requires knowledge of the probability 
distribution (also called the {\em sampling distribution}) 
of the statistic.
The sampling distribution can either be calculated 
analytically (if the statistic is sufficiently simple)
or via Monte Carlo simulations, which effectively
construct a histogram of the values of the statistic
by simulating many independent realizations of the data.
Given a statistic and its sampling distribution, one
can then calculate either {\em confidence intervals}
for parameter estimation or {\em p-values} for 
hypothesis testing.
(These will be discussed in more detail below.)
Note that a potential problem with frequentist statistical
inference is that the sampling distribution depends on 
data values that were {\em not} actually observed, 
which is related to how the experiment was carried out 
{\em or might have been} carried out.
The so-called {\em stopping problem} of frequentist 
statistics is an example of such a problem~\cite{Howson-Urbach:2006}.

\subsubsection{Frequentist hypothesis testing}
\label{s:freq-hypothesis-testing}

Suppose, as a frequentist, you want to test the hypothesis 
$H_1$ that a gravitational-wave signal, 
having some fixed but unknown amplitude $a>0$,
is present in the data.
Since you cannot assign probabilities to hypotheses or 
to parameters like $a$ as a frequentist, 
you need to introduce instead an 
alternative (or {\em null}) hypothesis $H_0$, which, 
for this example, is the hypothesis that there is no 
gravitational-wave signal in the data
(i.e., that $a=0$).
You then argue for $H_1$ by arguing {\em against} $H_0$,
similar to proof by contradiction in mathematics.
Note that $H_1$ is a {\em composite} hypothesis since
it depends on a range of values of the unknown parameter $a$.
It can be written as the union, $H_1=\cup_{a>0} H_a$, 
of a set of simple hypotheses $H_a$
each corresponding to a single fixed value of the 
parameter $a$.

\begin{figure}[h!tbp]
\begin{center}
\includegraphics[angle=0,width=.7\columnwidth]{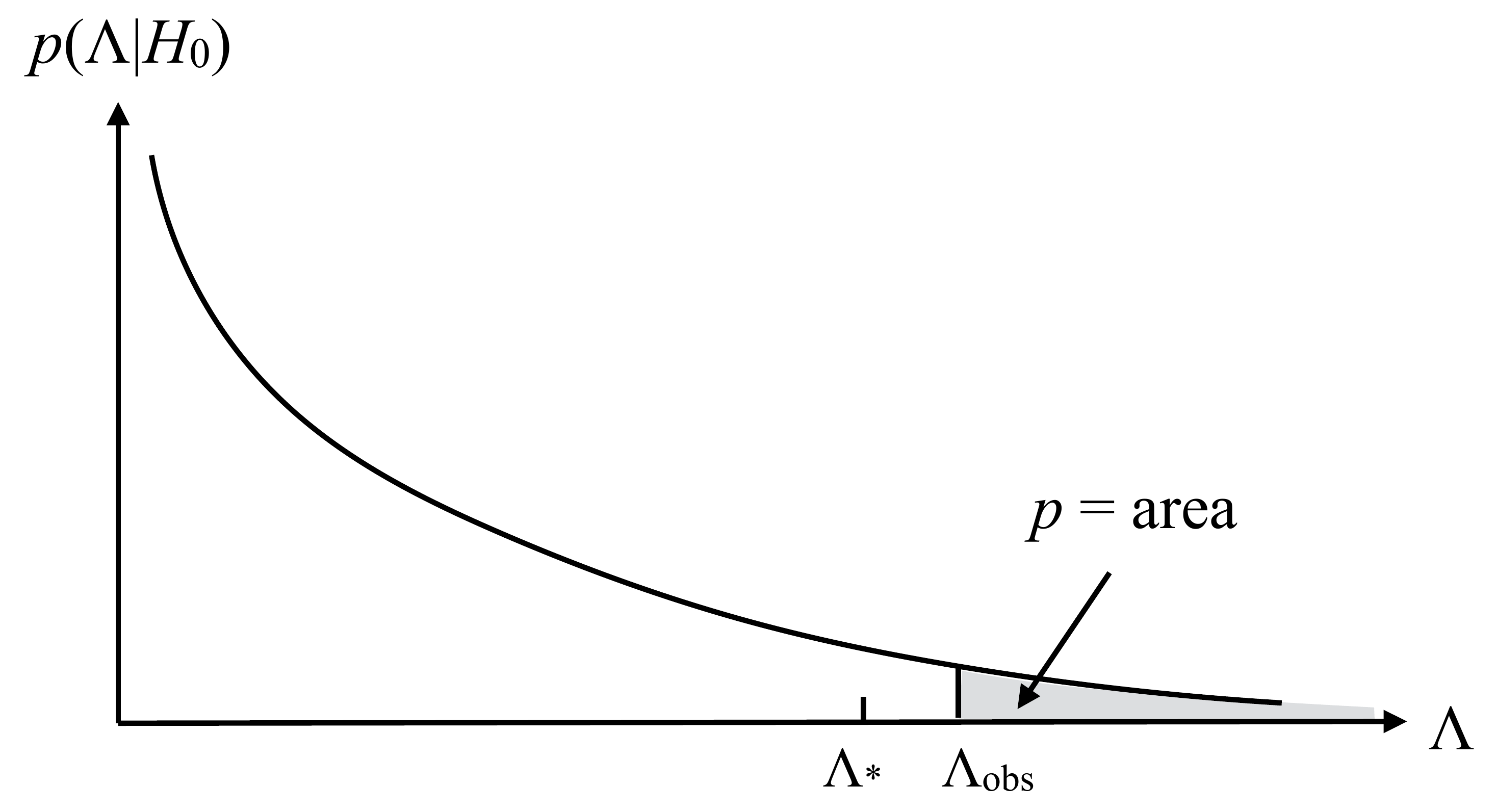}
\caption{Definition of the $p$-value (or significance)
for frequentist hypothesis testing.
The value of $p$ equals the area under the probability
distribution $p(\Lambda|H_0)$ for $\Lambda\ge\Lambda_{\rm obs}$.}
\label{f:pvalue}
\end{center}
\end{figure}

To rule either in favor or against $H_0$, you construct 
a statistic $\Lambda$, called a {\em test} or 
{\em detection statistic}, 
on which the statistical test will be based.
As mentioned above, you will need to calculate analytically 
or via Monte Carlo simulations the sampling distribution 
for $\Lambda$ under the assumption that the null hypothesis is 
true, $p(\Lambda|H_0)$.
If the observed value of $\Lambda$ lies far out in the tails of 
the distribution, then the data are most likely not
consistent with the assumption of the null hypothesis,
so you reject $H_0$ 
(and thus accept $H_1$) at the $p*100$\% level, where 
\be
p\equiv {\rm Prob}(\Lambda>\Lambda_{\rm obs}|H_0)
\equiv\int_{\Lambda_{\rm obs}}^\infty  p(\Lambda|H_0)\,d\Lambda\,.
\ee
This is the so-called {\em $p$-value} 
(or {\em significance}) of the test;
it is illustrated graphically in Figure~\ref{f:pvalue}.
The $p$-value required to reject the null hypothesis 
determines a {\em threshold} $\Lambda_*$, above
which you reject $H_0$ and accept $H_1$ (e.g., claim
a detection).
It is related to the {\em false alarm probability} for 
the test as we explain below.

The above statistical test is subject to two types of
errors:
(i) type I or {\em false alarm} errors, which arise 
if the data are such that you reject the null hypothesis 
(i.e., $\Lambda_{\rm obs}>\Lambda_*$) when it is actually true, and
(ii) type II or {\em false dismissal} errors, which 
arise if the data are such that you accept the null 
hypothesis (i.e., $\Lambda_{\rm obs}<\Lambda_*$) 
when it is actually false.
The false alarm probability $\alpha$ and 
false dismissal probability $\beta(a)$ are given 
explicitly by 
\begin{align}
\alpha &\equiv {\rm Prob}(\Lambda>\Lambda_*|H_0)\,,
\\
\beta(a) &\equiv {\rm Prob}(\Lambda<\Lambda_*|H_a)\,,
\end{align}
where $a$ is the amplitude of the gravitational-wave
signal, assumed to be present 
under the assumption that $H_1$ is true.
To calculate the false dismissal probability
$\beta(a)$, one needs the sampling distribution 
of the test statistic assuming the presence of
a signal with amplitude $a$.

Different test statistics are judged according to 
their false alarm and false dismissal probabilities.
Ideally, you would like your statistical test to have 
false alarm and false dismissal probabilities that are 
both as small as possible.
But these two properties compete with one another
as setting a larger threshold value to minimize 
the false alarm probability will increase the false 
dismissal probability.
Conversely, 
setting a smaller threshold value to minimize the false
dismissal probability will increase the false alarm
probability.

In the context of gravitational-wave data analysis, 
the gravitational-wave community is (at least initially)
reluctant to falsely claim detections.
Hence the false alarm probability is set to some very low value.
The best statistic then is the one that minimizes 
the false dismissal probability 
(i.e., maximizes detection probability)
for fixed false alarm.
This is the {\em Neyman-Pearson criterion}.
For medical diagnosis, on the other hand, a doctor
is very reluctant to falsely dismiss an illness.
Hence the false dismissal probability will be set to some 
very low value.
The best statistic then is the one which minimizes 
the false alarm probability for fixed false dismissal.

\subsubsection{Frequentist detection probability}
\label{s:freq-DP}

\begin{figure}[h!tbp]
\begin{center}
\includegraphics[angle=0,width=.7\columnwidth]{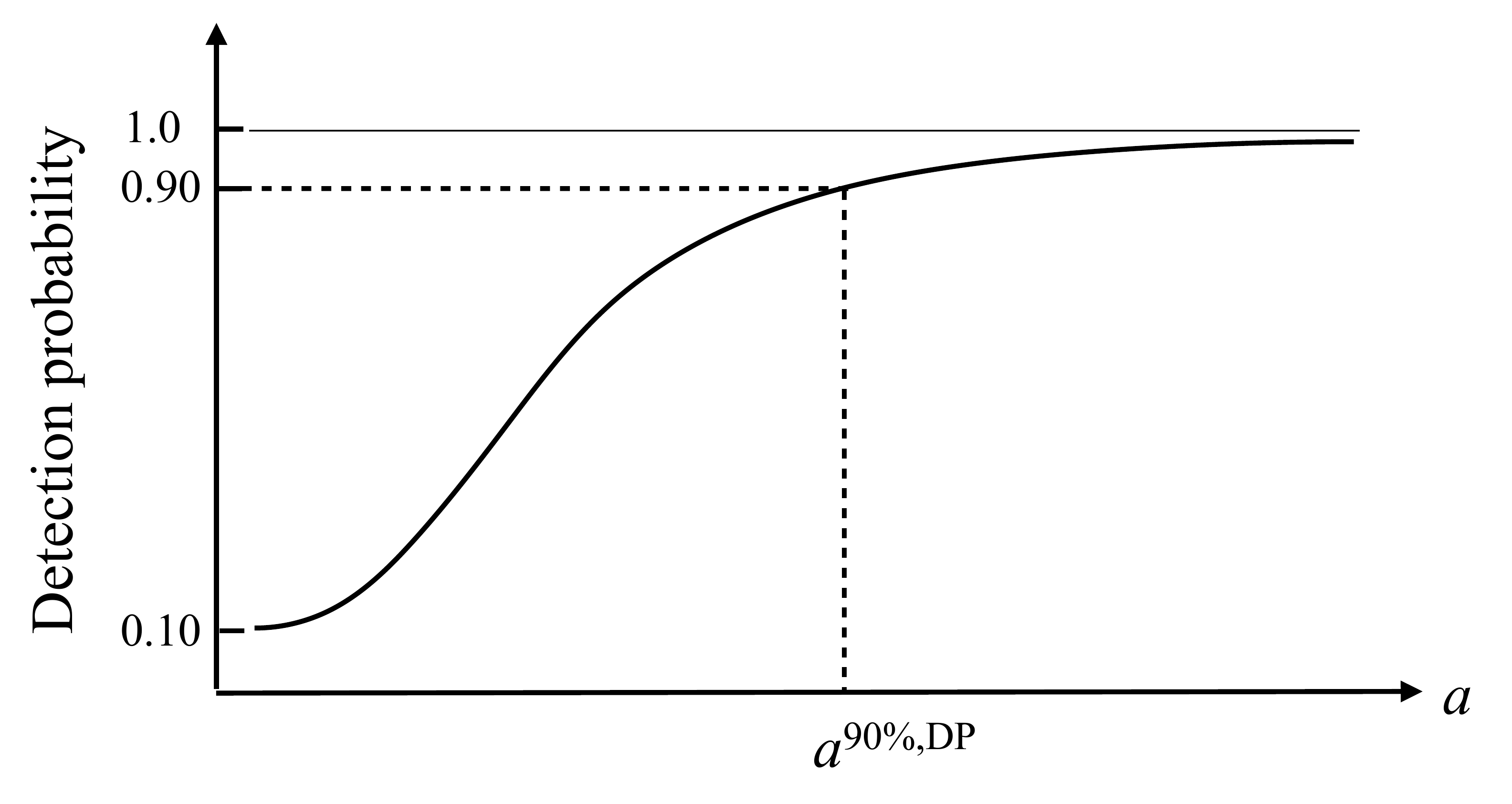}
\caption{Detection probability as a function 
of the signal amplitude for a false alarm probability
equal to $10\%$.
The value of $a$ needed for 90\% detection probability 
is indicated by the dashed vertical line and is denoted
by $a^{90\%,{\rm DP}}$.}
\label{f:frequentist-DP}
\end{center}
\end{figure}

The value $1-\beta(a)$ is called the 
{\em detection probability} or {\em power} 
of the test. 
It is the fraction of times that the test 
stastic $\Lambda$ correctly identifies the presence of a 
signal of amplitude $a$ in the data, for a fixed
false alarm probability $\alpha$ (which sets the
threshold $\Lambda_*$).
A plot of detection probability versus signal
strength is often used 
to show how strong a signal has to be in order 
to detect it with a certain probability.
Since detection probability does not depend on 
the observed data---it depends only on the sampling 
distribution of the test statistic and a choice 
for the false alarm probability---detection probability 
curves are often used as a {\em figure-of-merit} for 
proposed search methods for a signal.
Figure~\ref{f:frequentist-DP} shows a detection 
probability curve, with the value of $a$ needed 
to be detectable with 90\% frequentist probability 
indicated by the dashed vertical line.
We will denote this value of $a$ by $a^{90\%,{\rm DP}}$.
Note that as the signal amplitude goes to zero,
the detection probability reduces to the false
alarm probability $\alpha$, which for this example
was chosen to be $0.10$. 
%

\subsubsection{Frequentist upper limits}
\label{s:freq-UL}

\begin{figure}[h!tbp]
\begin{center}
\includegraphics[angle=0,width=.7\columnwidth]{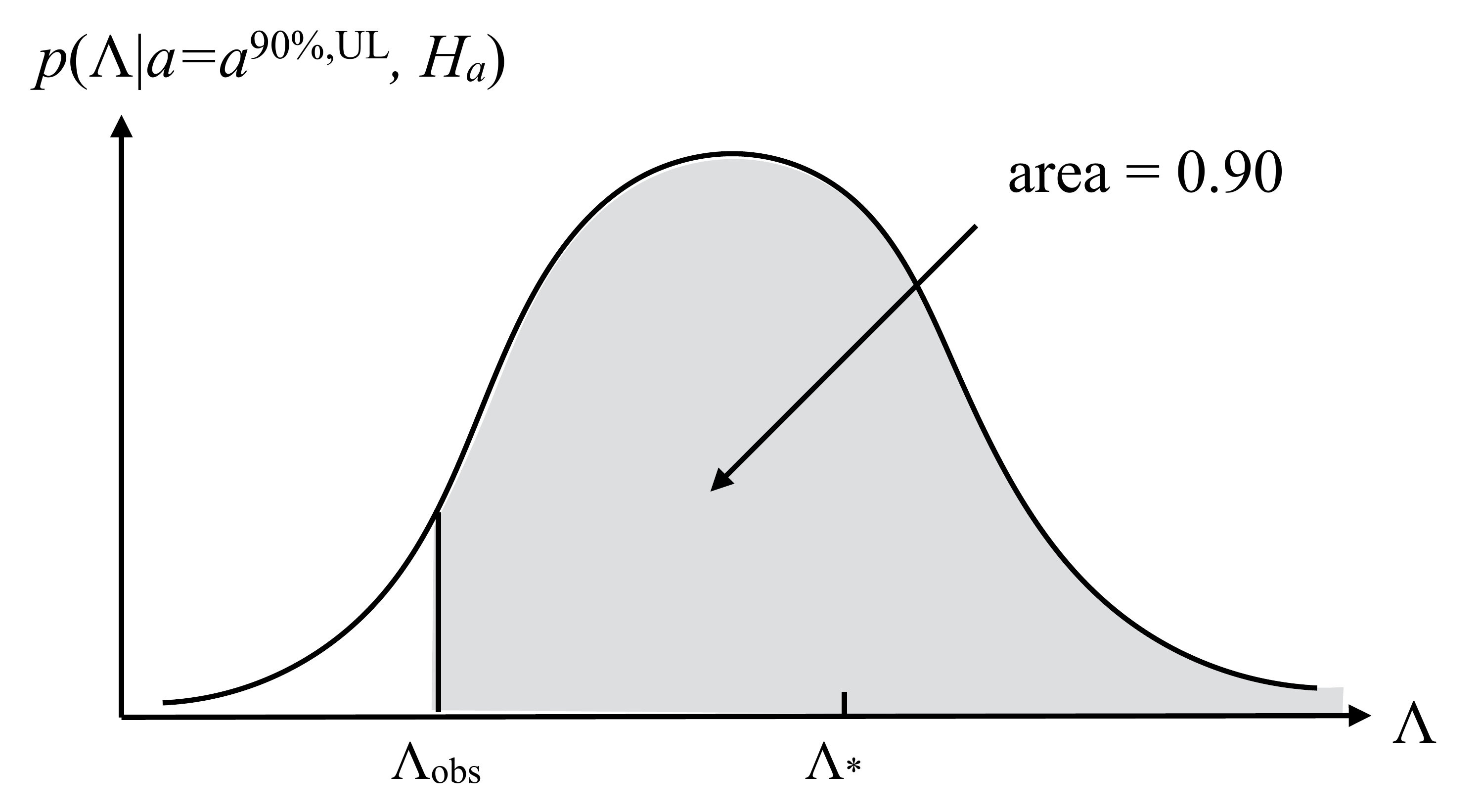}
\caption{Graphical representation of a frequentist 90\%
confidence level upper limit.
When $a=a^{90\%,{\rm UL}}$, the probability of obtaining a value 
of the detection statistic $\Lambda\ge \Lambda_{\rm obs}$ is equal 
to $0.90$.}
\label{f:frequentist-UL}
\end{center}
\end{figure}

In the absence of a detection (i.e., if the observed value of the 
test statistic is less than the detection threshold $\Lambda_*$), 
one can still set a bound (called an {\em upper limit}) 
on the strength of the signal that one was trying to detect.
The upper limit depends on the observed value of the
test statistic, $\Lambda_{\rm obs}$, and a choice of confidence 
level, $\mr{CL}$, interpreted in the frequentist framework
as the long-run relative occurence for a set of 
repeated identical experiments. 
For example, one defines the 90\% confidence-level upper limit 
$a^{90\%,{\rm UL}}$ as the minimum value of $a$ for which 
$\Lambda\ge \Lambda_{\rm obs}$ at least 90\% of the time:
\be
{\rm Prob}(\Lambda\ge \Lambda_{\rm obs}| a\ge a^{90\%,{\rm UL}}, H_a)
\ge 0.90\,.
\ee
In other words, if the signal has an amplitude $a^{90\%,{\rm UL}}$ 
or higher, we would have detected it in at least 
90\% of repeated observations.
A graphical representation of a frequentist upper limit is given
in Figure~\ref{f:frequentist-UL}.
%

\subsubsection{Frequentist parameter estimation}
\label{s:freq-parameter-estimation}

\begin{figure}[h!tbp]
\begin{center}
\includegraphics[angle=0,width=.7\columnwidth]{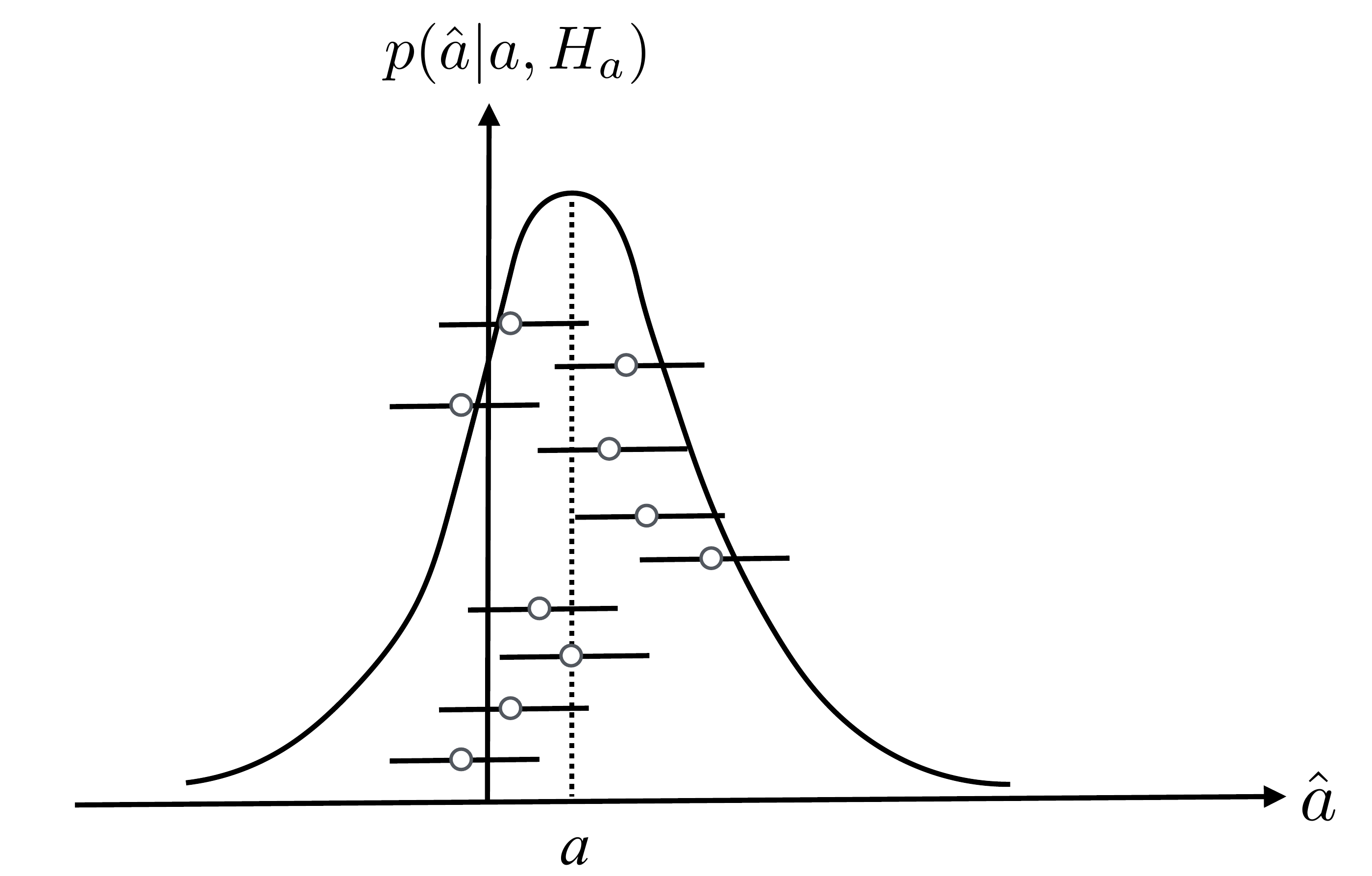}
\caption{Definition of the frequentist confidence interval
for parameter estimation.
Each circle and line represents a measured interval
$[\hat a-\Delta, \hat a+\Delta]$.
The set of all such intervals will contain the true value 
of the parameter $a$ (indicated here by the dotted vertical 
line) $\mr{CL}*100\%$ of the time, where $\mr{CL}$ is the confidence level.}
\label{f:frequentist-intervals}
\end{center}
\end{figure}

The frequentist prescription for estimating the value
of a particular parameter $a$, like the amplitude
of a gravitational-wave signal, is slightly different 
than the method used to claim a detection.
You need to first construct a statistic 
(called an {\em estimator}) 
$\hat a$ of the parameter $a$ you are interested in.
(This might be a maximum-likelihood estimator of 
$a$, but other estimators can also be used.)
You then calculate its sampling distribution 
$p(\hat a|a, H_a)$.
Note that statements like 
\be
{\rm Prob}(a-\Delta < \hat a < a+\Delta)=0.95\,,
\ee
which one constructs from $p(\hat a|a,H_a)$ make sense in the 
frequentist framework, since $\hat a$ is a random variable.
Although the above inequality can be rearranged to yield
\be
{\rm Prob}(\hat a-\Delta < a < \hat a+\Delta)=0.95\,,
\ee
this should {\em not} be interpreted as a statement about
the probability of $a$ lying within a particular interval
$[\hat a-\Delta,\hat a+\Delta]$, since $a$ is not a random variable.
Rather, it should be interpreted as a probabilistic
statement about the {\em set of intervals} 
$\{[\hat a-\Delta,\hat a+\Delta]\}$
for all possible values of $\hat a$.
Namely, in a set of many repeated experiments, 0.95 is
the fraction of the intervals that will contain the 
true value of the parameter $a$.
Such an interval is called a 
{\em $95\%$ frequentist confidence interval}.
This is illustrated graphically in 
Figure~\ref{f:frequentist-intervals}.

It is important to point out that an estimator can 
sometimes take on a value of the parameter that is 
{\em not physically allowed}.
For example, if the parameter $a$ denotes the amplitude
of a gravitational-wave signal (so physically $a\ge 0$), 
it is possible for $\hat a <0$ for a particular realization 
of the data. 
Note that there is nothing mathematically wrong with 
this result.
Indeed, the sampling distribution for $\hat a$ 
specifies the probability of obtaining such values
of $\hat a$.
It is even possible to have a confidence interval 
$[\hat a-\Delta, \hat a+\Delta]$ all of whose values
are unphysical, especially 
if one is trying to detect a weak signal in noise.
Again, this is mathematically allowed, but it is a 
little awkward to report a frequentist 
confidence interval that is completely unphysical.
We shall see that within the Bayesian framework 
unphysical intervals and unphysical posteriors 
never arise, as a simple consequence of including a 
prior distribution on the parameter that requires $a > 0$.

\subsubsection{Unified approach for frequentist upper limits
and confidence intervals}

Frequentists also have a way of avoiding unphysical or empty 
confidence intervals, which at the same time {\em unifies} the 
treatment of upper limits for null results and two-sided 
intervals for non-null results.
This procedure, developed by Feldman and Cousins~\cite{Feldman-Cousins:1998},
also solves the problem that the choice of an upper limit or 
two-sided confidence interval leads to intervals that
do not have the proper coverage (i.e., the probability that 
an interval contains 
the true value of a parameter does not match the stated
confidence level) if the choice of reporting an upper limit 
or two-sided confidence interval is {\em based on the data}
and not decided upon before performing the experiment.

\begin{figure}[h!tbp]
\begin{center}
\includegraphics[angle=0,width=.5\columnwidth]{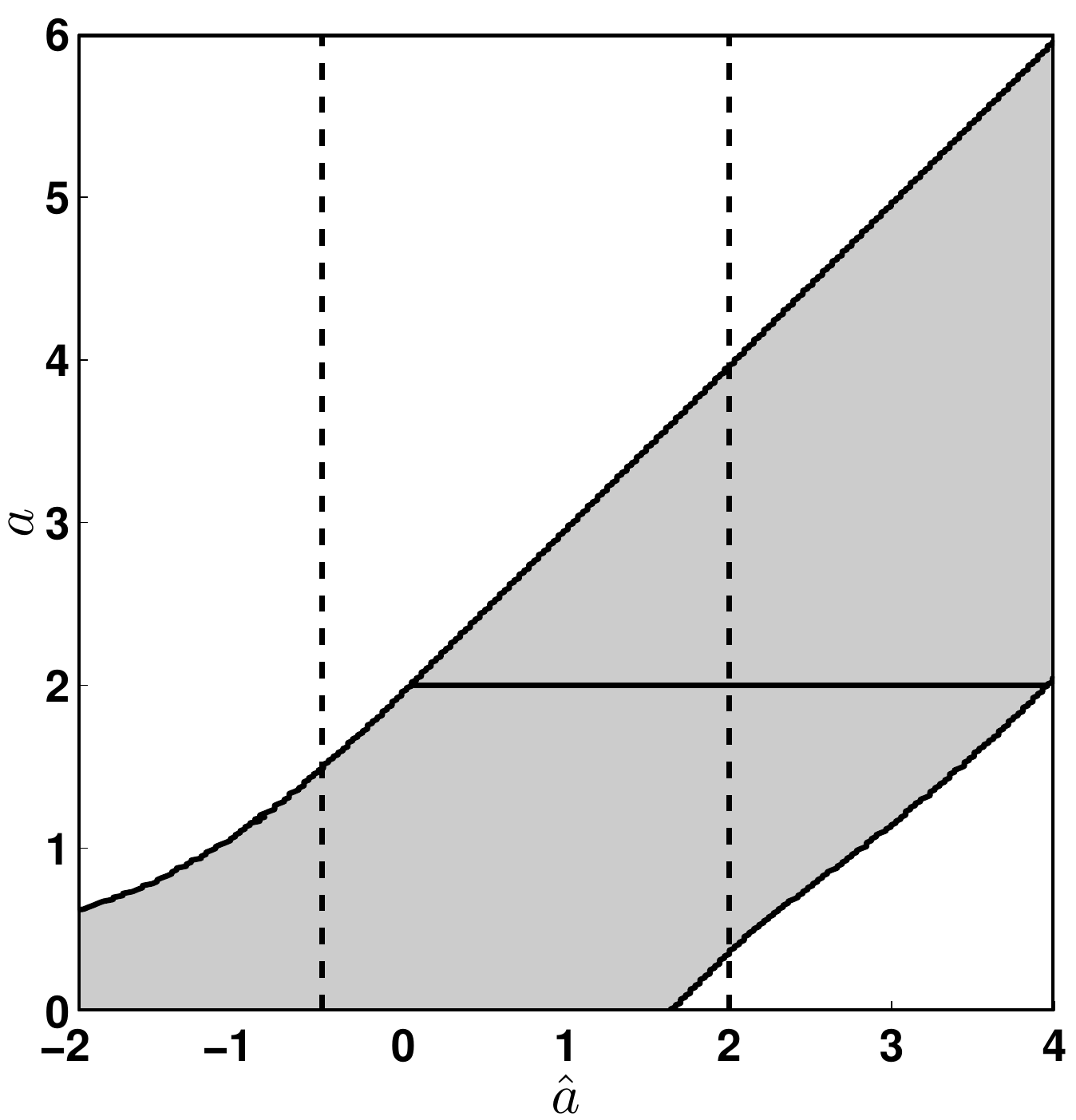}
\caption{Confidence belt for 95\% confidence-level 
intervals for a Gaussian distribution with mean $a> 0$.
(The values for $a$ and $\hat a$ are given here 
in units of $\sigma$.)
The solid horizontal line shows the acceptance interval for 
$a=2.0$.
The two dashed vertical lines correspond to two different 
observed values for the estimator $\hat a$:
$\hat a=-0.5$, which has a 95\% confidence-level upper limit $a\le 1.5$; and
$\hat a=2$, which has a 95\% confidence-level two-sided interval $a\in [0.35, 3.95]$.}
\label{f:feldmancousins}
\end{center}
\end{figure}

The basic idea underlying this unified approach to frequentist 
intervals is a new specification (or {\em ordering}) 
of the values of the random variable to include in the 
acceptance intervals for an unknown parameter.
If we let $a$ denote the parameter whose 
value we are trying to determine, and $\hat a$ be an 
estimator of $a$ with sampling distribution 
$p(\hat a|a,H_a)$,
then the choice of acceptance intervals becomes,
for each value of $a$, how do we choose 
$[\hat a_1, \hat a_2]$ such that 
\be
{\rm Prob}(\hat a_1 < \hat a < \hat a_2)
\equiv 
\int_{\hat a_1}^{\hat a_2} p(\hat a|a, H_a)\,d\hat a
=\mr{CL}\,,
\label{e:acceptance_interval}
\ee
where $\mr{CL}$ is the confidence level, e.g., $\mr{CL}=0.95$.
The ordering priniciple proposed by Feldman and 
Cousins~\cite{Feldman-Cousins:1998} is based on the
ranking function
\be
R(\hat a|a) \equiv 
\frac{p(\hat a|a, H_a)}{p(\hat a|a, H_a)\big|_{a=a_{\rm best}}}\,,
\label{e:ranking_function}
\ee
where $a_{\rm best}$ is the value of the parameter $a$
that maximizes the sampling distribution $p(\hat a|a,H_a)$
for a given value of $\hat a$.
The prescription then for constructing the acceptance 
intervals is to find, for each allowed value of $a$, 
values of $\hat a_1$ and $\hat a_2$ such that 
$R(\hat a_1|a)=R(\hat a_2|a)$ and for which 
(\ref{e:acceptance_interval}) is satisfied.
The set of all such acceptance intervals for different 
values of $a$ forms a {\em confidence belt} in the 
$\hat aa$-plane, which is then used to construct an 
upper limit or a two-sided confidence interval for a 
particular observed value of the estimator $\hat a$, 
as explained below and illustrated in
Figure~\ref{f:feldmancousins}.

As a specific example, let us suppose that $\hat a$ is
Gaussian-distributed about $a$ with variance $\sigma^2$:
\be
p(\hat a|a,H_a) = \frac{1}{\sqrt{2\pi}\sigma}
e^{-\frac{1}{2}\frac{(\hat a-a)^2}{\sigma^2}}\,,
\ee
and that the unknown parameter $a$ represents the 
amplitude of a signal, so that $a > 0$.
(Recall that it is possible, however, for the 
estimator $\hat a$ to take on negative values.)
Then $a_{\rm best}=\hat a$ if $\hat a > 0$, while
$a_{\rm best} = 0$ if $\hat a \le 0$, for which
\be
p(\hat a|a, H_a)\Big|_{a=a_{\rm best}} = 
\left\{
\begin{array}{lr}
\frac{1}{\sqrt{2\pi}\sigma}\,, & \hat a> 0\\
\frac{1}{\sqrt{2\pi}\sigma} 
\exp\left[-\frac{1}{2}\frac{\hat a^2}{\sigma^2}\right]\,,
& \hat a \le 0
\end{array}
\right.
\ee
and
\be
R(\hat a|a) = 
\left\{
\begin{array}{lr}
\exp\left[-\frac{1}{2}\frac{(\hat a-a)^2}{\sigma^2}\right]\,,
& \hat a> 0\\
\exp\left[-\frac{1}{2}\frac{(-2\hat a a + a^2)}{\sigma^2}\right]\,,
& \hat a\le 0
\end{array}
\right.\,.
\ee
The confidence belt constructed from this ranking 
function is shown in Figure~\ref{f:feldmancousins}.
The solid horizontal line at $a=2$ shows the corresponding
95\% confidence-level acceptance interval for this ranking function.
The two dashed vertical lines correspond to two different 
observed values for the estimator $\hat a$, leading to a 
95\% confidence-level upper limit and two-sided interval, respectively.

\subsection{Bayesian inference}
\label{s:bayesian}

In the following subsections, we again describe parameter
estimation and hypothesis testing, but this time from the
perspective of Bayesian inference.

\subsubsection{Bayesian parameter estimation}
\label{s:bayesian-parameter-estimation}

In Bayesian inference, a parameter, e.g., $a$, is estimated in 
terms of its posterior distribution, $p(a|d)$, in light of 
the observed data $d$.
As discussed in the introduction to this section, the posterior
$p(a|d)$ can be calculated from the likelihood $p(d|a)$ and
the prior probability distribution $p(a)$ using Bayes' theorem
\be
p(a\vert d) = \frac{p(d\vert a)p(a)}{p(d)}\,.
\label{e:bayes}
\ee
The posterior distribution tells you everything you need to know 
about the parameter, although you might sometimes want to reduce 
it to a few numbers---e.g., its mode, mean, standard deviation, etc.

\begin{figure}[h!tbp]
\begin{center}
\includegraphics[angle=0,width=.7\columnwidth]{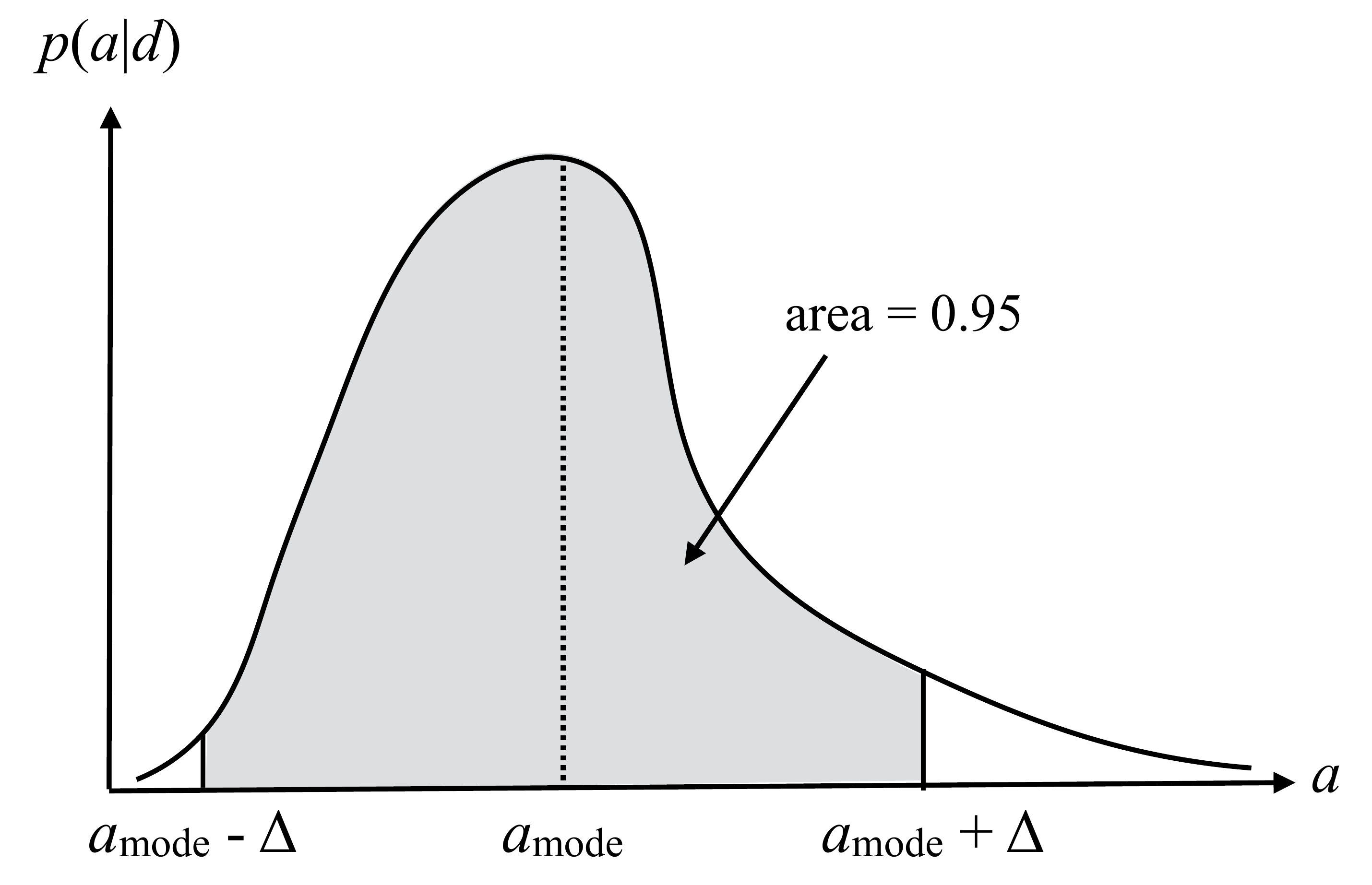}
\caption{Definition of a Bayesian credible interval for parameter 
estimation.
Here we construct a symmetric 95\% credible interval centered
on the mode of the distribution.}
\label{f:bayesian-interval}
\end{center}
\end{figure}

Given a posterior distribution $p(a|d)$, a Bayesian confidence 
interval (often called a {\em credible interval} given the Bayesian 
interpretation of probability as degree of belief, or state of 
knowledge, about an event) is simply defined in terms of the area 
under the posterior between one parameter value and another.
This is illustrated graphically in Figure~\ref{f:bayesian-interval},
for the case of a 95\% symmetric credible interval, centered on the 
mode of the distribution $a_{\rm mode}$.
If the posterior distribution depends on two parameters $a$ and 
$b$, but you really only care about $a$, then you can obtain the 
posterior distribution for $a$ by marginalizing the 
joint distribution $p(a,b|d)$ over $b$:
\be
p(a\vert d) = \int db\>p(a,b\vert d) = \int db\>p(a\vert b,d) p(b)\,,
\label{e:marginalize-nuisance}
\ee
where the second equality follows from the relationship
between joint probabilities and conditional probabilities,
e.g., $p(a|b,d) p(b) = p(a,b|d)$.
Variables that you don't particularly care about (e.g., the
variance of the detector noise as opposed to the strength of
a gravitational-wave signal) are called {\em nuisance parameters}.
Although nuisance parameters can be handled in a straight-forward
manner using Bayesian inference, they are problematic to deal 
with (i.e., they are a nuisance!)~in 
the context of frequentist statistics.
The problem is that marginalization doesn't make sense to a 
frequentist, for whom parameters cannot be assigned probability 
distributions.

The interpretation of Bayes' theorem (\ref{e:bayes}) is that our prior
knowledge is updated by what we learn from the data, as measured by
the likelihood, to give our posterior state of knowledge.  The amount
learned from the data is measured by the information gain
\begin{equation}
I = \int da\>
p( a \vert d) \log\left(\frac{p( a \vert d)}{p(a)}\right)\,.
\end{equation}
Using a natural logarithm gives the information in {\em nats}, while
using a base 2 logarithm gives the information in {\em bits}. If the
data tells us nothing about the parameter, then 
$p(d\vert a) = {\rm  constant}$, which implies $p(a\vert d)=p(a)$ 
and thus $I=0$.

\subsubsection{Bayesian upper limits}
\label{s:bayes-UL}

\begin{figure}[h!tbp]
\begin{center}
\includegraphics[angle=0,width=.7\columnwidth]{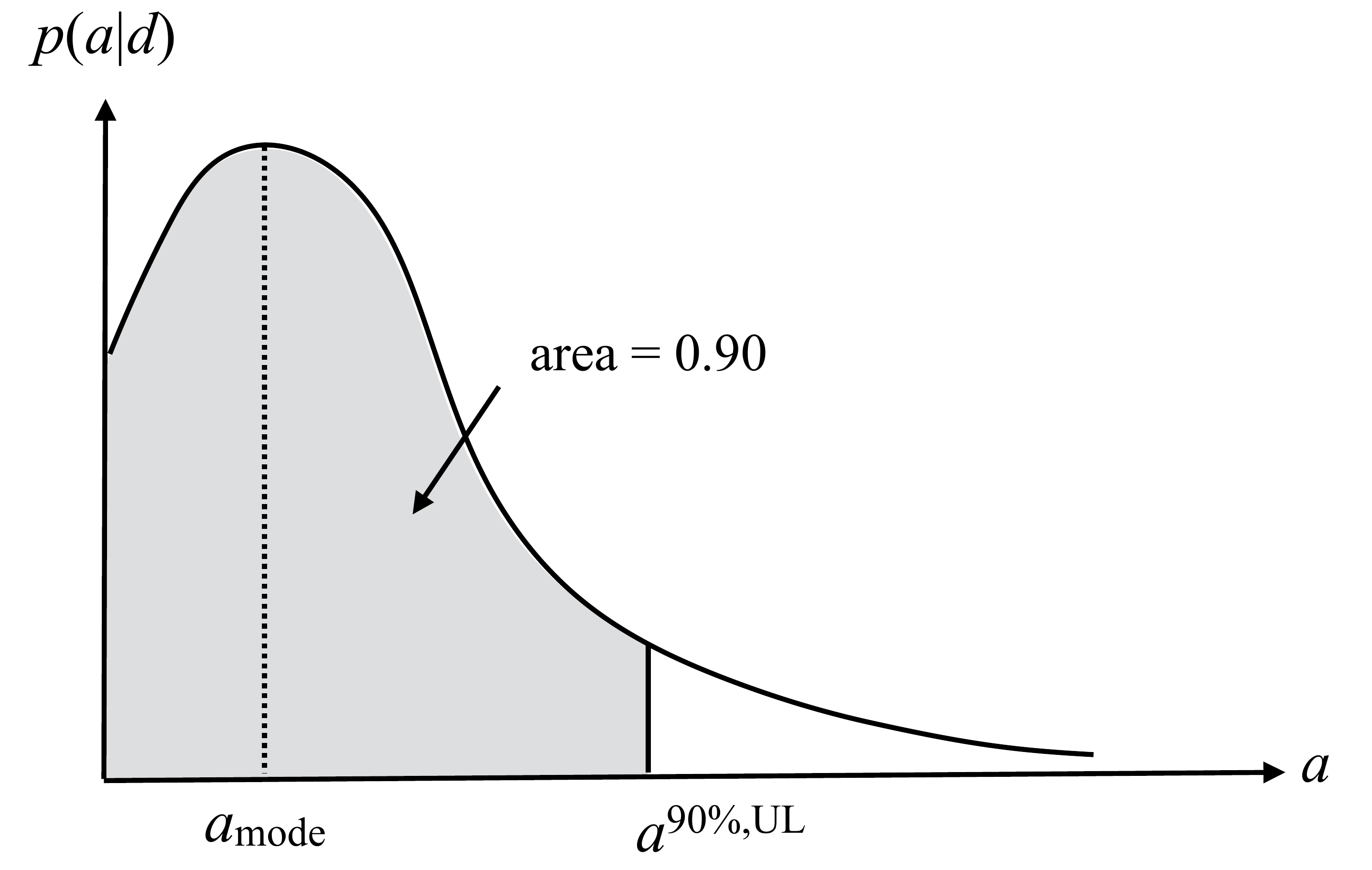}
\caption{Bayesian 90\% credible upper limit for the parameter $a$.}
\label{f:bayesian-UL}
\end{center}
\end{figure}

A Bayesian upper limit is simply a Bayesian credible interval 
for a parameter with the lower end point of the interval set 
to the smallest value that the parameter can take.
For example, the Bayesian 90\% upper limit
on a parameter $a> 0$ is defined by:
\be
{\rm Prob}(0 < a < a^{90\%,{\rm UL}}| d) = 0.90\,,
\ee
where probability is interpreted as degree of belief, 
or state of knowledge, that 
the parameter $a$ has a value in the indicated range.
One usually sets an upper limit on a parameter when the mode of 
the distribution for the parameter being estimated is not 
sufficiently displaced from zero, as shown in Figure~\ref{f:bayesian-UL}. 
%

\subsubsection{Bayesian model selection}
\label{s:bayes-model-selection}

Bayesian inference can easily be applied to multiple models or 
hypotheses, each with a different set of parameters.
In what follows, we will denote the different models by 
${\cal M}_\alpha$, where the index $\alpha$ runs over the 
different models, and the associated set of parameters by the vector
$\vec{\theta}_\alpha$.
The joint posterior distribution for the parameters $\vec{\theta}_\alpha$ 
is given by
\begin{equation}
\label{full_bayes}
p( \vec{\theta}_\alpha \vert d, {\cal M}_\alpha) 
= \frac{ p( d \vert \vec{\theta}_\alpha, {\cal M}_\alpha) 
p(\vec{\theta}_\alpha \vert {\cal M}_\alpha)}{p(d \vert {\cal M}_\alpha)} \, ,
\end{equation}
and the model evidence is given by
\begin{equation}
p(d \vert {\cal M}_\alpha) = 
\int p( d\vert \vec{\theta}_\alpha, {\cal M}_\alpha) 
p(\vec{\theta}_\alpha  \vert {\cal M}_\alpha) \, 
d\vec{\theta}_\alpha \,,
\label{e:model_evidence}
\end{equation}
where we marginalize over the parameter values associated 
with that model.
The posterior probability for model ${\cal M}_\alpha$ is given by Bayes' theorem as
\begin{equation}
p({\cal M}_\alpha \vert d ) = \frac{p(d \vert {\cal M}_\alpha) p({\cal M}_\alpha)}{p(d)}\,,
\end{equation}
where the normalization constant $p(d)$ involves a sum over all possible models:
\begin{equation}
p(d ) =  \sum_\alpha p(d \vert {\cal M}_\alpha) p({\cal M}_\alpha)\,.
\end{equation}
Since the space of all possible models is generally unknown, the sum
is usually taken over the subset of models being considered. The
normalization can be avoided by considering the posterior 
odds ratio between two models:
\begin{equation}
{\cal O}_{\alpha\beta}(d) 
= \frac{p({\cal M}_\alpha \vert d )}{p({\cal M}_\beta \vert d )}  
= \frac{p({\cal M}_\alpha)}{p({\cal M}_\beta)}\,
\frac{ p(d \vert {\cal M}_\alpha)}{p(d \vert {\cal M}_\beta)}\,.
\end{equation}
The first ratio on the right-hand side of the above equation
is the {\em prior} odds ratio for models
$\alpha,\beta$, while the second term is the evidence ratio, or 
{\em Bayes factor},
\begin{equation}
{\cal B}_{\alpha\beta}(d)
\equiv \frac{p(d \vert {\cal M}_\alpha)}{p(d \vert {\cal M}_\beta)} \, .
\end{equation}
The prior odds ratio is often taken to equal unity, but this is not
always justified. For example, the prior odds that a signal is
described by general relativity versus some alternative theory of
gravity should be much larger than unity given the firm theoretical and
observational footing of Einstein's theory.

While the foundations of Bayesian inference were laid out by Laplace
in the 1700s, it did not see widespread use until the late
20th century with the advent of practical implementation schemes
and the development of fast electronic computers. Today, Monte Carlo
sampling techniques, such as Markov Chain Monte Carlo (MCMC) and
Nested Sampling, are used to sample the posterior and estimate the
evidence~\cite{Skilling:2006, Gair-et-al:2010}. 
Successfully applying these techniques is something of an
art, but in principle, once the likelihood and prior have been written
down, the implementation of Bayesian inference is purely mechanical.
Calculating the likelihood and choosing a prior will be discussed in 
some detail in Section~\ref{s:likelihood+prior}.

\subsection{Relating Bayesian and frequentist detection statements}
\label{s:relating-freq-bayes}

It is interesting to compare the Bayesian model selection
calculation discussed above to 
frequentist hypothesis testing based on the 
{\em maximum-likelihood ratio}.
For concreteness, let us assume that we have two models 
${\cal M}_0$ (noise-only) and ${\cal M}_1$ (noise plus
gravitational-wave signal), with parameters
$\vec\theta_n$ and $\{\vec\theta_n,\vec\theta_h\}$,
respectively.
The frequentist detection statistic will be defined in
terms of the ratio of the maxima of the likelihood 
functions for the two models:
\be
\Lambda_{\rm ML}(d)
\equiv \frac
{\max_{\vec\theta_n} \max_{\vec\theta_h} p(d\vert\vec\theta_n,\vec\theta_h,{\cal M}_1)}
{\max_{\vec\theta_n'} p(d\vert \vec\theta_n',{\cal M}_0)}\,.
\label{e:likelihood_ratio}
\ee
As described above,
the Bayes factor calculation also involves a ratio
of two quantities, the model evidences $p(d\vert {\cal M}_1)$
and $p(d\vert {\cal M}_0)$,
but instead of maximizing over the parameters,
we marginalize over the parameters:
\be
{\cal B}_{10}(d)
=\frac
{\int d\vec\theta_n\int d\vec\theta_h\>
p(d\vert \vec\theta_n,\vec\theta_h,{\cal M}_1) p(\vec\theta_n,\vec\theta_h\vert {\cal M}_1)}
{\int d\vec\theta_n'\> p(d\vert \vec\theta_n',{\cal M}_0)p(\vec\theta_n'\vert {\cal M}_0)}\,.
\label{e:posterior_ratio}
\ee
These two expressions can be related using Laplace's
approximation to individually approximate the model
evidences $p(d\vert {\cal M}_1)$ and $p(d\vert {\cal M}_0)$.
This approximation is valid when the data are 
{\em informative}---i.e., when the likelihood functions
are peaked relative to the joint prior probability 
distributions of the parameters.
For an arbitrary model ${\cal M}$ with parameters 
$\vec\theta$,
the Laplace approximation yields:
\be
\int d\vec \theta\>
p(d|\vec\theta, {\cal M})p(\vec\theta|{\cal M})
\simeq
p(d|\vec\theta_{\rm ML},{\cal M})
\frac{\Delta V_{\cal M}}{V_{\cal M}}\,,
\ee
where $\vec\theta_{\rm ML}\equiv\vec\theta_{\rm ML}(d)$ 
maximizes the likelihood with respect to variations of 
$\vec\theta$ given the data $d$;
$\Delta V_{\cal M}$ is the characteristic spread of the 
likelihood function around its maximum
(the volume of the uncertainty ellipsoid for the parameters); 
and $V_{\cal M}$ is the total parameter space volume of the 
model parameters.
Applying this approximation to models ${\cal M}_0$ and
${\cal M}_1$ in (\ref{e:posterior_ratio}), we obtain
\be
{\cal B}_{10}(d)
\simeq \Lambda_{\rm ML}(d)\frac{\Delta V_1/V_1}{\Delta V_0/V_0}\,,
\ee
or, equivalently,
\be
2\ln {\cal B}_{10}(d)
\simeq 2\ln\left(\Lambda_{\rm ML}(d)\right)
+ 2\ln\left(\frac{\Delta V_1/V_1}{\Delta V_0/V_0}\right)\,.
\label{e:2lnBF_laplace}
\ee
The second term on the right-hand side of the above equation 
is negative and penalizes 
models that require a larger parameter space volume than
necessary to fit the data.
This is basically an {\em Occam penalty factor}, which
prefers the simpler of two models that fit the data equally well.
The first term has the interpretation of being the squared 
signal-to-noise ratio of the data, assuming an additive signal
in Gaussian-stationary noise, and it can be used as an alternative 
frequentist detection statistic in place of $\Lambda_{\rm ML}$.

Table~\ref{t:bayesfactors} from \cite{Kass-Raftery:1995} 
gives a range of Bayes factors and 
their interpretation in terms of the strength of the evidence 
in favor of one model relative to another.
The precise levels at which one considers the evidence 
to be ``strong" or ``very strong" is rather subjective.
But recent studies~\cite{Cornish-Sampson:2016, Taylor-et-al:2016b}
in the context of pulsar timing have been trying to 
make this correspondence a bit firmer, using {\em sky} 
and {\em phase scrambles} to effectively 
destroy signal-induced spatial correlations between pulsars while 
retaining the statistical properties of each individual dataset.
This is similar to doing time-slides for LIGO analyses, 
which are used to assess the significance of a detection.
\begin{table}
\centering
\begin{tabular}{c c c}
\toprule
${\cal B}_{\alpha\beta}(d)$ & $2\ln {\cal B}_{\alpha\beta}(d)$ &
Evidence for model ${\cal M}_\alpha$ relative to ${\cal M}_\beta$ \\
\midrule
$<1$ & $<0$ & Negative (supports model ${\cal M}_\beta$) \\
1--3 & 0--2 & Not worth more than a bare mention \\ 
3--20 & 2--6 & Positive \\ 
20--150 & 6--10 & Strong \\ 
$>150$ & $>10$ & Very strong \\ 
\bottomrule
\end{tabular}
\caption{Bayes factors and their
interpretation in terms of the strength of the evidence 
in favor of one model relative to the other.
Adapted from \cite{Kass-Raftery:1995}.}
\label{t:bayesfactors}
\end{table}

Taylor~et~al.~\cite{Taylor-et-al:2016b} even go so far as to perform a {\em
hybrid} frequentist-Bayesian analysis, doing Monte Carlo simulations:
(i) over different noise-only realizations, and 
(ii) over different sky and phase scrambles, which null the correlated signal. 
These simulations produce different null {\em distributions}
for the Bayes factor, similar to a null-hypothesis distribution for a
frequentist detection statistic (in this case, the log of the Bayes factor).
The significance of the measured Bayes factor is then its corresponding 
$p$-value with respect to one of these null distributions.  The utility of 
such a hybrid analysis is its ability to better assess the significance of 
a detection claim, especially when there might be questions about the 
suitability of one of the models (e.g., the noise model) used in the 
construction of a likelihood function.

\subsection{Simple example comparing Bayesian and frequentist analyses}
\label{s:example-bayes-freq}

To further illustrate the relationship between Bayesian and 
frequentist analyses, we consider in this section a very 
simple example---a constant signal with amplitude 
$a>0$ in white, Gaussian noise (zero mean, variance $\sigma$):
\be
d_i = a + n_i\,,
\qquad i=1,2,\cdots, N\,,
\ee
where the index $i$ labels the individual samples of the data.
The likelihood functions for the noise-only and 
signal-plus-noise models ${\cal M}_0$ and ${\cal M}_1$ are 
thus simple Gaussians:
\begin{align}
p(d|{\cal M}_0) &= \frac{1}{(2\pi)^{N/2}\sigma^N}
e^{-\frac{1}{2\sigma^2}\sum_{i=1}^N d_i^2}\,,
\label{e:pdM0}
\\
p(d|a, {\cal M}_1) &= \frac{1}{(2\pi)^{N/2}\sigma^N}
e^{-\frac{1}{2\sigma^2}\sum_{i=1}^N (d_i - a)^2}\,.
\label{e:pdM1}
\end{align}
We will assume that the value of $\sigma$ is known a~priori.
Thus, the noise model has no free parameters,
while the signal model has just one parameter,
which is the amplitude of the signal that we are trying 
to detect.
We will choose our prior on $a$ to be flat over the 
interval $(0,a_{\rm max}]$, so $p(a)=1/a_{\rm max}$.

It is straight-forward exercise to check that the 
maximum-likelihood estimator of the amplitude $a$ is 
given by the sample mean of the data:
\be
\hat a
\equiv a_{\rm ML}(d) 
= \frac{1}{N}\sum_{i=1}^N d_i
\equiv \bar d\,.
\ee
This is is an unbiased estimator of $a$ 
and has variance $\sigma_{\hat a}^2 =\sigma^2/N$ 
(the familiar variance of the sample mean).
Thus, the sampling distribution of $\hat a$ is simply
\be
p(\hat a|a, {\cal M}_1) = 
\frac{1}{\sqrt{2\pi}\sigma_{\hat a}}
e^{-\frac{1}{2\sigma_{\hat a}^2}(\hat a-a)^2}\,,
\label{e:pahat}
\ee
where $\hat a$ can take on either positive or negative 
values (even though $a>0$).

To compute the posterior distribution $p(a|d,{\cal M}_1)$ 
for the Bayesian analysis, we first note that
\be
\sum_{i=1}^N (d_i - a)^2 = N( {\rm Var}[d] + (a-\hat{a})^2 )\,.
\ee
The model evidence $p(d | {\cal M}_1)$ is then given by
\be
p(d| {\cal M}_1)
=\frac{e^{- \frac{{\rm Var}[d]} {2 \sigma_{\hat a}^2 }}
\left[
{\rm erf}\left(\frac{a_{\rm max}-\hat a}{\sqrt{2} \sigma_{\hat a}}\right)+
{\rm erf}\left(\frac{\hat{a}}{\sqrt{2} \sigma_{\hat a}}\right)
\right]} 
{2   a_{\rm max} \sqrt{N} (2\pi)^{(N-1)/2} \sigma^{(N-1)}}\,,
\label{e:pdM1_marg}
\ee
and the posterior distribution is given by
\be
p(a|d,{\cal M}_1) 
= \frac{1}{\sqrt{2\pi} \sigma_{\hat a}}   
e^{-\frac{(a-\hat a)^2}{2\sigma_{\hat a}^2}}  
2\left[
{\rm erf}\left(\frac{a_{\rm max}-\hat a}{\sqrt{2} \sigma_{\hat a}}\right)+
{\rm erf}\left(\frac{\hat{a}}{\sqrt{2} \sigma_{\hat a}}\right)
\right]^{-1}\,.
\label{e:pa|d}
\ee
Note that this is simply a {\em truncated}
Gaussian on the interval $a\in(0,a_{\rm max}]$, with mean
$\hat a$ and variance $\sigma_{\hat a}^2$.

The above calculation shows that $\hat a$ is a {\em sufficient statisitic}
for $a$.
This means that the posterior distribution 
for $a$ can be written simply in terms of $\hat a$, 
in lieu of the individual samples $d\equiv\{d_1, d_2, \cdots, d_N\}$.
The Bayes factor 
\be
{\cal B}_{10}(d) = \frac{p(d| {\cal M}_1)}{p(d| {\cal M}_0)}\,,
\ee
is given by
\be
{\cal B}_{10}(d) = 
e^{\frac{ \hat a^2}{2 \sigma_{\hat a}^2}}
\left( \frac{\sqrt{2\pi} \sigma_{\hat a}}{a_{\rm max}}\right)
\frac{1}{2}
\left[
{\rm erf}\left(\frac{a_{\rm max}-\hat a}{\sqrt{2} \sigma_{\hat a}}\right)+
{\rm erf}\left(\frac{\hat{a}}{\sqrt{2} \sigma_{\hat a}}\right)
\right]\,.
\ee
In the limit where $\hat a$ is tightly peaked away from $0$ 
and $a_{\rm max}$, the Bayes factor simplifies to
\be
{\cal B}_{10}(d) \simeq 
e^{\frac{ \hat a^2}{2 \sigma_{\hat a}^2} }  
\left(\frac{  \sqrt{2\pi} \sigma_{\hat a} }{  a_{\rm max}  } \right)\, .
\label{e:BF_approx_example}
\ee
If we take the frequentist detection statistic to be
twice the log of the maximum-likelihood ratio, 
$\Lambda(d) \equiv 2\ln \Lambda_{\rm ML}(d)$, then
\be
\Lambda(d) = \frac{ \hat a^2}{\sigma_{\hat a}^2}
= \frac{\bar d^2}{\sigma^2/N}
\equiv \rho^2\,,
\ee
which is just the 
squared signal-to-noise ratio of the data. 
Furthermore, 
taking twice the log of the approximate Bayes factor
in (\ref{e:BF_approx_example}) gives
\be
2\ln {\cal B}_{10}(d) \simeq 
\Lambda(d) +
2\ln\left(\frac{  \sqrt{2\pi} \sigma_{\hat a} }{  a_{\rm max}  } \right)\, ,
\label{e:bayesapprox_toymodel}
\ee
where the first term is just the frequentist detection statistic 
and second term expresses the Occam penalty.
This last result is consistent with the general relation
(\ref{e:2lnBF_laplace}) discussed in the previous subsection.

\begin{figure}[h!tbp]
\begin{center}
\includegraphics[angle=0,width=.7\columnwidth]{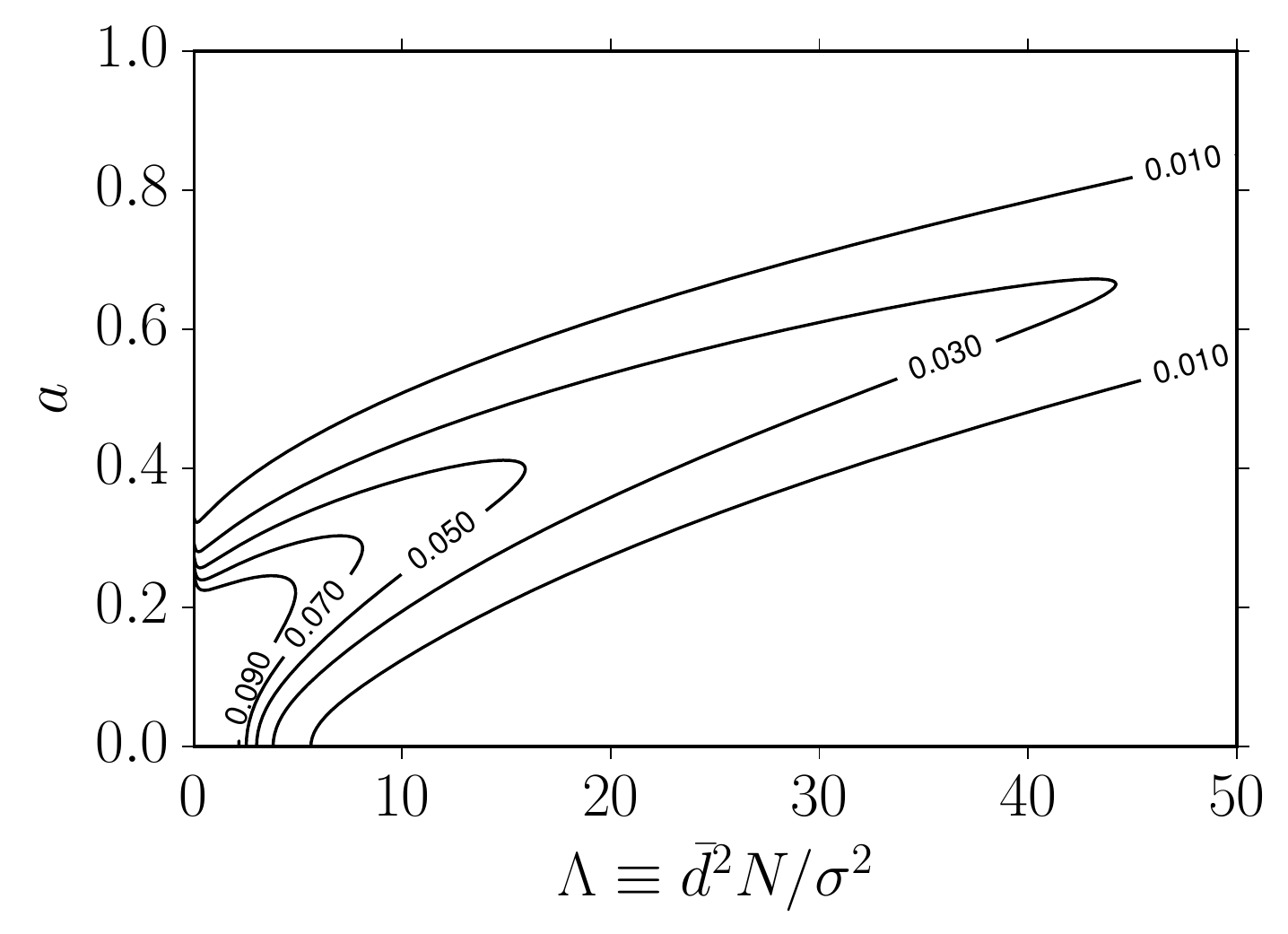}
\caption{Equal-probability contour plot 
for the frequentist detection statistic $\Lambda\equiv \bar d^2 N/\sigma^2$ 
for a signal with amplitude $a>0$.
The contours correspond to the values
$p(\Lambda|a, {\cal M}_1)=0.01$, 0.03, 0.05, 0.07, and 0.09.}
\label{f:example-plambdaA}
\end{center}
\end{figure}

The statistical distribution of the frequentist detection 
statistic can be found in closed form for this simple example.
Since a linear combination of Gaussian random variables
is also Gaussian-distributed,
$\Lambda$ is the {\em square} of a (single) Gaussian 
random variable $\rho=\bar d\sqrt{N}/\sigma$.
Moreover,
since $\rho$ has mean $\mu\equiv a\sqrt{N}/\sigma$ and 
unit variance, the sampling distribution for $\Lambda$ 
in the presence of a signal is a 
{\em noncentral chi-squared} distribution with one 
degree of freedom and non-centrality parameter 
$\lambda \equiv \mu^2 = a^2 N/\sigma^2$:
\be
p(\Lambda|a, {\cal M}_1) =\frac{1}{2}
e^{-(\Lambda+\lambda)/2}\left(\frac{\Lambda}{\lambda}\right)^{-1/4}
I_{-1/2}(\sqrt{\lambda\Lambda})\,,
\ee
where $I_{-1/2}$ is a modified Bessel function of the first
kind of order $-1/2$.
In the absence of a signal (i.e., when $a$ and hence $\lambda$
are equal to zero), $\Lambda$ is given by an (ordinary) 
chi-squared distribution with one degree of freedom:
\be
p(\Lambda|{\cal M}_0) 
=\frac{1}{\sqrt{2}\Gamma(1/2)}\Lambda^{-1/2}e^{-\Lambda/2}\,,
\ee
where $\Gamma$ is the gamma function.
Substituting explicit expressions for 
$I_{-1/2}(\sqrt{\lambda\Lambda})$ and $\Gamma(1/2)$, we find:
\begin{align}
p(\Lambda|{\cal M}_0) 
&=\frac{1}{\sqrt{2\pi\Lambda}}e^{-\Lambda/2}\,,
\label{e:pLambda0}
\\
p(\Lambda|a, {\cal M}_1) 
&=\frac{1}{\sqrt{2\pi\Lambda}}
\frac{1}{2}\left[
e^{-\frac{1}{2}(\sqrt{\Lambda}-\sqrt{\lambda})^2}
+e^{-\frac{1}{2}(\sqrt{\Lambda}+\sqrt{\lambda})^2}
\right]\,.
\label{e:pLambdaA}
\end{align}
An equal-probability contour plot of the sampling distribution 
of the detection statistic is shown in 
Figure~\ref{f:example-plambdaA}.
The fact that we are able to write down {\em analytic} 
expressions for the sampling distributions for the detection 
statistic $\Lambda$ is due to the simplicity of the
signal and noise models.
For more complicated real-world problems, these distributions
would need to be generated {\em numerically} using fake signal
injections and time-shifts to produce many different 
realizations of the data (signal plus noise) from which one 
can build up the distributions.

It is also important to point out that $\Lambda$ is {\em not}
a sufficient statistic for $a$, due to the fact that 
$\Lambda$ involves the {\em square} of the maximum-likelihood 
estimate $\hat a$---i.e., $\Lambda = \hat a^2 N/\sigma^2$.
Thus, we cannot take $p(\Lambda|a,{\cal M}_1)$ conditioned on 
$\Lambda$ (assuming a flat prior on $a$ from $[0,a_{\rm max}]$)
to get the posterior distribution for $a$ given $d$, since we
would be missing out on data samples that give negative values 
for $\hat a$.
Another way to see this is to start with 
$p(\Lambda|a,{\cal M}_1)$ given by (\ref{e:pLambdaA}), and 
then make a change of variables 
from $\Lambda$ to $\hat a$ using the general transformation relation
\be
p_Y(y)\, dy = p_X(x)\, dx
\quad\Rightarrow\quad
p_X(x) = \left[p_Y(y)\, |f'(x)|\right]_{y= f(x)}\,.
\ee
This leads to
\be
\tilde p(\hat a|a, {\cal M}_1) 
= \frac{1}{\sqrt{2\pi}\sigma_{\hat a}} 
\left[
e^{-\frac{1}{2\sigma_{\hat a}^2}(\hat a- a)^2}
+e^{-\frac{1}{2\sigma_{\hat a}^2}(\hat a + a)^2}
\right]\,,
\label{e:pahat_alt}
\ee
which is properly normalized for $\hat a>0$,
but differs from (\ref{e:pahat}) due to the second 
term involving $\hat a+a$.
Thus, we need to construct $p(a|d)$ from 
(\ref{e:pahat})---and {\em not} from 
(\ref{e:pahat_alt})---if we want the posterior 
to have the proper dependence on $a$.

\subsubsection{Simulated data}

\begin{figure}[h!tbp]
\begin{center}
\includegraphics[angle=0,width=.49\columnwidth]{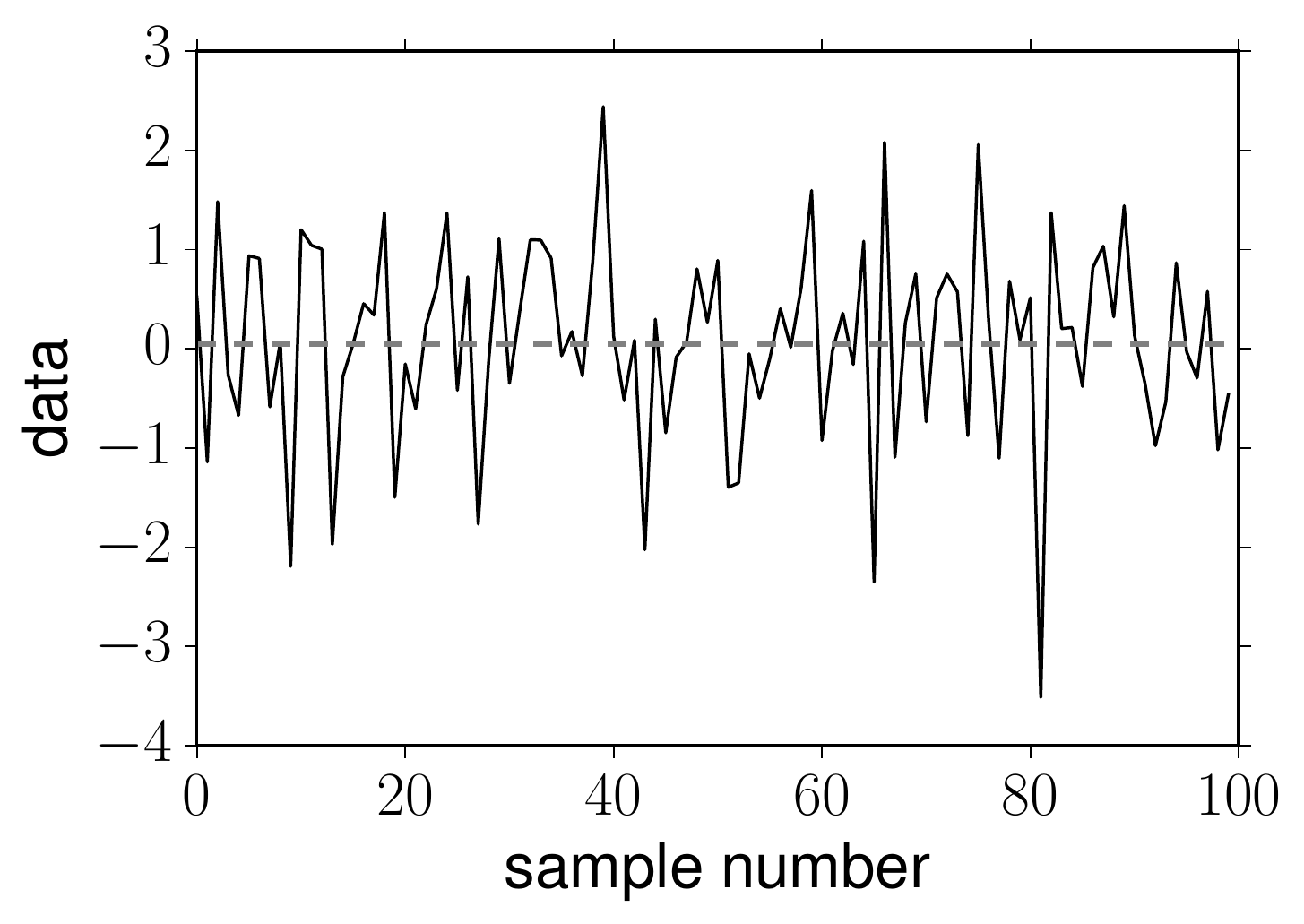}
\includegraphics[angle=0,width=.49\columnwidth]{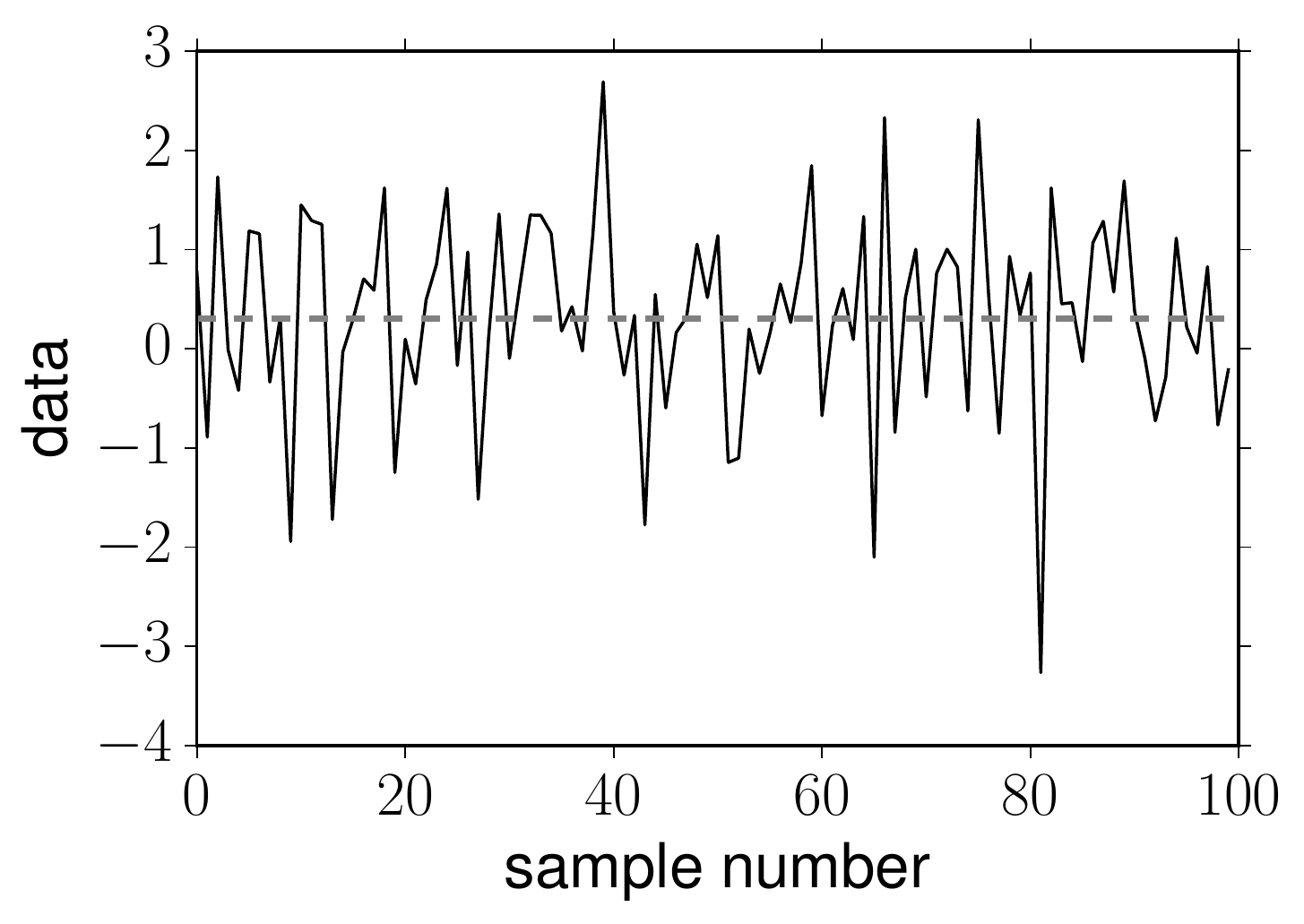}
\caption{Examples of simulated data for weak (left panel) and 
strong (right panel) signals injected into the 
data---$a_0=0.05$ and 0.3, respectively.}
\label{f:example-simulated-data}
\end{center}
\end{figure}

For our example, we will take $N=100$ samples,
$\sigma =1$, and $a_{\rm max} =1.0$.
We also simulate data with injected signals
having amplitudes $a_0=0.05$ and $0.3$, respectively.
Since the expected signal-to-noise ratio, $a\sqrt{N}/\sigma$, 
is given by $0.5$ and $3.0$,  these injections correspond to 
{\em weak} and (moderately) {\em strong} signals.
Single realizations of the data for the two different
injections are shown in Figure~\ref{f:example-simulated-data}.
The noise realization is the same for the two injections.

\subsubsection{Frequentist analysis}
\label{s:example-frequentist}

\begin{figure}[h!tbp]
\begin{center}
\includegraphics[angle=0,width=.49\columnwidth]{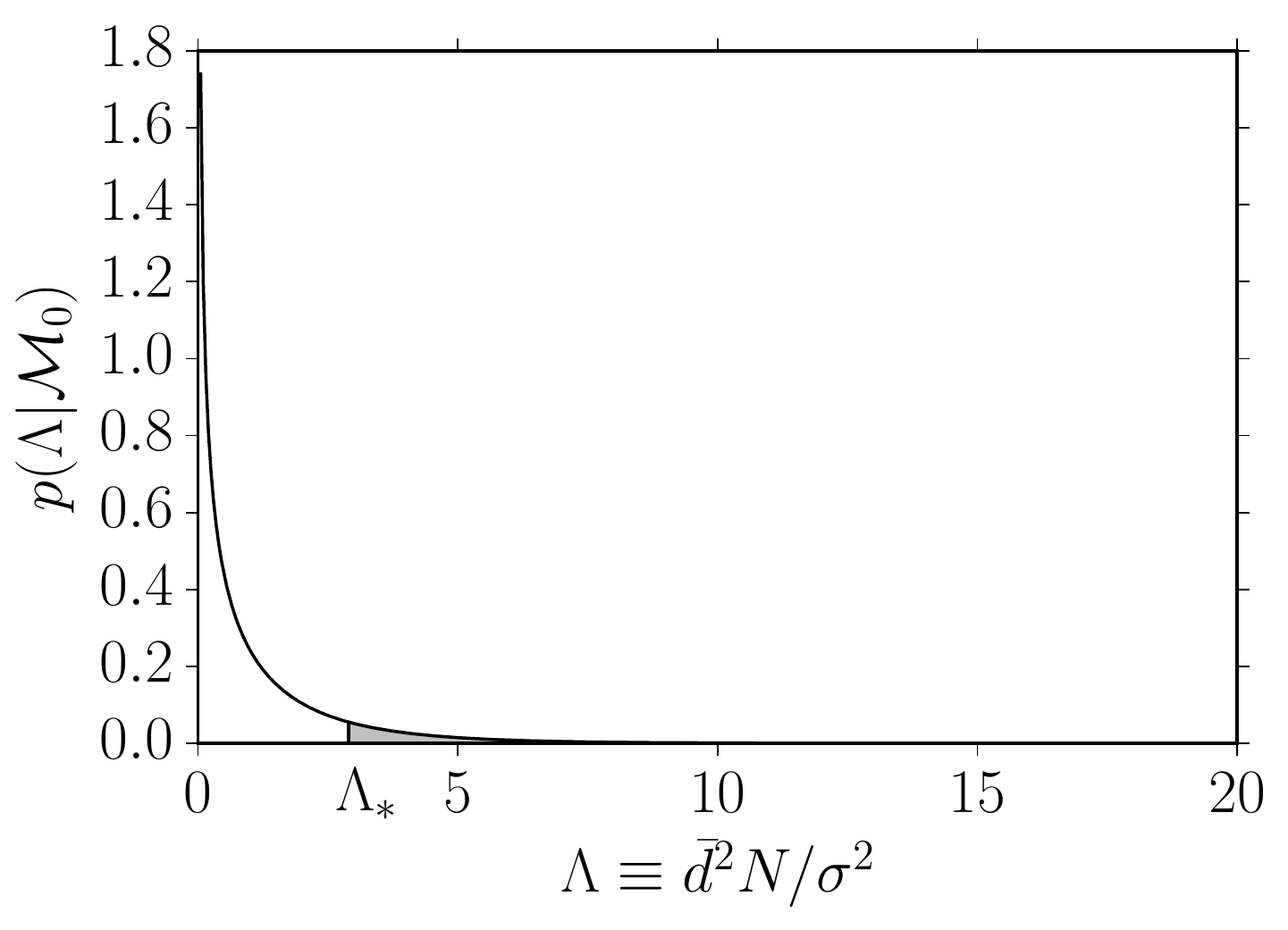}
\includegraphics[angle=0,width=.49\columnwidth]{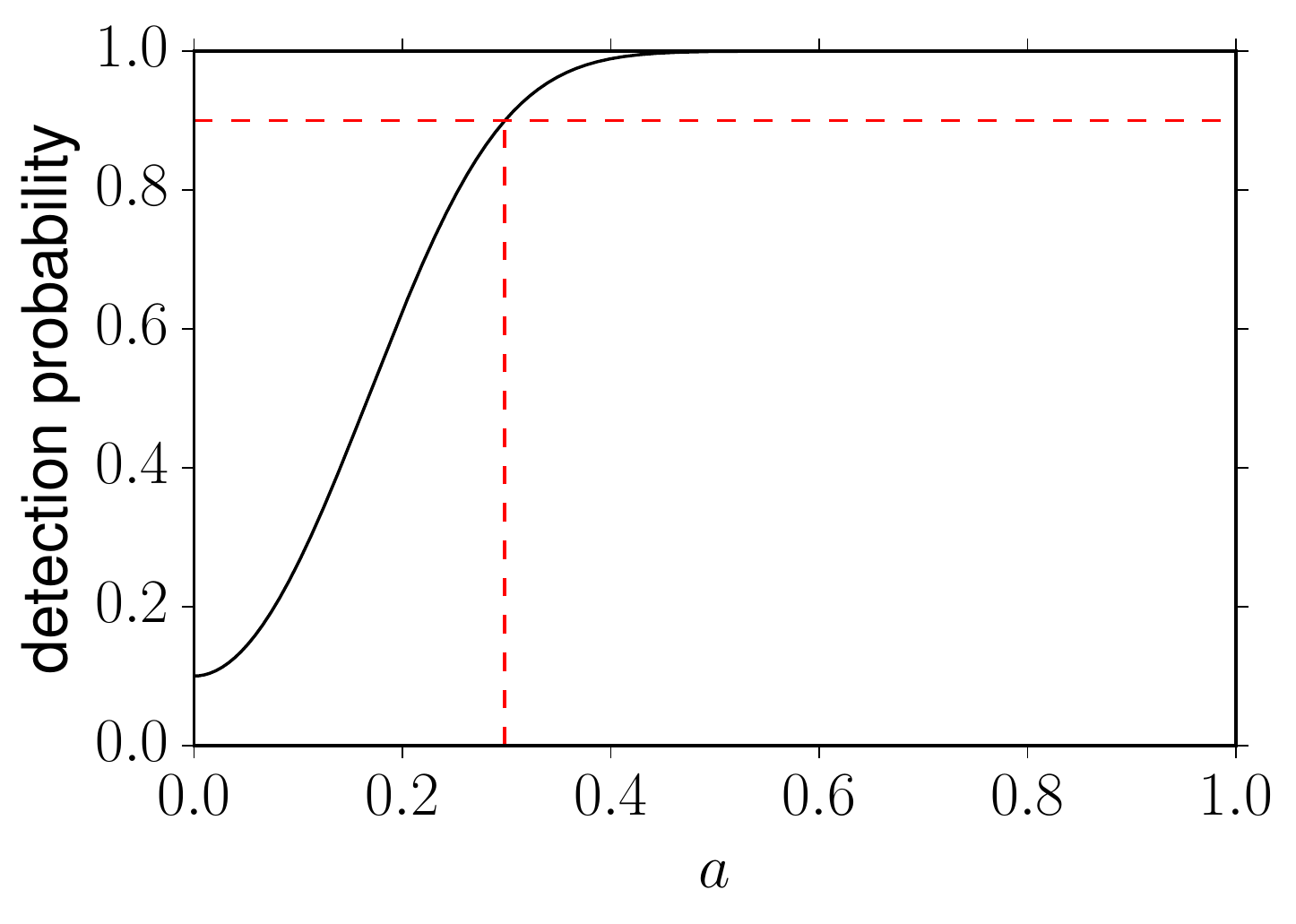}
\caption{Left panel: Probability distribution for the frequentist
detection statistic $\Lambda$ for the noise-only model.
The threshold value of the statistic for 10\% false alarm
probability is $\Lambda_*=2.9$.
Right panel: Detection probability as a function of the amplitude $a$.
The value of the amplitude needed for 90\% confidence detection
probability with $10\%$ false alarm probability is 
$a^{90\%,{\rm DP}} = 0.30$.} 
\label{f:example-DP}
\end{center}
\end{figure}

Given the values for $N$, $\sigma$, and the probability 
distributions (\ref{e:pLambda0}) and (\ref{e:pLambdaA})
for the frequentist detection 
statistic $\Lambda$, we can calculate the detection threshold 
for fixed false alarm probability $\alpha$ (which we will 
take to equal 10\%), and the corresponding 
detection probability as a function of the amplitude $a$.  
The detection threshold turns out to equal $\Lambda_* = 2.9$ 
(so 10\% of the area under the probability distribution 
$p(\Lambda|{\cal M}_0)$ is for $\Lambda\ge \Lambda_*$).
The value of the amplitude $a$ needed for 90\% confidence 
detection probability with 10\% false alarm probability is 
given by $a^{90\%,{\rm DP}}=0.30$.
(These results for the detection threshold and detection 
probability do {\em not} depend on the particular realizations 
of the simulated data.)
The corresponding curves are shown in Figure~\ref{f:example-DP}.

\begin{figure}[h!tbp]
\begin{center}
\includegraphics[angle=0,width=.49\columnwidth]{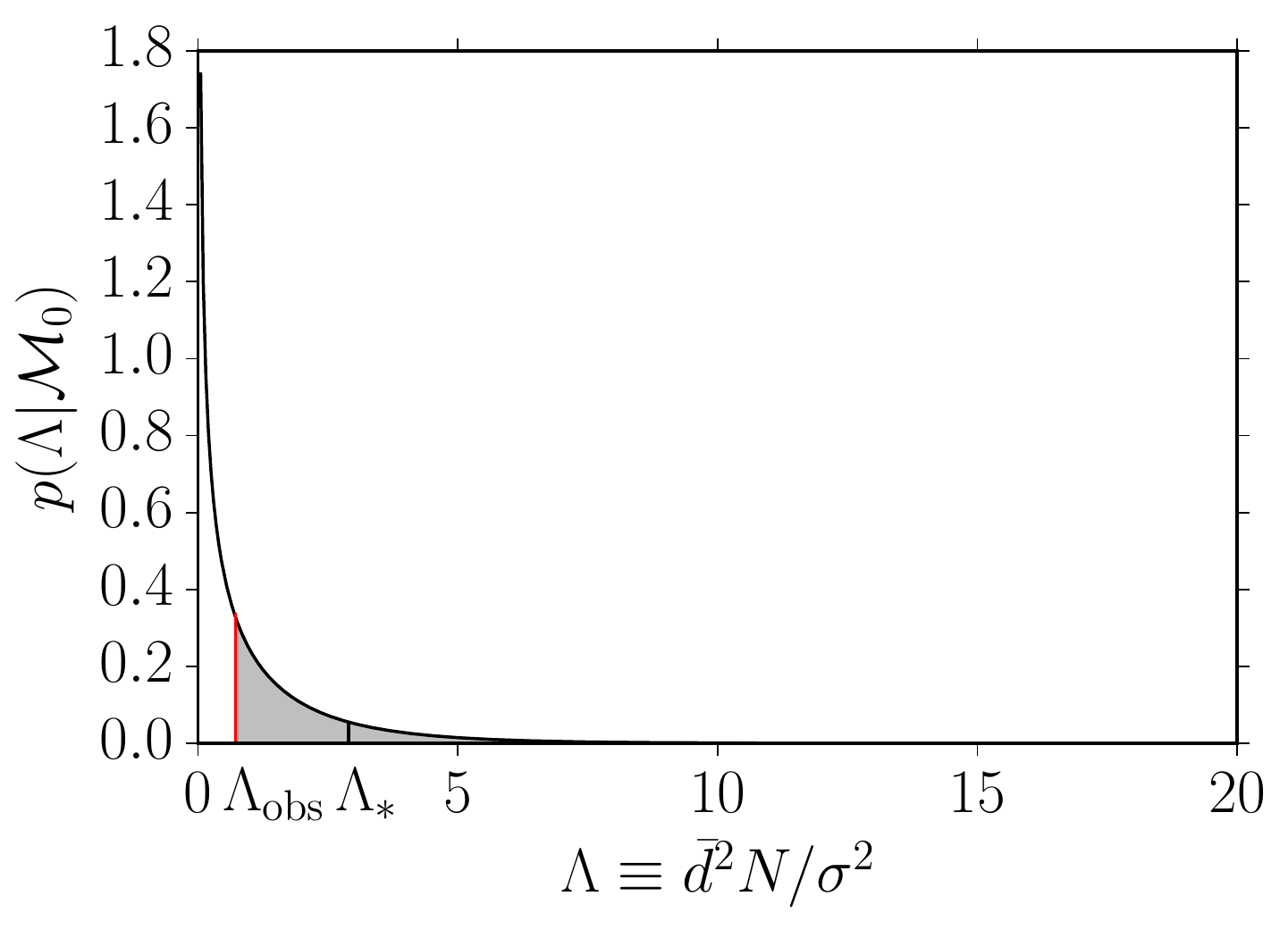}
\includegraphics[angle=0,width=.49\columnwidth]{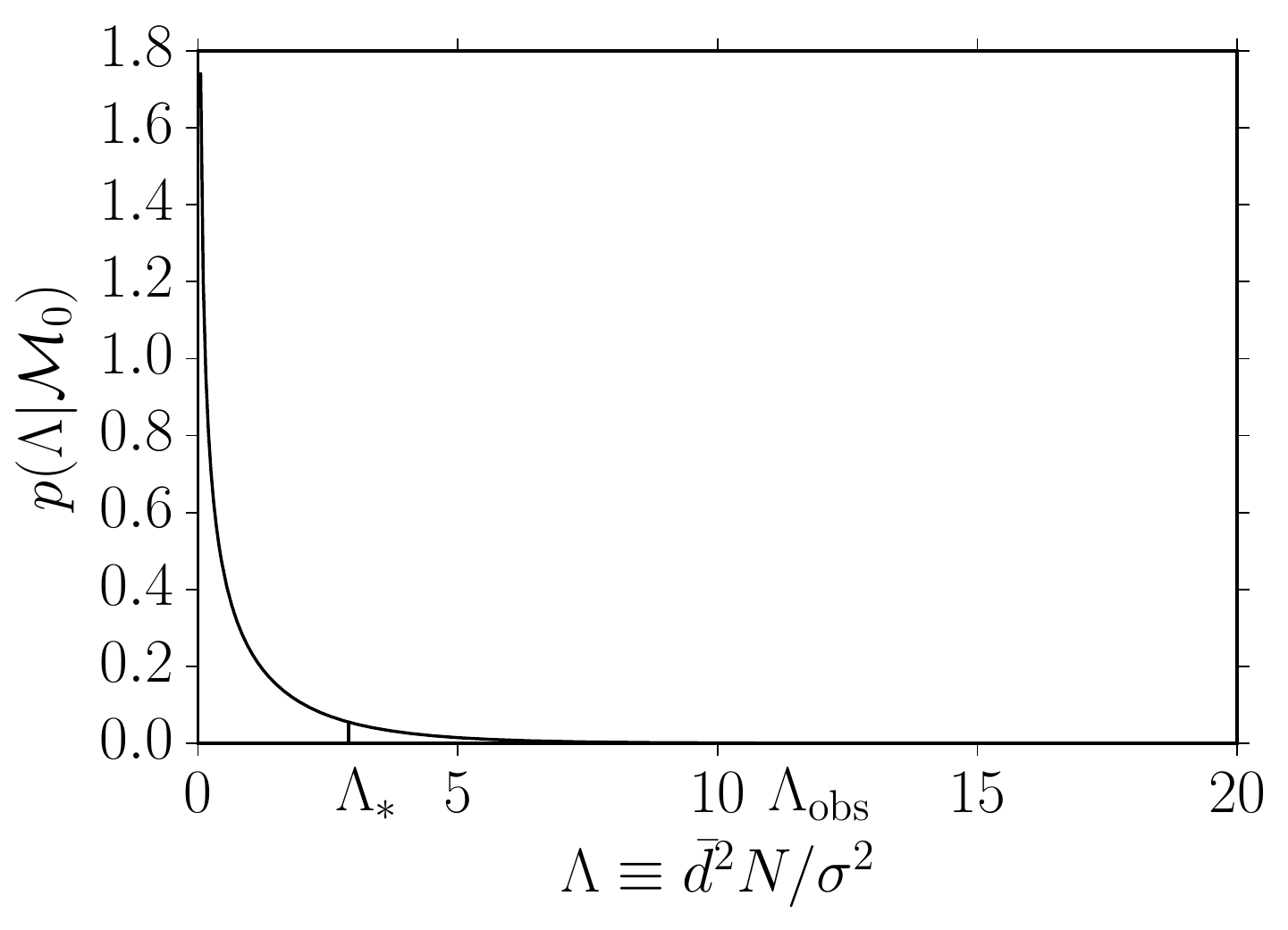}
\caption{Graphical representation of the $p$-value calculation
for the weak (left panel) and strong (right panel) injections.
For the weak injection, $\Lambda_{\rm obs}=0.72$ is marked by the red
vertical line, with corresponding $p$-value $0.45$.
For the strong injection, $\Lambda_{\rm obs}=11.2$ is sufficiently 
large that the corresponding red vertical line is not visible on 
this graph.
The $p$-value for the strong injection is $9.0\times 10^{-4}$.}
\label{f:example-pvalue}
\end{center}
\end{figure}

The sample mean of the data for the two simulations are
given by $\bar d = 0.085$ and $0.335$, respectively.
Since $\hat a = \bar d$, these are also the values of 
the maximum-likelihood estimator of the amplitude $a$.
The corresponding values of the detection statistic 
are $\Lambda_{\rm obs} = 0.72$ and $11.2$ for the two
injections, and have $p$-values equal to 0.45 and 
$9.0\times 10^{-4}$, as shown in Figure~\ref{f:example-pvalue}.
The 95\% frequentist confidence interval is given simply by
$[\hat a-2\sigma_{\hat a},\hat a+2\sigma_{\hat a}]$, 
since $\hat a$ is Gaussian-distributed, and has values 
$[-0.11,0.29]$ and $[0.14, 0.54]$, 
respectively.
These intervals contain the true value of the amplitudes
for the two injections, $a_0=0.05$ and 0.3.

\begin{figure}[h!tbp]
\begin{center}
\includegraphics[angle=0,width=.49\columnwidth]{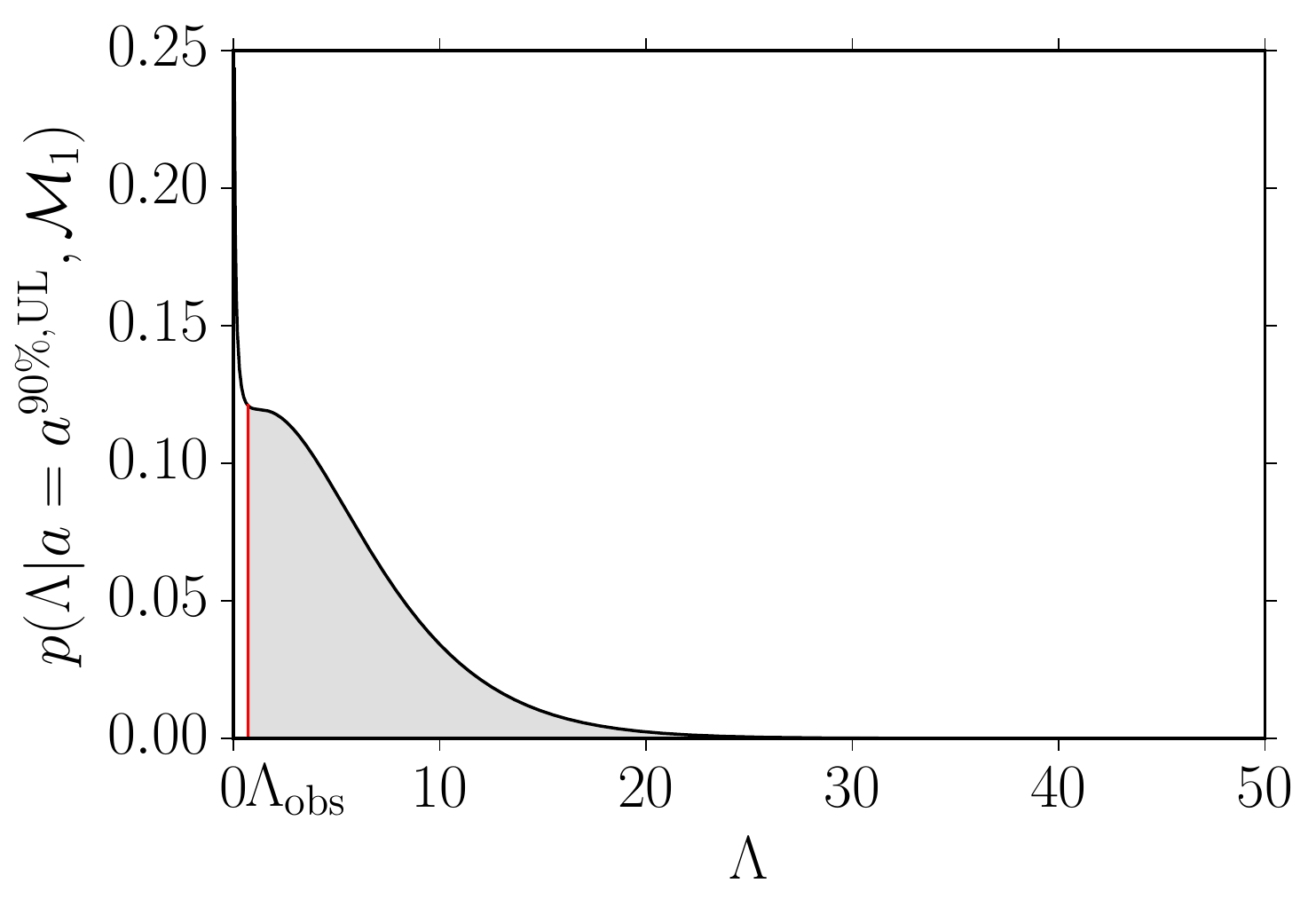}
\includegraphics[angle=0,width=.49\columnwidth]{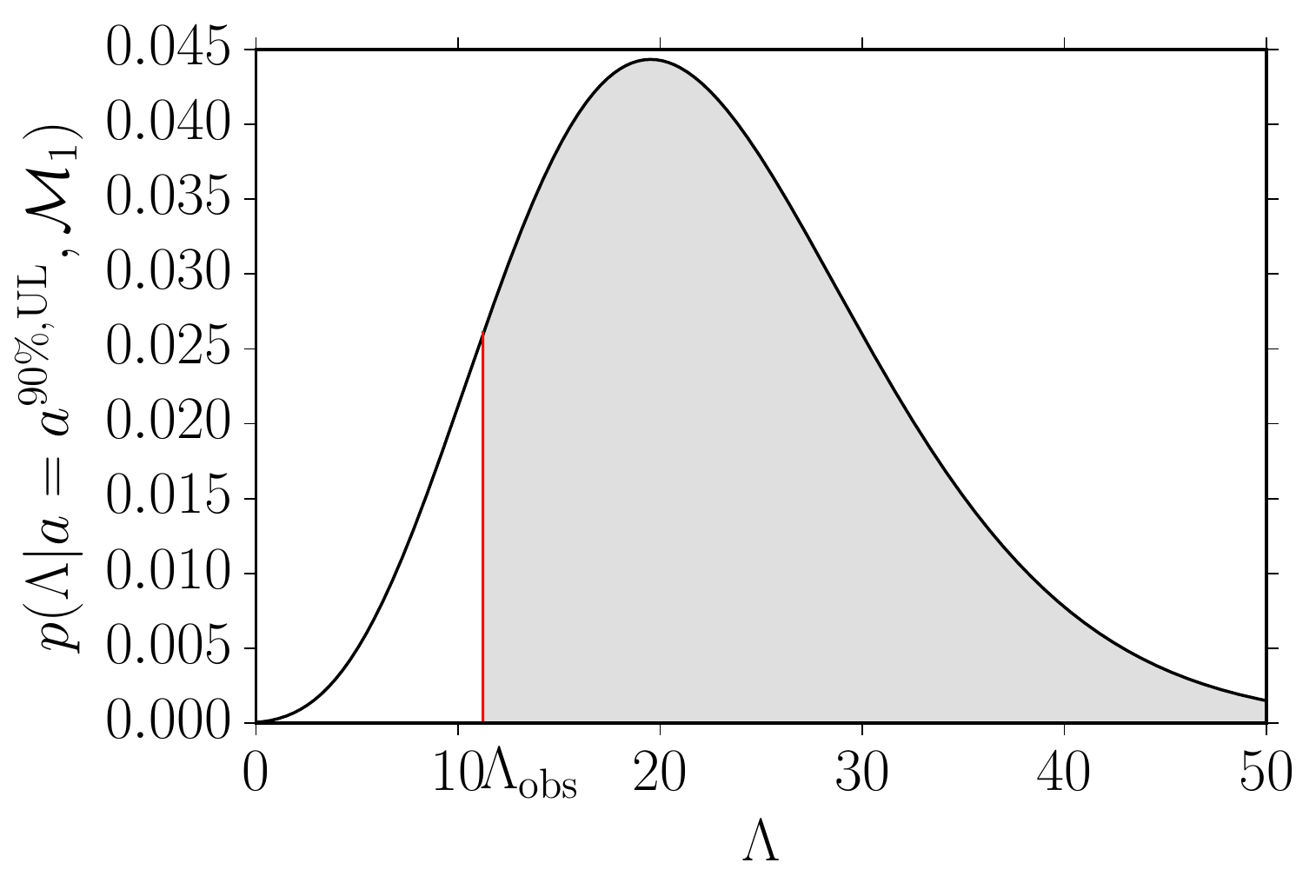}
\caption{Probability distributions for the frequentist detection statistic
$\Lambda$, conditioned on the value of the amplitude $a$ 
for which the probability of obtaining 
$\Lambda\ge \Lambda_{\rm obs}$ is equal to 0.90.
These define the 90\% confidence-level frequentist upper limits
$a^{90\%,{\rm UL}}= 0.20$ and 0.46, respectively.
The red vertical lines mark the value of $\Lambda_{\rm obs}$
for the weak (left panel, $\Lambda_{\rm obs}=0.72$) and strong 
(right panel, $\Lambda_{\rm obs}=11.2$) injections.}
\label{f:example-frequentist-UL}
\end{center}
\end{figure}

The 90\% confidence-level frequentist upper limits are 
$a^{90\%,{\rm UL}}= 0.20$ and 0.46, respectively.
Figure~\ref{f:example-frequentist-UL} shows the probability
distributions for the detection statistic $\Lambda$ conditioned
on these upper limit values for which the probability 
of obtaining $\Lambda\ge \Lambda_{\rm obs}$ is equal to 0.90.
%

\subsubsection{Bayesian analysis}
\label{s:example-bayesian}

\begin{figure}[h!tbp]
\begin{center}
\includegraphics[angle=0,width=.49\columnwidth]{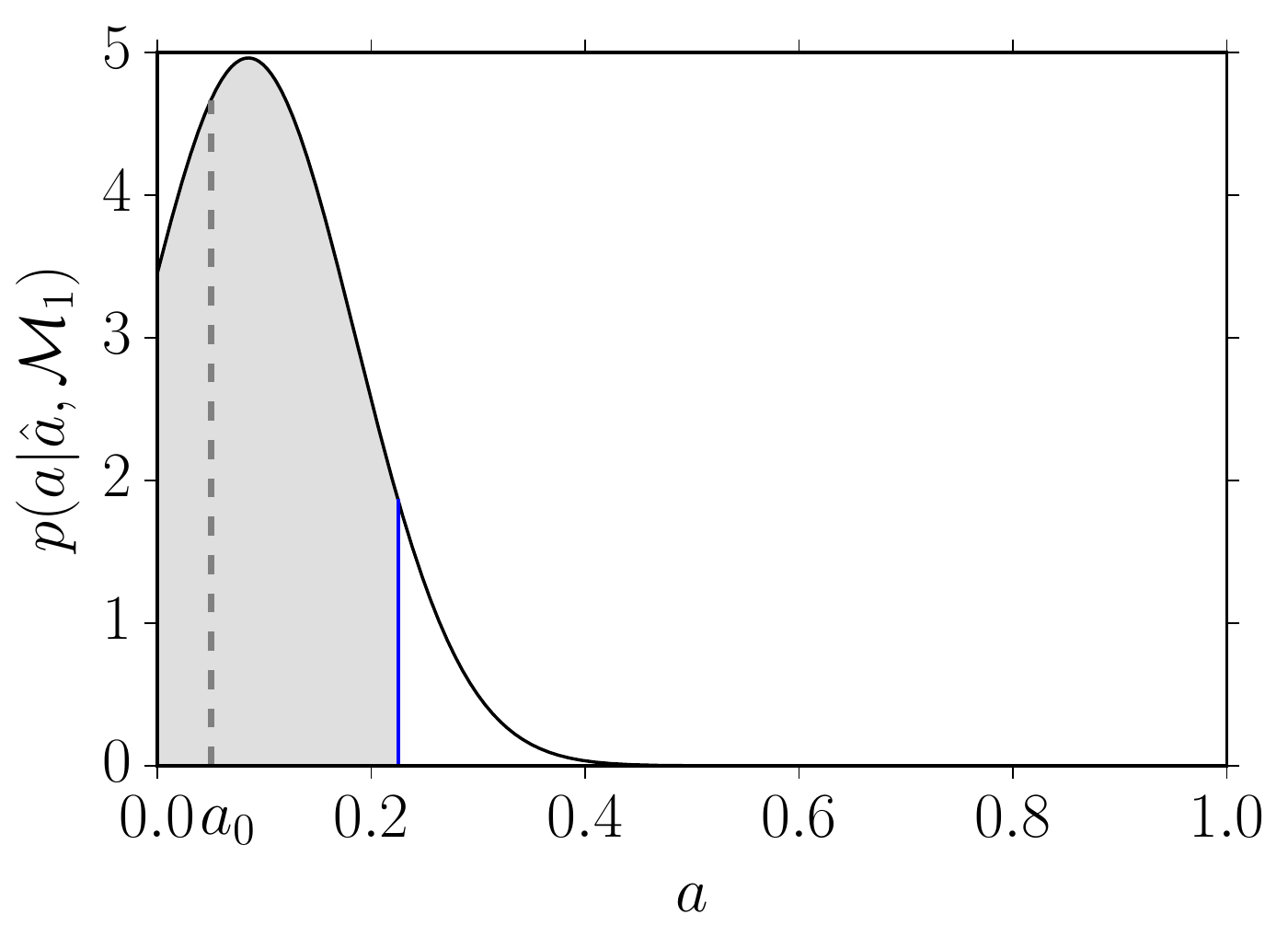}
\includegraphics[angle=0,width=.49\columnwidth]{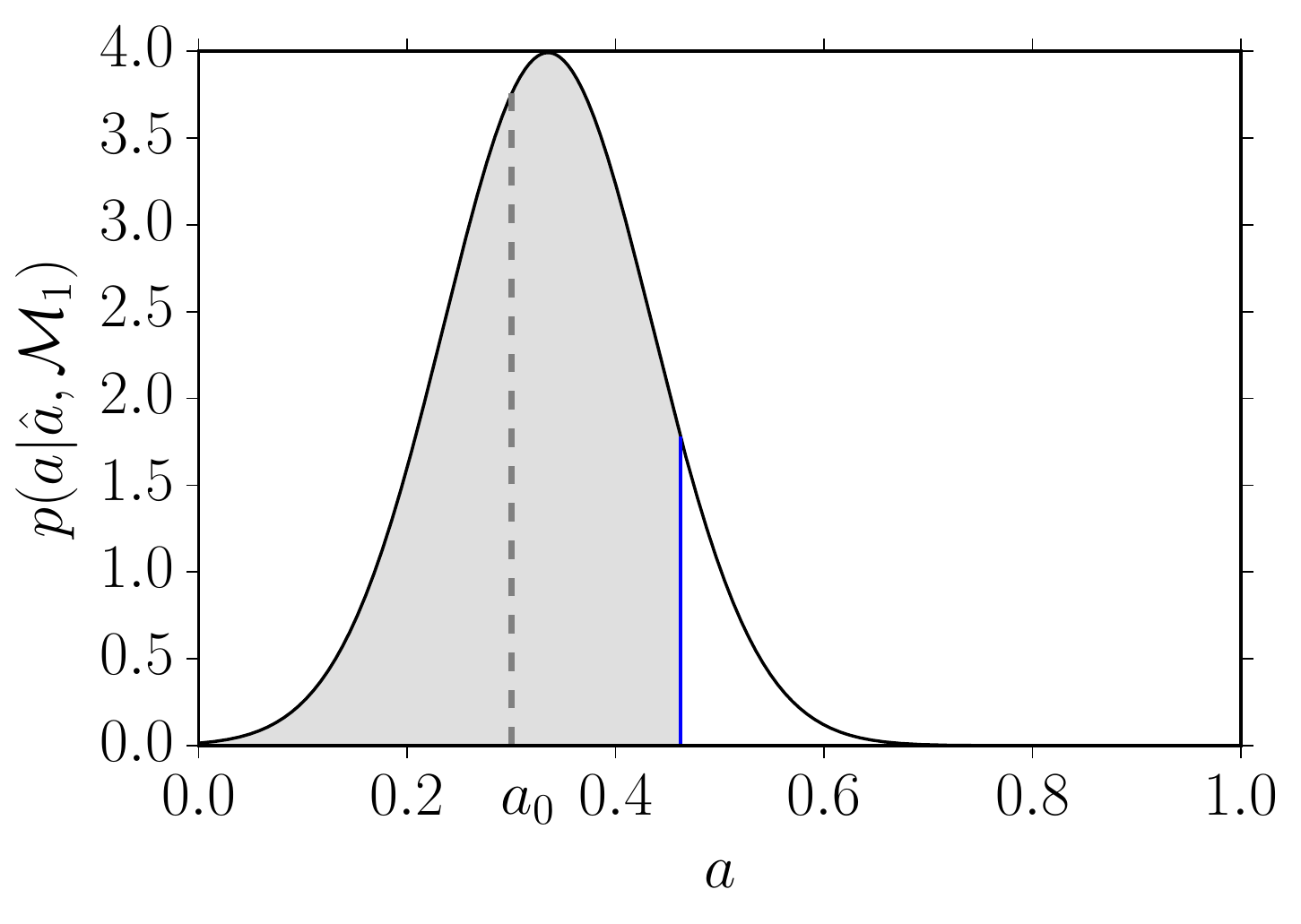}
\includegraphics[angle=0,width=.49\columnwidth]{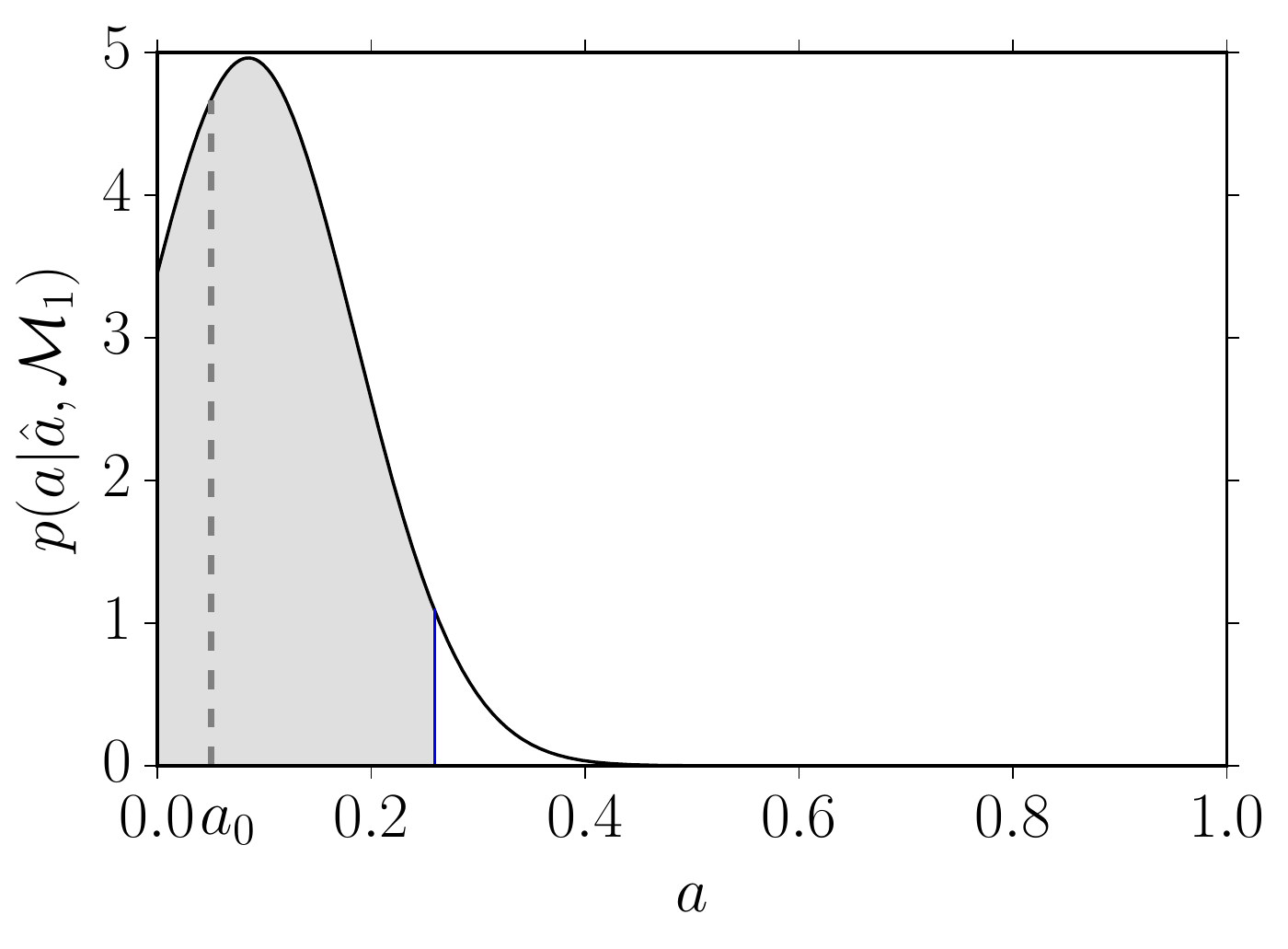}
\includegraphics[angle=0,width=.49\columnwidth]{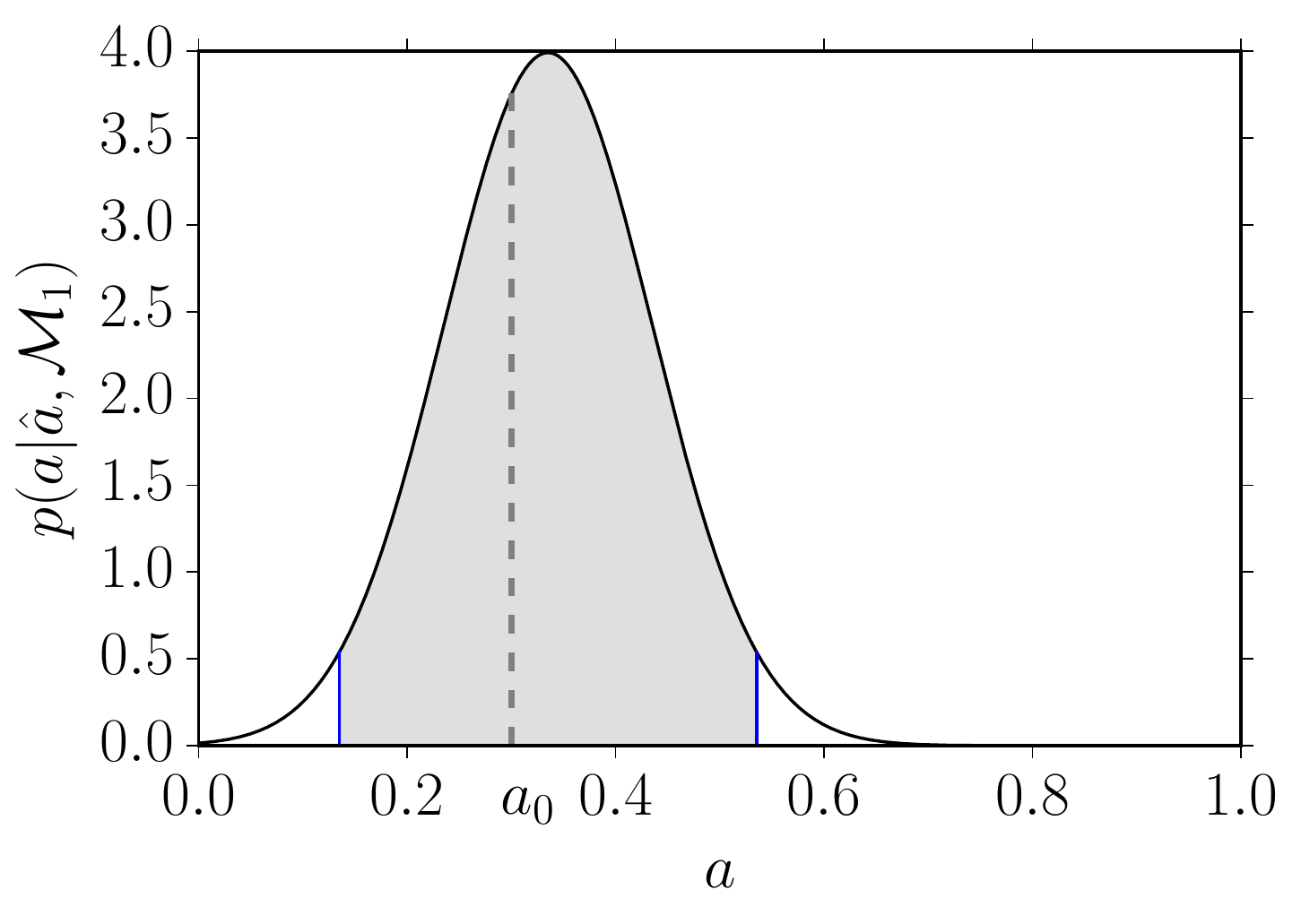}
\caption{Posterior distributions for the amplitude $a$ given the 
value of the maximum-likelihood estimator $\hat a$.
The left two panels are for the weak injection;
the right two panels are for the strong injection.
The top two plots illustrate the graphical construction of
Bayesian 90\% upper limits for the two injections;
the bottom two plots illustate the graphical construction of
the Bayesian 95\% credible intervals.
The dashed vertical lines indicate the values of the injected signal
amplitude $a_0$, which equal 0.05 and 0.3, respectively.}
\label{f:example-bayesian-posterior}
\end{center}
\end{figure}

The results of the Bayesian analysis for the two different injections
are summarized in Figure~\ref{f:example-bayesian-posterior}.
The plots show the posterior distribution for the amplitude $a$ given
the value of the maximum-likelihood estimator $\hat a$, which (as we
discussed earlier) is a sufficient statistic for the data $d$.
Recall that the posterior for $a$ for this example is 
simply a truncated Gaussian from 0 to $a_{\rm max}$ 
centered on $\hat a$, which could be negative, see~(\ref{e:pa|d}).
The left two panels show the graphical construction of the 
Bayesian 90\% upper limit and 95\% credible interval for 
the amplitude $a$ for the weak injection,
$a^{90\%,{\rm UL}}=0.23$ and $[0,0.26]$.
The right two panels show similar plots for the strong 
injection, $a^{90\%,{\rm UL}}=0.46$ and $[0.14,0.54]$.

Finally, the Bayes factor for the signal-plus-noise model ${\cal M}_1$ 
relative to the noise-only model ${\cal M}_0$ can be calculated by 
taking the ratio of the marginalized likelihood $p(d|{\cal M}_1)$
given by (\ref{e:pdM1_marg}) to $p(d|{\cal M}_0)$ given by (\ref{e:pdM0}).
Doing this, we find
$2 \ln {\cal B}_{10}=-2.2$ and $9.2$ for the weak and strong signal
injections, respectively.
The Laplace approximation to this quantity is
given by (\ref{e:bayesapprox_toymodel}), with values
$-2.0$ and $8.5$, respectively.
%
%

\subsubsection{Comparision summary}

Table~\ref{t:example-bayesfreq} summarizes the numerical 
results for the frequentist and Bayesian analyses.
We see that the frequentist and Bayesian 90\% upper limits and 95\% 
intervals numerically agree for the strong injection, 
but differ slightly for the weak injection.
The interpretation of these results is different, 
of course, for a frequentist and a Bayesian, given their
different definitions of probability.
But for a moderately strong signal in noisy data, 
we expect both approaches to yield a confident 
detection as they have for this simple example.
\begin{table}[htbp]
  \centering
  \begin{tabular}{lcccc}
    \toprule
     \multicolumn{1}{c}{} & \multicolumn{2}{c}{(weak injection, $a_0=-0.05$)} 
   & \multicolumn{2}{c}{(strong injection, $a_0=0.3$)}\\
     & Frequentist & Bayesian & Frequentist & Bayesian \\
    \midrule
    Detection threshold ($\Lambda_*$) & 2.9 & --- & 2.9 & --- \\
    Detection statistic ($\Lambda_{\rm obs}$) & 0.72 & --- & 11.2 & --- \\
    $p$-value & 0.45 & --- & $9.0\times 10^{-4}$ & --- \\
    90\% upper limit & $0.20$ & $0.23$ & $0.46$ & $0.46$ \\
    95\% interval & $[-0.11,0.29]$ & $[0, 0.26]$ & $[0.14,0.54]$ & $[0.14,0.54]$ \\
    ML estimator ($\hat a$) & $0.085$ & $0.085$ & $0.335$ & $0.335$ \\
    Bayes factor ($2\ln {\cal B}_{10}$) & --- & $-2.2$ & --- & $9.2$ \\ 
    Laplace approximation & --- & $-2.0$ & --- & $8.5$ \\ 
    \bottomrule
  \end{tabular}
  \caption{Tabular summary of the frequentist and Bayesian analysis results 
for the simulated data (both weak and strong injections).  
A dash indicates that a particular quantity 
is not relevant for either the frequentist or Bayesian analysis.}
  \label{t:example-bayesfreq}
\end{table}

\subsection{Likelihoods and priors for gravitational-wave searches}
\label{s:likelihood+prior}

To conclude this section on statistical inference, 
we discuss some issues related to calculating the likelihood 
and choosing a prior in the context of searches for
gravitational-wave signals using a network of 
gravitational-wave detectors.

\subsubsection{Calculating the likelihood}
\label{s:likelihood}

Defining the likelihood function (for either a frequentist or Bayesian
analysis) involves understanding the instrument response
and the instrument noise.  The data collected by gravitational-wave 
detectors comes in a variety of forms. For ground-based 
interferometers such as LIGO and Virgo, the data comes from the error
signal in the differential arm-length control system, which is
non-linearly related to the laser phase difference, which in turn is
linearly related to the gravitational-wave strain.  For pulsar timing
arrays, the data comes from the arrival times of radio pulses (derived
from the folded pulse profiles), which must be corrected using a
complicated timing model that takes into account the relative motion
of the telescopes and the pulsars, along with the spin-down of the
pulsars, in addition to a variety of propagation effects. The timing
residuals formed by subtracting the timing model from the raw arrival
times contain perturbations due to gravitational waves
integrated along the line of sight to the pulsar. For future space-based 
gravitational-wave detectors such as LISA, the data will be
directly read out from phase meters that perform a heterodyne
measurement of the laser phase. Synthetic combinations of these phase
read outs (chosen to cancel laser phase noise) are then linearly
proportional to the gravitational-wave strain.

Since gravitational waves can be treated as small perturbations to the
background geometry, the time delays or laser phase/frequency shifts
caused by a gravitational wave can easily be computed. These idealized
calculations have then to be related to the actual observations, either
by propagating the effects through an instrument response model, or,
alternatively, inverting the response model to convert the measured
data to something proportional to the gravitational-wave strain. 
(For example, most LIGO analyses work with the calibrated strain, rather
than the raw differential error signal.) 
If we assume that the
gravitational-wave signal and the instrument noise are linearly
independent, then the data taken at time $t$ can be written as
\begin{equation} 
d(t) = h(t) + n(t)\,,
\end{equation}
where $h(t)$ is shorthand for the gravitational-wave metric 
perturbations $h_{ab}(t,\vec x)$ 
convolved with the instrument response function and converted into
the appropriate quantity---phase shift, time delay, differential arm
length error, etc.  
(A detailed calculation of $h(t)$ and the associated detector
response functions will be given in Section~\ref{s:responsefunctions}.) 
As mentioned above, the data $d(t)$ may be the quantity that is
measured directly, or, more commonly, some quantity that is derived
from the measurements such as timing residuals or calibrated
strain. In any analysis, it is important to marginalize over the model
parameters used to make the conversion from the raw data. 

The likelihood of observing $d(t)$ is found by demanding that the residual
\be
r(t) \equiv d(t) - \bar{h}(t)\,,
\ee
be consistent with a draw from the noise distribution $p_n(x)$:
\begin{equation} 
 p( d(t) \vert \bar{h}(t)) = p_n(r(t))=p_n(d(t) - \bar{h}(t) )\,.
\end{equation}
Here $\bar h(t)$ is our model%
\footnote{Since the model $\bar h(t)$ will differ from the actual $h(t)$,
we use an overbar for the model to distinguish the two.}
for the gravitational-wave signal.
The likelihood of observing a collection of discretely-sampled data
$d \equiv \{ d_1, d_2, \cdots , d_N\}$, where 
$d_i\equiv d(t_i)$, is then given by $p( d\vert \bar{h}) = p_n(r)$,
where $r\equiv\{r_1, r_2, \cdots, r_N\}$ with $r_i\equiv r(t_i)$.
Since instrument noise is due to a large number of small disturbances
combined with counting noise in the large-number limit, the 
central limit theorem suggests that the noise distribution can be approximated
by a multi-variate normal (Gaussian) distribution:
\begin{equation} \label{gauss_likelihood}
 p( d\vert \bar{h}) =  
\frac{1}{\sqrt{{\rm det} (2\pi C_n)}}\,  
e^{-\frac{1}{2}\sum_{i,j} r_i \left(C_n^{-1}\right)_{ij} r_j} \, ,
\end{equation}
where $C_n$ is the
noise correlation matrix, with components
\be
(C_n)_{ij} = \langle n_i n_j\rangle -\langle n_i\rangle\langle n_j\rangle\,.
\ee
If the noise is stationary, then the
correlation matrix only depends on the lag $\vert t_i - t_j\vert$, and
the matrix $C_n$ can be (approximately)
diagonalized by transforming to the Fourier domain, where $r_i$ should 
then be interpreted as $\tilde{r}(f_i)$
(see Appendix~\ref{s:discrete-continuous-probability} for a more 
careful treatment of discrete probability distributions in the 
time and frequency domain).
In practice, the noise observed in
most gravitational-wave experiments is neither stationary nor
Gaussian (Section~\ref{s:complications} and Appendix~\ref{s:real}), 
but (\ref{gauss_likelihood}) still serves as a good starting
point for more sophisticated treatments.  The Gaussian likelihood
(\ref{gauss_likelihood}) immediately generalizes for a network of
detectors:
\begin{equation} 
\label{net_likelihood}
 p( d\vert \bar{h}) =  \frac{1}{\sqrt{{\rm det} (2\pi C_n)}}
e^{-\frac{1}{2} \sum_{Ii,Jj}r_{Ii} \left(C_n^{-1}\right)_{Ii,Jj} r_{Jj}} \, ,
\end{equation}
where $I$, $J$ labels the detector, and $i$, $j$
labels the discrete time or frequency sample for the 
corresponding detector. 
Note here that the parameters $\vec{\theta}$ appearing in 
(\ref{full_bayes}) are the
individual time or frequency samples $\bar{h}_i$.

\subsubsection{Choosing a prior}
\label{s:prior}

For Bayesian inference, it is also necessary to define a model ${\cal M}$ 
for the gravitational-wave signal, which is done by placing
a prior $p(\bar{h} \vert {\cal M})$ on the samples $\bar{h}_i$. In
some cases, a great deal is known about the signal model, such as when
approximate solutions to Einstein's equations provide waveform
templates.  In that case the prior can be written as
\begin{equation}
p(\bar{h} \vert {\cal M}) = 
\delta(\bar{h}  - \bar{h}(\vec{\theta}, {\cal M})) \, p(\vec{\theta} \vert {\cal M})  \, .
\end{equation}
Marginalizing over $\bar{h}$ converts the posterior $p(\bar{h}
\vert d)$ to a posterior distribution for the signal parameters
$p(\vec{\theta}\vert d, {\cal M})$. In other cases, such as for 
short-duration bursts associated with certain violent astrophysical events,
much less is known about the possible signals and weaker priors have
to be used. Models using wavelets, which have finite time-frequency
support, and priors that favor connected concentrations of power in
the time-frequency plane are commonly used for these ``unmodeled burst''
searches. At the other end of the spectrum from deterministic point
sources are the statistically-isotropic stochastic backgrounds that
are thought to be generated by various processes in the early
Universe, or through the superposition of a vast number of weak
astrophysical sources. In the case of Gaussian stochastic signals, 
the prior for a signal 
$\bar h=(\bar h{}_+(\hat n),\bar h{}_\times(\hat n))$
coming from direction $\hat{n}$ direction has the form
\begin{equation}
p(\bar{h}\vert {\cal M}) =   \frac{1}{2\pi S_h}  
e^{-(\bar h{}^2_+(\hat{n})+\bar h{}^2_\times(\hat{n}))/2 S_h}\,,
\end{equation}
where $S_h$ is the power spectrum of the background.
As we shall show in Section~\ref{s:corr}, marginalizing over 
$\bar{h}$ converts the posterior $p(\bar{h} \vert d)$ to a posterior 
$p(S_h \vert d, {\cal M})$ for $S_h$.

\section{Correlations}
\label{s:corr}

\begin{quotation}
Correlation is not cause, it is just a `music of chance'.
{\em Siri Hustvedt}
\end{quotation}

\noindent
Stochastic gravitational waves are indistinguishable from 
unidentified instrumetal noise in a single detector,
but are correlated between pairs of detectors in ways 
that differ, in general, from instrumental noise.
Cross-correlation methods basically use the random output 
of one detector as a template for the other, taking into
account the physical separation and relative orientation 
of the two detectors.
In this section, we introduce cross-correlation
methods in the context of both frequentist and
Bayesian inference, analyzing in detail a simple toy problem
(the data are ``white" and we ignore complications that 
come from the separation and relative orientation of the 
detectors---this we discuss in detail in Section~\ref{s:geom}).
We also briefly discuss possible alternatives to 
cross-correlation methods, e.g., using a null channel 
as a noise calibrator.

The basic idea of using cross-correlation to search for 
stochastic gravitational-waves can be found in several early 
papers~\cite{Grishchuk:1976, Hellings-Downs:1983, Michelson:1987, Christensen:PhD, 
Christensen:1992, Flanagan:1993}.
The derivation of the likelihood function in 
Section~\ref{s:unify} follows that of~\cite{Cornish-Romano:2013}; parts of 
Section~\ref{s:ML_statistic_derivation} are also discussed 
in~\cite{Allen:2002jw, Drasco-Flanagan:2003}.

\subsection{Basic idea}
\label{s:basic_idea}

The key property that allows one to distinguish a 
a stochastic gravitational-wave background from 
instrumental noise is that the
gravitational-wave signal is correlated across 
multiple detectors while instrumental noise typically 
is not.
To see this, consider the simplest possible example,
i.e., a single sample of data from two colocated and 
coaligned detectors:
\be
\begin{aligned}
d_1 &= h + n_1\,,
\\
d_2 &= h + n_2\,.
\label{e:data-singlesample}
\end{aligned}
\ee
Here $h$ denotes the common gravitational-wave 
signal and $n_1$, $n_2$ the noise in the two detectors.
To cross correlate the data, we simply form the
product of the two samples, 
$\hat C_{12}\equiv d_1 d_2$.
The expected value of the correlation is then
\be
\langle \hat C_{12}\rangle
=\langle d_1 d_2\rangle
=
\langle h^2\rangle+
\langle n_1 n_2\rangle+
\cancelto{0}{\langle h n_2\rangle}+
\cancelto{0}{\langle n_1 h\rangle}
=
\langle h^2\rangle+
\langle n_1 n_2\rangle\,,
\ee
since the gravitational-wave signal and the instrumental 
noise are uncorrelated.
If the instrumental noise in the two detectors are also
uncorrelated, then 
\be
\langle n_1 n_2\rangle=0\,,
\ee
which implies
\be
\langle \hat C_{12}\rangle
= \langle h^2\rangle
\equiv S_h\,.
\ee
This is just the variance (or power) of the stochastic
gravitational-wave signal.
So by cross-correlating data in two (or more) detectors, 
we can extract the common gravitational-wave component.

We have assumed here that there is no cross-correlated
noise (instrumental or environmental).
If there is correlated noise, then the simple 
procedure describe above needs to be augmented.
This will be discussed in more detail in Section~\ref{s:correlatednoise}.

\subsection{Relating correlations and likelihoods}
\label{s:unify}

The cross-correlation approach arises naturally from a standard 
likelihood analysis if we adopt a Gaussian stochastic template for the signal. 
Revisiting the example from the previous section, let's assume
that the detector noise is Gaussian-distributed with variances
$S_{n_1}$ and $S_{n_2}$.
Then the likelihood function for the data $d\equiv (d_1,d_2)$ for the 
noise-only model ${\cal M}_0$ is simply
\be
p(d | S_{{n_1}}, S_{{n_2}}, {\cal M}_0) 
= \frac{1}{2\pi \sqrt{S_{{n_1}} S_{{n_2}}}} 
\, \exp\left[-\frac{1}{2}\left(
\frac{d_1^2}{S_{n_1}}+\frac{d_2^2}{S_{n_2}}
\right)\right]\,. 
\label{e:nlike}
\ee
For the signal-plus-noise model ${\cal M}_1$, we have
\be
p(d | S_{{n_1}}, S_{{n_2}}, \bar{h}, {\cal M}_1) 
= \frac{1}{2\pi \sqrt{S_{{n_1}} S_{{n_2}}}} 
\, \exp\left[-\frac{1}{2}\left\{
\frac{(d_1-\bar h)^2}{S_{n_1}}+\frac{(d_2-\bar h)^2}{S_{n_2}}
\right\}\right]\,,
\label{e:slike0}
\ee
where the 
gravitational-wave signal $\bar h$ is assumed to be a Gaussian 
random deviate with probability distribution
\be
p(\bar h|S_h, {\cal M}_1) 
= \frac{1}{\sqrt{2 \pi S_h}}\,  
\exp\left[-\frac{1}{2}\frac{\bar h^2}{S_h}\right] \, .
\ee
In most applications we are not interested in the value of $\bar h$, 
but rather the power $S_h$. 
Marginalizing over $\bar h$, the likelihood takes the form
\be
p(d | S_{n_1}, S_{n_2}, S_h,{\cal M}_1) = \frac{1}{\sqrt{{\rm det}(2\pi C)}} 
e^{-\frac{1}{2} \sum_{I,J=1}^2 d_I \left(C^{-1}\right)_{IJ} d_J}\,,
\label{e:slike}
\ee
where
\be
C = \left[
\begin{array}{cc}
S_{n_1} +S_h & S_h
\\
S_h & S_{n_2} +S_h
\\
\end{array}
\right]\,.
\ee
Maximizing the likelihood with respect to $S_h$, $S_{n_1}$ and $S_{n_2}$ yields the 
maximum-likelihood estimators
\be
\begin{aligned}
&\hat S_h  = d_1 d_2 = \hat C_{12}\,,
\\
&\hat S_{n_1}  =  d_{1}^2 - d_{1} d_{2}\,,
\\
&\hat S_{n_2}  =  d_{2}^2 - d_{1} d_{2}\,.
\end{aligned}
\label{e:freq_estimators1d}
\ee
Thus, the cross-correlation statistic $\hat C_{12}$ is the
maximum-likelihood
estimator for a Gaussian stochastic gravitational wave template with zero mean and variance $S_h$.

\subsection{Extension to multiple data samples}

The extension to multiple data samples 
\be
\begin{array}{ll}
d_{1i} = h_i + n_{1i}\,, & i=1,2,\cdots, N\,,
\\
d_{2i} = h_i + n_{2i}\,,\ &  i=1,2,\cdots, N\,,
\end{array}
\label{e:data-Nsamples}
\ee
is fairly straightforward.
In the following two subsections, we consider
the cases where the detector 
noise and stochastic signal are either:
(i) both {\em white} (i.e., the data are 
uncorrelated between time samples) or 
(ii) both {\em colored}
(i.e., allowing for correlations in time).
The white noise example will be analyzed in 
more detail in Sections~\ref{s:ML_statistic_derivation}--\ref{s:freq-bayes-corr}.

\subsubsection{White noise and signal}
\label{s:white}

If the detector noise and stochastic signal are both white, 
then the likelihood functions for the data $d\equiv \{d_{1i};d_{2i}\}$,
are simply {\em products} of the likelihoods 
(\ref{e:nlike}) and (\ref{e:slike})
for the individual data samples.
We can write these product likelihoods as single multivariate
Gaussian distributions:
\be
p(d|S_{n_1},S_{n_2},{\cal M}_0) 
=\frac{1}{\sqrt{\det(2\pi C_n)}}\, e^{-\frac{1}{2} d^T C_n^{-1} d}\,,
\label{e:nlikelihood}
\ee
\be
p(d|S_{n_1},S_{n_2},S_h,{\cal M}_1) 
=\frac{1}{\sqrt{\det(2\pi C)}}\, e^{-\frac{1}{2} d^T C^{-1} d}\,,
\label{e:likelihood_marginalized}
\ee
where
\be
C_n 
= \left[
\begin{array}{cc}
S_{n_1}\,\unit_{N\times N} & \zero_{N\times N} 
\\
\zero_{N\times N} & S_{n_2}\,\unit_{N\times N}
\\
\end{array}
\right]\,,
\ee
\be
C 
= \left[
\begin{array}{cc}
(S_{n_1} +S_h)\,\unit_{N\times N} & S_h\,\unit_{N\times N} 
\\
S_h\,\unit_{N\times N} & (S_{n_2} +S_h)\,\unit_{N\times N}
\\
\end{array}
\right]\,.
\label{e:C_marginalized}
\ee
The arguments in the exponential have the form
\be
d^T C_n^{-1} d
= \sum_{I,J=1}^2\sum_{i,j=1}^N
d_{Ii} \left(C^{-1}_n\right)_{Ii,Jj} d_{Jj}\,,
\ee
and similarly for $d^T C^{-1} d$.
The maximum-likelihood estimators for this case are:
\be
\begin{aligned}
&\hat S_h \equiv \frac{1}{N}\sum_{i=1}^N d_{1i} d_{2i}\,,
\\
&\hat S_{n_1} \equiv \frac{1}{N}\sum_{i=1}^N d_{1i}^2 - \frac{1}{N}\sum_{i=1}^N d_{1i} d_{2i}\,,
\\
&\hat S_{n_2} \equiv \frac{1}{N}\sum_{i=1}^N d_{2i}^2 - \frac{1}{N}\sum_{i=1}^N d_{1i} d_{2i}\,.
\end{aligned}
\label{e:freq_estimators}
\ee
Note that these are just averages of the single-datum 
estimators (\ref{e:freq_estimators1d}) 
over the $N$ independent data samples.

A couple of remarks are in order:
(i) It is easy to show that the expectation values of the
estimators are the true values of the parameters
$S_h$, $S_{n_1}$, $S_{n_2}$.
It is also fairly straightforward to calculate the variances
of the estimators.
In particular, 
\be
{\rm Var}(\hat S_h)
\equiv \langle \hat S_h^2\rangle-\langle\hat S_h\rangle^2 
=\frac{1}{N}\left[
S_{n_1}S_{n_2} + S_h(S_{n_1} + S_{n_2}) + 2 S_h^2
\right]\,.
\ee
Note that this expression reduces to 
${\rm Var}(\hat S_h)\approx S_{n_1}S_{n_2}/N$
in the weak-signal limit, $S_h\ll S_{n_I}$, for $I=1,2$.
(ii) If we simply maximized the likelihood with respect
to variations of $S_h$, treating the noise variances
$S_{n_1}$ and $S_{n_2}$ as {\em known} parameters, then 
the frequentist estimator of $S_h$ would also include
{\em auto-correlation} terms for each detector:
\begin{multline}
\hat S_h = \frac{1}{(S_{n_1}+S_{n_2})^2}
\left[
2S_{n_1}S_{n_2} \frac{1}{N}\sum_{i=1}^N d_{1i} d_{2i} 
\right.
\\
\left.
+ S_{n_2}\left(\frac{1}{N}\sum_{i=1}^N d_{1i}^2 - S_{n_1}\right)
+ S_{n_1}\left(\frac{1}{N}\sum_{i=1}^N d_{2i}^2 - S_{n_2}\right)
\right]\,.
\end{multline}
In practice, however, the noise variances are not known well enough
to be able to extract useful information from the auto-correlation terms;
they actually worsen the performance of the simple cross-correlation 
estimator when the uncertainty in $S_{n_1}$ or $S_{n_2}$ is greater 
than or equal to $S_h$.

\subsubsection{Colored noise and signal}
\label{s:colored}

For the case where the detector noise and stochastic signal 
are colored, it simplest to work in the frequency domain,
since the Fourier components are {\em independent} of one 
another.
(This assumes that the data are {\em stationary}, so that 
there is no preferred origin of time.)
Assuming multivariate Gaussian distributions as before,
the variances $S_{n_1}$, $S_{n_2}$, and $S_h$ generalize
to {\em power spectral densitites}, which are functions
of frequency defined by
\be
\langle \tilde n_I(f)\tilde n^*_I(f')\rangle
= \frac{1}{2}\delta(f-f')\, S_{n_I}(f)\,,
\qquad
\langle \tilde h(f)\tilde h^*(f')\rangle
= \frac{1}{2}\delta(f-f')\, S_h(f)\,,
\label{e:psd-1sided}
\ee
where $I=1,2$ and tilde denotes Fourier transform.%
\footnote{Our convention for Fourier transform is
$\tilde h(f)= \int_{-\infty}^\infty dt \>e^{-i2\pi f t} h(t)$.}
The factor of $1/2$ in (\ref{e:psd-1sided})
is for {\em one-sided} power spectra, for which
the integral of the power spectrum over 
{\em positive} frequencies equals the variance of the data:
\be
{\rm Var}[h] 
=\int_0^\infty df\> S_h(f)\,.
\ee
This is just the continuous version of {\em Parseval's theorem},
see e.g., (\ref{e:parseval-continuous}).
For $N$ samples of discretely-sampled data from each 
of two detectors $I=1,2$ (total duration $T$), 
the likelihood function for a Gaussian stochastic 
signal template 
becomes \cite{Allen-et-al:2002, Cornish-Romano:2013}:
\be
p(d|S_{n_1}, S_{n_2}, S_h, {\cal M}_1)
= \prod_{k=0}^{N/2-1} \frac{1}{{\rm det}(2\pi \tilde C(f_k))}
e^{-\frac{1}{2}\sum_{I,J} \tilde d_I^*(f_k) \left(\tilde C(f_k)^{-1}\right)_{IJ}\tilde d_J(f_k)}\,,
\label{e:likelihood-colored}
\ee
where
\be
\tilde C(f) 
= \frac{T}{4}\left[
\begin{array}{cc}
S_{n_1}(f) +S_h(f) & S_h(f)
\\
S_h(f) & S_{n_2}(f) +S_h(f)
\\
\end{array}
\right]\,.
\label{e:Ctilde}
\ee
Here $k=0,1, \cdots, N/2-1$ labels the discrete positive frequencies.
There is no square root of the determinant in the 
denominator of (\ref{e:likelihood-colored})
since the volume element for the probability density involves both the
real and imaginary parts of the Fourier transformed data
(Appendix~\ref{s:discrete-continuous-probability}).

We do not bother to write down the 
maximum-likelihood estimators of the signal and noise 
power spectral densities for this particular example.
We will return to this problem in 
Section~\ref{s:optimal_filtering},
where we discuss the {\em optimally-filtered} 
cross-correlation statistic for isotropic stochastic 
backgrounds.
There one assumes a particular spectral {\em shape}
for the gravitational-wave power spectral density,
and then simply estimates its overall amplitude.
That simplifies the analysis considerably.

\subsection{Maximum-likelihood detection statistic}
\label{s:ML_statistic_derivation}

Let's return to the example discussed in
Section~\ref{s:white}, which consists of $N$ samples 
of data in each of two detectors, having uncorrelated
white noise and a common white stochastic signal.
As described in Section~\ref{s:relating-freq-bayes},
one can calculate a frequentist detection statistic 
based on 
the {\em maximum-likelihood ratio}:
\be
\Lambda_{\rm ML}(d)
\equiv \frac
{\max_{S_{n_1}, S_{n_2}, S_h} p(d\vert S_{n_1}, S_{n_2}, S_h, {\cal M}_1)}
{\max_{S_{n_1}, S_{n_2}} p(d\vert S_{n_1}, S_{n_2}, {\cal M}_0)}\,.
\ee
Substituting
(\ref{e:nlikelihood}) and (\ref{e:likelihood_marginalized}) for 
the likelihood functions and performing the maximizations yields
\be
\Lambda_{\rm ML}(d) =
\left[1-\frac{\hat S_h^2}{\hat S_1\hat S_2}\right]^{-N/2}\,,
\ee
where
\be
\begin{aligned}
\hat S_1 &\equiv \frac{1}{N}\sum_{i=1}^N d_{1i}^2=\hat S_{n_1} + \hat S_h\,,
\qquad
\hat S_2 &\equiv \frac{1}{N}\sum_{i=1}^N d_{2i}^2=\hat S_{n_2} + \hat S_h\,.
\end{aligned}
\ee
Note that the these estimators involve only 
{\em autocorrelations} of the data.
In the absence of a signal, 
they are maximum-likelihood estimators of the noise variances 
$S_{n_1}$ and $S_{n_2}$.
But in the presence of a signal, 
they are maximum-likelihood estimators of the {\em combined} 
variances $S_1\equiv S_{n_1}+S_h$ and
$S_2\equiv S_{n_2}+S_h$.

Recall that for comparison with Bayesian model selection
calculations, it is convenient to define the frequentist 
statistic $\Lambda(d)$ as twice the logarithm of the 
maximum-likelihood ratio:
\be
\Lambda(d) 
\equiv 2\ln\left(\Lambda_{\rm ML}(d)\right)
= -N \ln\left[1-\frac{\hat S_h^2}{\hat S_1\hat S_2}\right]\,.
\label{e:Lambda_corr_example}
\ee
In the limit that the stochastic gravitational-wave
signal is {\em weak} compared to the detector noise---i.e., 
$S_h\ll S_{n_I}$, for $I=1,2$---the above expression
reduces to
\be
\Lambda(d) 
\simeq \frac{\hat S_h^2}{\hat S_1\hat S_2/N} 
\simeq \frac{\hat S_h^2}{\hat S_{n_1}\hat S_{n_2}/N}\,.
\label{e:Lambda_weak}
\ee
This is just the squared signal-to-noise ratio of the 
cross-correlation statistic.
Note also that $\hat S_h^2/\hat S_1\hat S_2$ is the 
normalized cross-correlation (i.e., {\em coherence}) 
of the data from the two detectors.
It is a measure of how well the data in detector 2 
{\em matches} that in detector 1.

From (\ref{e:freq_estimators}), we see that $\Lambda(d)$ 
is a ratio of the square of a sum of products 
of Gaussian random variables to the product of a sum of squares 
of Gaussian random variables.
This is a sufficiently complicated expression that 
we will estimate the distribution of $\Lambda(d)$ 
{\em numerically}, doing fake signal injections into many 
realizations of simulated noise to build up the sampling
distribution. 
We do this explicitly in 
Section~\ref{s:freq-bayes-corr}, when we compare the 
frequentist and Bayesian correlation methods for this 
example.

\subsection{Bayesian correlation analysis}
\label{s:bayes-corr}

Compared to the frequentist cross-correlation analysis described 
above, a Bayesian analysis is conceptually much simpler.
One simply needs the likelihood functions 
$p(d|S_{n_1},S_{n_2},{\cal M}_0)$ and  
$p(d|S_{n_1},S_{n_2},S_h,{\cal M}_1)$ 
given by (\ref{e:nlikelihood}) and 
(\ref{e:likelihood_marginalized}), and joint
prior probability distributions for the signal 
and noise parameters.
For our example, we will assume that the signal and
noise parameters are
{\em statistically independent} of one another so that the 
joint prior distributions factorize into a product of 
priors for the individual parameters.
We use Jeffrey's priors for the individual noise variances:
\be
p_I(S_{n_I}) \propto 1/S_{n_I}\,,
\qquad
I=1,2\,,
\ee
and a flat%
\footnote{A flat prior for $S_h$ yields 
more conservative (i.e., larger) upper limits for 
$S_h$ than a Jeffrey's prior, since there 
is more prior weight at larger values of $S_h$
for a flat prior than for a Jeffrey's prior.}
prior for the signal variance:
\be
p(S_h) =  {\rm const}\,.
\ee
Then, using Bayes' theorem (\ref{full_bayes}), 
we obtain the joint posterior distribution:
\be
\begin{aligned}
p(S_{n_1},S_{n_2},S_h|d,{\cal M}_1)
&=\frac{p(d|S_{n_1},S_{n_2},S_h,{\cal M}_1)
p(S_{n_1},S_{n_2},S_h|{\cal M}_1)}
{p(d|{\cal M}_1)}
\\
&\propto
p(d|S_{n_1},S_{n_2},S_h,{\cal M}_1)
\frac{1}{S_{n_1}}\frac{1}{S_{n_2}}\,,
\end{aligned}
\ee
where $p(d|{\cal M}_1)$ is the evidence (or marginalized 
likelihood) for the signal-plus-noise model ${\cal M}_1$.
(Similar expressions can be written down for the noise-only 
model ${\cal M}_0$.)
The marginalized posterior distributions for the 
signal and noise parameters are given by marginalizing 
over the other parameters.
For example,
\be
p(S_h|d,{\cal M}_1)
\propto 
\int \frac{dS_{n_1}}{S_{n_1}}\>
\int \frac{dS_{n_2}}{S_{n_2}}\>
p(d|S_{n_1}, S_{n_2}, S_h, {\cal M}_1)
\ee
for the signal variance $S_h$.

Correlations enter the Bayesian analysis via the covariance 
matrix $C$ that appears in the 
likelihood function $p(d|S_{n_1},S_{n_2}, S_h,{\cal M}_1)$.
The covariance matrix for the data includes the cross-detector 
signal correlations, as we saw in (\ref{e:C_marginalized}).
So although one does not explicitly construct a 
cross-correlation statistic in the Bayesian framework, 
cross correlations do play an important role in the calculations.

\subsection{Comparing frequentist and Bayesian cross-correlation methods}
\label{s:freq-bayes-corr}

\begin{figure}[h!]
\begin{center}
\includegraphics[angle=0,width=0.49\columnwidth]{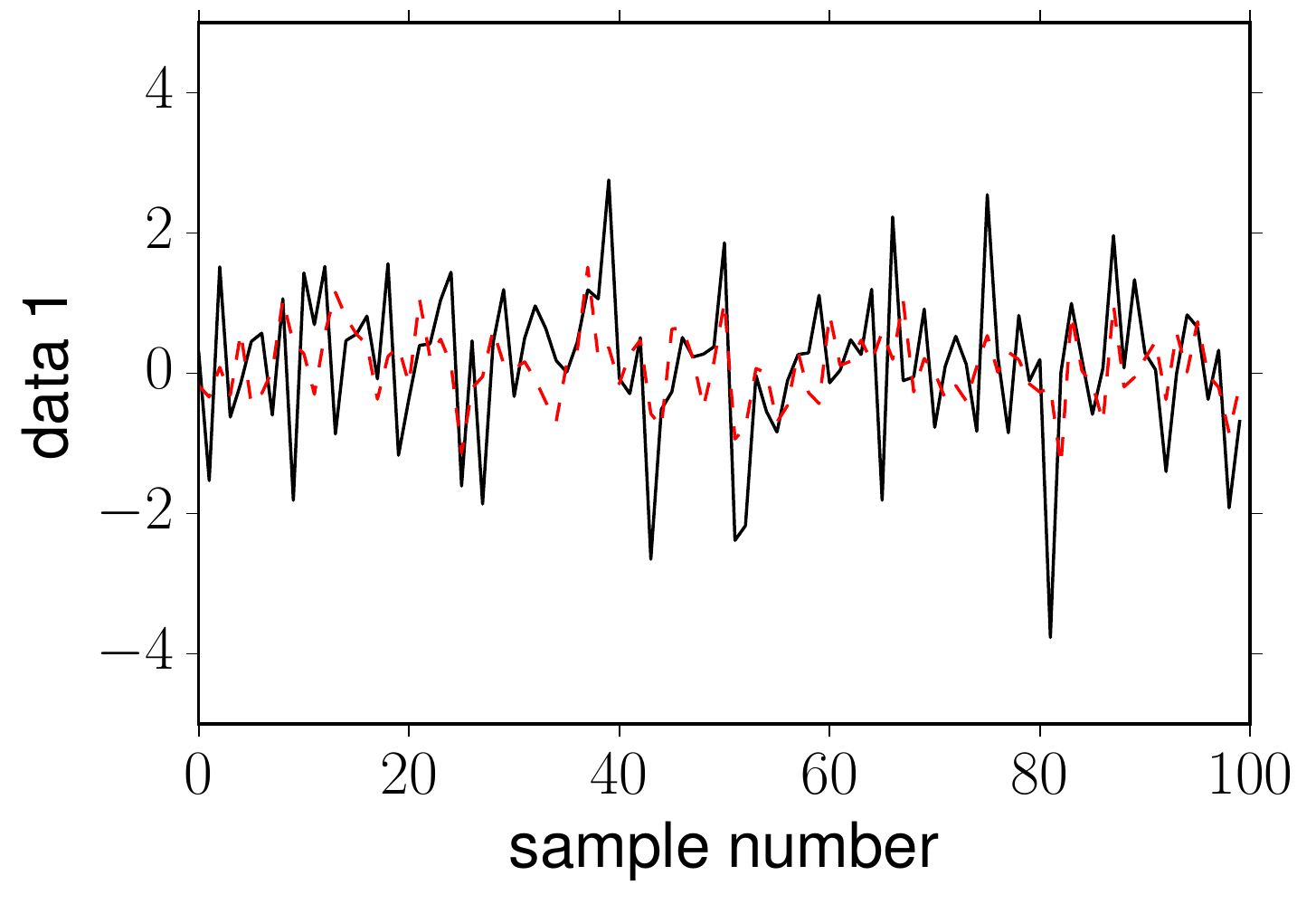}
\includegraphics[angle=0,width=0.49\columnwidth]{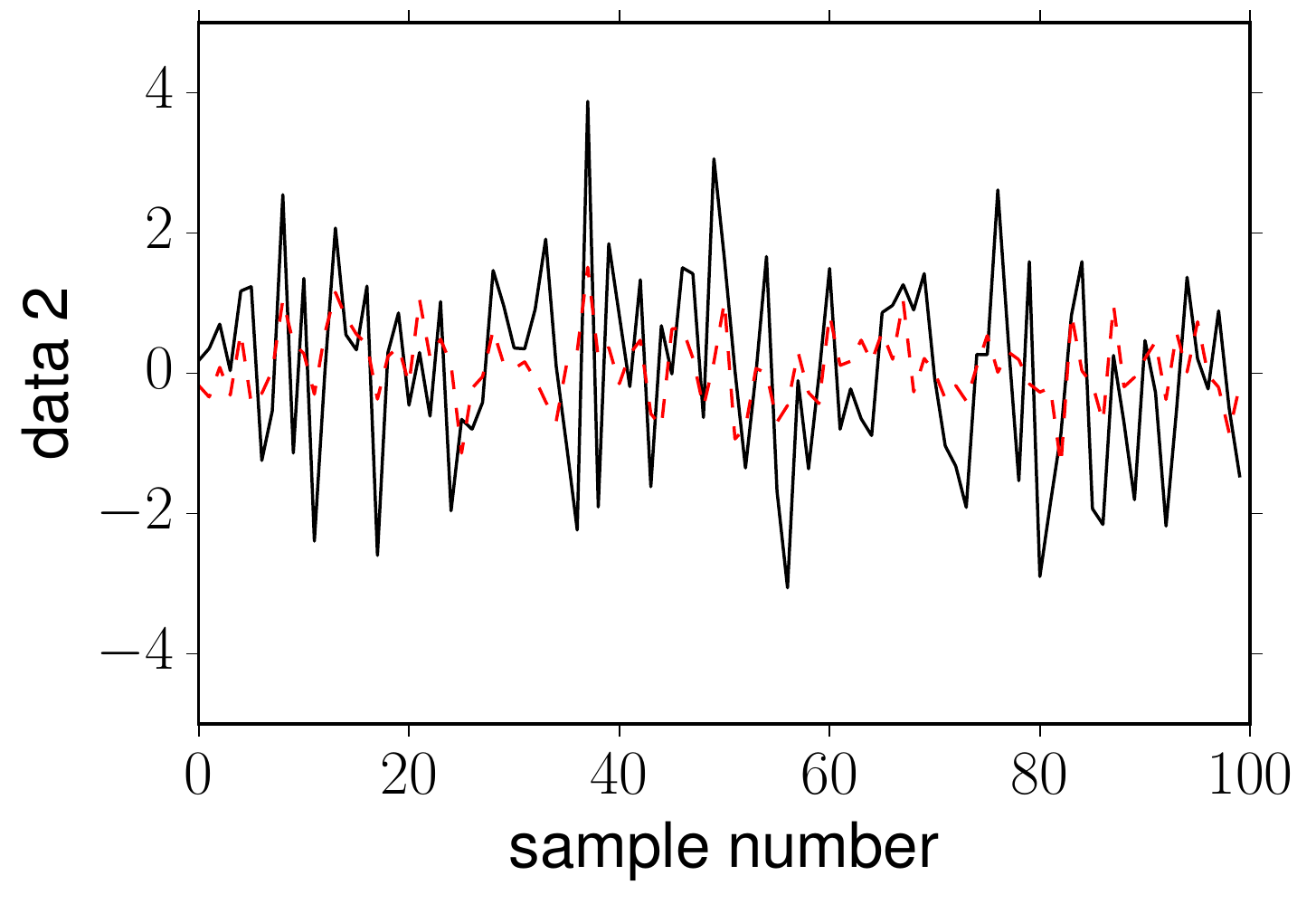}
\caption{Simulated data in the two detectors.
The detector output is shown by the black curves; 
the common stochastic signal is shown by the red dashed 
curves.}
\label{f:example-correlation-data}
\end{center}
\end{figure}

To explicitly compare the frequentist and Bayesian methods for
handling cross-correlations, we simulate data for the 
white noise, white signal example that we have been 
discussing in the previous subsections.
The particular realization of data that we generate has 
$N=100$ samples with 
$S_{n_1} =1$, $S_{n_2}=1.5$, and $S_h=0.3$.
Plots of the simulated data in the two detectors are given
in Figure~\ref{f:example-correlation-data}.

\subsubsection{Frequentist analysis}

The frequentist maximum-likelihood estimators 
(\ref{e:freq_estimators}) are very easy
to calculate.  
For the simulated data they have values:
\be
\hat S_{n_1} = 0.78\,,
\quad
\hat S_{n_2} = 1.46\,,
\quad
\hat S_h = 0.40\,.
\label{e:freq_estimators_example}
\ee
In addition
\be
\Lambda_{\rm ML}(d) = 44\,,
\quad
\Lambda(d)\equiv 2\ln\left(\Lambda_{\rm ML}(d)\right) = 7.6\,.
\ee
The weak-signal approximation to $\Lambda(d)$,
given by (\ref{e:Lambda_weak}), is 
significantly larger (having a value of $14$),
since the injected stochastic signal for this case 
was relatively strong, with the injected $S_h$ 
equal to $0.3 S_{n_1}$ and $0.2 S_{n_2}$.
In addition, for this realization of data, the 
signal variance was overestimated while both noise 
variances were underestimated, leading to a much 
larger value than the nominal squared signal-to-noise 
ratio of $6$.

As mentioned previously, the form 
(\ref{e:Lambda_corr_example}) of the detection
statistic $\Lambda(d)$ is sufficiently complicated
that it was simplest to resort to numerical 
simulations to estimate its sampling distribution,
$p(\Lambda|S_{n_1}, S_{n_2}, S_h, {\cal M}_1)$.
We took 50 values for each of
$S_{n_1}$, $S_{n_2}$, and $S_h$ 
in the interval $[0,3]$, and then for each
of the corresponding $50^3$ points in parameter space, 
we generated $10^4$ realizations of the data,
yielding $10^4$ values of $\Lambda(d)$.
By histogramming these values for each point
in parameter space, we were able to estimate 
the probability density function (and also the
cumulative distribution function) for $\Lambda$.

\begin{figure}[h!]
\begin{center}
\includegraphics[angle=0,width=0.9\columnwidth]{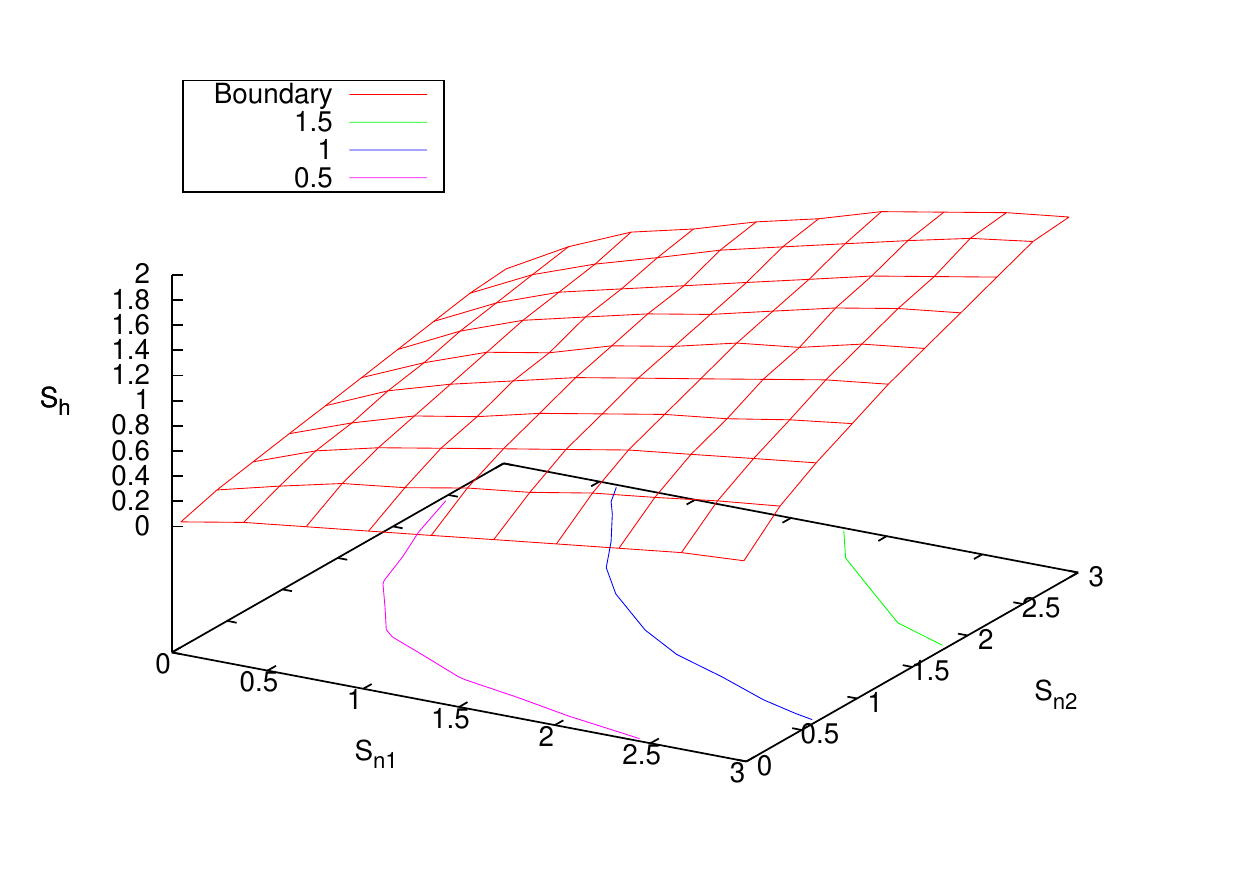}
\caption{Frequentist 90\% confidence-level exclusion
and inclusion regions for the simulated data with
$\Lambda_{\rm obs} = 7.6$.
The 90\% exclusion region ${\cal E}_{90\%}$ lies above
the red surface; 
the 90\% inclusion region ${\cal I}_{90\%}$ lies below
the red surface.
The green, blue and magenta curves are projections of
the $S_h=1.5, 1.0, 0.5$ level surfaces of the boundary
onto the $(S_{n_1}, S_{n_2})$ plane.}
\label{f:exclusion90}
\end{center}
\end{figure}

Figure~\ref{f:exclusion90} shows the frequentist
90\% confidence-level {\em exclusion} and 
{\em inclusion} regions for our simulated data 
with $\Lambda_{\rm obs}=7.6$.
The 90\% confidence-level exclusion region 
${\cal E}_{90\%}$ lies {\em above} the red surface; 
it consists of
points $(S_{n_1},S_{n_2},S_h)$ satisfying
\be
{\rm Prob}\left(\Lambda\ge \Lambda_{\rm obs}|
(S_{n_1}, S_{n_2}, S_h)\in {\cal E}_{90\%}\right)
\ge 0.90\,.
\ee
The region {\em below} the red surface is 
the 90\% confidence-level inclusion region 
${\cal I}_{90\%}$.
Note that construction of these regions is such
that the {\em true} values of the parameters 
$S_{n_1}$, $S_{n_2}$, and $S_h$ have a 90\%
frequentist probability of lying in ${\cal I}_{90\%}$.
This generalizes, to multiple parameters, the 
definition of the frequentist 
90\% confidence-level upper-limit for a single 
parameter, which was discussed
in detail in Section~\ref{s:freq-UL}.
Note that it is not correct to simply ``cut'' the 
surface using the maximum-likelihood point 
estimates $\hat S_{n_1}= 0.78$ and $\hat S_{n_2}=1.46$
to obtain a single value for $S_h^{90\%, {\rm UL}}$.
One needs to include the whole region in order
to get the correct frequentist coverage.

A similar procedure can be used
to estimate sampling distributions for the
frequentist maximum-likelihood estimators
$\hat S_{n_1}$, $\hat S_{n_2}$, and $\hat S_h$.
From these distributions, one can then 
calculate e.g., frequentist 95\% confidence-level 
exclusion and inclusion {\em regions} for the
given point estimates.
For example, 
$(S_{n_1}, S_{n_2}, S_h)\in {\cal I}_{95\%}$
for the observed point estimate 
$\hat S_{h,{\rm obs}}$ if and only if
$\hat S_{h,{\rm obs}}$ is contained in the 
symmetric 95\% confidence-level interval centered
on the mode of the probability distribution
$p(\hat S_h|S_{n_1}, S_{n_2}, S_h, {\cal M}_1)$.
These regions again generalize to 
multiple parameters the definition of a 
frequentist confidence {\em interval} for a 
single parameter, which was discussed in 
detail in Section~\ref{s:freq-parameter-estimation}.
They will be different, in general, for 
the different maximum-likelihood estimators.
But in order to move on to the Bayesian analysis 
for this example, we will leave the explicit 
construction of these regions to the interested reader.

\subsubsection{Bayesian analysis}

For the Bayesian analysis of this example, we 
limit ourselves to calculating the Bayes factor
$2 \ln {\cal B}_{10}(d)$ comparing the 
noise-only and signal-plus-noise models 
${\cal M}_0$ and ${\cal M}_1$, 
as well as the posterior distributions for the 
three parameters $S_h$, $S_{n_1}$, and $S_{n_2}$.
Following the procedure described above in
Section~\ref{s:bayes-corr} we find, for this
particular realization of data,
\be
{\cal B}_{10}=10\,,
\quad
2\ln {\cal B}_{10}(d) = 4.6\,.
\ee
This Bayes factor corresponds to 
{\em positive} evidence 
(see Table~\ref{t:bayesfactors})
in favor of a correlated stochastic signal in the data.

\begin{figure}[htbp!]
\begin{center}
\includegraphics[angle=0,width=0.7\columnwidth]{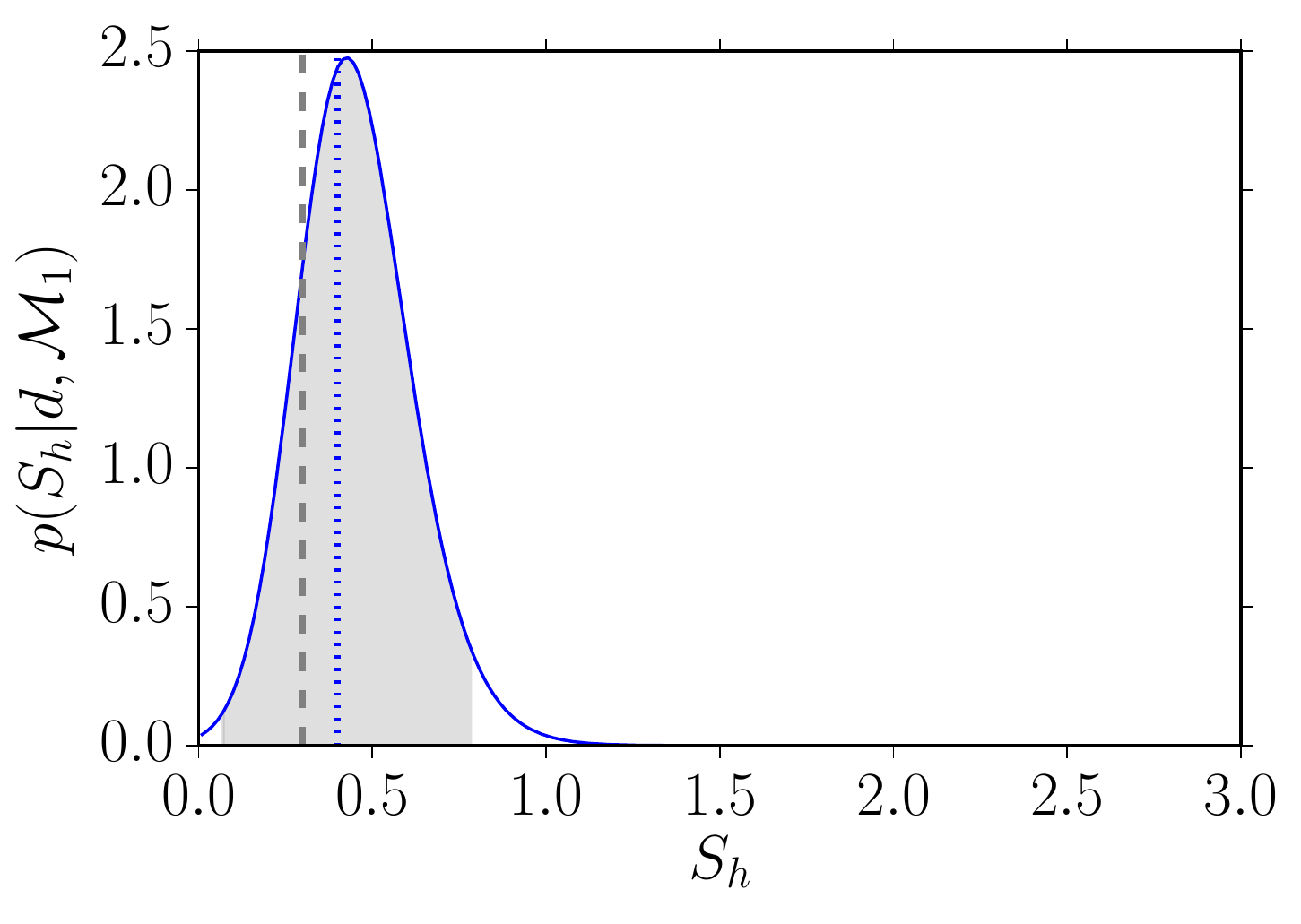}
\caption{Marginalized posterior distribution for the stochastic
signal variance $S_h$ for the signal-plus-noise model 
${\cal M}_1$.
The actual value of $S_h$ used for the 
simulation is shown by the grey dashed vertical line.
The 95\% Bayesian credible interval centered on
the mode of the distribution is the grey-shaded region.
For comparison, the frequentist maximum-likelihood estimator of 
$S_h$ is shown by the blue dotted vertical line.}
\label{f:example-correlation-posteriorh}
\end{center}
\end{figure}

Figure~\ref{f:example-correlation-posteriorh} shows the
marginalized posterior $p(S_h|d,{\cal M}_1)$ 
for the stochastic signal variance given the data $d$
and signal-plus-noise model ${\cal M}_1$.
The peak of the posterior lies close the frequentist
maximum-likelihood estimator $\hat S_h=0.40$ 
(blue dotted vertical line), and easily contains the injected 
value in its 95\% Bayesian credible interval (grey shaded region).
\begin{figure}[htbp!]
\begin{center}
\includegraphics[angle=0,width=0.49\columnwidth]{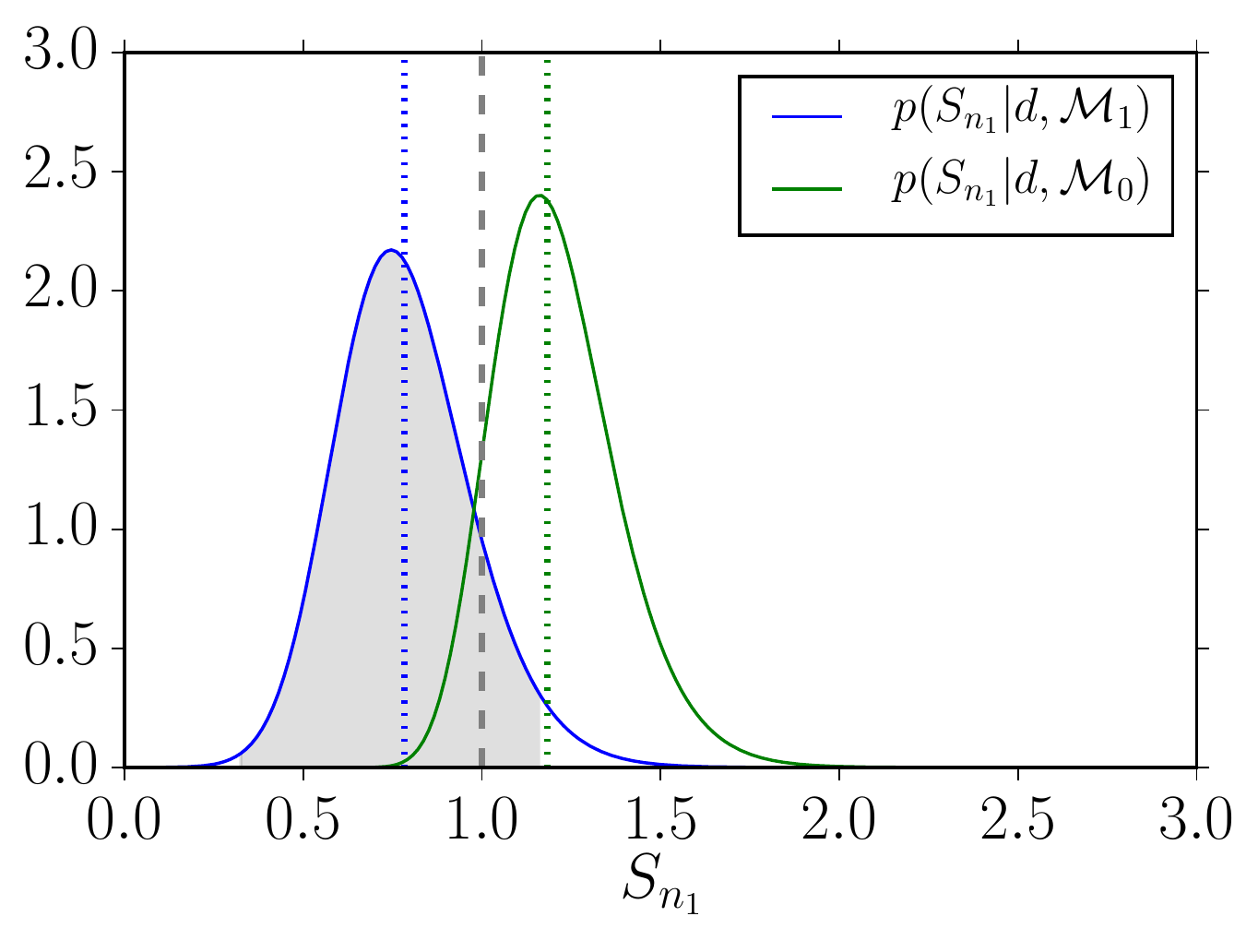}
\includegraphics[angle=0,width=0.49\columnwidth]{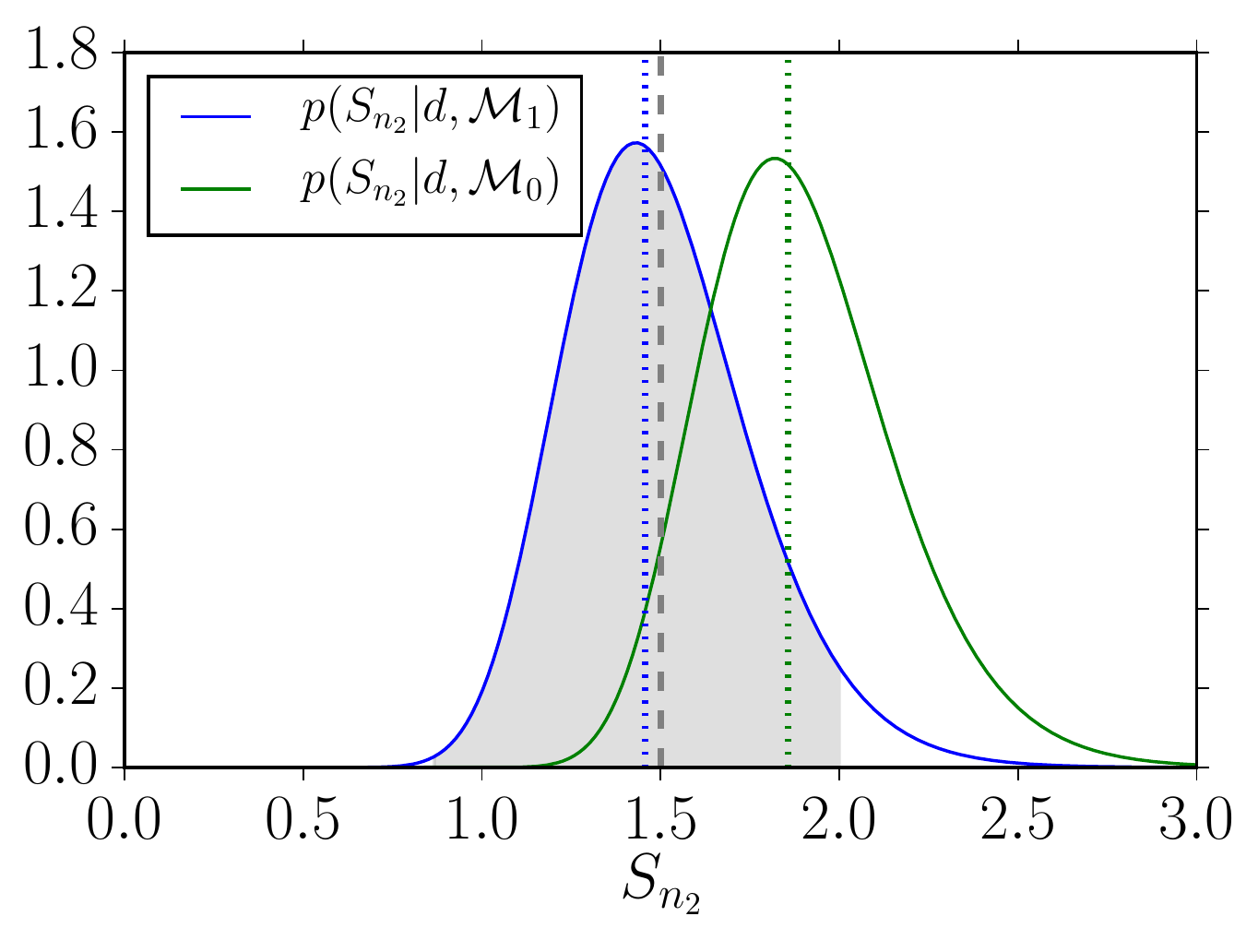}
\caption{Marginalized posterior distributions for the 
detector noise variances $S_{n_1}$ (left panel) and 
$S_{n_2}$ (right panel) for the signal-plus-noise model 
${\cal M}_1$ (blue curves) and the noise-only model ${\cal M}_0$ 
(green curves), respectively.
The actual values of $S_{n_1}$ and $S_{n_2}$ used for the 
simulation are shown by the grey dashed vertical lines.
The 95\% Bayesian credible intervals for the signal-plus-noise 
model ${\cal M}_1$ are the grey-shaded regions.
For comparison, 
the frequentist estimators of $S_{n_1}$ and $S_{n_2}$ 
for the two models are shown by the (blue and green) dotted 
vertical lines.}
\label{f:example-correlation-posteriorn}
\end{center}
\end{figure}
Figure~\ref{f:example-correlation-posteriorn} shows 
similar plots for the marginalized posteriors for the
noise variances $S_{n_1}$ and $S_{n_2}$ for both the
signal-plus noise model ${\cal M}_1$ (blue curves)
and the noise-only model ${\cal M}_0$ (green curves).
For comparison, the frequentist maximum-likelihood
estimators 
$\hat S_{n_1}, \hat S_{n_2}=0.78, 1.46$ and
$1.18, 1.86$ for the two
models are shown by the corresponding (blue and
green) dotted vertical lines.
Again, the peaks of the Bayesian posterior distributions
lie close to these values.
The 95\% Bayesian credible intervals for 
$S_{n_1}$ and $S_{n_2}$ for the signal-plus-noise model
${\cal M}_1$ are also shown (grey shaded region).
These intervals easily contain the injected values for
these two parameters.

\subsection{What to do when cross-correlation methods aren't available}
\label{s:null}

Cross-correlation methods can be used whenever one 
has two or more detectors that respond to a common 
gravitational-wave signal.
The beauty of such methods is that even 
though a stochastic background is another source of 
``noise" in a single detector, 
the common signal components in multiple detectors 
combine coherently when the data from pairs of 
detectors are multiplied together and summed, 
as described in Section~\ref{s:basic_idea}.
But with only a single detector, searches for a 
stochastic background need some other way to 
distinguish the signal from the noise---e.g.,
a difference between the spectra of the noise and 
the gravitational-wave signal, or the modulation 
of an anisotropic signal due to the motion of the 
detector (as is expected for the confusion-noise from 
galactic compact white dwarf binaries for LISA).
Without some way of differentiating instrumental noise
from gravitational-wave ``noise", there is no hope of 
detecting a stochastic background.

As a simple example, suppose that we have $N$ samples 
of data from each of two detectors $I=1,2$ 
(which we will call {\em channels} in what follows), 
but let's assume that the second channel 
is {\em insensitive} to the gravitational-wave signal:
\be
\begin{array}{ll}
d_{1i}=h_i + n_{1i}\,, & \quad i=1,2,\cdots, N\,,
\\
d_{2i}=n_{2i}\,, & \quad i=1,2,\cdots, N\,.
\label{e:data_nullchannel}
\end{array}
\ee
Then if we make the same assumptions as before 
for the signal and the noise, 
it follows that the likelihood function 
for the data $d\equiv \{d_{1i};d_{2i}\}$ is given by
\be
p(d|S_{n_1}, S_{n_2}, S_h, {\cal M}_1)=
\frac{1}{\sqrt{\det(2\pi C)}}
e^{-\frac{1}{2}d^T C^{-1} d}\,,
\label{e:LF}
\ee
where
\begin{equation}
C=
\left[
\begin{array}{cc}
(S_{n_1}+S_h)\,\unit_{N\times N} & \zero_{N\times N} \\
\zero_{N\times N} & S_{n_2}\,\unit_{N\times N} \\
\end{array}
\right]
\end{equation}
is the covariance matrix of the data.
Since the off-diagonal blocks of the covariance matrix 
are identically zero, it is clear that  
we will not be able to use the cross-correlation methods 
developed in the previous sections.
So we need to do something else if we are going to 
extract the gravitational-wave signal from the noise.

\subsubsection{Single-detector excess power statistic}

If we knew $S_{n_1}$ a~priori, then 
we could construct an 
{\em excess power} statistic from the autocorrelated
data to estimate the signal variance:
\begin{equation}
\hat S_h \equiv \frac{1}{N}\sum_{i=1}^N d_{1i}^2-S_{n_1}\,.
\label{e:ML1}
\end{equation}
(This is effectively how Penzias and Wilson discovered
the CMB~\cite{1965ApJ...142..419P}; 
they observed excess antenna noise that they
couldn't attribute to any other known source of noise.)
But as mentioned at the end of Section~\ref{s:white},
typically we do not know the detector noise well enough to 
use such a statistic, since the uncertainty in $S_{n_1}$ 
is much greater than the variance of the gravitational-wave
signal that we are trying to detect.
This is definitely the case for ground-based detectors like 
LIGO, Virgo, etc.
An exception to this ``rule" will probably be the predicted 
{\em foreground} signal from galactic white-dwarf 
binaries in the LISA band.
For frequencies below a few mHz, the gravitational-wave
confusion noise from these binaries is expected to dominate
the LISA instrument noise~\cite{Hils-et-al:1990, 
Bender-Hils:1997, Hils-Bender:2000, Nelemans-et-al:2001}.

\subsubsection{Null channel method}

If it were possible to make an {\em off-source} measurement 
using detector 1, then we could estimate the noise 
variance $S_{n_1}$ directly from the detector output, 
free of contamination from gravitational waves.
Using this noise estimate, $\hat S_{n_1}$, we could 
then define our excess power statistic as
\begin{equation}
\hat S_h \equiv \frac{1}{N}\sum_{i=1}^N d_{1i}^2
-\hat S_{n_1}\,.
\label{e:ML1_alt}
\end{equation}
Unfortunately, such off-source measurements are not
possible, since you cannot shield a gravitational-wave 
detector from gravitational waves.
However, in certain cases one can construct a particular 
{\em combination} of the data (called
a {\em null channel}) for which the response to gravitational 
waves is strongly suppressed.
The {\em symmetrized Sagnac} combination of the data for 
LISA~\cite{Tinto-et-al:2001, Hogan-Bender:2001} is one such example.

So let us assume that channel 2 for our example is such 
a null channel,
and let us also assume that there is some 
relationship between the noise in the two 
channels---e.g., $S_{n_1} = aS_{n_2}$,
with $a>0$.
(For colored noise, the variances would be replaced
by power spectra and $a$ would be replaced by a
function of frequency---i.e., a {\em transfer function}
relating the noise in the two channels.)
To begin with, we will also assume that 
$a$ is {\em known}.
Then the data from the second channel can be used 
as a {\em noise calibrator} for the first channel.
The frequentist estimators for this scenario are:
\be
\begin{aligned}
&\hat S_{n_2} = \frac{1}{N}\sum_{i=1}^N d_{2i}^2\,,
\\
&\hat S_{n_1} = a\hat S_{n_2}\,,
\\
&\hat S_h = \frac{1}{N}\sum_{i=1}^N d_{1i}^2 - \hat S_{n_1}\,.
\end{aligned}
\ee
These are the maximum-likelihood estimators of the 
signal and noise parameters, derived from the 
likelihood (\ref{e:LF}) with $S_{n_1}$ replaced
by $aS_{n_2}$.
In the Bayesian framework, the relation 
$S_{n_1}= a S_{n_2}$ is encoded 
in the joint prior probability distribution
\be
p(S_{n_1},S_{n_2}) = \delta(S_{n_1}-aS_{n_2})p_2(S_{n_2})\,,
\ee
which eliminates $S_{n_1}$ as an independent variable.
The marginalized posterior distribution for the signal
variance $S_h$, 
assuming a flat prior $p_h(S_h)={\rm const}$, 
is then
\be
p(S_h|d) \propto
\int dS_{n_2}\> 
p(d|S_{n_1}=aS_{n_2},S_{n_2},S_h)
p_2(S_{n_2})\,.
\ee
In the more realistic scenario where the transfer 
function $a$ is not known a~priori, but is 
described by its own prior probability distribution 
$p_a(a)$, we have
\be
p(S_{n_1}, S_{n_2}, a) 
= \delta(S_{n_1}- aS_{n_2})p_a(a)p_2(S_{n_2})
\ee
and
\be
p(S_h|d) \propto
\int da
\int dS_{n_2}\> 
p(d|S_{n_1}=aS_{n_2},S_{n_2},S_h)
p_a(a)
p_2(S_{n_2})\,.
\ee
This integral can be done numerically given priors for 
$S_{n_2}$ and $a$.

\begin{figure}[htbp]
\begin{center}
\includegraphics[angle=0,width=0.7\columnwidth]{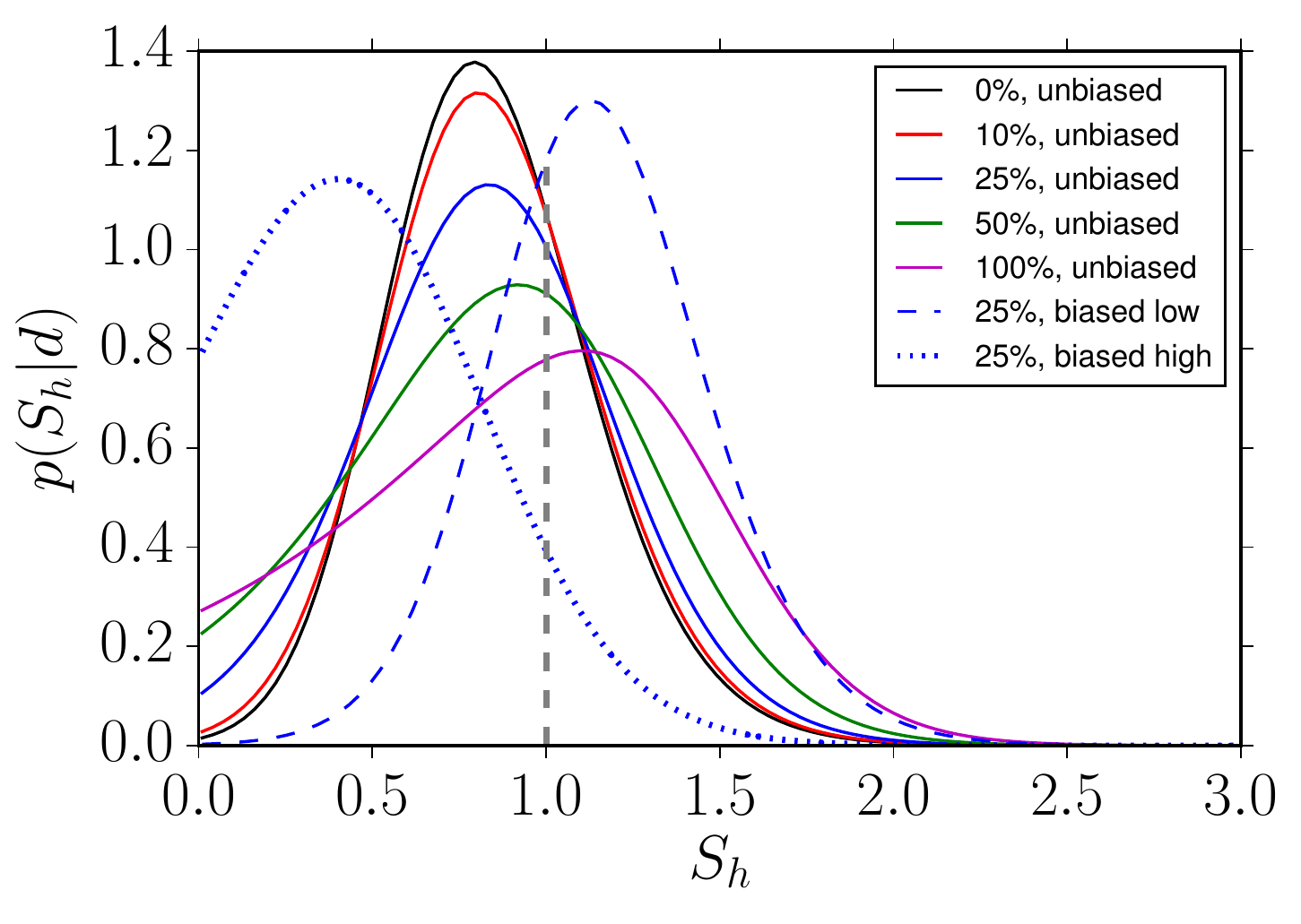}
\caption{Posterior distributions for $S_h$
for the null channel analysis, corresponding to different priors 
for the parameter $a$, which relates the instrumental noise variances 
in the two channels.
The labels ``$p$\%, unbiased" correspond to $A=a_0$ and $\Sigma=1+p/100$;
the labels ``25\%, biased low (or high)" correspond to $A=0.67 a_0$ 
(or $1.5 a_0$) and $\Sigma = 1.25$.
The vertical grey dashed line corresponds to the injected value of $S_h$.}
\label{f:PD2}
\end{center}
\end{figure}

To help illustrate the above discussion,
Figure~\ref{f:PD2} shows plots of several different
posterior distributions for $S_h$, corresponding 
to different choices for the prior distribution $p_a(a)$.
For these plots, we chose a {\em Jeffrey's prior} for 
$S_{n_2}$:
\be
p_2(S_{n_2})\propto 1/S_{n_2}\,,
\ee
and a {\em log-normal} prior for $a$:
\be
p(a|\mu,\sigma) 
= \frac{1}{a}\frac{1}{\sqrt{2\pi}\sigma}
e^{-\frac{1}{2}\frac{(\ln a-\mu)^2}{\sigma^2}}\,.
\ee
The different curves 
correspond to different values of $\mu$ and $\sigma$:
\be
\begin{aligned}
&\mu\equiv \ln A \,,\qquad A = a_0, 0.67a_0, 1.5a_0\,,
\\
&\sigma \equiv \ln\Sigma\,,
\qquad \Sigma = 1, 1.1, 1.25, 1.5, 2\,,
\end{aligned}
\ee
where $a_0$ denotes the nominal (true) value of $a$.
Note that $A=0.67a_0$ and $1.5a_0$ correspond to priors for 
$a$ that are {\em biased} away from its true value $a=a_0$.
Note also that 68\% of the prior distribution is contained in 
the region $a\in[A/\Sigma, A\Sigma]$
(so $\Sigma=1$ corresponds to a delta-function prior---i.e., 
no uncertainty in $a$).
The particular realization that we used consisted of $N=100$ 
samples of data (\ref{e:data_nullchannel}) with $S_h=1$, 
$S_{n_2}=1$, and $S_{n_1} = a_0 S_{n_2}$ with $a_0=1$.
Note that for the biased priors for $a$
(associated with the dashed and dotted curves in Figure~\ref{f:PD2}),
an under (over) estimate in $a$ corresponds to over (under) 
estimate in $S_h$, as $S_h$ is effectively the 
difference between the estimated variance in channel 1 and 
$a$ times the estimated variance in channel 2.
For this particular realization of the data, the mode of the 
``0\%, unbiased" posterior for $S_h$ is about 20\% less than the 
injected value, $S_h=1$.
On average, they would agree.
%

\section{Geometrical factors}
\label{s:geom}

\begin{quotation}
There is geometry in the humming of the strings, there is music in the spacing of the spheres.  
{\em Pythagoras}
\end{quotation}

\noindent
In the previous sections, we ignored many details regarding 
detector response and detector geometry.
We basically assumed that the detectors were {\em isotropic}, 
responding equally well to all gravitational waves,
regardless of the waves' directions of propagation, 
frequency content, and polarization.
We also ignored any loss in sensitivity in the
correlations between data from two or more detectors, 
due to the separation and relative orientation of 
the detectors.
But these details {\em are} important if we want to design
optimal (or near-optimal) data analysis algorithms to 
search for gravitational waves.
To specify the likelihood function, for example, requires 
models not only for the gravitational-wave signal and
instrument noise, but also for the reponse of the 
detectors to the waves that a source produces.

In this section, we fill in these details.
We first discuss the response of a single detector 
to an incident gravitational wave.
We then show how these non-trivial detector responses 
manifest themselves in the 
correlation between data from two or more detectors.
The results are first derived in a general setting making 
no assumption, for example, about the wavelength of a 
gravitational wave to the 
characteristic size of a detector.
The general results are then specialized, as appropriate, 
to the case of ground-based and space-based laser 
interferometers, spacecraft Doppler tracking, and 
pular timing arrays.
We conclude this section by discussing how the
motion of a detector relative to the gravitational-wave 
source affects the detector response.

The approach we take in this section is similar in spirit 
to that of~\cite{Hellings:1991}, attempting to unify the 
treatment of detector response functions and correlation 
functions across different gravitational-wave detectors.
Readers interested in more details about the effect of 
detector geometry on the correlation of data from two
or more detectors should see the original papers by 
Hellings and Downs \cite{Hellings-Downs:1983} for pulsar
timing arrays, and Flanagan \cite{Flanagan:1993} and 
Christensen \cite{Christensen:PhD, Christensen:1992} 
for ground-based laser interferometers.

\subsection{Detector response}
\label{s:detectorresponse}

Gravitational waves are time-varying perturbations
to the background geometry of spacetime.
Since gravitational waves induce time-varying changes 
in the separation between two freely-falling 
objects (so-called test masses), 
gravitational-wave detectors are designed
to be as sensitive as possible to this changing
separation.
For example, a resonant bar detector acts 
like a giant tuning fork, which is set 
into oscillation
when a gravitational wave of the natural frequency
of the bar is incident upon it.
These oscillations produce a stress against the
equilibrium electromagnetic forces that exist 
within the bar.
The stress (or oscillation) is measured by a strain 
gauge (or accelerometer), indicating the presence of a
gravitational wave.
The response for a bar detector is thus the 
fractional change in length of the bar, 
$h(t) = \Delta l(t)/l$, induced by the wave.
Since the length of the bar is typically much smaller
than the wavelength of a gravitational wave at the 
bar's resonant frequency,
the response is most easily computed using the 
geodesic deviation equation~\cite{MTW:1973} for the 
time-varying tidal field.

In this article, we will focus our attention on 
{\em beam} detectors, which use electromagnetic radiation to monitor 
the separation of two or more freely-falling objects.
Spacecraft Doppler tracking, pulsar timing arrays, and
ground- and space-based laser interferometers 
(e.g., LIGO-like and LISA-like detectors) are all 
examples of beam detectors, which can be used to 
search for gravitational waves (see, e.g., Section 4.2 
in \cite{Sathyaprakash-Schutz:2009}).

\subsubsection{Spacecraft Doppler tracking}

\begin{figure}[h!tbp]
\begin{center}
\includegraphics[angle=0,width=.4\columnwidth]{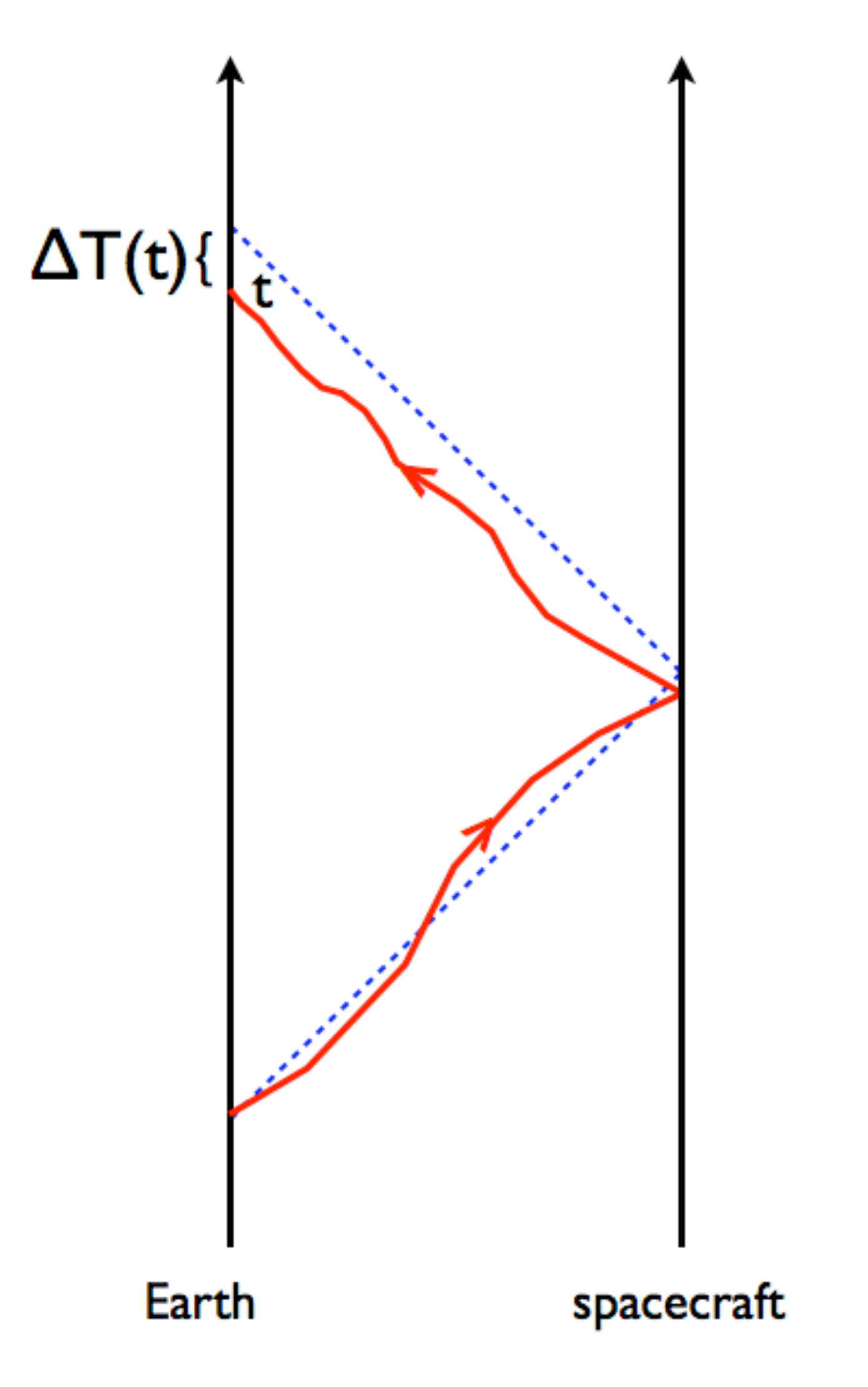}
\caption{A spacetime diagram representation of 
$\Delta T(t)$ for a
two-way spacecraft Doppler tracking measurement.
Time increases vertically upward.
The vertical arrows are spacetime worldlines for
the Earth and a spacecraft.
The measurement is made at time $t$.
The blue dotted line shows the trajectory of 
a pulse of electromagnetic radiation in the absence 
of a gravitational wave;
the red solid line shows the trajectory in the
presence of a gravitational wave.}
\label{f:twowayresponse}
\end{center}
\end{figure}

For spacecraft Doppler tracking, pulses of 
electromagnetic radiation are sent from one test 
mass (e.g., a radio transmitting tower on Earth) 
to another (e.g., the Cassini probe), and 
then bounced back (or coherently transponded) 
from the second test mass to the first.
From the arrival times of the returning 
pulses, one can calculate the 
fractional change in the frequency of the emitted 
pulses induced by a gravitational wave.
The detector response for such a measurement is
thus 
\be
h_{\rm doppler}(t)\equiv 
\frac{\Delta\nu(t)}{\nu_0}=\frac{d\Delta T(t)}{dt}\,,
\label{e:rdoppler}
\ee
where $\Delta T(t)$ is the deviation
of the round-trip travel time of a pulse 
away from the value it would have had at time $t$ 
in the absence of the gravitational wave.
A schematic representation of $\Delta T(t)$ for
spacecraft Doppler tracking is given in
Figure~\ref{f:twowayresponse}.

\subsubsection{Pulsar timing}

\begin{figure}[h!tbp]
\begin{center}
\includegraphics[angle=0,width=.4\columnwidth]{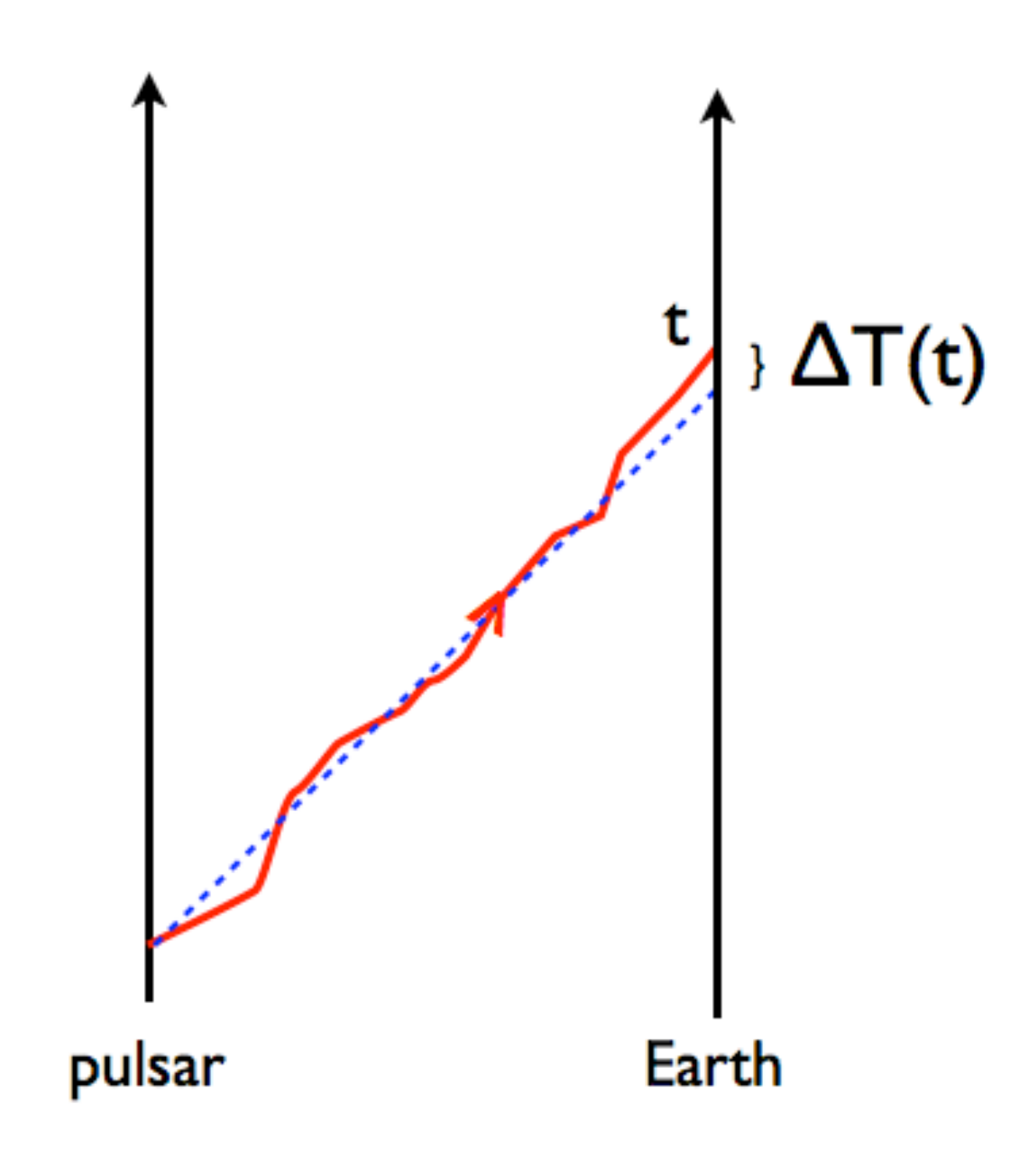}
\caption{A spacetime diagram representation of 
$\Delta T(t)$ for a
(one-way) pulsar timing residual measurement.
Time increases vertically upward.
The vertical arrows are spacetime worldlines for
a pulsar and a detector on Earth.
The measurement is made at time $t$.
The blue dotted line shows the trajectory of 
the radio pulse in the absence of a gravitational wave;
the red solid line shows the trajectory in the
presence of a gravitational wave.}
\label{f:onewayresponse}
\end{center}
\end{figure}

Pulsar timing is even simpler in the sense 
that we only have 
{\em one-way} transmission of electromagnetic
radiation (i.e., radio pulses are emitted by 
a pulsar and received by a radio antenna on Earth).
The response for such a system is simply the
timing residual
\be
h_{\rm timing}(t) = \Delta T(t)\,,
\label{e:rtiming}
\ee
which is the difference between the measured
time of arrival of a radio pulse and 
the expected time of arrival of the pulse
(as determined 
from a detailed timing model for the pulsar)
due to the presence of a gravitational wave.
A schematic representation of $\Delta T(t)$ for
a pulsar timing measurement is given in 
Figure~\ref{f:onewayresponse}.

\subsubsection{Laser interferometers}

\begin{figure}[h!tbp]
\begin{center}
\includegraphics[angle=0,width=.5\columnwidth]{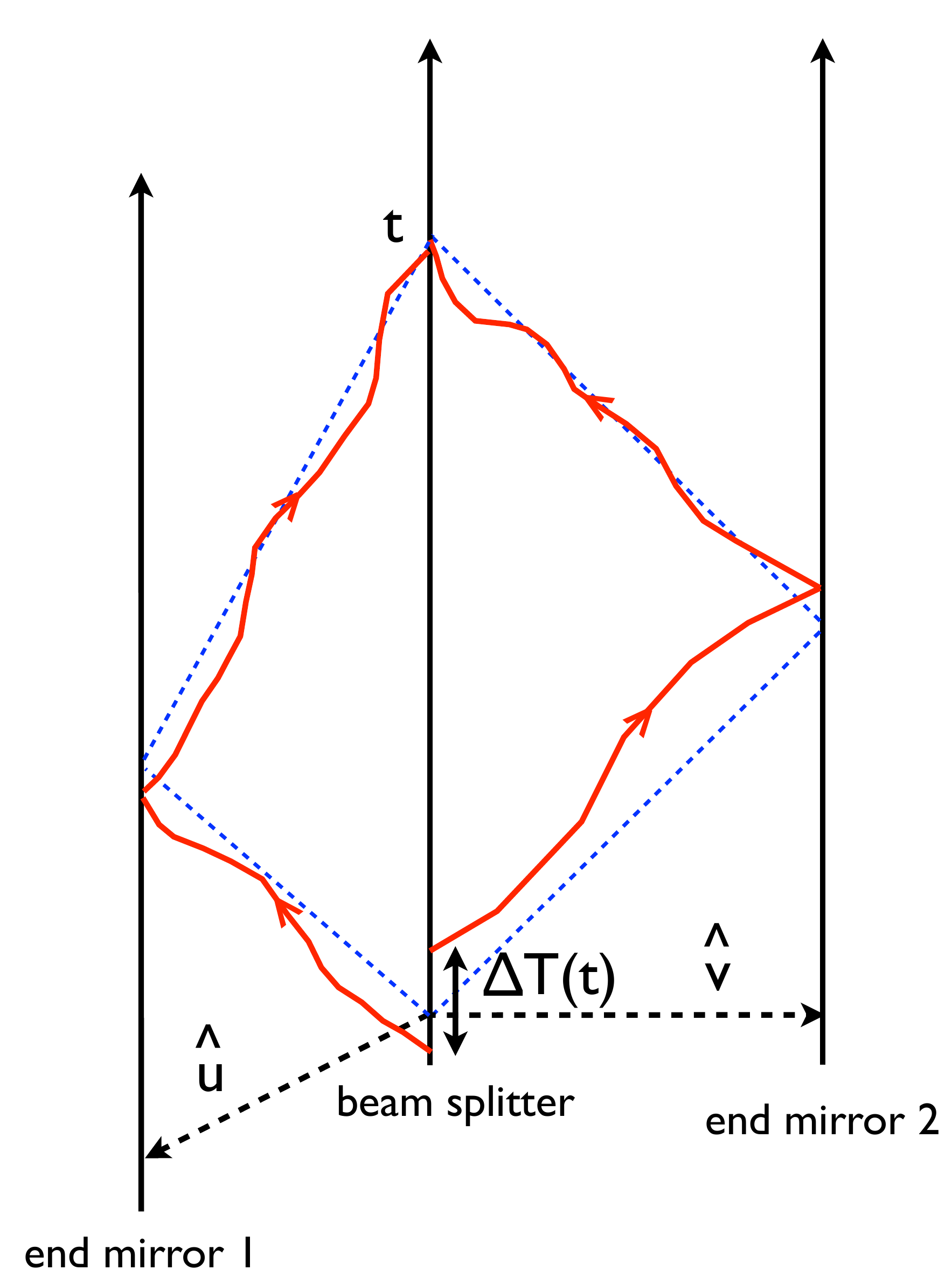}
\caption{A spacetime diagram representation of 
$\Delta T(t)$ for an equal-arm Michelson interferometer.
Time increases vertically upward.
The vertical arrows are spacetime worldlines for
the beam splitter and two end mirrors.
The blue dotted lines show the trajectory of the 
laser light in the two arms of the interferometer 
in the absence of a gravitational wave;
the red solid lines show the trajectory in the
presence of a gravitational wave. 
The black dotted arrows, labeled $\hat u$ and $\hat v$, 
show the orientation of the two arms, from beam
splitter to end mirrors, at $t=0$, assuming an opening
angle of $90^\circ$.}
\label{f:interferometerresponse}
\end{center}
\end{figure}

For laser interferometers like
LIGO or LISA, the detector response is the 
phase difference in the laser light sent 
down and back the two arms of the interferometer.
Again, the phase difference can be calculated 
in terms of the change in the round-trip travel 
time of the laser light from one test 
mass (e.g., the beam splitter) to another 
(e.g., one of the end test masses).
If we consider an equal-arm Michelson 
interferometer with unit vectors 
$\hat u$ and $\hat v$ pointing from the
beam splitter to the end masses in each of the
arms, then
\be
h_{\rm phase}(t) \equiv \Delta\Phi (t) 
= 2\pi\nu_0 \Delta T(t)\,,
\label{e:rphase}
\ee
where 
$\Delta T(t)\equiv T_{{\hat u},{\rm rt}}(t)
-T_{{\hat v},{\rm rt}}(t)$ 
is the difference of
the round-trip travel times, 
and $\nu_0$ is the frequency of the laser light.
(See Figure~\ref{f:interferometerresponse}.)
Alternatively, one often writes the interferometer 
response as a {\em strain} measurement 
in the two arms
\be
h_{\rm strain}(t)
\equiv\frac{\Delta L(t)}{L} 
=\frac{\Delta T(t)}{2L/c}\,,
\label{e:rstrain}
\ee
where
$\Delta L(t)\equiv L_{\hat u}(t)-L_{\hat v}(t)$ is
the difference of the proper lengths of the two
arms (having unperturbed length $L$), 
and $\Delta T(t)$ is the difference in round-trip
travel times as before.
Thus, interferometer phase and strain response
are simply related to one another.

Calculation of $\Delta T(t)$ for beam detectors
is most simply carried out in the transverse-traceless
gauge%
\footnote{See~\cite{Creighton-et-al:2009, Koop-Finn:2014}
for an alternative derivation of the response of a detector
to gravitational waves, which is done in terms of the 
curvature tensor and not the metric perturbations.} 
\cite{MTW:1973, Schutz:1985, Hartle:2002}
since the unperturbed separation $L$ of the two test 
masses can be larger than or comparable to the wavelength 
$\lambda\equiv c/f$ of an incident 
gravitational wave having frequency $f$.
This is definitely the case for pulsar timing
where $L$ is of order a few kpc, and for 
spacecraft Doppler tracking where $L$ is of 
order tens of AU.
It is also the case for space-based detectors
like LISA ($L=5\times 10^6 {\rm\ km}$) for 
gravitational waves with frequencies around a
tenth of a Hz.
On the other hand, for Earth-based detectors 
like LIGO ($L=4~{\rm km}$),
$L\ll \lambda$ is a good approximation 
below a few kHz.
Thus, the approach that we will take in the 
following subsections is to calculate the 
detector response in general, not making 
any approximation {\em a~priori} regarding the 
relative sizes 
of $\lambda=c/f$ and $L$.
To recover the standard expressions 
(i.e., in the long-wavelength or small-antenna limit)
for Earth-based detectors like LIGO
will be a simple matter of taking the limit 
$fL/c$ to zero.
For reference,
Table~\ref{t:beamdetectors} summarizes the characteristic
properties (i.e., size, characteristic frequency, 
sensitivity band, etc.)~of different beam detectors.
\begin{table}[htbp]
\centering
\begin{tabular}{lccccc}
\toprule
Beam detector & $L$ (km) & $f_*$ (Hz) & $f$ (Hz) & $f/f_*$ & 
Relation
\\
\midrule
Ground-based  & $\sim 1$ & $\sim 10^5$ & 
$10 - 10^4$ & $10^{-4} - 10^{-1}$ & 
$f\ll f_*$ 
\\
interferometer & & & & & 
\\
Space-based & $\sim 10^6$ & $\sim 10^{-1}$ & 
$10^{-4} - 10^{-1}$ & $10^{-3} -  1$ &
$f\lesssim f_*$ 
\\
interferometer & & & & & 
\\
Spacecraft Doppler & $\sim 10^9$ & $\sim 10^{-4}$ & 
$10^{-6} - 10^{-3}$ & $10^{-2} - 10$ & $f\sim f_*$
\\
tracking & & & & & 
\\
Pulsar timing & $\sim 10^{17}$ & $\sim 10^{-12}$ & 
$10^{-9} - 10^{-7}$ & $10^3 - 10^5$ & $f\gg f_*$ 
\\
\bottomrule
\end{tabular}
\caption{Characteristic properties of different beam detectors:
column 2 is the arm length or characteristic size of the detector
(tens of AU for spacecraft Doppler tracking; 
a few kpc for pulsar timing);
column 3 is the frequency corresponding to the characteristic
size of the detector, $f_*\equiv c/L$;
columns 4 and 5 are the frequencies at which the detector is
sensitive in units of Hz and units of $f_*$, respectively; and
column 6 is the relationship between $f$ and $f_*$.}
\label{t:beamdetectors}
\end{table}
%

\subsection{Calculation of response functions and antenna patterns}
\label{s:responsefunctions}

Gravitational waves are weak.
Thus, the detector response is {\em linear} in the metric perturbations 
$h_{ab}(t,\vec x)$ describing the wave, and can be written
as the convolution of the metric perturbations $h_{ab}(t,\vec x)$
with the {\em impulse response} $R^{ab}(t,\vec x)$ of the detector:
\be
h(t) = (\mb{R}*\mb{h})(t,\vec x)
\equiv \int_{-\infty}^\infty d\tau\int d^3y\,
R^{ab}(\tau,\vec y)
h_{ab}(t-\tau,\vec x-\vec y)\,,
\ee
where $\vec x$ is the location of the measurement at time $t$.
In terms of a plane-wave expansion (\ref{e:planewave}) of the metric 
perturbations, we have
\begin{align}
h(t) 
= \int_{-\infty}^\infty df
\int d^2\Omega_{\hat n}
R^{ab}(f,\hat n)
h_{ab}(f,\hat n)
e^{i2\pi f t}\,,
\end{align}
or, in the frequency domain,
\be
\tilde h(f) = 
\int d^2\Omega_{\hat n}
R^{ab}(f,\hat n)
h_{ab}(f,\hat n)\,,
\ee
where%
\footnote{\label{fn:response-phase}Some authors
\cite{Christensen:PhD, Christensen:1992, Flanagan:1993, 
Allen-Romano:1999, Cornish-Larson:2001, Finn-et-al:2009},
including us in the past, have defined the 
response function $R^{ab}(f,\hat n)$ {\em without}
the factor of $e^{i 2\pi f\hat n\cdot\vec x/c}$.
If one chooses coordinates so that the 
measurement is made at $\vec x=\vec 0$, then 
these two definitions agree.
Just be aware of this possible 
difference when reading the literature.
To distinguish the two definitions, we will use the 
symbol $\bar R^{ab}(f,\hat n)$ to denote the expression
without the exponential term, i.e.,
$R^{ab}(f,\hat n) = 
e^{i 2\pi f\hat n\cdot\vec x/c}\bar R^{ab}(f,\hat n)$.}
\be
R^{ab}(f,\hat n)
=e^{i2\pi f\hat n\cdot\vec x/c}
\int_{-\infty}^\infty {\rm d}\tau
\int {\rm d}^3 y\> 
R^{ab}(\tau,\vec y)\,
e^{-i2\pi f(\tau+\hat n\cdot\vec y/c)}\,.
\ee
Further specification of the response function
depends on the choice of gravitational-wave
detector as well as on the basis tensors used to 
expand $h_{ab}(f,\hat n)$, as we shall see below
and in the following subsections.

For example, if we work in the polarization basis, 
with expansion coefficients $h_A(f,\hat n)$, 
where $A=\{+,\times\}$, then
\be
\tilde h(f) =
\int d^2\Omega_{\hat n}
\sum_A
R^A(f,\hat n)h_A(f,\hat n)\,,
\label{e:responseRA}
\ee
with
\be
R^A(f,\hat n) = R^{ab}(f,\hat n)e^A_{ab}(\hat n)\,.
\label{e:RA}
\ee
If we work instead in the tensor spherical harmonic basis, 
with expansion coefficients $a^P_{(lm)}(f)$, 
where $P=\{G,C\}$, then
\be
\tilde h(f) =
\sum_{(lm)}
\sum_P
R^P_{(lm)}(f)a^P_{(lm)}(f)\,,
\label{e:responseRP}
\ee
with
\be
R^P_{(lm)}(f) 
= \int d^2\Omega_{\hat n}\>
R^{ab}(f,\hat n)Y^{P}_{(lm)ab}(\hat n)\,.
\label{e:RP}
\ee
Note that in the polarization basis the response function
$R^A(f,\hat n)$ is the detector response to a sinusoidal 
plane-wave with frequency $f$, coming from direction 
$\hat n$, and having polarization $A=+,\times$.
Plots of $|R^A(f,\hat n)|$ for fixed frequency $f$ 
are {\em antenna beam patterns} for gravitational waves
with polarization $A$.
A plot of
\be
{\cal R}(f,\hat n)\equiv
\left(|R^+(f,\hat n)|^2+ |R^\times(f,\hat n)|^2\right)^{1/2}
\label{e:beam}
\ee
for fixed frequency $f$ is the beam pattern for
an {\em unpolarized} gravitational wave---i.e., a 
wave having statistically equivalent $+$ and $\times$ polarization components.

Since the previous subsection showed that the
response of all beam detectors can be 
written rather simply in terms of the change in
the light-travel time of an electromagnetic 
wave propagating between two test masses, 
we now calculate $\Delta T(t)$ in various 
scenarios and use the resulting expressions to 
read-off the response functions $R^{ab}(f,\hat n)$ 
for the different detectors.
We also make plots of various antenna patterns.

\subsubsection{One-way tracking}
\label{s:one-way-tracking}
\begin{figure}[h!tbp]
\begin{center}
\includegraphics[angle=0,width=.4\columnwidth]{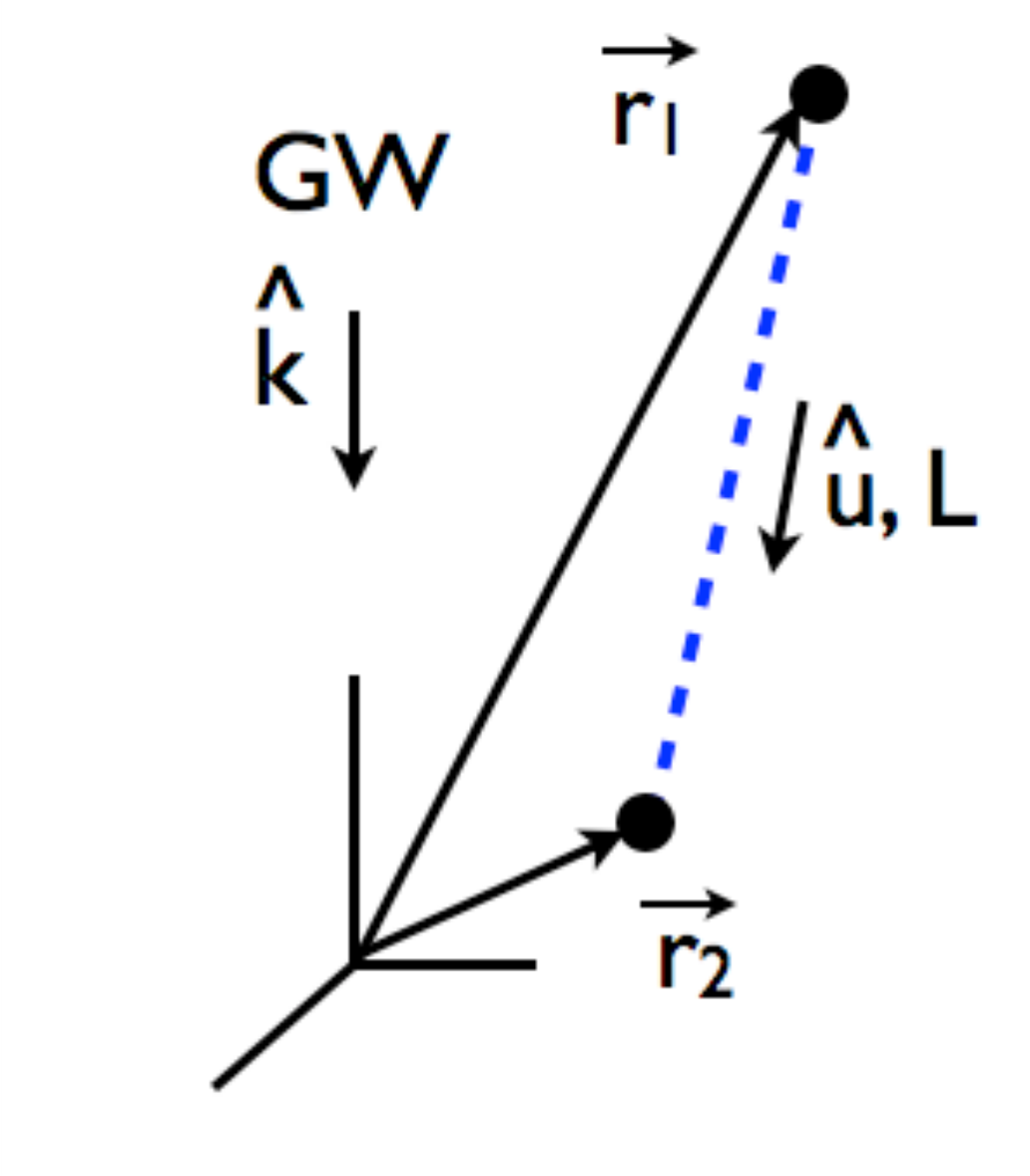}
\caption{Geometry for calculating the change in the photon  
propagation time from $\vec r_1$ to $\vec r_2=r_1+L\hat u$
in the presence of a plane gravitational wave propagating
in direction $\hat k$.}
\label{f:oneway}
\end{center}
\end{figure}

Consider two test masses located at position vectors 
$\vec r_1$ and $\vec r_2=\vec r_1+L\hat u$, 
respectively, in the presence of a plane gravitational wave 
propagating in direction $\hat k = -\hat n$, 
as shown in Figure~\ref{f:oneway}.
Then the change in the light-travel time for a photon 
emitted at $\vec r_1$ and received at $\vec r_2$ at 
time $t$ is given by~\cite{Estabrook-Wahlquist:1975}:
\be
\Delta T(t)
=\frac{1}{2c} u^a u^b \int_{s=0}^L ds\, 
h_{ab}(t(s),\vec x(s))\,,
\ee
where the 0th-order expression for the photon
trajectory can be used in $h_{ab}$:
\begin{align}
t(s) = (t-L/c) + s/c\,,
\quad
\vec x(s) = \vec r_1 + s\hat u\,.
\end{align}
Since $h_{ab}(t,\vec x) = h_{ab}(t+\hat n\cdot \vec x/c)$
for a plane wave, it is relatively easy to do the integral.
The result is
\begin{eqnarray}
{\Delta T(t)}
&=&
\int_{-\infty}^\infty df
\frac{1}{2}
u^a u^b
h_{ab}(f,\hat n)\,
\nonumber
\\
&&
\qquad\qquad
\frac{1}{i 2\pi f}\,
\frac{1}{1+\hat n\cdot \hat u}\,
\left[
e^{i 2\pi f(t_2+\hat n\cdot \vec r_2/c)}-
e^{i 2\pi f(t_1+\hat n\cdot \vec r_1/c)}
\right]
\label{e:deltaT1}
%
%
\\
&=&
\int_{-\infty}^\infty df
\frac{1}{2}
u^a u^b
h_{ab}(f,\hat n)\,
e^{i 2\pi f (t+\hat n\cdot\vec r_2/c)}
\nonumber
\\
&&
\qquad\qquad
\frac{1}{i 2\pi f}\,
\frac{1}{1+\hat n\cdot \hat u}\,
\left[1-e^{-\frac{i2\pi f L}{c}(1+\hat n\cdot\hat u)}\right]\,,
\label{e:deltaT2}
\end{eqnarray}
where we factored out 
$e^{i 2\pi f (t+\hat n\cdot\vec r_2/c)}$, corresponding to
the time and location of the measurement, to get the last line.
Note that the two terms in square brackets in 
(\ref{e:deltaT1}) correspond to sampling 
the gravitational-wave phase at photon reception 
(location $\vec r_2$ at time $t_2\equiv t$) 
and photon emission 
(location $\vec r_1$ at time $t_1\equiv t-L/c$), respectively.  
In the context of pulsar timing, these two terms 
are called the {\em Earth term} and {\em pulsar term}, 
respectively.

\begin{figure}[h!tbp]
\begin{center}
\includegraphics[angle=0,width=.6\columnwidth]{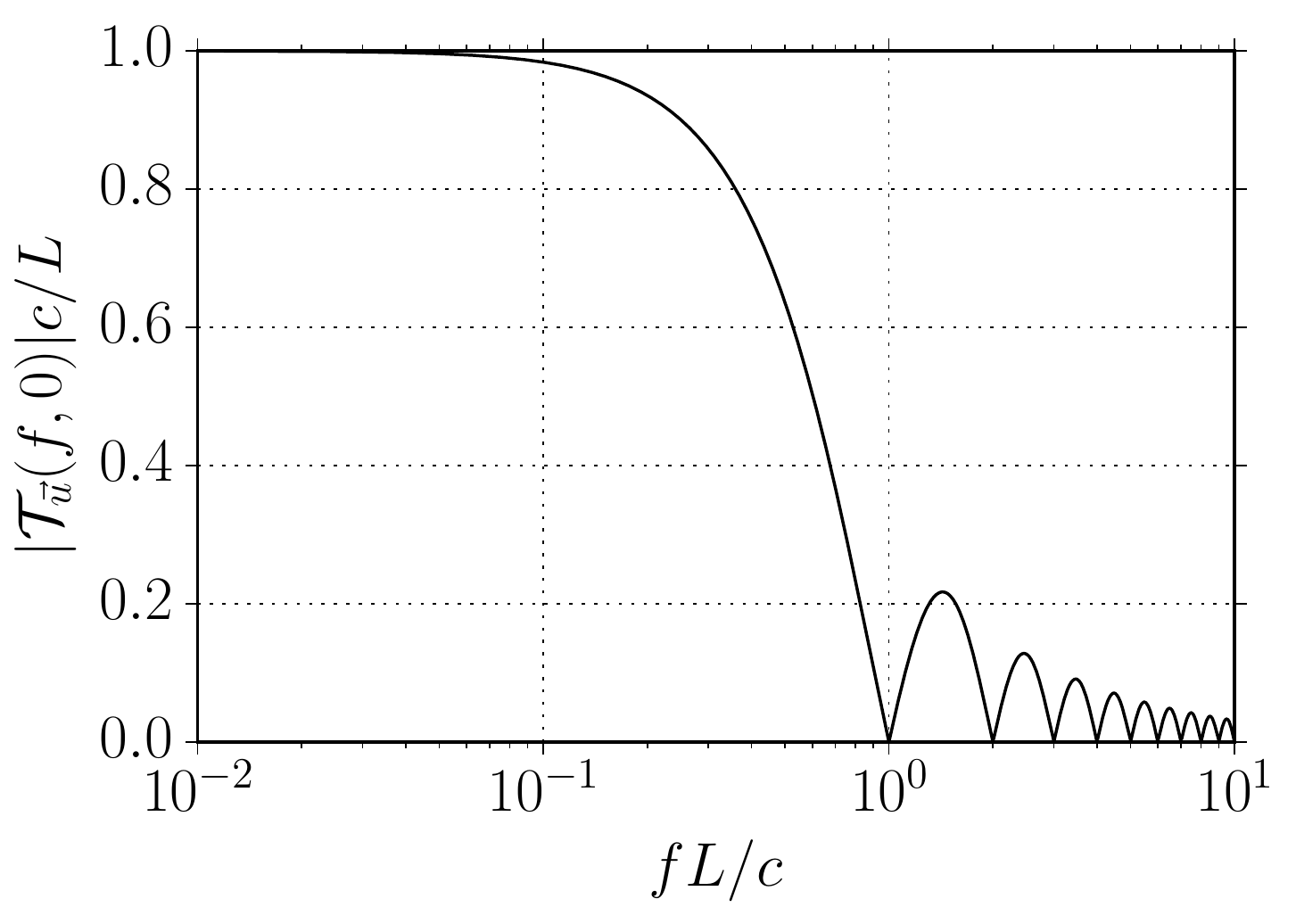}
\caption{Magnitude of the one-way tracking 
timing transfer funtion $|{\cal T}_{\vec u}(f,0)|$
for normal incidence of the gravitational wave,
plotted on a logarithmic frequency scale.
Nulls in the transfer function occur at frequencies equal to
integer multiples of $c/L$.}
\label{f:timingtransfer}
\end{center}
\end{figure}

From Equation (\ref{e:deltaT2}), we can read-off the 
response function for a timing residual measurement,
$h_{\rm timing}(t)\equiv \Delta T(t)$.
It is
%
%
\be
R^{ab}_{\rm timing}(f,\hat n)
=\frac{1}{2}
u^a u^b\,
{\cal T}_{\vec u}(f,\hat n\cdot\hat u)
e^{i2\pi f\hat n\cdot\vec r_2/c}\,,
\label{e:onewayRAtiming}
\ee
where
\be
\begin{aligned}
{\cal T}_{\vec u}(f,\hat n\cdot\hat u)
&\equiv \frac{1}{i 2\pi f}\,
\frac{1}{1+\hat n\cdot \hat u}\,
\left[1-e^{-\frac{i2\pi f L}{c}(1+\hat n\cdot\hat u)}\right]
\\
&= \frac{L}{c}
e^{-\frac{i\pi fL}{c}(1+\hat n\cdot\hat u)}\,
{\rm sinc}\left(\frac{\pi fL}{c}[1+\hat n\cdot\hat u]\right)
\end{aligned}
\ee
is the {\em timing transfer function} for one-way 
photon propagation along $\vec u=L\hat u$.
(Here ${\rm sinc}\,x \equiv \sin x/x$.)
If we choose $\vec r_2$ to be the origin of coordinates, 
then ${\cal T}_{\vec u}(f,\hat n\cdot\hat u)$ 
contains all the frequency-dependence of the timing response.
For example, for normal incidence of the gravitational wave
($\hat n\cdot\hat u=0$),
$|{\cal T}_{\vec u}(f,0)| = (L/c)\,|{\rm sinc}(\pi fL/c)|$.
Figure~\ref{f:timingtransfer} is a plot of 
$|{\cal T}_{\vec u}(f,0)|$ versus frequency on a logarithmic
frequency scale.

If we choose instead to measure the fractional Doppler 
frequency shift of the incoming photons, 
then we need to differentiate the timing response 
with respect to $t$ as
indicated in (\ref{e:rdoppler}).
This simply pulls-down a factor of $i2\pi f$ from the
exponential in $\Delta T(t)$, leading to
\be
R^{ab}_{\rm doppler}(f,\hat n)
=
i 2\pi f\,
R^{ab}_{\rm timing}(f,\hat n)\,.
\label{e:doppler-timing}
\ee
Thus, the frequency-dependence of the Doppler frequency
response is $i2\pi f$ times the timing transfer function
${\cal T}_{\vec u}(f,\hat n\cdot \hat u)$.
All of the above remarks are relevant for pulsar 
timing and {\em one-way} spacecraft Doppler tracking.

\begin{figure}[h!tbp]
\begin{center}
\includegraphics[angle=0,width=.55\columnwidth]{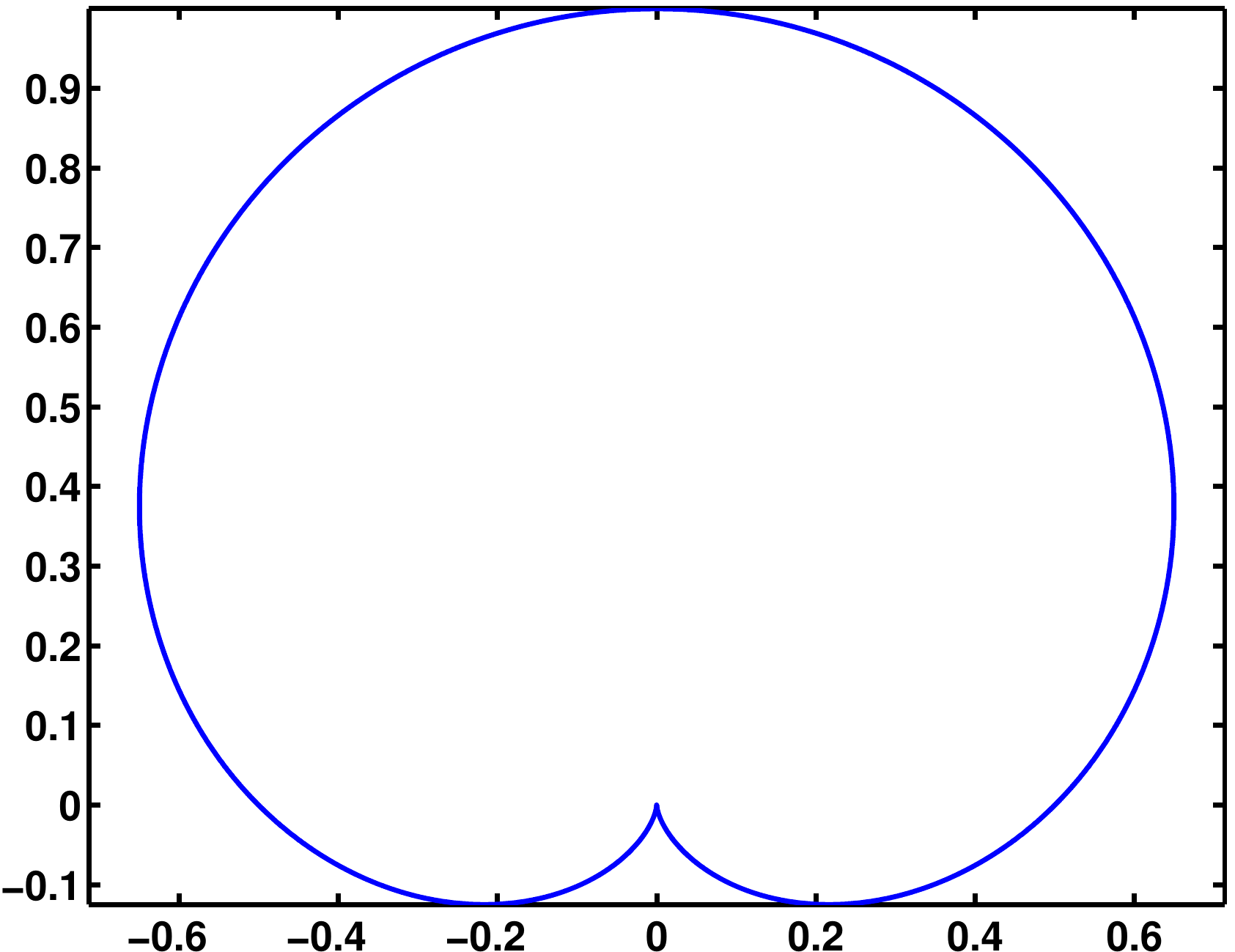} 
\caption{Antenna pattern for unpolarized gravitational
waves for a one-way tracking Doppler frequency measurement
with $\hat u =-\hat z$.
The gravitational waves propagate toward the origin.
The 3-d antenna pattern is axially symmetric around $\hat u$.}
\label{f:pulsarpeanut}
\end{center}
\end{figure}

\begin{figure}[h!tbp]
\begin{center}
\includegraphics[trim=3cm 6.5cm 3cm 3.5cm, clip=true, angle=0, width=0.49\textwidth]{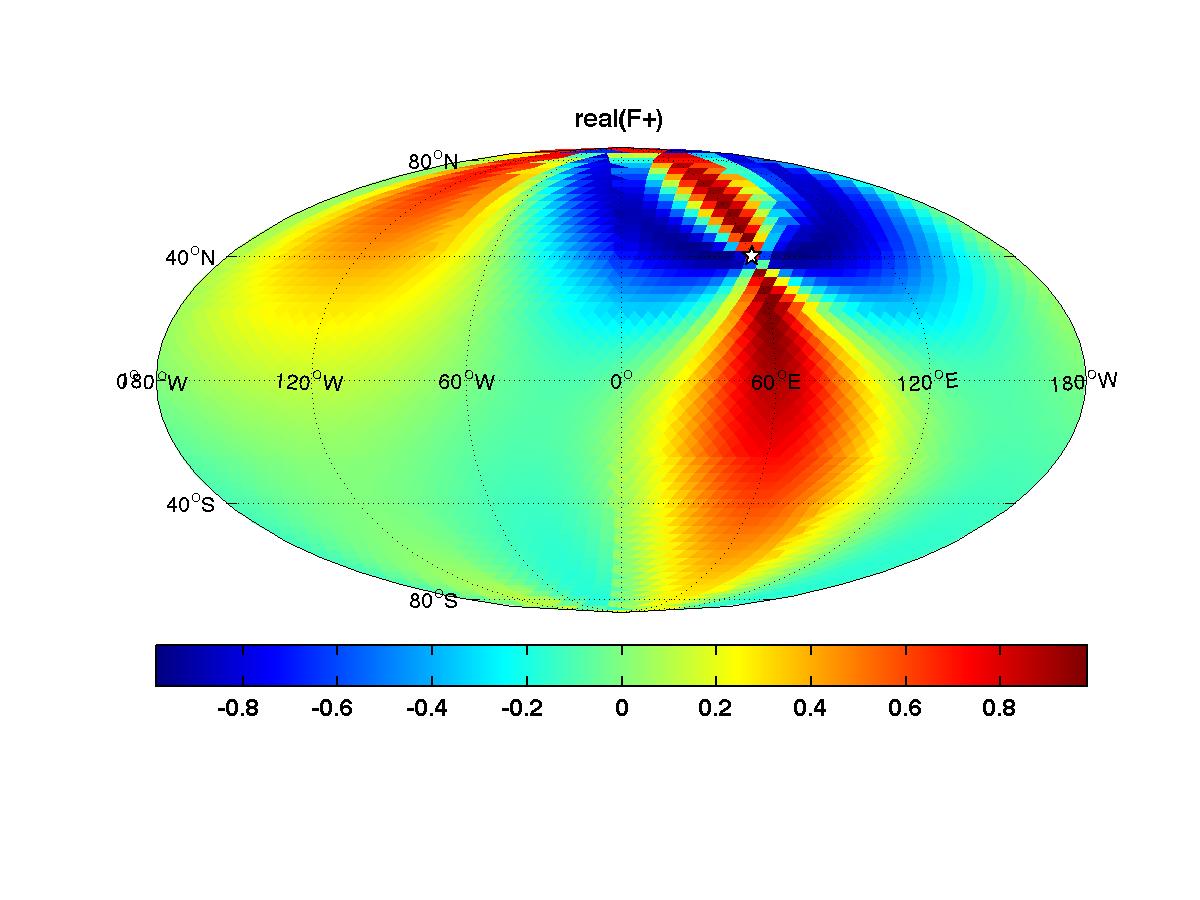}
\includegraphics[trim=3cm 6.5cm 3cm 3.5cm, clip=true, angle=0, width=0.49\textwidth]{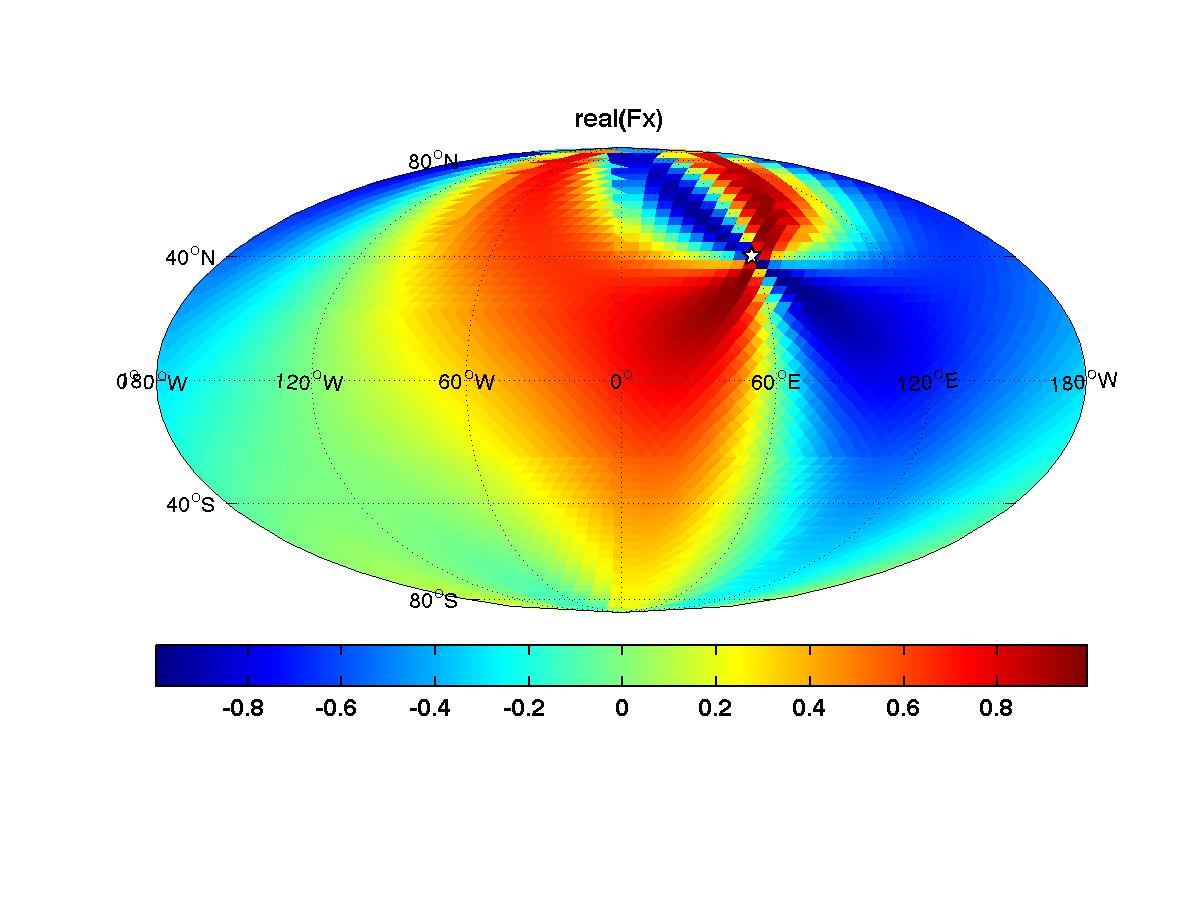}
\caption{Mollweide projections of the repsonse functions
$R_{\rm doppler}^+(\hat n)$, $R_{\rm doppler}^\times(\hat n)$, for one-way tracking 
Doppler frequency measurements 
corresponding to a pulsar located in the
direction of the white star 
$(\theta,\phi)=(50^\circ, 60^\circ)$.
The imaginary parts of both response functions are identically zero, so are not shown above.}
\label{f:RpRc_pulsar}
\end{center}
\end{figure}

In Figure~\ref{f:pulsarpeanut} we plot the 
antenna beam pattern (\ref{e:beam}) 
for unpolarized gravitational waves 
for a one-way tracking Doppler frequency measurement
(e.g., pulsar timing) with $\hat u=-\hat z$.
For this calculation, we chose $\vec r_2=0$ and ignored 
the exponential (i.e., `pulsar') term in the timing
transfer function, which yields
\be
R^A_{\rm doppler}(f,\hat n) =
\frac{1}{2} \frac{u^a u^b}{1+\hat u\cdot\hat n} 
e_{ab}^A(\hat n)
\qquad({\rm Earth\ term\ only})\,,
\label{e:pulsarresponse-earthonly}
\ee
for the $A=+,\times$ polarization modes.
Setting $\hat u=-\hat z$ and taking the gravitational waves 
to propagate inward (toward the origin), we find
\be
{\cal R}_{\rm doppler}(\hat n) 
= \frac{1}{2}(1+\cos\theta)\,,
\ee
which is axially symmetric around $\hat u$.
The response is maximum when the photon and 
the gravitational wave both propagate in the same
direction.
Figure~\ref{f:RpRc_pulsar} shows plots of the real parts of the individual 
polarization basis response functions (\ref{e:pulsarresponse-earthonly}),
represented as color bar plots on a Mollweide projection of the sky.
For this plot we chose the pulsar to be located in the
direction $(\theta,\phi) = (50^\circ,60^\circ)$.
(The direction $\hat p$ to the pulsar is given by $\hat p = -\hat u$.)
The imaginary parts of both response functions are identically
zero, so are not shown in the figure.

Making the same approximations as above, 
we can also calculate the corresponding Doppler-frequency
response functions for the gradient and curl
tensor spherical harmonic components 
$\{a^G_{lm}(f),a^C_{lm}(f)\}$
by performing the integration in (\ref{e:RP}).
As shown in \cite{Gair-et-al:2014}, this leads to%
\footnote{There is a factor of $(-1)^l$ difference
between $R^G_{(lm)}(f)$ in (\ref{e:RP_PTA}) and 
(92) in \cite{Gair-et-al:2014}.
The difference is due to the change in expressing the
response functions in terms of the direction to the
gravitational-wave source, $\hat n$, as opposed to 
the direction of gravitational-wave propagation, $\hat k=-\hat n$.
Appendix~\ref{s:translation} provides expressions
relating the response functions calculated using 
these two different conventions.}
\be
R^G_{(lm)}(f) = 2\pi\,{}^{(2)}\!N_l Y_{lm}(\hat p)\,,
\qquad
R^C_{(lm)}(f) = 0\,,
\label{e:RP_PTA}
\ee
where ${}^{(2)}\!N_l$ is given by (\ref{e:N_l}) and
$\hat p = -\hat u$ is the direction on the sky 
to a pulsar.
Note, somewhat surprisingly, that the curl response 
is {\em identically zero}.
We will discuss the consequences of this result
in more detail in Section~\ref{s:phase-coherent-PTA},
in the context of phase-coherent mapping of anisotropic
gravitational-wave backgrounds.

\subsubsection{Two-way tracking}

To calculate $\Delta T(t)$ for {\em two-way} 
spacecraft Doppler tracking,
we need to generalize the calculation of the previous 
subsection to include a return trip of the photon from 
$\vec r_2$ back to $\vec r_1$.
This can be done by simply summing the expressions
for the one-way timing residuals:
\be
\Delta T(t) =
\Delta T_{12}(t-L/c) + \Delta T_{21}(t)
\ee
where the subscripts on the $\Delta T$'s on
the right-hand side of the above equation 
indicate the direction of one-way photon 
propagation (e.g., $12$ indicates photon propagation from test
mass 1 to test mass 2), and the arguments of $\Delta T_{12}$
and $\Delta T_{21}$ indicate when the photon arrived at 
test mass 2 and test mass 1, respectively.
Doing this calculation leads to the following expression 
for the timing residual:
\be
\begin{aligned}
{\Delta T(t)}
&=
\int_{-\infty}^\infty df
\frac{1}{2}
u^a u^b
h_{ab}(f,\hat n)\,
\frac{1}{i 2\pi f}\,
\bigg[
\frac{1}{1-\hat n\cdot \hat u}\,
e^{i 2\pi f(t+\hat n\cdot \vec r_1/c)}
\\
&\qquad
-\frac{2\hat n\cdot\hat u}{1-(\hat n\cdot \hat u)^2}\,
e^{i 2\pi f(t-L/c+\hat n\cdot \vec r_2/c)}
-\frac{1}{1+\hat n\cdot \hat u}\,
e^{i 2\pi f(t-2L/c+\hat n\cdot \vec r_1/c)}
\bigg]\,,
\label{e:deltaTrt}
\end{aligned}
\ee
which has {\em three} terms corresponding to 
the final reception of the photon at $\vec r_1$ at time $t$,
the reflection of the photon at $\vec r_2$ at time $t-L/c$, and
the emission of the photon at $\vec r_1$ at time $t-2L/c$.
The timing response function is given by 
\be
R^{ab}_{\rm timing}(f,\hat n)
=\frac{1}{2}u^a u^b\,
{\cal T}_{{\vec u},{\rm rt}}(f,\hat n\cdot\hat u)
e^{i2\pi f\hat n\cdot \vec r_1/c}\,,
\label{e:twowayRAtiming}
\ee
where
\be
\begin{aligned}
{\cal T}_{{\vec u},{\rm rt}}(f,\hat n\cdot\hat u)
\equiv \frac{L}{c}
e^{-\frac{i 2\pi f L}{c}}\,
&\bigg[e^{-\frac{i\pi f L}{c}(1-{\hat n}\cdot{\hat u})}\,
{\rm sinc}\left(\frac{\pi f L}{c}[1+{\hat n}\cdot{\hat u}]\right)
\\
&\qquad
 +e^{\frac{i\pi f L}{c}(1+{\hat n}\cdot{\hat u})}\,
{\rm sinc}\left(\frac{\pi f L}{c}[1-{\hat n}\cdot{\hat u}]\right)
\bigg]
\end{aligned}
\ee
is the timing transfer function for {\em two-way} (or roundtrip) 
photon propagation along $\vec u$ and back.
For normal incidence, the magnitude of the timing transfer 
function is given by
$|{\cal T}_{{\vec u},{\rm rt}}(f,0)|=
(2L/c)|{\rm sinc}(2\pi fL/c)|$, 
which is  identical to the expression for one-way tracking 
with $L/c$ replaced by $2L/c$.
We also note that if we choose the origin of coordinates to
be at $\vec r_1$ (which we can always do for a single
detector), and if the frequency $f$ is such that $fL/c\ll 1$, 
then the timing response simplifies to
\be
R^{ab}_{\rm timing}(f,\hat n) =
u^a u^b\,\frac{L}{c}
\qquad
({\rm for}\ fL/c\ll 1)\,.
\label{e:RAuLW}
\ee
We will use the terminology {\em small-antenna limit}
(instead of {\em long-wavelength limit}) for this 
type of limit, since it avoids an ambiguity that might
arise if we want to compare three or more length scales.
For example, if we have two detectors that are physically
separated and the wavelength of a gravitational wave 
is {\em large} compared to the size of 
each  detector but {\em small} compared to the separation 
of the detectors, we would be in the long-wavelength
limit with respect to detector size but in the 
short-wavelength limit with respect to detector separation.
(This is actually the case for the current network
of ground-based interferometers.)
The terminology {\em small-antenna, large-separation limit} 
is more appropriate for this case.

\subsubsection{Michelson interferometer}

\begin{figure}[h!tbp]
\begin{center}
\includegraphics[angle=0,width=.4\columnwidth]{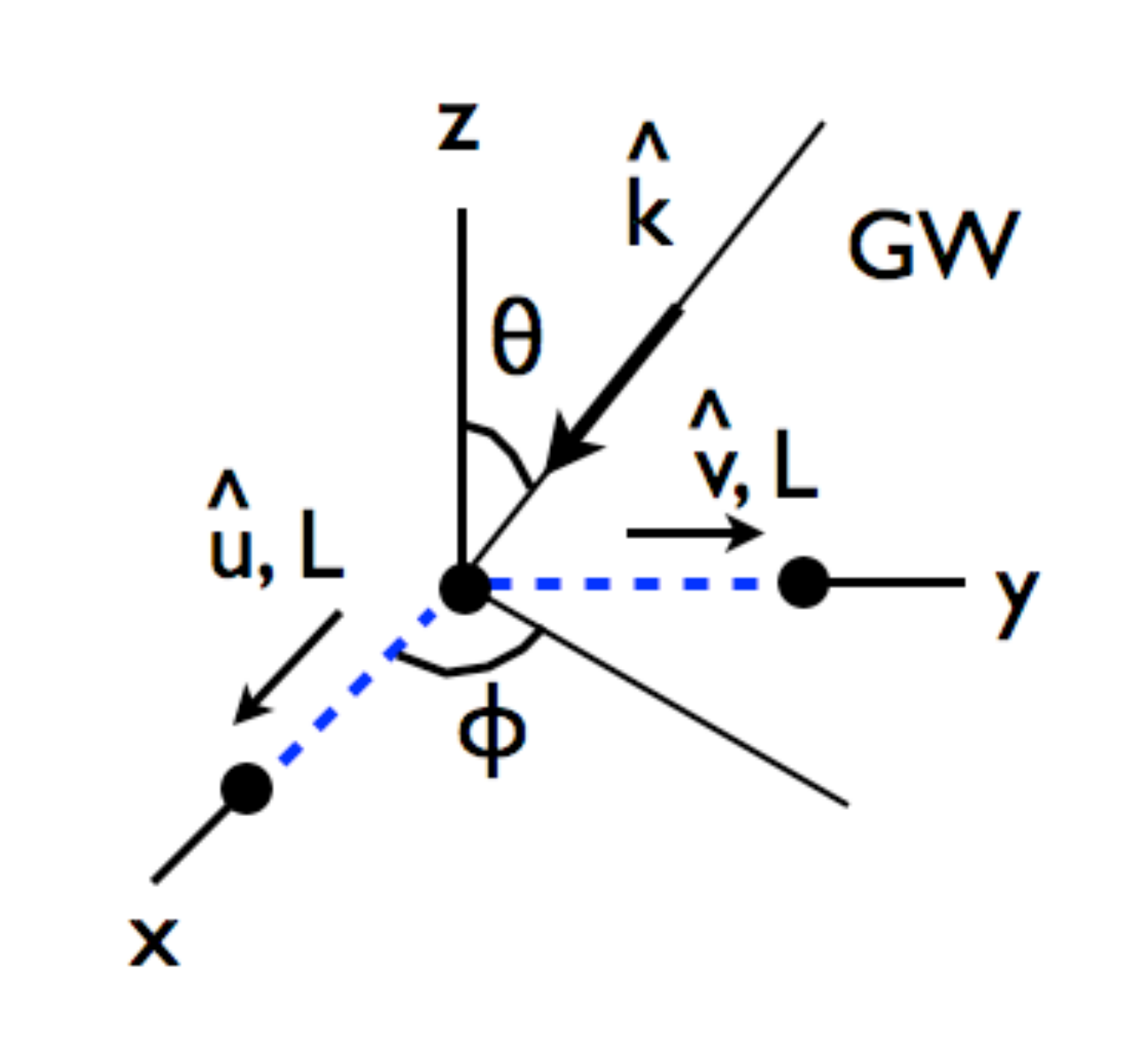}
\caption{Geometry for calculating the difference in
the round-trip light-travel times in the two arms 
of a Michelson interferometer:
$\hat k=-\hat n$ is the direction of propagation for a plane
gravitational wave;
$\hat u$ and $\hat v$ are unit vectors that point
from the vertex of the interferometer (e.g., the 
beam splitter) to the two end masses; and
$L$ denotes the lengths of each of the arms in the 
absence of a gravitational wave.}
\label{f:interferometer}
\end{center}
\end{figure}

\begin{figure}[h!tbp]
\begin{center}
\begin{tabular}{c}
\includegraphics[angle=0,width=\columnwidth]{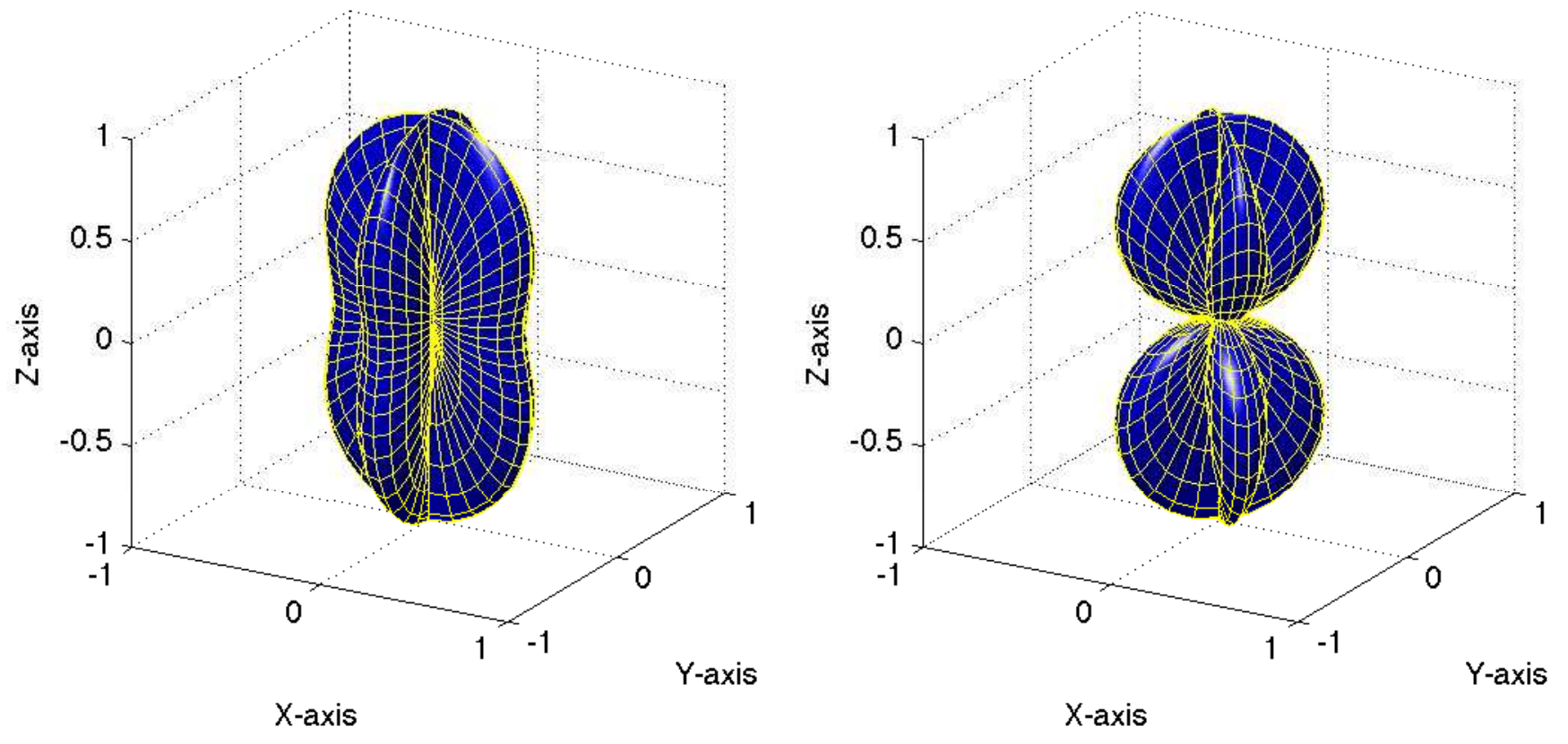}
\\
\includegraphics[angle=0,width=\columnwidth]{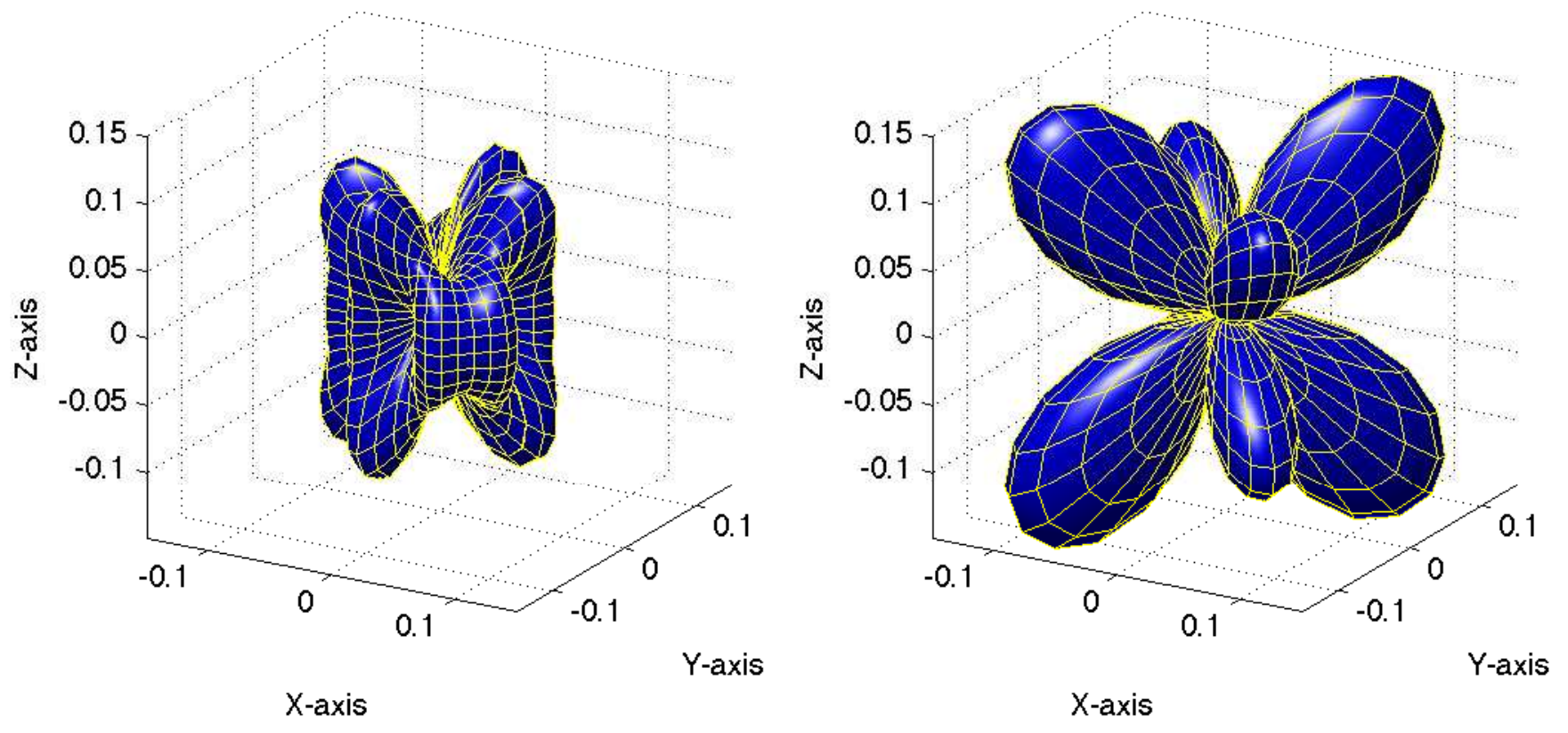}
\end{tabular}
\caption{Antenna patterns for 
Michelson interferometer strain response
$|R^+_{\rm strain}|$ and 
$|R^\times_{\rm strain}|$
evaluated in the small-antenna limit,
$f=0$ (top two plots)
and at the free-spectral range frequency,
$f=c/(2L)$
(bottom two plots).
The interferometer arms point in the $\hat x$ 
and $\hat y$ directions.
Note the change in the scale of the axes between the
top and bottom two plots.}
\label{f:antenna+x}
\end{center}
\end{figure}

\begin{figure}[h!tbp]
\begin{center}
\includegraphics[angle=0,width=\columnwidth]{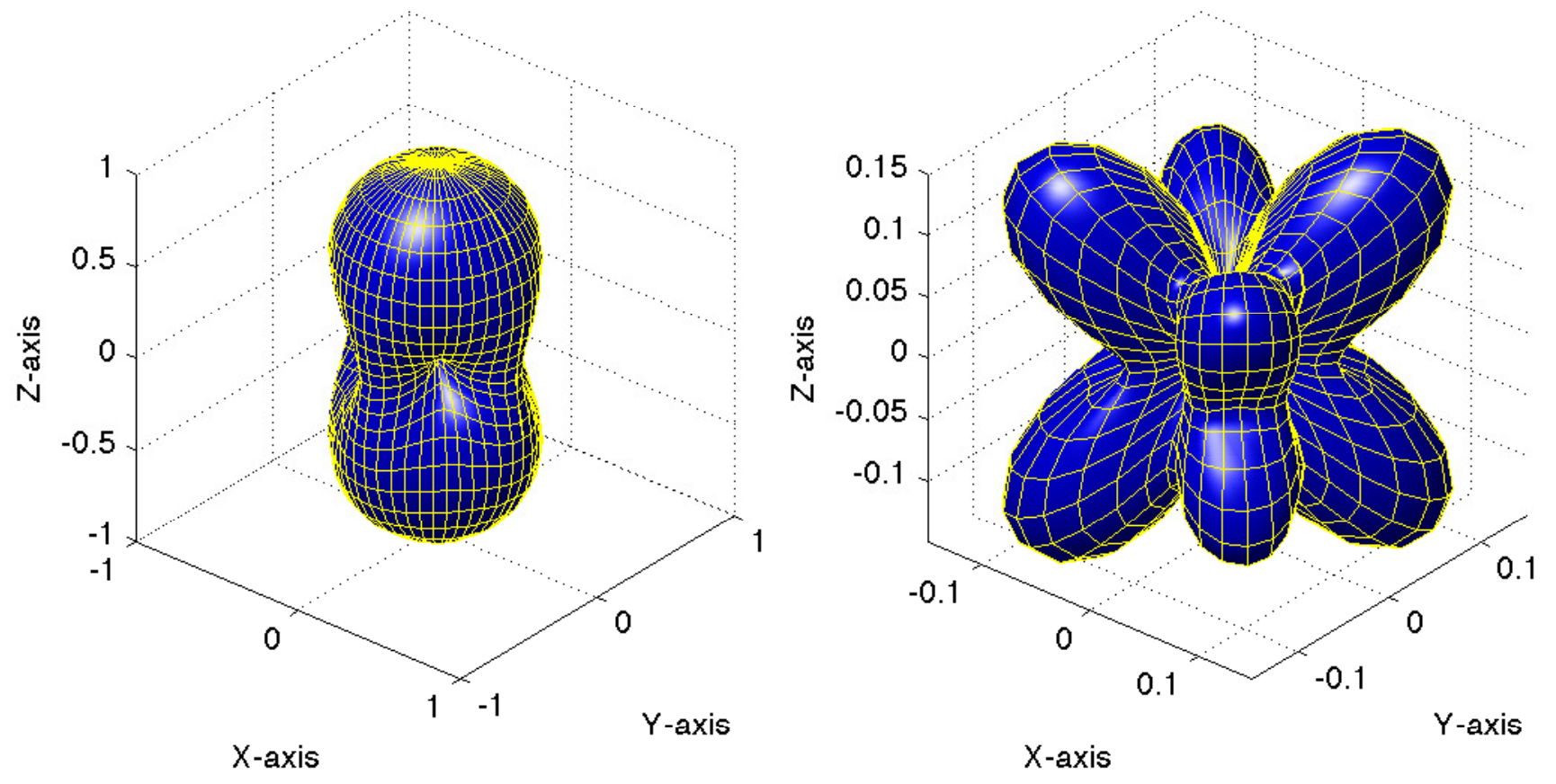}
\caption{Antenna pattern for 
Michelson interferometer strain response
to unpolarized gravitational waves 
evaluated in the small-antenna limit,
$f=0$ (left plot)
and at the free-spectral range frequency,
$f=c/(2L)$
(right plot).
The interferometer arms point in the $\hat x$ 
and $\hat y$ directions.
Note the change in the scale of the axes between the
two plots.}
\label{f:peanuts}
\end{center}
\end{figure}

\begin{figure}[h!tbp]
\begin{center}
\begin{tabular}{c}
\includegraphics[angle=0,width=.7\columnwidth]{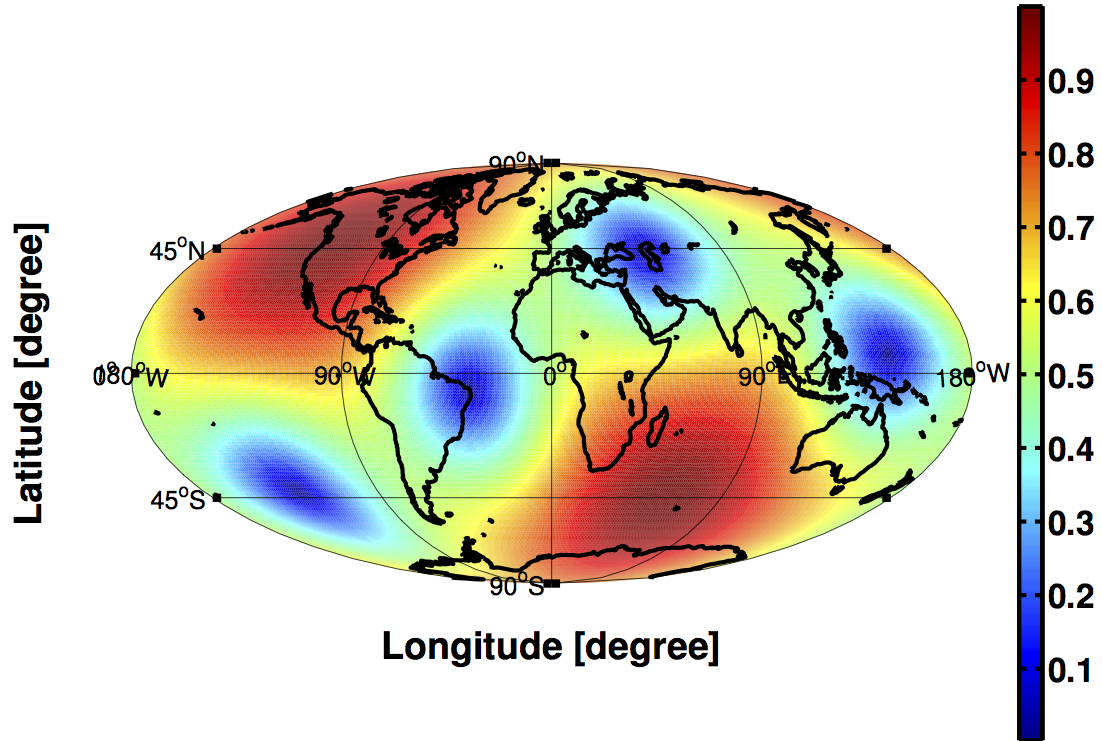}
\\
\includegraphics[angle=0,width=.7\columnwidth]{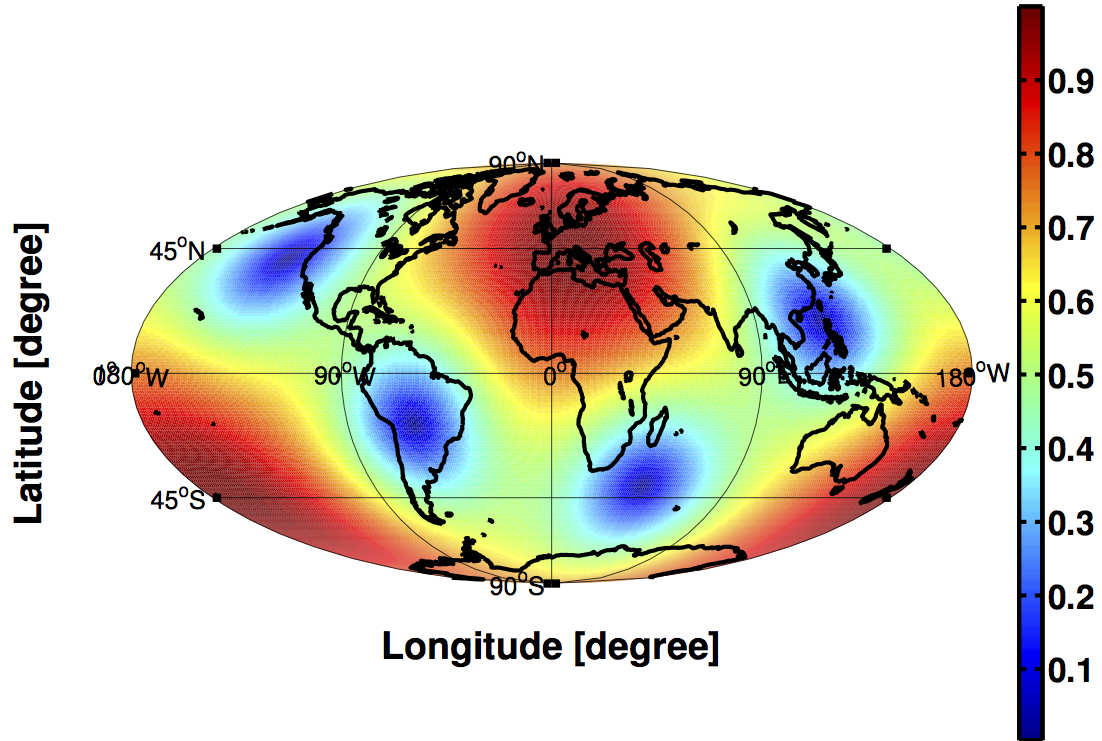}
\end{tabular}
\caption{Antenna patterns for the 
strain response to unpolarized gravitational waves 
for the LIGO Hanford (top panel) and 
Virgo (bottom panel) interferometers
evaluated in the small-antenna limit.
The antenna patterns are represented as colorbar plots 
on a Mollweide projection of the Earth.
Note that the maxima of the antenna patterns 
(the centers of the red regions) are
directly above (and below) the location of the two 
interferometers---in Hanford, WA and Cascina, Italy, respectively.
The blue regions correspond to the minima of the 
antenna patterns---i.e., the `dimples' in the left panel 
plot of Figure~\ref{f:peanuts}.}
\label{f:LIGOVirgo-peanuts}
\end{center}
\end{figure}

For an equal-arm Michelson interferometer, the
timing residual that we calculate is 
the difference in the round-trip 
light-travel times down and back each of the arms.
(See Figure~\ref{f:interferometer}.)
If we let $\vec u$ and $\vec v$ denote the 
vectors pointing from e.g., the beam splitter
to the two end mirrors for LIGO, or from one 
spacecraft to the other two spacecraft for LISA,
then%
\footnote{Although Figure~\ref{f:interferometer}
shows $\hat u$ and $\hat v$ making right angles with
one another, the following calculation is valid
for $\hat u$ and $\hat v$ separated by an {\em arbitrary}
angle.}
\be
\Delta T(t)\equiv 
T_{{\vec u},{\rm rt}}(t)-
T_{{\vec v},{\rm rt}}(t)
=
\Delta T_{{\vec u},{\rm rt}}(t)-
\Delta T_{{\vec v},{\rm rt}}(t)\,,
\ee
where the last equality is valid for an 
equal-arm interferometer.
But we just calculated these 
single-arm round-trip 
$\Delta T$'s in the previous section.
Thus, the timing response of an 
equal-arm Michelson is simply
\begin{align}
R^{ab}_{\rm timing}(f,\hat n)
&=\frac{1}{2}
\left[
u^a u^b\,{\cal T}_{{\vec u},{\rm rt}}(f,\hat n\cdot\hat u)-
v^a v^b\,{\cal T}_{{\vec v},{\rm rt}}(f,\hat n\cdot\hat v)
\right]\,,
\label{e:michRAtiming}
\end{align}
where we have chosen the origin of coordinates to be at
the vertex of the interferometer.
The phase and strain responses of a Michelson 
are related to the timing response by constant
multiplicative factors, cf.\ (\ref{e:rphase}) and 
(\ref{e:rstrain}), so that
\be
\begin{aligned}
R^{ab}_{\rm phase}(f,\hat n)
&= 2\pi\nu_0 
R^{ab}_{\rm timing}(f,\hat n)\,,
\\
R^{ab}_{\rm strain}(f,\hat n)
&= 
R^{ab}_{\rm timing}(f,\hat n)/(2L/c)\,,
\end{aligned}
\ee
where $\nu_0$ is the frequency of the laser.
Note that in the small-antenna limit, 
which is valid for the LIGO detectors 
below a few kHz, the strain response is
given by
\be
R^{ab}_{\rm strain}(f,\hat n) =
\frac{1}{2}(u^a u^b-v^a v^b)\,
\qquad
({\rm for}\ fL/c\ll 1)\,.
\label{e:RAuvLW}
\ee
Plots of the antenna patterns for the strain 
response to $A=+,\times$ 
polarized gravitational waves are given in
Figure~\ref{f:antenna+x}, for both the 
small-antenna limit (where we simply set $f=0$) 
and at the {\em free-spectral range} of the 
interferometer, $f = f_{\rm fsr}\equiv c/(2L)$.
Similar plots of the antenna patterns for 
unpolarized gravitational waves are given
in Figure~\ref{f:peanuts}.
In Figure~\ref{f:LIGOVirgo-peanuts} we 
show colorbar plots of the antenna patterns 
for the strain response to unpolarized gravitational 
waves for the LIGO Hanford and Virgo interferometers 
(located in Hanford, WA and Cascina, Italy, respectively), 
again evaluated in the small-antenna limit. 

We can also calculate the strain response of an interferometer
to the gradient and curl tensor spherical harmonic components
$\{a^G_{(lm)}(f), a^C_{(lm)}(f)\}$ by performing the integration 
in (\ref{e:RP}).
As shown in Appendix~E of \cite{Gair-et-al:2014}, this leads to
\be
R^G_{(lm)}(f) =
\delta_{l2}\frac{4\pi}{5}\sqrt{\frac{1}{3}}
\left[Y_{2m}(\hat u) - Y_{2m}(\hat v)\right]\,,
\qquad
R^C_{(lm)}(f) = 0\,,
\label{e:RP_IFO}
\ee
for an interferometer in the small-antenna limit, where the 
vertex is at the origin of coordinates,
and $\hat u$, $\hat v$ are unit vectors pointing
in the direction of the interferometer arms.
Similar to (\ref{e:RP_PTA}) for pulsar timing, 
the curl response is again identically zero.
We will discuss the consequences of this result
in more detail in Section~\ref{s:phase-coherent-IFO},
in the context of phase-coherent mapping of 
anisotropic gravitational-wave backgrounds.

\subsection{Overlap functions}
\label{s:overlap}

\noindent
As mentioned in Section~\ref{s:corr},
a stochastic gravitational-wave background manifests 
itself as a non-vanishing correlation between the data 
taken by two or more detectors.
This correlation differs, in general, from that due
to instrumental noise, allowing us to distinguish between 
a stochastic gravitational-wave signal and other noise
sources.
In this section, we calculate the expected correlation 
due to a gravitational-wave background,
allowing for non-trivial detector response functions and 
non-trivial detector geometry.
Interested readers can find more details in 
\cite{Hellings-Downs:1983, Christensen:PhD, Christensen:1992, 
Flanagan:1993, Finn-et-al:2009}.

\subsubsection{Definition}

Let $d_I$ and $d_J$ denote the data taken by two 
detectors labeled by $I$ and $J$.
In the presence of a gravitational wave, these data will 
have the form
\be
\begin{aligned}
d_I &= h_I + n_I\,,
\\
d_J &= h_J + n_J\,,
\end{aligned}
\ee
where $h_{I,J}$ denote the response of detectors $I$, $J$ 
to the gravitational wave, and $n_{I,J}$ denote the 
contribution from instrumental noise.
If the instrumental noise in the two detectors
are uncorrelated with one another, it follows that the 
expected correlation of the data is just the expected
correlation of the detector responses,
$\langle d_I d_J\rangle = \langle h_I h_J\rangle$.
If we also assume that the gravitational wave is due 
to a stationary, Gaussian, isotropic, and unpolarized
stochastic background, then
\be
\langle h_I(t) h_J(t')\rangle
=\frac{1}{2}\int_{-\infty}^\infty df\>
e^{i 2\pi f(t-t')}\Gamma_{IJ}(f)S_h(f)\,,
\label{e:rIrJtime}
\ee
where $S_h(f)$ is the one-sided 
strain power spectral density of the gravitational-wave 
background, computed from the expectation values of
the Fourier components of the metric perturbations
(\ref{e:iso_hh}), and
\be
\Gamma_{IJ}(f)
\equiv
\frac{1}{8\pi}
\int d^2\Omega_{\hat n} 
\sum_A
R_I^A(f,\hat n)R_J^{A}{}^*(f,\hat n)
\label{e:GammaIJ}
\ee
is the so-called {\em overlap function} for the
two detectors $I$, $J$ written in terms of the
polarization-basis response function $R^A_{I,J}(f,\hat n)$,%
\footnote{Recall from Footnote~\ref{fn:response-phase}
that the phase factors 
$e^{i 2\pi f\hat n\cdot \vec x_{I,J}/c}$
are already contained in our definition of the 
response functions $R^A_{I,J}(f,k)$.
If we explicitly display this dependence then
\be
\Gamma_{IJ}(f)
\equiv
\frac{1}{8\pi}
\int d^2\Omega_{\hat n} 
\sum_A
\bar R_I^A(f,\hat n)\bar R_J^{A}{}^*(f,\hat n)
e^{i 2\pi f\hat n\cdot(\vec x_I -\vec x_J)/c}\,,
\nonumber
\label{e:GammaIJ-alt}
\ee
where $\bar R^A_{I,J}(f,\hat n) \equiv \bar R^{ab}_{I,J}(f,\hat n)e^A_{ab}(\hat n)$.
One often sees this latter expression for $\Gamma_{IJ}(f)$ 
in the literature.}
where $A=\{+,\times\}$.
In terms of the tensor spherical harmonic-basis
response functions $R^P_{I,J(lm)}(f)$, we would have
\be
\Gamma_{IJ}(f)
=
\frac{1}{8\pi}
\sum_{(lm)}
\sum_P
R_{I(lm)}^P(f)R_{J(lm)}^{P*}(f)\,,
\label{e:GammaIJ-spherical}
\ee
%
where $P=\{G,C\}$ for the gradient and curl tensor spherical harmonic 
components.

\subsubsection{Interpretation}

The overlap function $\Gamma_{IJ}(f)$ 
quantifies the reduction in sensitivity of the
cross-correlation to a stochastic gravitational-wave
background due to the non-trivial response of the 
detectors and their separation and orientation
relative to one another.
This meaning of the overlap function is most easily 
seen in the frequency domain,
where (\ref{e:rIrJtime}) becomes
\be
\langle\tilde h_I(f) \tilde h_J^*(f')\rangle
=
\frac{1}{2}\delta(f-f')\, \Gamma_{IJ}(f)S_h(f)\,.
\label{e:rIrJfreq}
\ee
This implies
\be
\tilde C_{h_I h_J}(f) = \Gamma_{IJ}(f) S_h(f)\,,
\ee
where $\tilde C_{h_I h_J}(f)$ is the (one-sided)
cross-spectrum of the response in the two detectors.
Thus, $\Gamma_{IJ}(f)$ can be interpreted as the 
transfer function between gravitational-wave 
strain power $S_h(f)$ and detector response cross-power
$\tilde C_{h_I h_J}(f)$.

Expression (\ref{e:GammaIJ}) for the overlap 
function involves four length scales:
the lengths of the two detectors, 
$L_I$ and $L_J$, 
which appear in the response functions $R^A_{I,J}(f,\hat n)$;
the separation of the detectors, 
$s\equiv |\vec x_I-\vec x_J|$, 
which appears in the exponential factor;
and the wavelength of the gravitational waves,
$\lambda = c/f$. 
In general, one has to evaluate the integral in 
(\ref{e:GammaIJ}) {\em numerically}, due to the 
non-trivial frequency dependence of 
the response functions.
However, as we shall see in Section~\ref{s:orf-examples},
in certain limiting cases of the ratio of these length scales, 
we can do the integral {\em analytically} 
and obtain relatively simple expressions for the 
overlap function in terms of spherical Bessel
or trigonometric functions.
This is the case for ground-based interferometers,
which operate in the {\em small-antenna limit}---i.e., 
$fL/c\ll 1$ for both detectors, even though the separation 
can be large compared to the wavelength, $fs/c \gtrsim 1$.
It is also the case for pulsar timing arrays, 
which operate in the {\em large-antenna, 
small-separation limit}, 
since $fL/c\gg 1$ for each pulsar and $fs/c\ll 1$
for different radio receivers on Earth.
(The Earth effectively resides at the solar system barycenter
relative to the wavelength of the gravitational waves relevant
for pulsar timing.)

\subsubsection{Normalization}

It is often convenient to define a {\em normalized}
overlap function $\gamma_{IJ}(f)\propto \Gamma_{IJ}(f)$
by requiring that $\gamma_{IJ}(0)=1$ for two 
detectors that are co-located and co-aligned.
For the strain response of two identical equal-arm 
Michelson interferometers, this leads to the relation
\be
\gamma_{IJ}(f) = \frac{5}{\sin^2\beta}\,\Gamma_{IJ}(f)
\label{e:gamma_normalized}
\ee
where $\beta$ is the opening angle between the two arms
($\pi/2$ for LIGO and $\pi/3$ for LISA).

\subsubsection{Auto-correlated response}
\label{s:autocorrelatedresponse}

\begin{figure}[h!tbp]
\begin{center}
\includegraphics[angle=0,width=.6\columnwidth]{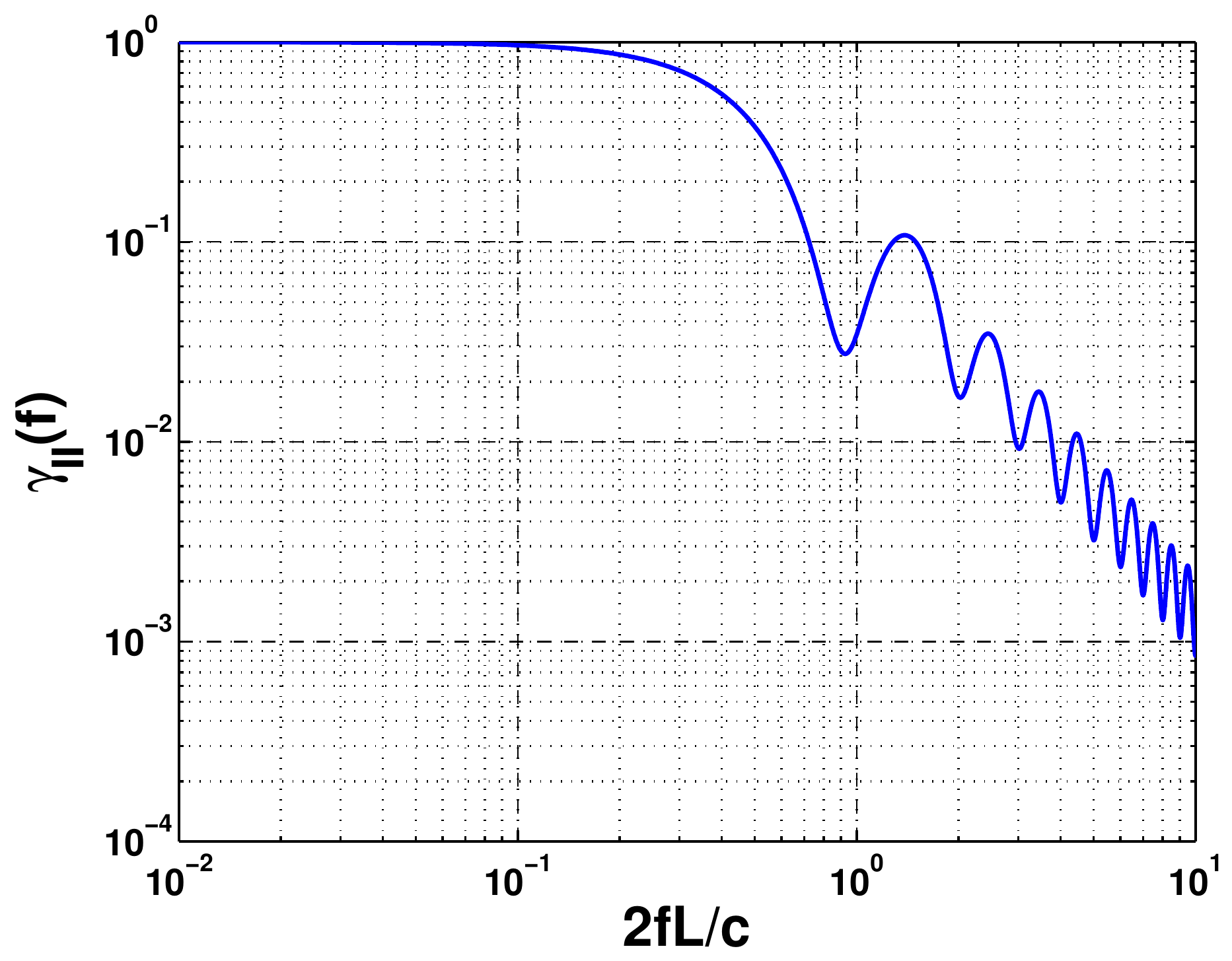}
\caption{A plot of the normalized transfer function
$\gamma_{II}(f)$ for the strain response of an equal-arm 
Michelson interferometer.
The dips in the transfer function occur around 
integer multiples of $c/(2L)$.}
\label{f:gammaII}
\end{center}
\end{figure}

\begin{figure}[h!tbp]
\begin{center}
\includegraphics[angle=0,width=.6\columnwidth]{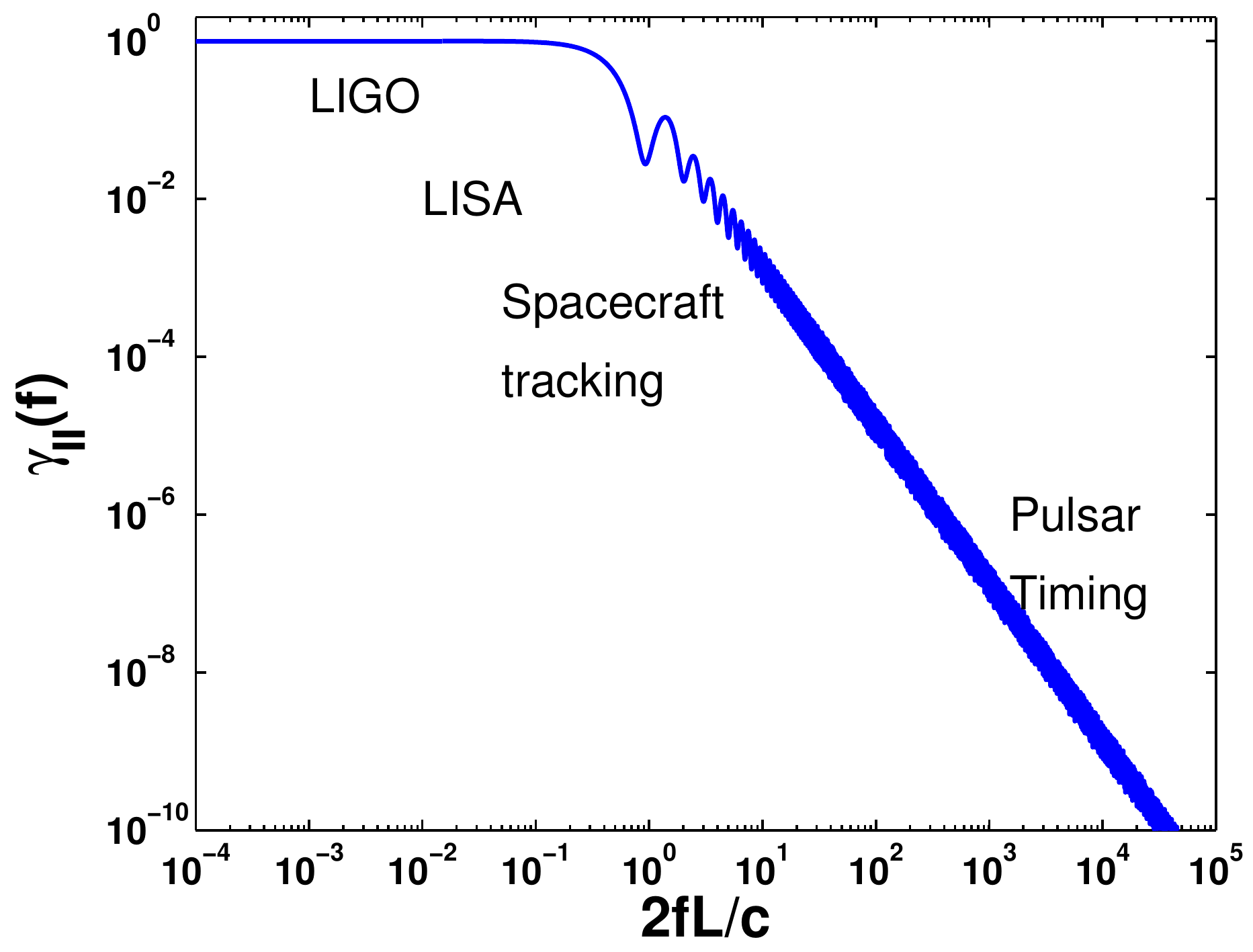}
\caption{An extension of Figure~\ref{f:gammaII} to 
lower and higher frequencies, and plotted on a log-log scale.
The position of the labels show the relative 
location of the frequency bands for gravitational-wave 
searches using ground-based interferometers like LIGO, 
space-based interferometers like LISA, 
spacecraft Doppler tracking, 
and pulsar timing arrays, expressed in units of $c/(2L)$.
See also Table~\ref{t:beamdetectors} for more details.}
\label{f:gammaII_extended}
\end{center}
\end{figure}

To obtain the {\em auto-correlated} response of a 
{\em single} detector, we can simply set $I=J$ 
in the previous expressions.
This means that the gravitational-wave strain 
power $S_h(f)$ and
the detector response power $P_{h_I}(f)$ 
in detector $I$ are related by
\be
P_{h_I}(f) = \Gamma_{II}(f) S_h(f)\,,
\label{e:Ph}
\ee
where
\be
\Gamma_{II}(f) =\frac{1}{8\pi}
\int d^2\Omega_{\hat n} 
\sum_A |R_I^A(f,\hat n)|^2\,.
\label{e:GammaII}
\ee
Note that $\Gamma_{II}(f)$ 
is just the square of the antenna pattern 
for the response to unpolarized gravitational waves
{\em integrated over the whole sky}.
A plot of the normalized transfer function 
$\gamma_{II}(f)$ for the strain response 
of an equal-arm Michelson interferometer
is shown in Figure~\ref{f:gammaII}.
Compared to Figure~\ref{f:timingtransfer} for the 
timing transfer function $|{\cal T}_{\vec u}(f,0)|$ 
for one-way photon propagation evaluated at normal 
incidence of the gravitational wave,
we see that the relevant frequency scale for an 
equal-arm Michelson is $c/(2L)$ (as opposed to $c/L$) 
due to the round-trip motion of the photons.
Also, the hard nulls in Figure~\ref{f:timingtransfer} 
have been softened into {\em dips} due to averaging of 
the waves over the whole sky.
The high-frequency `bumps' for $\gamma_{II}(f)$ are
lower than those for $|{\cal T}_{\vec u}(f,0)|$ due 
to the squaring of $|R^A_I(f,\hat n)|$ which enters
into the defintion of $\Gamma_{II}(f)$ (and $\gamma_{II}(f)$).
Figure~\ref{f:gammaII_extended} is an extended 
version of Figure~\ref{f:gammaII}, 
with the appropriate frequency ranges 
for ground-based interferometers (like LIGO), 
space-based interferometers (like LISA), 
spacecraft Doppler tracking, and 
pulsar timing searches indicated on the plot.
See also Table~\ref{t:beamdetectors} for more
details.
%
\subsection{Examples of overlap functions}
\label{s:orf-examples}

\subsubsection{LHO-LLO overlap function}
\label{s:LHO-LLO-example}

As mentioned above, Earth-based interferometers like LIGO 
operate in the small-antenna limit where $fL/c\ll 1$.
This implies that the associated response functions are 
well-approximated by the expression in (\ref{e:RAuvLW}).
If we denote the unit vectors along the two arms 
of one Earth-based interferometer by $\hat u_1$ and $\hat v_1$,
and the corresponding unit vectors of a second Earth-based
interferometer by $\hat u_2$ and $\hat v_2$, 
then the strain responses in the two interferometers are simply
\be
\begin{aligned}
R^A_{1,{\rm strain}}(f,\hat n) &\simeq 
D_1^{ab}\,e^A_{ab}(\hat n)
e^{i2\pi f\hat n\cdot\vec x_1/c}\,,
\\
R^A_{2,{\rm strain}}(f,\hat n) &\simeq 
D_2^{ab}\,e^A_{ab}(\hat n)
e^{i2\pi f\hat n\cdot\vec x_2/c}\,,
\label{e:IFOstrainresponse-x0}
\end{aligned}
\ee
where
\be
D_1^{ab} \equiv \frac{1}{2}\left(u_1^a u_1^b-v_1^a v_1^b\right)\,,
\qquad
D_2^{ab} \equiv \frac{1}{2}\left(u_2^a u_2^b-v_2^a v_2^b\right)\\,
\ee
and $\vec x_1$ and $\vec x_2$ denote the vertices of
the two interferometers.
The tensors $D_1^{ab}$, $D_2^{ab}$ defined above 
are called {\em detector tensors};
they are symmetric and trace-free with respect to their
$ab$ indices.
In terms of the detector tensors, the overlap function 
becomes
\be
\Gamma_{12}(f)=D_1^{ab} D_2^{cd}
\Gamma_{abcd}(\Delta \vec x)\,,
\label{e:Gamma_in_terms_of_Gamma_abcd}
\ee
where
\be
\Gamma_{abcd}(\Delta \vec x) 
\equiv \int d^2\Omega_{\hat n}\sum_A
e_{ab}^A(\hat n)e_{cd}^A(\hat n)\,
e^{-i2\pi f\hat n\cdot\Delta\vec x/c}
\label{e:Gamma_abcd}
\ee
and $\Delta\vec x\equiv\vec x_2-\vec x_1$ is the 
separation vector connecting the two vertices.
We will also define:
\be
\alpha\equiv 2\pi fs/c\,,
\qquad
s\equiv|\Delta\vec x|\,,
\qquad
\hat s\equiv \Delta\vec x/s\,.
\ee
Thus, in the small-antenna limit,
the orientation-dependence of the overlap function
$\Gamma_{12}(f)$ is encoded
in the detector tensors $D_1^{ab}$, $D_2^{ab}$,
while the separation-dependence is encoded in 
$\Gamma_{abcd}(\Delta\vec x)$.

Note that $\Gamma_{abcd}$ is a tensor which is symmetric 
under the interchanges 
$a\leftrightarrow b$,
$c\leftrightarrow d$, and
$ab\leftrightarrow cd$;
it is also trace-free with respect to the $ab$ and $cd$ index pairs.
The most general expression that we construct for 
$\Gamma_{abcd}(\Delta\vec x)$ given $\delta_{ab}$, $s_a$, 
and its symmetry properties is:
\begin{multline}
\Gamma_{abcd}(\Delta\vec x)
= A(\alpha)\delta_{ab}\delta_{cd} 
+ B(\alpha)(\delta_{ac}\delta_{bd} + \delta_{bc}\delta_{ad})
+ C(\alpha)(\delta_{ab}s_c s_d + \delta_{cd}s_a s_b)
\\
+ D(\alpha)(\delta_{ac}s_b s_d + \delta_{ad} s_b s_c + \delta_{bc}s_a s_d + \delta_{bd}s_a s_c)
+ E(\alpha) s_a s_b s_c s_d\,.
\label{e:Gamma_abcd_general}
\end{multline}
By contracting the above expression with tensors of the form 
$\delta^{ab}\delta^{cd}$, 
$(\delta^{ac}\delta^{bd} + \delta^{bc}\delta^{ad})$,
$\cdots$,
$s^a s^b s^c s^d$, we obtain a linear system of equations 
for $A, B, \cdots, E$, which we can solve in terms of 
scalar integrals involving contractions of the products 
of the polarization tensors, $e^A_{ab}(\hat n)e^A_{cd}(\hat n)$,
with various combinations of $\delta^{ab}$ and $s^a$.
As shown in \cite{Flanagan:1993, Allen-Romano:1999}, these integrals
can be done {\em analytically}, leading to
\be
\left[
\begin{array}{c}
A(\alpha)\\
B(\alpha)\\
C(\alpha)\\
D(\alpha)\\
E(\alpha)\\
\end{array}
\right]
= \frac{1}{2\alpha^2}
\left[
\begin{array}{rrr}
-5\alpha^2 &  10\alpha & 5 \\
 5\alpha^2 & -10\alpha & 5 \\
 5\alpha^2 & -10\alpha & -25 \\
-5\alpha^2 &  20\alpha & -25 \\
 5\alpha^2 & -50\alpha & 175 \\
\end{array}
\right]
\left[
\begin{array}{c}
j_0(\alpha)\\
j_1(\alpha)\\
j_2(\alpha)\\
\end{array}
\right]\,,
\ee
where $j_0(\alpha)$, $j_1(\alpha)$, and $j_2(\alpha)$ 
are the standard spherical Bessel functions \cite{Abramowitz-Stegun:1972}.
With these explicit expressions for $A, B,\cdots, E$ in hand, 
all that is left to do is to contract the right-hand side of 
(\ref{e:Gamma_abcd_general}) with $D_1^{ab} D_2^{cd}$ to obtain 
$\Gamma_{12}(f)$.
If we only assume that the detector tensors are symmetric,%
\footnote{This is needed, for example, to calculate the overlap 
functions for an array of seismometers in the small-antenna 
limit~\cite{Coughlin-Harms:2014}.
For this case, the detector tensors are simply 
$D_I^{ab} \equiv u_I^a u_I^b$, where
$\hat u_I$ is a unit vector pointing along the sensitive
direction of the $I$th seismometer.}
then all terms contribute~\cite{Coughlin-Harms:2014}:
\be
\begin{aligned}
\Gamma_{12}(f) 
&=A(\alpha)\mr{Tr}\,(D_1)\mr{Tr}\,(D_2) +2B(\alpha)D_1^{ab}D_{2ab} 
\\
&\qquad
+C(\alpha)\left[\mr{Tr}\,(D_1)D_2^{ab} + \mr{Tr}\,(D_2)D_1^{ab}\right]s_a s_b
\\
&\qquad\quad+4D(\alpha) D_1^{ab} D_{2a}{}^c s_b s_c
+E(\alpha)D_1^{ab} D_2^{cd} s_a s_b s_c s_d\,.
\end{aligned}
\ee
For symmetric, {\em trace-free} detector tensors, as is the case 
for ground-based interferometers, there is no contribution from 
the $A$ and $C$ terms.
Thus, in the small-antenna limit, the overlap function for the 
strain response of two equal-arm Michelson interferometers can be 
written as a sum of the first three spherical Bessel functions
with coefficients that depend on the product of the frequency
and separation of the two detectors.
(The analytic expression for the overlap function can also be 
derived using (\ref{e:GammaIJ-spherical}), 
which involves the tensor spherical harmonic repsonse functions.
A detailed derivation using these response functions is given 
in \cite{Romano-et-al:2015}.)
 
\begin{figure}[h!tbp]
\begin{center}
\begin{tabular}{cc}
\includegraphics[angle=0,width=.5\columnwidth]{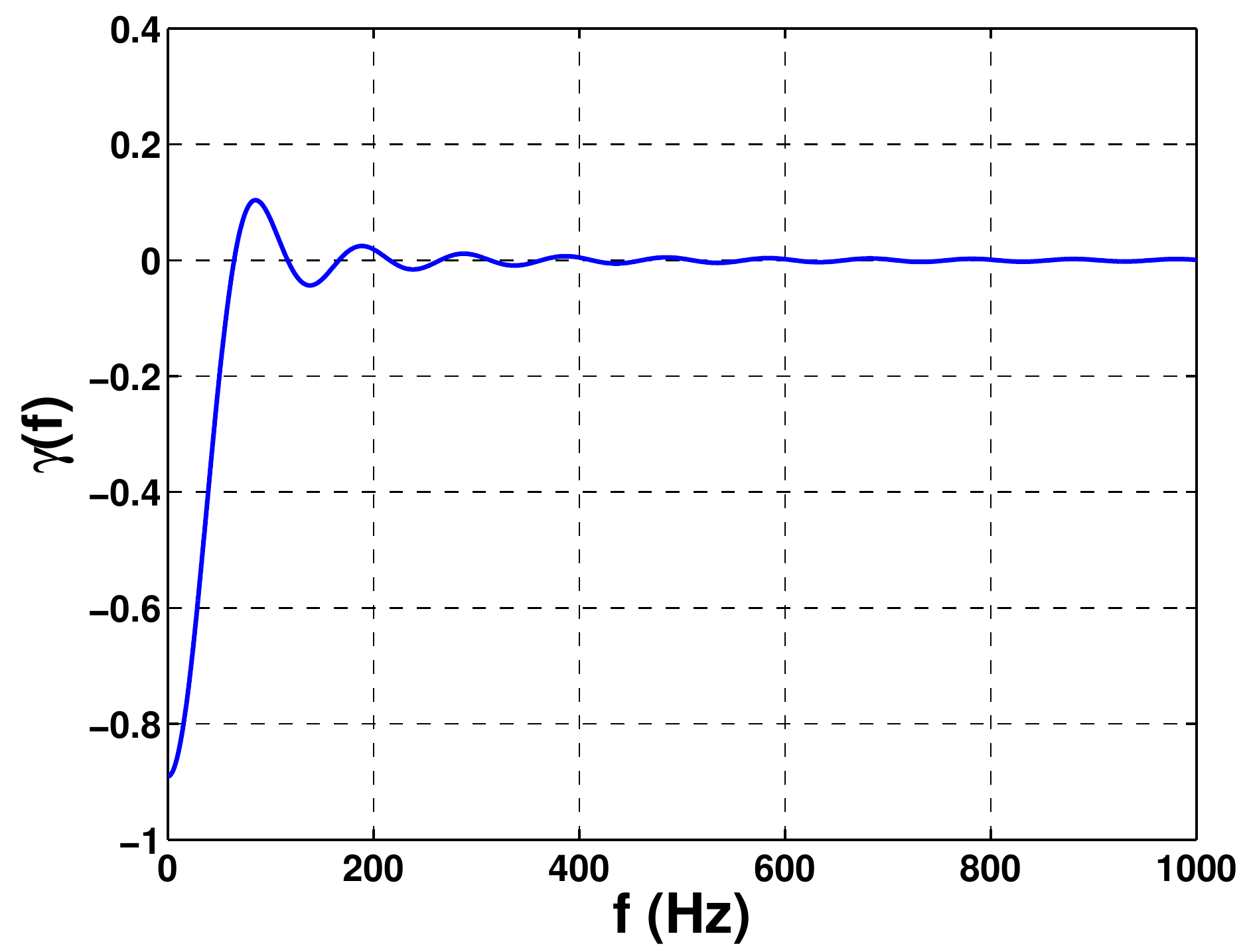}
\includegraphics[angle=0,width=.5\columnwidth]{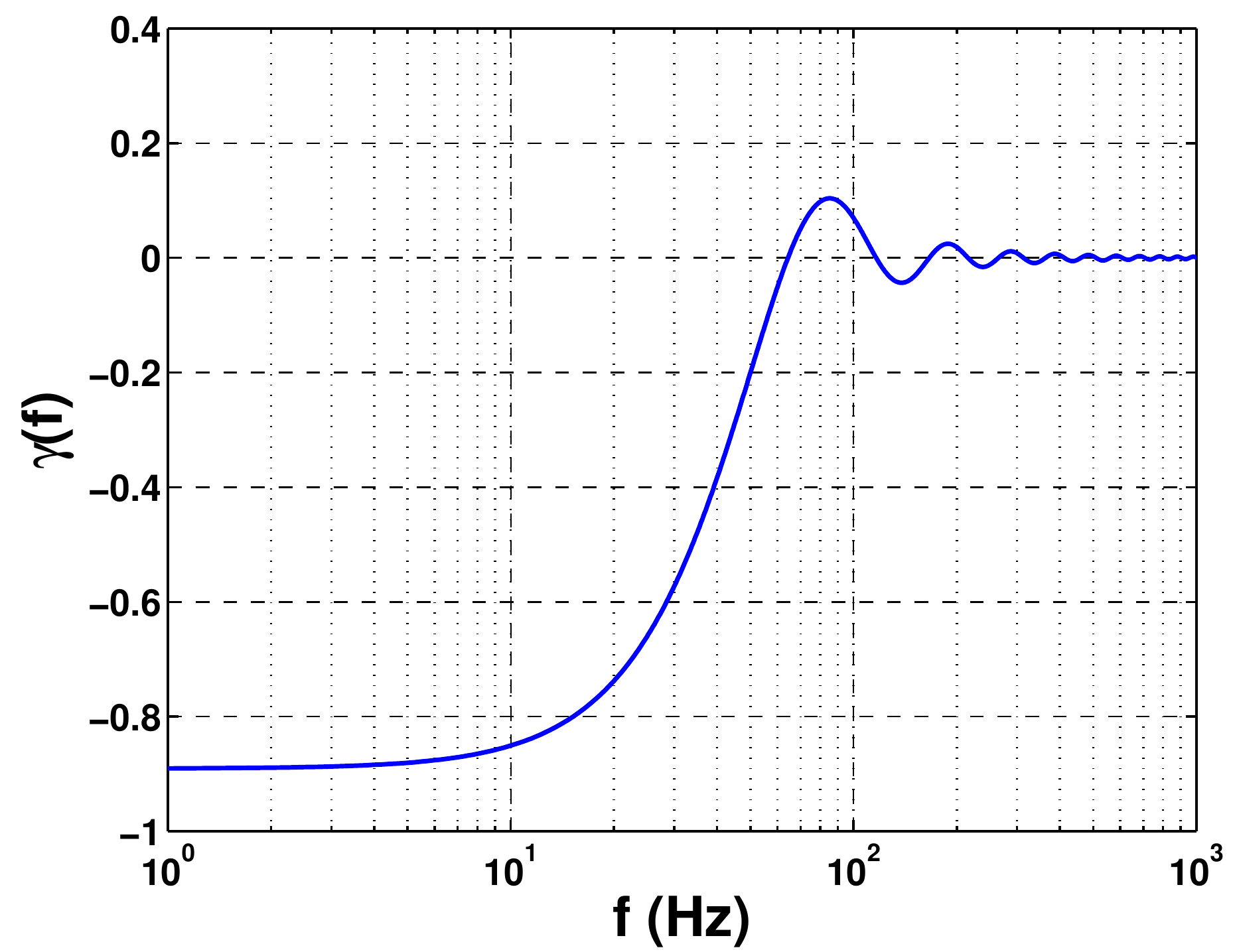}
\end{tabular}
\caption{Overlap function for the LIGO Hanford-LIGO Livinston
cross-correlation in the small-antenna limit.
Left panel: linear frequency scale. 
Right panel: logarithmic frequency scale.}
\label{f:overlapHL}
\end{center}
\end{figure}

Figure~\ref{f:overlapHL} is a plot of the normalized
overlap function
for the strain response of the 4-km LIGO interferometers
in Hanford, WA and Livingston, LA.
There are several things to note about the plot:
(i) The overlap function is negative as $f\rightarrow 0$.
This is because the arms of the Hanford and Livingston 
interferometers are rotated by $90^\circ$ with 
respect to one another.
(ii) The magnitude of the overlap function at
$f=0$ is less than unity---i.e., $|\gamma_{HL}(0)|=0.89$, 
even though the overlap function was normalized.
This is because the planes of the Hanford and 
Livingston interferometers are not identical;
these two detectors are separated by $27.2^\circ$ as 
seen from the center of the Earth.
(iii) The first zero of the overlap function 
occurs just above 60~Hz.
This is roughly equal to $c/(2s)=50~{\rm Hz}$, 
where $s= 3000~{\rm km}$ is 
the separation between the two interferometers.
Note that $f=c/(2s)$ is the frequency of a 
gravitational wave that has a wavelength equal 
to twice the separation of the two sites.
For lower frequencies, the two interferometers
will be driven (on average) by the same 
positive (or negative) part of the incident
gravitational wave.
For slightly higher frequencies, one interferometer 
will be driven by the positive (or negative) part of the incident
wave, while the other interferometer will be driven
by the negative (or positive) part.
The zeros of the overlap function correspond to 
the transitions between the in-phase and out-of-phase
excitations of the two interferometers.
%

\subsubsection{Big-Bang Observer overlap function}
\label{s:BBO-example}

\begin{figure}[h!tbp]
\begin{center}
\includegraphics[angle=0,width=.6\columnwidth]{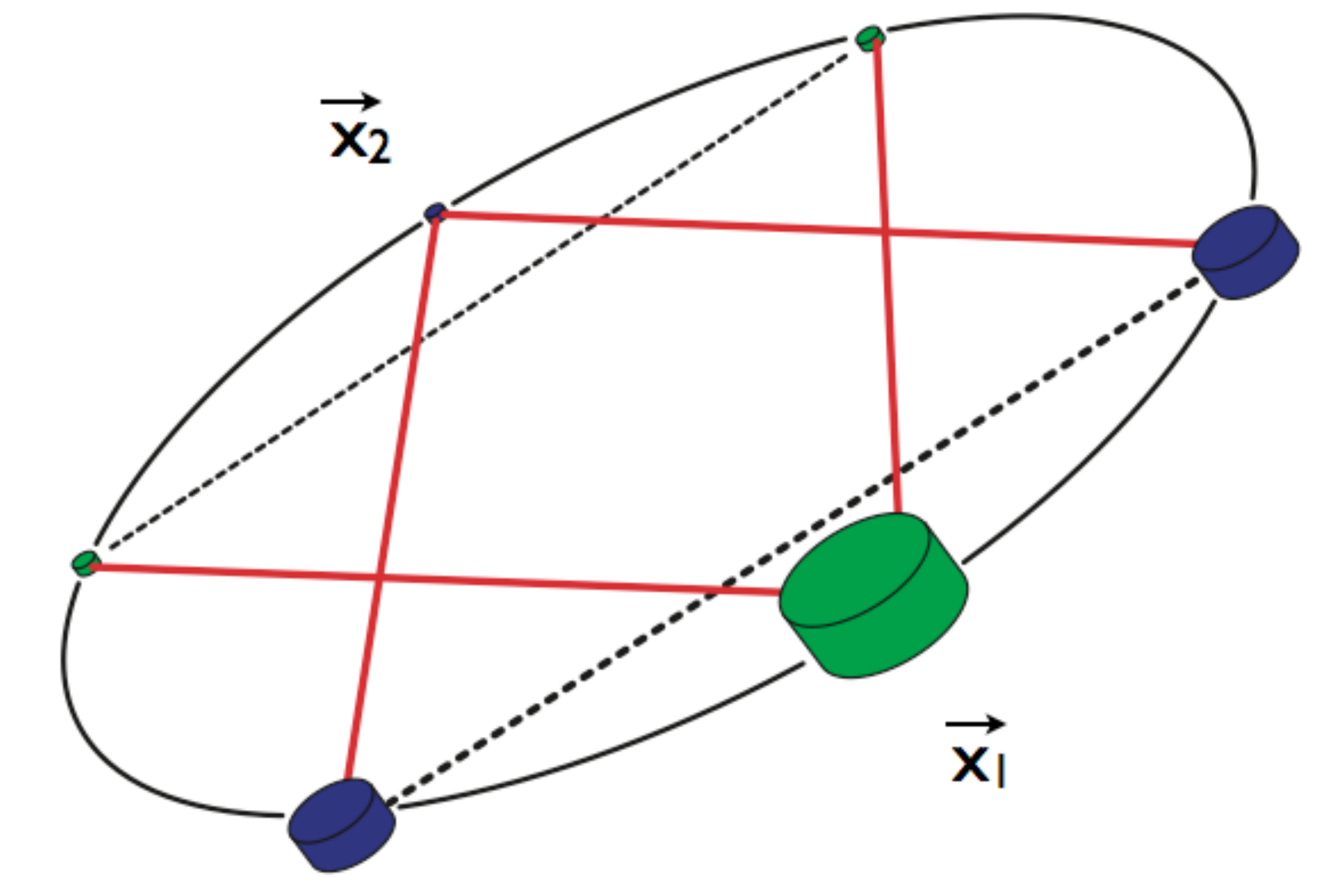}
\caption{Hexagram configuration for the cross-correlation 
of two LISA-like detectors, relevant for the proposed
Big-Bang Observer space mission.
Spacecraft, which house lasers and freely-falling 
test masses, are located at each vertex of the hexagram.
The vectors $\vec x_1$ and $\vec x_2$ denote the vertices 
of two equal-arm Michelson interferometers, with opening 
angle $\beta =60^\circ$.
Image reproduced with permission from \cite{Cornish-Larson:2001},
copyright by IOP.}
\label{f:BBO}
\end{center}
\end{figure}

\begin{figure}[h!tbp]
\begin{center}
\includegraphics[angle=0,width=.6\columnwidth]{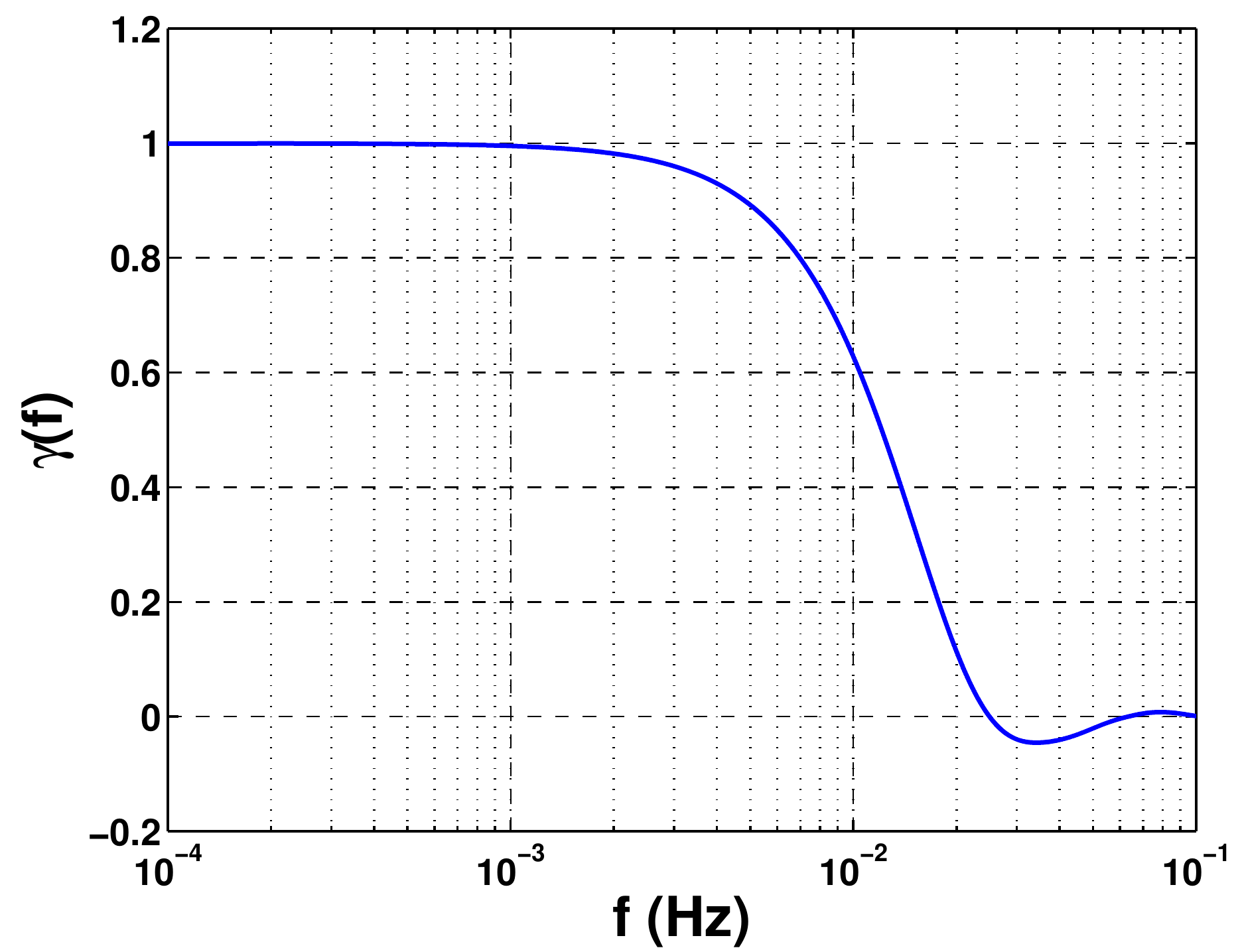}
\caption{Plot of the normalized overlap function for 
strain response for the hexagram configuration shown 
in Figure~\ref{f:BBO}.}
\label{f:BBOoverlap}
\end{center}
\end{figure}

As a second example, we consider the overlap function between
two LISA-like constellations oriented in a hexagram
(i.e., `six-pointed star') configuration as shown in 
Figure~\ref{f:BBO}.
This is one of the configurations being considered
 for the Big-Bang Observer (BBO), which is
a proposed space mission designed to detect or put stringent 
limits on a cosmologically-generated
gravitational-wave background~\cite{Phinney-et-al:2004}.
The arm lengths of the two interferometers, with
vertices $\vec x_1$ and $\vec x_2$, are taken to be
$L=5\times 10^6~{\rm km}$.
The opening angle for the two interferometers is
$\beta = 60^\circ$.
For this example, we calculate the normalized overlap 
function for strain response numerically, since the 
small-antenna limit is not valid for the high-frequency
end of the sensitivity band.
A plot of the normalized overlap function is given in 
Figure~\ref{f:BBOoverlap}.
%

\subsubsection{Pulsar timing overlap function (Hellings and Downs curve)}
\label{s:example-HD}

\begin{figure}[h!tbp]
\begin{center}
\includegraphics[angle=0,width=.6\columnwidth]{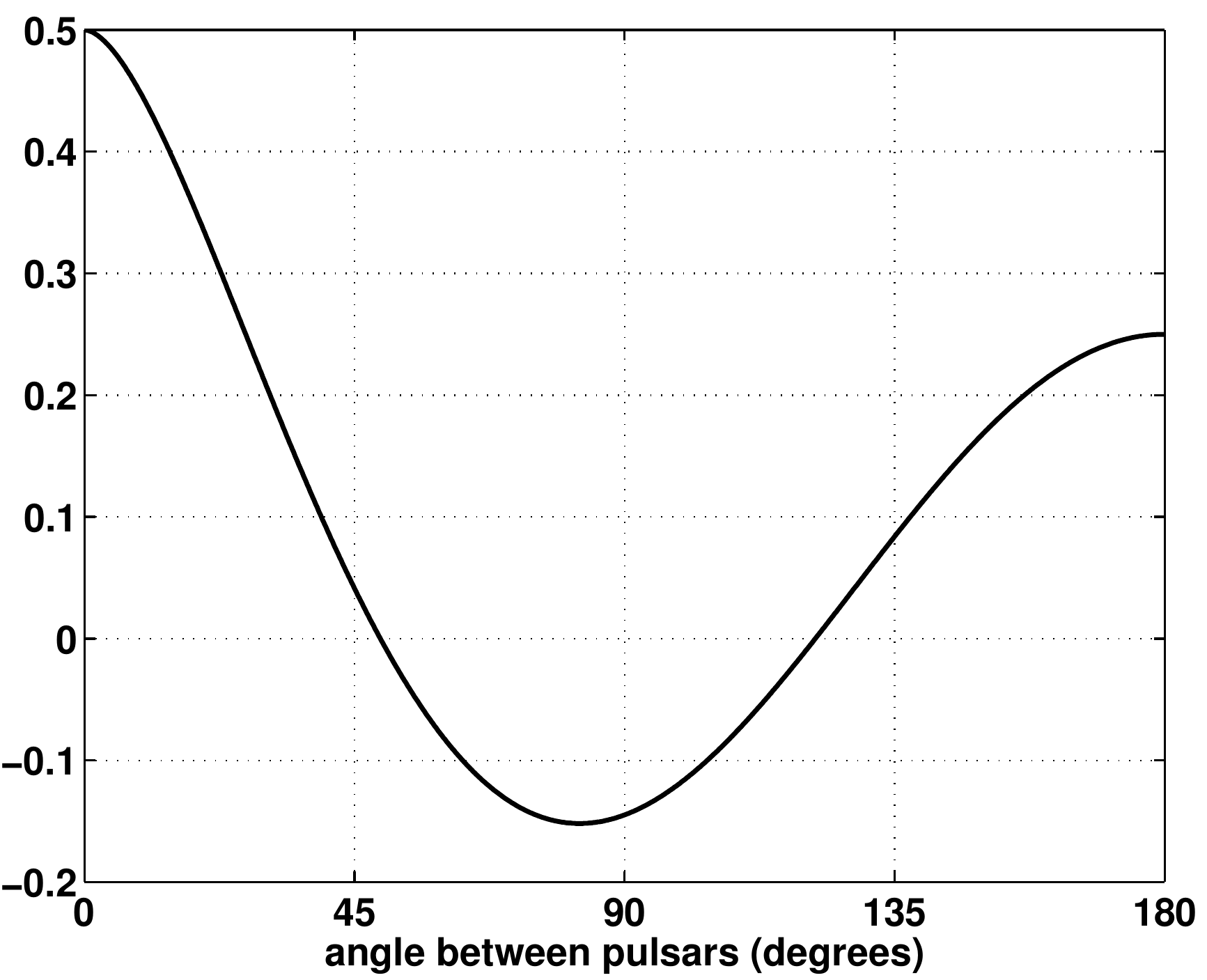}
\caption{Plot of the Hellings and Downs curve 
as a function of the angular separation between 
two distinct pulsars.}
\label{f:HDcurveExact}
\end{center}
\end{figure}

As our final example, we consider the overlap function
for timing residual measurements from an array of $N$ pulsars,
labeled by index $I=1,2,\cdots, N$.
Each pulsar defines a one-way tracking beam 
detector with the position of pulsar $I$ at 
$\vec p_{I}$ and the postion of detector $I$ 
(i.e., a radio receiver on Earth) by $\vec x_{I}$.
For convenience, we will take the origin of coordinates 
to lie at the solar system barycenter. 
Since the diameter of the Earth ($\sim 10^4~{\rm km}$)
and its distance from the Sun ($\sim 10^8~{\rm km}$)
are both small compared to the wavelength of 
gravitational waves relevant for pulsar timing 
($\lambda = c/f\sim 10^{13}~{\rm km}$),
we can effectively set 
$\vec x_{I}\approx \vec x_J\approx \vec 0$
in the argument of the exponential term that enters 
expression (\ref{e:GammaIJ}) for the overlap function.
Thus,
\begin{multline}
\Gamma_{IJ}(f)=
\frac{1}{(2\pi f)^2}\,
\int d^2\Omega_{\hat n} \sum_A
\frac{1}{2}
u_I^a u_I^b
e_{ab}^A(\hat n)\,
\frac{1}{2}
u_J^c u_J^d
e_{cd}^A(\hat n)
e^{i2\pi f\hat n\cdot\cancel{(\vec x_I - \vec x_J)}/c}
\,\times
\\
\frac{1}{1+\hat n\cdot \hat u_I}\,
\frac{1}{1+\hat n\cdot \hat u_J}\,
\left[1-\cancel{e^{-i\frac{2\pi f L_I}{c}(1+\hat n\cdot\hat u_I)}}\right]
\left[1-\cancel{e^{+i\frac{2\pi f L_J}{c}(1+\hat n\cdot\hat u_J)}}\right]\,,
\end{multline}
where the unit vectors $\hat u_I$, $\hat u_J$ are defined
by $\vec x_I= \vec p_I + L_I\hat u_I$, where $L_I$ is the
distance to pulsar $I$.
But since $\vec x_I\approx \vec 0$, it follows that 
$\hat u_I$ and $\hat u_J$ are just unit vectors pointing
from the location of pulsars $I$ and $J$ {\em toward} the 
solar system barycenter.
For distinct pulsars ($I\ne J$), we can ignore the 
exponential terms in the square brackets, 
since $fL/c\gg 1$ for $L\sim 1~{\rm kpc}\ (=3\times 10^{16}~{\rm km})$ 
implies that
$e^{-i 2\pi fL_I(1+\hat n\cdot \hat u_I)/c}$
and its product with the corresponding term for 
pulsar $J$ are rapidly varying functions of $\hat n$ 
and do not contribute
significantly when integrated over the whole sky 
\cite{Hellings-Downs:1983, Anholm-et-al:2009}.
(For a single pulsar ($I=J$), the product of the two 
exponential terms equals 1 and hence cannot be ignored.)
With these simplifications, the integral can be done 
analytically \cite{Hellings-Downs:1983, Anholm-et-al:2009, Jenet-Romano:2015}.
The result is
\be
\Gamma_{IJ}(f)
=
\frac{1}{(2\pi f)^2}\,
\frac{1}{3}\,\chi(\zeta_{IJ})\,,
\ee
where
\be
\chi(\zeta_{IJ})\equiv
\frac{3}{2}\left(\frac{1-\cos\zeta_{IJ}}{2}\right)
\ln
\left(\frac{1-\cos\zeta_{IJ}}{2}\right)
-\frac{1}{4}
\left(\frac{1-\cos\zeta_{IJ}}{2}\right)
+\frac{1}{2}
+\frac{1}{2}\delta_{IJ}\,,
\label{e:HD_normalized}
\ee
and $\zeta_{IJ}$ is the angle between the two pulsars
$I$ and $J$ relative to the solar system barycenter.
(For Doppler frequency measurments, the overlap function
is {\em independent} of frequency,
$\Gamma_{IJ}=\chi(\zeta_{IJ})/3$.)
$\chi(\zeta)$ is the {\em Hellings and Downs} 
function~\cite{Hellings-Downs:1983};
it depends only on the angular separation of a pair of pulsars.
The normalization was chosen so that for a single pulsar, 
$\chi(0)=1$
(for two {\em distinct} pulsars occupying the same 
angular position on the sky, $\chi(0)=0.5$).
A plot of the Hellings and Downs curve is given in 
Figure~\ref{f:HDcurveExact}.

\begin{figure}[h!tbp]
\begin{center}
\includegraphics[angle=0,width=.6\columnwidth]{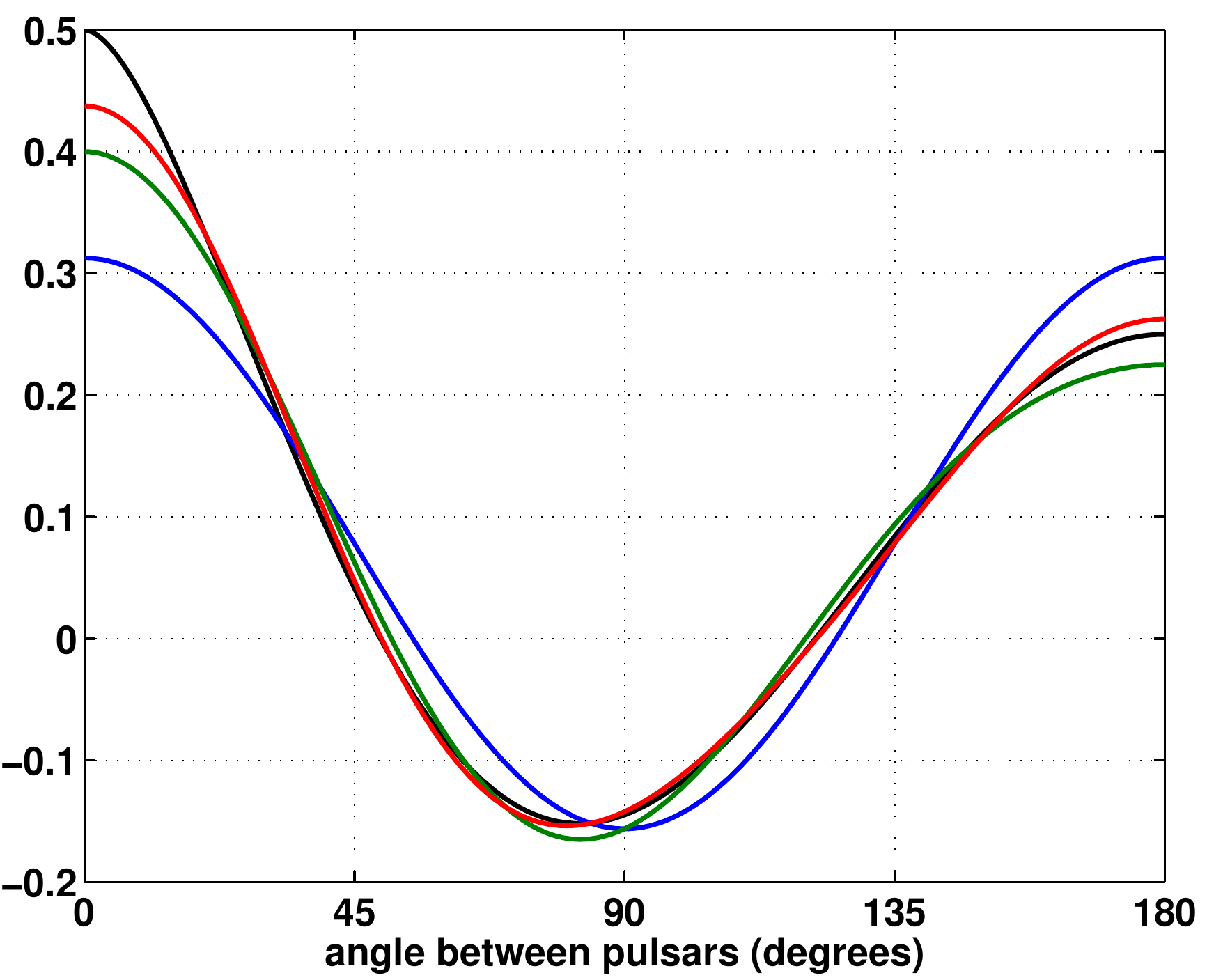}
\caption{Comparison of the exact expression of the 
Hellings and Downs curve (black) with Legendre series 
approximations for different values of $l_{\rm max}$.
The blue, green, and red curves correspond to 
$l_{\rm max} = 2$, 3, and 4, respectively.}
\label{f:HDcurveApprox}
\end{center}
\end{figure}

A couple of remarks are in order:
(i) The Hellings and Downs curve 
is {\em independent} of frequency; 
it is a function of the {\em angle} $\zeta$ 
between different 
pulsar pairs.
This contrasts with the overlap functions for the 
two LIGO interferometers and for BBO given in 
Figures~\ref{f:overlapHL} and \ref{f:BBOoverlap}.
These overlap functions were calculated for a 
fixed pair of detectors; they are functions
instead of the {\em frequency} of the gravitational wave.
(ii) The value of the Hellings and Downs
function $\chi(\zeta_{IJ})$ for a pair of pulsars $I$, $J$ 
can be written as a Legendre series in the cosine of the
angle between the two pulsars.
This follows immediately if one uses 
(\ref{e:GammaIJ-spherical}) for the overlap function
and (\ref{e:RP_PTA}) for the pulsar timing response 
functions in the tensor spherical harmonic basis.
As shown in \cite{Gair-et-al:2014}:
\be
\chi(\zeta_{IJ}) = \frac{3}{4}\sum_{l=2}^\infty ({}^{(2)}\!N_l)^2 
(2l+1)P_l(\hat p_I\cdot\hat p_j)\,,
\label{e:HD_legendre_series}
\ee
where $\hat p_I$ and $\hat p_J$ are unit vectors 
that point in the directions to the two pulsars.
A Legendre series expansion out to $l_{\rm max}=4$
(i.e., only three terms) gives very good agreement with 
the exact expression for the Hellings and Downs function, 
except for very small angular separations.
This is illustrated in Figure~\ref{f:HDcurveApprox}.
%

\subsection{Moving detectors}
\label{s:moving-detectors}

\begin{figure}[h!tbp]
\begin{center}
\includegraphics[angle=0,width=.7\columnwidth]{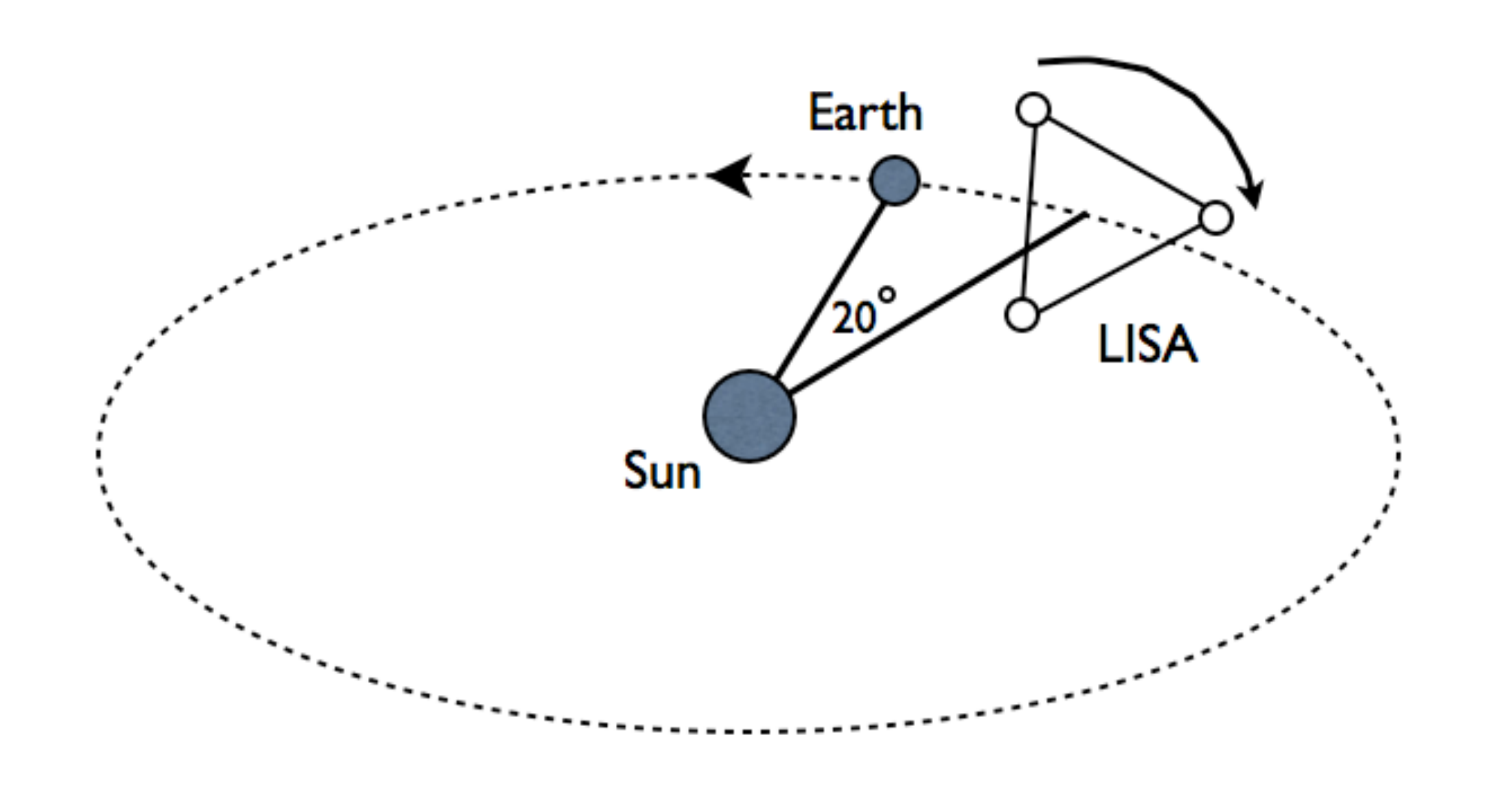}
\caption{Original LISA configuration:
The center-of-mass of the equilateral-triangle configuration
of spacecrafts orbits the Sun in a circle of radius
1~AU, $20^\circ$ behind Earth, while the configuration 
`cartwheels' in retrograde motion about the center-of-mass, 
also with a period of one year.
(Figure adapted from \cite{Cornish-Larson:2001}.)}
\label{f:LISAorbit}
\end{center}
\end{figure}

So far, we have ignored any time-dependence in the detector 
response introduced by the motion of the detectors relative 
to the gravitational-wave source.
In general, this relative motion produces a {\em modulation} 
in both the {\em amplitude} and the {\em phase} of the 
response of a detector to a monochromatic, plane-fronted 
gravitational wave \cite{Cutler:1998}.
For Earth-based interferometers like LIGO, the modulation is 
due to both the Earth's daily rotation and yearly orbital 
motion around the Sun.
For space-based interferometers like LISA, the modulation is 
due to the motion of the individual spacecraft as they 
orbit the Sun with a period of one year.
For example, for the original LISA design, three 
spacecraft fly in an equilateral-triangle configuration 
around the Sun.
The center-of-mass (or guiding center) of the configuration
moves in a circular orbit of radius 1~AU, at an angle of
$20^\circ$ behind Earth,
while the configuration `cartwheels' in retrograde motion about 
the guiding center, also with a period of one year (see 
Figure~\ref{f:LISAorbit}).
%

\subsubsection{Monochromatic plane waves}
\label{s:moving-detectors-plane}

The phase modulation of a monochromatic plane wave 
will have contributions from both the time-varying 
{\em orientation} of the detector as well as the
detector's {\em translational} motion relative the source.
The time-varying orientation leads to changes in the 
response of the detector to the $+$ and $\times$
polarization components of the wave, 
$|R^+ h_+|$ and $|R^\times h_\times|$.
The translational motion leads to a Doppler shift
in the observed frequency of the wave, which is 
proportional to $v/c$ times the nominal frequency, 
where $v$ is velocity of the detector relative to 
the source:
\be
\Delta_{\rm D} f 
= \frac{1}{2\pi} \frac{d\varphi_{\rm D}(t)}{dt}
=-f\hat n\cdot \vec v(t)/c\,.
\ee
For example, for a monochromatic source with 
$f=100~{\rm Hz}$ observed by ground-based detectors 
like LIGO, the Earth's daily rotational motion 
($v\approx 500~{\rm m/s}$) produces a Doppler
shift of order $\sim\! 10^{-4}~{\rm Hz}$, 
while the Earth's yearly orbital motion 
($v\approx 3\times 10^4~{\rm m/s}$), produces a
shift of order $\sim\! 10^{-2}~{\rm Hz}$.
A matched-filter search for a sinusoid must take 
this latter modulation into account,
as the frequency shift is larger than the width of a 
frequency bin for a typical search for such a signal.

\subsubsection{Stochastic backgrounds}
\label{s:moving-detectors-stoch}

For stochastic gravitational-wave backgrounds, 
things are slightly more complicated as the signal 
is an incoherent sum of sinusoidal plane waves having 
different amplitudes, frequencies, and phases,
and coming from different directions on the sky 
(\ref{e:planewave}).
But since the signal is {\em broad-band}, the Doppler
shift associated with the phase modulation of the 
individual component plane waves is not important, 
as the gravitational-wave signal power is (at worst)
shuffled into nearby bins.%
\footnote{Actually, the bin size for a typical
LIGO search for a stochastic background is 
{\em larger} than the $\sim\! 10^{-2}~{\rm Hz}$ 
Doppler shift 
due to the Earth's orbital motion around the Sun.}
On the other hand,
the amplitude modulation of the signal, 
due to the time-varying orientation of a detector, 
{\em can} be significant if the background is 
{\em anisotropic}---i.e., stronger coming from 
certain directions on the sky than from others.
(We will discuss searches for anisotropic 
backgrounds in detail in Section~\ref{s:anisotropic}.)
As the lobes of the antenna pattern sweep through
the ``hot" and ``cold" spots of the anisotropic
background, the amplitude of the signal is modulated
in time.

\begin{figure}[h!tbp]
\begin{center}
\includegraphics[angle=0,width=.7\columnwidth]{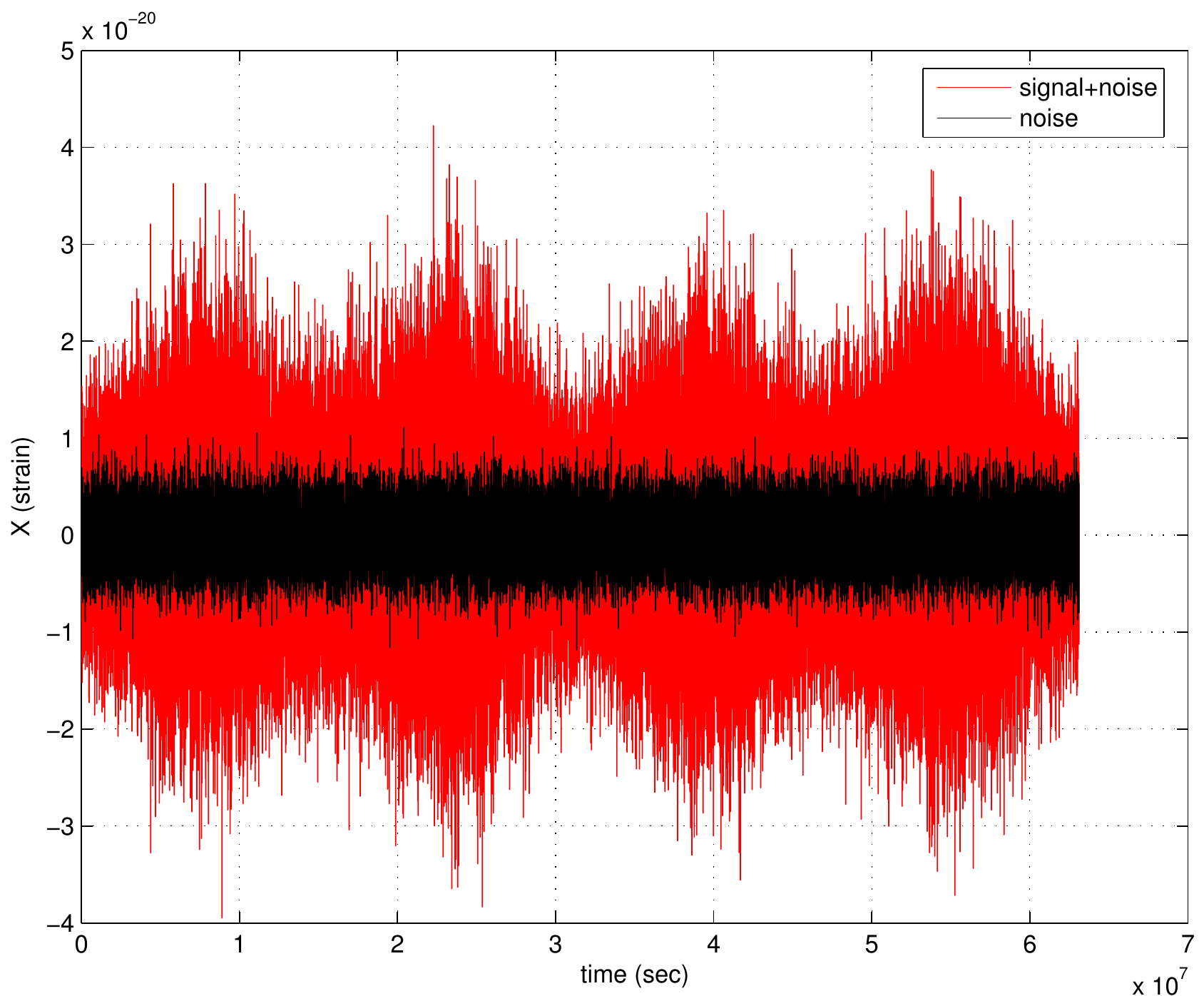}
\caption{The time-domain output of a particular
Michelson combination, $X(t)$, of the LISA data over a 
two-year period.
The contribution from the detector noise is shown 
in black.
The combined output, consisting of both detector
noise and the confusion noise from the Galactic
population of compact white-dwarf binaries, is shown
in red.
The modulation in the amplitude is due to the 
time-varying orientation of the LISA constellation
as it performs a `cart-wheel' in its 1-year orbit
around the Sun (Figure~\ref{f:LISAorbit}).
The amplitude of the output is largest when the 
main lobes of LISA's antenna pattern points in the 
general direction of the galactic center. 
(Data provided by Matt Benacquista.)}
\label{f:cyclostationary}
\end{center}
\end{figure}

Figure~\ref{f:cyclostationary} shows the expected 
time-domain output of a particular Michelson
combination, $X(t)$, of the LISA data over a two-year
period.
The combined signal (red) consists of both detector
noise (black) and the confusion-limited 
gravitational-wave signal from the galactic 
population of compact white-dwarf binaries.
At freuquencies $\sim 10^{-4} - 10^{-3}~{\rm Hz}$,
which corresponds to the lower end of 
LISA's sensitivity band, 
the contribution from these binaries dominates the 
detector noise.
The modulation of the detector output is clearly 
visible in the figure.
The peaks in amplitude are more than 50\% larger
than the minimima; they repeat on a 6~month time 
scale, as expected from LISA's yearly orbital 
motion around the Sun (Figure~\ref{f:LISAorbit}).

\begin{figure}[h!tbp]
\begin{center}
\includegraphics[angle=0,width=.8\columnwidth]{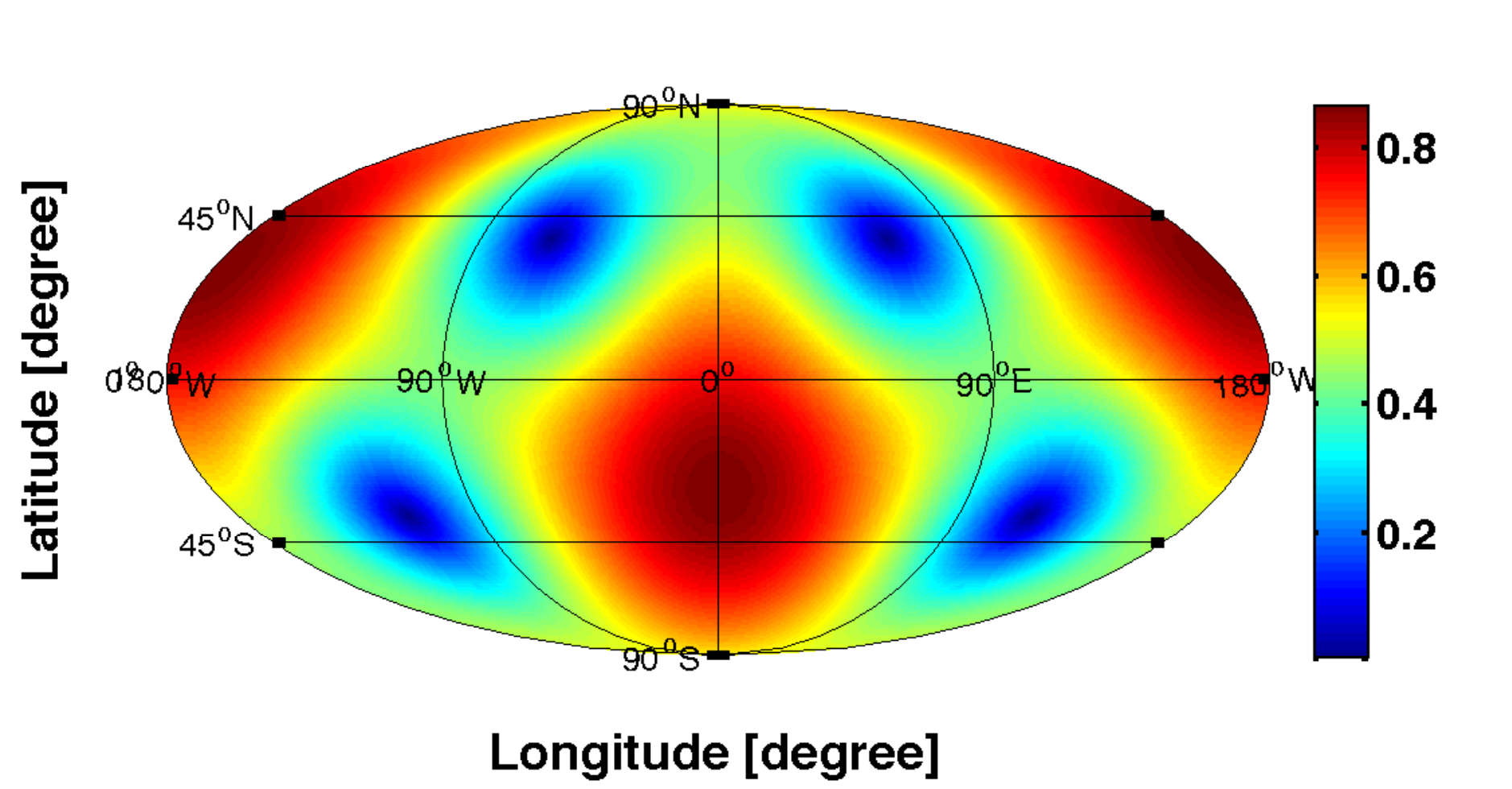}
\caption{A single frame of an animation showing 
the time evolution of the LISA antenna pattern,
represented as a colorbar plot on a Mollweide
projection of the sky in ecliptic coordinates.
Maxima (minima) of the antenna pattern are shown 
by the red (blue) regions. 
The full animation corresponds to a period of
1~year.
To view the animation, please go to the online version 
of this review article at 
\url{http://www.livingreviews.org/}.}
\label{f:lisaPeanutAnimation}
\end{center}
\end{figure}

Figure~\ref{f:lisaPeanutAnimation} is a single frame
of an animation showing the time evolution of the 
LISA antenna pattern, represented as as a colorbar plot on a 
Mollweide projection of the sky in ecliptic coordinates.
The peaks in the detector output that we saw earlier
in Figure~\ref{f:cyclostationary} correspond 
to those times when the maxima of the antenna 
pattern point in the general direction of the galactic
center, $({\rm lon},{\rm lat})=(-93.3^\circ, -5.6^\circ)$
in ecliptic coordinates.%
\footnote{In equatorial coordinates, the  galactic 
center is located at
$({\rm ra},{\rm dec}) = (-6^{\rm h} 15^{\rm m}, -29^\circ)$.}
The motion of the LISA constellation was taken 
from \cite{Cutler:1998}, and
the antenna pattern was calculated for the 
$X$-Michelson combination of the LISA data, 
assuming the small-antenna approximation for 
the interferometer response functions.
The full animation corresponds to LISA's orbital
period of 1~year.
Go to \url{http://www.livingreviews.org/} to view
the animation.
%

\subsubsection{Rotational and orbital motion 
of Earth-based detectors}
\label{s:rotational-orbital}

As mentioned above, given the broad-band nature of a
stochastic signal, the Doppler shift associated with 
the motion of a detector does not play an important
role for stochastic background searches.
This means that we can effectively ignore the velocity
of a detector, and treat its motion as {\em quasi-static}.
So, for example, the motion of a single Earth-based 
detector like LIGO can be thought of as 
synthesizing a {\em set} of {\em static} virtual 
detectors located along an approximately circular ring 1~AU 
from the solar system barycenter \cite{Romano-et-al:2015}.
Each virtual detector in this set observes the 
gravitational-wave background from a different spatial
location and with a different orientation.

As described in \cite{Romano-et-al:2015}, the relevant time-scale 
for a set of virtual detectors is the time over which 
measurements made by the different virtual detectors 
are {\em correlated} with one another.
Basically, we want two neighboring virtual detectors to be 
spaced far enough apart that they provide 
{\em independent} information about the background.
For a gravitational wave of frequency $f$, the minimal
separation corresponds to $|\Delta \vec x|\approx \lambda/2$, 
where $\lambda = c/f$ is the wavelength 
of the gravitational wave.
For smaller separations, the two detectors will 
be driven in coincidence (on average), 
as discussed in item (iii) at the very end of 
Section~\ref{s:LHO-LLO-example}.
Writing $|\Delta \vec x|= v\Delta t$ and solving 
for $\Delta t$ yields
\be
\Delta t\approx \frac{\lambda}{2v} = \frac{c}{2vf}\equiv t_{\rm corr}\,,
\ee
where $t_{\rm corr}$ is the correlation time-scale.
For $\Delta t \lesssim t_{\rm corr}$, the measurements taken 
by the two virtual detectors will be correlated with 
one another;
for $\Delta t\gtrsim t_{\rm corr}$ the measurements will
be uncorrelated with one another.

As a concrete example, let us consider a gravitational 
wave having frequency $f=100~{\rm Hz}$, and 
calculate the correlation time scale for the Earth's 
rotational and orbital motion, treated independently.
Since $v\approx 500~{\rm m/s}$ for daily rotation
and $v\approx 3\times 10^4~{\rm m/s}$ for orbital 
motion, we get
\be
\begin{array}{ll}
t_{\rm corr}\approx 3000 ~{\rm s}
& ({\rm rotational\ motion})\,,
\\
t_{\rm corr}\approx 50 ~{\rm s}
& ({\rm orbital\ motion})\,.
\end{array}
\label{e:tcorr_values}
\ee
Thus, the orbital motion of the Earth around the Sun will 
more rapidly synthesize a large network of independent detectors 
from the motion of a single detector, compared to just
rotational motion.

\begin{figure}[h!tbp]
\begin{center}
\includegraphics[angle=0, width=.45\textwidth]{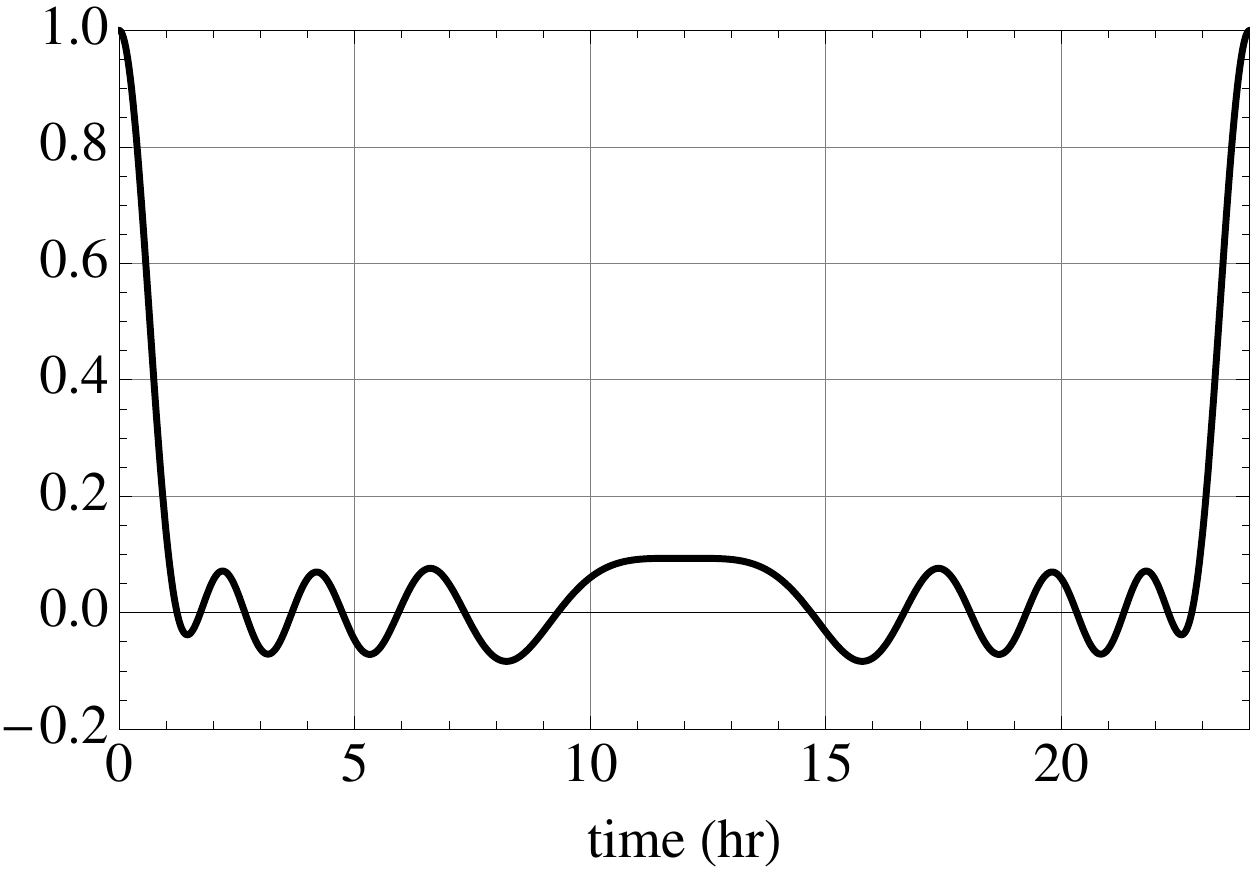}
\hspace{0.04\textwidth}
\includegraphics[angle=0, width=.45\textwidth]{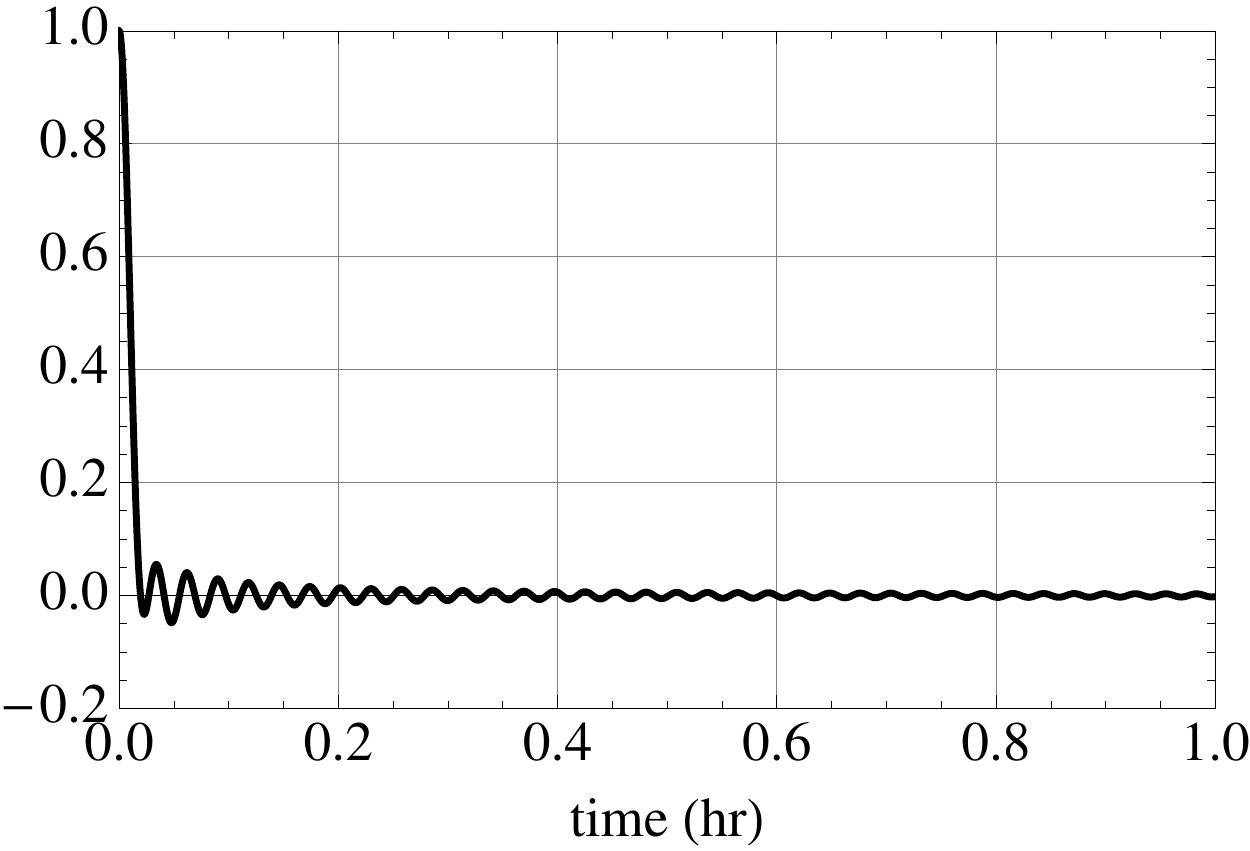}
\caption{Overlap function at $f=100~{\rm Hz}$ for two
virtual interferometers as a function of time. 
The left-hand plot is for a set of virtual interferometers 
located on Earth's equator, associated with Earth's
daily rotational motion.
The right-hand plot is for a set of virtual interferometers
at $1~{\rm AU}$ from the SSB, associated with
Earth's yearly orbital motion.
The first zero-crossing times in these two plots are 
consistent with the correlation times given in 
(\ref{e:tcorr_values}).
Image reproduced with permission from \cite{Romano-et-al:2015},
copyright by APS.}
\label{f:overlap_vs_time}
\end{center}
\end{figure}

We can confirm these approximate results by plotting 
the overlap function at 
$f=100~{\rm Hz}$ for two virtual interferometers 
synthesized by the Earth's rotational and orbital 
motion as function of time.
This is done in Figure~\ref{f:overlap_vs_time}, 
assuming an isotropic and unpolarized stochastic
background, and using the small-antenna approximation
to calculate the detector response functions.
The left-hand plot is for a set of virtual 
interferometers synthesized by the daily 
rotation of a detector located on the 
Earth's equator, with no orbital motion.
The center of the Earth is fixed at the solar system barycenter, 
and the virtual interferometers have one arm pointing North
and the other pointing East.
One sees from the plot that the virtual interferometers
decorrelate on a timescale of roughly an hour,
consistent with (\ref{e:tcorr_values}), 
and recorrelate after 24~hrs when the original detector
returns to its starting position.
The right-hand plot is for a set of virtual 
interferometers
at $1~{\rm AU}$ from the solar system barycenter, associated with
Earth's yearly orbital motion.
There is no rotational motion for this case, as the 
interferometers are located at the center of the 
Earth in its orbit around the Sun, with the orientation 
of the interferometer arms unchanged by the orbital motion.
Here we see that the virtual interferometers 
decorrelate on a timescale of roughly 1~minute,
again consistent with (\ref{e:tcorr_values}).
They will recorrelate only after 1~year (not shown
on the plot).
Since the orbital velocity of the Earth is much
larger than the velocity of a detector on the surface of the
Earth due to the Earth's daily rotational motion, 
the virtual interferometers
associated with orbital motion build up a larger
separation and decorrelate on a much shorter time scale.

We will return to this idea of using the motion of a 
detector to synthesize a set of static virtual detectors 
when we discuss a {\em phase-coherent} approach 
for mapping anisotropic gravitational-wave backgrounds in 
Section~\ref{s:phase-coherent}.

\section{Optimal filtering}
\label{s:optimal_filtering}

\begin{quotation}
Filters are for cigarrettes and coffee.
{\em Cassandra Clare}
\end{quotation}

\noindent
Optimal filtering, in its most simple form, is a method 
of combining data so as to extremize some quantity of interest.
The optimality criterion depends on the particular 
application, but for signal processing, one typically
wants to:
(i) maximize the detection probability for a fixed
rate of false alarms, 
(ii) maximize the signal-to-noise ratio of some test statistic, or 
(iii) find the minimal variance, unbiased estimator of 
some quantity.
Finding such optimal combinations plays a key role in 
both Bayesian and frequentist approaches to statistical inference
(Section~\ref{s:inference}),
and it is an important tool for every data analyst. 
For a Bayesian, the optimal combinations are often
implicitly contained in the likelihood function, while 
for a frequentist, optimal filtering is usually more 
explicit, as there is much more freedom in the construction 
of a statistic.

In this section, we give several simple examples of 
optimal (or matched) filtering for deterministic signals,
and we then show how the standard optimally-filtered
cross-correlation statistic~\cite{Allen:1997, Allen-Romano:1999}
for an Gaussian-stationary, unpolarized, isotropic 
gravitational-wave background
can be derived as a matched-filter statistic for the
expected cross-correlation.
This derivation of the optimally-filtered 
cross-correlation statistic differs from the standard 
derivation given e.g., in~\cite{Allen:1997}, but it illustrates a 
connection between searches for deterministic and stochastic 
signals, which is one of the goals of this review article.

\subsection{Optimal combination of independent measurements}

As a simple explicit example, suppose we have $N$ 
{\em independent} measurements 
\begin{equation}
d_i = a+n_i\,,
\qquad i=1,2,\cdots,N\,,
\end{equation}
where $a$ is some astrophysical parameter that we want
to estimate and $n_i$ are (independent) noise terms.
Assuming the noise has zero mean and known 
variance $\sigma_i^2$
(which can be different from measurement to measurement),
it follows that 
\begin{equation}
\langle d_i\rangle = a\,,
\qquad
\mr{Var}(d_i) \equiv \langle d_i^2\rangle - \langle d_i\rangle^2
=\sigma_i^2\,.
\end{equation}
The goal is to find a linear combination of the data
\begin{equation}
\hat a\equiv\sum_i \lambda_i d_i
\label{e:linear_combination}
\end{equation}
that is optimal in the sense of being an 
{\em unbiased, minimal variance} estimator of $a$.
Unbiased (i.e., $\langle\hat a\rangle=a$) implies
\begin{equation}
\sum_i \lambda_i=1\,,
\label{e:constraint}
\end{equation}
while minimum variance implies
\begin{equation}
\mr{Var}(\hat a)
\equiv
\sigma_{\hat a}^2
=\sum_i\lambda_i^2 \sigma_i^2
=\mr{minimum}\,.
\label{e:variance}
\end{equation}
Since (\ref{e:constraint}) is a constraint that must hold 
when we minimize the variance, we can use Lagrange's method 
of undetermined multipliers~\cite{Boas:2006}
and minimize instead
\begin{equation}
f(\lambda_i,\Lambda)
\equiv \sum_i\lambda_i^2 \sigma_i^2
+\Lambda\left(1-\sum_i\lambda_i\right)
\end{equation}
with respect to both $\lambda_i$ and $\Lambda$.
The final result is:
\begin{equation}
\lambda_i = \left(\sum_j \frac{1}{\sigma_j^2}\right)^{-1}
\frac{1}{\sigma_i^2}
\end{equation}
so that
\begin{equation}
\hat a = \left(\sum_j \frac{1}{\sigma_j^2}\right)^{-1}
\sum_i \frac{d_i}{\sigma_i^2}\,.
\label{e:ahat_filt}
\end{equation}
Thus, the linear combination is a {\em weighted average} 
that gives less weight to the noiser measurements
(i.e., those with large variance $\sigma_i^2$).
The variance of the optimal combination is
\begin{equation}
\sigma_{\hat a}^2
=\left(\sum_j \frac{1}{\sigma_j^2}\right)^{-1}\,.
\label{e:Var(ahat)}
\end{equation}
If the individual variances happen to be equal 
(i.e., $\sigma_i^2\equiv\sigma^2$), then the above
expressions reduce to 
$\hat a = N^{-1}\sum_i d_i$ and
$\sigma_{\hat a}^2 = \sigma^2/N$,
which are the standard formulas for the sample mean and
the reduction in the variance for $N$ independent and
identically-distributed measurements as we saw in 
Section~\ref{s:example-bayes-freq}.

The above results can also be derived by maximizing the
likelihood function
\begin{equation}
p(d|a,\sigma_1^2, \sigma_2^2,\cdots,\sigma_N^2)=
\frac{1}{(2\pi)^{N/2} \sqrt{\sigma_1^2\sigma_2^2\cdots\sigma_N^2}}
\exp\left[-\frac{1}{2}\sum_{i=1}^N\frac{(d_i-a)^2}{\sigma_i^2}\right]
\end{equation}
with respect to the signal parameter $a$, assuming that the 
noise terms $n_i$ are Gaussian-distributed and independent of 
one another.
In fact, similar to what we showed in Section~\ref{s:example-bayes-freq},
one can rewrite the argument of the exponential so that
\begin{equation}
p(d|a,\sigma_1^2, \sigma_2^2,\cdots,\sigma_N^2)
\propto 
\exp\left[-\frac{1}{2}\frac{(a-\hat a)^2}{\sigma_{\hat a}^2}\right]\,,
\end{equation}
where $\hat a$ and $\sigma_{\hat a}^2$ are given by
(\ref{e:ahat_filt}) and (\ref{e:Var(ahat)}), respectively.
From this expression, it immediately follows that $\hat a$ 
maximizes the likelihood, and also the posterior distribution
of $a$, if the prior for $a$ is flat. 

\subsection{Correlated measurements}

Suppose the $N$ measurements $d_i$ are {\em correlated}, so 
that the covariance matrix $C$ has non-zero elements 
\begin{equation}
C_{ij} 
\equiv
\langle d_i d_j\rangle 
-\langle d_i\rangle\langle d_j\rangle
\end{equation}
when $i\ne j$.
Again, we want to find a linear combination (\ref{e:linear_combination})
that is unbiased and has minimum variance
\begin{equation}
\sigma_{\hat a}^2
=\sum_i\sum_j\lambda_i\lambda_j C_{ij}\,.
\end{equation}
By following the same Lagrange multiplier procedure 
described in the previous subsection, one can show that 
the optimal estimator is 
\begin{equation}
\hat a = \left(\sum_k\sum_l \left(C^{-1}\right)_{kl}\right)^{-1}
\sum_i \sum_j \left(C^{-1}\right)_{ij} d_i\,.
\label{e:Ahat-correlated}
\end{equation}
Thus, the weighting factors $1/\sigma_i^2$ of the 
previous subsection are replaced by 
$\sum_j (C^{-1})_{ij}$.
Note that for uncorrelated measurements, 
$C_{ij}=\delta_{ij}\sigma_i^2$, so the above expression
for $\hat a$ reduces to that found previously in (\ref{e:ahat_filt}).

NOTE: Although (\ref{e:Ahat-correlated}) shows how to 
optimally combine data that are correlated with one 
another, it turns out that for most practical purposes one 
can get by using expressions like 
(\ref{e:ahat_filt}) and (\ref{e:matched-filter}) below, 
which are valid for {\em uncorrelated} data.
This is because the values of the Fourier transform of 
a stationary random process are uncorrelated for 
different frequency bins.
Basically, the Fourier transform is a rotation in data space
to a basis in which the covariance matrix is diagonal;
this is called a {\em Karhunen-Loeve transformation}.
(See also Appendix~\ref{s:discrete-continuous-probability}.)
This is one of the reasons why much of signal processing is
done in the frequency domain.

\subsection{Matched filter}

Suppose that the astrophysical signal is not constant but also 
has a `shape' $h_i$ so that
\begin{equation}
d_i = a h_i + n_i\,,
\qquad
i=1,2,\cdots,N\,.
\end{equation}
We will assume that the $h_i$ are known, so that the only
unknown signal parameter is $a$.
We will also assume that the different measurements are 
independent, as will be the case for a stationary random 
process in the frequency domain.
Since $\langle d_i\rangle = a h_i$ is not a constant, the 
analysis of the previous subsection does not immediately apply.
However, if we simply rescale $d_i$ by $h_i$, we obtain 
a new set of measurements	
\begin{equation}
\bar d_i \equiv d_i/h_i
\end{equation}
for which
\begin{equation}
\langle \bar d_i\rangle = a\,,
\qquad
\mr{Var}(\bar d_i) 
\equiv\bar\sigma_i^2
=\sigma_i^2/h_i^2\,,
\end{equation}
so that the previous analysis {\em is} now valid.
Thus,
\begin{equation}
\hat a 
=\left(\sum_j \frac{1}{\bar\sigma_j^2}\right)^{-1}
\sum_i \frac{\bar d_i}{\bar\sigma_i^2}
=\left(\sum_j \frac{h_j^2}{\sigma_j^2}\right)^{-1}
\sum_i \frac{h_i d_i}{\sigma_i^2}
\label{e:matched-filter}
\end{equation}
is the optimal estimator of $a$.

The above expression for $\hat a$ is often called a 
{\em matched filter}~\cite{Wainstein-Zubakov:1971}
since the data $d_i$ are projected onto the expected 
signal shape $h_i$ (as well as weighted 
by the inverse of the noise variance $\sigma_i^2$).
The particular combination
\begin{equation}
Q_i \equiv h_i/\sigma_i^2
\end{equation}
multiplying $d_i$ is the {\em optimal filter} for 
this analysis.%
\footnote{For correlated measurements,
$Q_i = \sum_j(\bar C^{-1})_{ij}/h_i$ where 
$\bar C^{-1}$ is the inverse of the re-scaled covariance matrix
$\bar C_{ij}\equiv C_{ij}/(h_i h_j)$.}
When there are many possible candidate signal shapes,
one constructs a {\em template bank}---i.e.,
a collection of possible shapes against which the data
compared.
By normalizing each of the templates so that
$\sum_i (h_i^2/\sigma_i^2)=1$, the signal-to-noise 
ratio of the matched filter 
\begin{equation}
\hat\rho(h)\equiv\sum_i \frac{h_i d_i}{\sigma_i^2}\,,
\end{equation}
or its square, can be used as a frequentist 
detection statistic.
That is, the maximum value of $\hat\rho(h)$ over the
space of templates $\{h_i\}$ is compared against
some threshold $\rho_*$ (chosen so that the false
alarm probability is below some acceptable value).
If the maximum signal-to-noise ratio exceeds the 
threshold, then one claims detection of the signal
with a certain level of confidence.  
The shape of the detected signal is that which 
corresponds to the maximum matched-filter 
signal-to-noise ratio.

\subsection{Optimal filtering for a stochastic background}

As noted by Fricke~\cite{Fricke:2006}, the above results 
can be used to derive the optimal cross-correlation statistic 
for the stochastic background search.
(A more standard derivation can be found e.g., 
in \cite{Allen:1997}.)
To see this, consider a cross-correlation search for a
Gaussian-stationary, unpolarized, isotropic gravitational-wave 
background using two detectors having uncorrelated noise.
Let $T$ be the total observation time of the measurement.
In the frequency domain, the measurements are given by
the values of the complex-valued cross-correlation
\begin{equation}
x(f)=\tilde d_1(f)\tilde d_2^*(f)
\end{equation}
where $\tilde d_I(f)$, $I=1,2$ are the Fourier transforms
of the time-series output of the two detectors:
\be
\begin{aligned}
d_1(t) &=h_1(t) + n_1(t)\,,
\\
d_2(t) &=h_2(t) + n_2(t)\,.
\end{aligned}
\ee
The $x(f)$ for different frequencies correspond to the 
measurements $d_i$ of the previous subsections.
Since we are assuming uncorrelated detector noise,
\begin{equation}
\langle x(f)\rangle
=\langle \tilde h_1(f)\tilde h_2^*(f)\rangle
=\frac{T}{2}\Gamma_{12}(f)S_h(f)\,,
\label{e:<h1h2>}
\end{equation}
where $S_h(f)$ is the power spectral density of the
stochastic background signal, and $\Gamma_{12}(f)$ is 
the overlap function for the two detectors.%
\footnote{The last equality in (\ref{e:<h1h2>}) 
follows from (\ref{e:rIrJfreq}) with the Dirac delta 
function $\delta(f-f')$ replaced by its finite-time
version $\delta_T(f-f') = T\mr{sinc}[\pi (f-f')T]$,
which equals $T$ when $f=f'$.}
In the weak-signal limit, the covariance matrix
is dominated by the diagonal terms:
\be
\begin{aligned}
C_{ff'}
&\equiv
\langle x(f)x^*(f')\rangle - \langle x(f)\rangle\langle x^*(f')\rangle
\\
&\approx 
\langle \tilde n_1(f)\tilde n^*_1(f')\rangle
\langle \tilde n^*_2(f)\tilde n_2(f')\rangle
\\
&=\frac{T}{4} P_{n_1}(f)P_{n_2}(f)\,\delta(f-f')\,,
\end{aligned}
\ee
where $P_{n_I}(f)$ are the 1-sided power spectral densities
of the noise in the two detectors:
\begin{equation}
\langle \tilde n_I(f)\tilde n^*_I(f')\rangle =
\frac{1}{2}P_{n_I}(f)\,\delta(f-f')\,.
\end{equation}
Thus, in this approximation
\begin{equation}
\int_{-\infty}^\infty df'\> (C^{-1})_{ff'} \approx
\frac{4}{T}\frac{1}{P_{n_1}(f) P_{n_2}(f)}\,.
\label{e:weighting}
\end{equation}
Now, suppose we are searching for a stochastic background
with a power-law spectrum 
\begin{equation}
\Omega_\mr{gw}(f)=\Omega_\beta\left(\frac{f}{f_{\mr{ref}}}\right)^\beta\,,
\end{equation}
whose amplitude $\Omega_\beta$ we would like to estimate.
Then, according to (\ref{e:Sh-Omega_gw}), 
\begin{equation}
S_h(f) = 
\frac{3 H_0^2}{2\pi^2} \frac{\Omega_\beta}{f_{\rm ref}^3}
\left(\frac{f}{f_{\mr{ref}}}\right)^{\beta-3}
= \Omega_\beta H_\beta(f)\,,
\end{equation}
where
\be
H_\beta(f)\equiv 
\frac{3 H_0^2}{2\pi^2} \frac{1}{f_{\rm ref}^3}
\left(\frac{f}{f_{\mr{ref}}}\right)^{\beta-3}\,.
\ee
Using the above form of $S_h(f)$ and (\ref{e:<h1h2>}),
we see that
\begin{equation}
\frac{T}{2}\Gamma_{12}(f)H_\beta(f)
\quad\longleftrightarrow\quad
h_i
\label{e:stoch-shape}
\end{equation}
is the expected signal `shape' $h_i$ in the notation
of the previous subsection.
Given (\ref{e:weighting}) and (\ref{e:stoch-shape}),
it is now a simple matter to show that
\begin{equation}
\hat\Omega_\beta
={\cal N}\int_{-\infty}^\infty df\>
\frac{\Gamma_{12}(f)H_\beta(f)}{P_{n_1}(f) P_{n_2}(f)}
\tilde d_1(f)\tilde d_2^*(f)\,,
\label{e:Omega_betahat}
\end{equation}
where
\begin{equation}
{\cal N}
\equiv\left[ \frac{T}{2}
\int_{-\infty}^\infty df\>
\frac{\Gamma^2_{12}(f)H_\beta^2(f)}{P_{n_1}(f) P_{n_2}(f)}\right]^{-1}\,.
\end{equation}
The variance and expected signal-to-noise ratio of the estimator 
$\hat\Omega_\beta$ are:
\be
\sigma_{\hat\Omega_\beta}^2
=\left[T
\int_{-\infty}^\infty df\>
\frac{\Gamma^2_{12}(f)H_\beta^2(f)}{P_{n_1}(f) P_{n_2}(f)}\right]^{-1}\,,
\label{e:varOmegabetahat}
\ee
and
\be
\rho = \sqrt{T}
\left[
\int_{-\infty}^\infty df\>
\frac{\Gamma^2_{12}(f) S_h^2(f)}{P_{n_1}(f) P_{n_2}(f)}
\right]^{1/2}\,.
\label{e:SNR-expected}
\ee
The combination
\begin{equation}
\tilde Q(f)\equiv{\cal N}
\frac{\Gamma_{12}(f)H_\beta(f)}{P_{n_1}(f) P_{n_2}(f)}
\label{e:Q_optfilter}
\end{equation}
multiplying $\tilde d_1(f)\tilde d_2^*(f)$ in (\ref{e:Omega_betahat})
is the standard optimal filter 
(see e.g.,~\cite{Allen:1997, Allen-Romano:1999}),
which was derived in those references for a flat spectrum, $\beta=0$.
The optimally-filtered cross-correlation statistic, denoted $S$ 
in~\cite{Allen:1997, Allen-Romano:1999}, is given by $S=\hat\Omega_0 T$.

\subsubsection{Optimal estimators for individual frequency bins}

As shown in ~\cite{Aasi-et-al:H1H2}, 
we can also construct estimators of the 
amplitude $\Omega_\beta$ of a power-law spectrum
using cross-correlation data for {\em individual} 
frequency bins, of width $\Delta f$, centered at 
{\em each} (positive) frequency $f$:
\be
\hat\Omega_\beta(f)\equiv
\frac{2}{T} \frac{\Re[\tilde d_1(f) \tilde d_2^*(f)]}
{\Gamma_{12}(f)H_\beta(f)}\,.
\label{e:hat_Omega_beta}
\ee
Note that these estimators are just the measured values of 
the cross-spectrum divided by the expected spectral shape 
of the cross-correlation due to a gravitational-wave 
background with spectral index $\beta$.
In the above expression,
$T$ is the duration of the data segments used in 
calculating the Fourier transforms 
$\tilde d_1(f)$, $\tilde d_2(f)$; and $\Gamma_{12}(f)$ is 
the overlap function for the two detectors.

In the absence of correlated noise, the above estimators
are optimal in the sense that they are unbiased estimators 
of $\Omega_\beta$ and have minimal variance for a single bin:
\be
\sigma^2_{\hat\Omega_\beta}(f)
\approx
\frac{1}{2T\Delta f}\frac{P_{n_1}(f) P_{n_2}(f)}
{\Gamma^2_{12}(f)H^2_\beta(f)}\,,
\ee
where we assumed the weak-signal limit to obtain the approximate
equality for the variance.
For a frequency band consisting of many bins of 
width $\Delta f$, we can optimally combine the individual
estimators $\hat\Omega_\beta(f)$ using the standard 
$1/\sigma^2$-weighting discussed earlier:
\be
\hat\Omega_\beta \equiv 
\frac{\sum_f \sigma^{-2}_{\hat\Omega_\beta}(f) \hat\Omega_\beta(f)}
{\sum_{f'} \sigma^{-2}_{\hat\Omega_\beta}(f')}\,,
\qquad
\sigma^{2}_{\hat\Omega_\beta}\equiv
\left[\sum_{f} \sigma^{-2}_{\hat\Omega_\beta}(f)\right]^{-1}\,.
\ee
The expressions for $\hat\Omega_\beta$ and $\sigma^2_{\hat\Omega_\beta}$
obtained in this way reproduce the standard optimal filter expressions
(\ref{e:Omega_betahat}) and (\ref{e:varOmegabetahat}) in the limit where 
$\Delta f\rightarrow df$ and the sums are replaced by integrals.

\subsubsection{More general parameter estimation}

The analyses in the previous two subsections 
take as given the spectral shape of an isotropic stochastic 
background, and then construct estimators of its overall amplitude.  
But it is also possible to construct estimators of {\em both} 
the amplitude and spectral index of the background.
One simply treats these as free parameters in the signal model
e.g., when constructing the likelihood function.
Interested readers should see \cite{Mandic-et-al:2012} for details.
 
\section{Anisotropic backgrounds}
\label{s:anisotropic}

\begin{quotation}
Sameness is the mother of disgust, variety the cure.
{\em Francesco Petrarch}
\end{quotation}

\noindent
An anisotropic background of gravitational radiation
has {\em preferred} directions on the sky---the associated 
signal is stronger coming from certain directions 
(``hot'' spots) than from others (``cold'' spots).
The anisotropy is produced primarily by sources 
that follow the local distribution of matter in the 
universe (e.g., compact white-dwarf binaries 
in our galaxy), as opposed to sources 
at {\em cosmological} distances (e.g., cosmic strings
or quantum fluctuations in the gravitational field
amplified by inflation \cite{Allen:1997, Maggiore:2000}),
which would produce an {\em isotropic} background.
This means that the measured distribution of 
gravitational-wave power
on the sky can be used to discriminate between 
cosmologically-generated backgrounds, 
produced in the very early Universe, and 
astrophysically-generated backgrounds, produced by 
more recent populations of astrophysical sources.
In addition, an anisotropic distribution of power 
may allow us to detect the gravitational-wave 
signal in the first place;
as the lobes of the antenna pattern of a detector sweep across
the ``hot'' and ``cold'' spots of the anisotropic 
distribution, the amplitude of the signal is modulated
in time, while the detector noise remains unaffected 
\cite{Adams-Cornish:2010}.

In this section, we describe several different 
approaches for searching for anisotropic backgrounds of 
gravitational waves:
The first approach (described in Section~\ref{s:allen-ottewill})
looks for modulations in the 
correlated output of a pair of detectors, at 
harmonics of the rotational or orbital frequency 
of the detectors (e.g., daily rotational
motion for ground-based detectors like LIGO, Virgo,
etc., or yearly orbital motion for space-based 
detectors like LISA).
This approach assumes a known distribution of 
gravitational-wave power ${\cal P}(\hat n)$, 
and filters the data so as to
maximize the signal-to-noise ratio of the harmonics
of the correlated signal.
The second approach (Section~\ref{s:ML})
constructs maximum-likehood
estimates of the gravitational-wave power on the sky
based on cross-correlated data from a network of 
detectors.
This approach produces sky maps of ${\cal P}(\hat n)$, 
analogous to sky maps of temperature anisotropy in
the cosmic microwave background radiation.
The third approach (Section~\ref{s:MLR_statistic})
constructs frequentist detection statistics for 
either an unknown or an assumed distribution of 
gravitational-wave power on the sky.
The fourth and final approach we describe
(Section~\ref{s:phase-coherent}) 
attempts to measure both the amplitude {\em and} phase of the 
gravitational-wave background at each point on the 
sky, making minimal assumptions about the statistical
properties of the signal.
This latter approach 
produces sky maps of the real and imaginary parts 
of the random fields $h_+(f,\hat n)$ and $h_\times(f,\hat n)$, 
from which the power in the background 
${\cal P}(\hat n) = |h_+|^2 + |h_\times|^2$ is just one 
of many quantities that can be estimated from the 
measured data.

Numerous papers have been written over the last 
$\approx 20$~years on the problem of detecting anisotropic 
stochastic backgrounds, starting with the seminal paper by 
Allen and Ottewill \cite{Allen-Ottewill:1997}, which
laid the foundation for much of the work that followed.
Readers interested in more details should see 
\cite{Allen-Ottewill:1997} regarding modulations of 
the cross-correlation statistic at harmonics of the 
Earth's rotational frequency;
\cite{Ballmer-PhD:2006, Ballmer:2006,
Mitra-et-al:2008, Thrane-et-al:2009, Mingarelli-et-al:2013,
Taylor-Gair:2013} for maximum-likelihood estimates
of gravitational-wave power;
~\cite{Thrane-et-al:2009, Talukder-et-al:2011} for
maximum-likelihood ratio detection statistics; and
\cite{Gair-et-al:2014, Cornish-vanHaasteren:2014, Romano-et-al:2015}
regarding phase-coherent mapping.
For results of actual analyses of initial LIGO data 
and pulsar timing data for anisotropic backgrounds,
see~\cite{Abadie-et-al:S5-anisotropic, Taylor-et-al:EPTA-anisotropic}
and Section~\ref{s:aniso-bounds}.

Note that we will not discuss in any detail methods to 
detect anisotropic backgrounds using space-based 
interferometers like LISA or the Big-Bang Observer (BBO).
As mentioned in Section~\ref{s:moving-detectors-stoch},
the confusion noise from the galactic population of compact 
white dwarf binaries is a guaranteed source of anisotropy
for such detectors.
At low frequencies, measurements made using a single LISA will 
be sensitive to only the $l=0,2,4$ components of the background,
while cross-correlating data from two independent LISA-type detectors
(as in BBO) will allow for extraction of the full range of multipole moments.
The proposed data analysis methods are similar to those that 
we will discuss in Sections~\ref{s:allen-ottewill} and \ref{s:ML},
but using the synthesized $A$, $E$, and $T$ data channels for a 
single LISA (see Section~\ref{s:singledetector}).
Readers should see
\cite{Giampieri-Polnarev:1997, Cornish:2001, Ungarelli-Vecchio:2001, 
Seto:2004, Seto-Cooray:2004,
Kudoh-Taruya:2005, Edlund-et-al:2005, Taruya-Kudoh:2005} 
for details.

\subsection{Preliminaries}
\label{s:preliminaries-anisotropic}

\subsubsection{Quadratic expectation values}

For simplicity, we will restrict our attention to 
Gaussian-stationary, unpolarized, anisotropic
backgrounds with quadratic expectation values given 
by (\ref{e:aniso_hh}):
\be
\langle h_A(f,\hat n) h_{A'}^*(f',\hat n')\rangle
=\frac{1}{4}{\cal P}(f,\hat n)
\delta(f-f')
\delta_{AA'}
\delta^2(\hat n,\hat n')\,,
\label{e:aniso_hh_new}
\ee
where
\be
S_h(f) = \int d^2\Omega_{\hat n}\> {\cal P}(f,\hat n)\,.
\ee
We will also assume that ${\cal P}(f,\hat n)$ 
factorizes 
\be
{\cal P}(f,\hat n) = \bar H(f){\cal P}(\hat n)\,,
\label{e:factorize}
\ee
so that the angular distribution of power on the sky 
is independent of frequency.
We will chose our normalization so that $\bar H(f_{\rm ref})=1$,
where $f_{\rm ref}$ is a reference frequency, typically taken
to equal 100~Hz for ground-based detectors.
We will also assume that the spectral shape $\bar H(f)$
is known, so that we only need to recover ${\cal P}(\hat n)$.
If we expand the power ${\cal P}(\hat n)$ in terms of 
spherical harmonics,
\be
{\cal P}(\hat n) = \sum_{l=0}^\infty \sum_{m=-l}^l {\cal P}_{lm} Y_{lm}(\hat n)\,,
\label{e:P(k)expansion}
\ee
then this normalization choice is equivalent to 
${\cal P}_{00} = S_h(f_{\rm ref})/\sqrt{4\pi}$, and has units
of $({\rm strain})^2 \, {\rm Hz}^{-1}\, {\rm sr}^{-1}$,
where ${\rm sr} \equiv {\rm rad}^2$ is one steradian.
Thus, ${\cal P}_{00}$ is a measure of the {\em isotropic}
component of the background, and sets the overall normalization
of the strain power spectral density $S_h(f)$.

\subsubsection{Short-term Fourier transforms}
\label{s:SFT}

Since the response of a detector changes as its antenna 
pattern sweeps across the ``hot'' and ``cold''
spots of an anisotropic distribution,
we will need to split the data taken by the detectors
into chunks of duration $\tau$, where $\tau$ is much 
greater than the light-travel time between any 
pair of detectors, 
but small enough that the detector response functions do not 
change appreciably over that interval.
(For Earth-based interferometers like LIGO, 
$\tau\sim 100~{\rm s}$ to 1000~s is appropriate.)
Each chunk of data $[t-\tau/2,t+\tau/2]$ 
will then be Fourier transformed 
over the duration $\tau$, yielding
\be
\tilde d_I(t;f) = \int_{t-\tau/2}^{t+\tau/2} dt'\> 
d_I(t') e^{-i2\pi ft'}\,.
\label{e:SFT}
\ee
This operation is often called a {\em short-term} Fourier 
transform.
Note that, in this notation, $t$ labels a {\em particular} 
time chunk, and is not a variable that is subsequently 
Fourier transformed.

\subsubsection{Cross-correlations}
\label{s:anisotropic-crosscorr}

For many of the approaches that map the distribution of
gravitational-wave  power, it is convenient to work with 
cross-correlated data from two detectors, evaluated at
the same time chunk $t$ and frequency $f$:
\be
\hat C_{IJ}(t; f) = \frac{2}{\tau}\tilde d_I(t;f)\tilde d_J^*(t;f)\,.
\label{e:CIJ}
\ee
The factor of 2 is a convention consistent
with the choice of one-sided power spectra.
Assuming uncorrelated detector noise and using 
expectation values given in (\ref{e:aniso_hh_new}), we find
\be
\langle \hat C_{IJ}(t; f)\rangle
=\bar H(f) \int d^2\Omega_{\hat n} \>
\gamma_{IJ}(t; f, \hat n){\cal P}(\hat n)\,,
\label{e:<CIJ>}
\ee
where 
\be
\gamma_{IJ}(t; f, \hat n) 
\equiv \frac{1}{2}\sum_A R^A_I(t; f, \hat n) R^{A*}_J(t; f, \hat n)\,.
\label{e:gamma_khat}
\ee
Note that up to a factor of $1/(4\pi)$, the function
$\gamma_{IJ}(t;, f,\hat n)$
is just the integrand of the isotropic 
overlap function $\Gamma_{IJ}(f)$ given by (\ref{e:GammaIJ}).
In what follows, we will drop the detector labels $IJ$ from
both $\hat C_{IJ}(t;f)$ and $\gamma_{IJ}(t;f,\hat n)$ 
when there is no chance for confusion.
 
\begin{figure}[h!tbp]
\begin{center}
\begin{tabular}{cc}
\includegraphics[angle=0,width=.5\columnwidth]{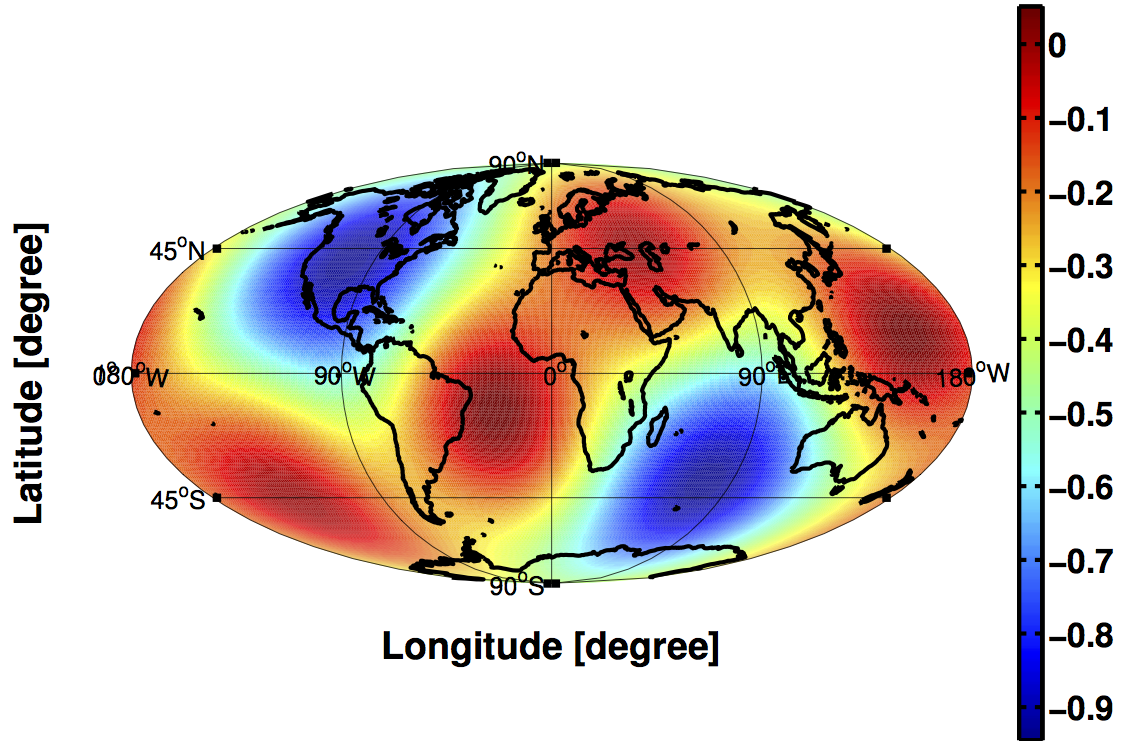}
\includegraphics[angle=0,width=.5\columnwidth]{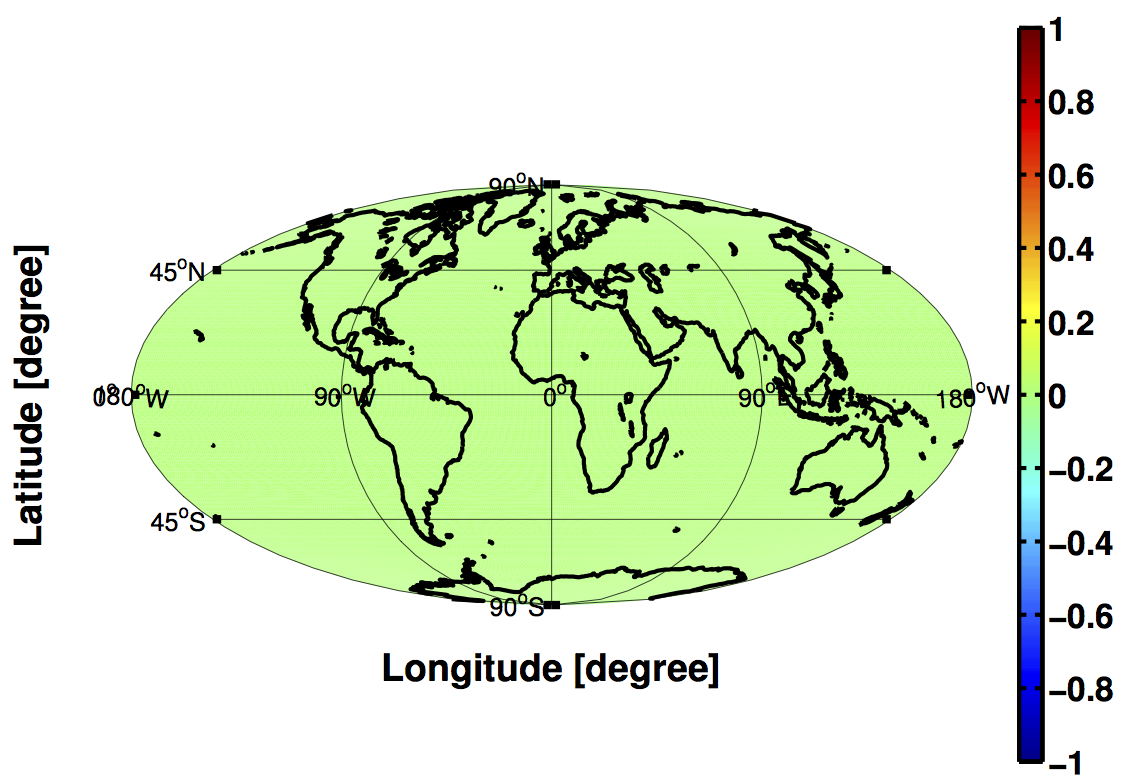}
\\
\includegraphics[angle=0,width=.5\columnwidth]{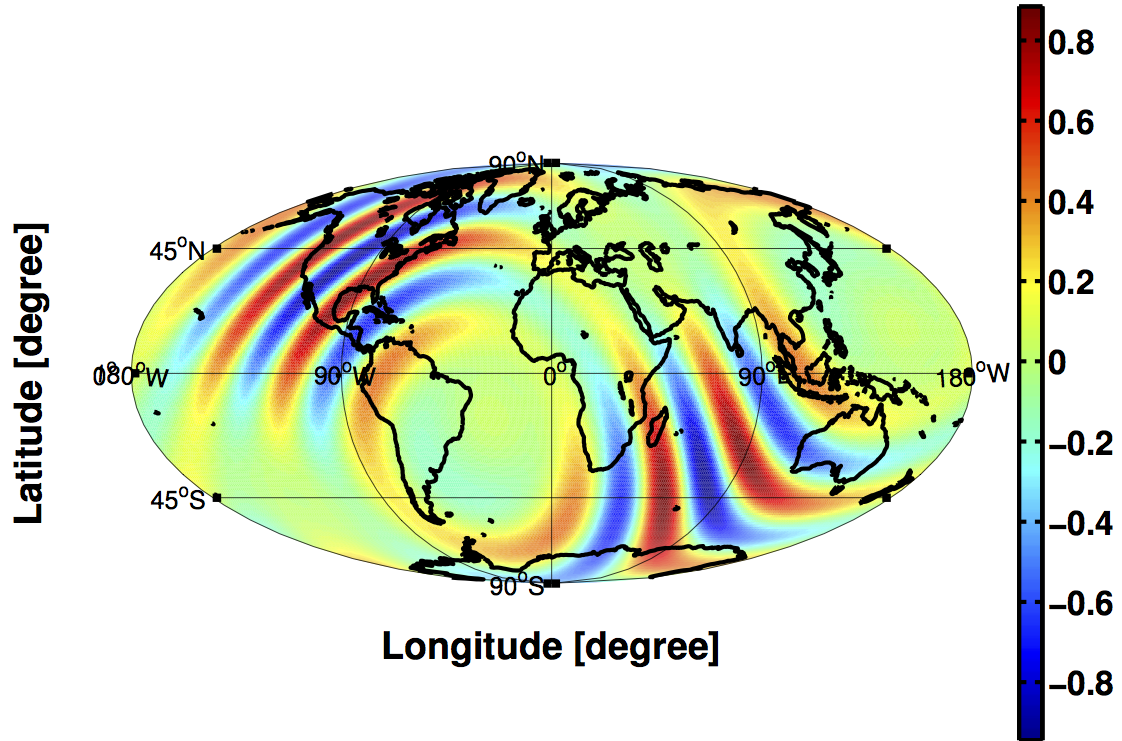}
\includegraphics[angle=0,width=.5\columnwidth]{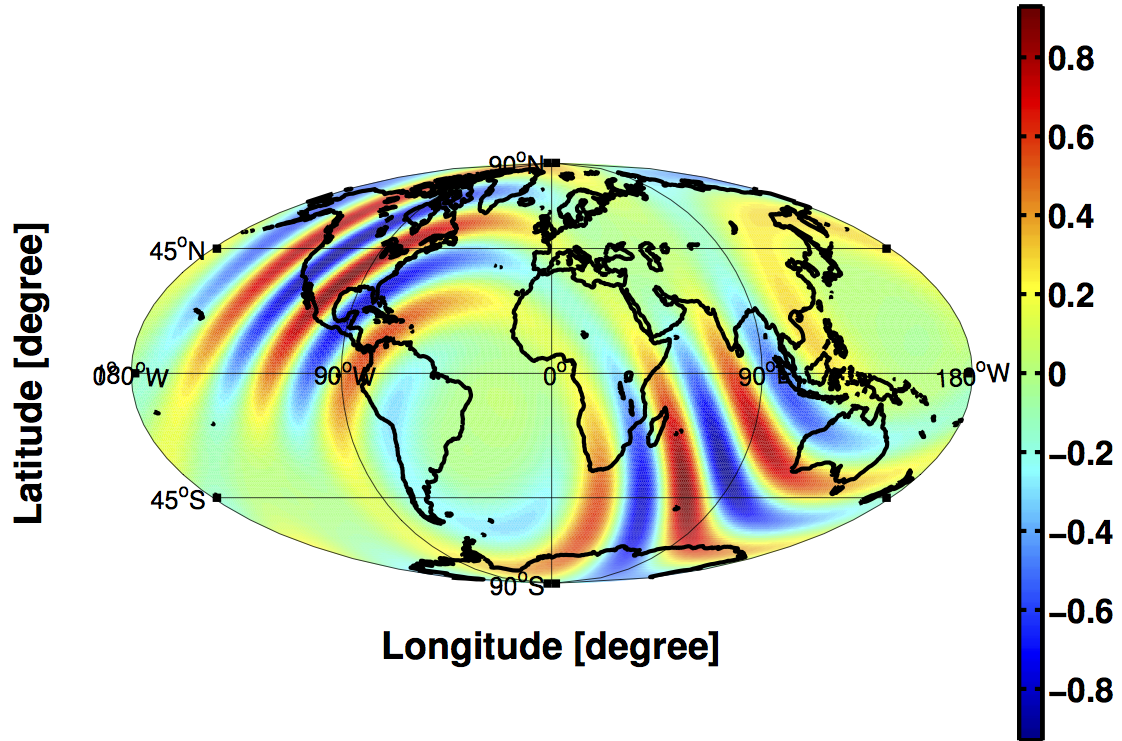}
\end{tabular}
\caption{Real and imaginary parts 
of $\gamma(f,\hat n)$ (appropriately normalized)
for the strain response of the 4-km LIGO Hanford and LIGO 
Livingston interferometers for $f=0~{\rm Hz}$ (top two plots)
and $f=200~{\rm Hz}$ (bottom two plots).
In the top left plot, note the large blue region in the 
vicinity of the two detectors, 
corresponding to the {\em anti-alignment} of the 
Hanford and Livingston interferometers---i.e., the arms of 
the two interferometers are rotated by $90^\circ$ with respect
to one another. 
As shown in the top right plot,
there is no imaginary component to the integrand of the overlap
function at 0~Hz.
The bottom two plots show multiple 
positive and negative oscillations (`lobes'), which come from
the exponential factor $e^{-i 2\pi f\hat n\cdot \Delta\vec x/c}$
of the product of the two response functions (\ref{e:IFOstrainresponse-x0}).
The location of the positive and negative lobes are shifted 
relative to one another for the real and imaginary parts.
The separation between the lobes depends inversely on the 
frequency.}
\label{f:overlapHL_0Hz_200Hz}
\end{center}
\end{figure}

Figure~\ref{f:overlapHL_0Hz_200Hz} shows maps of the real and
imaginary parts of 
$\gamma(t; f, \hat n)$ (appropriately normalized)
for the strain response of the 4-km LIGO Hanford
and LIGO Livingston interferometers evaluated at $f=0~{\rm Hz}$
(top two plots) and $f=200~{\rm Hz}$ (bottom two plots).
(In the Earth-fixed frame, the detectors don't move so there is
no time dependence to worry about.)
Note the presence of oscillations or `lobes' for the 
$f=200~{\rm Hz}$ plots, which come from the exponential factor 
$e^{-i2\pi f\hat n\cdot \Delta \vec x/c}$ of the product of the
two response functions (\ref{e:IFOstrainresponse-x0}).
For $f=0$, this factor is unity.

\begin{figure}[h!tbp]
\begin{center}
\includegraphics[trim=3cm 6.5cm 3cm 3.5cm, clip=true, angle=0,width=.19\columnwidth]{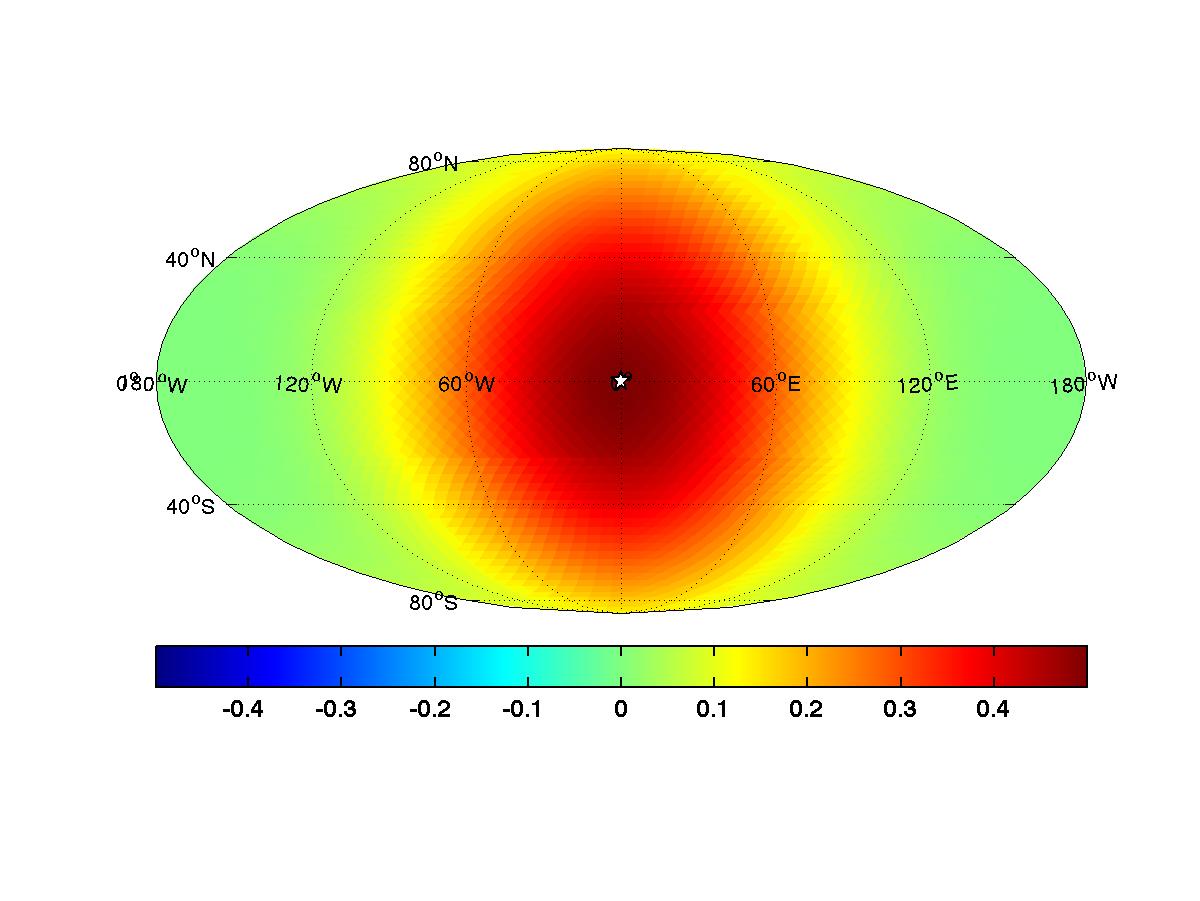}
\includegraphics[trim=3cm 6.5cm 3cm 3.5cm, clip=true, angle=0,width=.19\columnwidth]{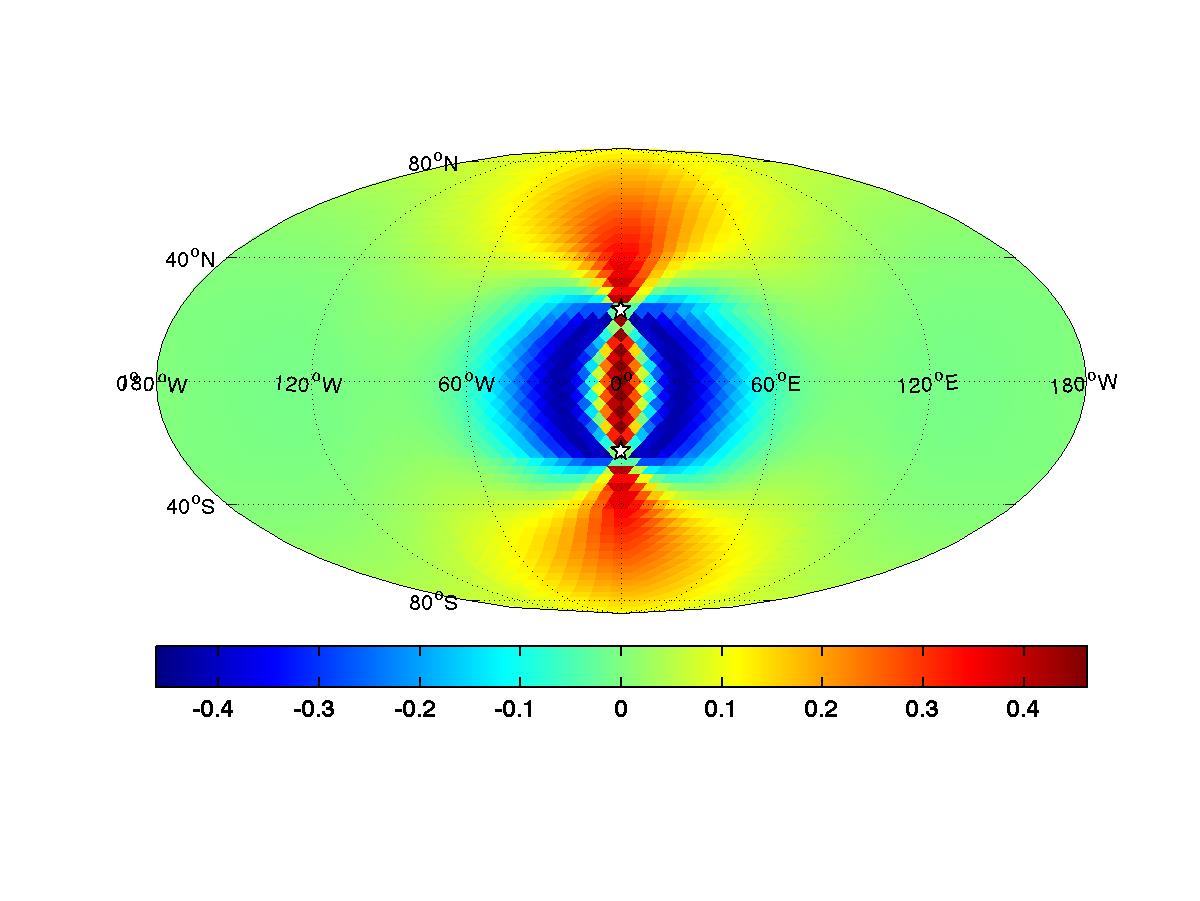}
\includegraphics[trim=3cm 6.5cm 3cm 3.5cm, clip=true, angle=0,width=.19\columnwidth]{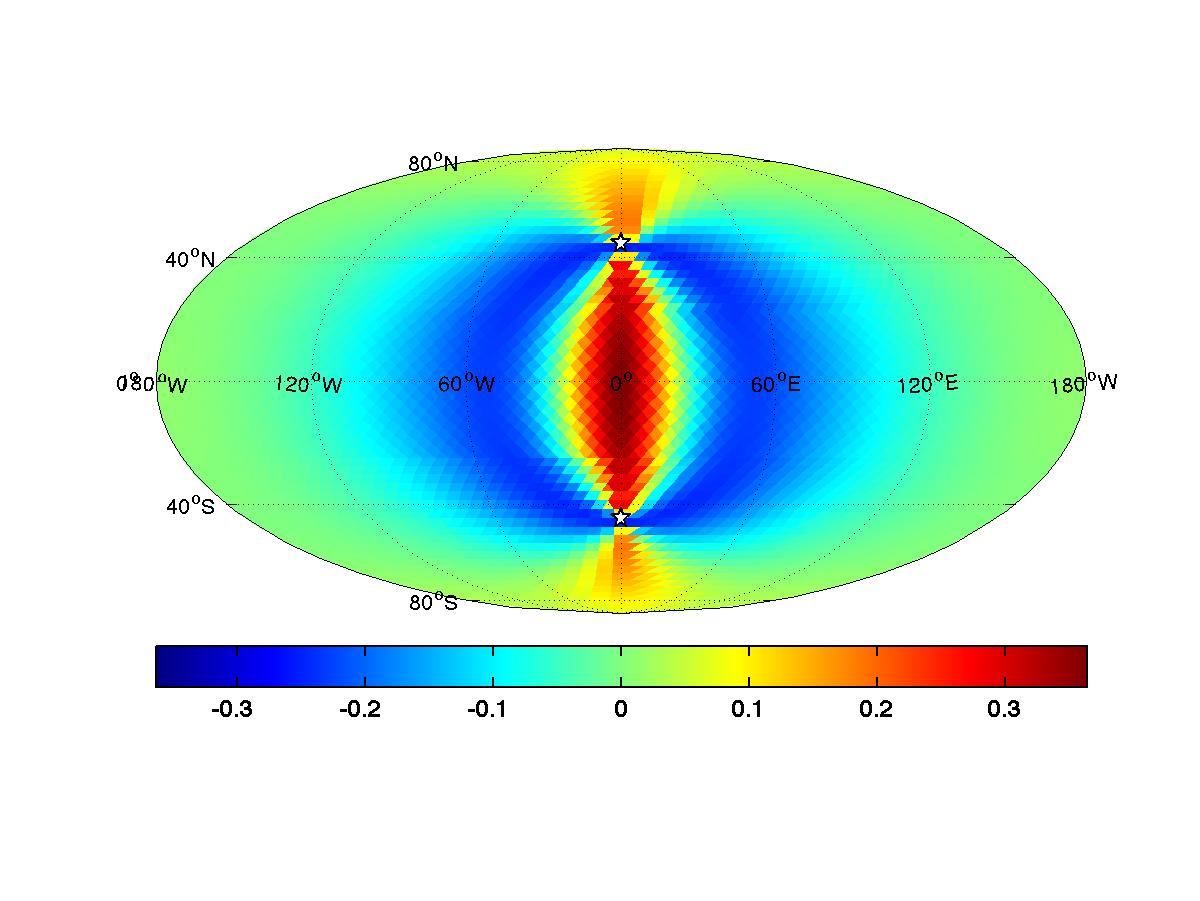}
\includegraphics[trim=3cm 6.5cm 3cm 3.5cm, clip=true, angle=0,width=.19\columnwidth]{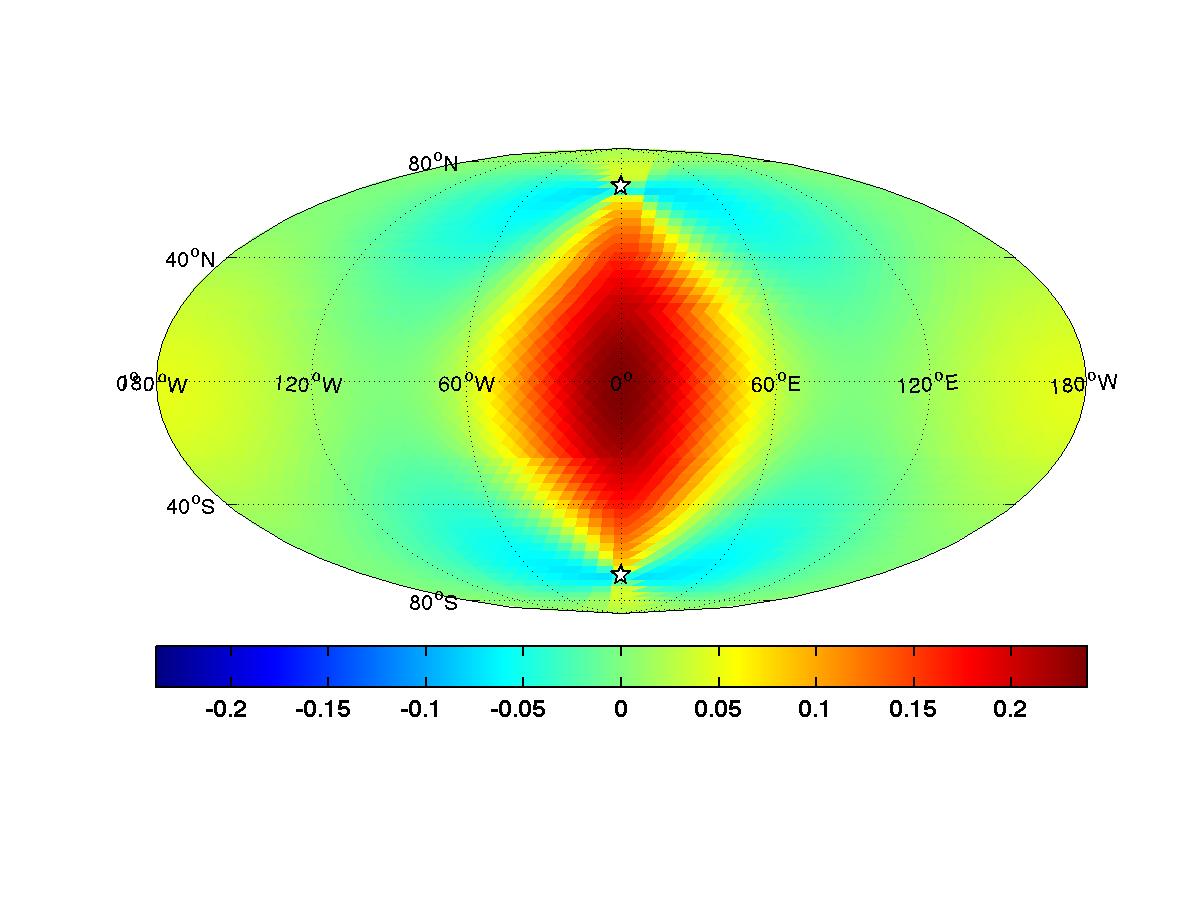}
\includegraphics[trim=3cm 6.5cm 3cm 3.5cm, clip=true, angle=0,width=.19\columnwidth]{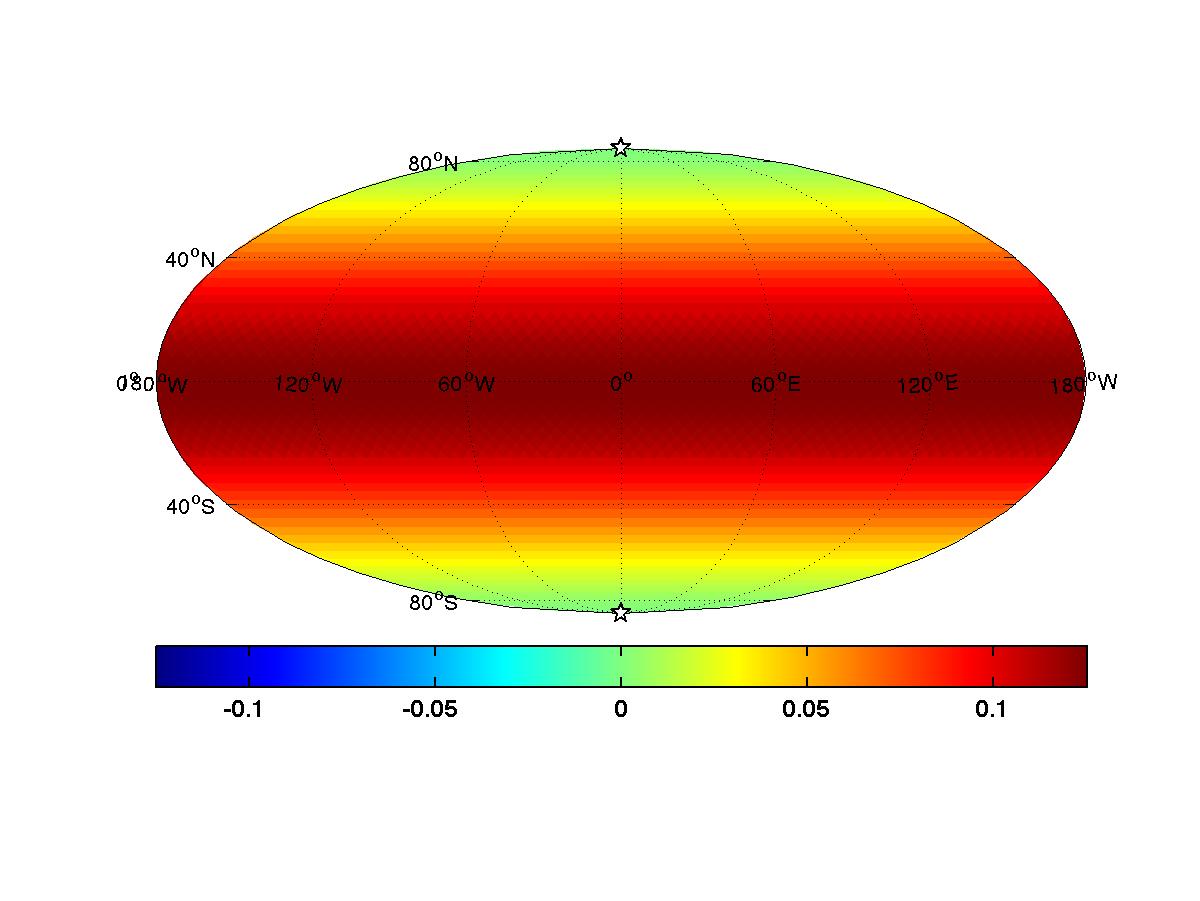}
\\
\includegraphics[angle=0,width=0.9\columnwidth]{HDcurveExact}
\caption{Top row:
Mollweide projections of $\gamma(\hat n)$ for pairs of pulsars 
separated on the sky by $\zeta=0^\circ$, $45^\circ$, $90^\circ$, $135^\circ$, $180^\circ$.
Reddish regions correspond to positive values of $\gamma(\hat n)$;
blueish regions correspond to negative values of $\gamma(\hat n)$.
Bottom:
Hellings and Downs curve as a function of the angular separation 
between two distinct pulsars.
The integral of the top plots over the whole sky equal the values of the 
Hellings and Downs curve for these angular separations.
(See also Figure~\ref{f:HDcurveExact}.)}
\label{f:HDintegrandMollweide}
\end{center}
\end{figure}

Figure~\ref{f:HDintegrandMollweide} is a similar plot, showing Mollweide projections
of $\gamma(t; f, \hat n)$ for the Earth-term-only Doppler frequency 
response (\ref{e:pulsarresponse-earthonly}) 
of pairs of pulsars separated on the sky by 
$\zeta=0^\circ$, $45^\circ$, $90^\circ$, $135^\circ$, $180^\circ$.
(There is no time dependence nor frequency dependence for 
these functions.)
The bottom panel is a plot of the Hellings and Downs curve as a function
of the angular separation between a pair of Earth-pulsar baselines.
By integrating the top plots over the whole sky (appropriately
normalized), one obtains the values of the Hellings and Downs 
curve for those angular separations.

\subsubsection{Spherical harmonic components of $\gamma(t; f, \hat n)$}
\label{s:gammaLM}

As first noted in \cite{Allen-Ottewill:1997}, the functions 
$\gamma(t; f,\hat n)$ defined above 
(\ref{e:gamma_khat}) play a very important role in searches 
for anisotropic backgrounds.
For a fixed pair of detectors at a fixed time $t$ and 
for fixed frequency $f$, these functions are scalar fields 
on the unit 2-sphere and hence can be expanded in terms of 
the ordinary spherical harmonics $Y_{lm}(\hat n)$:
\be
\gamma(t; f, \hat n) \equiv \sum_{l=0}^\infty \sum_{m=-l}^l
\gamma_{lm}(t;f) Y_{lm}^*(\hat n)\,,
\label{e:gammaLM_def1}
\ee
or, equivalently,
\be
\gamma_{lm}(t;f) \equiv \int d^2\Omega_{\hat n}\>
\gamma(t; f, \hat n)Y_{lm}(\hat n)\,.
\label{e:gammaLM_def2}
\ee
Note that this definition differs from (\ref{e:P(k)expansion}) 
for ${\cal P}_{lm}$ by a complex conjugation,
but agrees with the convention used in \cite{Allen-Ottewill:1997}.
In terms of the spherical harmonic components, it follows that
\be
\int d^2\Omega_{\hat n}\>\gamma(t;f,\hat n){\cal P}(\hat n)
= \sum_{l=0}^\infty\sum_{m=-l}^l \gamma_{lm}(t;f){\cal P}_{lm}\,,
\label{e:int_gamma_P}
\ee
as a consequence of the orthogonality of the $Y_{lm}(\hat n)$.
%
%
This expression enters (\ref{e:<CIJ>}) for the 
expected cross-correlation of the output in two detectors.
As explained in \cite{Allen-Ottewill:1997, Thrane-et-al:2009},
the time dependence of $\gamma_{lm}(t;f)$ is particularly 
simple:
\be
\gamma_{lm}(t;f) = \gamma_{lm}(0; f)\,e^{im2\pi t/T_{\rm mod}}\,,
\label{e:gammaLM-timedependence}
\ee
where $T_{\rm mod}$ is the relevant modulation period associated
with the motion of the detectors.
For example, for ground-based detectors like LIGO and Virgo,
$T_{\rm mod}=1$~sidereal day, since the displacement vector 
$\Delta\vec x(t) \equiv \vec x_2(t)-\vec x_1(t)$ connecting the 
vertices of the two interferometers (and which enters the expression
for the overlap function) traces out a cone on the sky 
with a period of one sidereal day.
If there is no time dependence, as is the case for pulsar timing, 
$T_{\rm mod}$ is infinite.

\medskip
\noindent{\bf Example: Earth-based interferometers}
\medskip

\begin{figure}[h!tbp]
\begin{center}
\includegraphics[width=.32\textwidth]{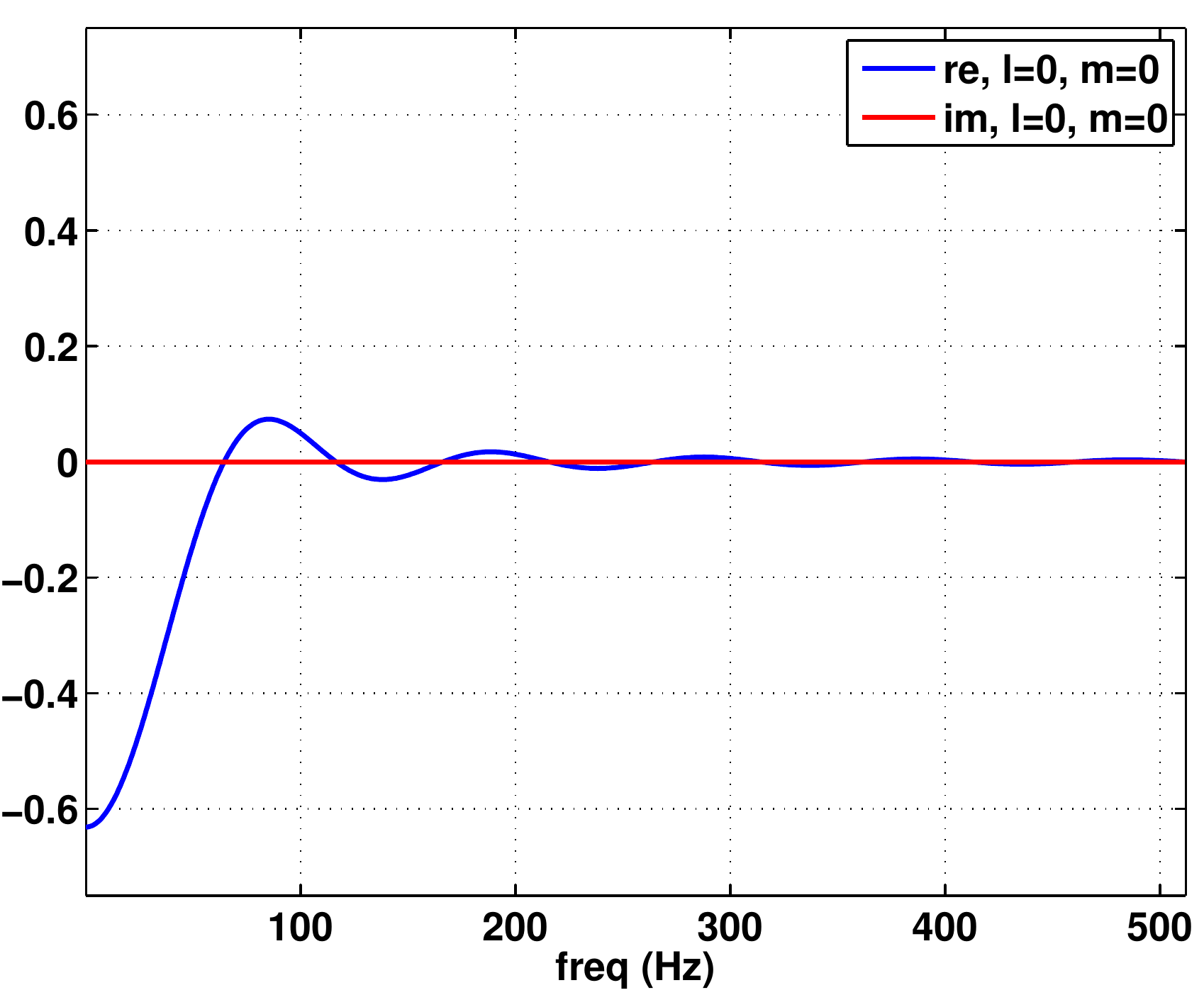}
\includegraphics[width=.32\textwidth]{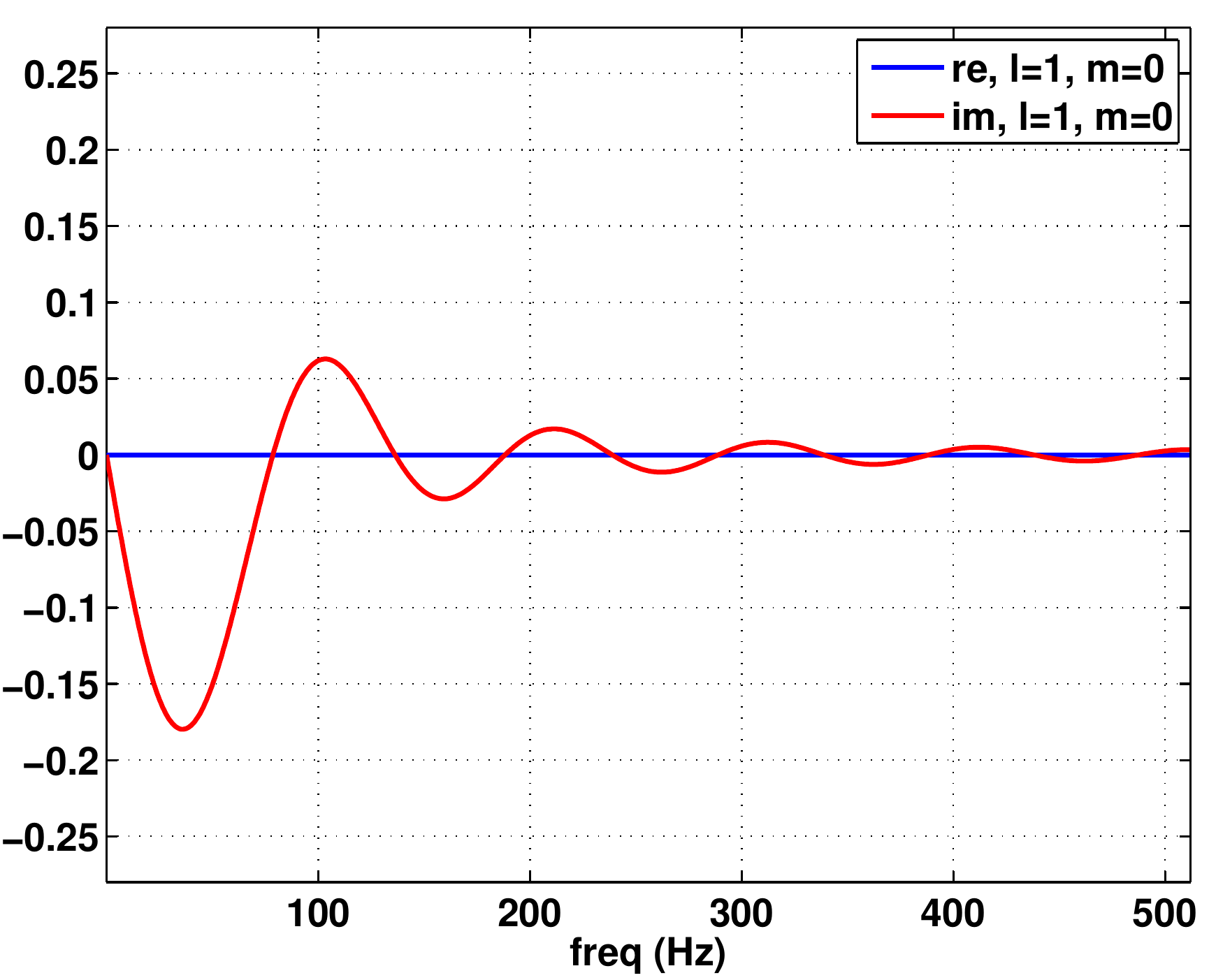}
\includegraphics[width=.32\textwidth]{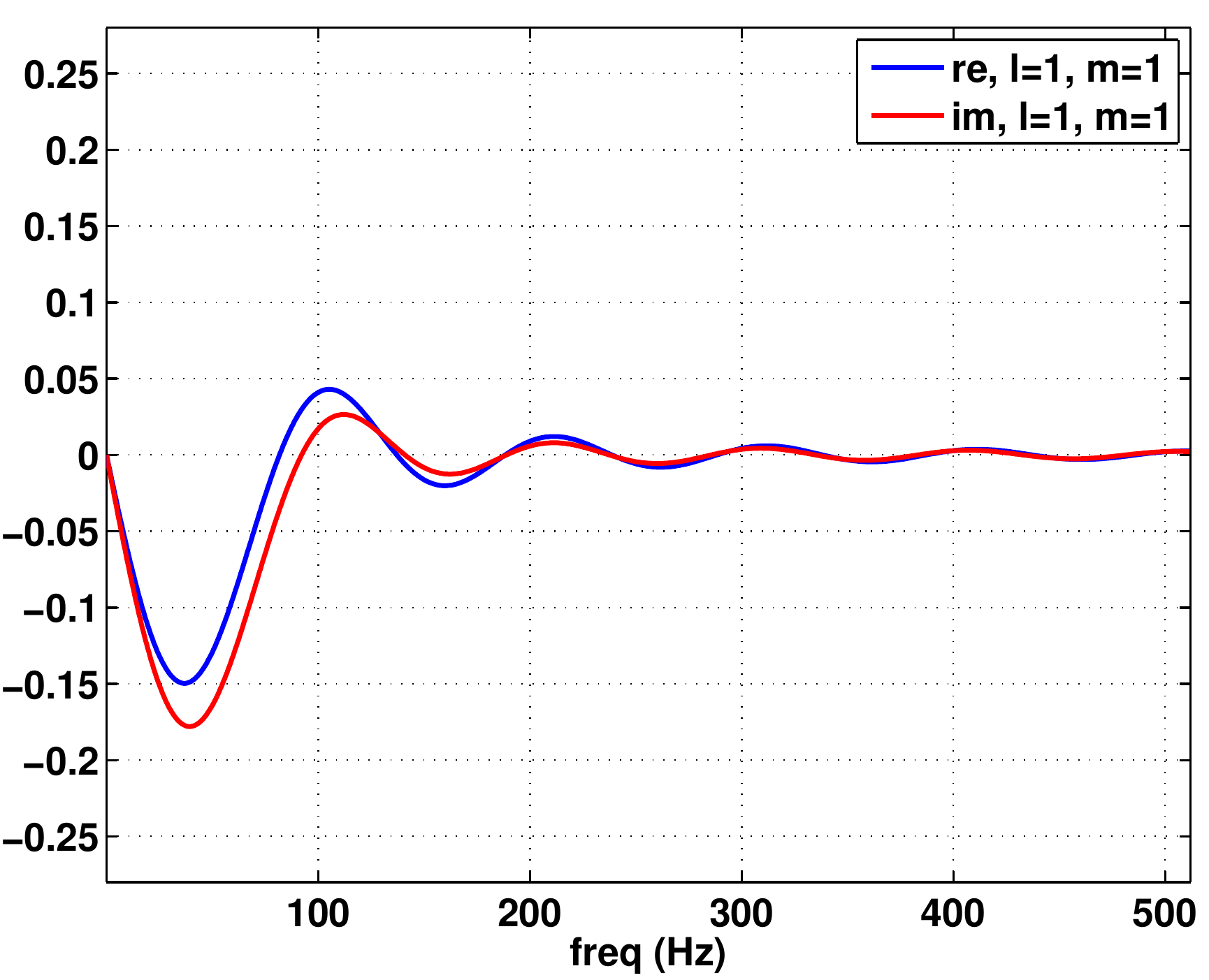}
\includegraphics[width=.32\textwidth]{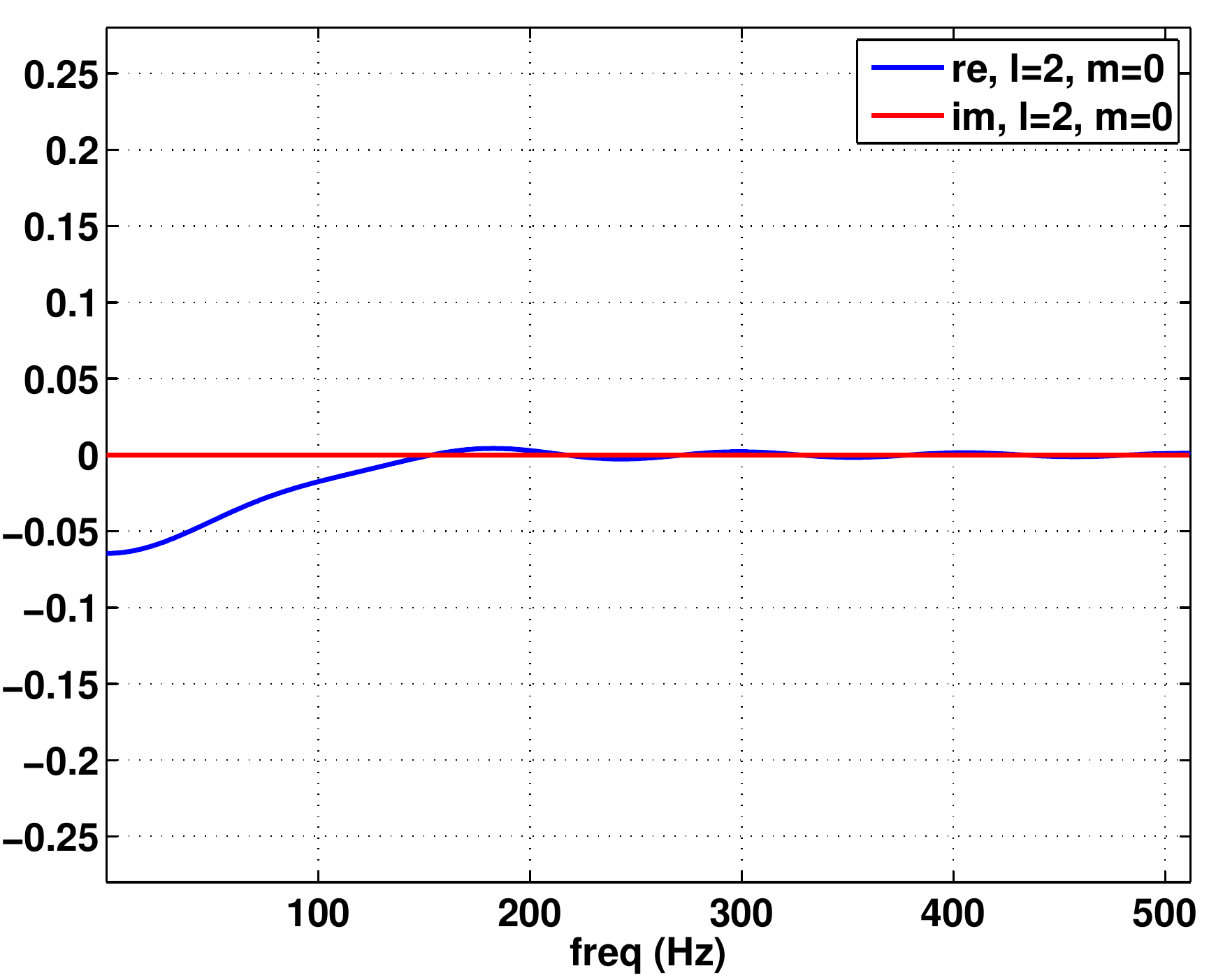}
\includegraphics[width=.32\textwidth]{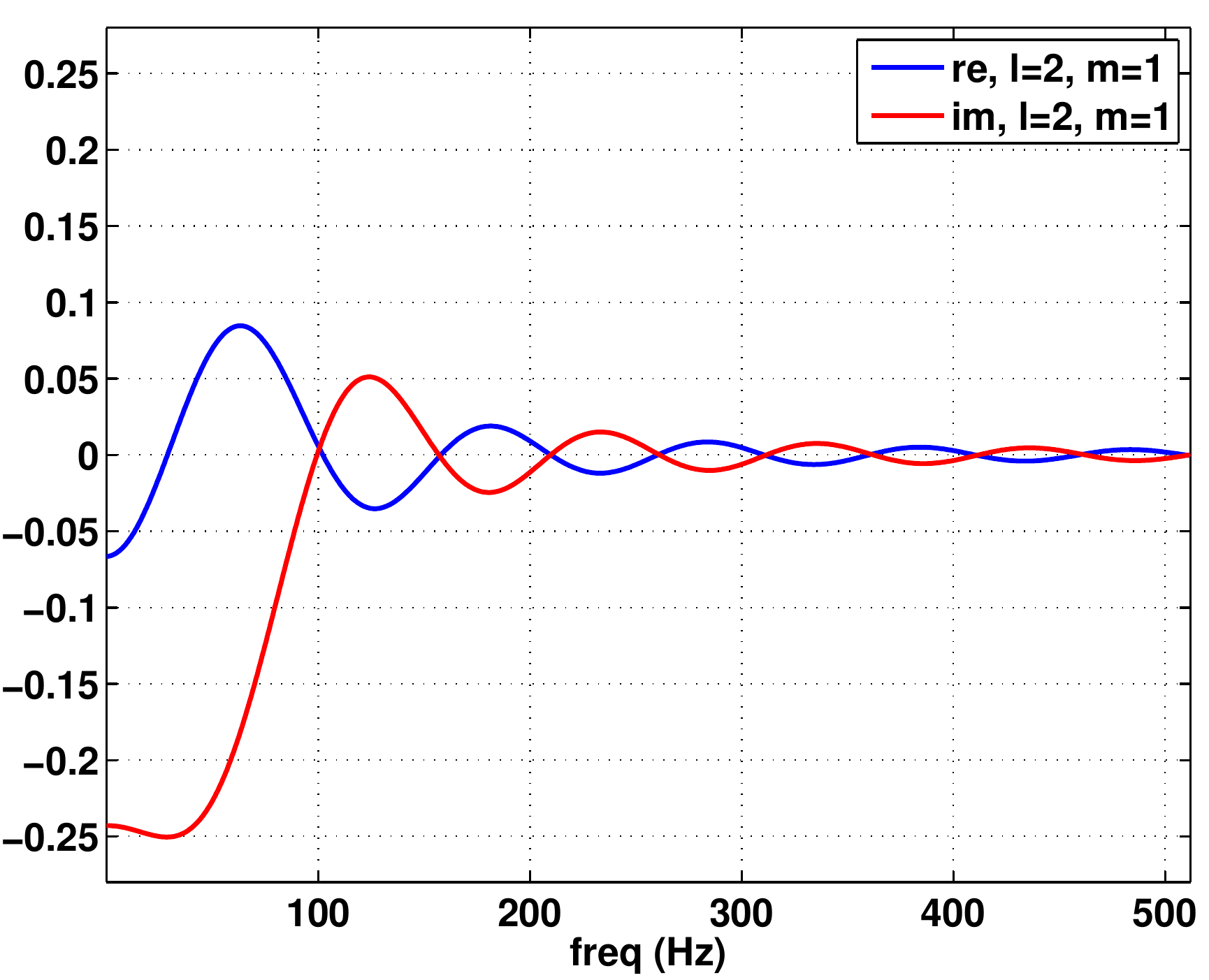}
\includegraphics[width=.32\textwidth]{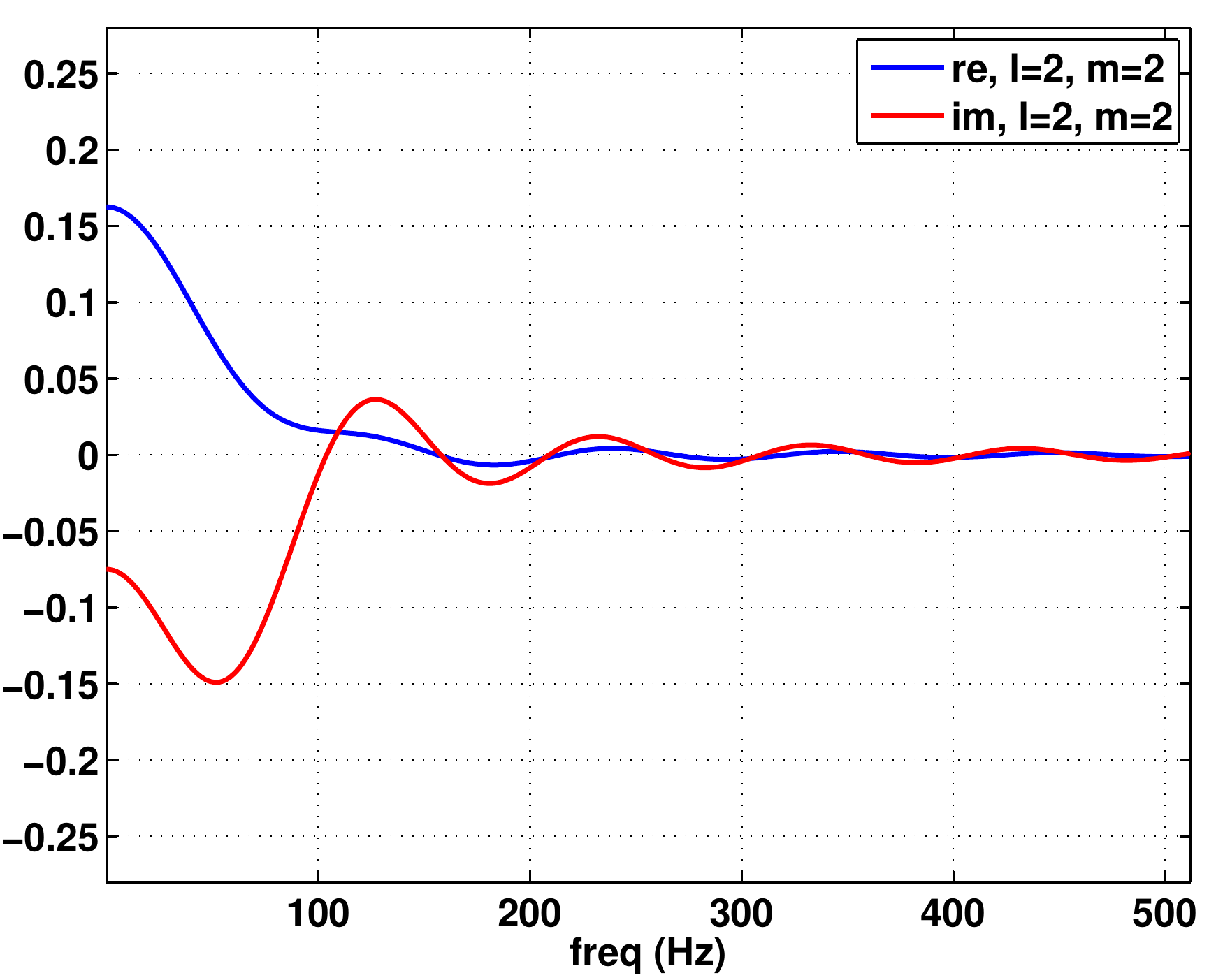}
\includegraphics[width=.32\textwidth]{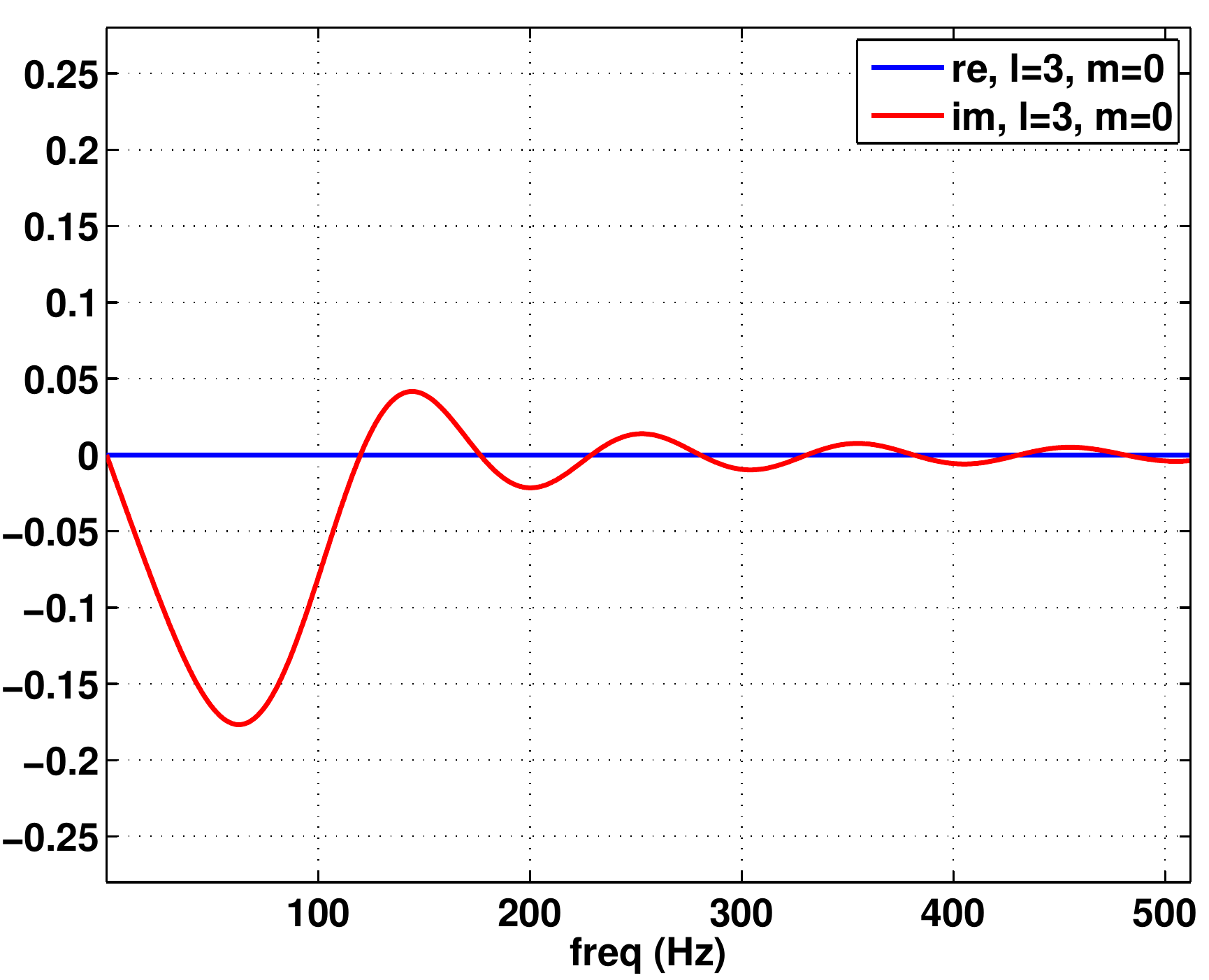}
\includegraphics[width=.32\textwidth]{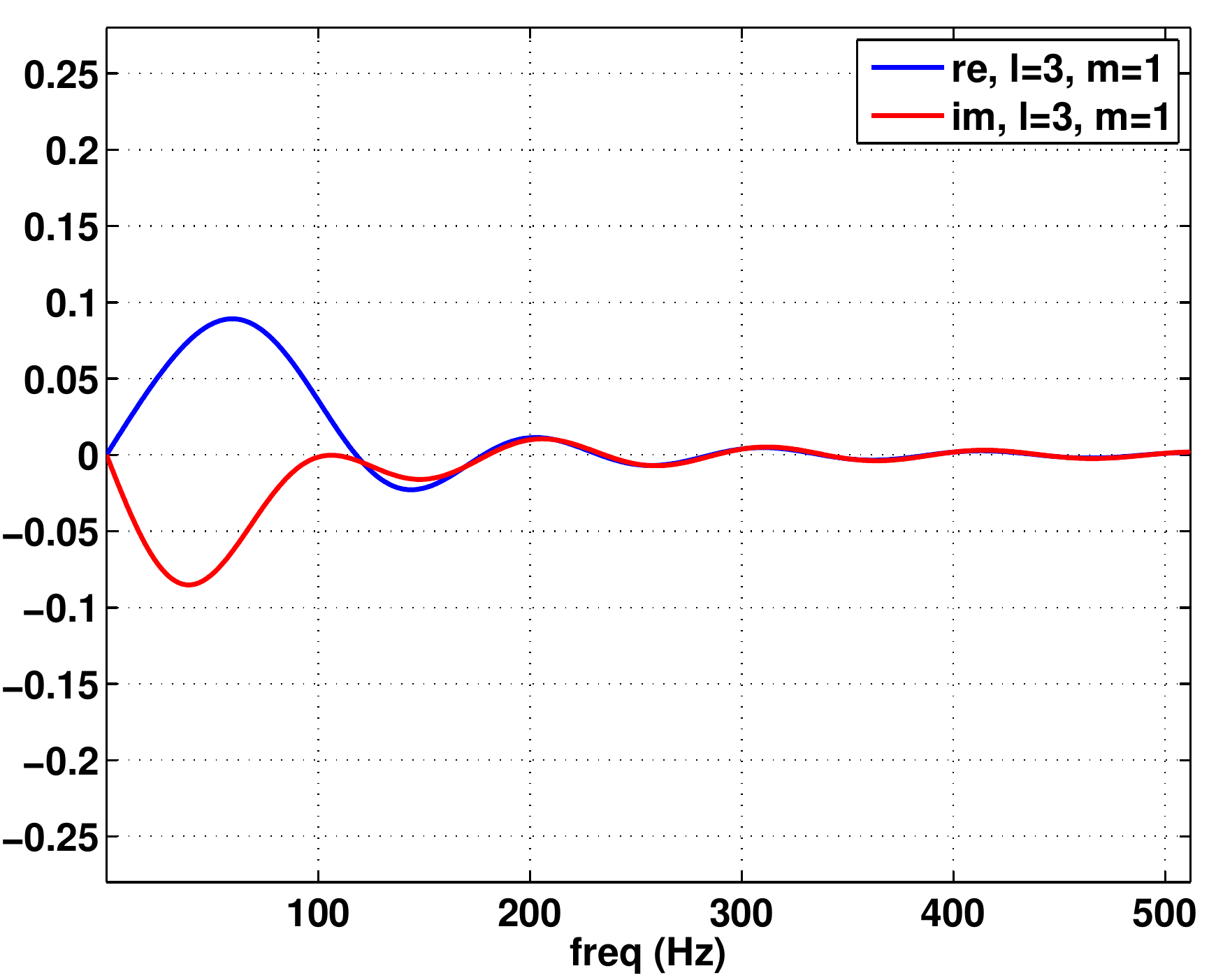}
\includegraphics[width=.32\textwidth]{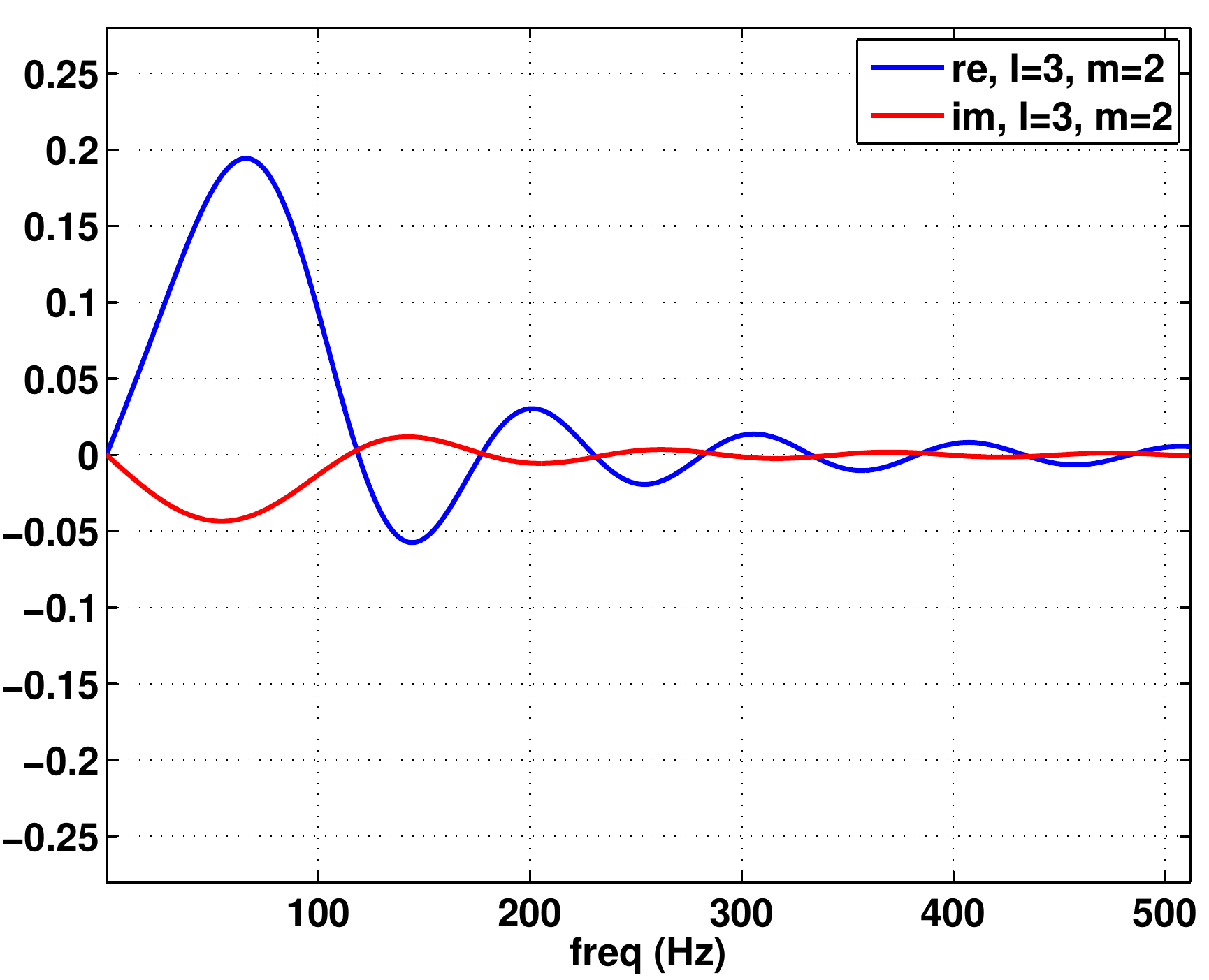}
\includegraphics[width=.32\textwidth]{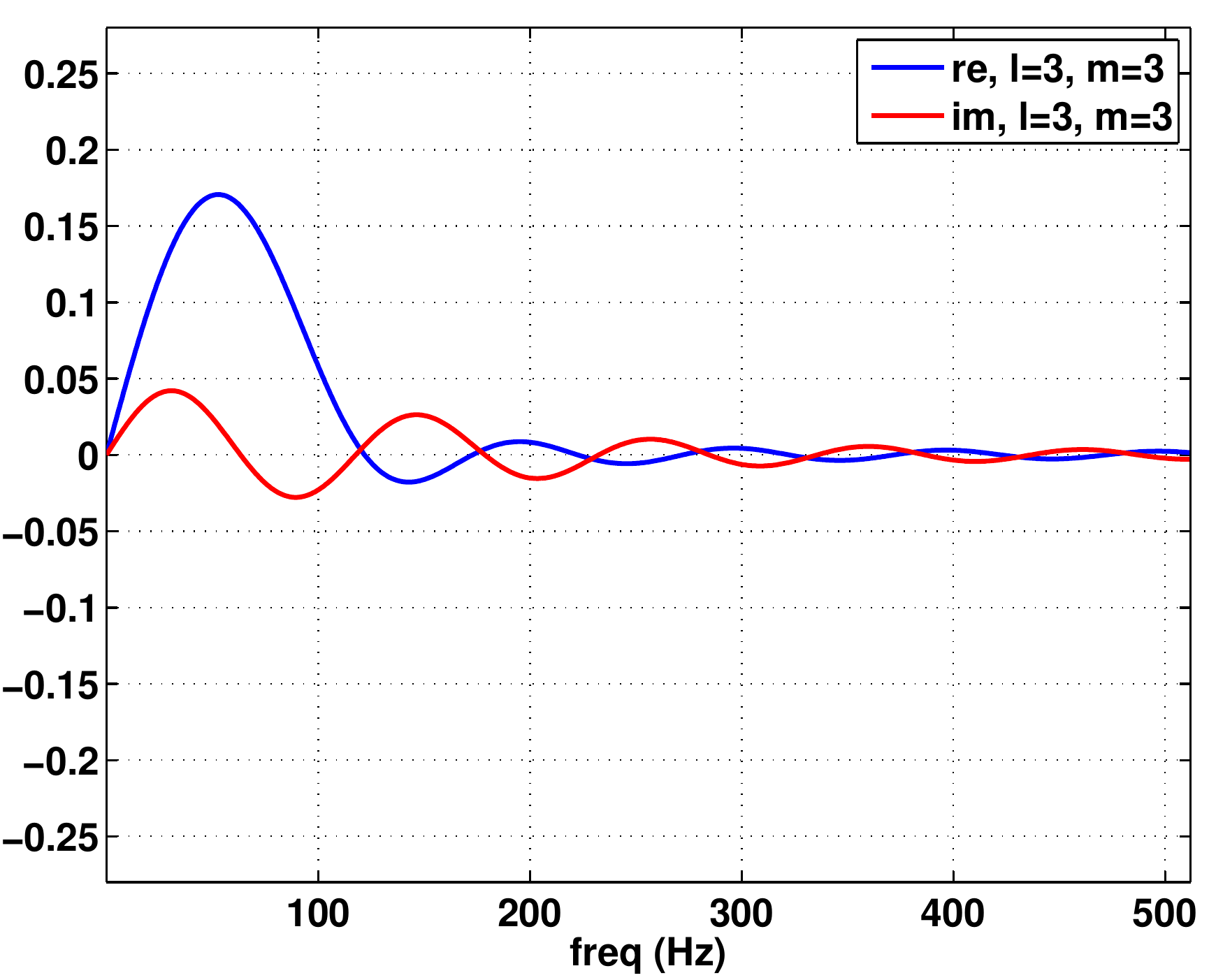}
\includegraphics[width=.32\textwidth]{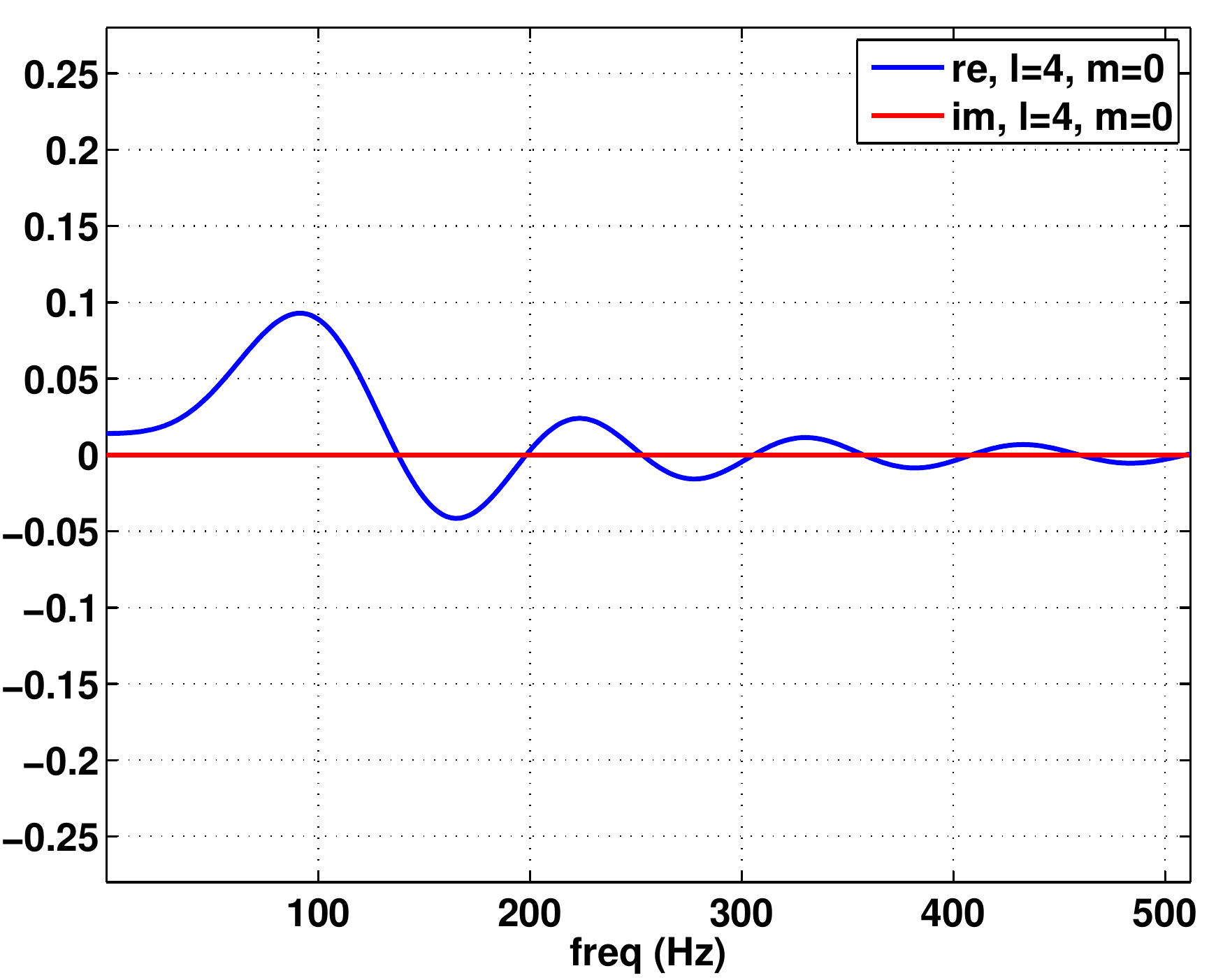}
\includegraphics[width=.32\textwidth]{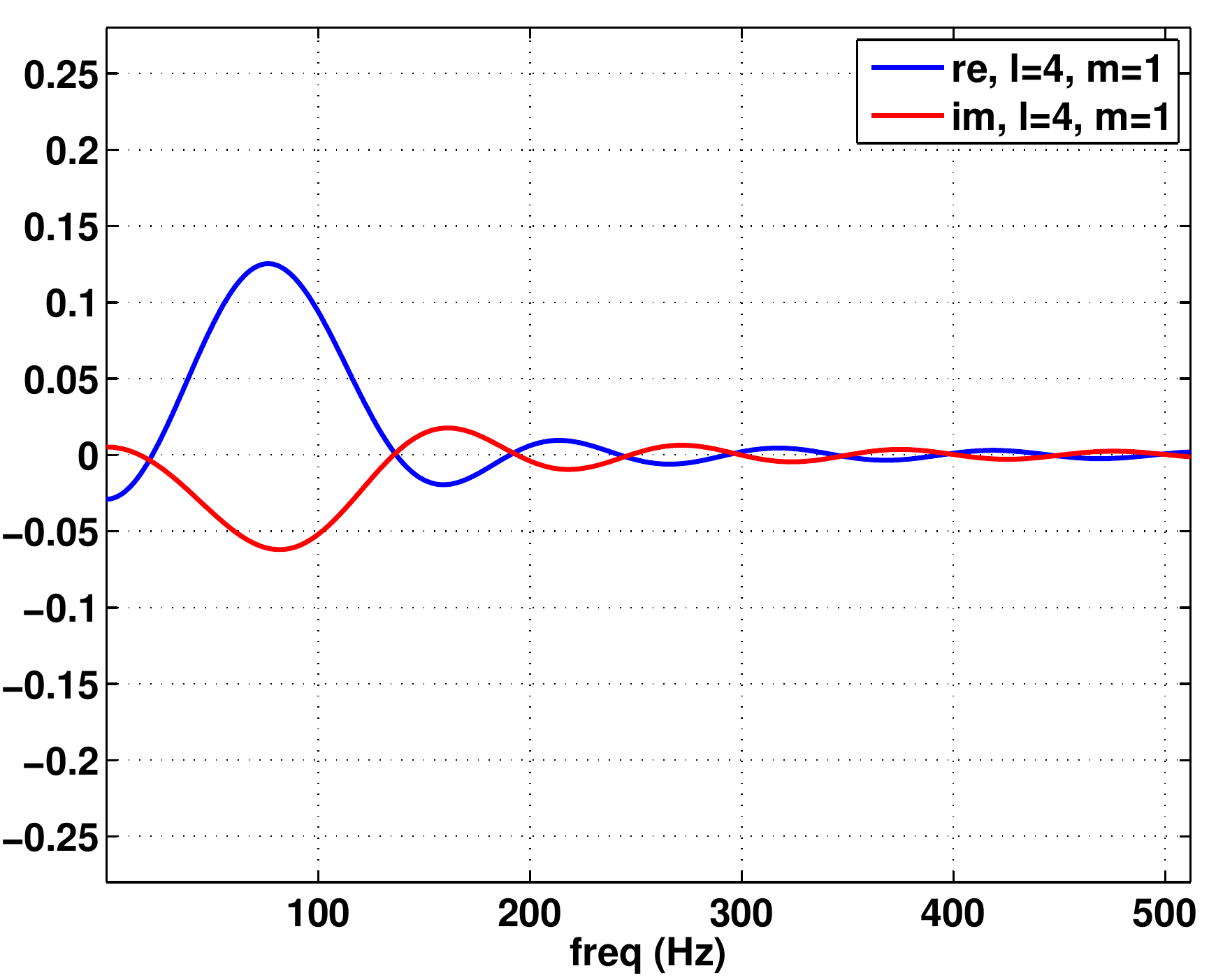}
\includegraphics[width=.32\textwidth]{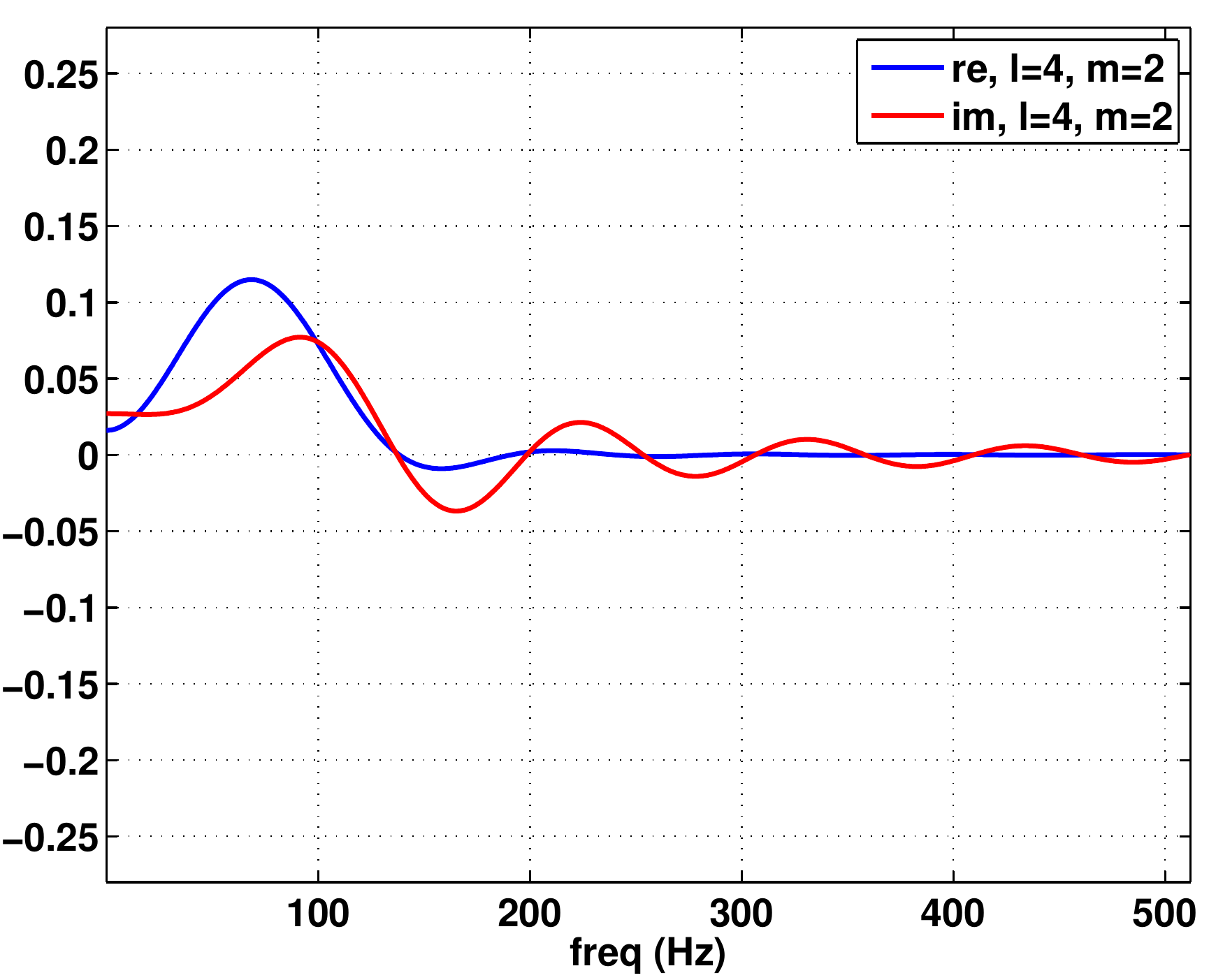}
\includegraphics[width=.32\textwidth]{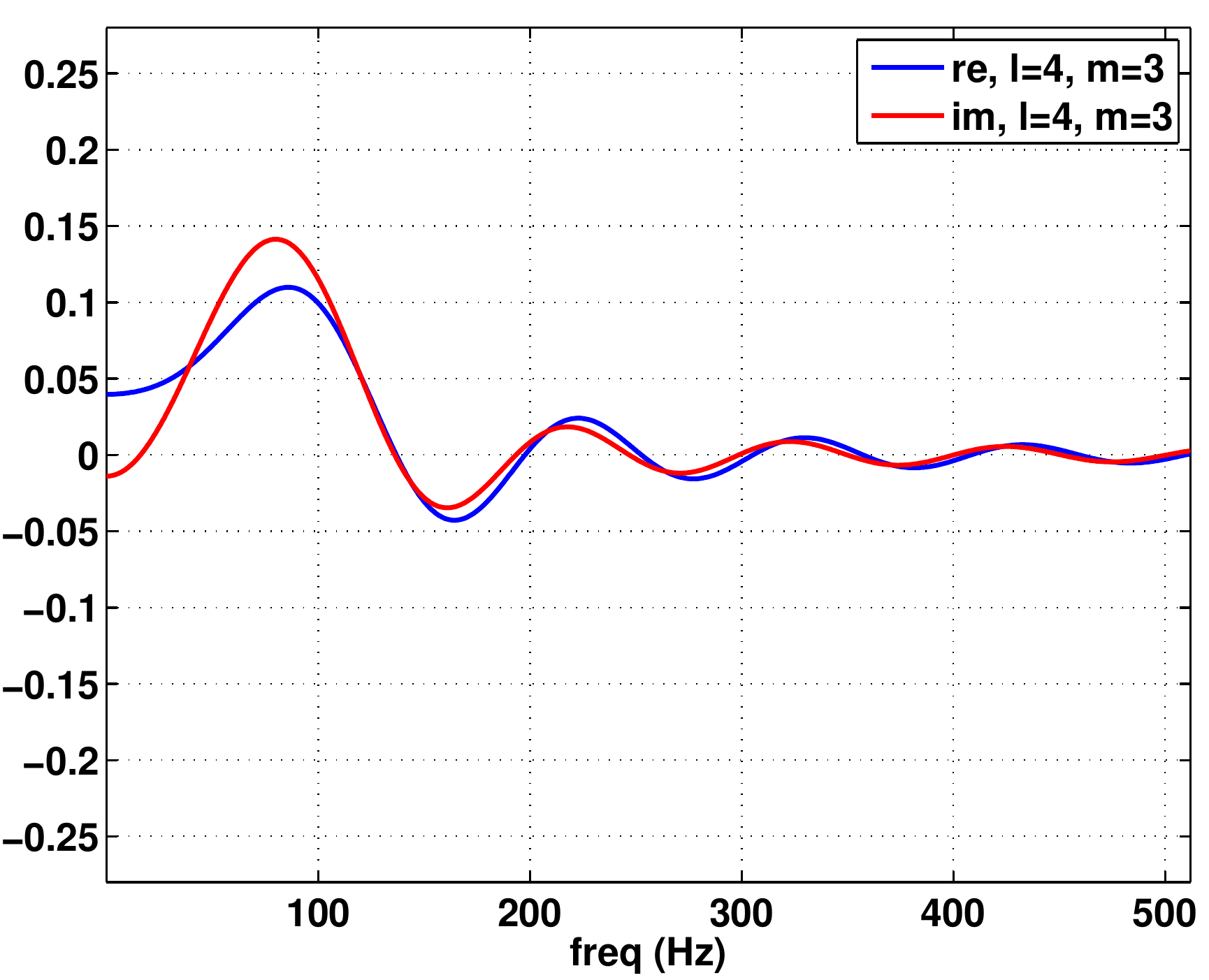}
\includegraphics[width=.32\textwidth]{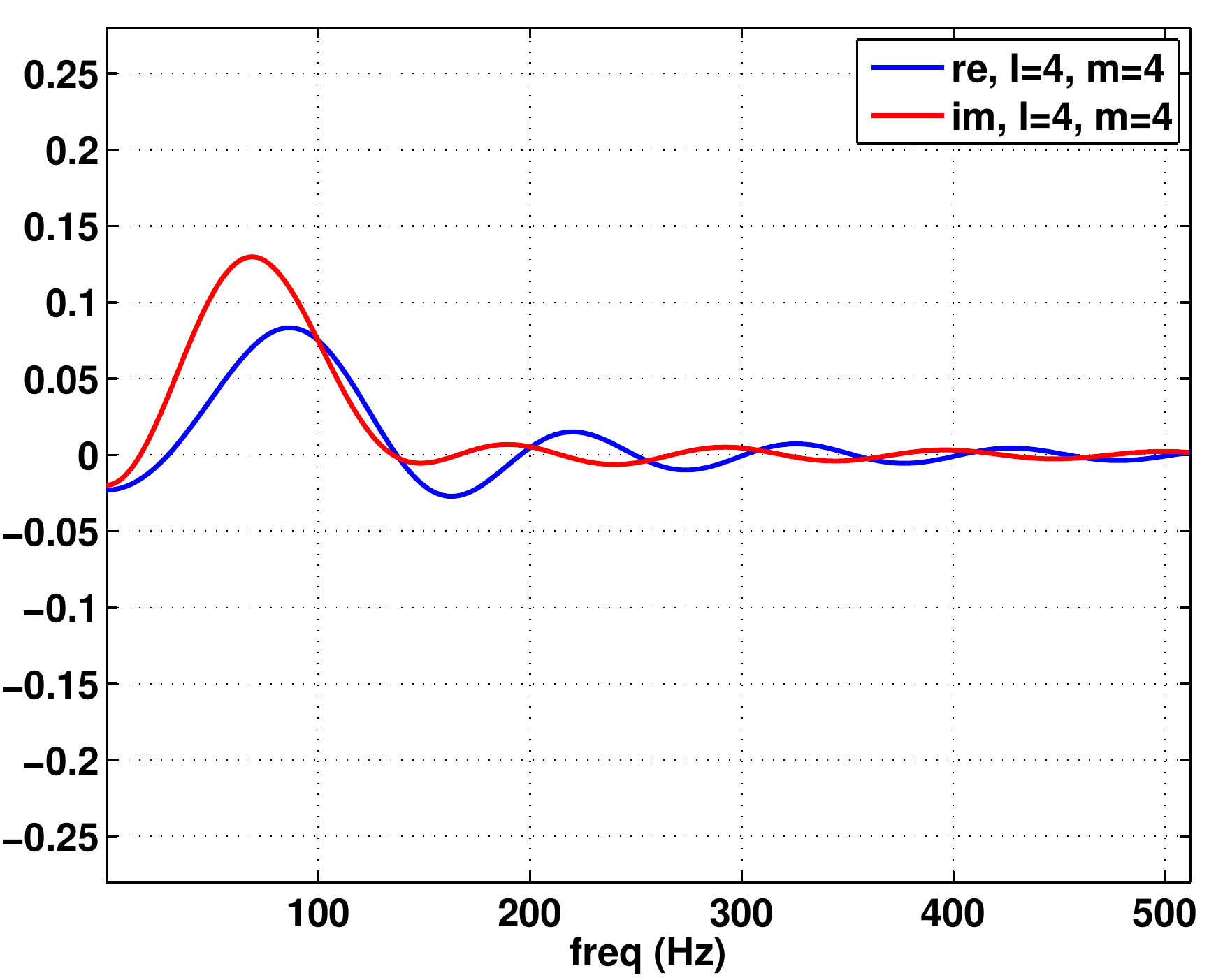}
\caption{Real and imaginary parts of the spherical
harmonic components $\gamma_{lm}(0;f)$ for the 
LIGO Hanford--LIGO Livingston detector pair.
Here we show plots for $l=0$, 1, 2, 3, 4 and $m\ge 0$.
For $m<0$, use (\ref{e:m<0}).}
\label{f:HLgammaLM}
\end{center}
\end{figure}

\noindent
As was also shown in \cite{Allen-Ottewill:1997}, one can derive 
{\em analytic} expressions for $\gamma_{lm}(t;f)$ for a pair of
Earth-based interferometers in the short-antenna limit.
If we set $t=0$, then $\gamma_{lm}(0;f)$ can be written as a
linear combination%
\footnote{The number of terms in the expansion is given by
$2+{\rm floor}(1+l/2)$.}
involving spherical Bessel functions,
$j_n(x)/x^n$ (for $l$ even) and  $j_n(x)/x^{n-1}$ (for $l$ odd),
where $x$ depends on the relative separation of the two 
detectors, $x \equiv 2\pi f|\Delta \vec x|/c$.
The coefficients of the expansions are complex numbers that 
depend on the relative orientation of the detectors.
Explicit expression for the first few spherical harmonic 
components for the LIGO Hanford--LIGO Livingston pair are given below:
\be
\begin{aligned}
\gamma_{00}(0;f)= 
&-0.0766 j_0(x)
-2.1528 j_1(x)/x
+2.4407 j_2(x)/x^2\,,
\\
\gamma_{10}(0;f)=
&- 0.0608i\,j_1(x)
- 2.6982i\,j_2(x)/x
+ 7.7217i\,j_3(x)/x^2\,,
\\
\gamma_{11}(0;f)=
&-(0.0519 + 0.0652i)j_1(x)
-(1.8621 + 1.0517i)j_2(x)/x
\\
&+(4.0108 - 2.4933i)j_3(x)/x^2\,,
\\
\gamma_{20}(0;f)=
&\ 0.0316 j_0(x)
-0.9612 j_1(x)/x
+10.9038 j_2(x)/x^2
-52.7905 j_3(x)/x^3\,,
\\
\gamma_{21}(0;f)=
&-(0.0669 - 0.0532i)j_0(x)
-(1.9647 - 2.6145i) j_1(x)
\\
&+(15.0524 -24.7604i)j_2(x)/x^2
-(36.5620 -50.7179i)j_3/x^3\,,
\\
\gamma_{22}(0;f)=
&-(0.0186 - 0.0807i) j_0(x)
+(1.2473 + 1.6858i) j_1(x)/x
\\
&-(12.2048 +12.5814i) j_2(x)/x^2
+(60.7859 +12.7191i) j_3(x)/x^3\,.
\end{aligned}
\ee
Note that the above numerical coefficients
do not agree with those in \cite{Allen-Ottewill:1997}
due to an overall normalization factor of 
$4\pi/5$ and 
phase $e^{im\phi}$, where $\phi = -38.52^\circ$ is 
the angle between the separation vector between
the vertices of the LIGO-Hanford and LIGO-Livingston
interferometers and the Greenwich meridian
\cite{Thrane-et-al:2009}.
Plots of the real and imaginary parts of 
$\gamma_{lm}(0;f)$ for $l=0$, 1, 2, 3, 4 and
$m\ge 0$ 
for the the LIGO Hanford-LIGO Livingston detector pair 
are given in Figure~\ref{f:HLgammaLM}.
For $m<0$, one can use the relation
\be
\gamma_{lm}(t; f) = (-1)^{l+m} \gamma_{l,-m}(t;f)\,,
\label{e:m<0}
\ee
which follows from the properties of the 
spherical harmonics $Y_{lm}(\hat n)$ (see Appendix~\ref{s:spinweightedY}).
Note that up to an overall normalization factor of 
$5/\sqrt{4\pi}$, the real part of 
$\gamma_{00}(0;f)$ is the Hanford-Livingston 
overlap function for an unpolarized, isotropic 
stochastic background, shown in 
Figure~\ref{f:overlapHL}.

\medskip
\noindent{\bf Example: Pulsar timing arrays}
\medskip

\begin{figure}[h!tbp]
\begin{center}
\subfigure[]{\includegraphics[width=.49\textwidth]{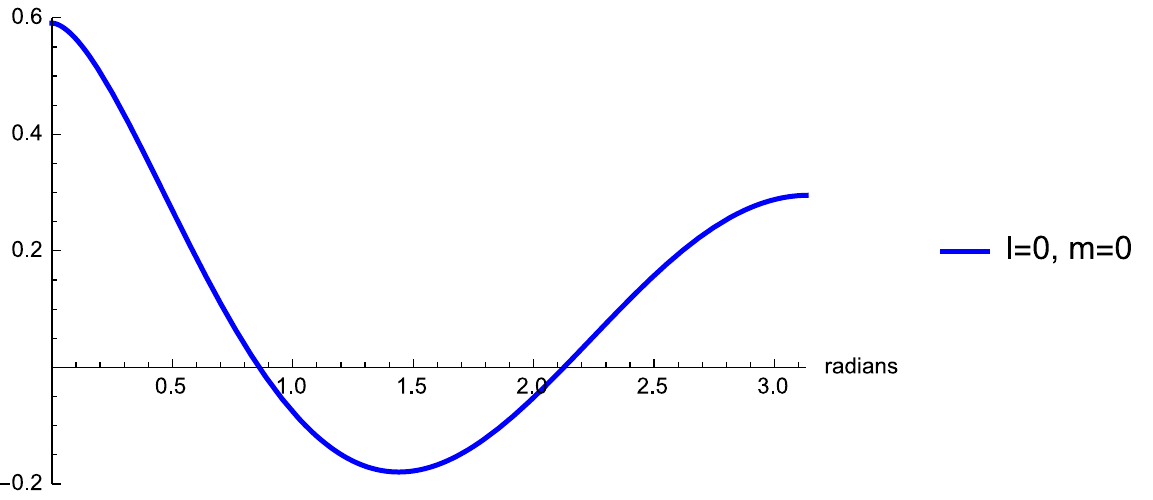}}
\subfigure[]{\includegraphics[width=.49\textwidth]{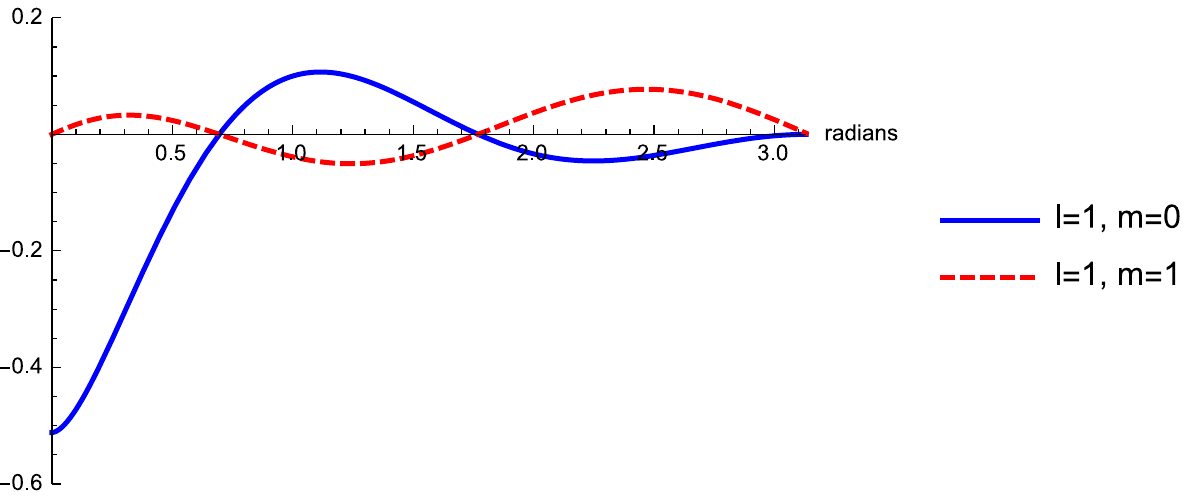}}
\subfigure[]{\includegraphics[width=.49\textwidth]{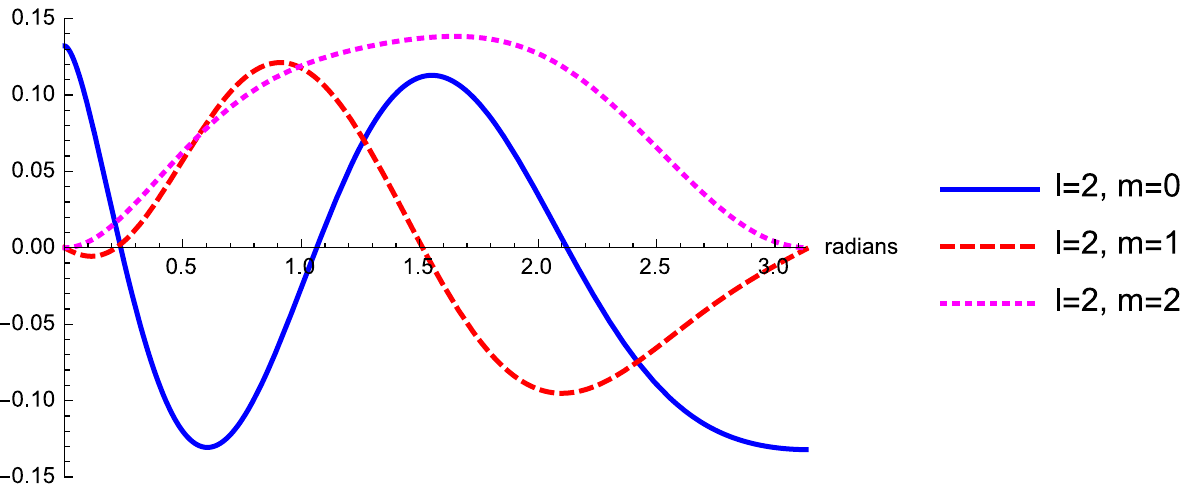}}
\subfigure[]{\includegraphics[width=.49\textwidth]{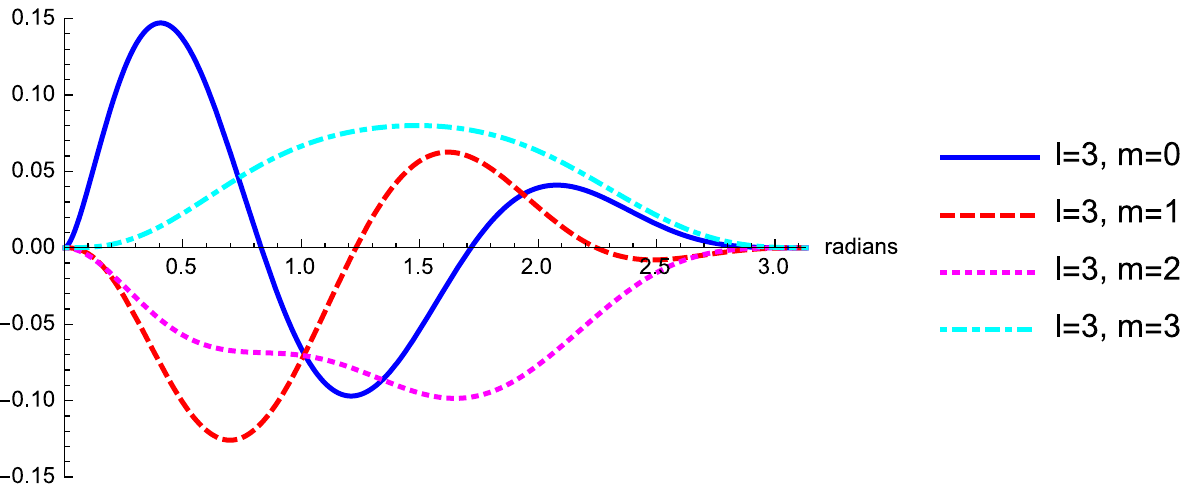}}
\subfigure[]{\includegraphics[width=.49\textwidth]{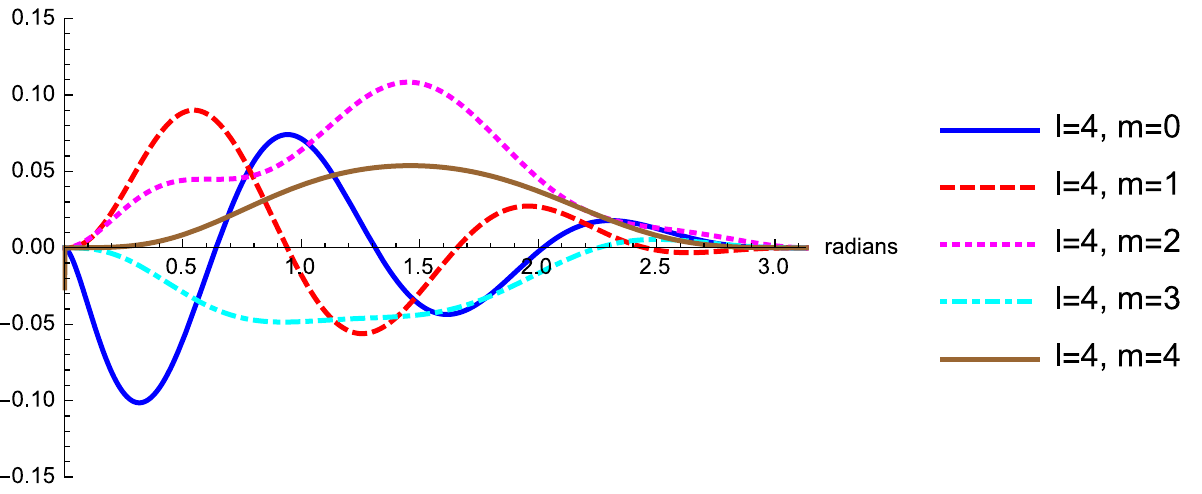}}
\subfigure[]{\includegraphics[width=.49\textwidth]{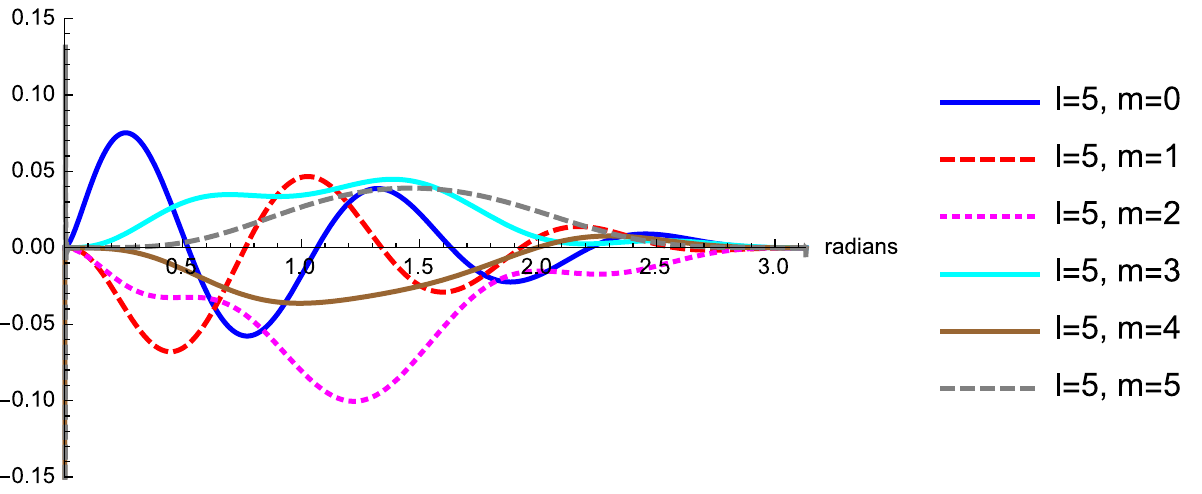}}
\caption{Spherical harmonic component functions
  $\gamma_{lm}(\zeta)$ for pulsar timing as a function 
  of the angle $\zeta$ between two distinct pulsars.
  Here we show plots for  $l=0,1,\cdots,5$ and $m\ge 0$.
  We used the Earth-term-only Doppler-frequency response
  (\ref{e:pulsarresponse-earthonly}) to calculate these functions.}
\label{f:PTAgammaLM}
\end{center}
\end{figure}

\noindent
In Figure~\ref{f:PTAgammaLM}, we show plots of the spherical 
harmonic components of $\gamma(t;f, \hat n)$ 
calculated using the Earth-term-only Doppler-frequency 
response functions (\ref{e:pulsarresponse-earthonly})
for pulsar timing.
Since there is no frequency or time-dependence for these
response functions, the spherical harmonic components of 
$\gamma(\hat n)$ depend only of the angular 
separation $\zeta$ between the two pulsars that 
define the detector pair.
As shown in \cite{Mingarelli-et-al:2013, Gair-et-al:2014},
these functions can be calculated analytically for {\em all} 
values of $l$ and $m$.
A detailed derivation with all the relevant formulae can 
be found in Appendix~E of \cite{Gair-et-al:2014};
there the calculation is done in a `computational' frame,
where one of the pulsars is located along the $z$-axis and 
the other is in the $xz$-plane, making an angle $\zeta$ with 
respect to the first.  
In this computational frame, all of the components 
$\gamma_{lm}(\zeta)$ are real.
Note that up to an overall normalization factor%
\footnote{The functions here are a factor of $1/2$ smaller
than those in Figure~8 in \cite{Gair-et-al:2014}, due to 
different definitions of $\gamma(t; f, \hat n)$.
Compare (115) in that paper to 
(\ref{e:gamma_khat}) and (\ref{e:gammaLM_def2}) above.}
of $3/\sqrt{4\pi}$, the function $\gamma_{00}(\zeta)$ is just 
the Hellings and Downs function for an unpolarized,
isotropic stochastic background, shown in Figure~\ref{f:HDcurveExact}.

\subsection{Modulations in the correlated output of two detectors}
\label{s:allen-ottewill}

For ground-based detectors like LIGO and Virgo, an anisotropic
gravitational-wave background will modulate the correlated 
output of a pair of detectors at harmonics of the Earth's 
rotational frequency.
It turns out that for an 
unpolarized, anisotropic background, the contribution 
to the $m$th harmonic of the correlation has a frequency 
dependence proportional to
\be
\bar H(f) \sum_{l=|m|}^\infty \gamma_{lm}(0;f){\cal P}_{lm}\,,
\ee
where ${\cal P}_{lm}$ are the spherical harmonic components of the
gravitational-wave power on the sky ${\cal P}(\hat n)$.
(We are assuming here that the spherical harmonic 
decomposition of ${\cal P}(\hat n)$ is with respect to a 
coordinate system whose $z$-axis points along the Earth's 
rotational axis.)
In this section, we derive the above result following the
presentation in \cite{Allen-Ottewill:1997} and construct
an optimal filter for the cross-correlation that maximizes the 
signal-to-noise ratio for the $m$th harmonic.
This was the first concrete approach that was proposed for 
detecting an anisotropic stochastic background.

\subsubsection{Time-dependent cross-correlation}

We start by writing down an expression (in the frequency
domain) for the correlated output of two ground-based detectors 
(e.g., LIGO Hanford and LIGO Livingston):
\be
\hat C(t) = 
\int_{-\infty}^\infty df\>
\tilde Q(t;f) \tilde d_1(t;f) \tilde d_2^*(t;f)\,,
\label{e:C(t)}
\ee
where $\tilde d_{1,2}(t;f)$ are (short-term) Fourier 
transforms (\ref{e:SFT}) centered around $t$, and 
where we have included a filter function $\tilde Q(t;f)$,
whose specific form we will specify later.
Since the cross-correlation is periodic with a period
$T_{\rm mod} = 1$~sidereal day (due to the motion of 
the detectors attached to the surface of the Earth),
we can expand $\hat C(t)$ as a Fourier series:
\be
\begin{aligned}
&\hat C(t) = \sum_{m=-\infty}^\infty 
\hat C_m e^{im 2\pi t/T_{\rm mod}}\,, 
%
%
\\
&\hat C_m = \frac{1}{T}\int_{0}^T dt\>
\hat C(t) e^{-im 2\pi t/T_{\rm mod}}\,.
\label{e:Cm}
\end{aligned}
\ee
Here $T$ is the total observation time, e.g., 1~sidereal year,
which we will assume for simplicity is an integer multiple of 
$T_{\rm mod}$. 

Assuming as usual that the detector noise is uncorrelated
across detectors, and using the expectation values (\ref{e:aniso_hh_new})
for an unpolarized, anisotropic background, we find
\be
\langle \hat C(t)\rangle 
=\frac{\tau}{2}\int_{-\infty}^\infty df\>
\tilde Q(t;f)\bar H(f)\sum_{l=0}^\infty\sum_{m=-l}^l \gamma_{lm}(t;f)
{\cal P}_{lm}\,, 
\ee
where $\gamma_{lm}(t;f)$ are the 
spherical harmonic components of $\gamma_{12}(t;f,\hat n)$.
(We have dropped the 12 indices to simplify the notation.)
Similarly, if we assume that the gravitational-wave signal 
is weak compared 
to the detector noise, and that the duration $\tau$ is 
also much larger than the correlation time of the 
detectors, then
\be
\langle \hat C(t)\hat C^*(t')\rangle 
-\langle \hat C(t)\rangle\langle \hat C^*(t')\rangle
\approx 
\frac{\tau}{4} \delta_{tt'}^2 \int_{-\infty}^\infty df\>
|\tilde Q(t;f)|^2 P_{n_1}(t;f) P_{n_2}(t;f)\,,
\ee
where $P_{n_I}(t;f)$ is the one-sided power spectral density
for the noise in detector $I=1,2$ centered around $t$.
These two results can now be cast in terms of the Fourier 
components $\hat C_m$ using (\ref{e:Cm}). 
Since (\ref{e:gammaLM-timedependence}) implies
\be
\frac{1}{T}\int_0^T dt\>
\gamma_{lm'}(t;f)
e^{-im2\pi t/T_{\rm mod}} 
= \delta_{mm'}\,\gamma_{lm}(0;f)\,,
\ee
we immediately obtain
\be
\langle \hat C_m\rangle 
=\frac{\tau}{2}\int_{-\infty}^\infty df\>
\tilde Q(t;f)\bar H(f)\sum_{l=|m|}^\infty\gamma_{lm}(0;f){\cal P}_{lm}\,,
\label{e:<Cm>}
\ee
where we used
\be
\sum_{l=0}^\infty \sum_{m=-l}^l = \sum_{m=-\infty}^\infty \sum_{l=|m|}^\infty\,.
\ee
Similarly,
\be
\langle \hat C_m\hat C^*_{m'}\rangle 
-\langle \hat C_m\rangle\langle \hat C^*_{m'}\rangle
\approx 
\delta_{mm'}\frac{1}{T}\left(\frac{\tau}{2}\right)^2
\int_{-\infty}^\infty df\>
|\tilde Q(t;f)|^2 P_{n_1}(t;f) P_{n_2}(t;f)
\ee
for the covariance of the estimators.

\subsubsection{Calculation of the optimal filter}

To determine the optimal form of the filter $\tilde Q(t;f)$
for the $m$th harmonic $\hat C_m$, we {\em maximize} the 
(squared) signal-to-noise:
\be
{\rm SNR}_m^2 
\equiv 
\frac{|\langle C_m\rangle|^2}
{\langle |\hat C_m|^2\rangle 
-|\langle \hat C_m\rangle|^2}
=\frac{T\left|
\int_{-\infty}^\infty df\>
\tilde Q(t;f)\bar H(f)\sum_{l=|m|}^\infty\gamma_{lm}(0;f){\cal P}_{lm}
\right|^2}
{\int_{-\infty}^\infty df\>
|\tilde Q(t;f)|^2 P_{n_1}(t;f) P_{n_2}(t;f)}\,.
\ee
The above expression can be written in a more suggestive
form if we introduce an {\em inner product}
on the space of complex-valued functions \cite{Allen:1997}: 
\be
(A,B)\equiv
\int_{-\infty}^\infty df\>
A(f)B^*(f) P_{n_1}(t;f) P_{n_2}(t;f)\,.
\ee
In terms of this inner product,
\be
{\rm SNR}_m^2 
=\frac{T\left|
\left(\tilde Q, \frac{\bar H}{P_{n_1}P_{n_2}}\sum_{l=|m|}^\infty
\gamma_{lm}{\cal P}_{lm}\right)\right|^2}
{(\tilde Q,\tilde Q)}\,.
\ee
But now the maximization problem is trivial, as it has been 
cast as a simple problem in vector algebra---namely to find the vector 
$\tilde Q$ that maximizes the ratio 
$|(\tilde Q, A)|^2/(\tilde Q,\tilde Q)$
for a {\em fixed} vector $A$.
But since this ratio is proportional to the squared cosine of 
the angle between $\tilde Q$ and $A$, it is maximized by 
choosing $\tilde Q$ {\em proportional} to $A$.
Thus,
\be
\tilde Q(t;f)
\propto
\frac{\bar H(f)}{P_{n_1}(t;f) P_{n_2}(t;f)}
\sum_{l=|m|}^\infty\gamma_{lm}(0;f){\cal P}_{lm}
\ee
is the form of the filter function that maximizes 
the SNR for the $m$th harmonic.

Note that this expression reduces to the standard
form of the optimal filter (\ref{e:Q_optfilter})
for an isotropic background,
${\cal P}_{lm} = \delta_{l0}\delta_{m0}{\cal P}_{00}$.
Note also that the optimal filter assumes knowledge of both the
spectral shape $\bar H(f)$ {\em and} the angular distribution
of gravitational-wave power on the sky, ${\cal P}_{lm}$.
So if one has some model for the expected anisotropy
(e.g., a dipole in the same direction as the cosmic
microwave background), then one can filter the 
cross-correlated data to be optimally sensitive to the 
harmonics $\hat C_m$ induced by that anisotropy.

\subsubsection{Inverse problem}

In \cite{Allen-Ottewill:1997}, there was no attempt to 
solve the {\em inverse problem}---that is, given the 
{\em measured values} of the correlation harmonics, 
how can one 
{\em infer} (or {\em estimate}) the components ${\cal P}_{lm}$?
The first attempt to solve the inverse problem was given
in \cite{Cornish:2001}, in the context of correlation 
measurements for both ground-based and space-based interferometers.
Further developments in solving the inverse problem
were given in subsequent
papers, e.g., \cite{Ballmer:2006, Ballmer-PhD:2006, Mitra-et-al:2008,
Thrane-et-al:2009}, which we explain in more detail
in the following subsections. 
Basically, these latter methods constructed frequentist maximum-likelihood 
estimators for the ${\cal P}_{lm}$, using singular-value decomposition
to `invert' the Fisher matrix (or point spread function), 
which maps the true gravitational-wave power distribution
to the measured distribution on the sky.

\subsection{Maximum-likelihood estimates of gravitational-wave power}
\label{s:ML}

In this section, we describe an approach for constructing
maximum-likehood estimates of the gravitational-wave
power distribution ${\cal P}(\hat n)$.
It is a solution to the inverse problem discussed at the end 
of the previous subsection.
But since a network of gravitational-wave detectors typically
does not have perfect coverage of the sky, the inversion 
requires some form of regularization, which we describe below.
The gravitational-wave radiometer and spherical harmonic decomposition
methods (Section~\ref{s:radiometer-SHD}) are the two main 
implementations of this approach, and have been used to analyze
LIGO science data \cite{Abadie-et-al:S5-anisotropic, LVC:O1-anisotropic}.

\subsubsection{Likelihood function and maximum-likelihood estimators}
\label{s:likelihood-MLestimators}

As shown in Section~\ref{s:anisotropic-crosscorr} the 
cross-correlated data from two detectors
\be
\hat C_{IJ}(t; f) = \frac{2}{\tau}\tilde d_I(t;f)\tilde d_J^*(t;f)
\label{e:CIJ_new}
\ee
has expectation values
\be
\langle \hat C_{IJ}(t; f)\rangle
=\bar H(f) \int d^2\Omega_{\hat n} \>
\gamma_{IJ}(t; f, \hat n){\cal P}(\hat n)\,.
\label{e:<CIJ>_new}
\ee
We can write this relation abstractly as a matrix equation
\be
\langle \hat C_{IJ}\rangle = M_{IJ}\,{\cal P}\,,
\label{e:<CIJ>_matrix}
\ee
where $M_{IJ}\equiv\bar H(f)\gamma_{IJ}(t;f,\hat n)$ and 
the matrix product is summation over directions $\hat n$ on the sky.
The covariance matrix for the cross-correlated data
is given by
\be
\begin{aligned}
N_{tf,t'f'}
&\equiv\langle \hat C_{IJ}(t; f) \hat C^*_{IJ}(t'; f')\rangle
-\langle \hat C_{IJ}(t; f)\rangle\langle \hat C^*_{IJ}(t'; f')\rangle
\\
&\approx \delta_{tt'}\delta_{ff'} P_{n_I}(t; f) P_{n_J}(t; f)\,,
\end{aligned}
\ee
where we have assumed as before that there is no 
cross-correlated detector noise, and that the 
gravitational-wave signal is weak compared to the
detector noise.

If we treat the detector noise and the gravitational-wave
spectral shape $\bar H(f)$ as known quantities
(or if we estimate the detector noise from the 
auto-correlated output of each detector), then we 
can write down a likelihood function for the cross-correlated 
data given the signal model (\ref{e:<CIJ>_matrix}).
Assuming a Gaussian-stationary distribution for the noise,
we have
\be
p(\hat C|{\cal P}) \propto
\exp\left[-\frac{1}{2}
(\hat C- M{\cal P})^\dagger N^{-1}
(\hat C- M{\cal P})\right]\,,
\label{e:likelihood_P(k)}
\ee
where we have temporarily dropped the $IJ$ indices for 
notational convenience.%
\footnote{The multiplications inside the exponential
are {\em matrix} multiplications---either summations 
over sky directions $\hat n$
or summations over discrete times and frequencies, $t$
and $f$.}
Since the gravitational-wave power distribution ${\cal P}$
enters quadratically in the exponential of the likelihood, 
we can immediately
write down the maximum-likelihood estimators of ${\cal P}$:
\be
\hat {\cal P}
=F^{-1} X\,,
\label{e:Phat}
\ee
where
\be
F\equiv M^\dagger N^{-1} M\,,
\qquad
X\equiv M^\dagger N^{-1} \hat C\,.
\label{e:F,X}
\ee
The (square) matrix $F$ is called the {\em Fisher information matrix}.
It is typically a singular matrix, since the response 
matrix $M=\bar H\gamma$ usually has {\em null} directions 
(i.e., anisotropic
distributions of gravitational-wave power that are mapped to 
zero by the detector response).
Inverting $F$ therefore requires some sort of regularization,
such as singular-value decomposition \cite{Press:1992}
(Section~\ref{s:svd_power}).
The vector $X$ is the so-called {\em dirty map}, as it 
represents the gravitational-wave sky as `seen' by 
a pair of detectors.
If the spectral shape $\bar H(f)$ that we used for our
signal model exactly matches that of the observed
background, then 
\be
\langle X\rangle 
= M^\dagger N^{-1} M \,{\cal P} 
= F\,{\cal P}\,.
\label{e:<X>}
\ee
Thus, even in the absence of noise, a point source 
${\cal P}(k) = \delta^2(\hat n,\hat n_0)$ does not 
map to a point source by the response of the detectors,
but it maps instead to $F_{\hat n\hat n_0}$.
This `blurring' or `spreading' of point sources is represented
by a {\em point spread function}, which is a characteristic
feature of any imaging system.
We give plots of point spread functions for both pulsar
timing arrays and ground-based interferometers in
Section~\ref{s:PSF_power}.

\subsubsection{Extension to a network of detectors}

The above results generalize to a {\em network} of detectors.
One simply replaces $X$ and $F$ in (\ref{e:Phat}) 
by their network expressions, which are simply sums of the
the dirty maps and Fisher matrices for each distinct
detector pair:
\be
X= \sum_I\sum_{J>I} X_{IJ}\,,
\qquad
F= \sum_I\sum_{J>I} F_{IJ}\,.
\ee
Explicit expressions for the dirty map and Fisher matrix 
for a network of detectors are:
\be
X \equiv
X_{\hat n} =
\sum_I\sum_{J>I}\sum_t\sum_f 
\gamma_{IJ}^*(t;f,\hat n)
\frac{\bar H(f)}{P_{n_I}(t;f)P_{n_J}(t,f)}
\hat C_{IJ}(t; f)\,,
\label{e:Xn-power}
\ee
and
\be
F \equiv
F_{\hat n\hat n'} =
\sum_I\sum_{J>I}\sum_t\sum_f 
\gamma_{IJ}^*(t;f,\hat n)
\frac{\bar H^2(f)}{P_{n_I}(t;f)P_{n_J}(t,f)}
\gamma_{IJ}(t; f,\hat n')\,.
\label{e:Fnn'-power}
\ee
Note that including more detectors in the network is
itself a form of regularization, as adding more detectors
typically means better coverage of the sky.
This tends to `soften' the singularities that may exist
when trying to deconvolve (i.e., invert) the detector response.

\subsubsection{Error estimates}

Using (\ref{e:<X>}) it follows that
$\hat{\cal P}$ is an unbiased estimator of ${\cal P}$:
\be
\langle \hat{\cal P}\rangle = {\cal P}\,.
\label{e:<Phat>}
\ee
Similarly, in the weak-signal approximation,
\be
\begin{aligned}
&\langle XX^\dagger\rangle - \langle X\rangle \langle X^\dagger\rangle
\approx F\,,
\\
&\langle \hat{\cal P}\hat{\cal P}^\dagger\rangle - 
\langle \hat{\cal P}\rangle \langle \hat{\cal P}^\dagger\rangle
\approx F^{-1}\,.
\label{e:cov(Phat)}
\end{aligned}
\ee
Thus, $F$ is the covariance matrix for the dirty map $X$, 
while $F^{-1}$ is the covariance matrix of the clean map $\hat{\cal P}$.
We will see below (Section~\ref{s:svd_power}) 
that regularization necessarily changes 
these results as one cannot recover modes of ${\cal P}$ 
to which the detector network is insensitive.
This introduces a bias in $\hat {\cal P}$, and changes the 
corresponding elements of the covariance matrix for $\hat{\cal P}$.

\subsubsection{Point spread functions}
\label{s:PSF_power}

\begin{figure}[h!tbp]
\begin{center}
\includegraphics[trim=3cm 4cm 3cm 2.5cm, clip=true, width=.32\textwidth]{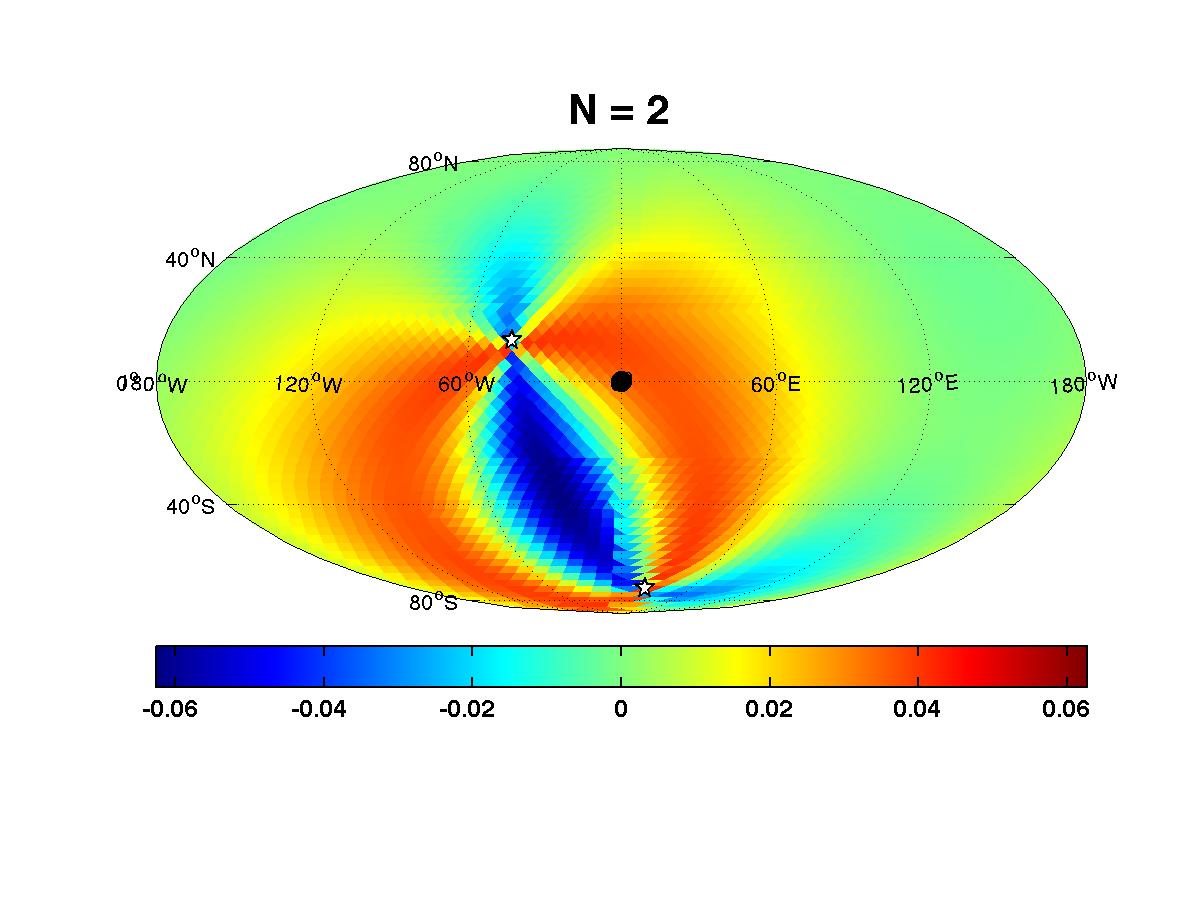}
\includegraphics[trim=3cm 4cm 3cm 2.5cm, clip=true, width=.32\textwidth]{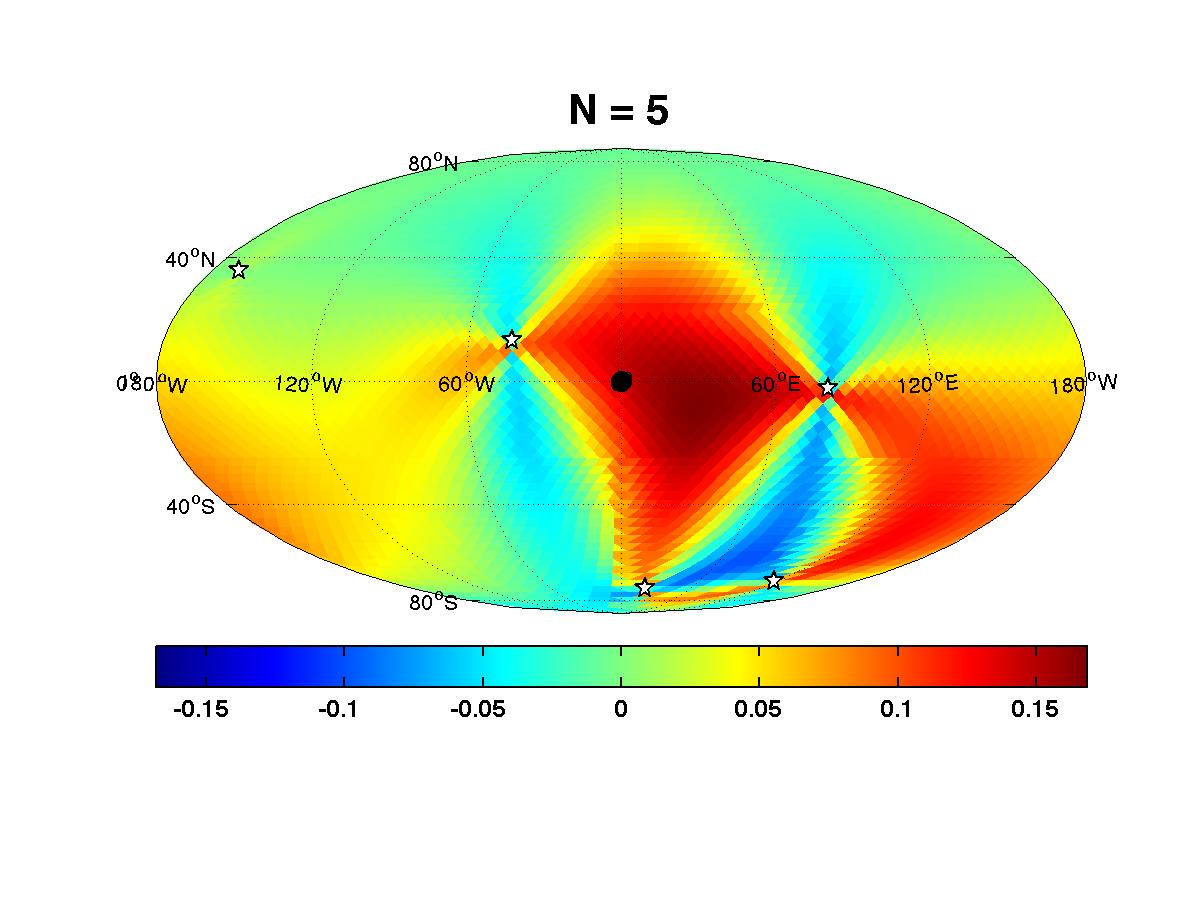}
\includegraphics[trim=3cm 4cm 3cm 2.5cm, clip=true, width=.32\textwidth]{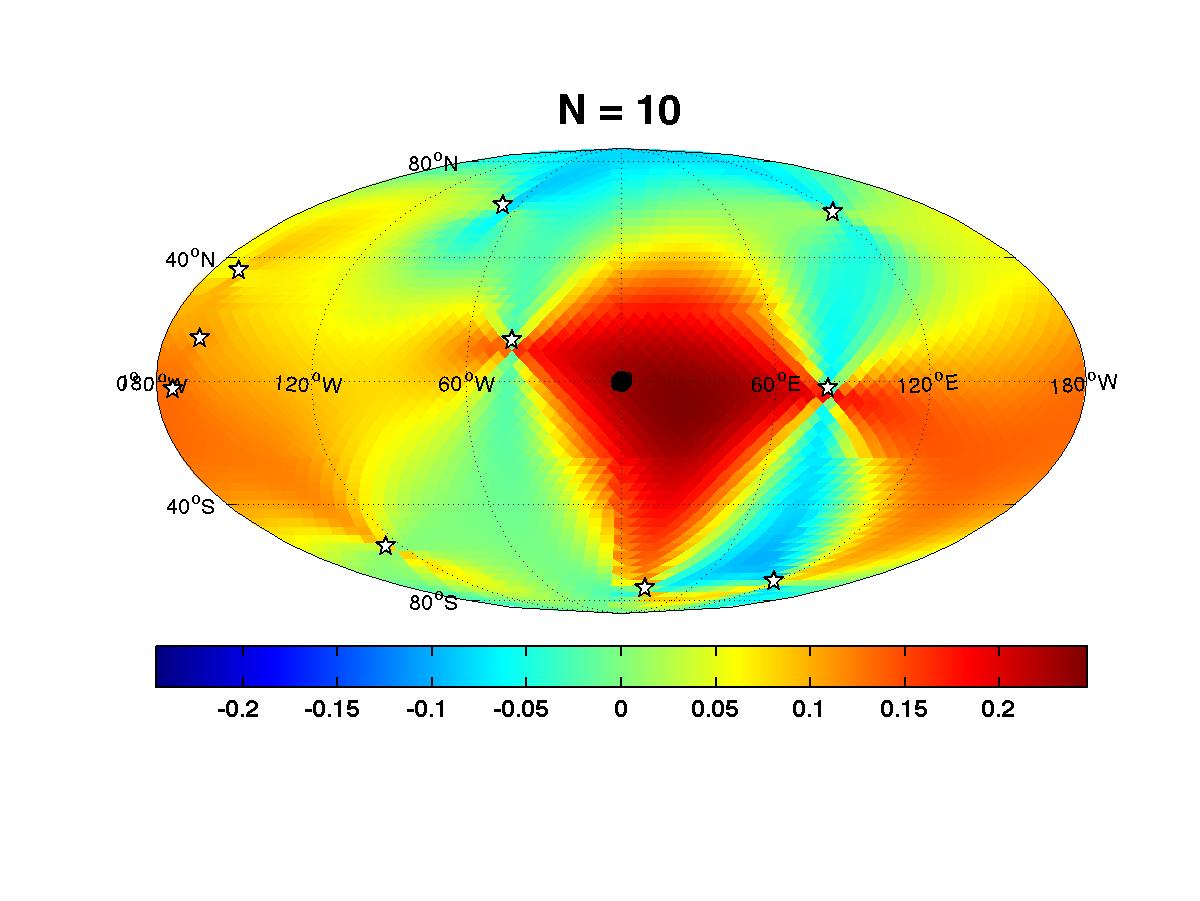}
\includegraphics[trim=3cm 4cm 3cm 2.5cm, clip=true, width=.32\textwidth]{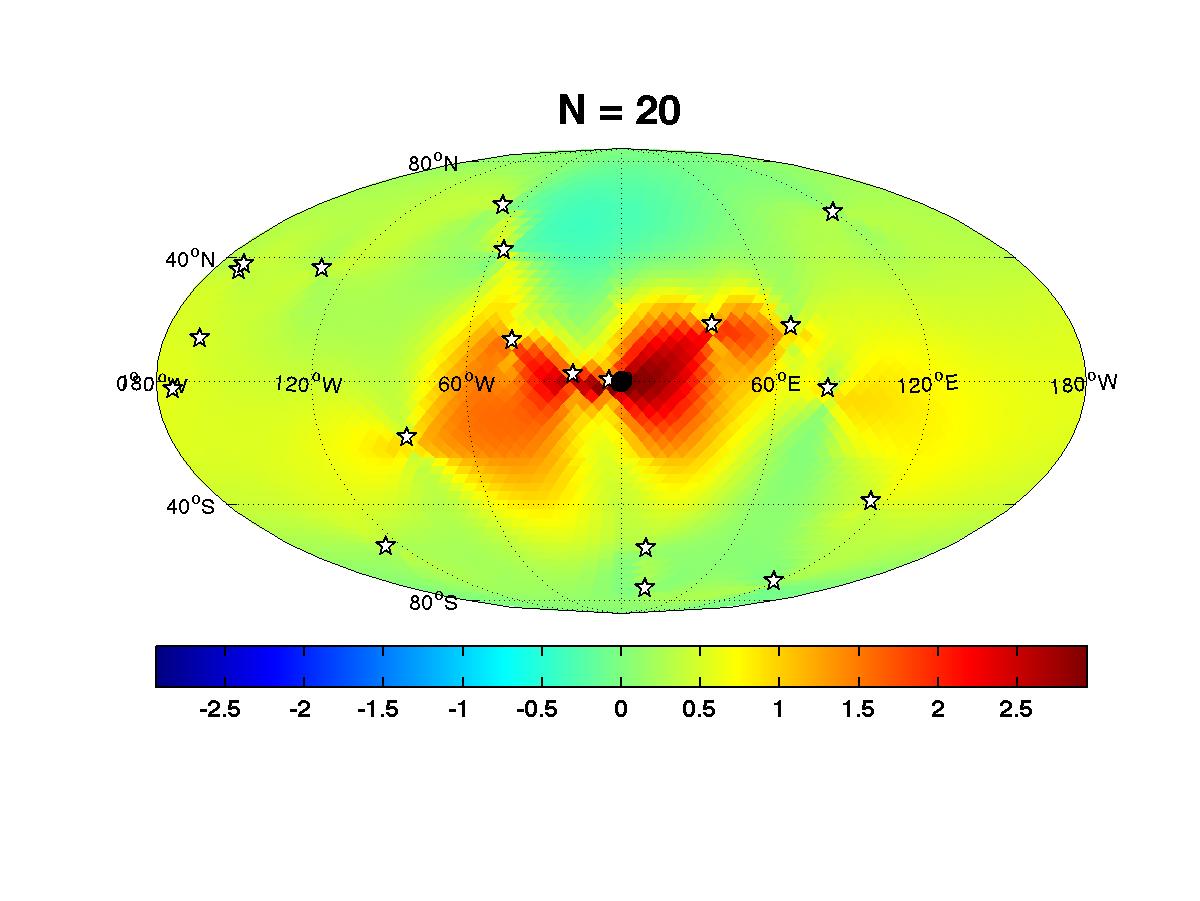}
\includegraphics[trim=3cm 4cm 3cm 2.5cm, clip=true, width=.32\textwidth]{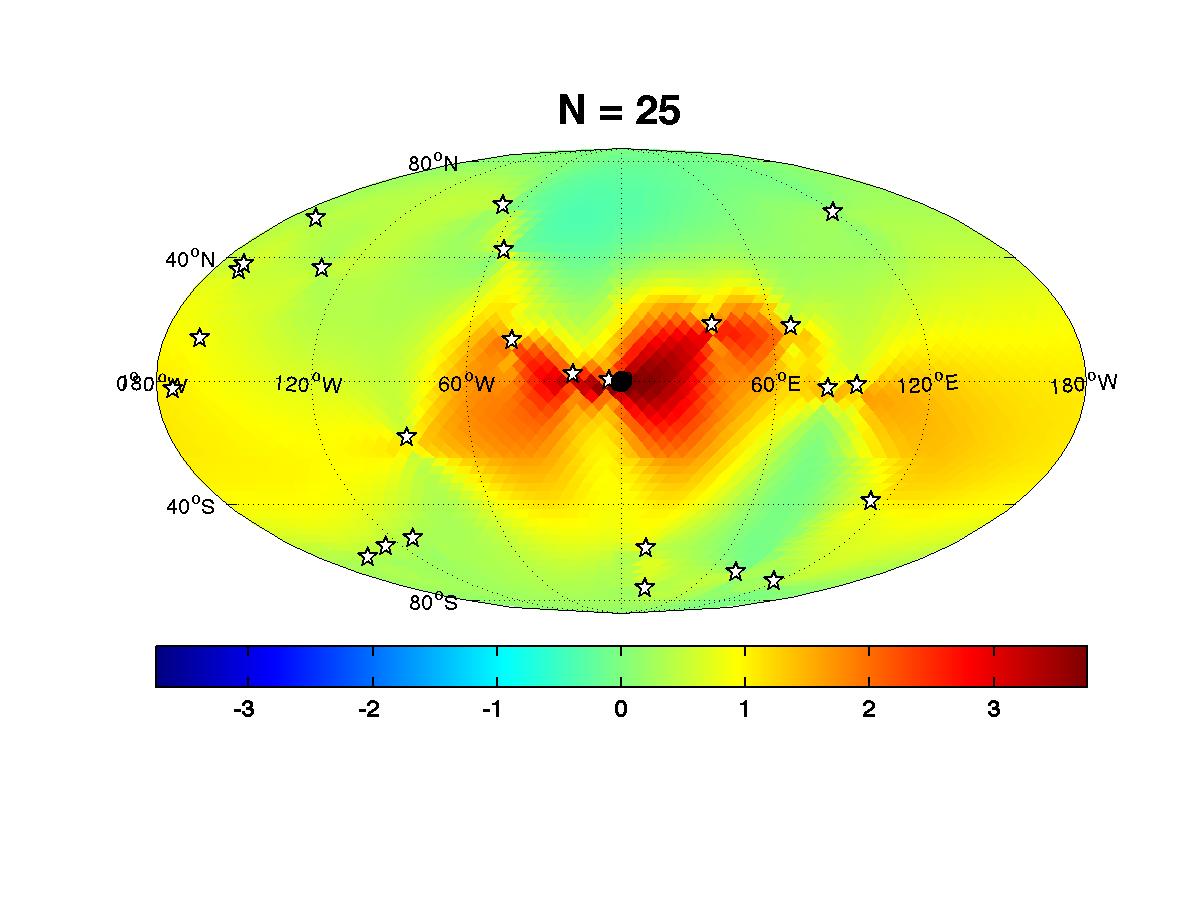}
\includegraphics[trim=3cm 4cm 3cm 2.5cm, clip=true, width=.32\textwidth]{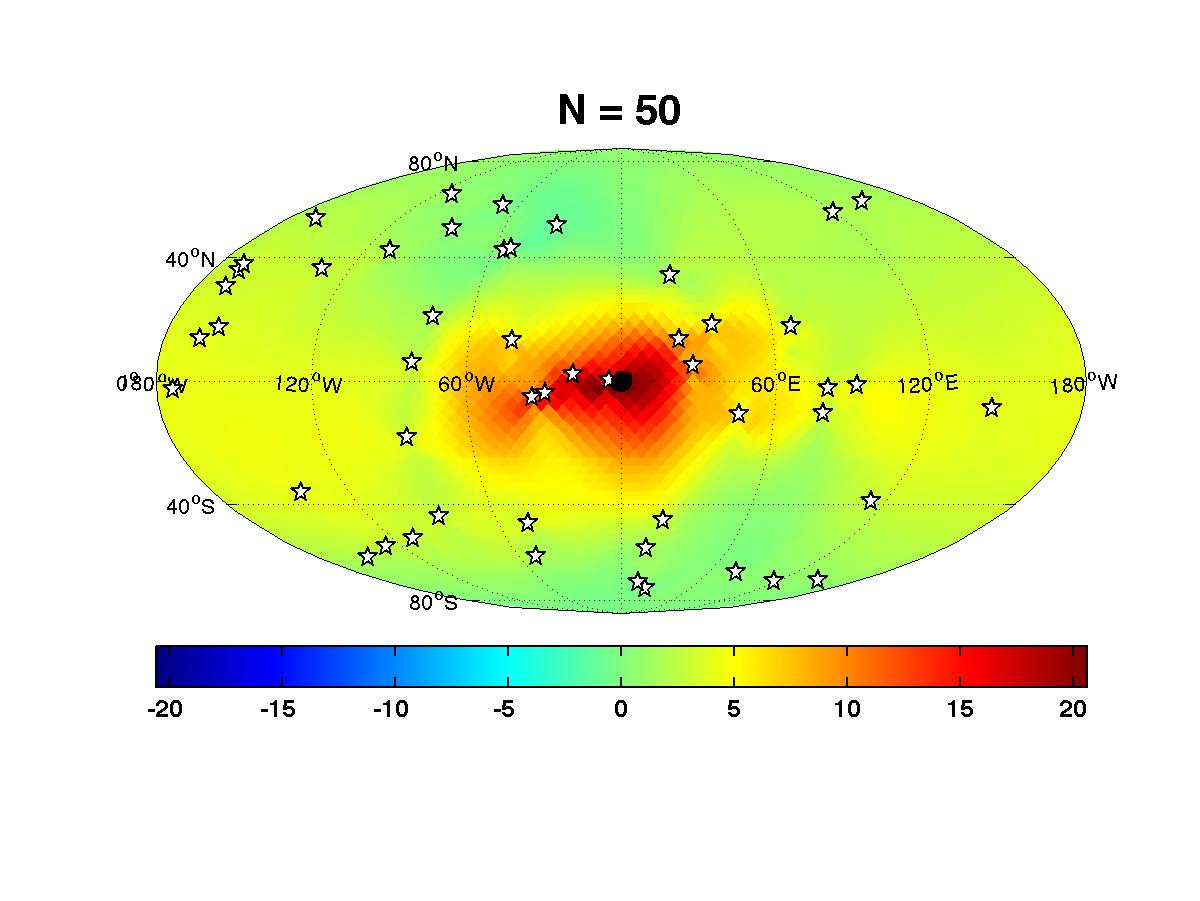}
\caption{Point spread functions for gravitational-wave
power for pulsar timing arrays consisting of 
$N=2$, 5, 10, 20, 25, 50 pulsars.
The point source is located at the center of the maps,
$(\theta,\phi)=(90^\circ, 0^\circ)$, indicated by a black dot.
The pulsar locations (indicated by white stars) are
randomly placed on the sky.
The point spread function becomes tighter as the number
of pulsars in the array increases.}
\label{f:PSFpulsar-random}
\end{center}
\end{figure}

\begin{figure}[h!tbp]
\begin{center}
\includegraphics[trim=3cm 4cm 3cm 2.5cm, clip=true, width=.49\textwidth]{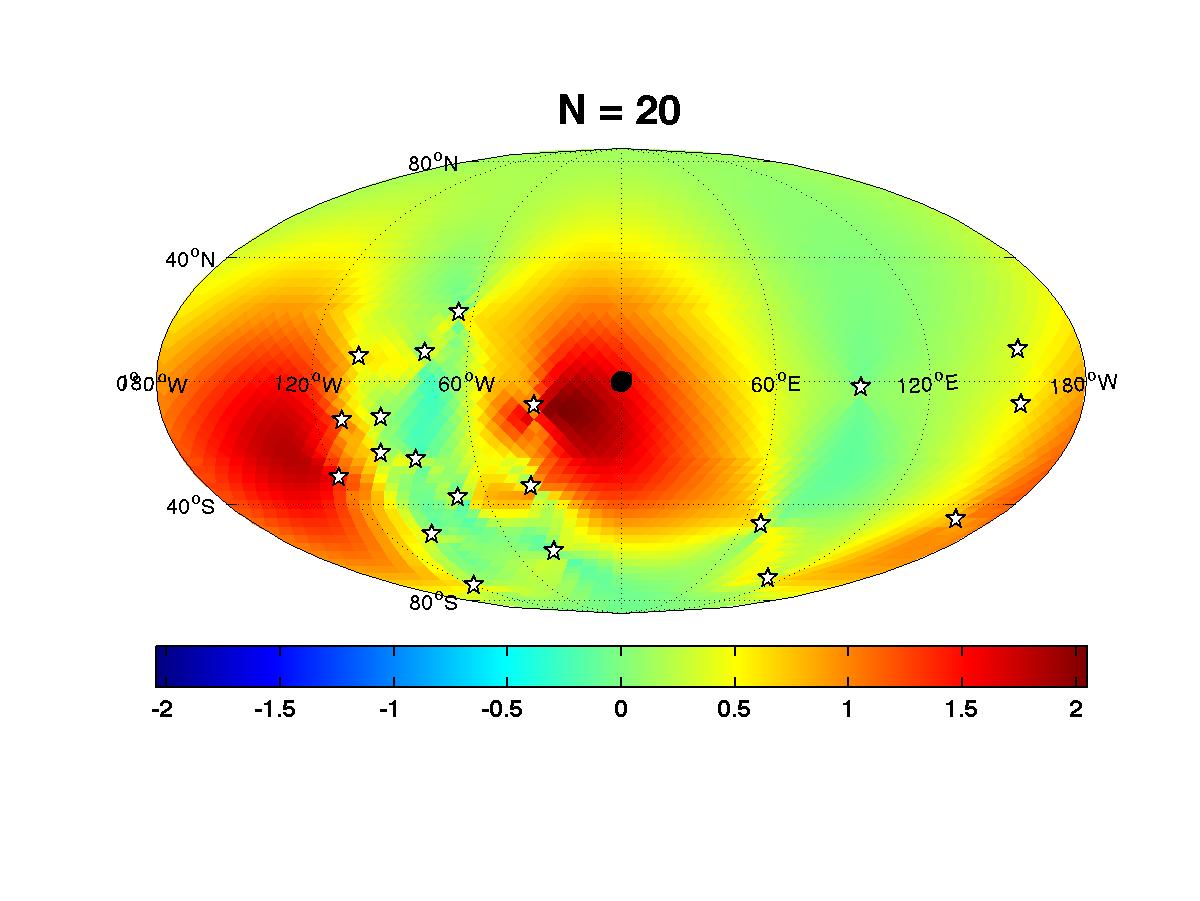}
\includegraphics[trim=3cm 4cm 3cm 2.5cm, clip=true, width=.49\textwidth]{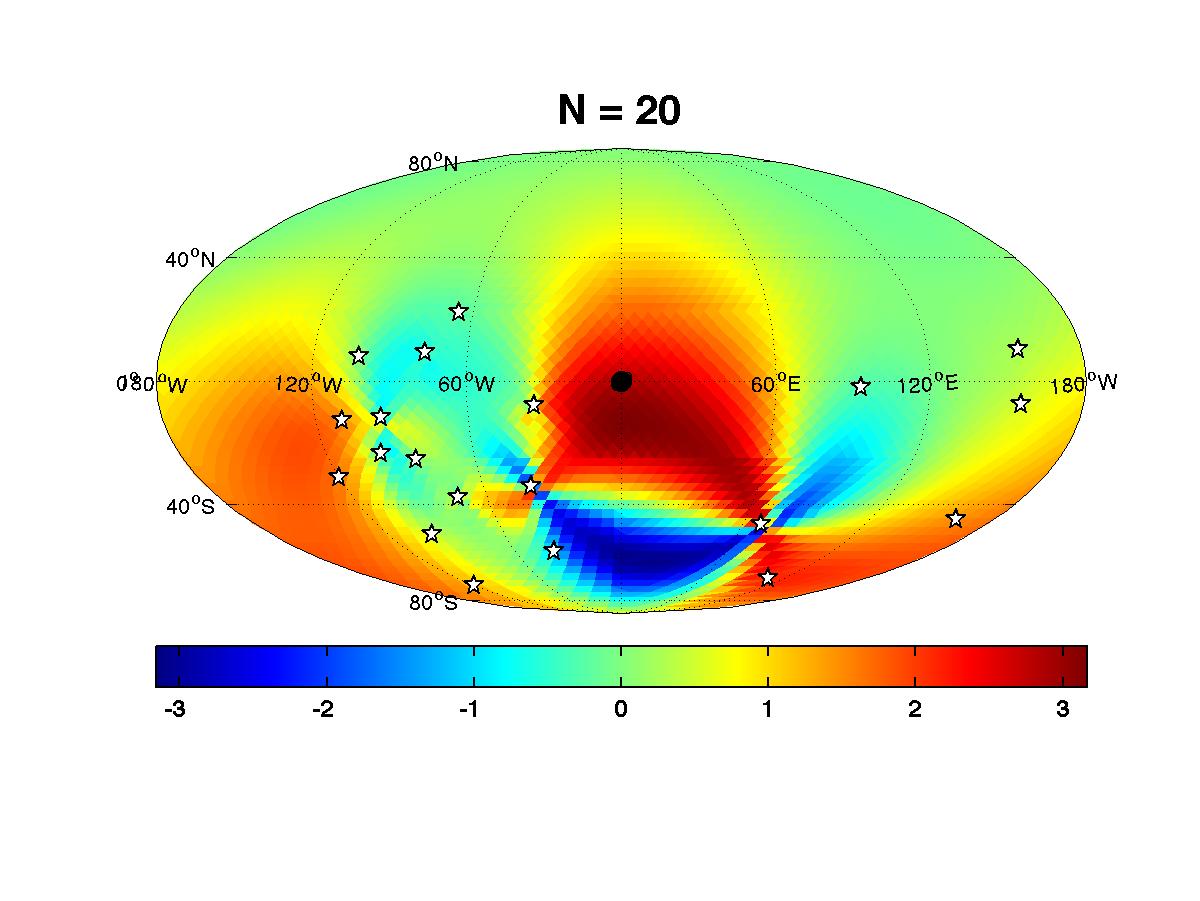}
\caption{Point spread functions for the array of $N=20$
pulsars listed in Table~\ref{t:pulsar_locations} for both 
{\em equal-noise} weighting (left panel) and 
{\em actual-noise} weighting (right panel), using the 
timing noise values in the second column of the Table.
The timing noise values were rescaled by an overall 
factor so that the maps for the two different weighting 
schemes could be meaningfully compared with one another.
The point source is located at the center of the maps,
indicated by a black dot.}
\label{f:PSFpulsar-actual}
\end{center}
\end{figure}

\begin{figure}[h!tbp]
\begin{center}
\includegraphics[trim=3cm 6.5cm 3cm 3.5cm, clip=true, width=.49\textwidth]{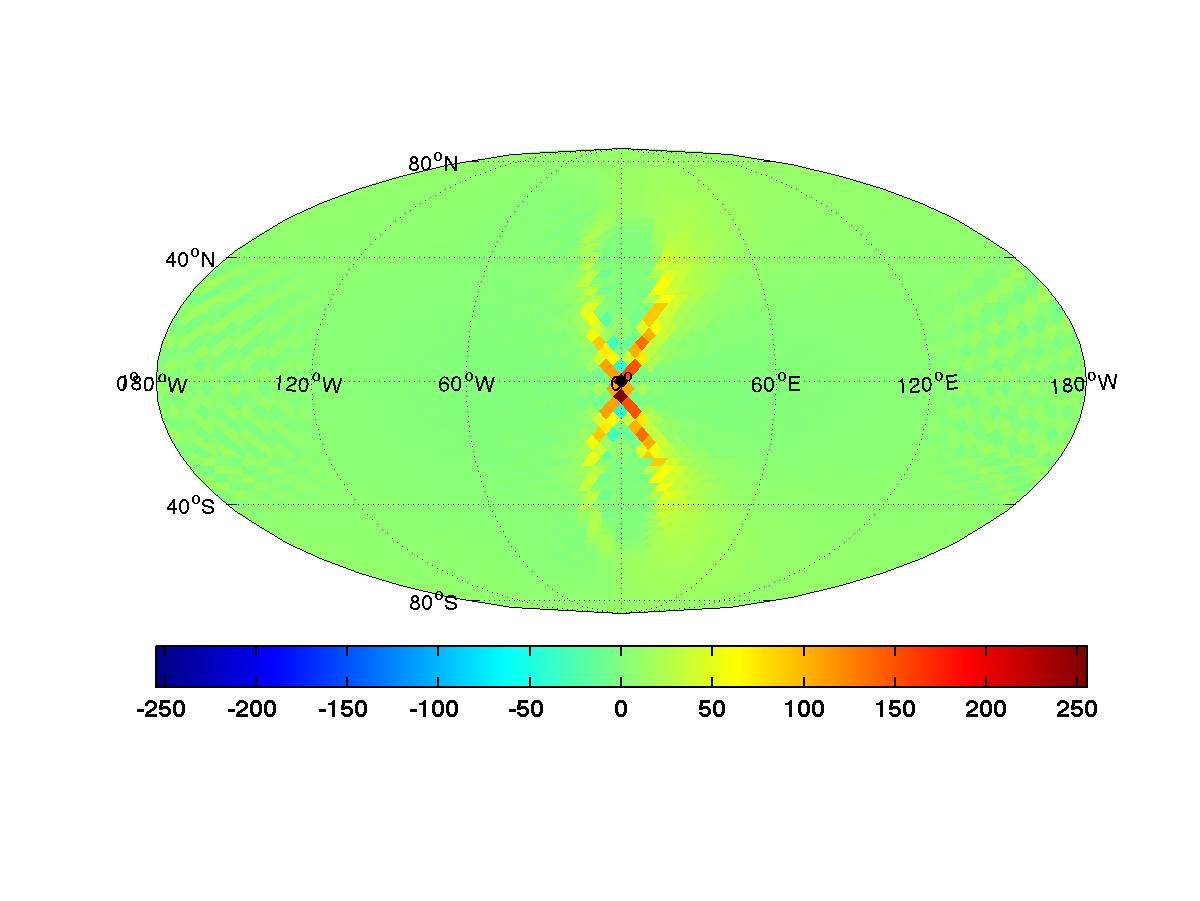}
\includegraphics[trim=3cm 6.5cm 3cm 3.5cm, clip=true, width=.49\textwidth]{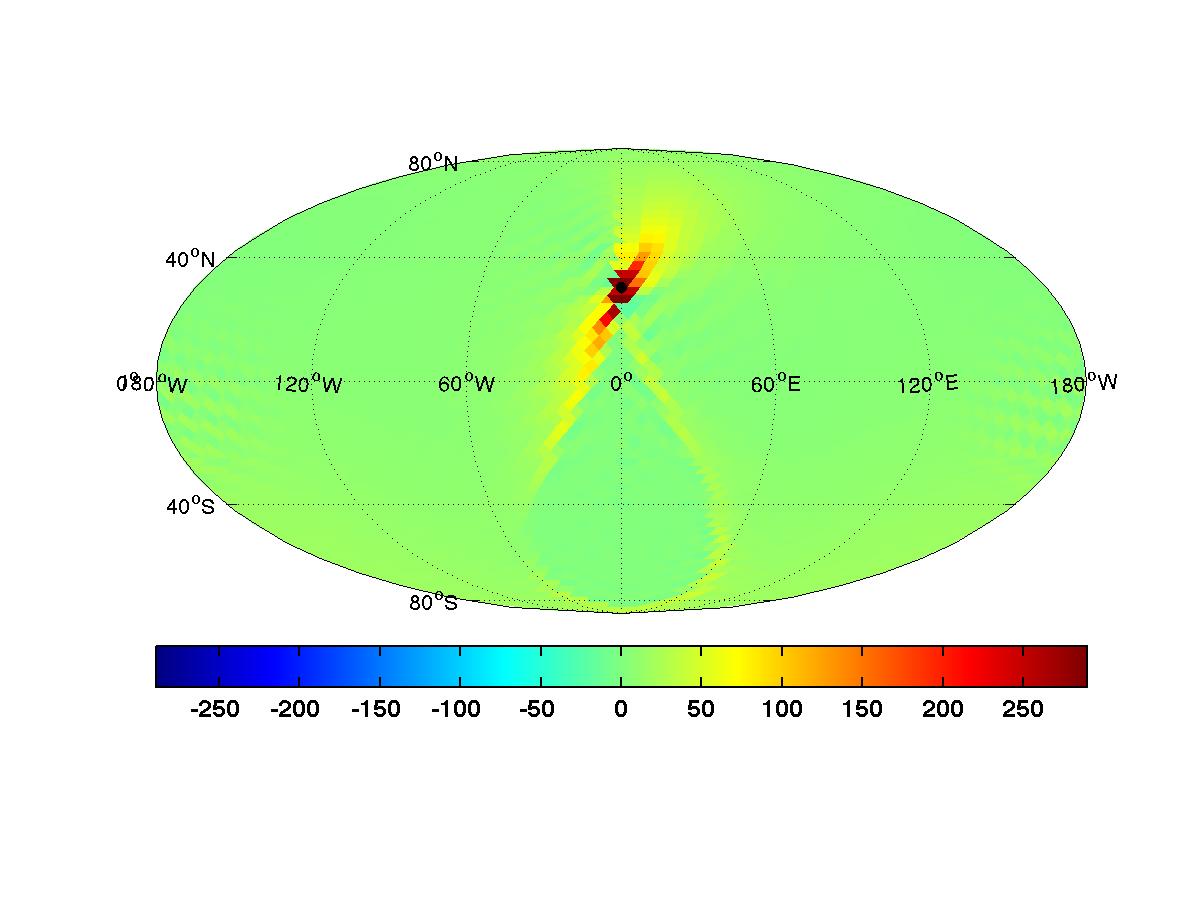}
\caption{Point spread functions for gravitational-wave power 
for the LIGO Hanford--LIGO Livingston detector pair.
Left panel: point source at the center of the map, $(\theta,\phi)=(90^\circ, 0^\circ)$.
Right panel: point source at $(\theta,\phi)=(60^\circ,0^\circ)$.}
\label{f:PSFIFO}
\end{center}
\end{figure}

As discussed in the previous section, the point spread
function for mapping gravitational-wave power is given 
by the components of the Fisher information matrix:
\be
{\rm PSF}_{\hat n_0}(\hat n) 
\equiv{\rm PSF}(\hat n, \hat n_0)
=F_{\hat n \hat n_0}\,.
\label{e:PSF}
\ee
Here $\hat n_0$ is the direction to the point source
and $\hat n$ is an arbitrary point on the sky.
In the following three figures 
(Figures~\ref{f:PSFpulsar-random}, \ref{f:PSFpulsar-actual}, \ref{f:PSFIFO})
we shows plots of 
point spread functions for both pulsar timing arrays
and the LIGO Hanford--LIGO Livingston detector pair.

\medskip
\noindent{\bf Example: Pulsar timing arrays}
\medskip

\noindent
Figure~\ref{f:PSFpulsar-random} shows plots of 
point spread functions for pulsar timing arrays consisting of 
$N=2$, 5, 10, 20, 25, 50 pulsars.
The point source is located at the center of the maps,
indicated by a black dot.
The pulsar locations (indicated by white stars) were
randomly-distributed on the sky, and we used equal-noise
weighting for calculating the point spread function.
One can see that the point spread function becomes tighter
as the number of pulsars in the array increases.
Figure~\ref{f:PSFpulsar-actual} are similar plots for an actual
array of $N=20$ pulsars given in Table~\ref{t:pulsar_locations}.
Note that the pulsar locations are concentrated in the 
direction of the galactic center, 
$({\rm ra},{\rm dec}) = (-6^{\rm h}15^{\rm m}, -29^\circ)$
in equatorial coordinates.
The point source is again located at the center of the maps, 
indicated by a black dot.
The left panel shows the point spread function calculated using
equal-noise weighting, while the right panel shows the point spread
function calculated using {\em actual-noise} weighting, 
based on the timing noise values given
in the second column of Table~\ref{t:pulsar_locations}.
Note that this latter plot is similar to the small-$N$ plots
in Figure~\ref{f:PSFpulsar-random}, being dominated by 
pulsars with low timing noise---in this particular case, 
J0437-4715 and J2124-3358, which have the lowest and third-lowest 
timing noise.
\begin{table}[t!]
\centering
\begin{tabular}{lclc}
\toprule
pulsar name & timing noise ($\mu$s) & pulsar name & timing noise ($\mu$s) \\
\midrule
J0437-4715  & 0.14 & J1730-2304  & 0.51 \\
J0613-0200  & 2.19 & J1732-5049  & 1.81 \\
J0711-6830  & 1.04 & J1744-1134  & 0.17 \\
J1022+1001  & 0.60 & J1824-2452  & 3.62 \\
J1024-0719  & 0.35 & J1909-3744  & 0.56 \\
J1045-4509  & 3.24 & J1939+2134  & 3.58 \\
J1600-3053  & 2.67 & J2124-3358  & 0.25 \\
J1603-7202  & 1.64 & J2129-5721  & 2.55 \\
J1643-1224  & 4.86 & J2145-0750  & 0.50 \\
J1713+0747  & 0.89 & B1855+0900  & 0.70 \\
\bottomrule
\end{tabular}
\caption{Actual pulsar locations and timing noise.
The pulsar name specifies its location:
the first four digits is right ascension (ra) 
in hours and minutes (hhmm);
the last four digits is declination (dec) 
in degrees and minutes (ddmm),
with the preceding $+$ or $-$ sign.
The rms timing noise is in microsec.}
\label{t:pulsar_locations}
\end{table}

\medskip
\noindent{\bf Example: Earth-based interferometers}
\medskip

\noindent
In Figure~\ref{f:PSFIFO} we plot point spread
functions for gravitational-wave power for the LIGO Hanford-LIGO 
Livingston pair of detectors.
The left-hand plot is for a point source located at the 
center of the map, $(\theta,\phi)=(90^\circ, 0^\circ)$, while
the right-hand plot is for a point source located at 
$(\theta,\phi)=(60^\circ, 0^\circ)$
(indicated by black dots).
We assumed equal white-noise power spectra for the two 
detectors, and we combined the contributions from 100 
discrete frequencies between 0 and 100~Hz, and 100 discrete
time chunks over the course of one sidereal day.
The point spread functions for the two different 
point source locations are shaped, respectively,
like a {\em figure-eight} with a bright region at 
the center of the figure-eight pattern, and a {\em tear drop} 
with a bright region near the top of the drop.
These results are in agreement with \cite{Mitra-et-al:2008} 
(see e.g., Figure~1 in that paper).
Provided one combines data over a full sidereal day, the 
point spread function is independent of the right ascension
(i.e., azimuthal) angle of the source.
Readers should see \cite{Mitra-et-al:2008} for more details, 
including a stationary phase approximation for calculating 
the point spread function.

\medskip
\noindent{\bf Angular resolution estimates}
\medskip

\noindent
There are ``rules of thumb" that can be used to estimate the 
angular resolution $\Delta\theta$ (or size of a 
point spread function) for an anisotropic stochastic background search.
For cross-correlations using ground-based interferometers 
like LIGO, Virgo, etc., the angular resolution of the detector
network can be estimated from the diffraction limit~\cite{Monnier:2003}:
\be
\Delta\theta \simeq \frac{\lambda}{2D} 
= \frac{c}{2fD}\,,
\label{e:deltaThetaIFO}
\ee
where $f$ is gravitational-wave frequency and $D$ is separation 
between a pair of detectors.
Thus, the larger the separation between detectors and the higher 
frequencies searched for, the better the angular resolution.
For a pulsar timing array consisting of $N$ pulsars, the corresponding
estimate is given by 
\be
\Delta\theta \simeq 180^\circ/l_{\rm max}
\simeq 180^\circ/\sqrt{N}\,,
\label{e:deltaThetaPTA}
\ee
where $l_{\rm max}$ is the maximum value of $l$ for a spherical
harmonic decomposition of the background having angular features 
of size $\Delta\theta$.
The last approximate equality follows from the fact that, at each frequency, 
one can extract at most $N$ (complex) pieces of information about 
the gravitational-wave background using an $N$-pulsar 
array~\cite{Boyle-Pen:2012, Cornish-vanHaasteren:2014, Gair-et-al:2014};
and those $N$ pieces of information correspond to the number of 
spherical harmonic components $(lm)$ out to $l_{\rm max}$, so 
$N\sim l_{\rm max}^2$.
(We will discuss this again in Section~\ref{s:basis_skies},
in the context of {\em basis skies} for a phase-coherent search for
anisotropic backgrounds.) 
Note that if we knew the distances to the pulsars in the array and used 
information from the pulsar-term contribution to the timing
residuals (\ref{e:deltaT2}), then $\Delta\theta$ for a 
pulsar timing array would have the same form as (\ref{e:deltaThetaIFO}), 
but with $D$ now representing the Earth-pulsar distance.
See~\cite{Boyle-Pen:2012} for details.

\subsubsection{Singular-value decomposition}
\label{s:svd_power}

Expression (\ref{e:Phat}) for the maximum-likelihood estimator 
$\hat{\cal P}$ involves the inverse of the Fisher matrix $F$.
But this is just a {\em formal} expression, as $F$ is typically 
a singular matrix, requiring some sort of regularization to invert.
Here we describe how {\em singular-value decomposition}~\cite{Press:1992}
can be used to `invert' $F$.
Since this a general procedure, we will frame our discussion in 
terms of an arbitrary matrix $S$.

Singular value decomposition factorizes an 
$n\times m$ matrix $S$ into the product of three matrices:
\be
S = U\Sigma V^\dagger\,,
\label{e:SVD}
\ee
where $U$ and $V$ are $n\times n$ and $m\times m$ unitary 
matrices, and $\Sigma$ is an $n\times m$ rectangular matrix 
with (real, non-negative) singular values $\sigma_k$ along
its diagonal, and with zeros everywhere else.
We will assume, without loss of generality, that the 
singular values are arranged from largest to smallest along
the diagonal.
We define the {\em pseudo-inverse} $S^+$ of $S$ as
\be
S^+ \equiv V\Sigma^+ U^\dagger\,,
\label{e:S+}
\ee
where $\Sigma^+$ is obtained by taking the reciprocal of 
each nonzero singular value of $\Sigma$, leaving all the
zeros in place, and then transposing the resulting matrix.
Note that when $S$ is a square matrix with non-zero 
determinant, then the pseudo-inverse $S^+$ is identical 
to the ordinary matrix inverse $S^{-1}$.
Thus, the pseudo-inverse of a matrix generalizes the 
notion of ordinary inverse to non-square or singular
matrices.

As a practical matter, it is important to note that 
if the nonzero singular values of $\Sigma$ vary over 
several orders of magnitude, it is usually necessary to 
first set to zero (by hand) all nonzero singular values $\le$ 
some minimum threshold value $\sigma_{\rm min}$ 
(e.g., $10^{-5}$ times that of the largest singular value).
Alternatively, we can set those very small singular 
values equal to the threshold value $\sigma_{\rm min}$.
This procedure helps to reduce the noise in the 
maximum-likelihood estimates, which is dominated by the 
modes to which we are least sensitive.

Returning to the gravitational-wave case, the above 
discussion means that all of the previous expressions 
for the inverse of the Fisher matrix, $F^{-1}$, should 
actually be written in terms of the pseudo-inverse $F^+$.
Thus,
\be
\hat{\cal P} = F^+ X\,,
\ee
which then implies
\be
\begin{aligned}
&\langle \hat{\cal P}\rangle = F^+ F\,{\cal P}\,,
\\
&\langle \hat{\cal P}\hat{\cal P}^\dagger\rangle - 
\langle \hat{\cal P}\rangle \langle \hat{\cal P}^\dagger\rangle
\approx F^+\,.
\label{e:bias_P}
\end{aligned}
\ee
So $\hat{\cal P}$ is actually a {\em biased} estimator 
of ${\cal P}$ if $F^+\ne F^{-1}$, as was discussed in 
\cite{Thrane-et-al:2009}.
 
\begin{figure}[h!tbp]
\begin{center}
\includegraphics[width=.6\textwidth]{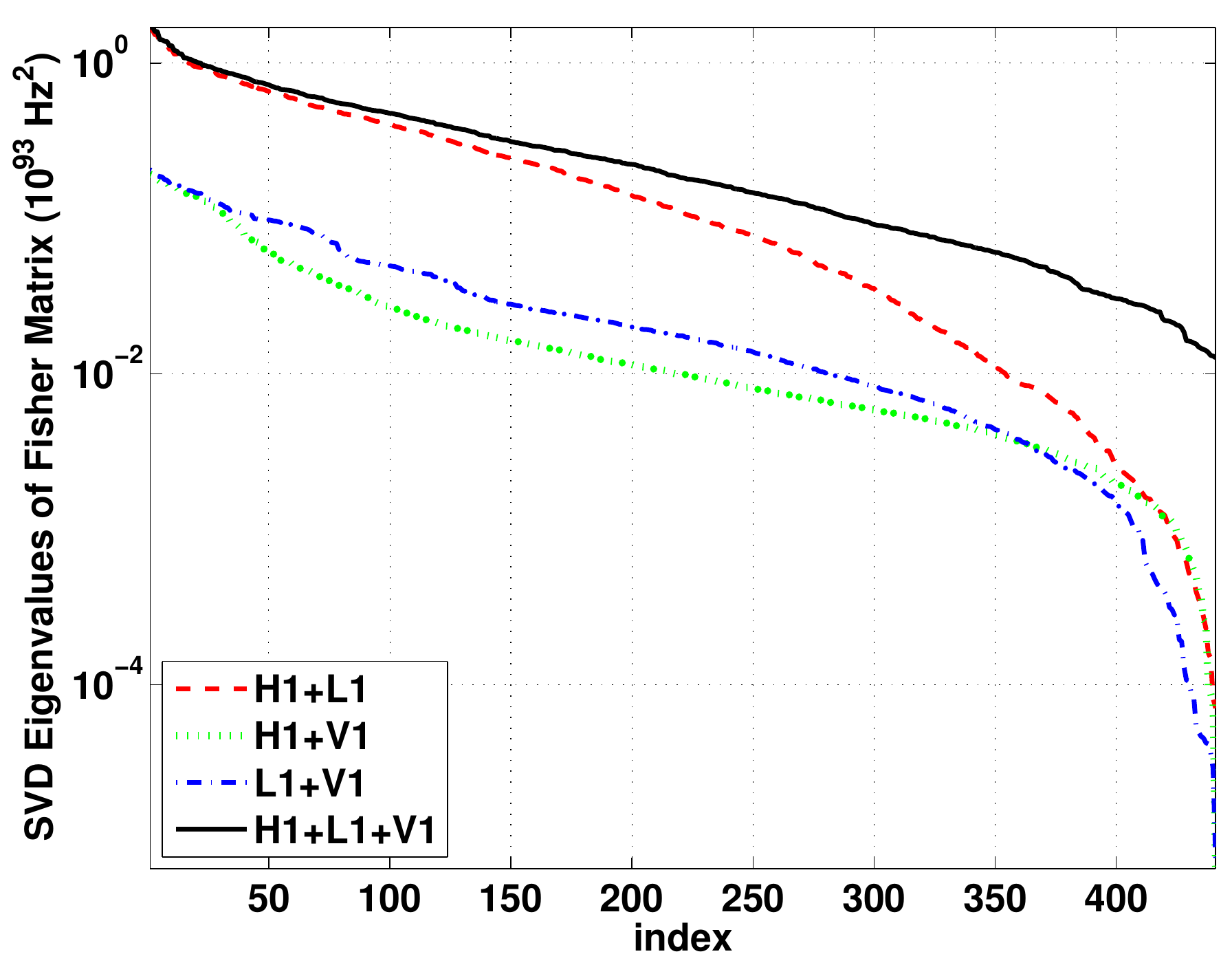}
\caption{Singular values of typical Fisher matrices
$F$ for different ground-based interferometer
detector pairs and a multibaseline detector network.
For this analysis there were $441$ total modes.
For each individual detector pair, some of the
singular values are (almost) null.
The multibaseline network has fewer null modes,
thus acting as a natural regularizer.
Image reproduced with permission from \cite{Thrane-et-al:2009},
copyright by APS.}
\label{f:singular_values}
\end{center}
\end{figure}

Figure~\ref{f:singular_values} is a plot of the singular 
values of typical Fisher matrices for 
different ground-based interferometer detector pairs
(Hanford--Livingston, Hanford--Virgo, Livingston--Virgo)
and a multibaseline detector network (Hanford--Livingston--Virgo).
For these examples, we chose to expand the gravitational-wave
power on the sky ${\cal P}(\hat n)$ and the integrand of 
the overlap functions $\gamma_{IJ}(t;f,\hat n)$ in terms 
of spherical harmonics out to $l_{\rm max}=20$.
(See Section~\ref{s:radiometer-SHD} for more details 
about the spherical harmonic decomposition method.)
This yields $(l_{\rm max}+1)^2 = 441$ modes of 
gravitational-wave sky that we would like to recover.
Note how the inclusion of more detectors to the network
reduces the dynamic range of the singular values of $F$, 
hence making the matrix less singular without any external 
form of regularization.
%

\subsubsection{Radiometer and spherical harmonic 
decomposition methods}
\label{s:radiometer-SHD}

The gravitational-wave radiometer~\cite{Ballmer-PhD:2006, Ballmer:2006,
Mitra-et-al:2008} and spherical harmonic decomposition 
methods~\cite{ Thrane-et-al:2009, Abadie-et-al:S5-anisotropic}
are two different ways of implementing the maximum-likelihood approach 
for mapping gravitational-wave power ${\cal P}(\hat n)$.
They differ primarily in their choice of signal model, and 
their approach for deconvolving the detector response from the
underlying (true) distribution of power on the sky.

\medskip
\noindent{\bf Gravitational-wave radiometer}
\medskip

\noindent
The radiometer method takes as its signal model a point source
characterized by a direction $\hat n_0$ and amplitude 
${\cal P}_{\hat n_0}$:
\be
{\cal P}(\hat n) = {\cal P}_{\hat n_0}\,\delta^2(\hat n,\hat n_0)\,.
\ee
It is applicable to an anisotropic gravitational-wave 
background dominated by a limited number of 
widely-separated point sources.
As the number of point sources increases or if two point
sources are sufficiently close to one another, the 
point spread function for the detector network will 
cause the separate signals to interfere with one another.
Thus, the radiometer method is not appropriate for 
diffuse backgrounds.
Moreover, by assuming that the signal is point-like, 
the radiometer method ignores correlations between 
neigboring pixels on the sky, 
effectively side-stepping the deconvolution problem.
Explicitly, the inverse of the Fisher matrix that appears 
in the maximum-likelihood estimator
$\hat{\cal P}=F^{-1} X$ is replaced by the 
inverse of the {\em diagonal element} 
$F_{\hat n\hat n}$ to obtain an estimate of the 
point-source amplitude at $\hat n$:
\be
\hat{\cal P}_{\hat n} = 
(F_{\hat n\hat n})^{-1} X_{\hat n}\,,
\label{e:eta_k}
\ee
where $X$ is the dirty map (\ref{e:F,X}).
Thus, the radiometer method estimates the strength of 
point sources at different points on the sky, 
{\em ignoring} any correlations between neighboring 
pixels.

Note that for a single pair of detectors $IJ$ the above 
estimator (\ref{e:eta_k}) is equivalent to an 
appropriately normalized cross-correlation statistic:
\be
\hat C_{IJ}(t; \hat n) 
\equiv \int_{-\infty}^\infty df\>
Q_{IJ}(t; f,\hat n) \tilde d_I(t;f) \tilde d_J^*(t;f)\,,
\label{e:radiometer-CCstat}
\ee
with filter function
\begin{equation}
Q_{IJ}(t;f,\hat n) \propto
\frac{\gamma_{IJ}(t;f,\hat n)\, \bar H(f)}
{P_{n_I}(t;f) P_{n_J}(t;f)}\,,
\label{e:radiometer-Q}
\end{equation}
where $\gamma_{IJ}$ is given by (\ref{e:gamma_khat}).
For a network of detectors, one recovers the
estimator $\hat{\cal P}_{\hat n}$ by summing the 
individual-baseline 
statistics (\ref{e:radiometer-CCstat}) 
over both time and distinct detector 
pairs, weighted by the inverse variances of the 
individual-baseline statistics.
See e.g., 
\cite{Ballmer-PhD:2006, Ballmer:2006, Mitra-et-al:2008} 
for more details.
 
\medskip
\noindent{\bf Spherical harmonic decomposition}
\medskip

\noindent
The spherical harmonic decomposition method is appropriate 
for extended anisotropic distributions on the sky, 
assuming a signal model for gravitational-wave power that 
includes spherical harmonic components up to some specified 
value of $l_{\rm max}$:
\be
{\cal P}(\hat n) = \sum_{l=0}^{l_{\rm max}}\sum_{m=-l}^l
{\cal P}_{lm} Y_{lm}(\hat n)\,.
\ee
The cutoff in the expansion at $l_{\rm max}$ 
corresponds to an angular scale 
$\Delta\theta \simeq 180^\circ/l_{\rm max}$.
The diffraction limit~\cite{Monnier:2003}:
\be
\Delta\theta \simeq \frac{\lambda}{2D} = \frac{c}{2f D}\,,
\label{e:diffractionlimit}
\ee
where $f$ is the maximum gravitational-wave frequency 
and $D$ is the separation between a pair of detectors,
sets an upper limit on the size of $l_{\rm max}$, 
since the detector network is not able to resolve 
features having smaller angular scales.
For example, for the LIGO Hanford--LIGO Livingston 
detector pair ($D= 3000\ {\rm km}$) and a stochastic
background having contributions out to 
$f\sim 500~{\rm Hz}$, we find $l_{\rm max} \lesssim 30$.
Alternatively, one can use Bayesian model selection
to determine the value of $l_{\rm max}$ that is most 
consistent with the data.

Since the spherical harmonic method targets extended
distributions of gravitational-wave power on the sky,
correlations between neighboring pixels 
or, equivalently, between different spherical harmonic 
components must be taken into account.
This is addressed by using singular-value decomposition 
as described in Section~\ref{s:svd_power} to `invert' 
the Fisher matrix.
By effectively ignoring those modes to which the detector 
network is insensitive, we can construct the 
pseudo-inverse $F^+$ to perform the deconvolution.
In terms of $F^+$, we have
\be
\hat{\cal P}_{lm} 
= \sum_{l'=0}^{l_{\rm max}}
\sum_{m'=-l'}^{l'} F^+_{lm,l'm'} X_{l'm'}
\ee
for the spherical harmonic components of the 
maximum-likelihood estimators $\hat{\cal P}$.
The sky map constructed from the $\hat{\cal P}_{lm}$ 
is called a `clean' map, since the inversion
removes the detector response from the `dirty'
map $X$.

\begin{figure}[h!tbp]
\begin{center}
\includegraphics[width=.49\textwidth]{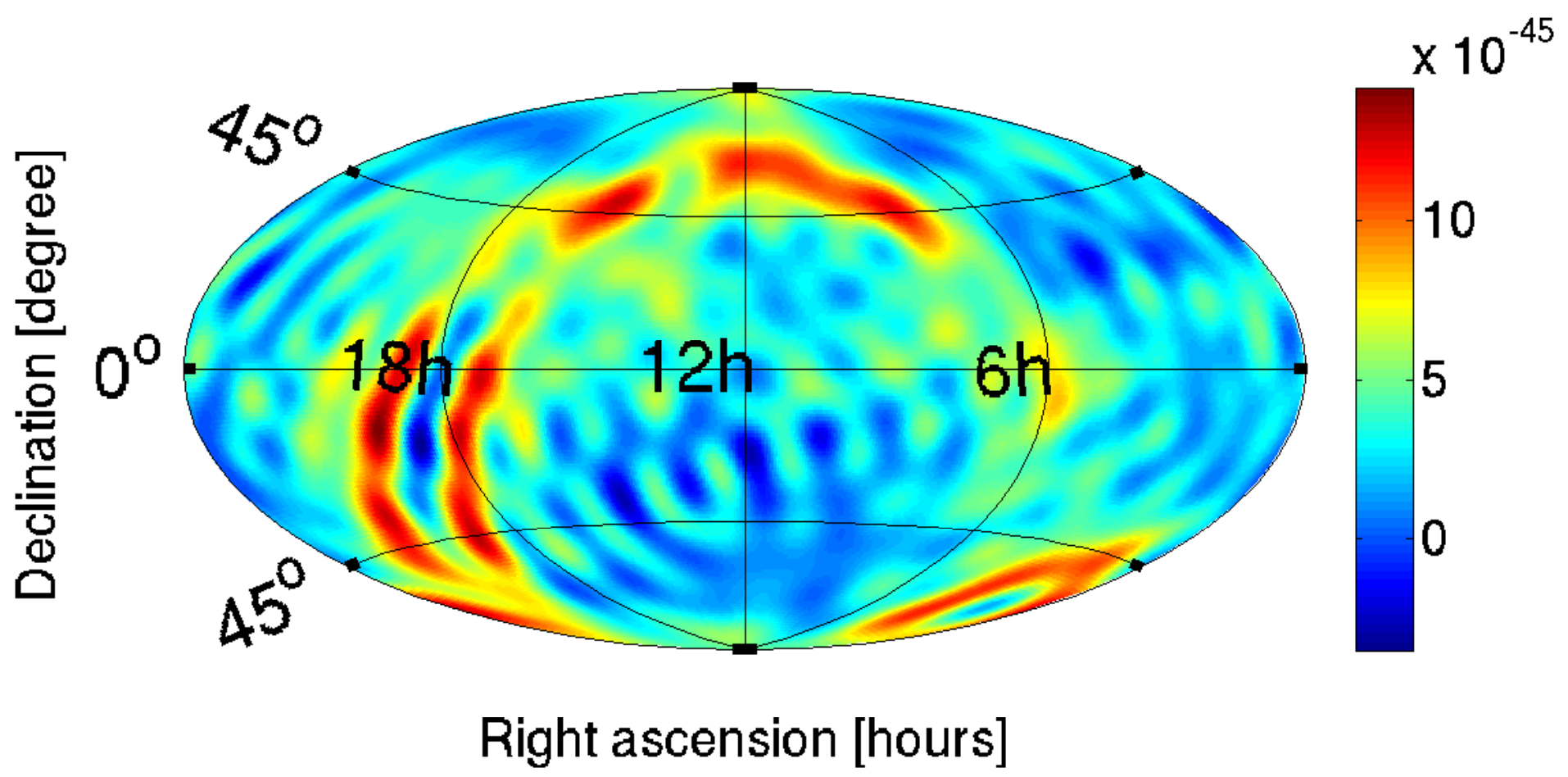}
\includegraphics[width=.49\textwidth]{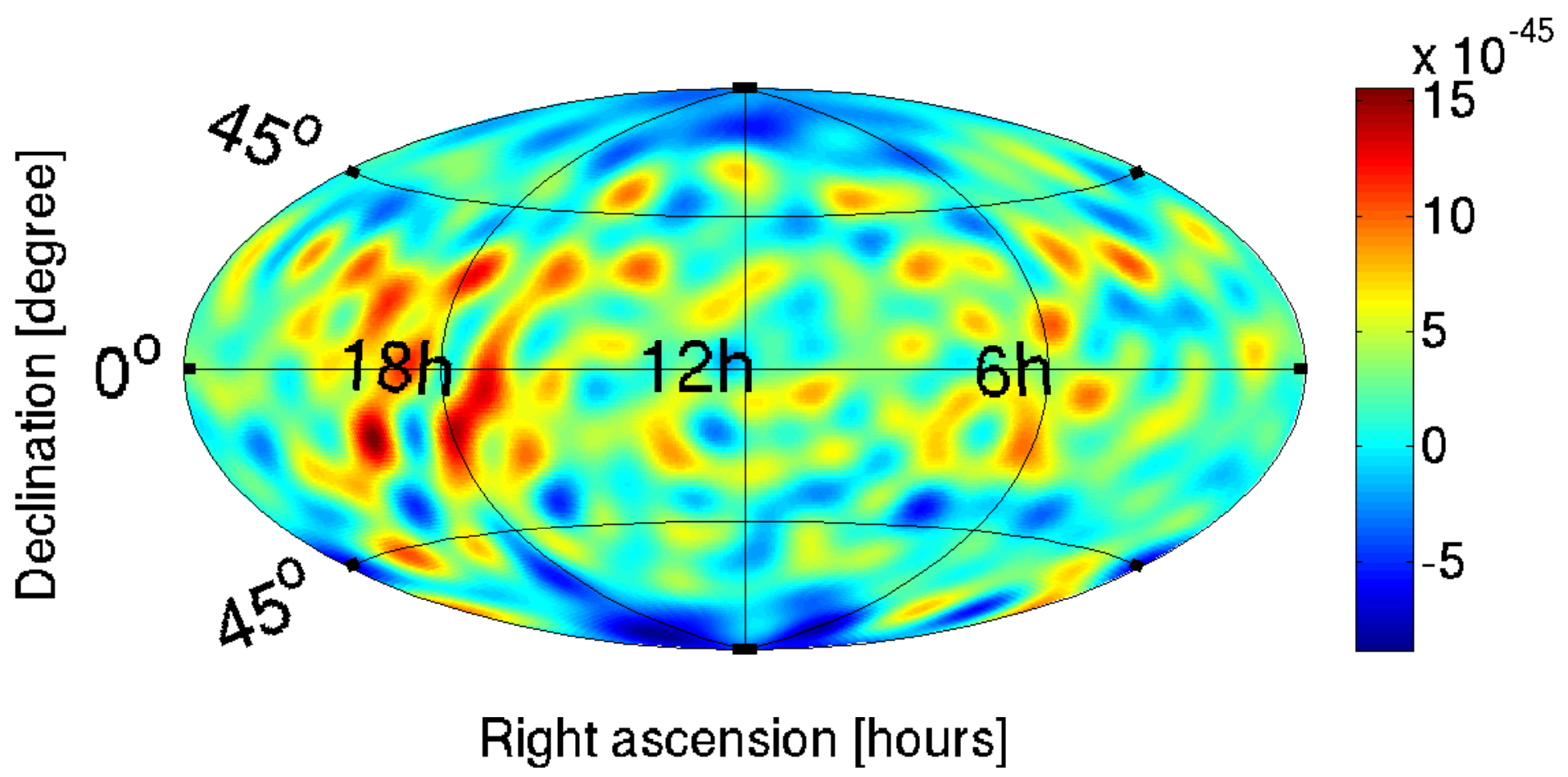}
\includegraphics[width=.49\textwidth]{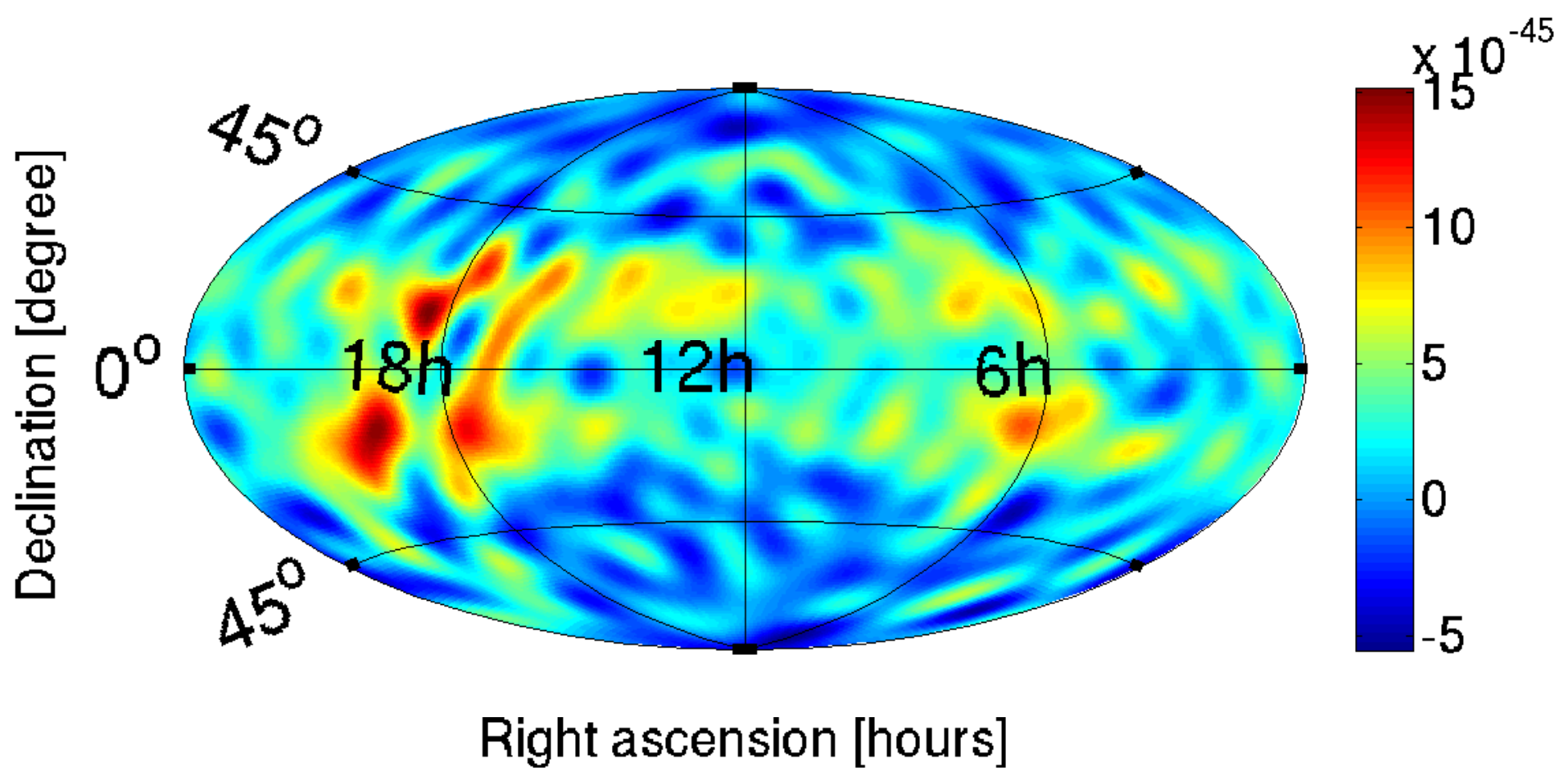}
\includegraphics[width=.49\textwidth]{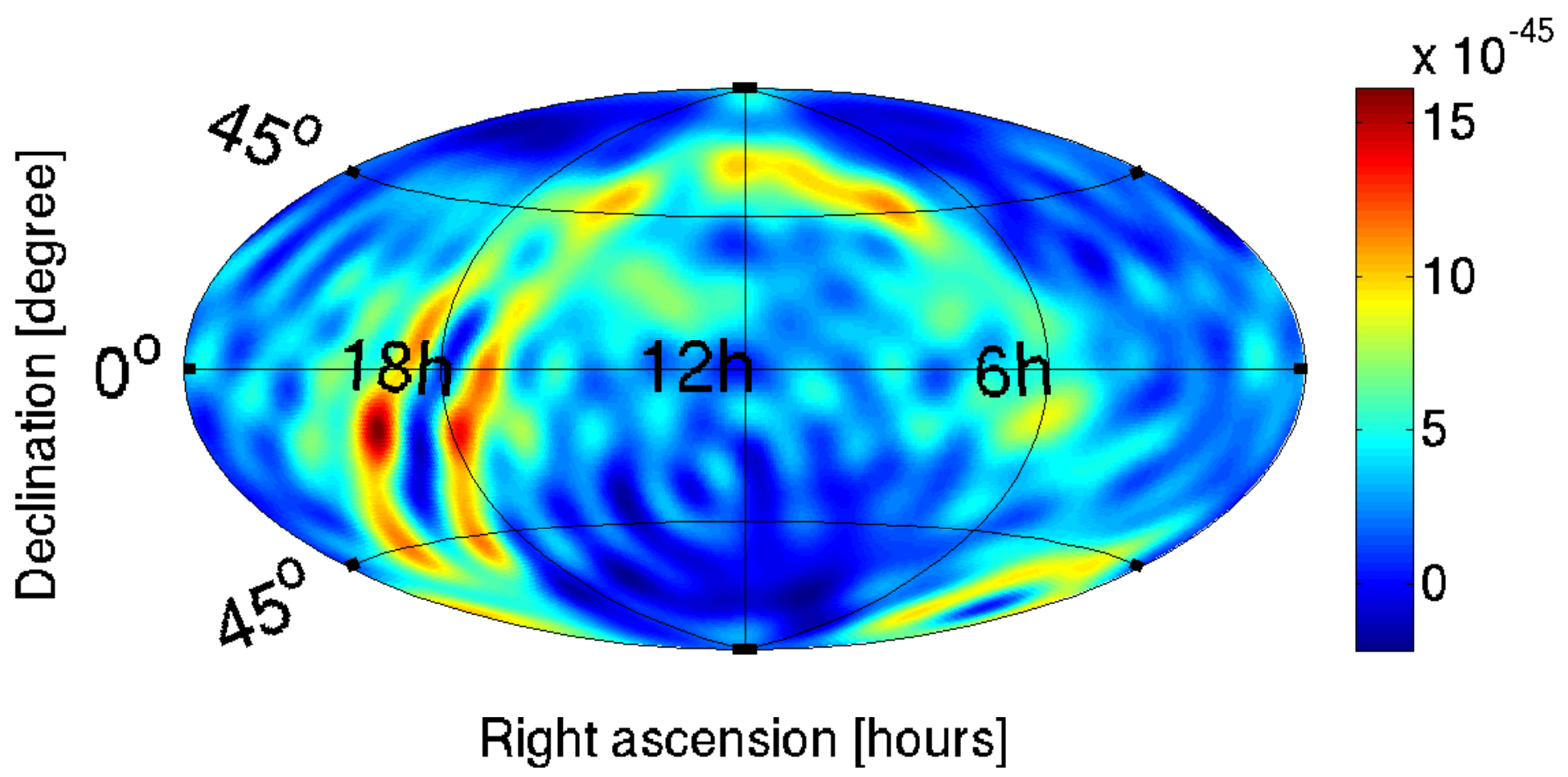}
\includegraphics[width=.49\textwidth]{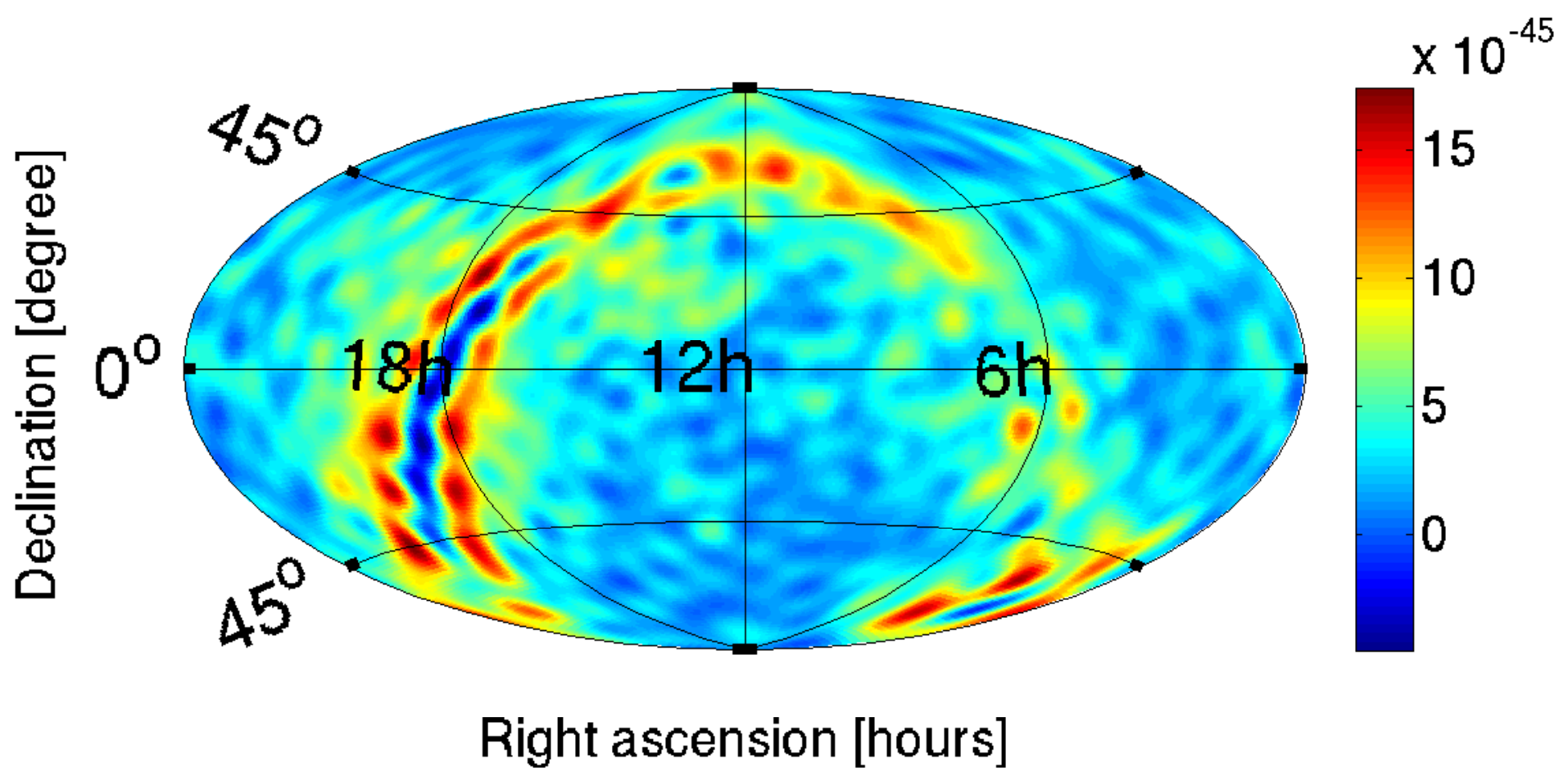}
\caption{Results of spherical harmonic 
  decomposition analyses performed using different 
  detector pairs and a multibaseline detector network.
  The simulated anisotropic power distribution is 
  shown in the bottom plot.
  Top row: Clean maps for the Hanford--Livingston and
  Hanford--Virgo detector pairs.
  Second row: Same as the top row, but for the 
  Livingston--Virgo detector pair and for the
  Hanford--Livingston--Virgo multibaseline detector network.
  For all maps $l_{\rm max}=20$.  
  Image reproduced with permission from \cite{Thrane-et-al:2009},
  copyright by APS.}
\label{f:multi}
\end{center}
\end{figure}

\begin{figure}[h!tbp]
\begin{center}
\includegraphics[width=.6\textwidth]{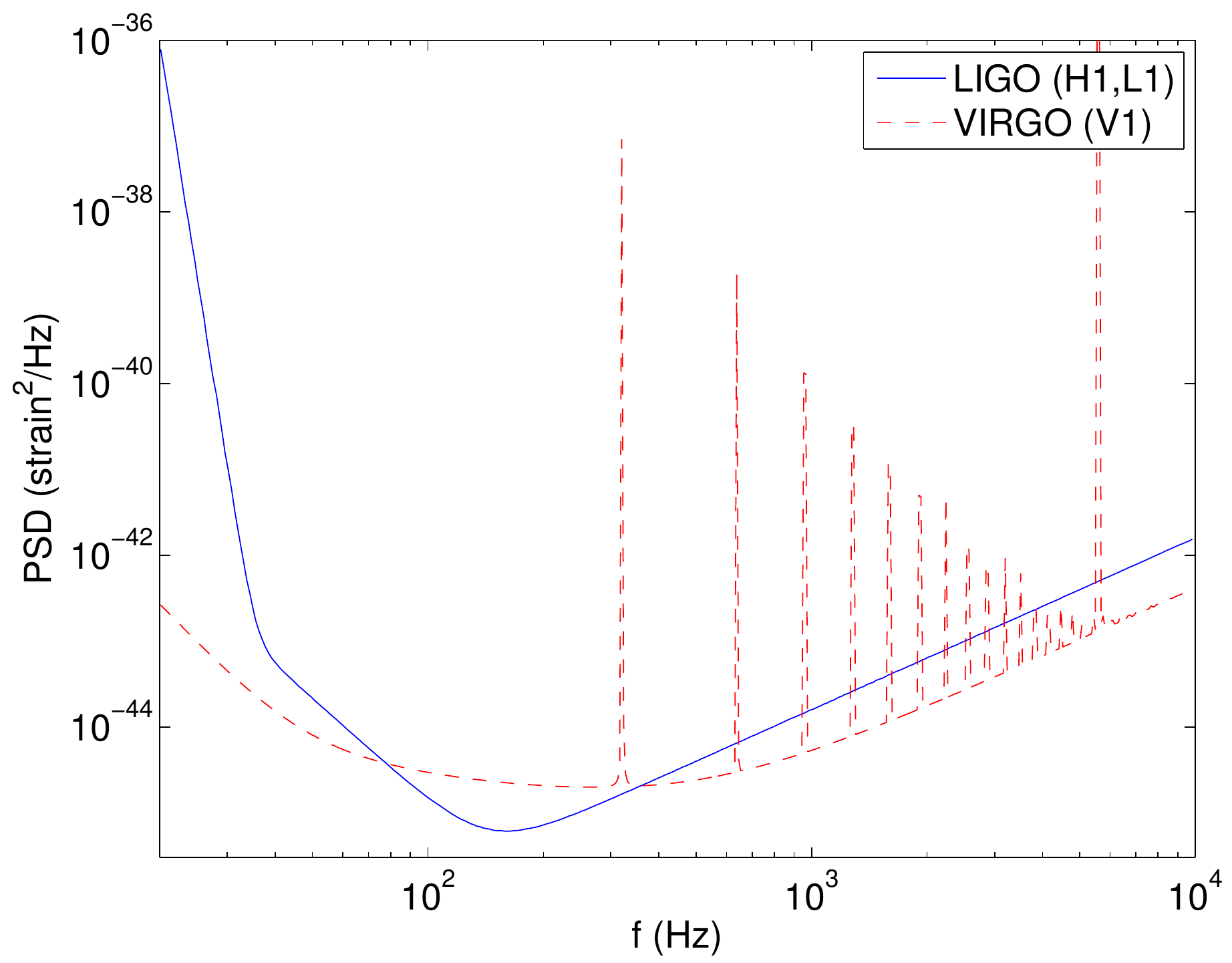}
\caption{The power spectral densities used for
the simulated detector noise for the injections
described in Figure~\ref{f:multi}.
Image reproduced with permission from \cite{Thrane-et-al:2009},
copyright by APS.}
\label{f:compPSDs}
\end{center}
\end{figure}

Figure~\ref{f:multi} shows clean maps produced by 
the spherical harmonic 
decomposition method for a simulated anisotropic 
background distributed along the galactic plane~\cite{Thrane-et-al:2009}.
The injected map is the bottom plot in the figure.
(All sky maps are in equatorial coordinates.)
The four maps shown in the top two rows of the 
figure correspond to analyses with different 
interferometer detector pairs 
(Hanford--Livingston, Hanford--Virgo, and Livingston--Virgo) 
and a multibaseline detector network
(Hanford--Livingston--Virgo).
Consistent with our findings in 
Figure~\ref{f:singular_values}, we see that 
the recovered map is best for the multibaseline 
network, whose Fisher matrix has singular values 
with the smallest dynamic range.
For the reconstructed maps, $F^+$ was calculated
by keeping 2/3 of all the eigenmodes (those
with the largest singular values), setting the 
remaing singular values equal to the minimum 
value $\sigma_{\rm min}$ of the modes that were kept.
For all cases, $l_{\rm max}=20$.
The anisotropic background was injected into 
simulated LIGO and Virgo detector noise (initial 
design sensitivity) whose power spectra are shown
in Figure~\ref{f:compPSDs}.
The overall amplitude of the signal was chosen to
be large enough that it was easily detectable in
1~sidereal day's worth of simulated data.
For additional details see \cite{Thrane-et-al:2009}.

\subsection{Frequentist detection statistics}
\label{s:MLR_statistic}

As discussed in Sections~\ref{s:relating-freq-bayes} and 
\ref{s:ML_statistic_derivation}, one can construct
a frequentist detection statistic $\Lambda_{\rm ML}(d)$
by taking the ratio of 
the maxima of the likelihood functions for the signal-plus-noise
model to the noise-only model.
The logarithm,
\be
\Lambda(d) \equiv 2\, \ln[\Lambda_{\rm ML}(d)]\,,
\ee
is the squared signal-to-noise ratio of the data.
If we calculate this quantity for an anisotropic background
${\cal P}(\hat n)$ using (\ref{e:likelihood_P(k)}) for the 
signal-plus-noise model, we find
\be
\Lambda(d) = \hat{\cal P}^\dagger F \hat{\cal P}\,,
\label{e:MLR_stat}
\ee
where $\hat{\cal P}$ are the maximum-likelihood estimators
of ${\cal P}$.
As described in Section~\ref{s:freq-hypothesis-testing},
one can use this statistic to do frequentist hypothesis
testing, comparing its observed value $\Lambda_{\rm obs}$
to a threshold $\Lambda_*$ to decide whether or not to claim 
detection of a signal.

The above detection statistic can be written in several
alternative forms:
\be
\Lambda(d) 
= \hat{\cal P}^\dagger F \hat{\cal P}
= X^\dagger F^{-1} X
= \frac{1}{2}\left(\hat{\cal P}^\dagger X + X^\dagger \hat{\cal P}\right)\,,
\ee
where $X$ is the `dirty' map, which is 
related to $\hat{\cal P}$ via 
$\hat{\cal P} = F^{-1} X$.
The last form suggests a standard matched-filter 
statistic:
\be
\lambda(d)
\equiv \frac{1}{2}\left(\bar{\cal P}^\dagger_{\rm model} X
+ X^\dagger \bar{\cal P}_{\rm model} \right)\,,
\ee
where $\bar{\cal P}_{\rm model}$ is an {\em assumed}
distribution of gravitational-wave power on the sky,
normalized such that
\be
\bar{\cal P}^\dagger_{\rm model} F
\bar{\cal P}_{\rm model}=1\,.
\ee
The above normalization is chosen so that if the 
true gravitational-wave background has the same 
spectral shape $\bar H(f)$ and the same angular distribution 
$\bar{\cal P}_{\rm model}$, then 
$\lambda(d)$ is an estimator of the overall
amplitude of the background.
In the absence of a signal, $\lambda(d)$
has zero mean and unit variance.

Such a matched-filter statistic was proposed in 
Appendix~C of \cite{Thrane-et-al:2009} and 
studied in detail in \cite{Talukder-et-al:2011}.
One nice property of this statistic is that it does
not require inverting the Fisher matrix.
Hence it avoids the inherent bias (\ref{e:bias_P})
and introduction of 
other uncertainties associated with the 
deconvolution process.
Indeed, if we are {\em given} a model of the expected 
anisotropy, $\lambda(d)$ 
is the {\em optimal} statistic for detecting its 
presence.
Thus, $\lambda(d)$ is especially good at 
detecting weak anisotropic signals.
See \cite{Talukder-et-al:2011} for more details.

\subsection{Phase-coherent mapping}
\label{s:phase-coherent}

Phase-coherent mapping is an approach that constructs
estimates of both the amplitude and phase of the 
gravitational-wave background at each point of the 
sky~\cite{Cornish-vanHaasteren:2014, Gair-et-al:2014,
Romano-et-al:2015}.
In some sense, it can be thought of as the 
``square root'' of the approaches described in the
previous subsections, which attempt to measure the 
distribution of gravitational-wave {\em power}
${\cal P}(\hat n) = |h_+|^2 + |h_\times^2|$.
The gravitational-wave signal can be characterized in 
terms of either the standard polarization basis 
components $\{h_+(f,\hat n), h_\times(f,\hat n)\}$ 
or the tensor spherical harmonic components 
$\{a^G_{(lm)}(f), a^C_{(lm)}(f)\}$.
In what follows we will restrict our attention the 
polarization basis components, although a similar
analysis can be carried out in terms of the spherical 
harmonic components~\cite{Gair-et-al:2014}.

\subsubsection{Maximum-likelihood estimators and Fisher matrix}
\label{s:likelihood_phase}

Unlike the previous approaches, which target
gravitational-wave power and hence use
cross-correlations (\ref{e:CIJ}) as their 
fundamental data product,
phase-coherent mapping works directly with the 
data from the individual detectors.
In terms of the short-term Fourier transforms defined
in Section~\ref{s:SFT}, we can write
\be
\tilde d_{I}(t;f)
=\int d^2\Omega_{\hat n}
\sum_A R^A_{I}(t; f,\hat n) h_A(f,\hat n)
+ \tilde n_{I}(t;f)\,,
\ee
where $I$ labels the different detectors, and 
$\tilde n_I(t;f)$ denotes the corresponding detector
noise.
Given our assumption (\ref{e:factorize}) that the spectral 
and angular dependence of the background 
factorize with known spectral function $\bar H(f)$,
we can rewrite the above equation as
\be
\tilde d_{I}(t; f)
=\int d^2\Omega_{\hat n}\>
\bar H^{1/2}(f) 
\sum_A R^A_{I}(t; f, \hat n)
h_A(\hat n)
+ \tilde n_{I}(t;f)\,,
\ee
so that the only unknowns are 
$\{h_+(\hat n), h_\times(\hat n)\}$
at different locations on the sky.
We will write this equation abstractly as
a matrix equation
\be
d = M a + n\,,
\ee
where 
\be
M \equiv 
\{\bar H^{1/2}(f) R_I^A(t;f,\hat n)\}\,,
\qquad
a \equiv \{h_A(\hat n)\}\,.
\label{e:M,a-defns}
\ee
The matrix multiplication corresponds to a sum over 
polarizations $A$ and directions $\hat n$ on the sky.
 
Assuming that the noise is uncorrelated across detectors, 
the noise covariance matrix is given by:
\be
\begin{aligned}
N_{Itf,I't'f'}
&\equiv \langle \tilde n_{I}(t;f)\tilde n^*_{I'}(t';f')\rangle
-\langle \tilde n_{I}(t;f)\rangle\langle \tilde n^*_{I'}(t';f')\rangle
\\
&=\frac{\tau}{2}\delta_{II'}\delta_{tt'}\delta_{ff'} P_{n_{I}}(t;f)\,,
\end{aligned}
\ee
where $P_{n_{I}}(t;f)$ is the one-sided power spectral density
of the noise in detector $I$ at time $t$.
Thus, we can write down a likelihood function for the data 
$d\equiv \{\tilde d_I(t;f)\}$ given $a$:
\be
p(d|a) \propto
\exp\left[ -\frac{1}{2}(d - Ma)^\dagger N^{-1} 
(d- Ma)\right]\,
\label{e:likelihood_a}
\ee
where the multiplications inside the exponential are
matrix multiplications, involving summations over detectors $I$,
times $t$, and frequencies $f$, or summations over 
polarizations $A$ and directions $\hat n$ on the sky. 
Note that (\ref{e:likelihood_a}) has exactly the same form 
as (\ref{e:likelihood_P(k)}), so the same general remarks 
made in Section~\ref{s:likelihood-MLestimators} apply here 
as well.
Namely, the maximum-likelihood estimators of $a$ are
\be
\hat a
=F^{-1} X\,,
\label{e:ahat}
\ee
where
\be
F\equiv M^\dagger N^{-1} M\,,
\qquad
X\equiv M^\dagger N^{-1} d\,,
\ee
are the Fisher matrices and `dirty' maps for this analysis.
(The definitions of $M$, $N$ here are different, of course, 
from those in Section~\ref{s:likelihood-MLestimators}.)
Explict expression for $X$ and $F$ are given below:
\be
X\equiv X_{A\hat n}
=\frac{2}{\tau}\sum_{I}\sum_t\sum_f
R^{A*}_{I}(t;f,\hat n) 
\frac{\bar H^{1/2}(f)}{P_{n_{I}}(f)}
\tilde d_{I}(t;f)\,,
\label{e:XAk}
\ee
and
\be
F\equiv F_{A\hat n,A'\hat n'}
=\frac{2}{\tau}\sum_{I}\sum_t\sum_f
R^{A*}_{I}(t;f,\hat n) 
\frac{\bar H(f)}{P_{n_{I}}(f)}
R^{A'}_{I}(t;f,\hat n')\,.
\label{e:FAkAk'}
\ee
Note that these expressions have an extra polarization index
$A$, compared to the corresponding expressions,
(\ref{e:Xn-power}) and (\ref{e:Fnn'-power}), 
for gravitational-wave power.

\subsubsection{Point spread functions}

The point spread function for the above analysis can
now be obtained by fixing values for both $A'$ and 
$\hat n'$, and letting $A$ and $\hat n$ vary.
Since there are two polarization modes ($+$ and $\times$),
there are actually {\em four} different point spread functions 
for each direction $\hat n'$ on the sky:
\be
{\rm PSF}_{AA'}(\hat n, \hat n') = F_{A\hat n,A'\hat n'}\,.
\label{e:PSF_phase}
\ee
These correspond to the $A=+,\times$ responses to the 
$A'=+,\times$-polarized point sources located in
direction $\hat n'$.

\begin{figure}[h!tbp]
\begin{center}
\includegraphics[trim=3cm 4cm 3cm 2.5cm, clip=true, width=.24\textwidth]{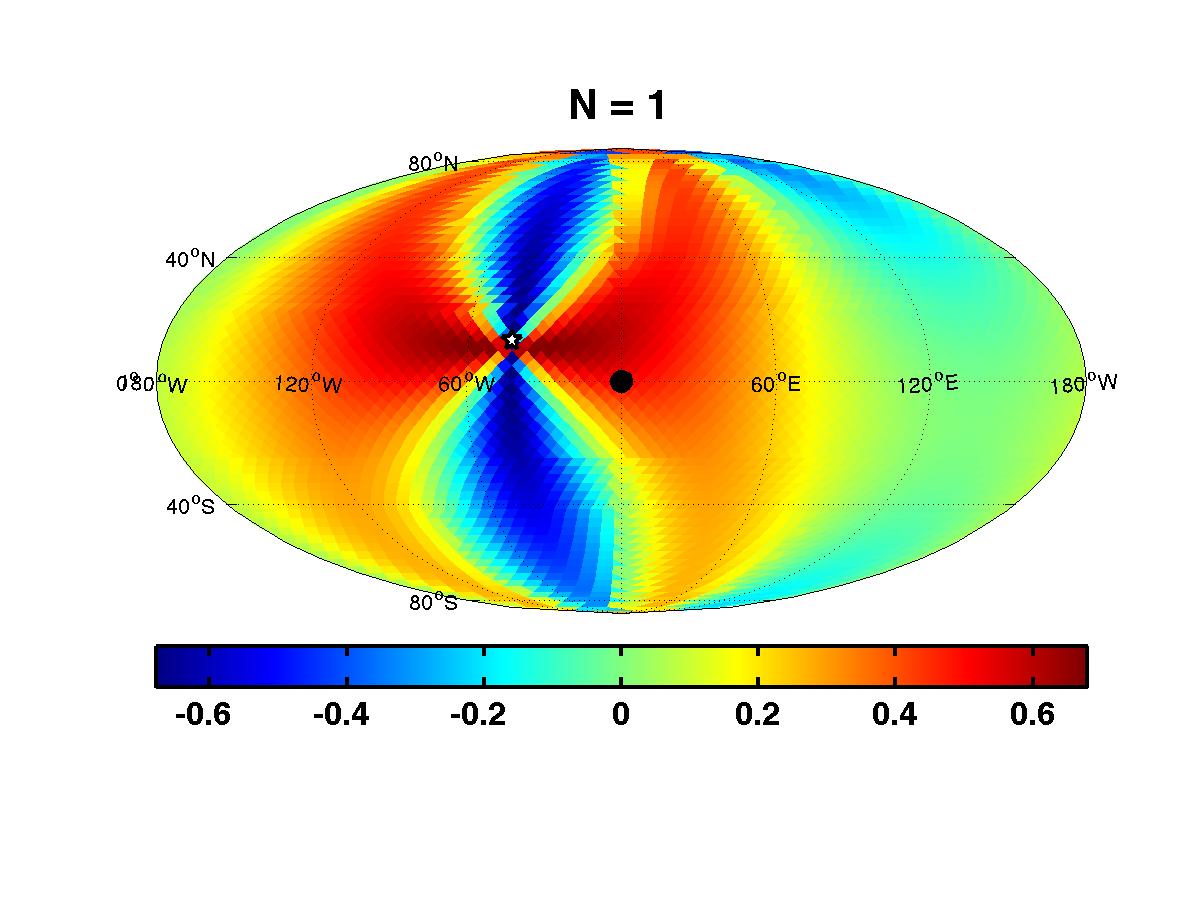}
\includegraphics[trim=3cm 4cm 3cm 2.5cm, clip=true, width=.24\textwidth]{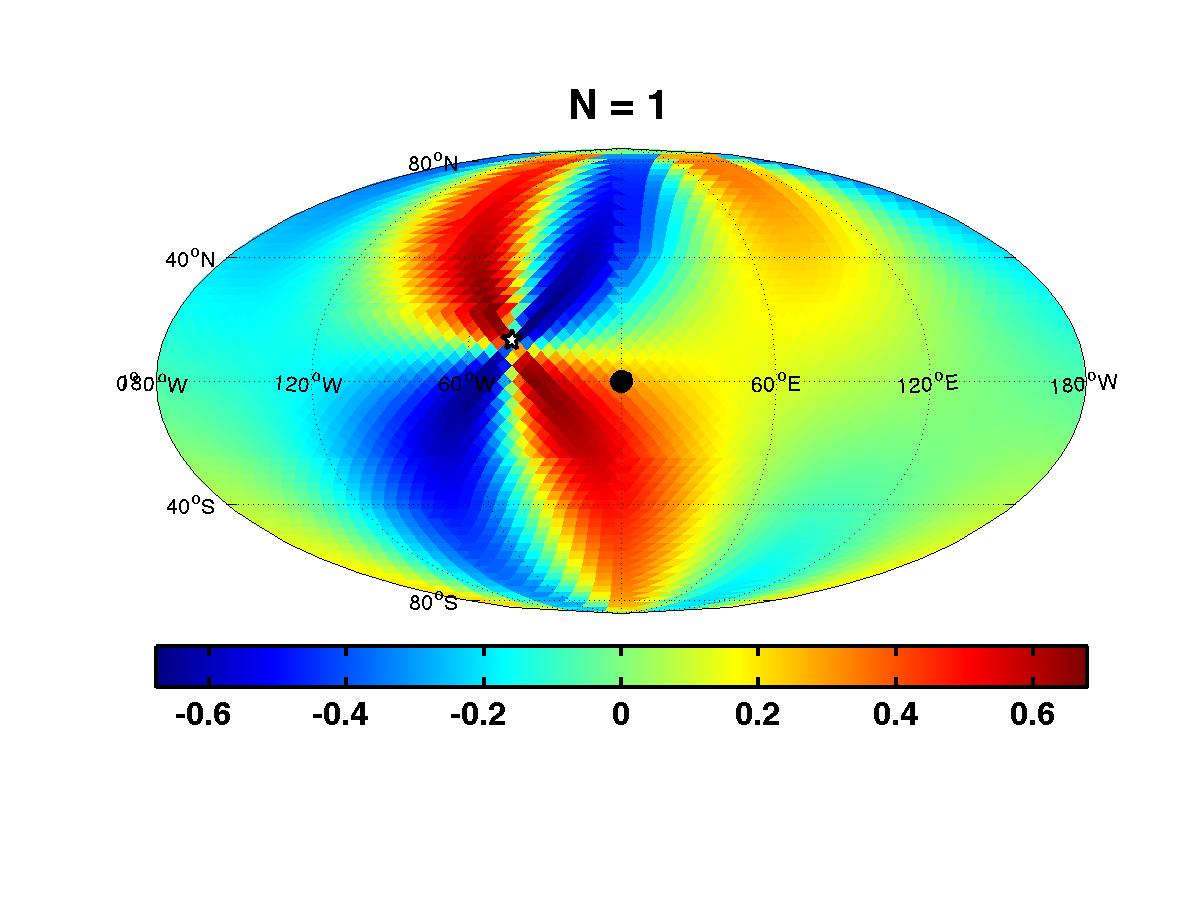}
\includegraphics[trim=3cm 4cm 3cm 2.5cm, clip=true, width=.24\textwidth]{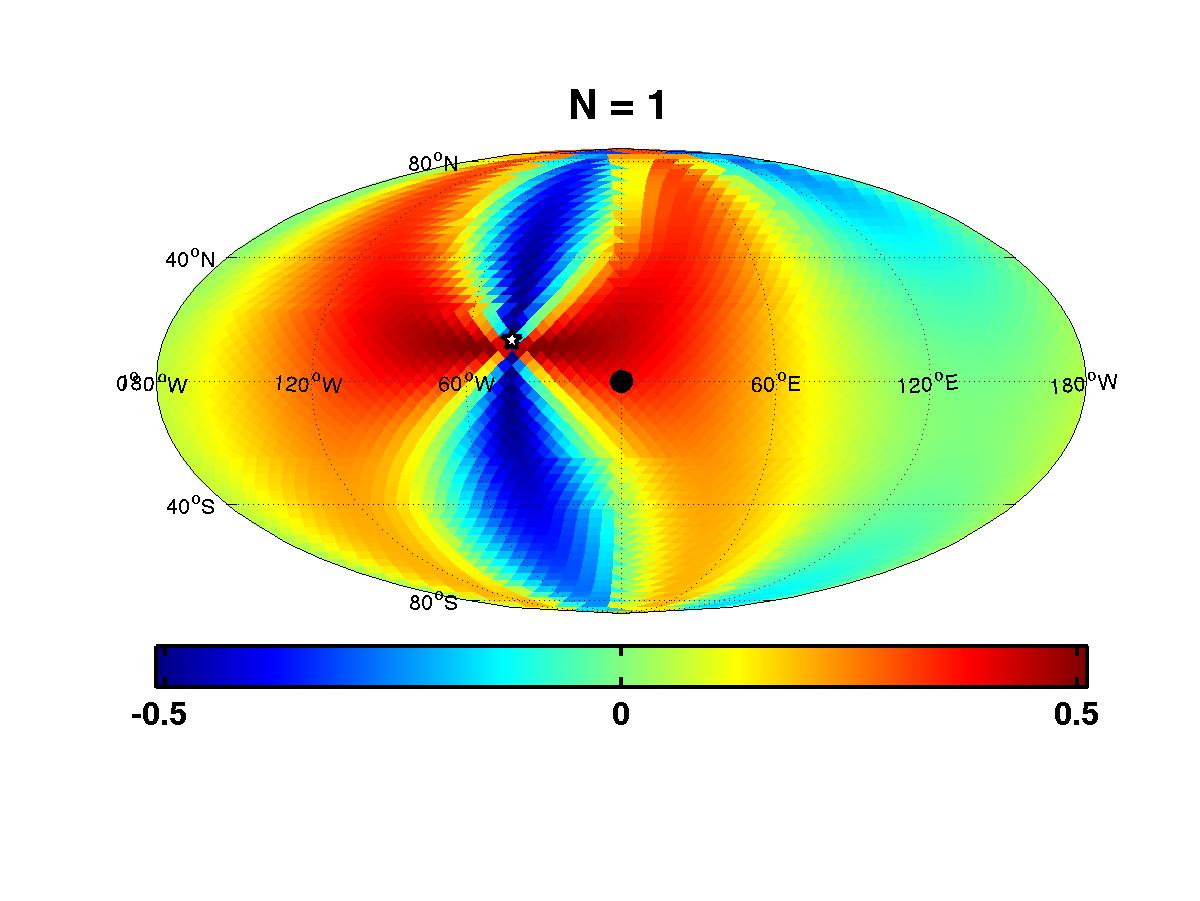}
\includegraphics[trim=3cm 4cm 3cm 2.5cm, clip=true, width=.24\textwidth]{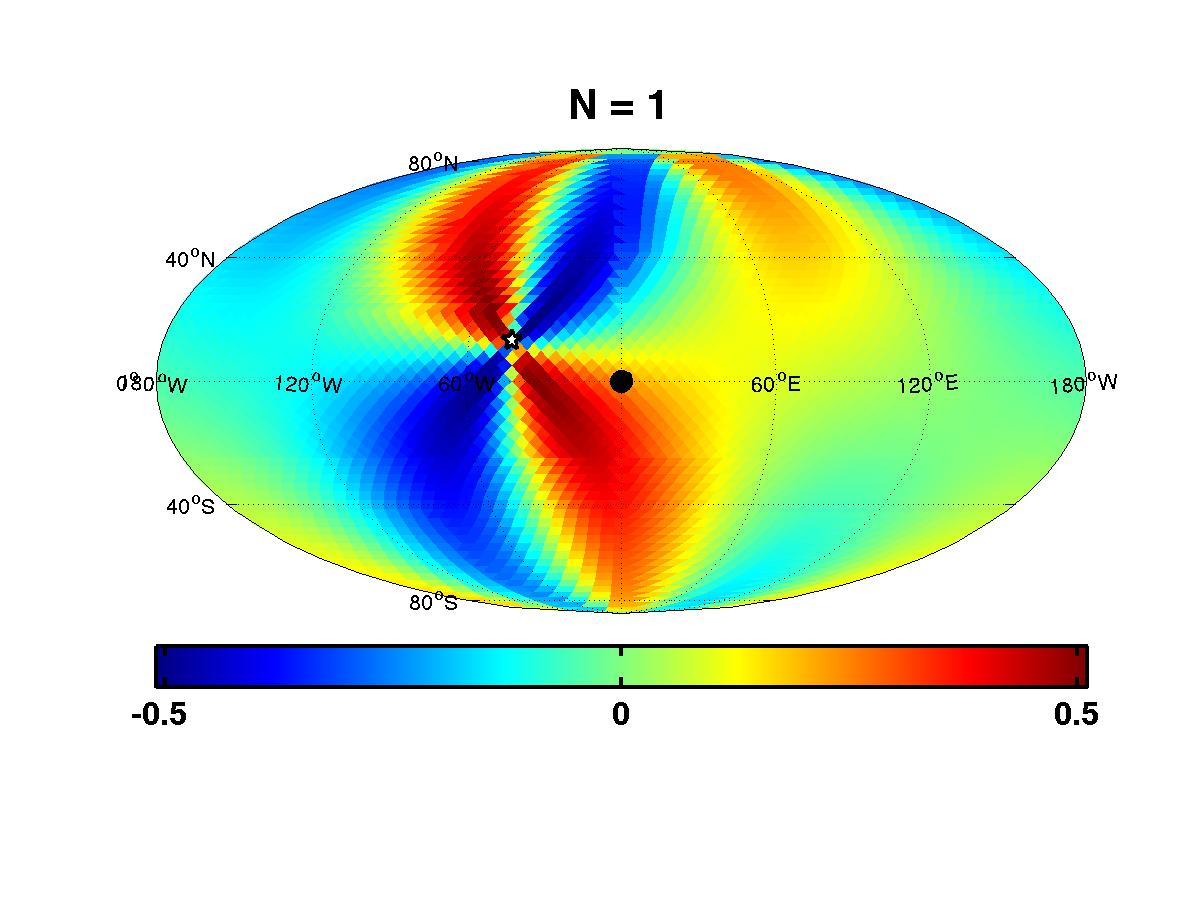}
\includegraphics[trim=3cm 4cm 3cm 2.5cm, clip=true, width=.24\textwidth]{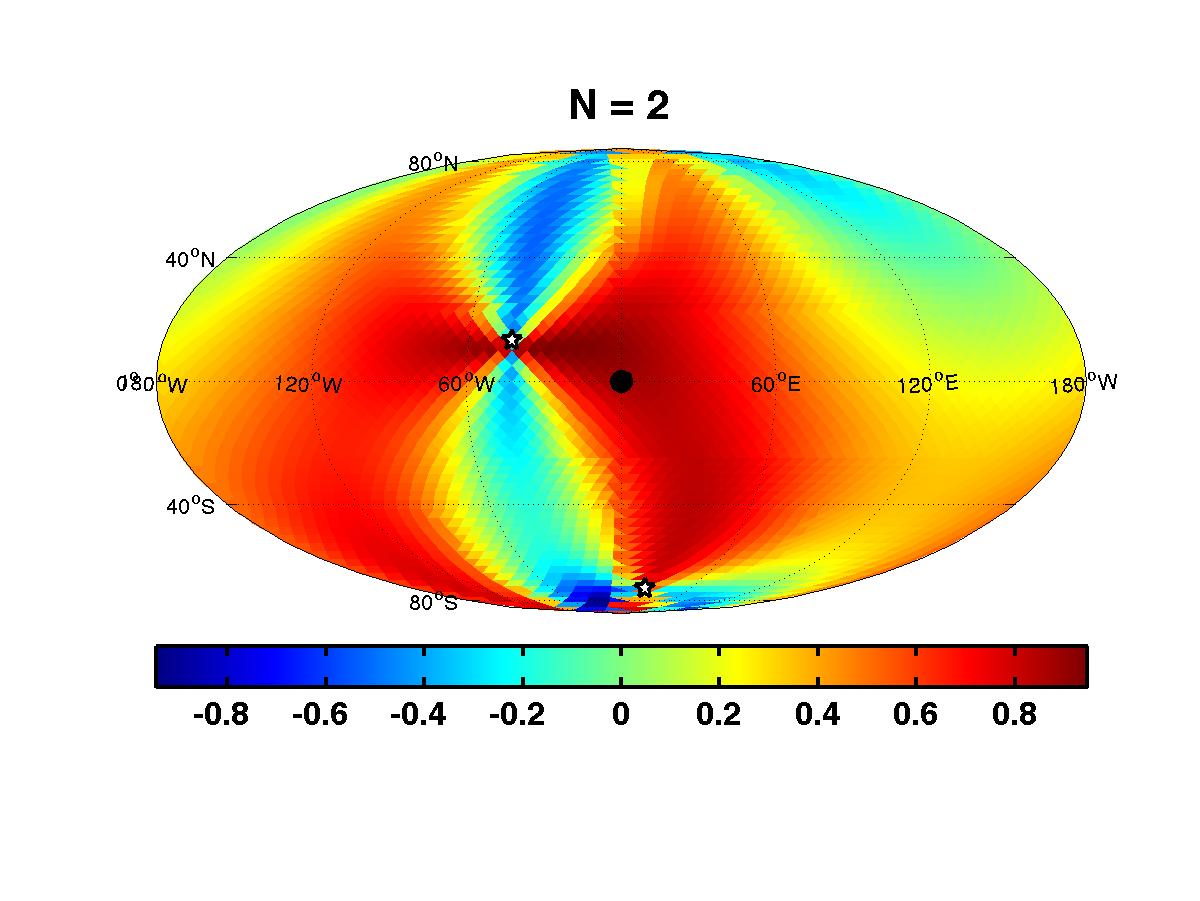}
\includegraphics[trim=3cm 4cm 3cm 2.5cm, clip=true, width=.24\textwidth]{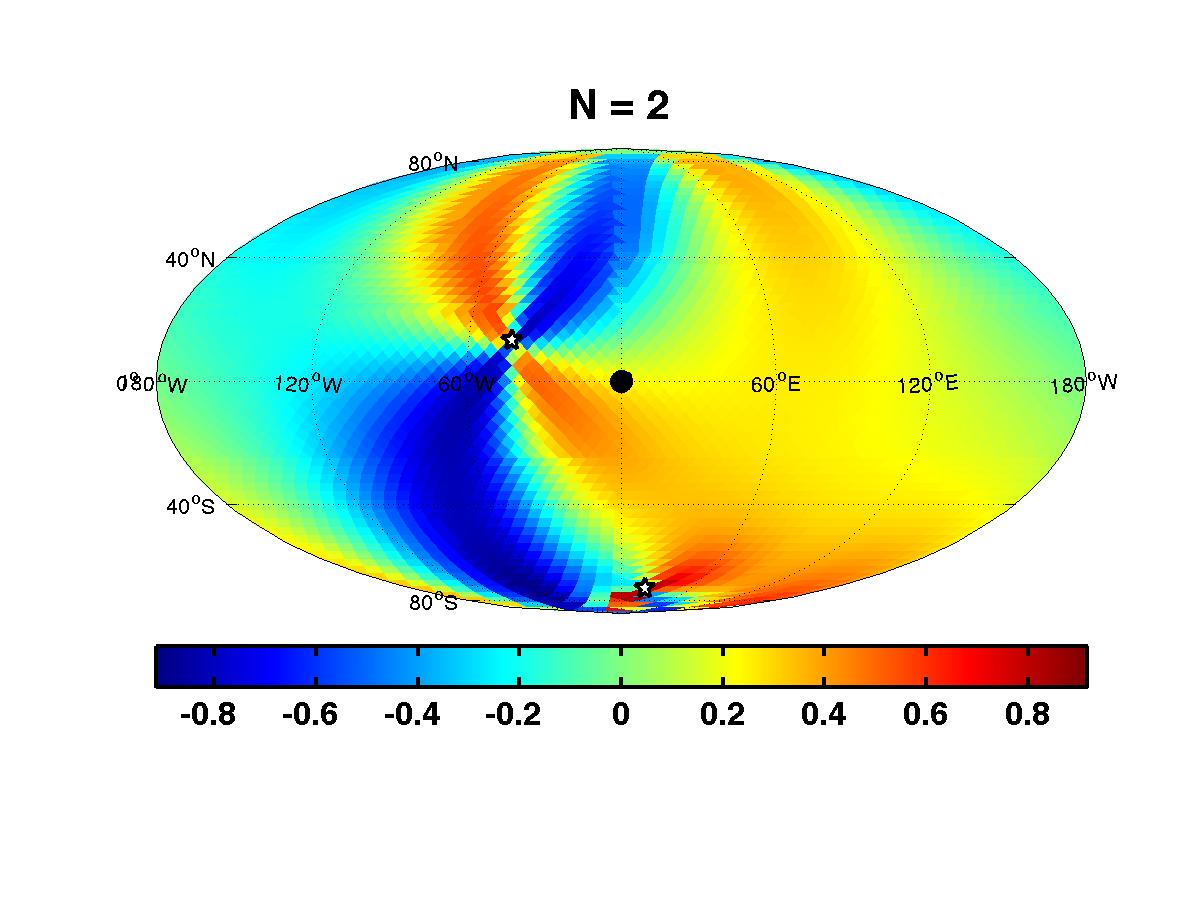}
\includegraphics[trim=3cm 4cm 3cm 2.5cm, clip=true, width=.24\textwidth]{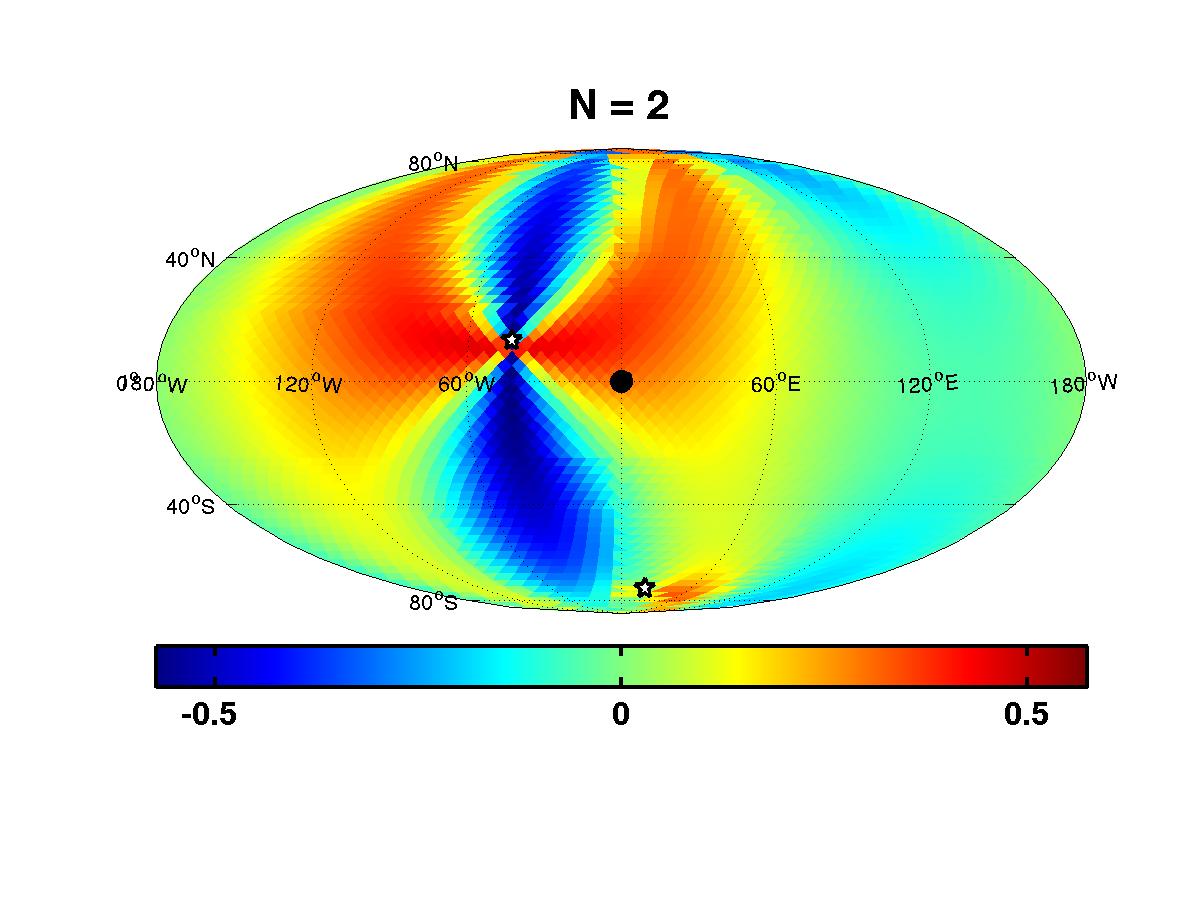}
\includegraphics[trim=3cm 4cm 3cm 2.5cm, clip=true, width=.24\textwidth]{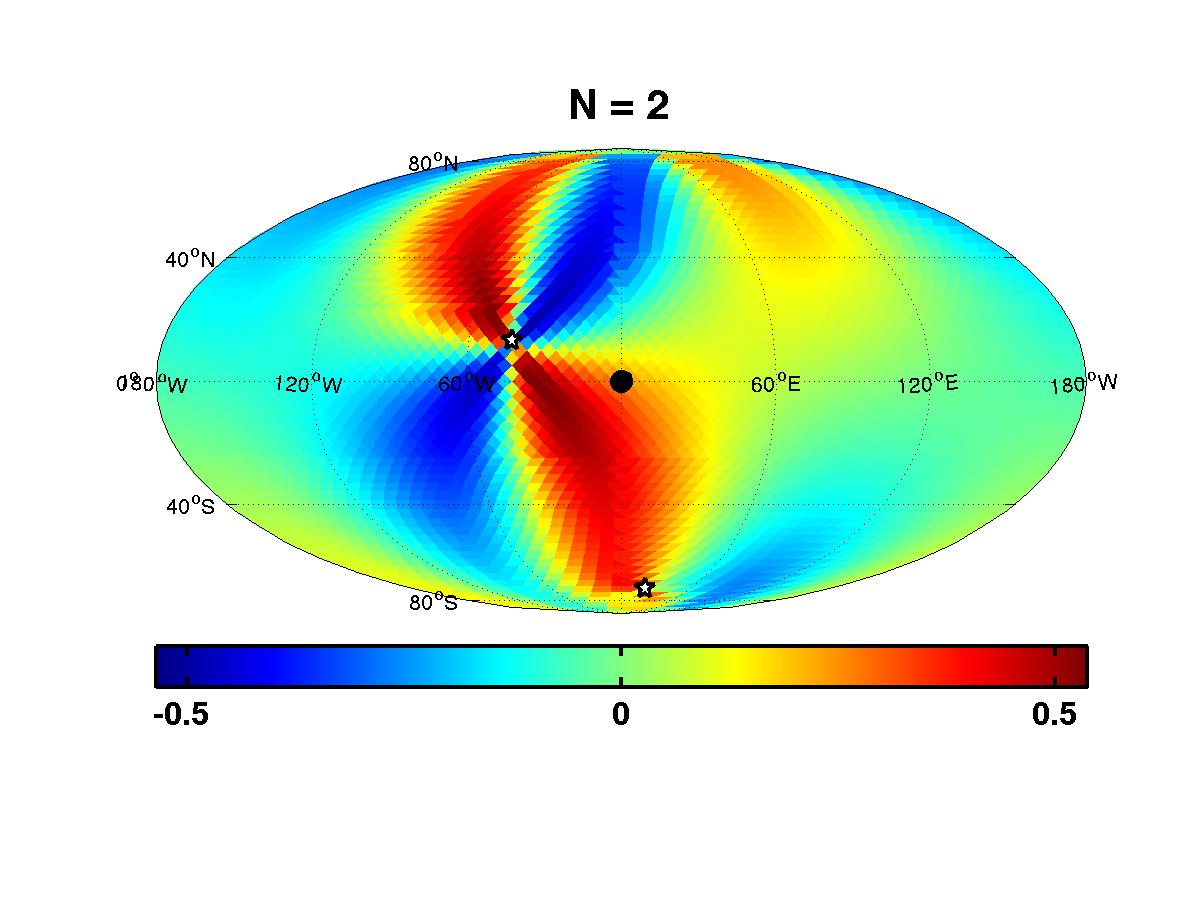}
\includegraphics[trim=3cm 4cm 3cm 2.5cm, clip=true, width=.24\textwidth]{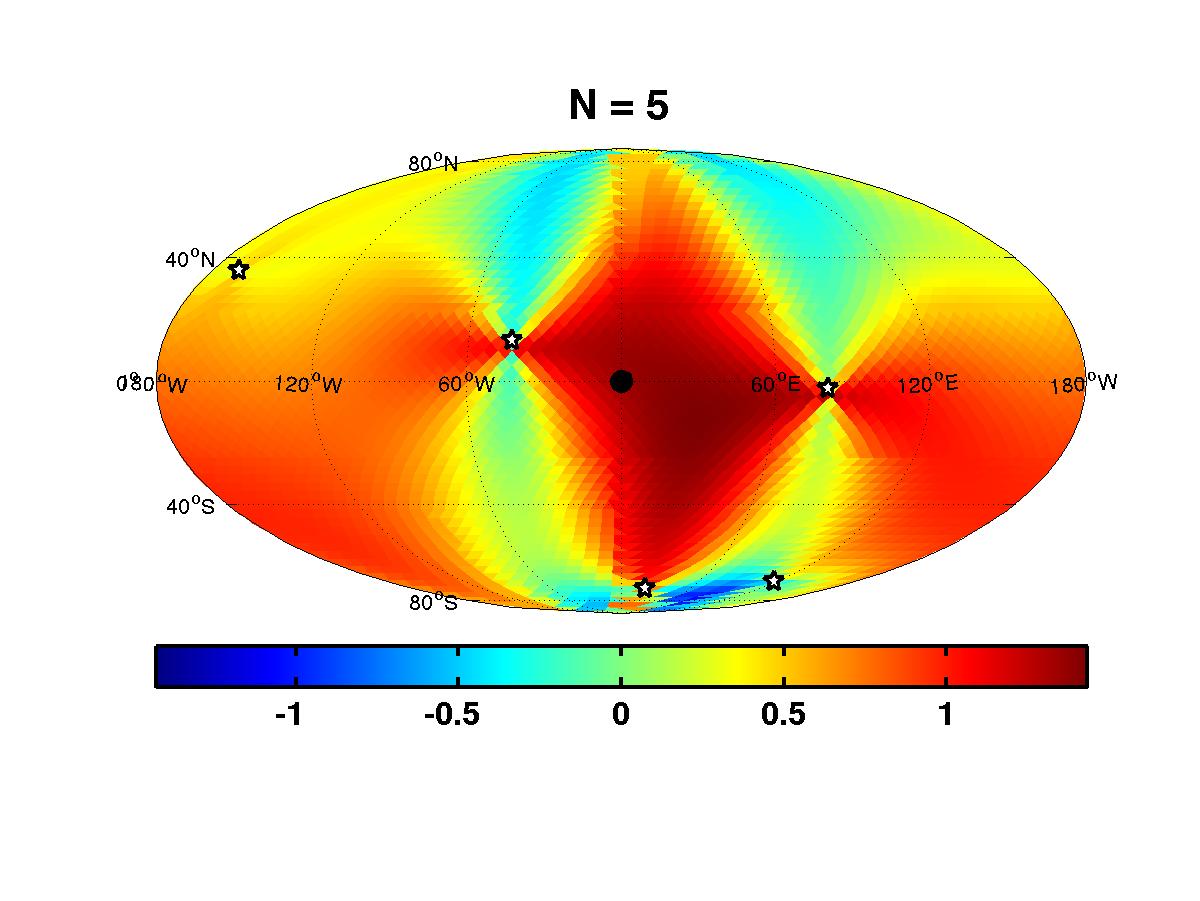}
\includegraphics[trim=3cm 4cm 3cm 2.5cm, clip=true, width=.24\textwidth]{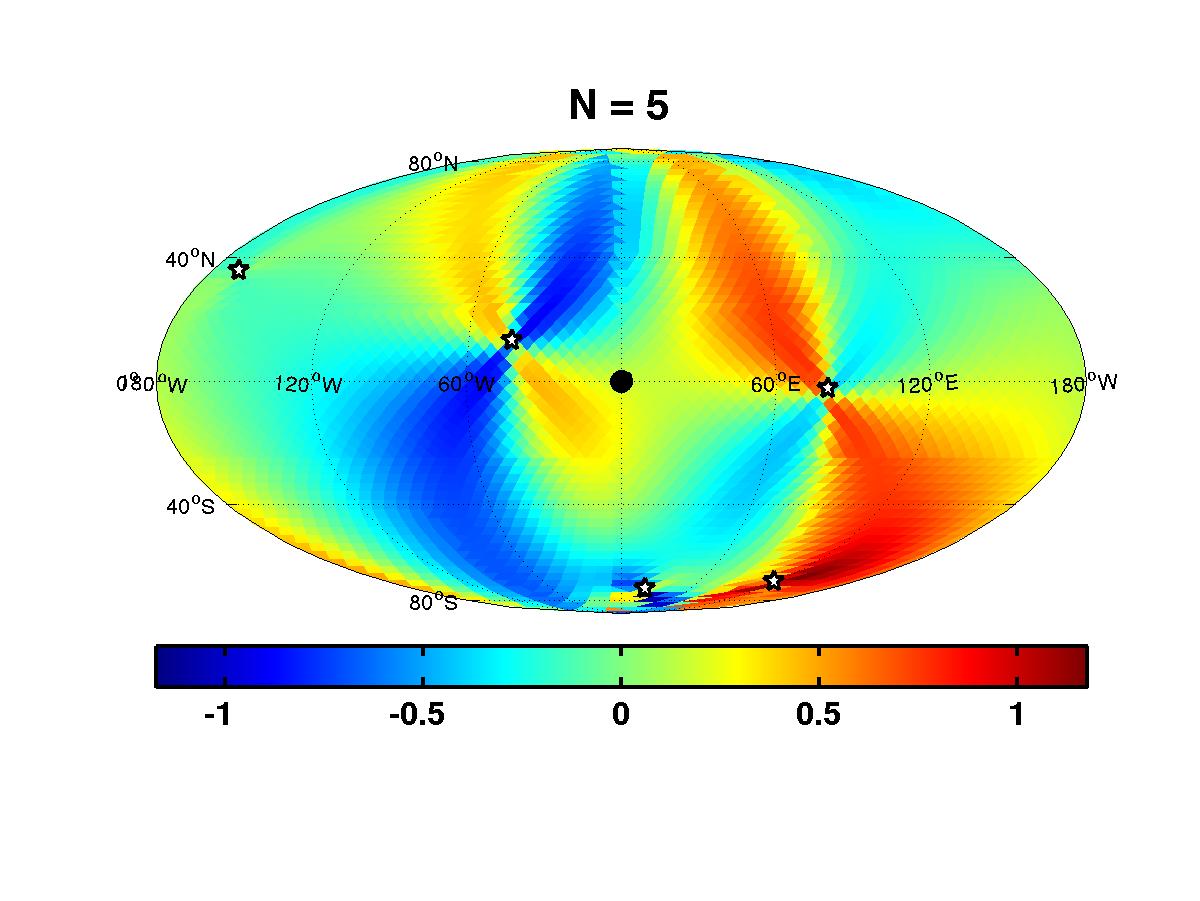}
\includegraphics[trim=3cm 4cm 3cm 2.5cm, clip=true, width=.24\textwidth]{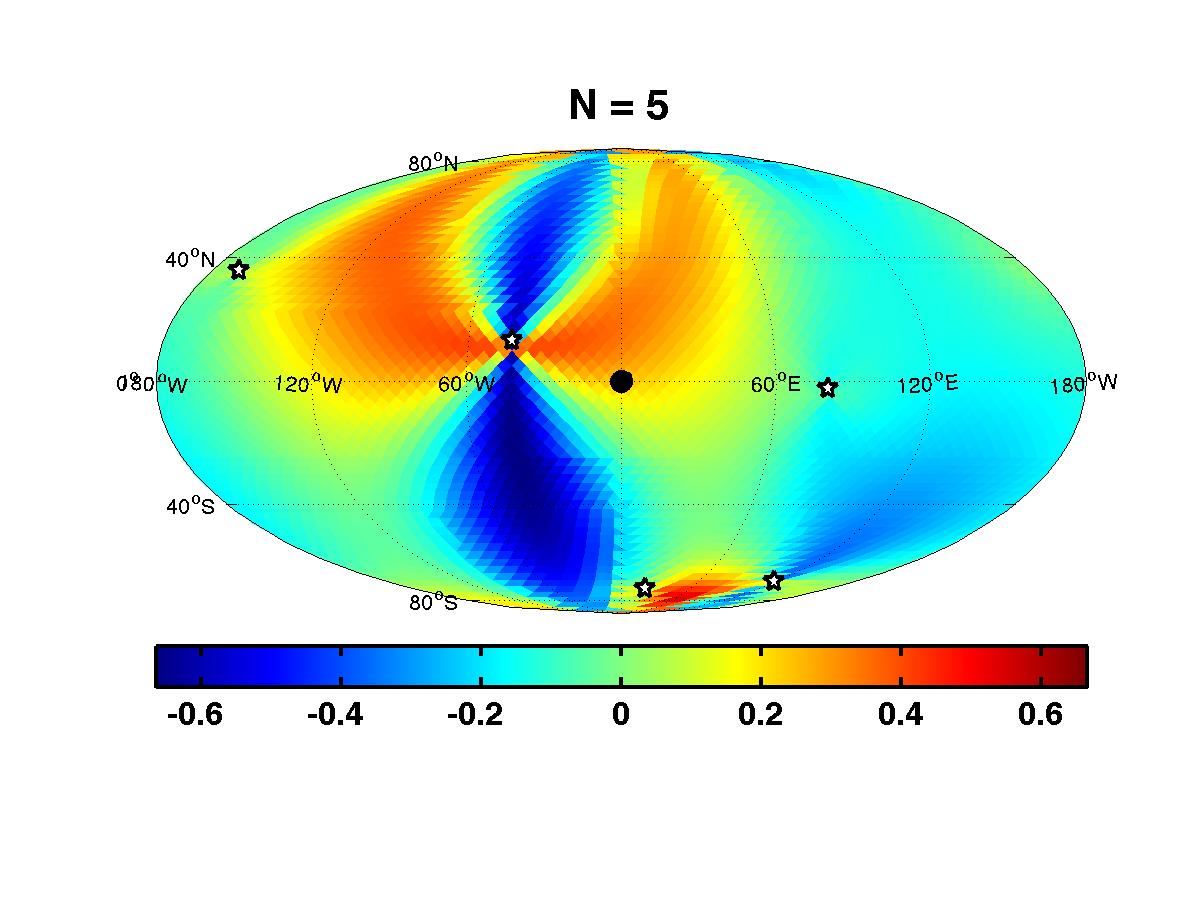}
\includegraphics[trim=3cm 4cm 3cm 2.5cm, clip=true, width=.24\textwidth]{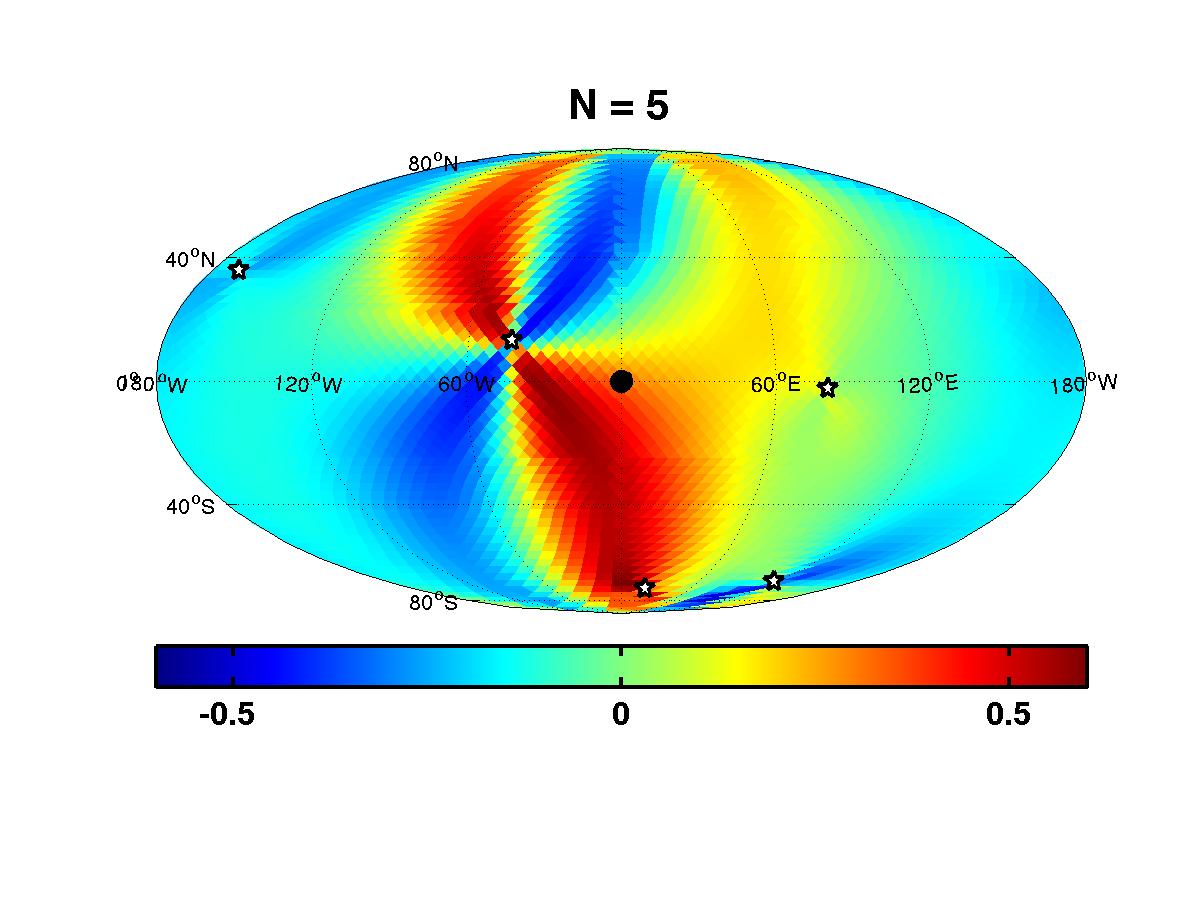}
\includegraphics[trim=3cm 4cm 3cm 2.5cm, clip=true, width=.24\textwidth]{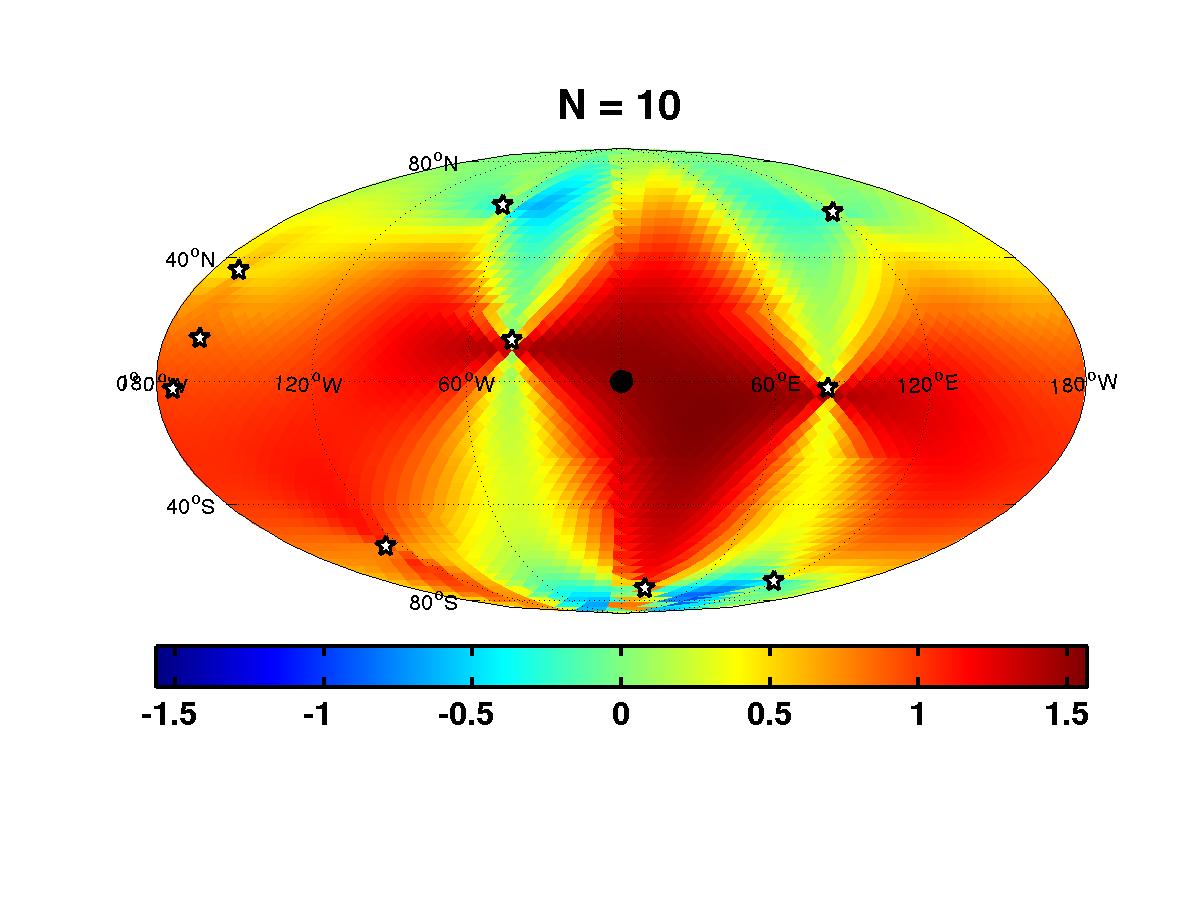}
\includegraphics[trim=3cm 4cm 3cm 2.5cm, clip=true, width=.24\textwidth]{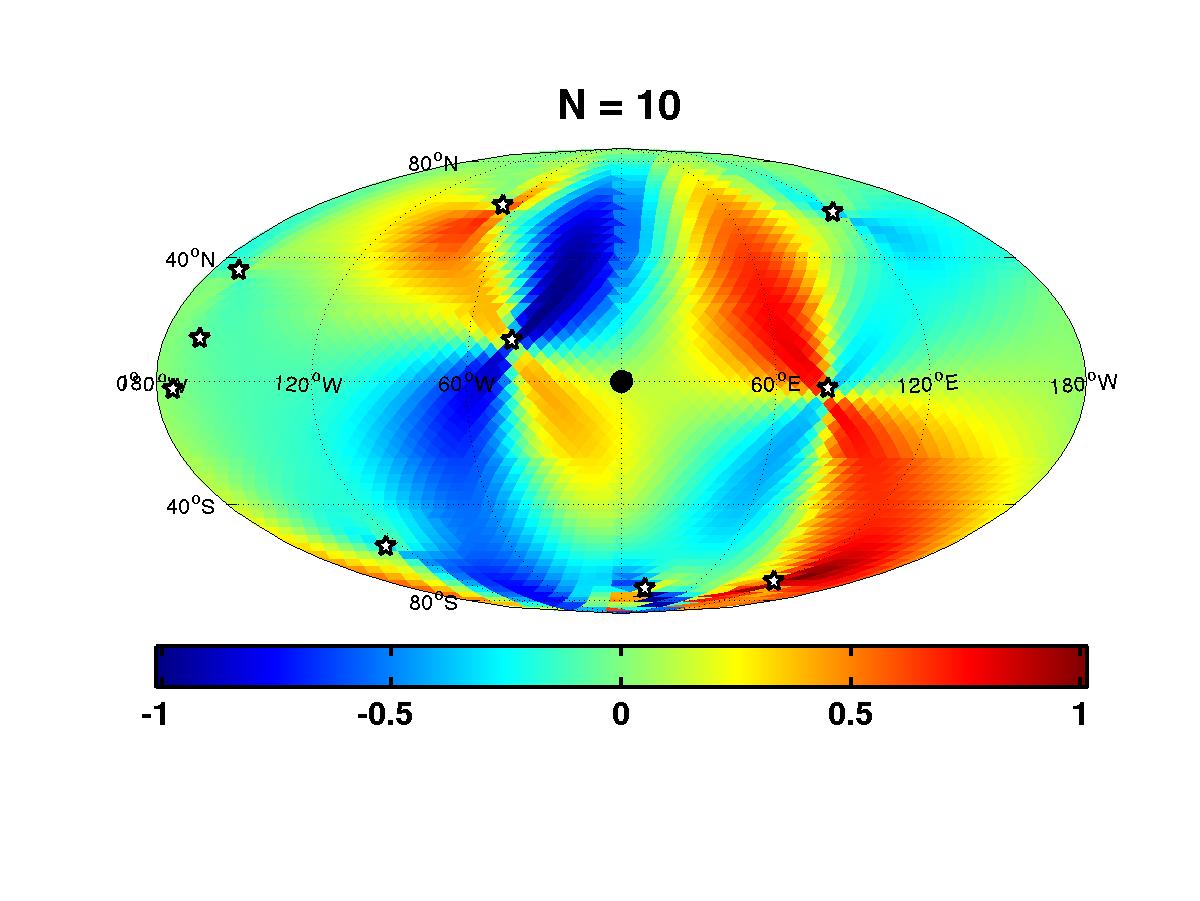}
\includegraphics[trim=3cm 4cm 3cm 2.5cm, clip=true, width=.24\textwidth]{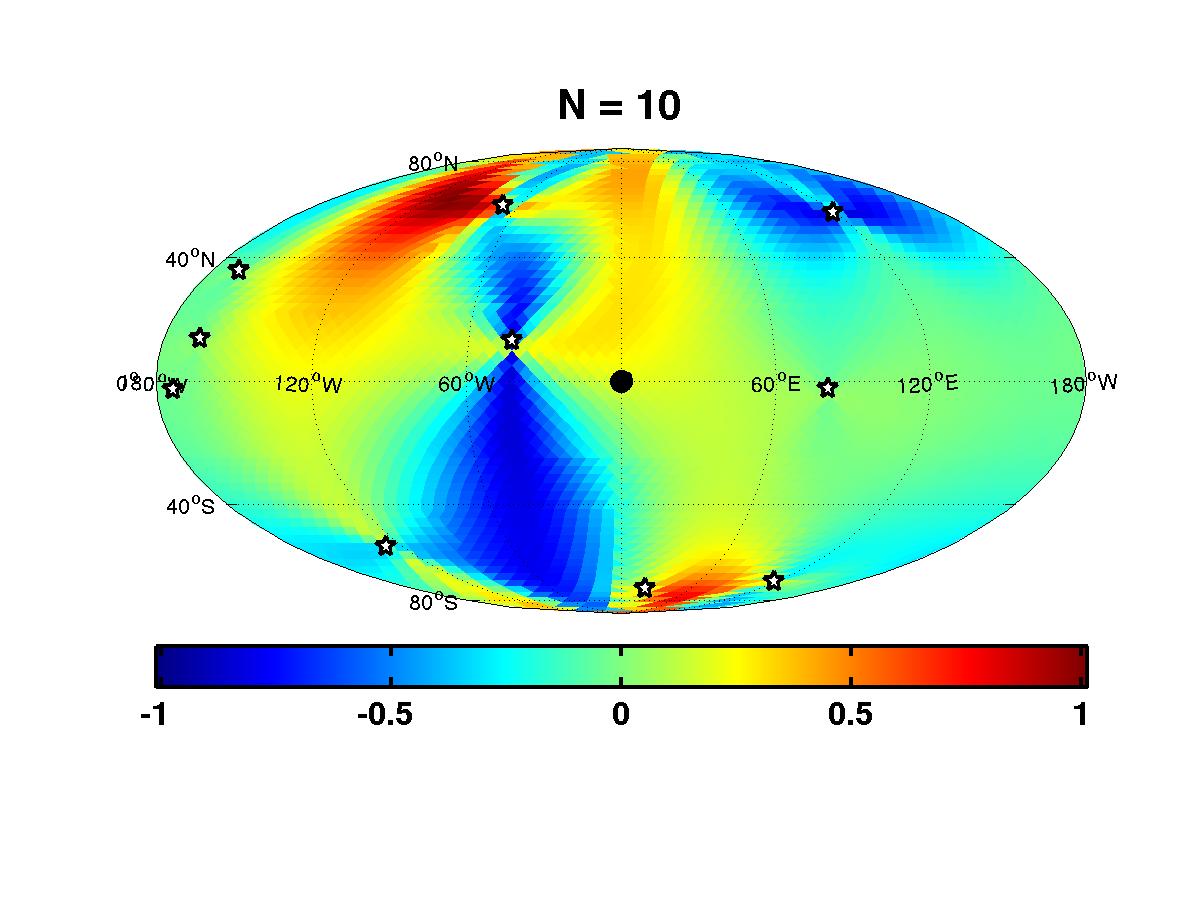}
\includegraphics[trim=3cm 4cm 3cm 2.5cm, clip=true, width=.24\textwidth]{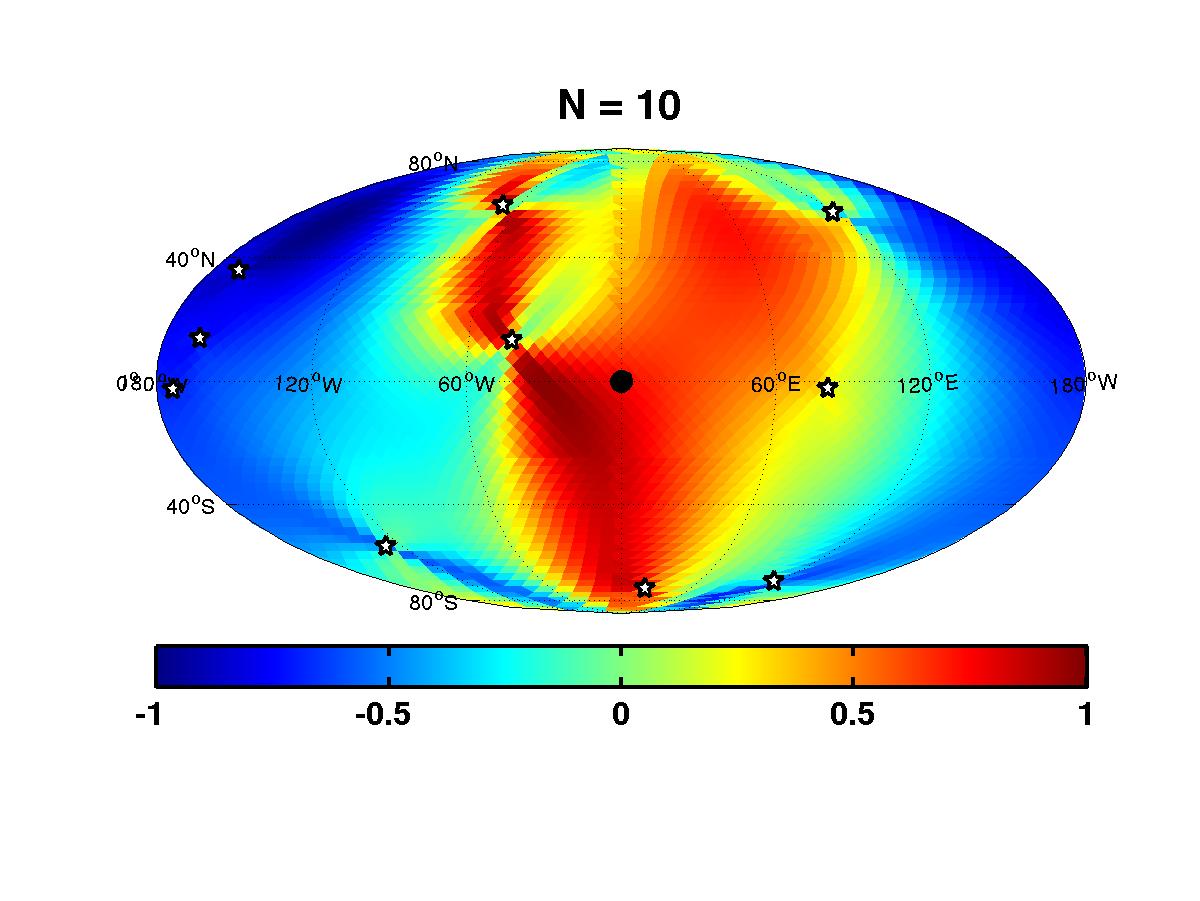}
\includegraphics[trim=3cm 4cm 3cm 2.5cm, clip=true, width=.24\textwidth]{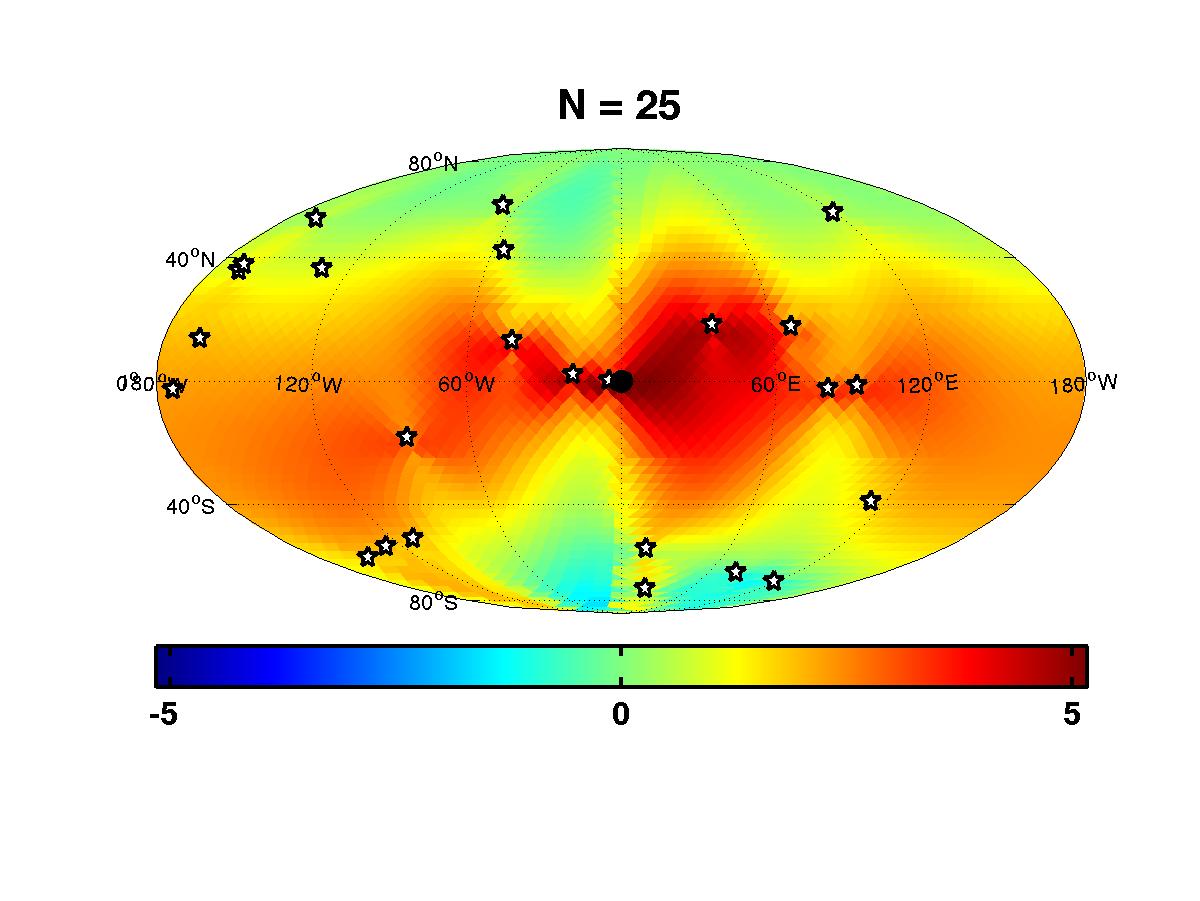}
\includegraphics[trim=3cm 4cm 3cm 2.5cm, clip=true, width=.24\textwidth]{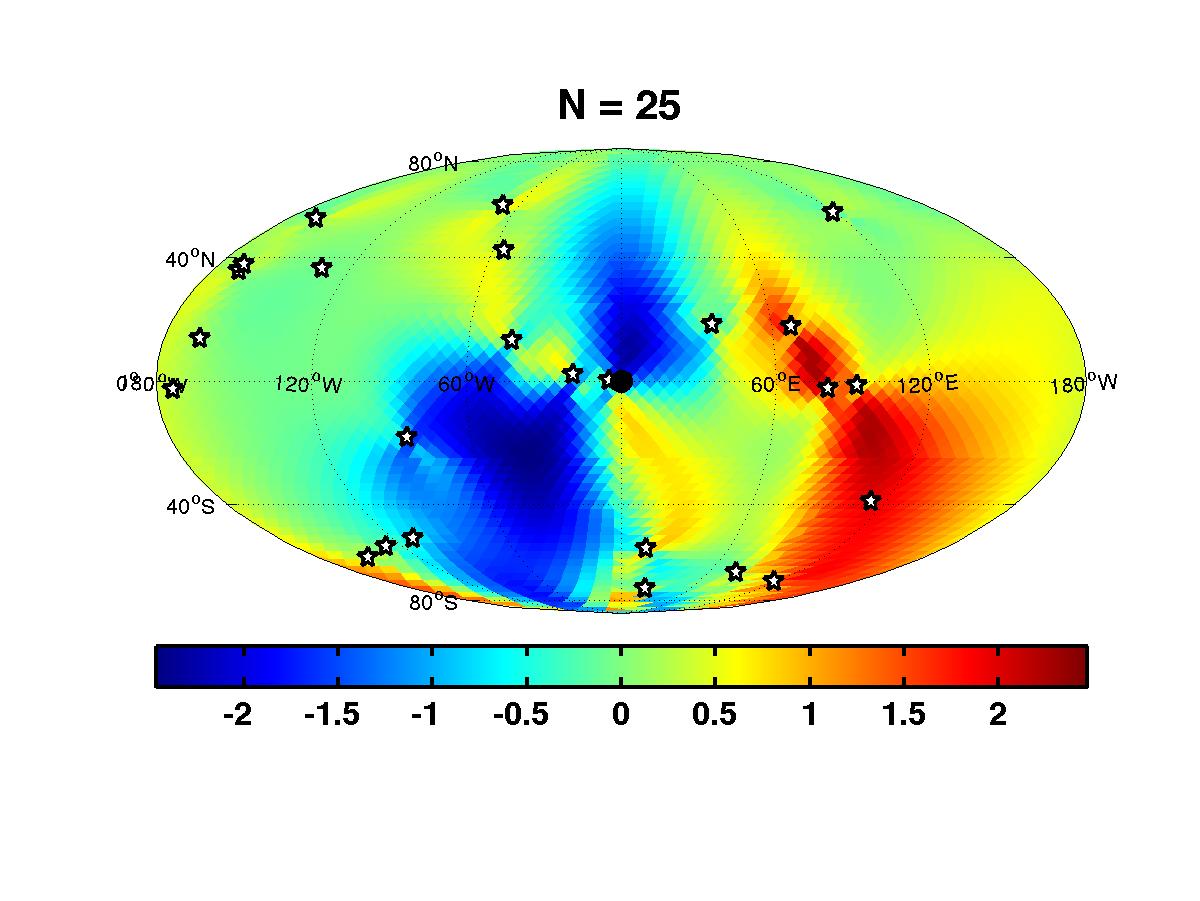}
\includegraphics[trim=3cm 4cm 3cm 2.5cm, clip=true, width=.24\textwidth]{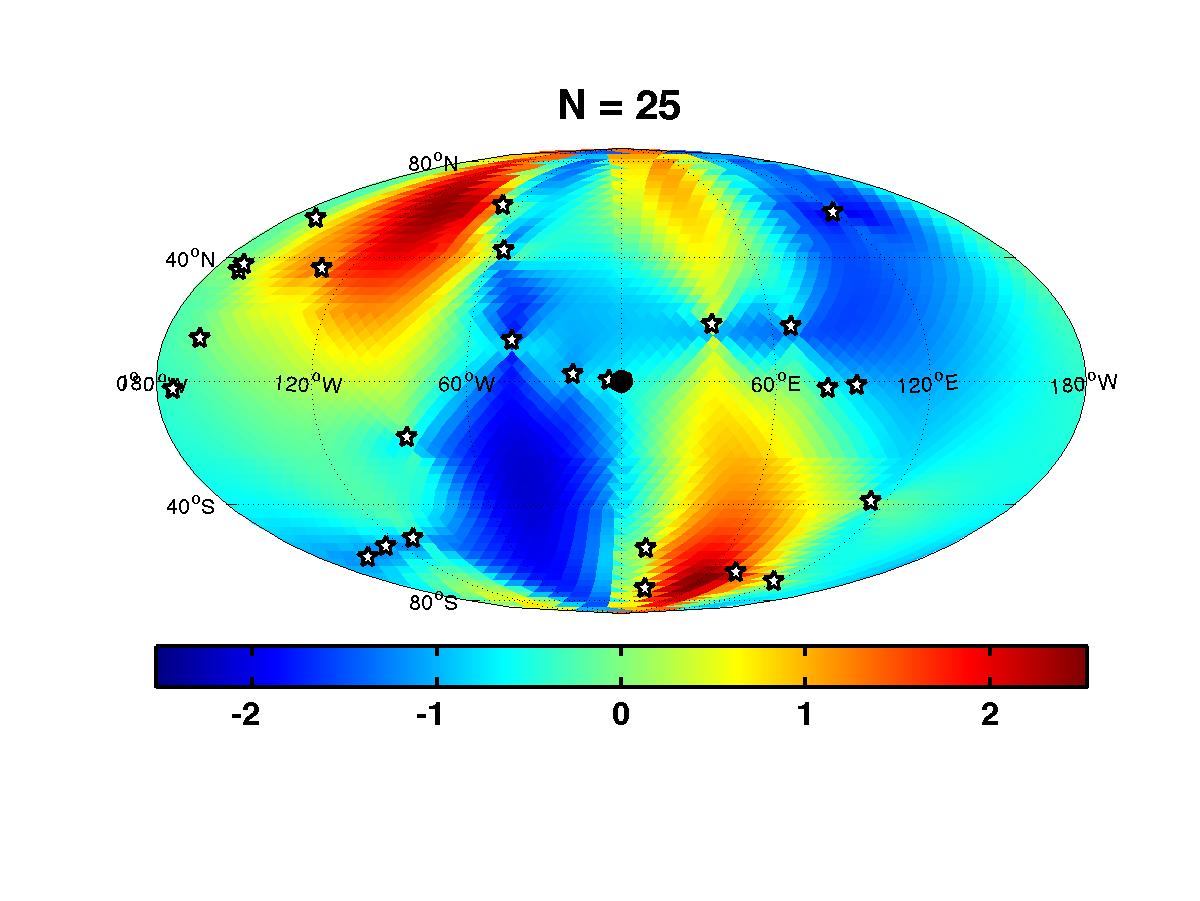}
\includegraphics[trim=3cm 4cm 3cm 2.5cm, clip=true, width=.24\textwidth]{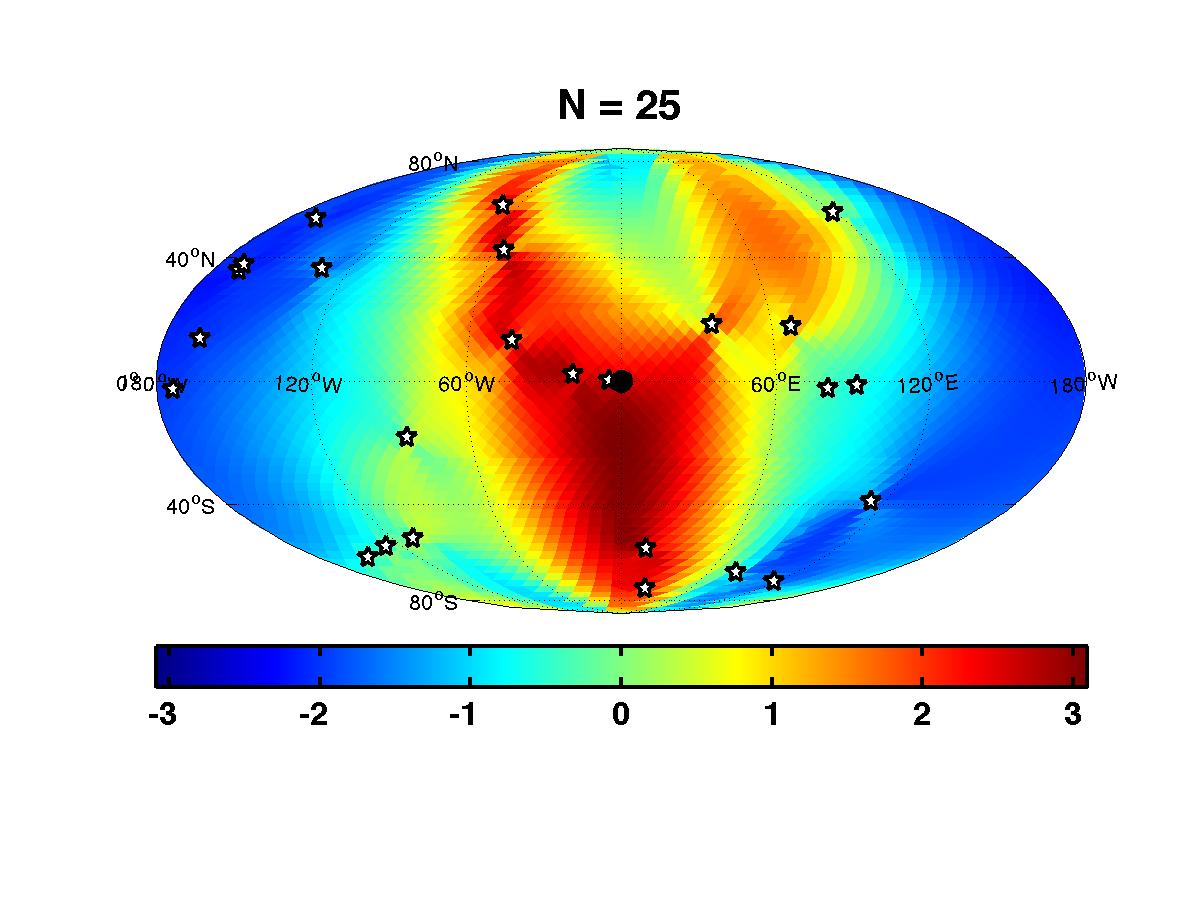}
\includegraphics[trim=3cm 4cm 3cm 2.5cm, clip=true, width=.24\textwidth]{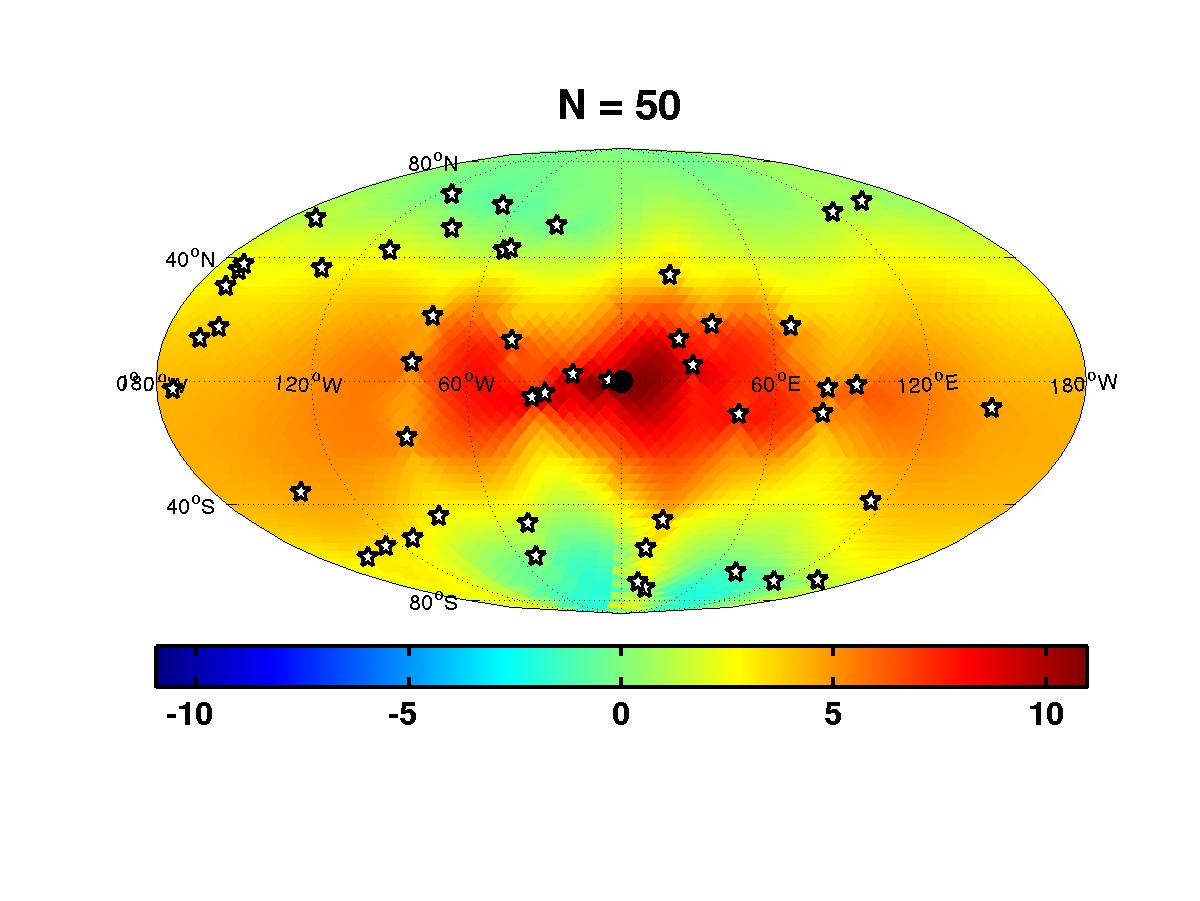}
\includegraphics[trim=3cm 4cm 3cm 2.5cm, clip=true, width=.24\textwidth]{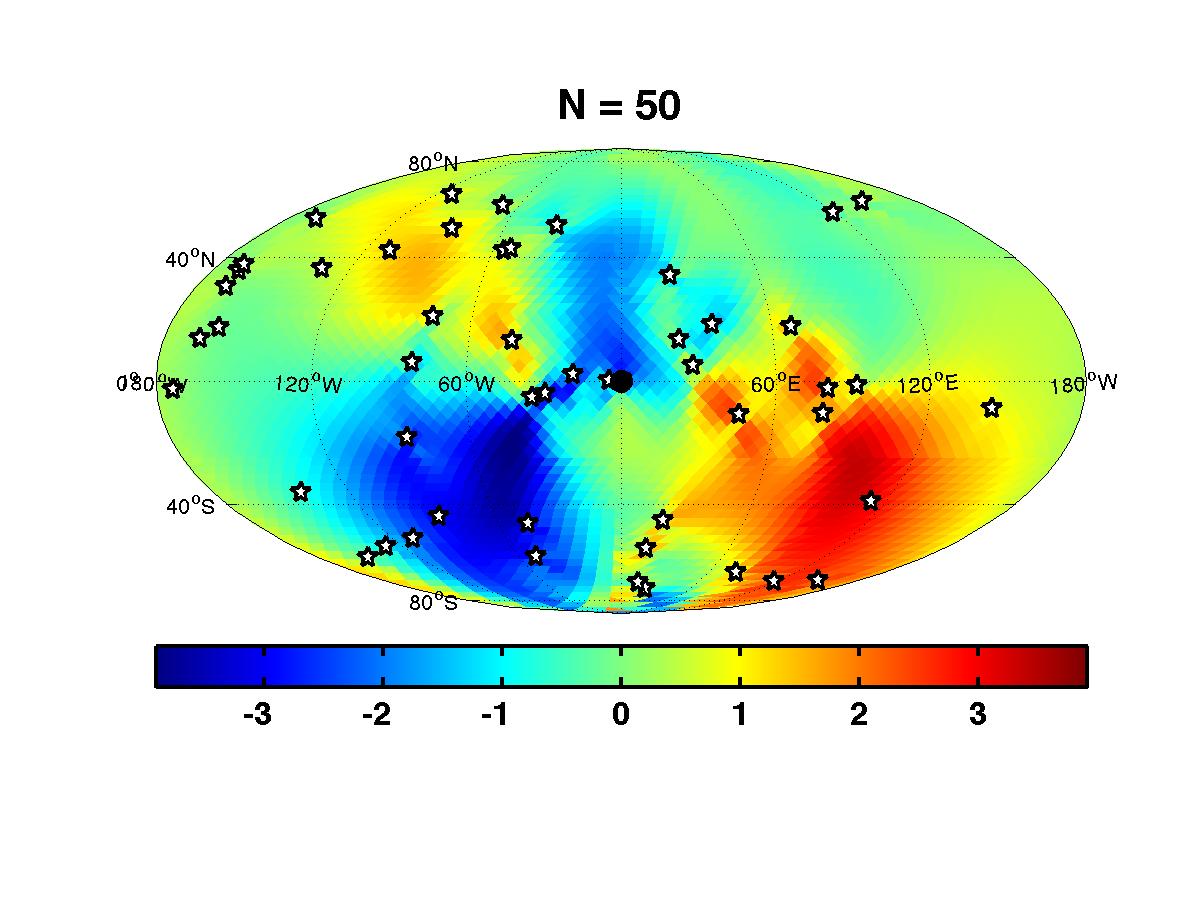}
\includegraphics[trim=3cm 4cm 3cm 2.5cm, clip=true, width=.24\textwidth]{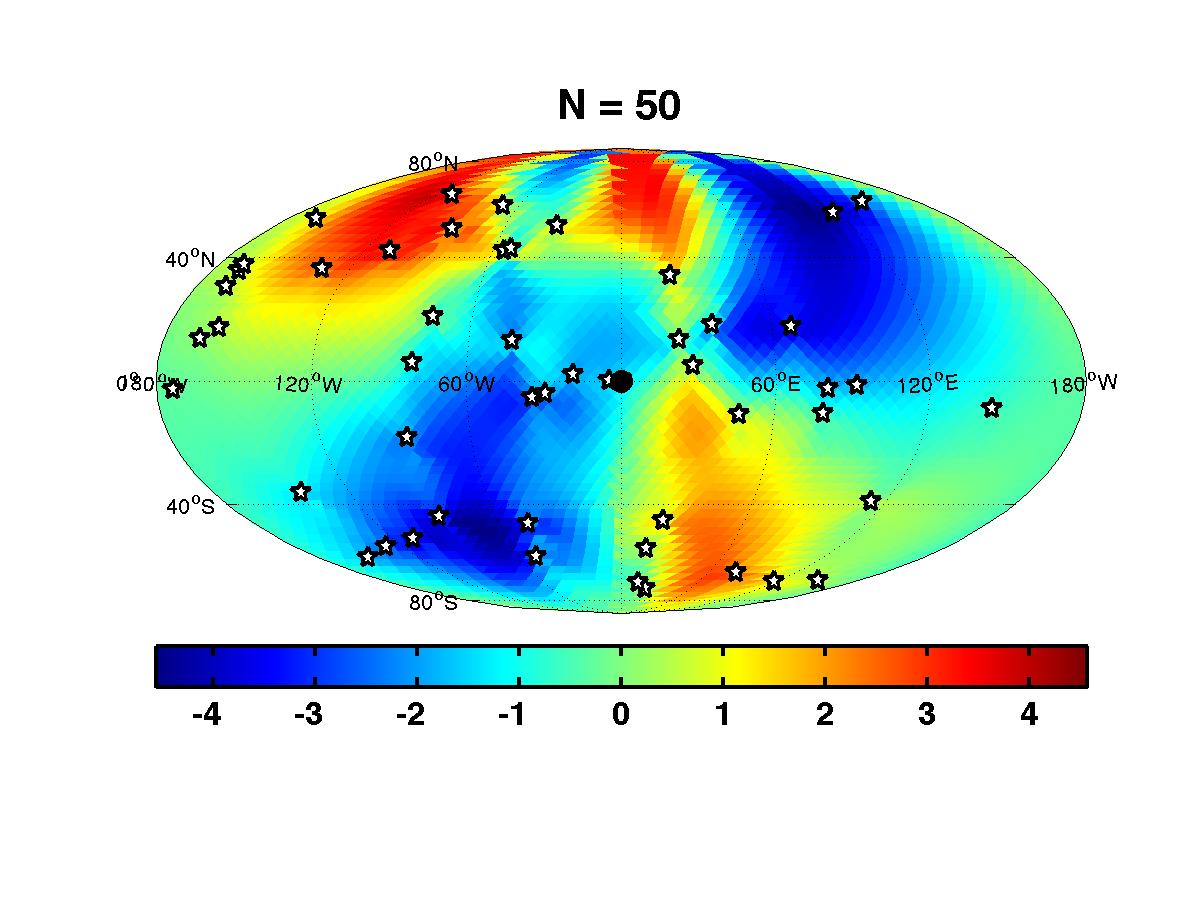}
\includegraphics[trim=3cm 4cm 3cm 2.5cm, clip=true, width=.24\textwidth]{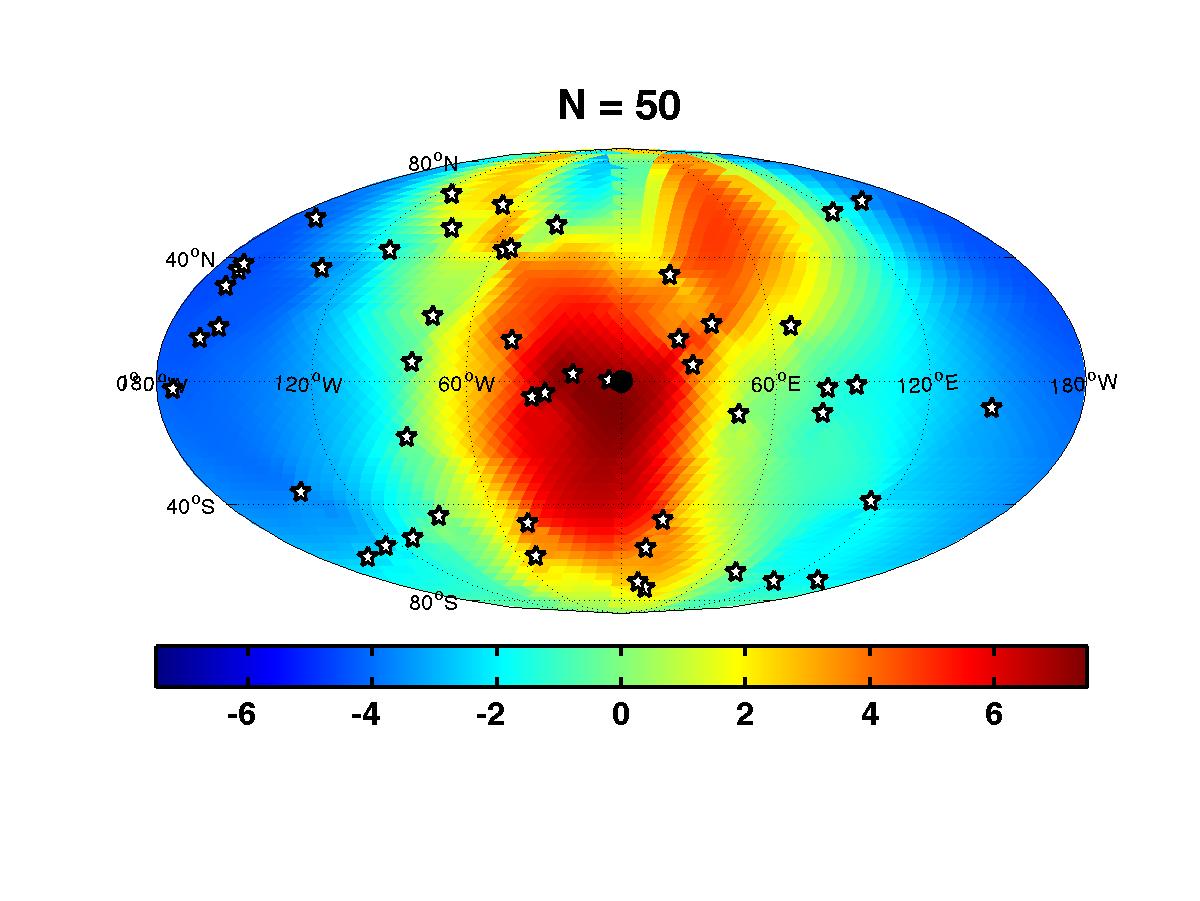}
\includegraphics[trim=3cm 4cm 3cm 2.5cm, clip=true, width=.24\textwidth]{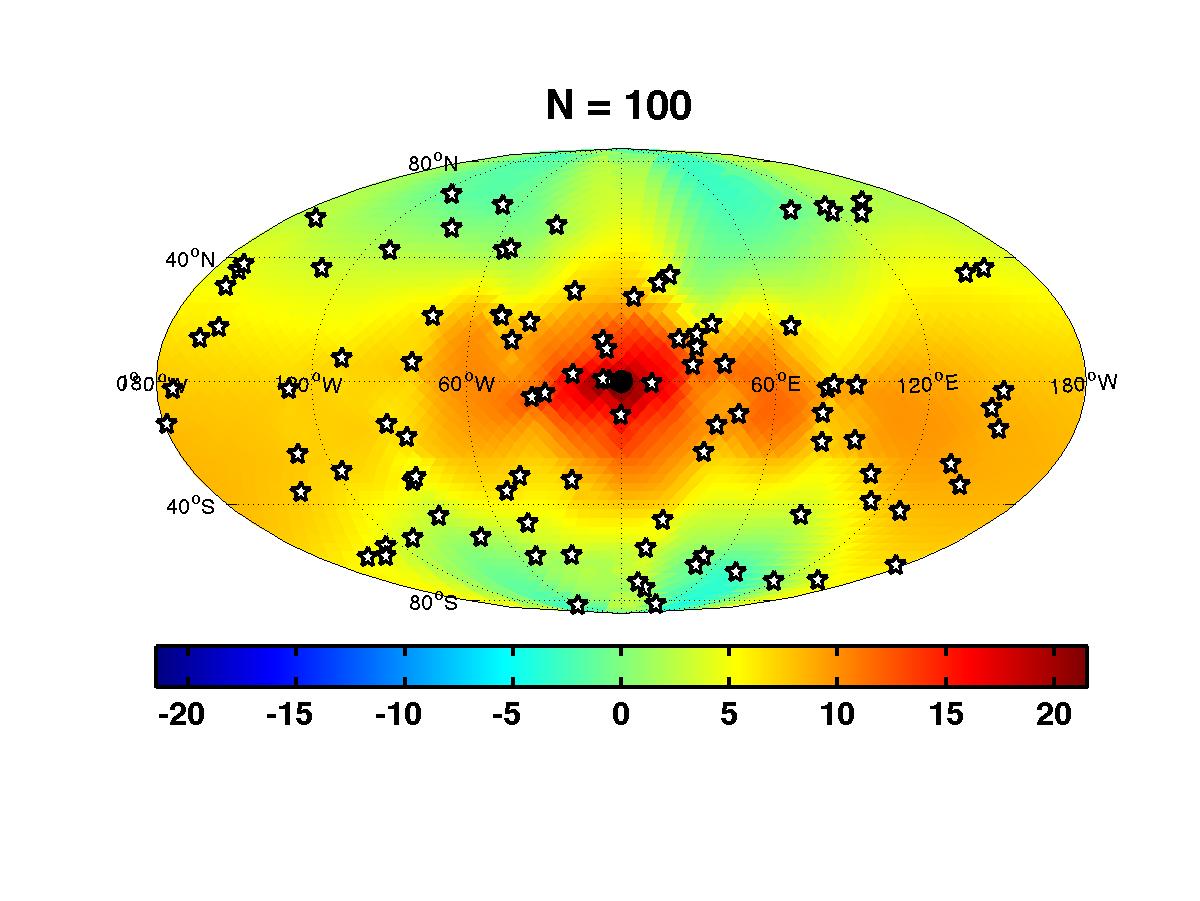}
\includegraphics[trim=3cm 4cm 3cm 2.5cm, clip=true, width=.24\textwidth]{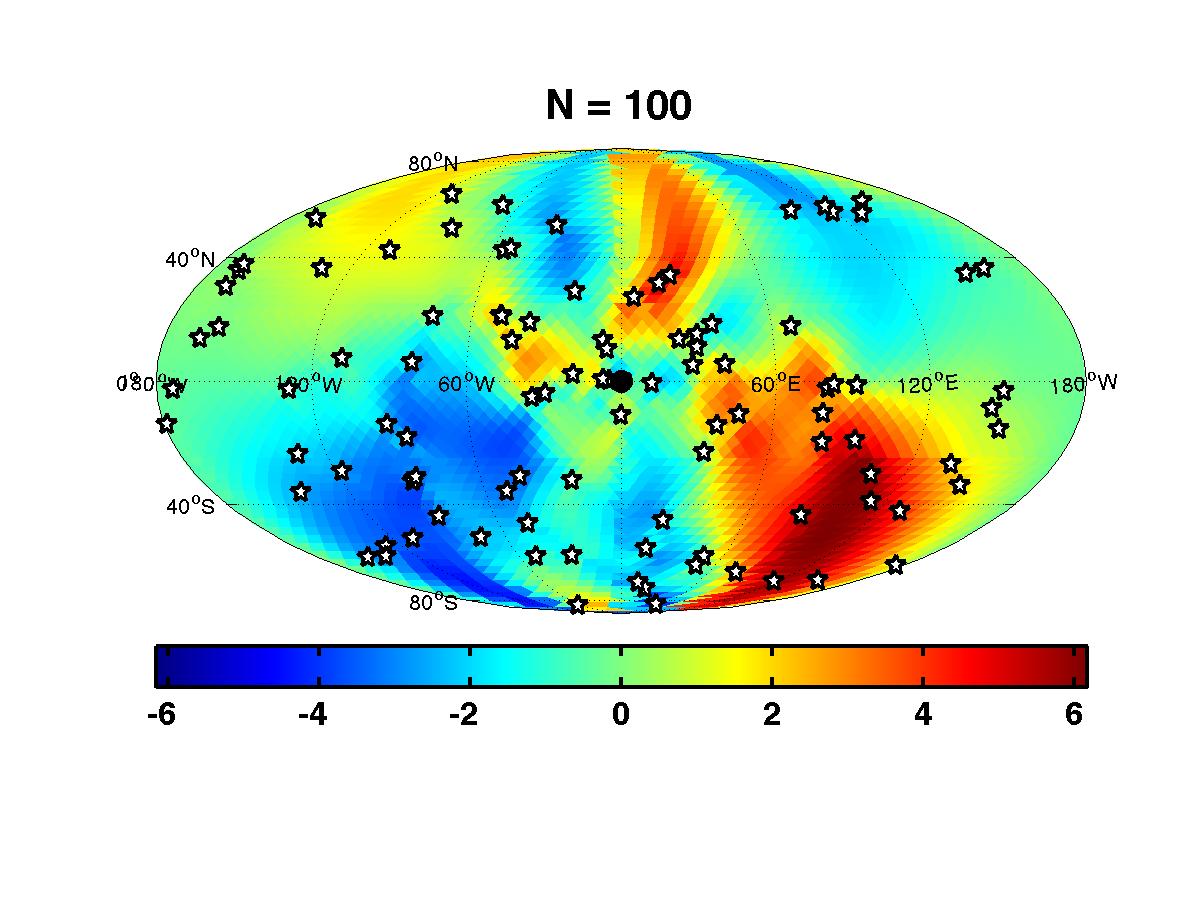}
\includegraphics[trim=3cm 4cm 3cm 2.5cm, clip=true, width=.24\textwidth]{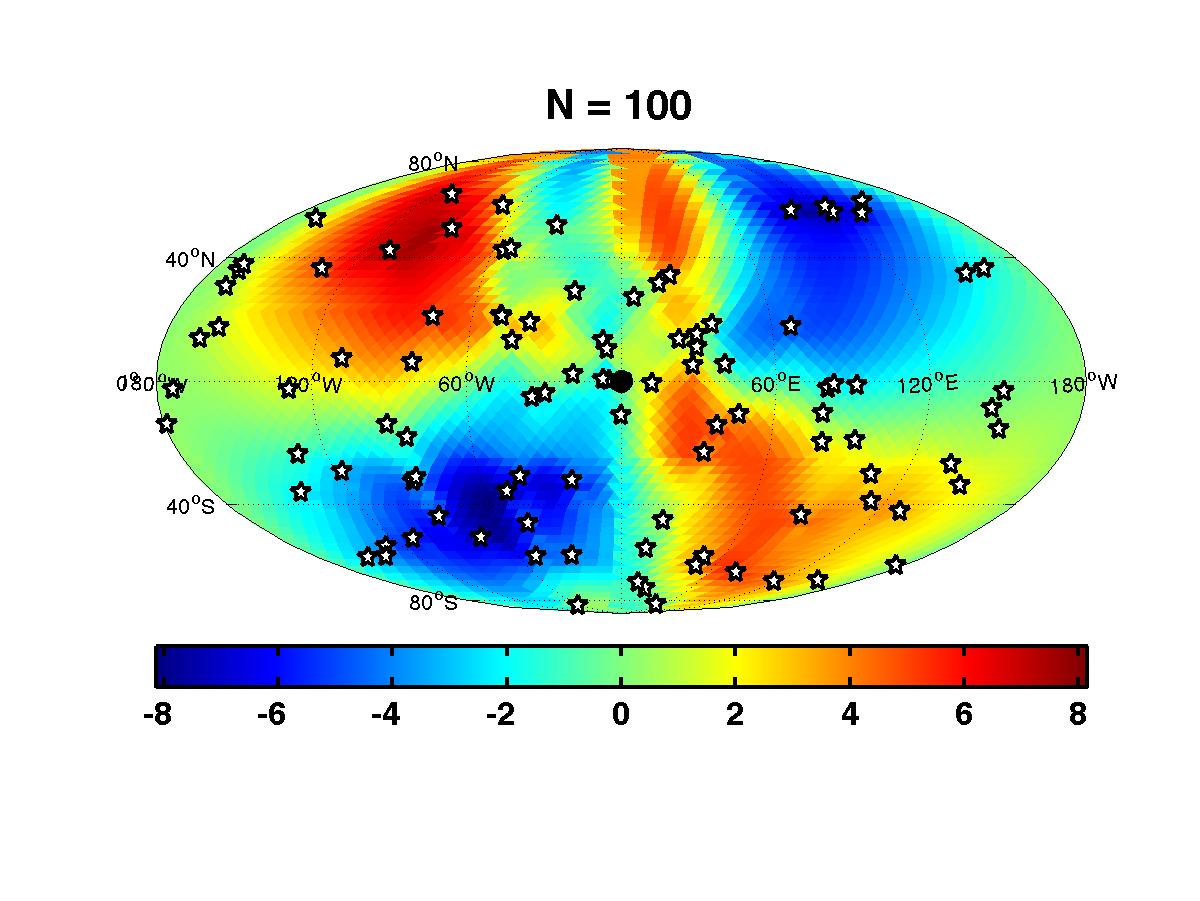}
\includegraphics[trim=3cm 4cm 3cm 2.5cm, clip=true, width=.24\textwidth]{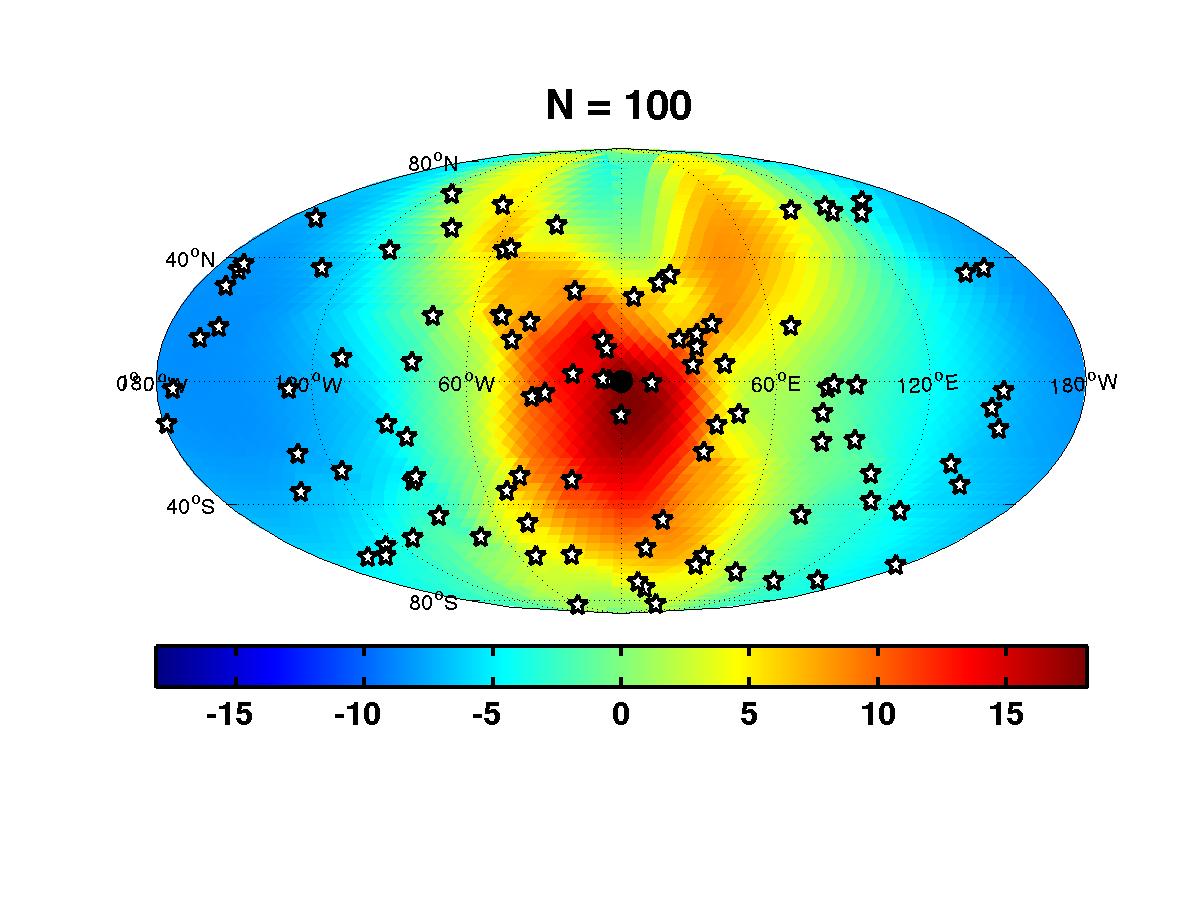}
\caption{Point spread functions for phase-coherent mapping,
for pulsar timing arrays consisting of 
$N=1$, 2, 5, 10, 25, 50, 100 pulsars.
The point source is located at the center of the maps,
$(\theta,\phi)=(90^\circ, 0^\circ)$, indicated by a black dot.
The pulsar locations (indicated by white stars) are
randomly placed on the sky.
Different rows correspond to different numbers of pulsars in the array.
Columns 1 and 2 correspond to the $+$ and $\times$ response of the 
pulsar timing array to a $+$-polarized point source;
columns 3 and 4 correspond to the $+$ and $\times$ response of the 
pulsar timing array to a $\times$-polarized point source.}
\label{f:PSFpulsar-phase}
\end{center}
\end{figure}

To illustrate the above procedure, we calculate point spread 
functions for phase-coherent mapping, for 
pulsar timing arrays consisting of $N=1$, 2, 5, 10, 25, 50,
100 pulsars.
Figure~\ref{f:PSFpulsar-phase} show plots of these point
spread functions.
The pulsars are randomly distributed over the sky (indicated by 
white stars), and the point source is located at the center of 
the maps (indicated by a black dot).
For simplicity, we assumed a single frequency bin, and used 
equal-noise weighting for 
calculating the point spread functions.
(In addition, there is no time dependence as the directions to 
the pulsars are fixed on the sky.)  
Different rows in the figure
correspond to different numbers of pulsars in the array.
Different columns correspond to different choices for $A$ and $A'$:
columns 1, 2 correspond to the $A=+,\times$ response of the 
pulsar timing array to an $A'=+$-polarized point source;
columns 3, 4 correspond to the $A=+,\times$ response of the 
pulsar timing array to an $A'=\times$-polarized point source.
Note that for $N=1$, the point spread functions are proportional to
either $R_I^+(\hat n)$ or $R_I^\times(\hat n)$ for that pulsar, 
producing maps similar to those shown in Figure~\ref{f:RpRc_pulsar}.
As $N$ increases the $++$ and $\times\times$ point spread functions
(columns 1 and 4) become tighter around the location of the point
source, which is at the center of the maps.
But since the $+$ and $\times$ polarizations are orthogonal, the 
$\times +$ and $+\times$ point spread functions (columns 2 and 3)
have values close to zero around the location of the point source.

\subsubsection{Singular value decomposition}
\label{s:svd_phase}

Just as we had to deconvolve the detector response in order to 
obtain the estimators $\hat{\cal P}$ for 
gravitational-wave power, 
we need to do the same for the estimators $\hat a$ for the
phase-coherent mapping approach.
Although we could use singular-value decomposition for the 
Fisher matrix $F$ given by (\ref{e:FAkAk'}), 
we will first {\em whiten} the data, which leads us 
directly to pseudo-inverse of the whitened response 
matrix $M$, (\ref{e:M,a-defns}).
This is the approach followed in 
\cite{Cornish-vanHaasteren:2014, Romano-et-al:2015}, 
and it leads to some interesting results regarding 
{\em sky-map basis vectors}, which we will describe in more detail 
in Section~\ref{s:basis_skies}.
An alternative approach involving the pseudo-inverse of
the unwhitened response matrix is given in 
\cite{Gair-et-al:2014} and Appendix~B of \cite{Romano-et-al:2015}.

To whiten the data, we start by finding the Cholesky
decomposition of the inverse noise covariance matrix
$N^{-1} = LL^\dagger$, where $L$ is a lower triangular
matrix.
The whitened data are then given by 
$\bar d = L^\dagger d$
(since this has unit covariance matrix),
and the whitened response matrix is given 
by $\bar M = L^\dagger M$.
In terms of these whitened quantities,
\be 
F=\bar M^\dagger \bar M\,,
\qquad
X= \bar M^\dagger \bar d\,,
\ee
implying
\be
\hat a = F^{-1} X = (\bar M^\dagger \bar M)^{-1} M^\dagger \bar d
\equiv \bar M^+ \bar d\,.
\ee
The last equality is a formal expression for the pseudo-inverse
$\bar M^+$ since $\bar M^\dagger \bar M$ is not necessarily
invertible.
But as shown in Section~\ref{s:svd_power} it is 
{\em always} possible to define the pseudo-inverse of
a matrix in terms of its singular-value decomposition.
Thus, given the singular-value decomposition:
\be
\bar M = U \Sigma V^\dagger\,,
\ee
we have
\be
\bar M^+ = V\Sigma^+ U^\dagger\,,
\ee
where $\Sigma^+$ is defined by the procedure described in Section~\ref{s:svd_power}.
Thus,
\be
\hat a = \bar M^+ \bar d = V\Sigma^+ U^\dagger \bar d\,.
\label{e:ahat_svd}
\ee
This is the expression we need to compute to calculate the 
maximum-likelihood
estimators $\hat a$ for the phase-coherent mapping approach.

\subsubsection{Basis skies}
\label{s:basis_skies}

The singular-value decomposition of $\bar M$ also 
has several nice geometrical properties.
For example, from (\ref{e:ahat_svd}), we see that 
the columns of $V$ corresponding to the non-zero
singular values of $\Sigma$ are {\em basis vectors} 
(which we will call {\em basis skies}) 
in terms of which $\hat a$ can be written as a 
linear combination.
Similarly, if write the whitened response to the 
gravitational-wave background as
\be
\bar M a = U\Sigma V^\dagger a\,,
\ee
then we see that the columns of $U$ corresponding 
to the non-zero singular values of $\Sigma$ can be 
interpreted as 
{\em range vectors} for the response.
To be more explicit, let $u_{(k)}$ and $v_{(k)}$ 
denote the $k$th columns of $U$ and $V$, and
let $r$ be the number of non-zero singular
values of $\Sigma$.
Then
\be
\begin{aligned}
\hat a 
&= \sum_{k=1}^r \sigma_k^{-1} (u_{(k)}\cdot \bar d)\, v_{(k)}\,,
\\
\bar M a 
&= \sum_{k=1}^r \sigma_k (v_{(k)}\cdot a)\, u_{(k)}\,,
\end{aligned}
\ee
where the dot product of two vectors $a$ and $b$ is
defined as $a\cdot b = a^\dagger b$.
If we further expand $\bar d = \bar M a + \bar n$ in
the first of these equations, then
\be
\hat a 
= \sum_{k=1}^r (v_{(k)}\cdot a) v_{(k)}
+\bar M^+ \bar n\,.
\ee
This last expression involves the projection of the 
true gravitational-wave sky $a$ onto the basis skies 
$v_{(k)}$ for only the non-zero singular 
values of $\Sigma$.

\begin{figure}[h!tbp]
\begin{center}
\includegraphics[trim=3cm 4.5cm 3cm 2.5cm, clip=true, width=.35\textwidth]{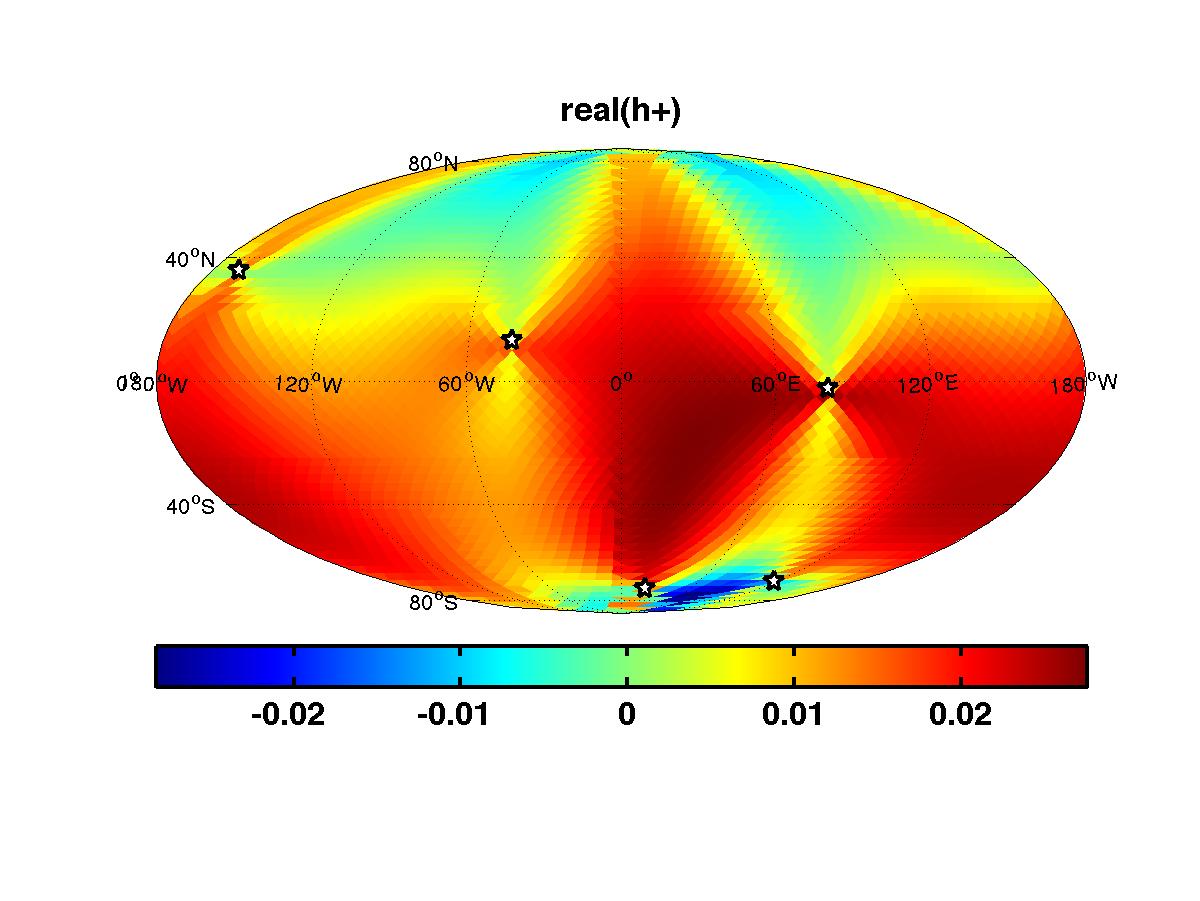}
\hspace{.05\textwidth}
\includegraphics[trim=3cm 4.5cm 3cm 2.5cm, clip=true, width=.35\textwidth]{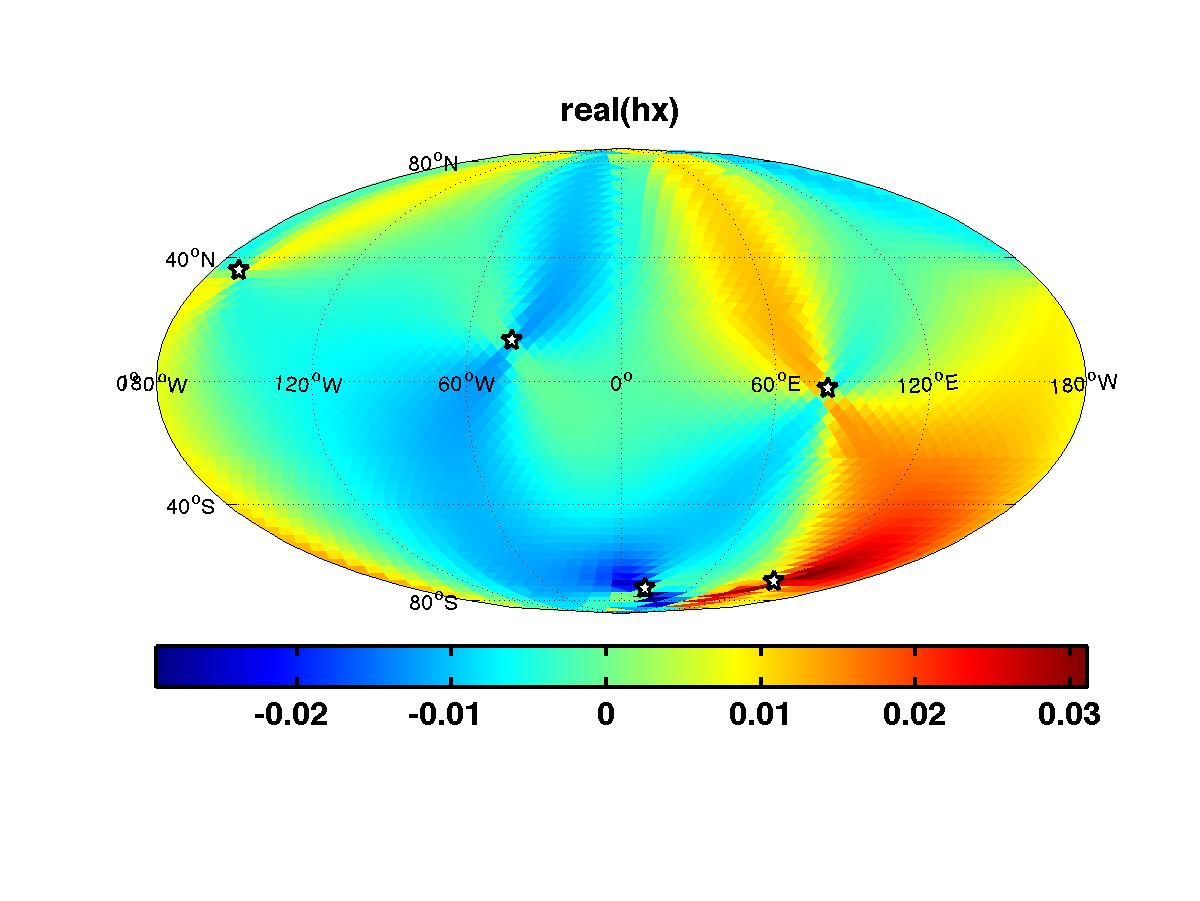}
\includegraphics[trim=3cm 4.5cm 3cm 2.5cm, clip=true, width=.35\textwidth]{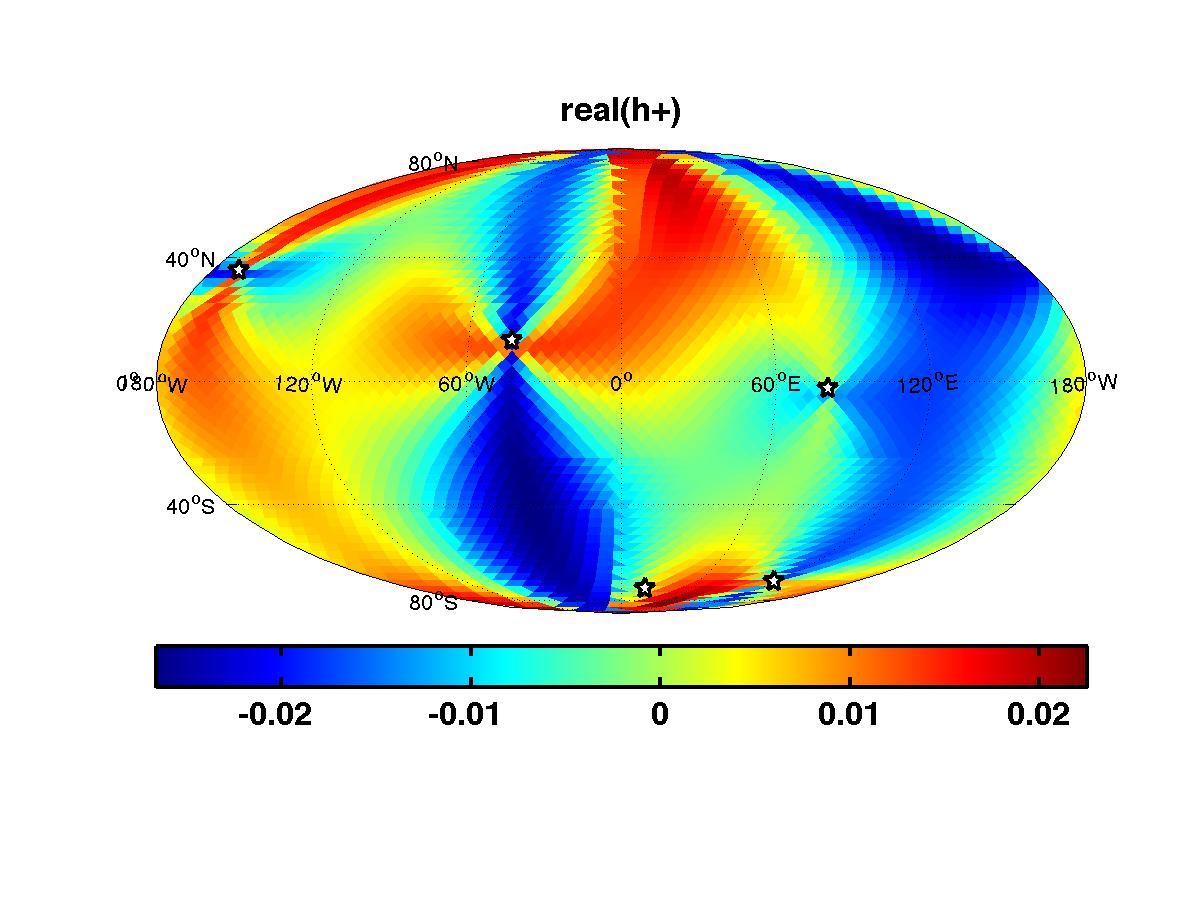}
\hspace{.05\textwidth}
\includegraphics[trim=3cm 4.5cm 3cm 2.5cm, clip=true, width=.35\textwidth]{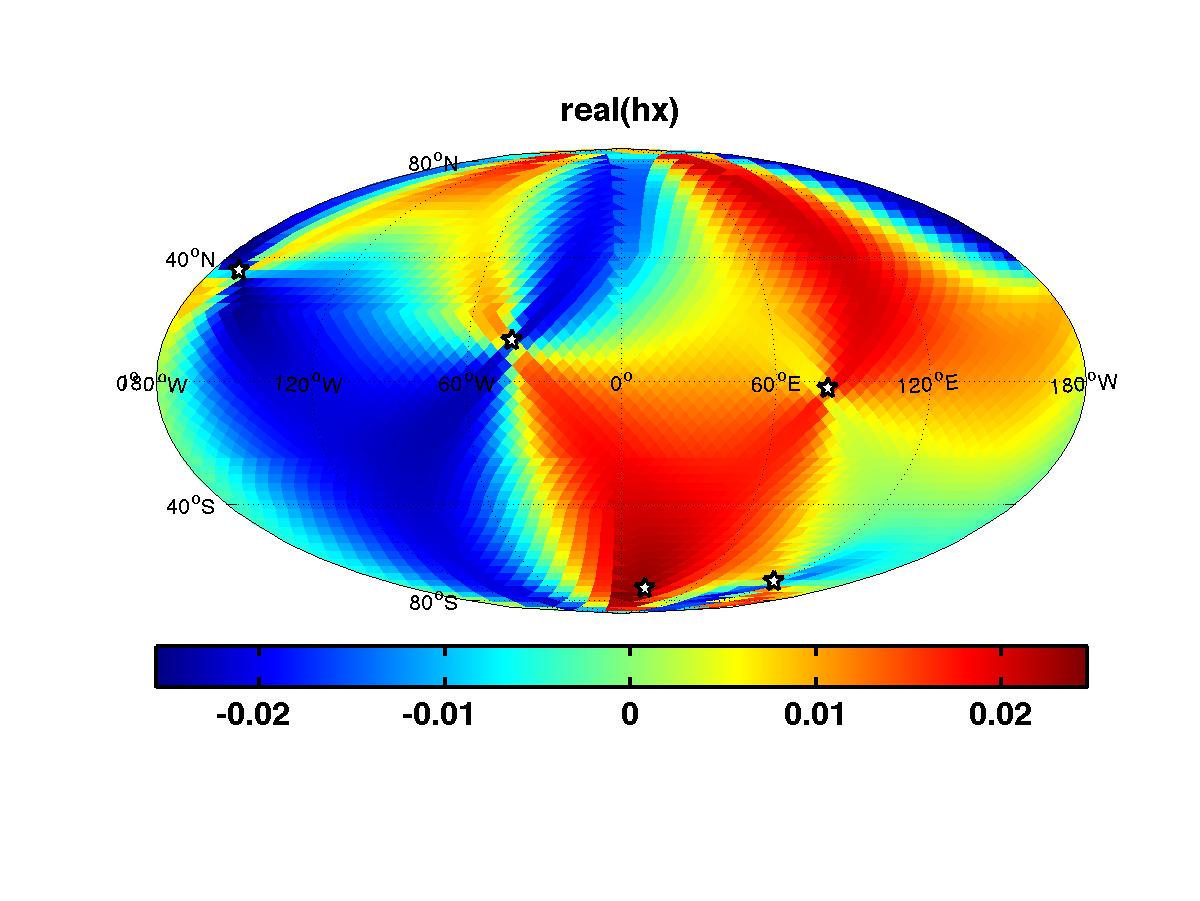}
\includegraphics[trim=3cm 4.5cm 3cm 2.5cm, clip=true, width=.35\textwidth]{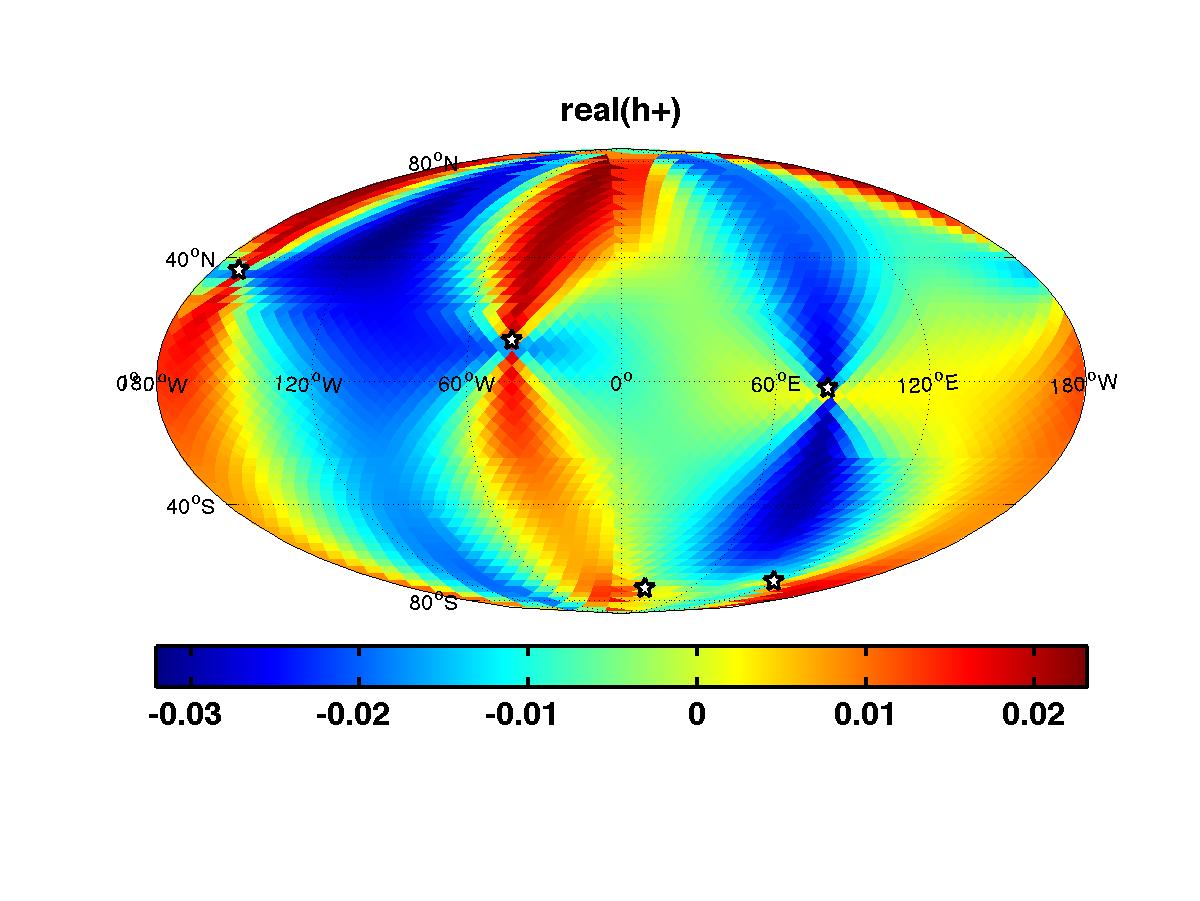}
\hspace{.05\textwidth}
\includegraphics[trim=3cm 4.5cm 3cm 2.5cm, clip=true, width=.35\textwidth]{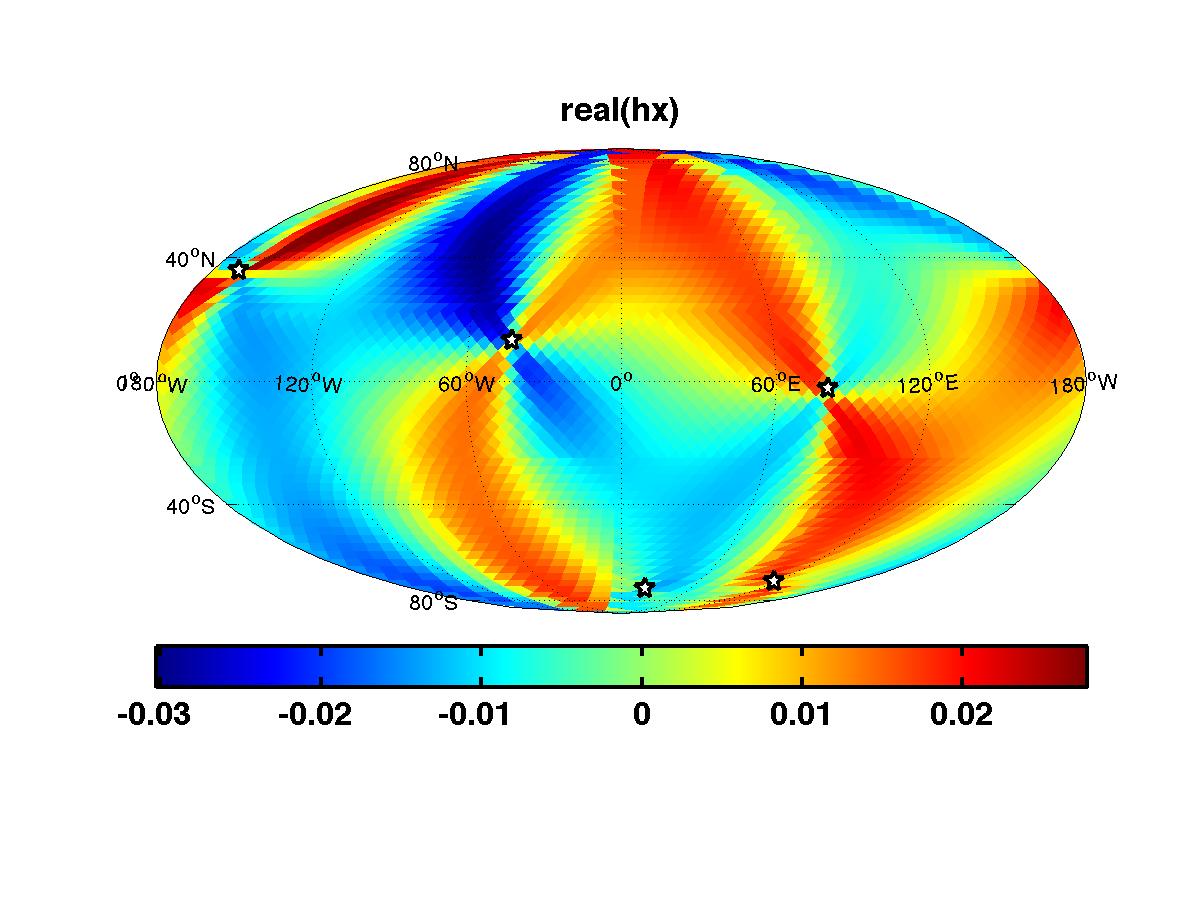}
\includegraphics[trim=3cm 4.5cm 3cm 2.5cm, clip=true, width=.35\textwidth]{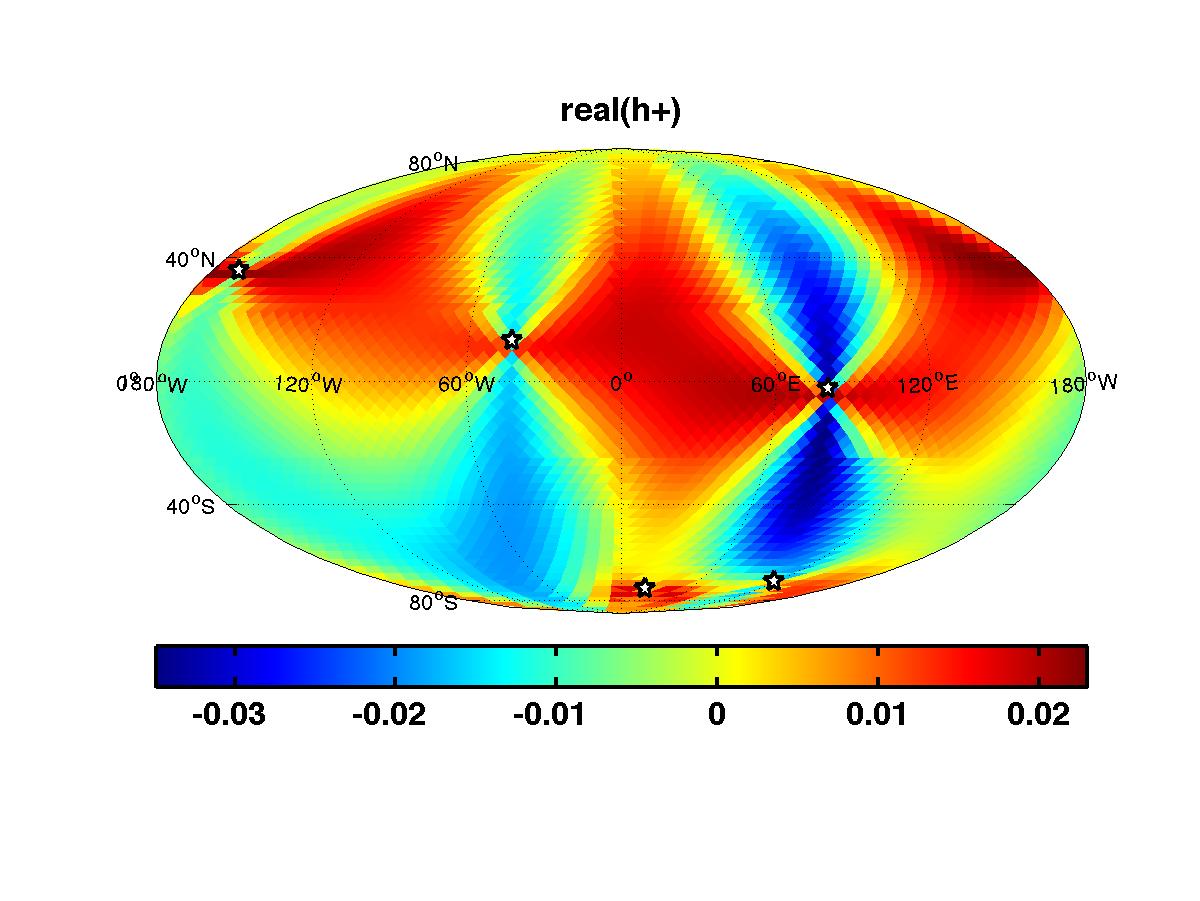}
\hspace{.05\textwidth}
\includegraphics[trim=3cm 4.5cm 3cm 2.5cm, clip=true, width=.35\textwidth]{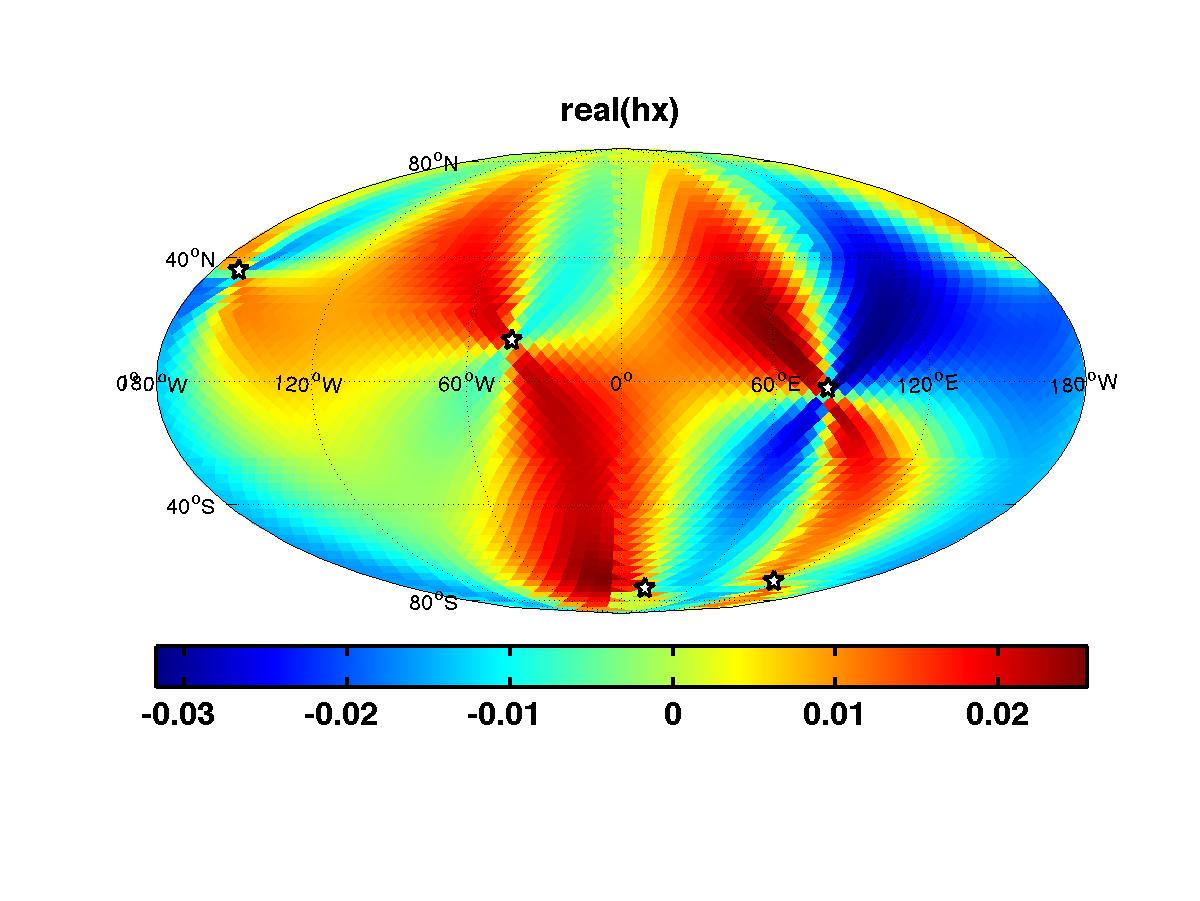}
\includegraphics[trim=3cm 4.5cm 3cm 2.5cm, clip=true, width=.35\textwidth]{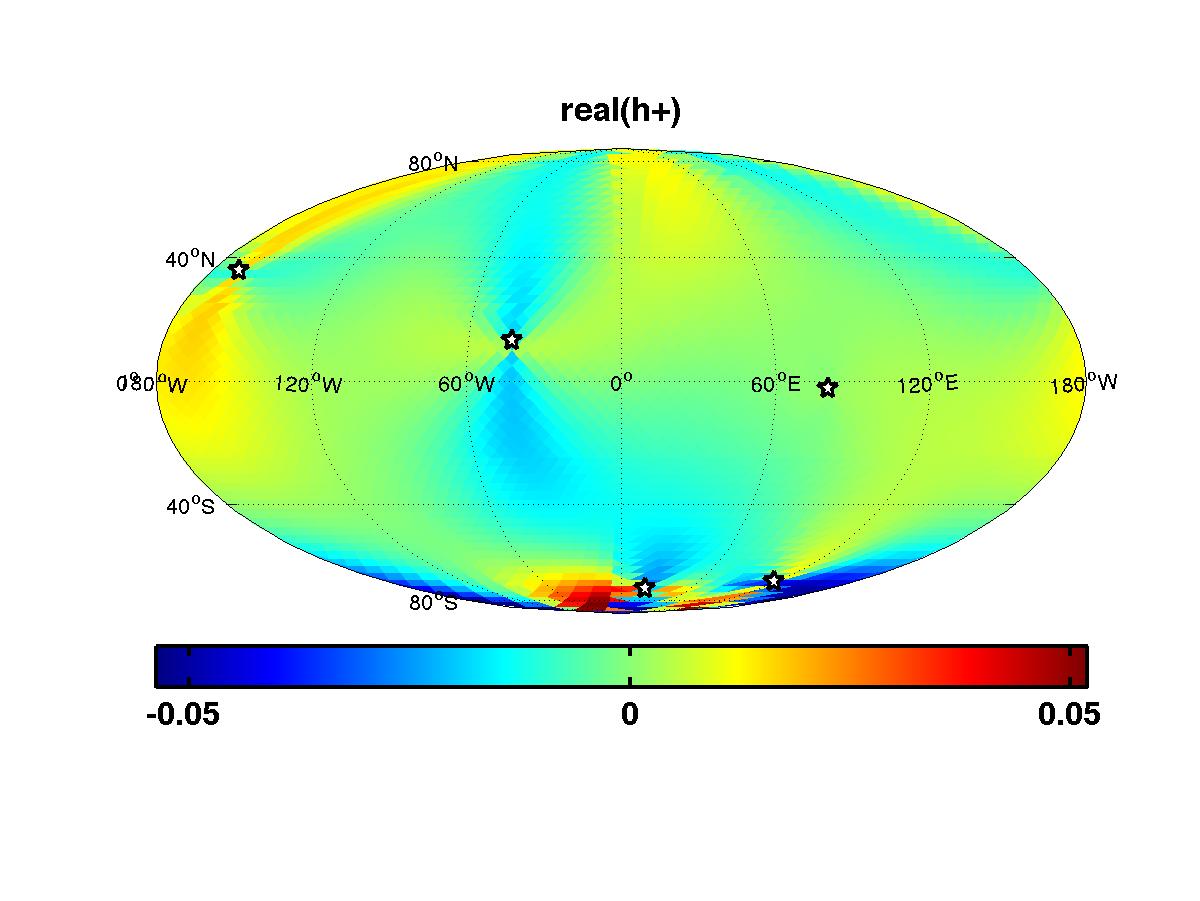}
\hspace{.05\textwidth}
\includegraphics[trim=3cm 4.5cm 3cm 2.5cm, clip=true, width=.35\textwidth]{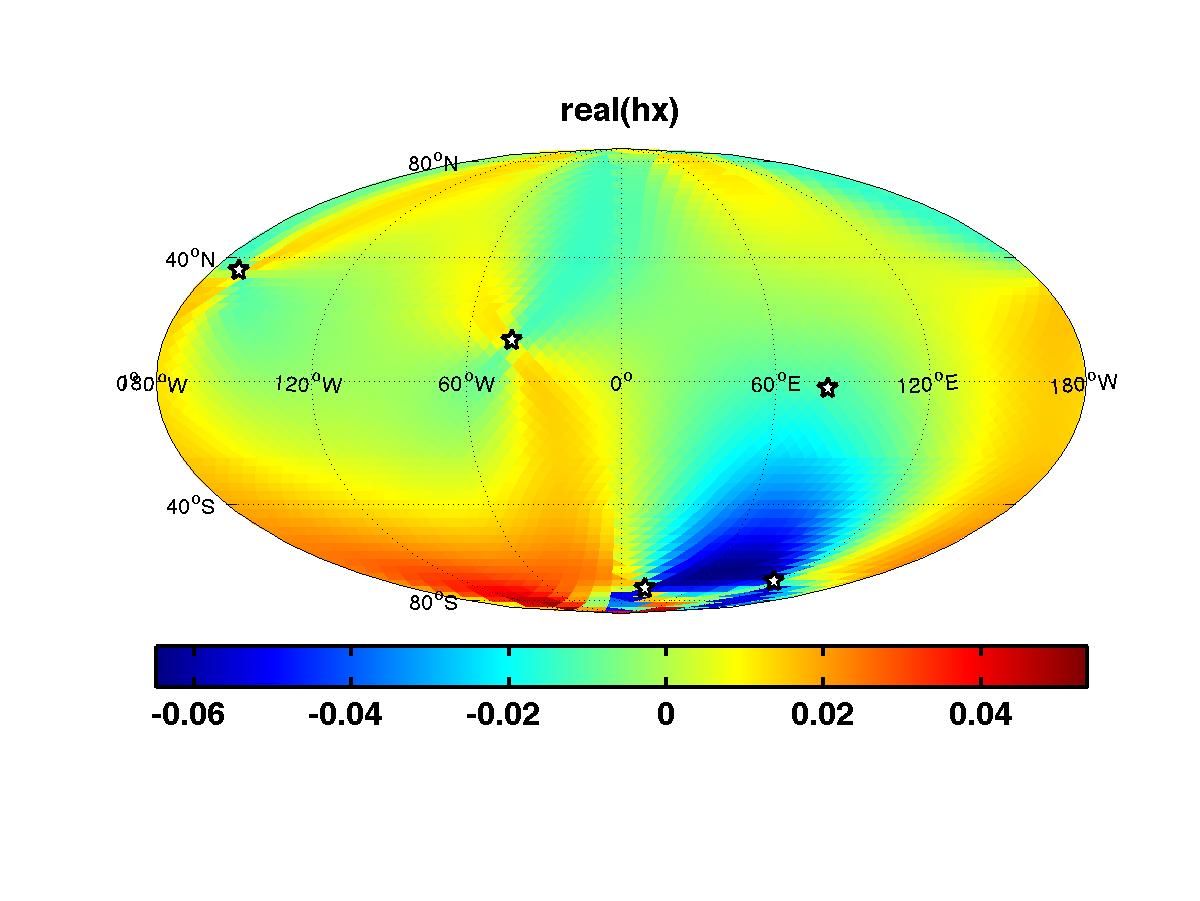}
\caption{The real parts of the $+$ and $\times$-polarization basis skies for 
pulsar timing array consisting of $N=5$ pulsars randomly distributed on the sky.
The imaginary components of the basis skies are identically zero.
The basis skies are shown in decreasing size of their singular values, from the 
top of the figure to the bottom.}
\label{f:basis_skies_PTA}
\end{center}
\end{figure}

In Figure~\ref{f:basis_skies_PTA}, we show plots of 
the real parts of the $+$ and $\times$-polarization 
basis skies
for a pulsar timing array consisting of $N=5$~pulsars
randomly distributed on the sky.
The imaginary components of the basis skies are 
identically zero, and hence are not shown in the figure.
The basis skies are shown in decreasing size of
their singular values, from top to bottom.
In general, if $N$ is the number of pulsars in the 
array, then the number of basis skies is $2N$
(the factor of 2 corresponding to the two polarizations,
$+$ and $\times$).
This means that one can extract at most $2N$ real pieces 
of information about the gravitational-wave background 
with an $N$-pulsar array.
This is typically fewer than the number of modes of 
the background that we would like to recover.

\subsubsection{Underdetermined reconstructions}

More generally, let's consider the case where the
total number of measured data points $n$ is less 
than the number of modes $m$ that we are trying to 
recover (so $n<m$),
or where there are certain modes of the 
gravitational-wave background (e.g., {\em null skies})
that our detector network is simply insensitive to.
Then, for both of these cases, the linear system 
of equations that we are trying to solve,
$\bar d = \bar M a$, is {\em underdetermined}---i.e.,
there exist multiple solutions for $a$, which 
differ from (\ref{e:ahat_svd}) by terms of the form
\be
a_{\rm null} = 
(\unit_{m\times m} - \bar M^+\bar M) a_{\rm arb}\,,
\ee
where $a_{\rm arb}$ is an {\em arbitrary}
gravitational-wave background.
(Note that $a_{\rm null}$ is an element of the 
{\em null space} of $\bar M$ as it maps to zero under 
the action of $\bar M$.)
Our solution for $\hat a$ given in (\ref{e:ahat_svd})
sets to zero those modes that we are insensitive to.
Our solution also sets to zero the variance of 
these modes.

In a Bayesian formulation of the problem, one needs
to specify prior probability distributions for the 
signal parameters, in addition to specifying the 
likelihood function (\ref{e:likelihood_a}).
For a mode of the background to which our detector 
network is insensitive, the marginalized posterior
for that mode will be the same as the prior, 
since the data are uniformative about this mode.
This is what one would expect for a mode that is 
unconstrained by the data, in contrast to setting 
the variance equal to zero
as we do with our maximum-likelihood reconstruction.
Basically, our maximum-likelihood reconstruction
does not attempt to say anything about the modes 
of the background for which we have no information.

\subsubsection{Pulsar timing arrays}
\label{s:phase-coherent-PTA}

The phase-coherent mapping approach was first developed in
the context of pulsar timing arrays~\cite{Cornish-vanHaasteren:2014, Gair-et-al:2014}.
In \cite{Cornish-vanHaasteren:2014}, the analysis was
done in terms of the standard polarization components
$a\equiv \{h_+(f,\hat n),h_\times(f,\hat n)\}$, similar to 
what we described above.
In \cite{Gair-et-al:2014}, the analysis was done 
in terms of the tensor spherical harmonic components
$a\equiv \{a_{(lm)}^G(f), a_{(lm)}^C(f)\}$.
Now recall from (\ref{e:RP_PTA}) that 
the Earth-term-only, Doppler-frequency response 
functions are given by 
\be
R^G_{(lm)}(f) = 2\pi {}^{(2)}\!N_l Y_{lm}(\hat p)\,,
\qquad
R^C_{(lm)}(f) = 0\,,
\label{e:RP_PTA_phase}
\ee
where $\hat p$ is the direction to an 
arbitrary pulsar.
Thus, the pulsar response to curl modes is 
identically zero.
This means that a pulsar timing array is 
blind to {\em half} of all possible 
modes of a gravitational-wave
background, regardless of how many pulsars 
there are in the array.
Note that this statement is not restricted to 
the tensor spherical harmonic analysis;
it is also true in terms of the standard $(+,\times)$
polarization components, since
$a^G_{(lm)}(f)$ and $a^C_{(lm)}(f)$ are 
linear combinations of $h_+(f,\hat n)$ and 
$h_\times(f,\hat n)$, see (\ref{e:aG+iaC}).
It is just that the insensivity of a pulsar 
timing array to half of the gravitational-wave
modes is {\em manifest} in the gradient and curl spherical 
harmonic basis for which (\ref{e:RP_PTA_phase}) 
is valid.

\begin{figure}[h!tbp]
\begin{center}
\subfigure[\ Total map (grad+curl)]
{\includegraphics[trim=3cm 6.5cm 3cm 3.5cm, clip=true, angle=0, width=0.32\textwidth]{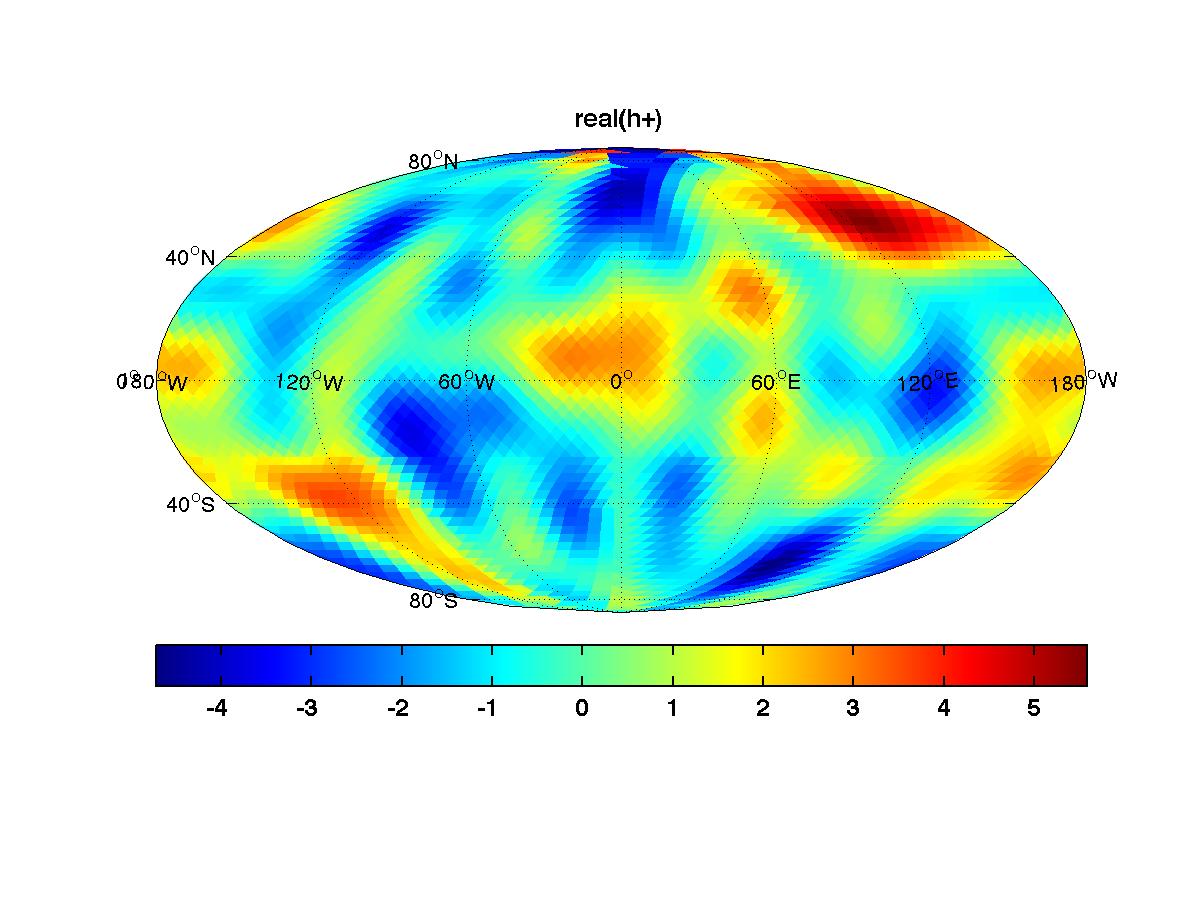}}
\subfigure[\ Gradient component]
{\includegraphics[trim=3cm 6.5cm 3cm 3.5cm, clip=true, angle=0, width=0.32\textwidth]{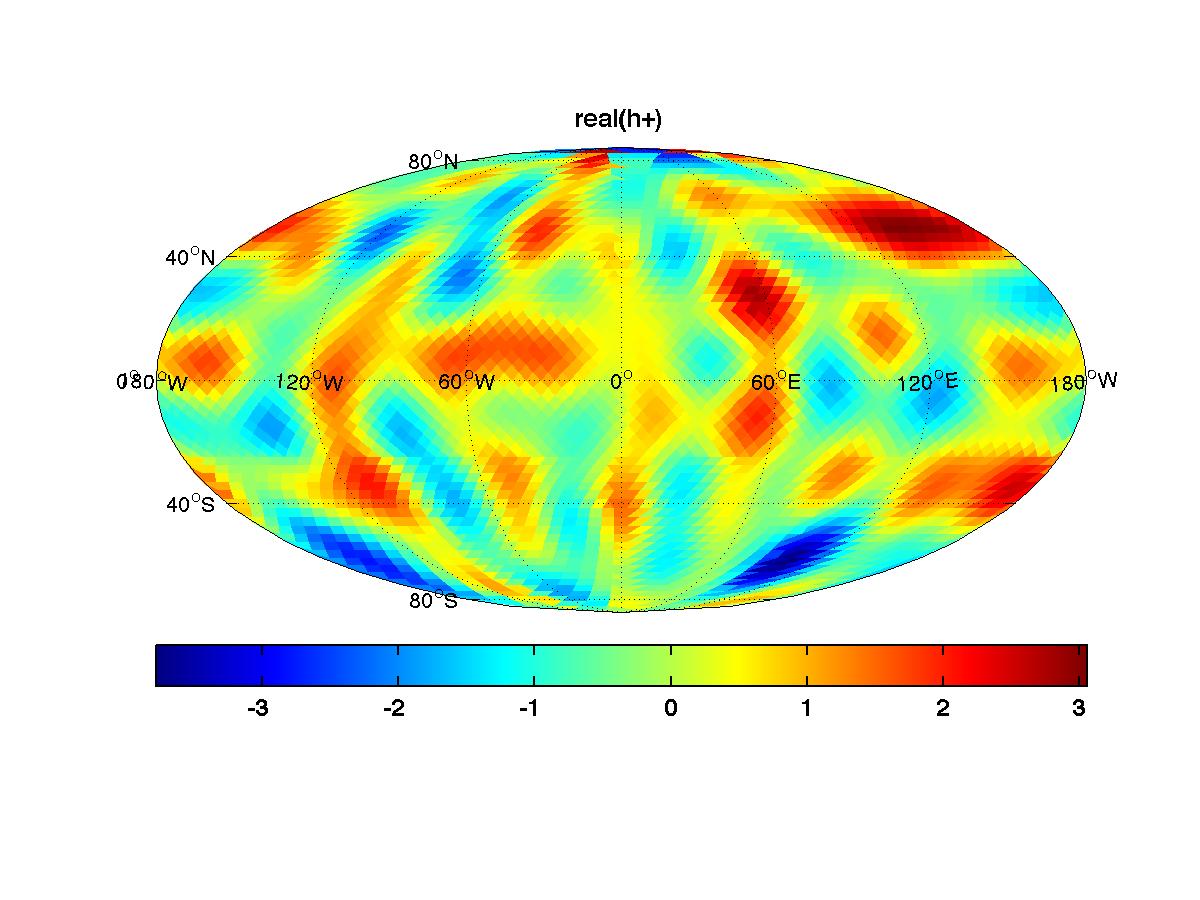}}
\subfigure[\ Curl component]
{\includegraphics[trim=3cm 6.5cm 3cm 3.5cm, clip=true, angle=0, width=0.32\textwidth]{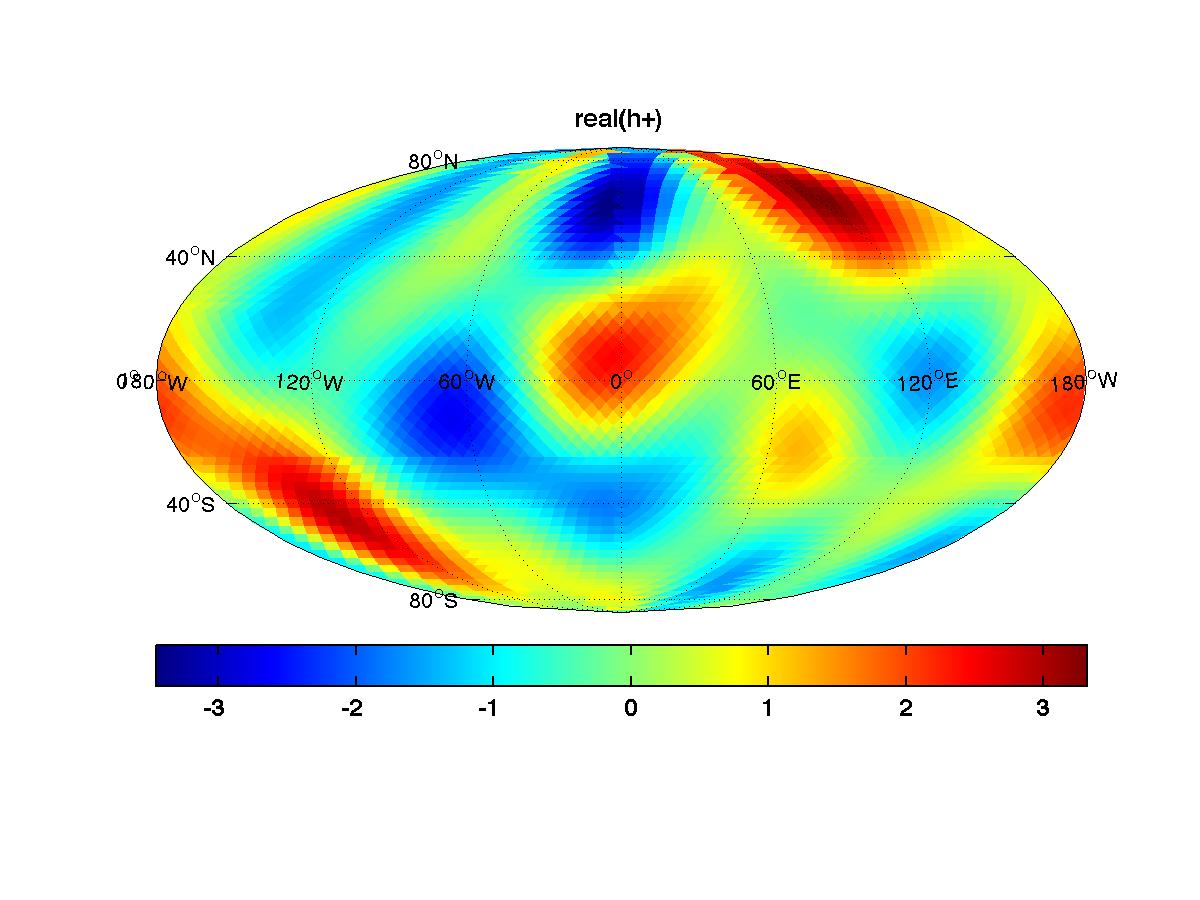}}
\subfigure[\ Gradient residual map]
{\includegraphics[trim=3cm 6.5cm 3cm 3.5cm, clip=true, angle=0, width=0.32\textwidth]{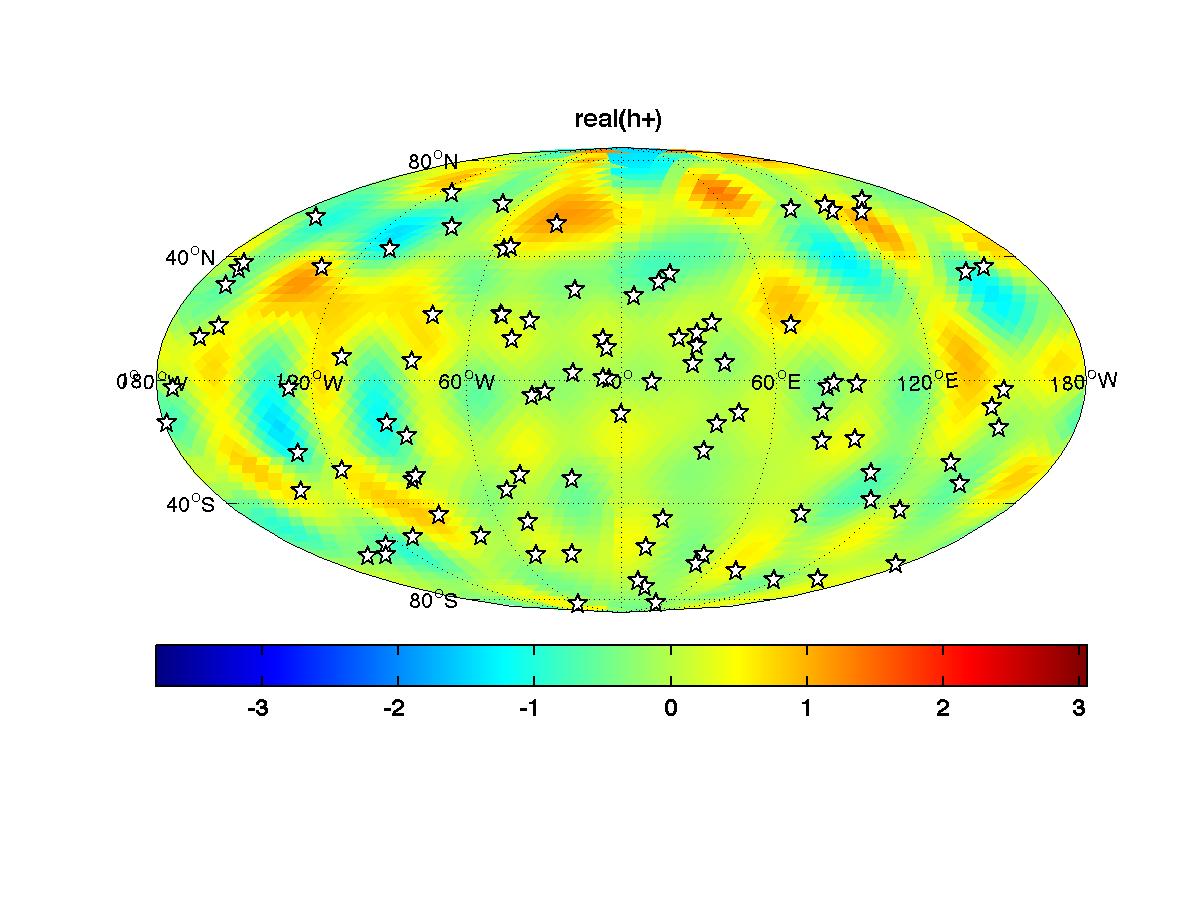}}
\subfigure[\ Max-likelihood recovered map]
{\includegraphics[trim=3cm 6.5cm 3cm 3.5cm, clip=true, angle=0, width=0.32\textwidth]{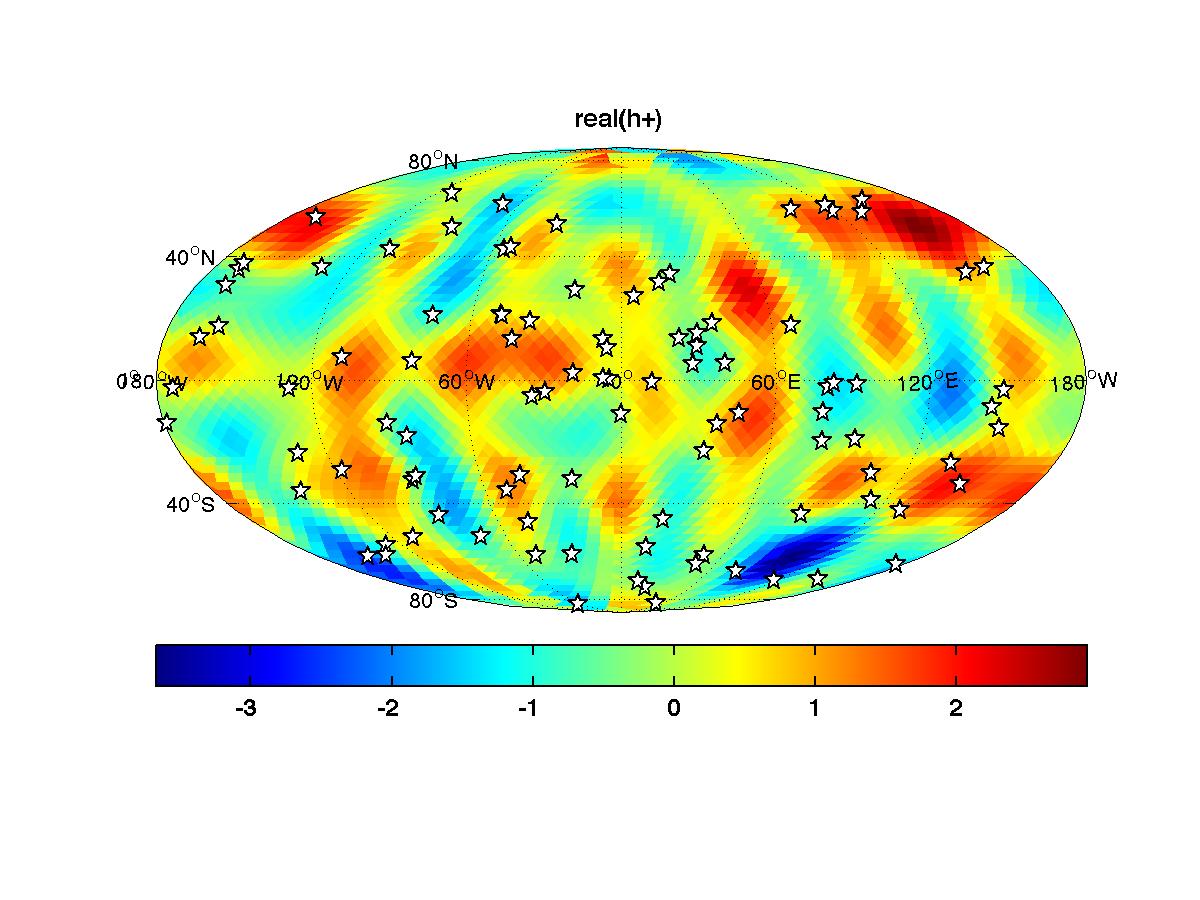}}
\subfigure[\ Total residual map]
{\includegraphics[trim=3cm 6.5cm 3cm 3.5cm, clip=true, angle=0, width=0.32\textwidth]{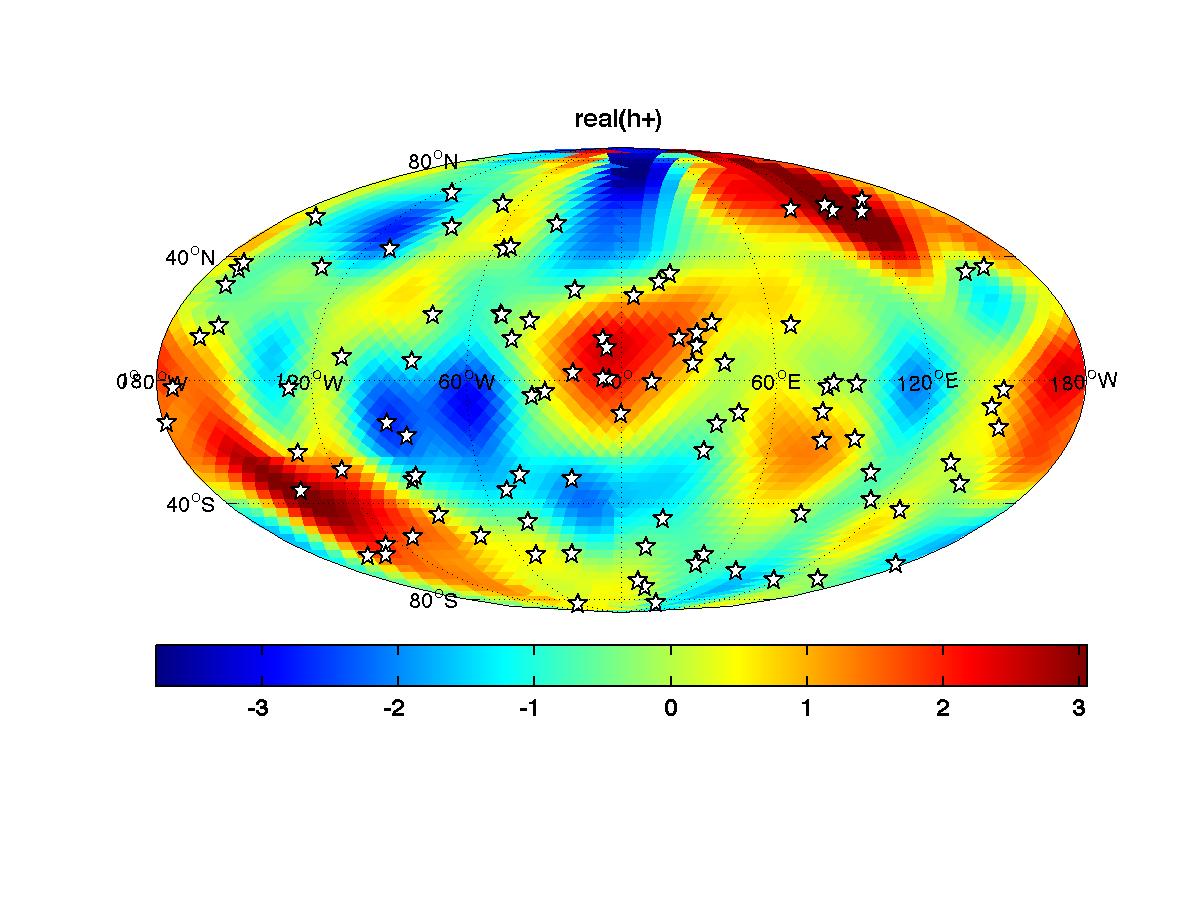}}
\caption{Mollweide projections of the real parts of
$h_+(\hat n)$ for the different components of the 
simulated background (panels a--c),
the maximum-likelihood recovered map for a 
pulsar timing array consisting of $N=100$ pulsars (panel e),
and the corresponding residual maps for the grad-component 
(panel d) and the total simulated background (panel f).
Sky maps of the imaginary part of $h_+(\hat n)$ and the 
real and imaginary parts of $h_\times(\hat n)$ are similar,
and hence are not shown in this figure.
Note that the maximum-likelihood recovered map 
most-closely resembles the
gradient component of the simulated background,
since a pulsar timing array is insensitive to 
the curl modes of a gravitational-wave background.
Image reproduced with permission from \cite{Gair-et-al:2014},
copyright by APS.}
\label{f:demo_comparisonskymapsPTA}
\end{center}
\end{figure}

To explicitly demonstrate that a pulsar timing 
array is insensitive to the curl-component of a 
gravitational-wave background, Gair et~al.~\cite{Gair-et-al:2014} 
constructed maximum-likelihood estimates of a simulated 
background containing both gradient and curl modes.
The total simulated background and its gradient
and curl components are shown in the top row (panels a--c) of 
Figure~\ref{f:demo_comparisonskymapsPTA}.
(Note that this is for a noiseless simulation so as 
not to confuse the lack of reconstructing
the curl component with the presence of detector noise.)
Panel e shows the maximum-likelihood recovered
map for a pulsar timing array consisting of
$N=100$ pulsars randomly distributed on the 
sky.
Panels d and f are residual maps obtained
by subtracting the maximum-likelihood recovered map 
from the gradient component and the total simulated
background, respectively.
Note that the maximum-likelihood recovered map 
resembles the gradient component of the background,
consistent with the fact that a pulsar timing
array is insenstive to the curl component of a 
gravitational-wave background.
The residual map for the gradient component (panel d)
is much cleaner than the residual map for the 
total simulated background (panel f), which 
has angular structure that closely resembles the
curl component of the background.
%

\subsubsection{Ground-based interferometers}
\label{s:phase-coherent-IFO}

The phase-coherent mapping approach can also be
applied to data taken by a network of 
ground-based interferometers~\cite{Romano-et-al:2015}.
Again the analysis can be performed in terms of either 
the standard $+$, $\times$ polarization components or 
the gradient and curl spherical harmonic components.
Recall from (\ref{e:RP_IFO}) that
\be
R^G_{(lm)}(f) = 
\delta_{l2}\frac{4\pi}{5}\sqrt{\frac{1}{3}}
\left[Y_{2m}(\hat u) - Y_{2m}(\hat v)\right]\,,
\qquad
R^C_{(lm)}(f) = 0\,,
\label{e:RP_IFO_phase}
\ee
for a ground-based interferometer in the 
small-antenna limit,
with its vertex at the origin, and with unit vectors 
$\hat u$, $\hat v$ 
pointing in the direction of the interferometer arms.
At first, one might think that these expressions
imply that a network of ground-based interferometers 
is also blind to the curl component of a 
gravitational-wave background.
But (\ref{e:RP_IFO_phase}) are valid only for 
interferometers with their vertices {\em at the origin} 
of coordinates.
Since a translation mixes gradient and curl components,
the response functions for an interferometer 
displaced from the origin by $\hat x_0$ are given 
by~\cite{Romano-et-al:2015}:
\be
\begin{aligned}
R^G_{(lm)}(f)
=&\sum_{m'=-2}^2 
\sum_{L=l-2}^{l+2}
\sum_{M=-L}^L 
F_{m'}(\hat u,\hat v)
4\pi (-i)^L j_L(\alpha)Y^*_{LM}(\hat x_0)
\frac{(-1)^{m'}}{2}
\left[(-1)^l+(-1)^L\right]
\\
&\times\sqrt{\frac{(2\cdot 2+1)(2l+1)(2L+1)}{4\pi}}
\left(\begin{array}{ccc}
2 & l & L\\
-m' & m & M 
\end{array}\right)
\left(\begin{array}{ccc}
2 & l & L\\
2 & -2 & 0 
\end{array}\right)\,,
\\
R^C_{(lm)}(f)
=&\sum_{m'=-2}^2 
\sum_{L=l-2}^{l+2}
\sum_{M=-L}^L 
F_{m'}(\hat u,\hat v)
4\pi (-i)^L j_L(\alpha)Y^*_{LM}(\hat x_0)
\frac{(-1)^{m'}}{2i}
\left[(-1)^l-(-1)^L\right]
\\
&\times\sqrt{\frac{(2\cdot 2+1)(2l+1)(2L+1)}{4\pi}}
\left(\begin{array}{ccc}
2 & l & L\\
-m' & m & M 
\end{array}\right)
\left(\begin{array}{ccc}
2 & l & L\\
2 & -2 & 0 
\end{array}\right)\,,
\label{e:RGRC_general}
\end{aligned}
\ee
where $\alpha \equiv 2\pi f|\vec x_0|/c$ 
and $j_L(\alpha)$ are spherical Bessel functions of order $L$.
Here
\be
F_m(\hat u, \hat v) \equiv
\frac{4\pi}{5}\sqrt{\frac{1}{3}}
\left[Y_{2m}(\hat u) - Y_{2m}(\hat v)\right]\,,
\ee
is shorthand for the particular combination 
of spherical harmonics that enter the expression for
$R^G_{(lm)}(f)$ in (\ref{e:RP_IFO_phase}).
The two expressions in parentheses $(\ )$ for each response
function are Wigner 3-$j$ symbols
(see for example~\cite{Wigner:1959, Messiah:1962}).
Note that the curl response is now non-zero, 
and both response functions depend on frequency 
via the quantity $\alpha$, which equals 
$2\pi$ times the number of radiation wavelengths 
between the origin and the vertex of the interferometer.
These expressions are valid in an arbitrary 
translated and rotated coordinate system, provided 
the angles for $\hat u$, $\hat v$, and $\hat x_0$ are 
calculated in the rotated frame.

\begin{figure}[hbtp!]
\begin{center}
\subfigure[Injected grad-only map]{\includegraphics[trim=3cm 4cm 3cm 2.5cm, clip=true, angle=0, width=0.49\textwidth]{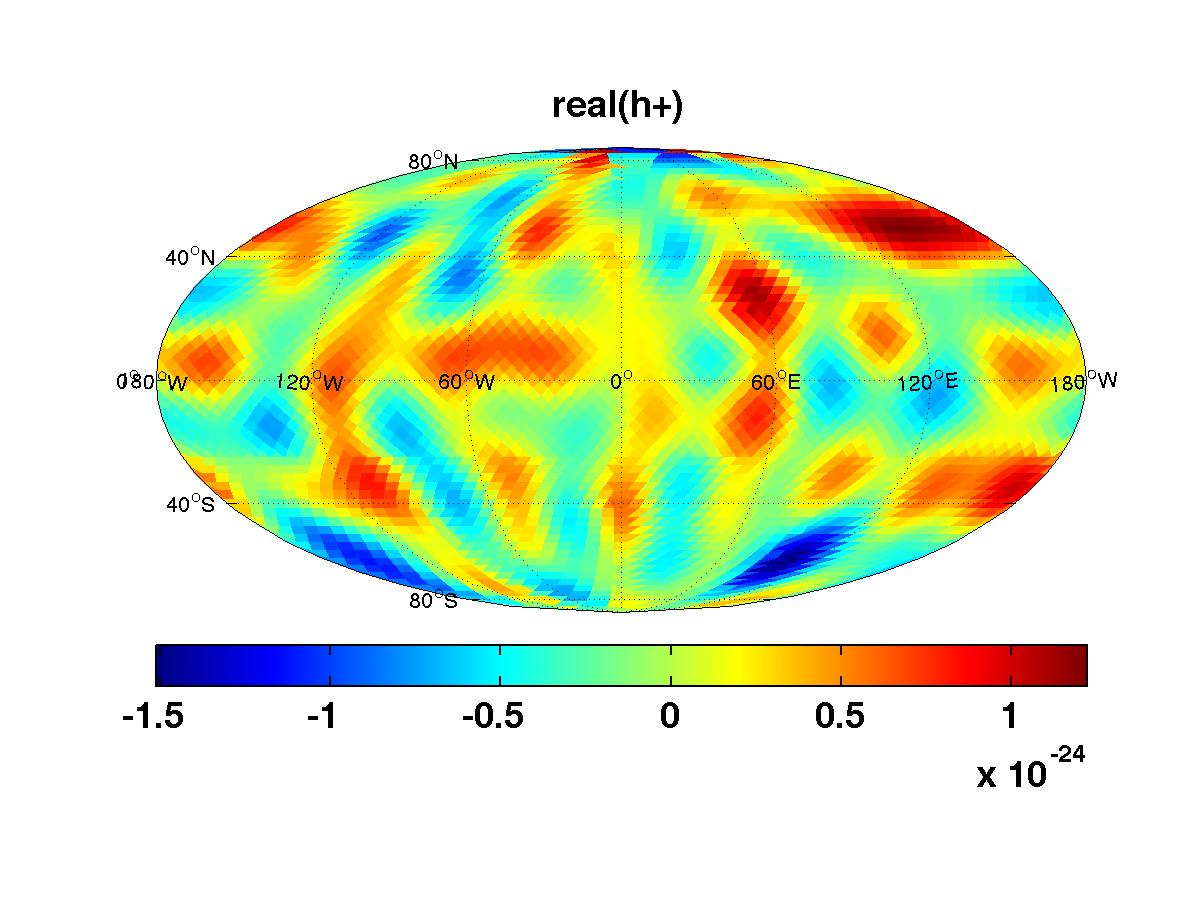}}
\subfigure[Injected curl-only map]{\includegraphics[trim=3cm 4cm 3cm 2.5cm, clip=true, angle=0, width=0.49\textwidth]{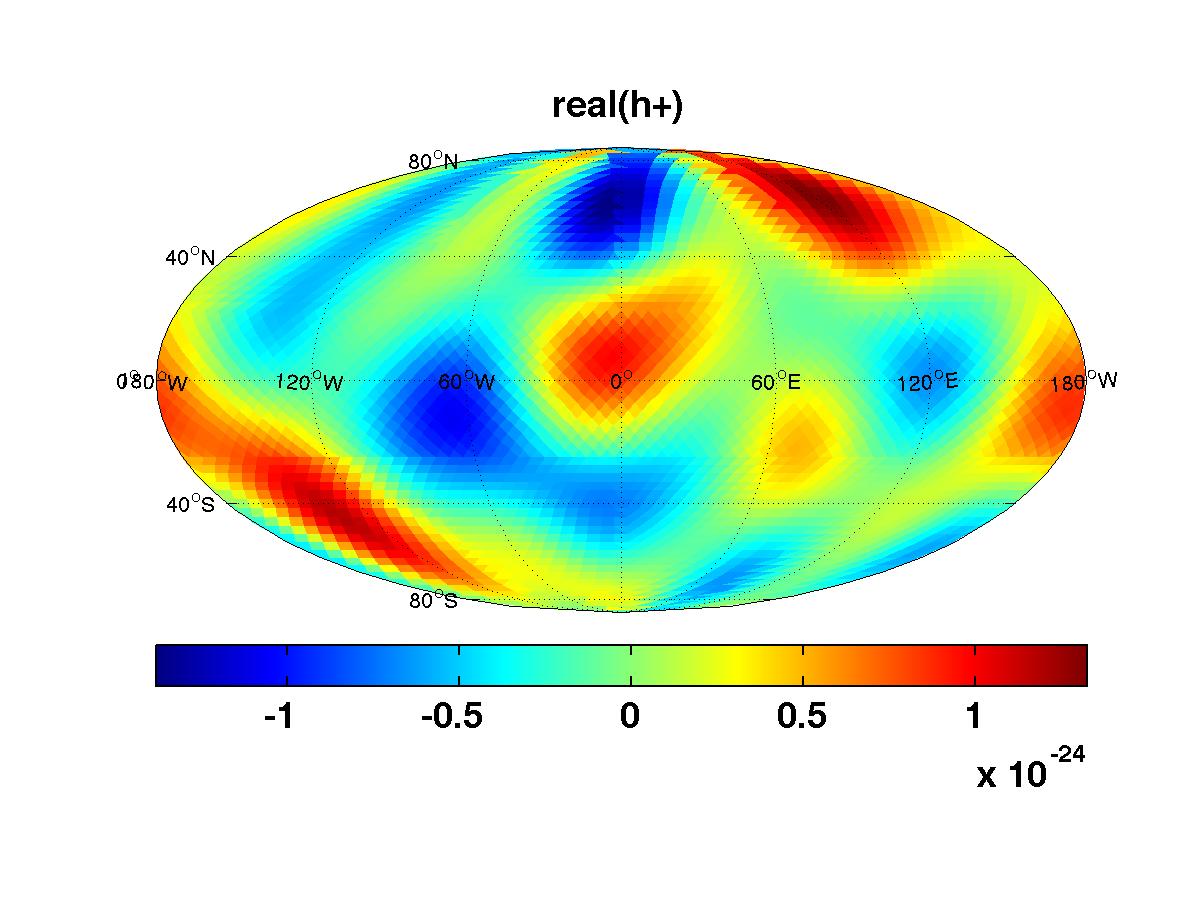}}
\subfigure[Max-likelihood recovered map]{\includegraphics[trim=3cm 4cm 3cm 2.5cm, clip=true, angle=0, width=0.49\textwidth]{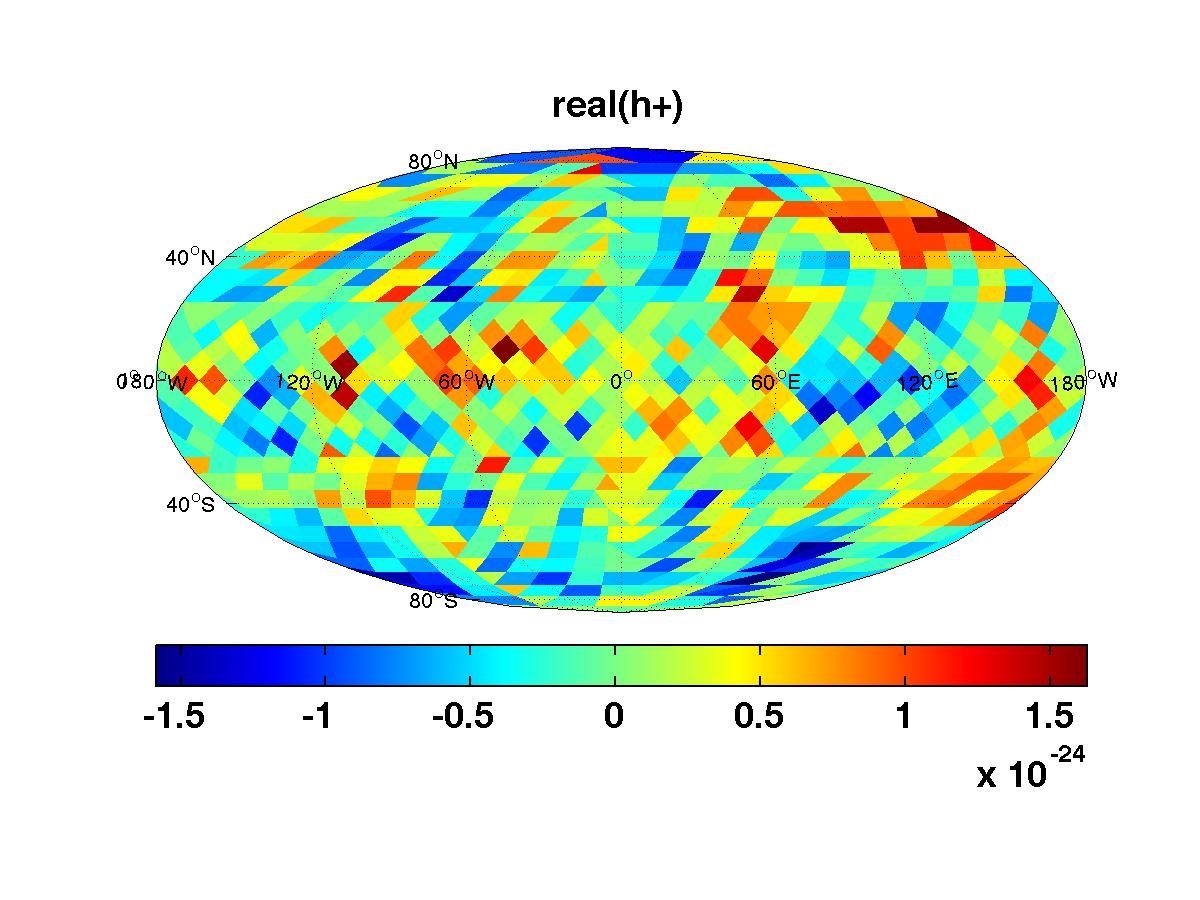}}
\subfigure[Max-likelihood recovered map]{\includegraphics[trim=3cm 4cm 3cm 2.5cm, clip=true, angle=0, width=0.49\textwidth]{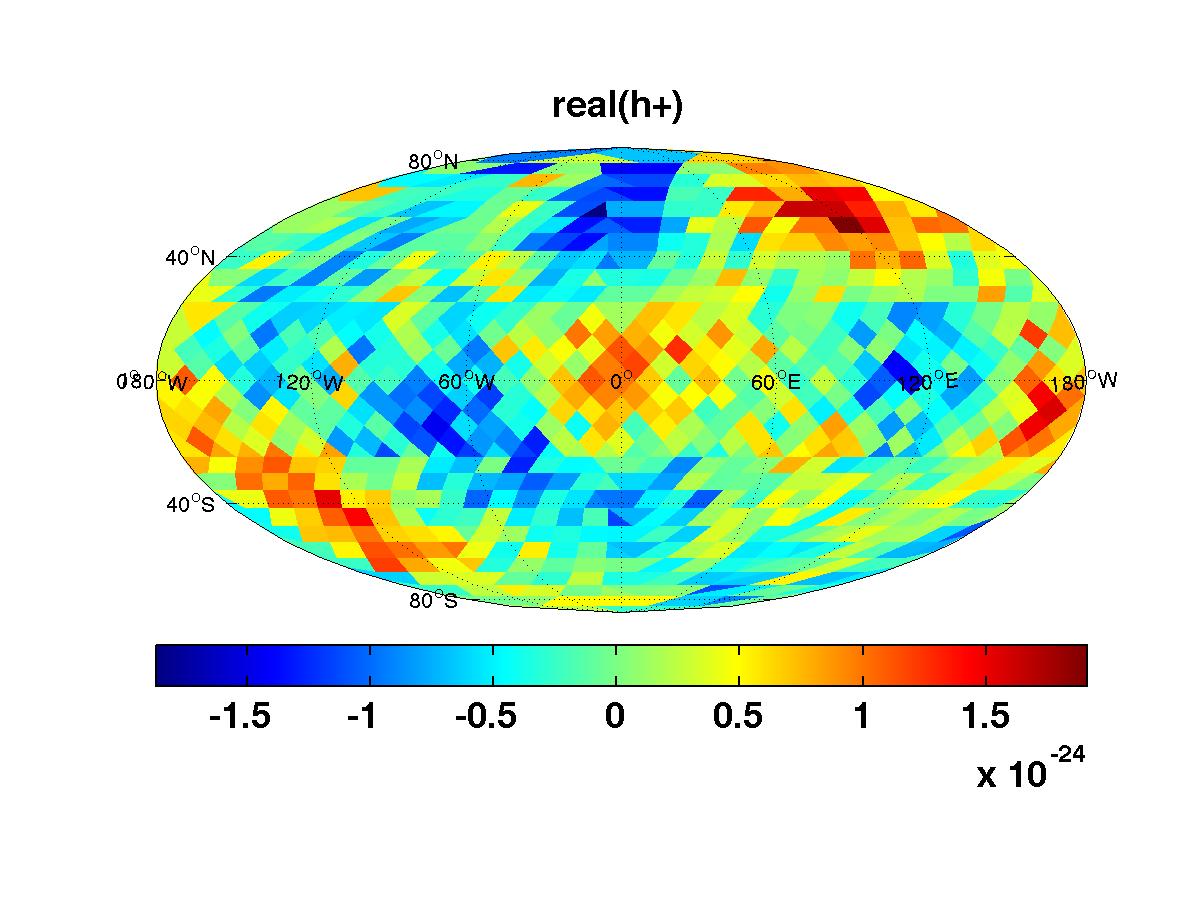}}
\caption{Mollweide projections of the real parts of $h_+(\hat n)$
for grad-only and curl-only anisotropic backgrounds injected
into noise and analysed using a 3-detector network 
of ground-based laser interferometers (Hanford-Livingston-Virgo).
The injected maps are shown in the top row;
the maximum-likelihood recovered maps are shown in the second row.
Sky maps of the imaginary part of $h_+(\hat n)$ and the real and 
imaginary parts of $h_\times(\hat n)$ are similar for both the
injections and the recovered maps, and hence are not shown in the figure.
Note that a network of ground-based interferometers is capable of
recovering both the gradient and curl components of a gravitational-wave
background, in contrast to a pulsar timing array 
(compare with Figure~\ref{f:demo_comparisonskymapsPTA}).
Image reproduced with permission from \cite{Romano-et-al:2015},
copyright by APS.}
\label{f:demo_comparisonskymapsIFO}
\end{center}
\end{figure}

Thus, the spatial separation of a network of 
ground-based interferometers, or of 
a single interferometer at different
times during its daily rotational and yearly orbital motion
around the Sun (Section~\ref{s:rotational-orbital}),
allows for recovery of both the gradient 
{\em and} curl components of a gravitational-wave background.
This is in contrast to a pulsar timing array, which 
is insensitive to the curl component, because one vertex 
of all the pulsar baselines are `pinned' to the solar system barycenter.
To illustrate this difference, 
we show in Figure~\ref{f:demo_comparisonskymapsIFO},
maximum-likelihood recovered sky maps for simulated grad-only
and curl-only anistropic backgrounds injected into noise for 
a 3-detector network of ground-based interferometers 
(Hanford-Livingston-Virgo).
The grad-only and curl-only backgrounds are the same as those
used for the simulated maps in
Figure~\ref{f:demo_comparisonskymapsPTA}.
In contrast to the recovered maps shown in that figure 
for the pulsar timing array, the maximum-likelihood maps 
(bottom row) for the network of ground-based interferometers 
reproduce the general angular structure of both the grad-only
{\em and} curl-only injected maps (shown in the top row).
(The noise for these injections degrades the 
recovery compared to the noiseless
injections in Figure~\ref{f:demo_comparisonskymapsPTA}.)
See \cite{Romano-et-al:2015} for more details and related
simulations.

\section{Searches for other types of backgrounds / signals}
\label{s:extensions}

\begin{quotation}
No idea is so outlandish that it should not be considered 
with a searching but at the same time a steady eye.
{\em Winston Churchill}
\end{quotation}

\noindent
Since stochastic gravitational-wave backgrounds come in 
many different ``flavors", one needs additional search 
methods that go beyond the standard ``vanilla" 
cross-correlation search for a Gaussian-stationary, 
unpolarized, isotropic signal 
(Sections~\ref{s:corr} and \ref{s:geom}) to extract 
the relevant information from the more exotic backgrounds.
In Section~\ref{s:anisotropic}, we discussed how to 
search for {\em anisotropic} signals, which are stronger
coming from certain directions on the sky than from others. 
In this section, we discuss search methods for
non-Gaussian signals (Section~\ref{s:nongaussian}),
circularly polarized backgrounds (Section~\ref{s:circular}),
and additional polarization modes predicted by alternative
(non-general-relativity)
metric theories of gravity (Sections~\ref{s:altpol}, 
\ref{s:altpolIFO}, \ref{s:altpolPTA}).
In Section~\ref{s:others}, we also briefly mention 
searches for other types of gravitational-wave signals, 
which are not really stochastic backgrounds, but
nonetheless can be searched for using the basic idea of
cross-correlation, which we developed in Section~\ref{s:corr}.
The majority of the search methods that we will 
describe here have been implemented ``across the band"---i.e.,
for ground-based interferometers, space-based 
interferometers, and pulsar timing arrays.
For these methods, we will highlight any significant 
differences in the implementations for the different 
detectors, if there are any.

Of course, we do not have enough time or space in this 
section to do justice for all of these methods.  
As such, readers are strongly encouraged to read the 
original papers for more details.
For non-Gaussian backgrounds,
see~\cite{Drasco-Flanagan:2003, Seto:2009, Thrane:2013, 
Martellini-Regimbau:2014, Cornish-Romano:2015};
for circular polarization, 
see~\cite{Seto-Taruya:2007, Seto-Taruya:2008, Kato-Soda:2016};
for polarization modes in alternative theories of gravity, 
see~\cite{Lee-et-al:2008, Nishizawa-et-al:2009, 
Chamberlin-Siemens:2012, Gair-et-al:2015};
and for the othe types of signals,
see~\cite{Thrane-et-al:2011, Messenger-et-al:2015}.

\subsection{non-Gaussian backgrounds}
\label{s:nongaussian}

In Section~\ref{s:when-stochastic}, we asked the 
question ``when is a gravitational-wave signal 
stochastic" to highlight the practical distinction 
between searches for 
deterministic and stochastic signals.
From an operational perspective, a signal is stochastic if 
it is best searched for using a stochastic signal model
(i.e., one defined in terms of probability distributions),
even if the signal is {\em intrinsically} deterministic,
e.g., a superposition of sinusoids.
This turns out to be the case if the signals are:
(i) {\em sufficiently weak} that they are individually 
unresolvable in a single detector, and hence can only 
be detected by 
integrating their correlated contribution across multiple 
detectors over an extended period of time, or
(ii) they are {\em sufficiently numerous} 
that they overlap in time-frequency space, 
again making them individually unresolvable, 
but producing a {\em confusion noise} that can be detected 
by cross-correlation methods.
If the rate of signals is large enough, the 
confusion noise
will be Gaussian thanks to the central limit theorem.
But if the rate or duty-cycle is small, then the resulting stochastic 
signal will be non-Gaussian and ``popcorn-like", as we 
discussed in Section~\ref{s:motivation}.
This is the type of signal that we expect from the 
population of binary black holes that produced GW150914
and GW151226; and it is the type of signal that we will focus on in 
the following few subsections.

\begin{figure}[htbp]
\begin{center}
\includegraphics[angle=0,width=0.49\columnwidth]{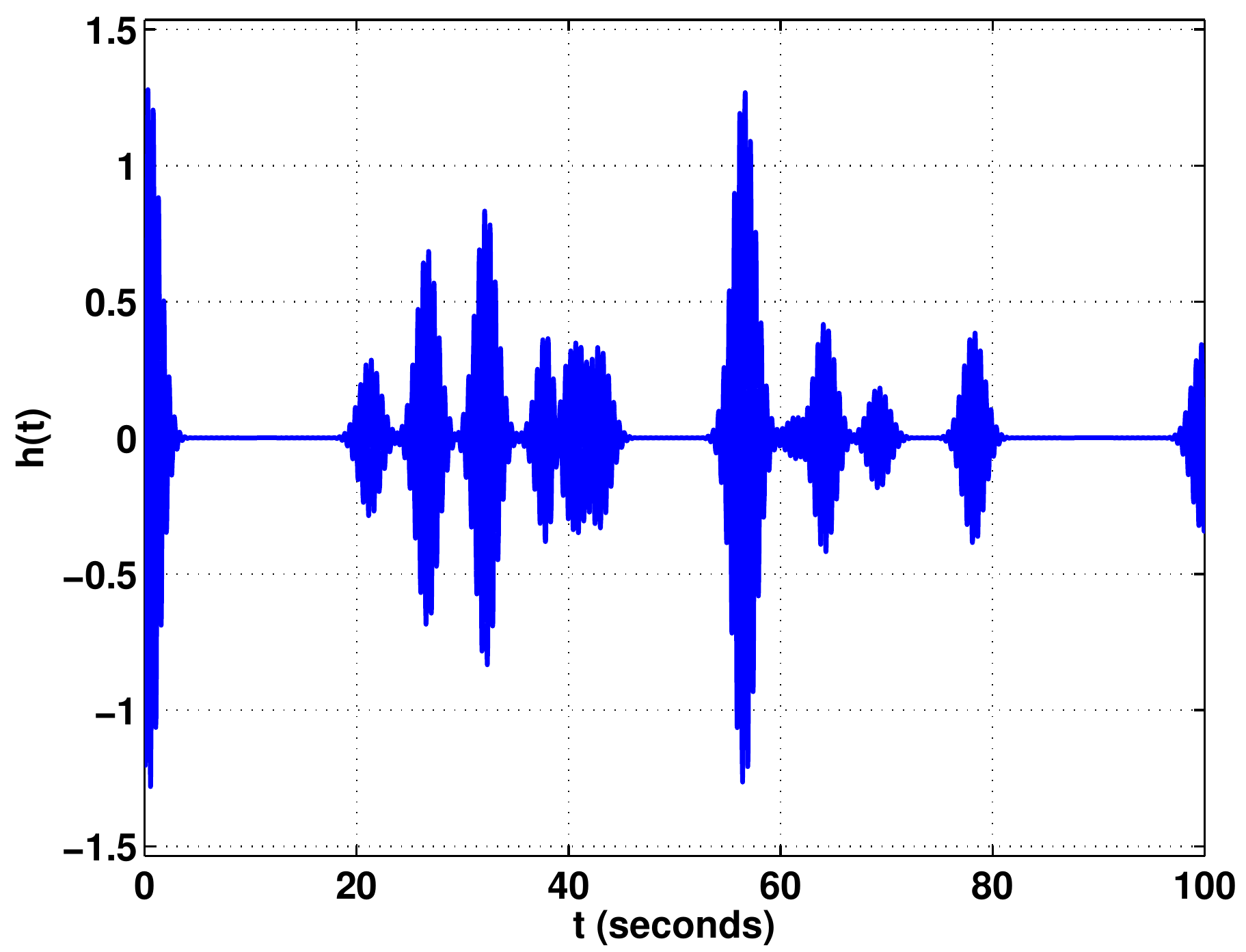}
\includegraphics[angle=0,width=0.49\columnwidth]{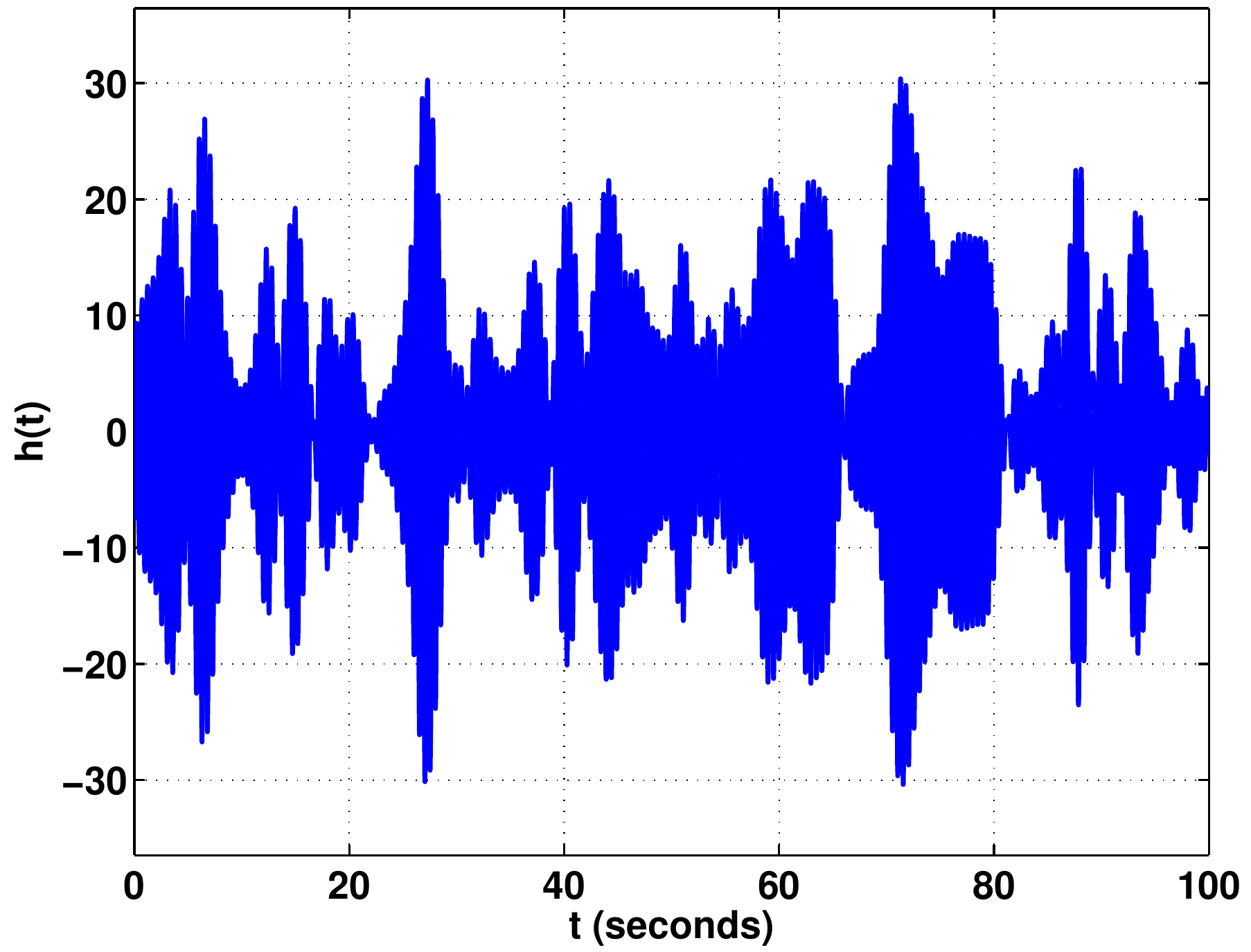}\\
\hspace{0.02true in}
\includegraphics[angle=0,width=0.45\columnwidth]{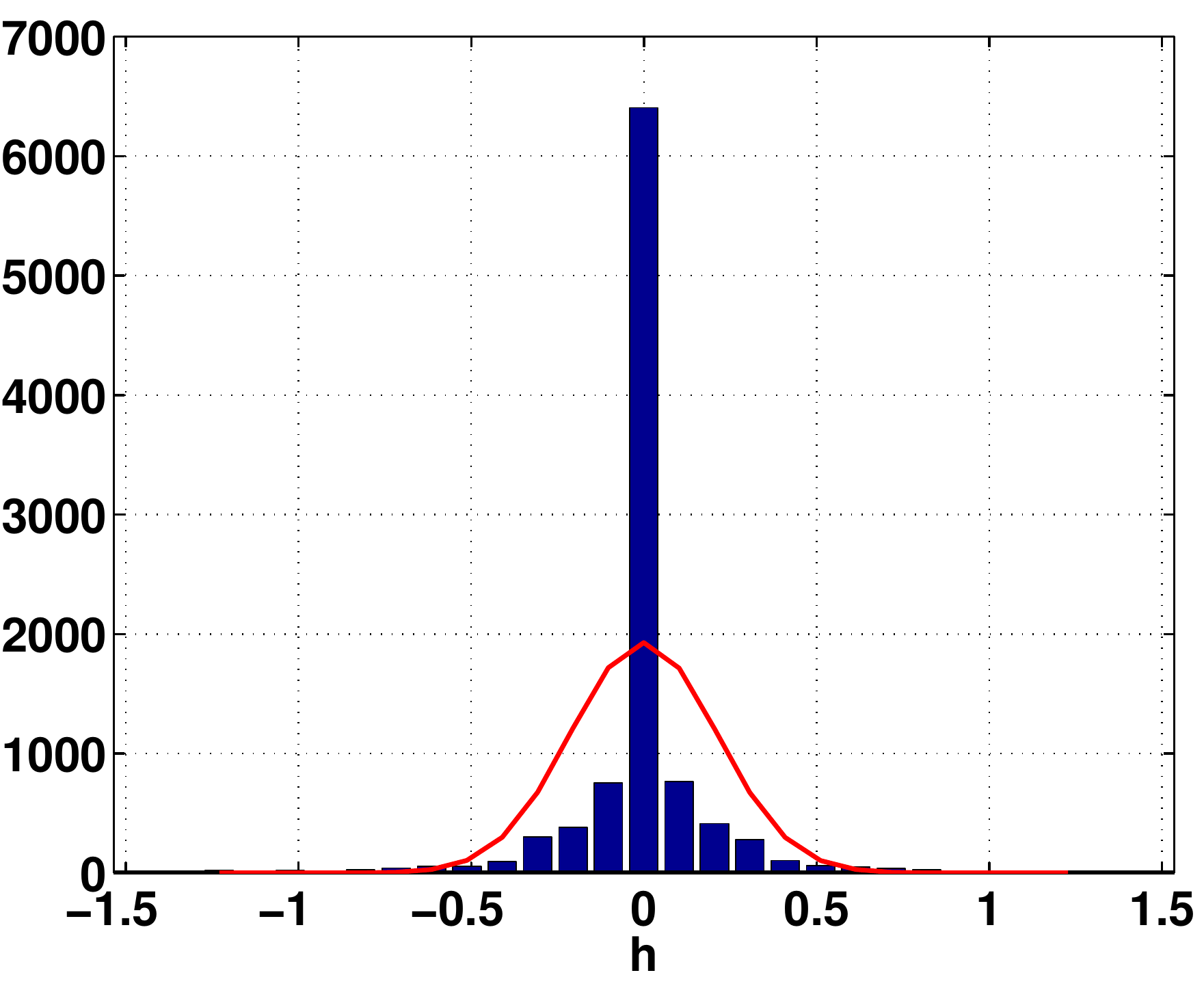}
\hspace{0.22true in}
\includegraphics[angle=0,width=0.45\columnwidth]{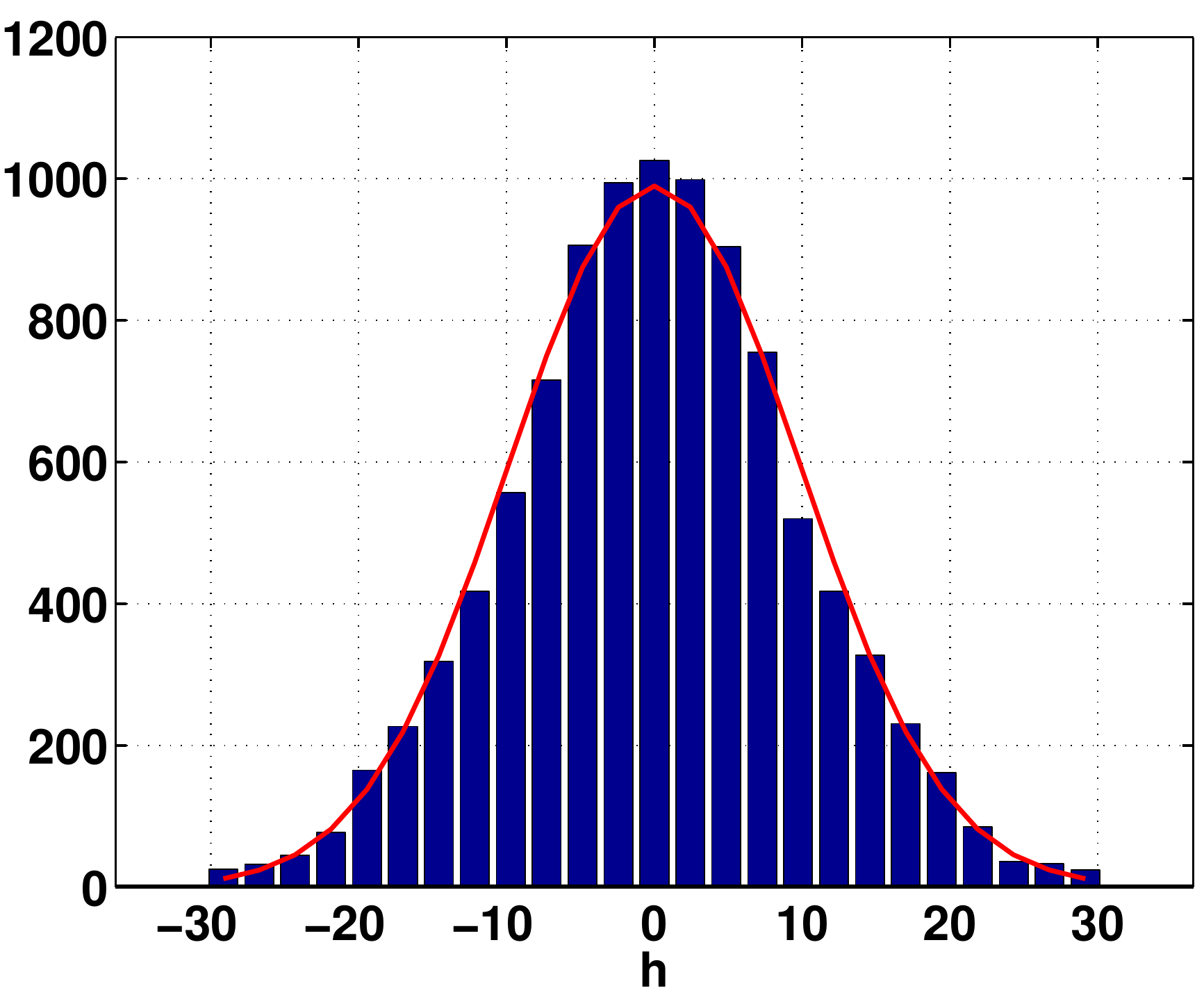}
\caption{Simulated toy-model signals and histograms for 
different duty cycles.
The left two panels correspond to 1 burst every 10~seconds (on average);
the right two panels correspond to 100 bursts every second (on average).
The red curves in the bottom two panels show the best-fit Gaussian 
distributions to the data.
Similar to Figure~1 from \cite{Thrane:2013}.}
\label{f:popcorn}
\end{center}
\end{figure}

Figure~\ref{f:popcorn} illustrates the above 
statements in the context of a simple toy-model signal
consisting of simulated sine-Gaussian bursts (each with
a width $\sigma_t= 1~{\rm s}$) 
having different rates or duty cycles.
The left two panels correspond to the case where 
there is 1 burst every 10~seconds (on average).
The probability distribution of the signal samples 
$h$ (estimated by the histogram in the 
lower-left-hand panel) is far from Gaussian for this case.
The right two panels correspond to 100 bursts
every second (on average), for which the probability distribution is
approximately Gaussian-distributed, as expected 
from the central limit theorem.
%

\subsubsection{non-Gaussian search methods -- overview}
\label{s:ng_search_methods}

There are basically two different approaches that one
can take to search for non-Gaussian stochastic signals:
(i) The first is to incorporate the non-Gaussianity
of the signal into the likelihood function by 
marginalizing over the 
appropriate signal model (Section~\ref{s:signal_models}).
Then given the likelihood, one can construct frequentist
detection statistics and estimators from the maximum-likelihood
ratio (\ref{e:likelihood_ratio}), or do Bayesian
model selection in the usual way (Section~\ref{s:inference}).
(ii) The second approach is to construct specific
frequentist statistics that targets the higher-order 
moments of the non-Gaussian distribution, and then use 
these statistics to do standard frequentist hypothesis 
testing and parameter estimation.
This approach is most simply cast in terms of the 
{\em skewness} and (excess) {\em kurtosis} of the 
distribution, which are the third and fourth-order 
{\em cumulants}, defined as follows:
If $X$ is a random variable with probability 
distribution $p_X(x)$, then the {\em moments} are
defined by (Appendix~\ref{s:basics}):
\be
\mu_n \equiv \langle X^n\rangle
= \int dx\> x^n p_X(x)\,,
\ee
and the {\em cumulants} by
\be
\begin{aligned}
c_1 &= \mu_1\,,
\\
c_2 &= \mu_2-\mu_1^2\,,
\\
c_3 &= \mu_3-3\mu_2\mu_1 + 2\mu_1^3\,,
\\
c_4 &= \mu_4-4\mu_3\mu_1 - 3\mu_2^2 + 12\mu_2\mu_1^2-6\mu_1^4\,,
\\
&\vdots
\end{aligned}
\ee
Note that $c_1$ and $c_2$ are just the {\em mean} $\mu$ 
and {\em variance} $\sigma^2$ of the distribution.
For a Gaussian distribution, $c_3=0, c_4=0, \cdots$.
For a distribution with zero mean, the above formulas
simplify to $c_1=0$, $c_2=\mu_2$, $c_3=\mu_3$, and 
$c_4 = \mu_4 - 3\mu_2^2$.
The higher-order-moment approach requires 
3rd or 4th-order 
correlation measurements (Section~\ref{s:fourth-order}).

\subsubsection{Likelihood function approach for non-Gaussian backgrounds}
\label{s:signal_models}

Fundamentally, searching for non-Gaussian stochastic
signals is no different than searching for a 
Gaussian stochastic signal.
In both cases one must:
(i) specify a signal model,
(ii) incorporate that signal model into a likelihood 
function or frequentist detection statistic/estimator, and 
(iii) then analyze the data to determine how likely 
it is that a signal is present.
It is the choice of signal model, of course, that 
determines what type of signal is being searched for.

The signal model is incorporated into the 
likelihood via marginalization over the signal
samples as discussed in Section~\ref{s:prior}.
Assuming Gaussian-stationary noise%
\footnote{What to do when the noise is non-stationary or
non-Gaussian is discussed in Sections~\ref{s:nonstationarynoise}
and \ref{s:nongaussiannoise}.}
with covariance matrix $C_n$, the probability of 
observing data $d$ in a network of detectors 
given signal model $\bar h$ is (\ref{net_likelihood}):
\be
p(d|\bar h, C_n) =
\frac{1}{\sqrt{\det(2\pi C_n)}}
e^{-\frac{1}{2}
\sum_{Ii,Jj}r_{Ii} \left(C_n^{-1}\right)_{Ii,Jj} r_{Jj}}\,,
\ee
where 
\be
r_{Ii}\equiv d_{Ii} - \bar h_{Ii}
\ee
are the residuals in detector $I$.
(The subscript $i$ labels either a time or frequency 
sample for the analysis, whichever is being used.)
Since one is often not interested in the 
particular values of $\bar h$, but rather the values
of the parameters $\vec\theta_h$ that describe the signal, 
one marginalizes over $\bar h$:
\be
p(d|\vec \theta_h,\vec\theta_n)
= \int d\bar h\> p(d|\bar h, C_n) p(\bar h|\vec\theta_h)\,.
\ee
This yields a likelihood function that depends
on the signal and noise parameters $\vec\theta_h$,
$\vec\theta_n\equiv C_n$.
It is this likelihood function that we then use for 
our statistical analysis.

Several different signal priors, which 
have been proposed in the literature, are given below.
For simplicity, we will consider the case where
the detectors are colocated and coaligned, and
have isotropic antenna patterns, so that the 
contribution from the signal is the same in each detector,
and is independent of direction on the sky.
For real analyses, these simplifications will need to 
be dropped, as is done e.g., in \cite{Thrane:2013}.
\medskip

\noindent
\underline{Gaussian signal prior}:
\be
p(\bar h| S_h)=
\frac{1}{(2\pi S_h)^{N/2}}\, e^{-\frac{1}{2 S_h}\sum_{i=1}^N \bar h_i^2}\,.
\label{e:gaussian_prior}
\ee
This is the standard prior that one uses for describing
a Gaussian-stochastic signal, and leads to the usual
Gaussian-stochastic cross-correlation detection 
statistic (Section~\ref{s:ML_statistic_derivation}).
\medskip

\noindent
\underline{Drasco and Flanagan~\cite{Drasco-Flanagan:2003} non-Gaussian signal prior}:
\be
p(\bar h| \xi, \alpha)=
\prod_{i=1}^N\left[
\xi\,\frac{1}{\sqrt{2\pi\alpha^2}}\, e^{-\bar h_i^2/2\alpha^2}
+(1-\xi)\,\delta(\bar h_i)
\right]\,.
\label{e:DF_prior}
\ee
This prior corresponds to 
Gaussian bursts occuring with probability $0\le \xi\le 1$
and with root-mean-square (rms) amplitude $\alpha$.
\medskip

\noindent
\underline{Mixture-Gaussian signal prior}:

\be
p(\bar h| \xi, \alpha, \beta)=
\prod_{i=1}^N\left[
\xi\,\frac{1}{\sqrt{2\pi\alpha^2}}\, e^{-\bar h_i^2/2\alpha^2}
+(1-\xi)\,\frac{1}{\sqrt{2\pi\beta^2}}\, e^{-\bar h_i^2/2\beta^2}
\right]\,.
\label{e:mixture_gaussian_prior}
\ee
The mixture-Gaussian signal prior is a 
non-Gaussian distribution, which reduces to the Gaussian 
signal prior in the limit $\xi\rightarrow 1$.
It reduces to the Drasco and Flanagan signal prior 
in the limit $\beta\rightarrow 0$.
\medskip

\noindent
\underline{Martellini and Regimbau~\cite{Martellini-Regimbau:2014} 
non-Gaussian signal prior}:
\be
p(\bar h| \xi, \alpha)=
\prod_{i=1}^N\left[
\xi\,p_{\rm NG}(\bar h_i)
+(1-\xi)\,\delta(\bar h_i)
\right]\,,
\label{e:MR_prior}
\ee
where
\be
p_{\rm NG}(\bar h_i) =
\frac{1}{\sqrt{2\pi\alpha^2}}\, e^{-\bar h_i^2/2\alpha^2}
\left[
1 
+\frac{c_3}{6\alpha^3}H_3\left(\frac{\bar h_i}{\alpha}\right)
+\frac{c_4}{24\alpha^4}H_4\left(\frac{\bar h_i}{\alpha}\right)
+\frac{c_3^2}{72\alpha^6}H_6\left(\frac{\bar h_i}{\alpha}\right)
\right]
\ee
is the 4th-order Edgeworth expansion~\cite{Martellini-Regimbau:2014} 
of a non-Gaussian distribution with 
third and fourth-order cumulants $c_3$ and $c_4$.
($H_n(x)$ denotes a Hermite polynomial of order $n$.)
The Edgeworth expansion is referenced off a
Gaussian probability distribution, and is 
thus said to be a {\em semi-parametric} representation of
a non-Gaussian distribution.
This prior reduces to the Drasco and Flanagan signal prior 
when $c_3=0$, $c_4=0$.
\medskip

\noindent
\underline{Multi-sinusoid signal prior}:
\be
\begin{aligned}
&p(\bar h|\vec \theta_h) 
= \delta\left(\bar h - \bar h(\vec\theta_h)\right)\,,
\\
&\bar h_i(\vec\theta_h) = \sum_{I=1}^M 
A_I \cos(2\pi f_I t_i - \varphi_I)\,.
\label{e:multisinusoid_prior}
\end{aligned}
\ee
This is a {\em deterministic} signal prior, corresponding to
the superposition of $M$ sinusoids with unknown amplitudes,
frequencies, and phases,
$\vec \theta_h = \{A_I, f_I, \varphi_I\vert I=1,2,\cdots, M\}$.
This was one of the signal models used in \cite{Cornish-Romano:2015} 
to investigate the question of when is a signal stochastic.
\medskip

\noindent
\underline{Superposition of finite-duration deterministic signals}:
\be
\begin{aligned}
&p(\bar h|\vec \theta_h) 
= \delta\left(\bar h - \bar h(\vec\theta_h)\right)\,,
\\
&\bar h_i(\vec\theta_h) = \sum_{I=1}^M 
A_I {\cal T}(t_i - t_I|\vec\theta_{\cal T})\,.
\label{e:finite_duration_prior}
\end{aligned}
\ee
Here, ${\cal T}(t|\vec\theta_{\cal T})$ is a normalized 
waveform (template) for some deterministic signal 
(e.g., a chirp from an inspiralling binary, 
a sine-Gaussian burst, a ringdown signal, $\cdots$)
described by parameters $\vec\theta_{\cal T}$
(e.g., chirp mass, correlation time, frequency, $\cdots$).
$A_I$ is the amplitude of the $I$th signal and 
$t_I$ is its arrival time.
Note that these signal waveforms can be {\em extended} 
in time, having a characteristic duration $\tau$.
Thus, this signal model is intermediate between the 
single-sample burst and multi-sinusoid signal models.
\medskip

\noindent
\underline{Generic likelihood for unresolvable signals}:
\medskip

\noindent
In~\cite{Thrane:2013}, Thrane writes down a generic
likelihood function for a non-Gaussian background formed 
from the superposition of signals which are 
individually unresolvable in a single detector.
The likelihood function:
\be
p(\hat\rho|\xi,\vec \theta_h, \vec \theta_n)
=\prod_i\left[
\xi\,S(\hat\rho_i|\vec\theta_h)
+(1-\xi)\,B(\hat\rho_i|\vec\theta_n)
\right]
\label{e:thrane}
\ee
is defined for a pair of detectors $I$, $J$, 
and takes as its fundamental data vector 
estimates of the signal-to-noise
ratio of the cross-correlated power in the 
two detectors:
\be
\hat \rho_i \equiv \hat\rho(t;f) 
= \sqrt{\tau\delta f}
\frac{\hat C_{IJ}(t;f)}{\sqrt{P_{n_I}(t;f) P_{n_J}(t;f)}}\,,
\ee
where
\be
\hat C_{IJ}(t;f) \equiv \frac{2}{\tau}\tilde d_I(t;f) \tilde d_J^*(t;f)\,.
\ee
Here $\tau$ is the duration of the short-term Fourier
transforms and $\delta f$ is the frequency resolution.
(Note that $\delta f$ can be greater than $1/\tau$ if 
one averages together neighboring frequency bins.)
The product over $i$ is over time-frequency pixels $tf$.
The functions $S$ and $B$ are probability distributions
for $\hat\rho_i$ for the signal and noise models, respectively.
These distributions are generic in the sense that they
are to be estimated using Monte Carlo simulations with
injected signals for the signal model $S$, and via 
time-slides on real data for the noise model $B$.
They need not be Gaussian for either the signal or the
detector noise.
The vectors $\vec\theta_h$ and $\vec\theta_n$ denote 
parameters specific to the signal and noise models.
Although the above likelihood function was not obtained 
by explicitly marginalizing over $\bar h$,
mathematically there is some signal prior and 
noise model which yields this likelihood upon marginalization.

\subsubsection{Frequentist detection statistic for non-Gaussian backgrounds}
\label{s:drasco-flanagan}

As discussed in Section~\ref{s:relating-freq-bayes},
given likelihood functions for the signal-plus-noise 
and noise-only models,
we can construct a frequentist detection statistic 
from either the 
maximum-likelihood ratio $\Lambda_{\rm ML}(d)$
given by (\ref{e:likelihood_ratio}), or twice its
logarithm, $\Lambda(d) \equiv 2\ln(\Lambda_{\rm ML}(d))$,
which has the 
interpretation of being the squared signal-to-noise
ratio of the relevant data.
For a white Gaussian stochastic signal in white 
Gaussian detector noise (assuming a pair of 
colocated and coaligned detectors), 
we showed in Section~\ref{s:ML_statistic_derivation}:
\be
\Lambda^{\rm G}_{\rm ML}(d)
=\left[1-\frac{\hat S_h^2}{\hat S_1\hat S_2}\right]^{-N/2}\,,
\qquad
\Lambda^{\rm G}(d) 
\approx \frac{\hat S_h^2}{\hat S_{n_1}\hat S_{n_2}/N}\,,
\ee
where $N$ is the number of samples, and where the last
approximate equality assumes that the gravitational-wave
signal is weak compared to the detector noise.
We have added the superscript G to indicate that this is 
for a Gaussian-stochastic signal model.

We can perform exactly the same calculations, 
making the same assumptions, for the likelihood functions 
constructed from {\em any} of the non-Gaussian signal 
priors given above (in Section~\ref{s:signal_models}).
These calculations have already been done for the 
Drasco-Flanagan and Martellini-Regimbau signal priors
\cite{Drasco-Flanagan:2003, Martellini-Regimbau:2014}.
The expressions that they find for 
the maximum-likelihood ratios 
$\Lambda_{\rm ML}^{\rm NG}(d)$ for their non-Gaussian
signal models are rather long and not particularly 
informative, so we do not
bother to write them down here (interested
readers should see (1.8) in \cite{Drasco-Flanagan:2003} 
and the last equation in \cite{Martellini-Regimbau:2014}.)
The values of the parameters that maximize the 
likelihood ratio are estimators of 
$\xi$, $\alpha$, $S_{n_1}$, $S_{n_2}$ for the 
Drasco and Flanagan signal model, and 
estimators of
$\xi$, $\alpha$, $c_3$, $c_4$, $S_{n_1}$, $S_{n_2}$
for the Martellini and Regimbau signal model.

\begin{figure}[h!tbp]
\begin{center}
\includegraphics[angle=0,width=0.6\columnwidth]{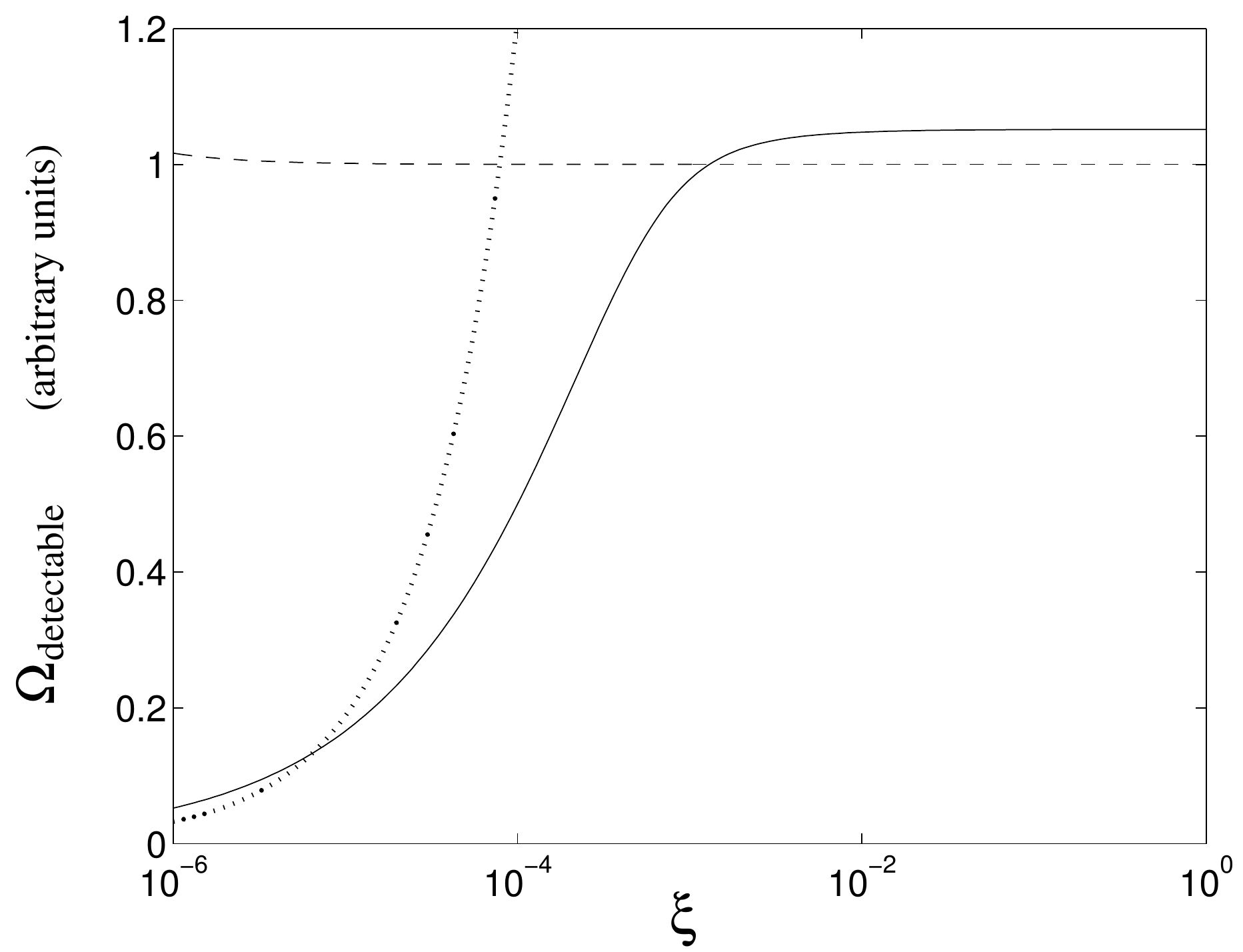}
\caption{The minimum detectable value 
of $\Omega_{\rm gw}$ as a function of the 
duty cycle $\xi$.
The solid line is the theoretical prediction for
the Drasco and Flanagan non-Gaussian 
maximum-likelihood statistic;
the dashed line is for the standard Gaussian-stochastic 
cross-correlation statistic; and the dotted line is 
for a single-detector burst statistic.
The number of data points used was $N=10^9$, 
and the false alarm and false dismissal probabilities
were both chosen to equal $0.01$.
Image reproduced with permission from \cite{Drasco-Flanagan:2003},
copyright by APS.}
\label{f:drasco-flanagan}
\end{center}
\end{figure}

To illustrate the performance of a non-Gaussian 
detection statististic,
we plot in Figure~\ref{f:drasco-flanagan}
the mimimum value of $\Omega_{\rm gw}$ 
($S_h$ in the notation above) 
necessary for detection as a function of the 
duty cyle $\xi$.
(The signal becomes Gaussian as $\xi\rightarrow 1$.)
The solid line is the theoretical prediction for
the Drasco and Flanagan
non-Gaussian maximum-likelihood statistic,
while the dashed line is the theoretical 
prediction for the standard Gaussian-stochastic 
cross-correlation statistic.
The dotted line is the theoretical prediction
for a single-detector {\em burst} statistic, 
which is just the maximum of the absolute value 
of the data samples in e.g., detector 1:
$\Lambda^{\rm B}(d) = \max_{i} |d_{1i}|$.
The false alarm and false dismissal probabilities
were both chosen to equal $0.01$ for this calculation.
From the figure one sees that for 
$\xi\gtrsim 10^{-3}$, the Gaussian-stochastic 
cross-correlation statistic performs best.
For smaller values of $\xi$, the 
non-Gaussian statistic is better.
In particular, for $\xi\sim 10^{-4}$.
there is a factor of $\sim\!2$ improvement
in the minimum detectable signal amplitude
if one uses the non-Gaussian maximum-likelihood
detection statistic. 

\begin{figure}[h!tbp]
\begin{center}
\includegraphics[angle=0,width=0.6\columnwidth]{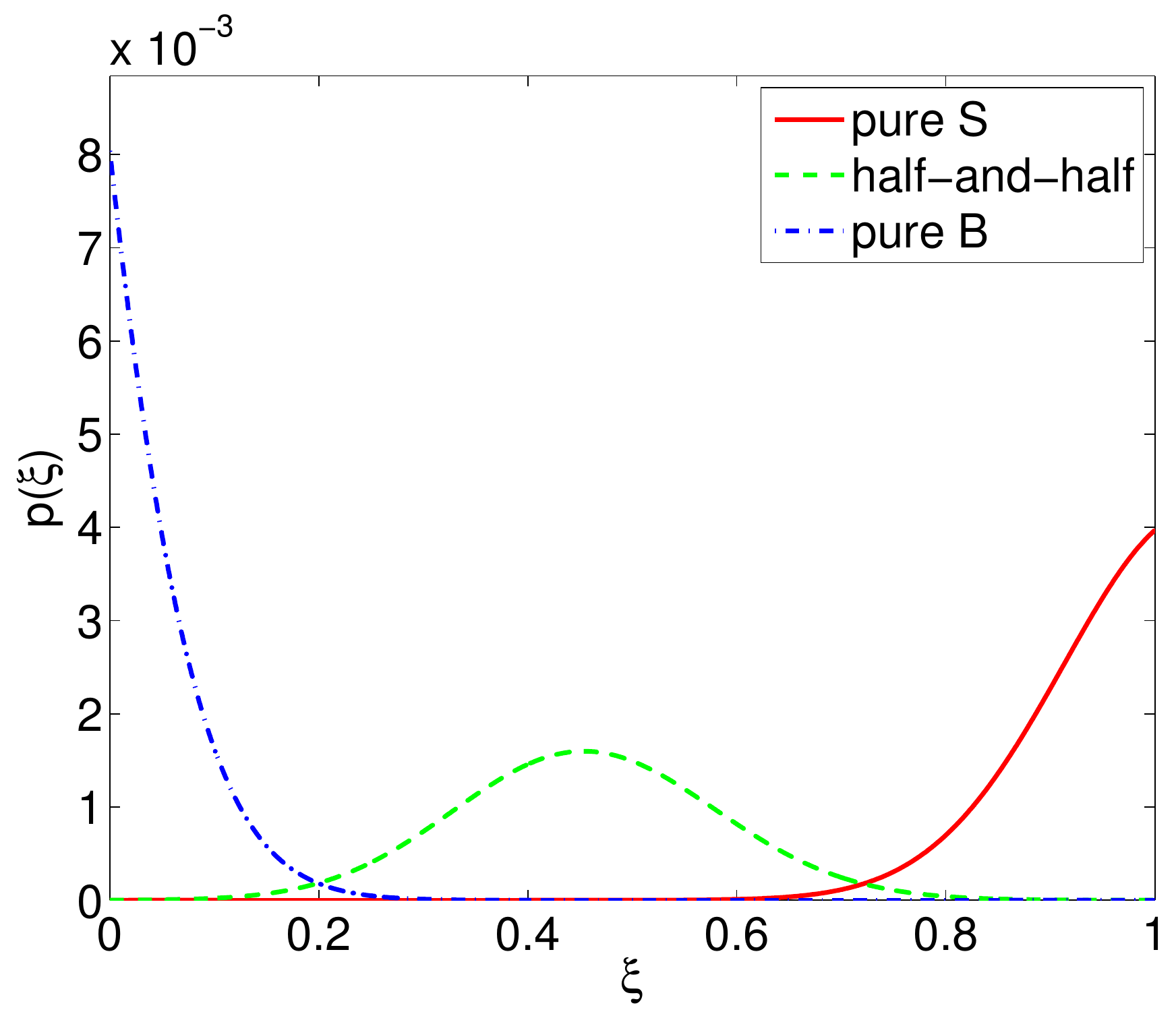}
\caption{Posterior distributions for the
duty cycle $\xi$ calculated for 
Monte Carlo simulations
having $\xi=0$ (dash-dot blue),
$\xi=1$ (solid red),
and $\xi=0.5$ (dashed green).
Image reproduced with permission from \cite{Thrane:2013},
copyright by APS.}
\label{f:thrane-duty}
\end{center}
\end{figure}

Figure~\ref{f:thrane-duty} is taken from \cite{Thrane:2013}
and shows posterior distributions for the duty cycle
$\xi$ calculated for Monte Carlo simulations 
corresponding to 
pure background $\xi=0$ (dash-dot blue), 
pure signal $\xi=1$ (solid red),
and an even mixture $\xi=0.5$ (dashed green).
These curves illustrate that the formalism in \cite{Thrane:2013}
can provide 
estimates of the duty cycle $\xi$ of the non-Gaussian background.
See \cite{Thrane:2013} for more details.
%

\subsubsection{Bayesian model selection}
\label{s:BF_nongaussian}

As an alternative to using frequentist detection statistics and 
estimators to search for potentially non-Gaussian signals, one 
can use Bayesian model selection to compare the noise-only model 
${\cal M}_0$ to different signal-plus-noise models ${\cal M}_1, {\cal M}_2, \cdots$.
This is a general procedure for Bayesian inference, which was 
discussed in Section~\ref{s:bayes-model-selection}.
As shown there, the posterior odds ratio between two different 
models ${\cal M}_\alpha$ and ${\cal M}_\beta$ can be written as
\begin{equation}
{\cal O}_{\alpha\beta}(d)
= \frac{p({\cal M}_\alpha \vert d )}{p({\cal M}_\beta \vert d )}
= \frac{p({\cal M}_\alpha)}{p({\cal M}_\beta)}\,
\frac{ p(d \vert {\cal M}_\alpha)}{p(d \vert {\cal M}_\beta)}\,,
\end{equation}
where the first ratio on the right-hand side is the 
{\em prior} odds for the two models, while the second term 
is the {\em Bayes factor}:
\begin{equation}
{\cal B}_{\alpha\beta}(d)
= \frac{p(d \vert {\cal M}_\alpha)}{p(d \vert {\cal M}_\beta)}\,,
\end{equation}
which is a ratio of model evidences:
\begin{equation}
p(d \vert {\cal M}_\alpha) =
\int p( d\vert \vec{\theta}_\alpha, {\cal M}_\alpha)
p(\vec{\theta}_\alpha  \vert {\cal M}_\alpha) \,
d\vec{\theta}_\alpha \,,
\end{equation}
and similarly for $p(d|{\cal M}_\beta)$.
If one assumes equal prior odds, then the posterior odds ratio
is just the Bayes factor, and we can use its value to rule 
in favor of one model or another 
(see Table~\ref{t:bayesfactors}).

The idea of using Bayesian model selection in the context
of searches for non-Gaussian stochastic backgrounds was
proposed by us in \cite{Cornish-Romano:2015}.
We considered a simple toy-problem consisting of 
simulated data in two colocated and coaligned detectors, 
having uncorrelated white Gaussian detector noise plus 
a gravitational-wave 
signal formed from the superposition of sinusoids 
having amplitudes 
drawn from an astrophysical population of sources.
Such a signal is effectively the 
{\em frequency-domain version} of the short-duration
time-domain bursts discussed in the previous subsections.
Five different models were considered:
\begin{itemize}

\i ${\cal M}_0$: 
noise-only model, consisting of uncorrelated white 
Gaussian noise in two detectors with unknown 
variances $\sigma_1^2$, $\sigma_2^2$.

\i ${\cal M}_1$: 
noise plus the Gaussian-stochastic signal model
defined by (\ref{e:gaussian_prior}).

\i ${\cal M}_2$: 
noise plus the mixture-Gaussian stochastic signal model
defined by (\ref{e:mixture_gaussian_prior}).

\i ${\cal M}_3$: 
noise plus the deterministic multisinusoid model 
defined by (\ref{e:multisinusoid_prior}).

\i ${\cal M}_4$:  
noise plus the deterministic multisinusoid signal model
plus the Gaussian-stochastic signal model.
This is a {\em hybrid} signal model that allows for 
both stochastic and determistic components for the 
signal.

\end{itemize}
Simulated data were generated by coadding sinusoidal
signals with amplitudes drawn from an astrophysical
model~\cite{2013MNRAS.433L...1S}, and phases and 
frequencies drawn uniformly across the range spanned 
by the data.
Gaussian-distributed white noise for the two detectors 
were then added to the signal data.
The amplitude of the signals were scaled so as to
produce a specified matched filter signal-to-noise
ratio per frequency bin.
Markov Chain Monte Carlo analyses were run to compare 
the noise-only model ${\cal M}_0$ to each of the four 
signal-plus-noise models ${\cal M}_1, \cdots, {\cal M}_4$.
Quantile intervals for the Bayes factors were estimated
from 256 independent realizations of the simulated 
data for each set of parameter values.
These intervals capture the fluctuation in the Bayes
factors that come from {\em different} realizations 
of the data; they are not uncertainties in the Bayes 
factors associated with different Monte Carlo 
simulations for a {\em single} realization, which 
were $\lesssim 10\%$.

\begin{figure}[htbp]
{\includegraphics[width=.49\textwidth]{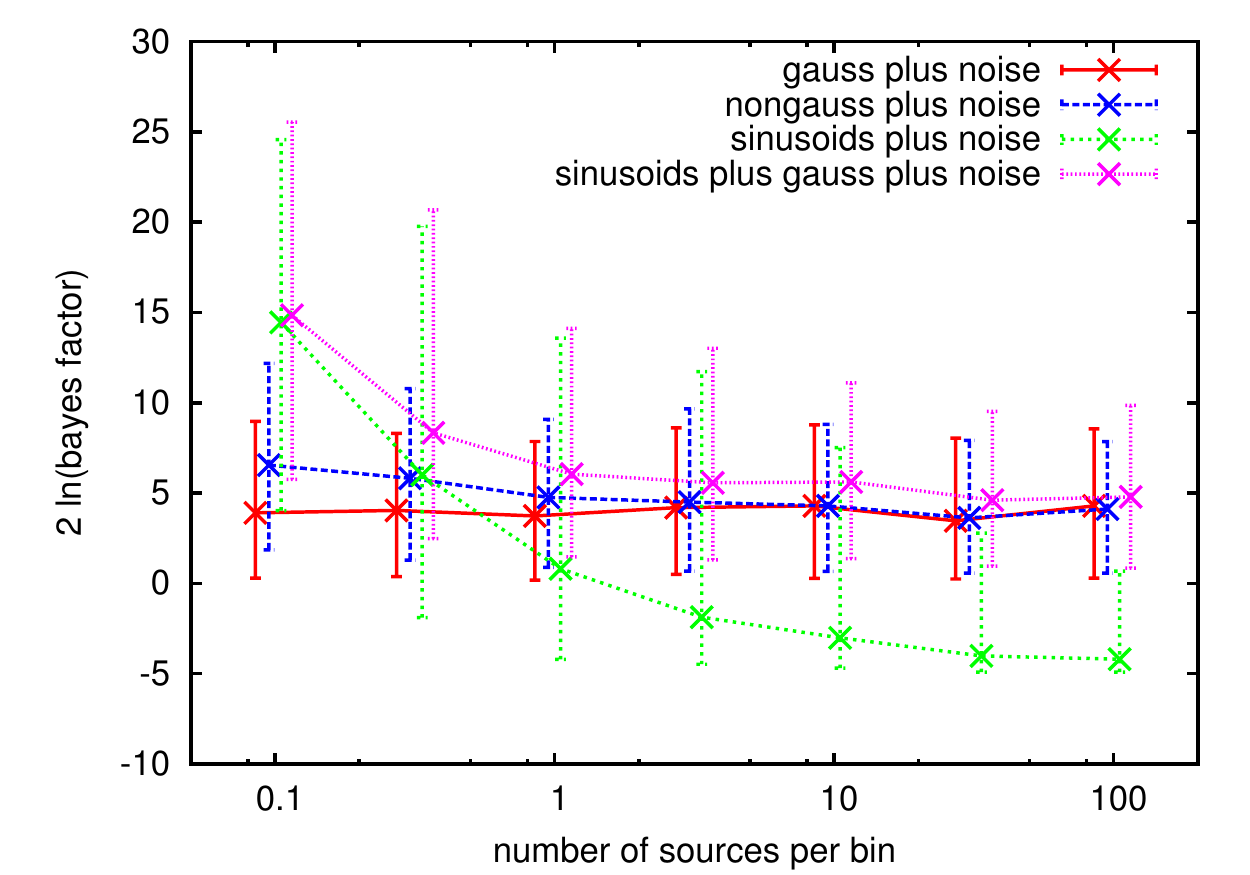}}
{\includegraphics[width=.49\textwidth]{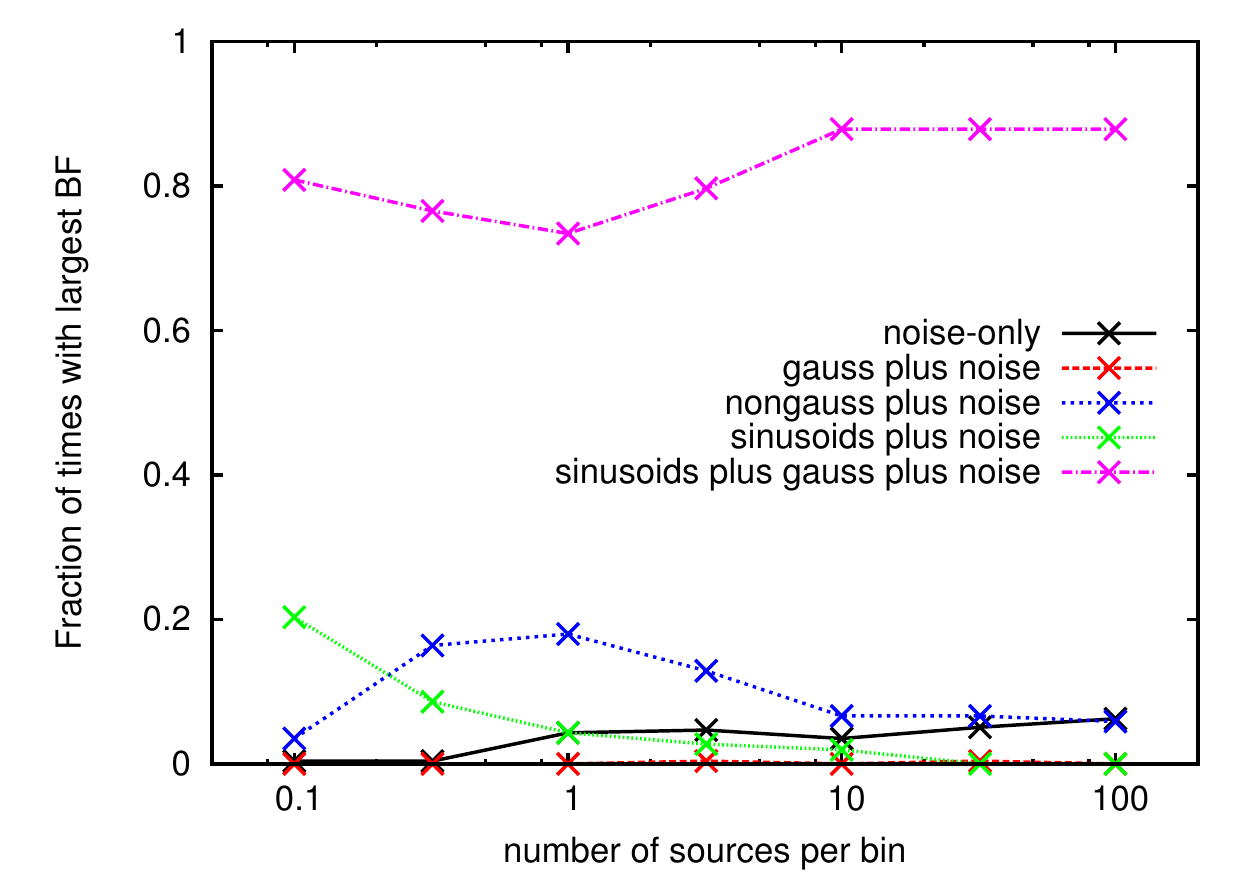}}
\caption{Left panel: Bayes factor 80\% quantile intervals for the
four different signal-plus-noise models relative to the noise-only
model as a function of the number of sources per bin.  
Right panel: Fraction of time that the different models had the largest Bayes 
factor for the different simulations.
Image reproduced with permission from \cite{Cornish-Romano:2015},
copyright by APS.}
\label{f:SdenNoiseSourceModel2}
\end{figure}

Figure~\ref{f:SdenNoiseSourceModel2} is a representative
plot taken from \cite{Cornish-Romano:2015}, comparing 
the different models.
The left panel shows the Bayes factors for the 
four different signal-plus-noise models 
relative to the noise-only model plotted as a 
function of the average number of sources per bin.
The right panel shows the fraction of time 
that the different models had the largest Bayes
factor for the different simulations.
The total number of bins was set to 32 for these
simulations and the SNR per bin was fixed at 2.
From these and other similar plots in 
\cite{Cornish-Romano:2015},
one can draw the general conclusion that 
deterministic models are generally favored for 
small source densities, 
a non-Gaussian stochastic model is preferred for 
intermediate source densities, and a Gaussian-stochastic
model is preferred for large source densities.
Given the large fluctuations in the Bayes factors
associated with different signal realizations, 
the boundaries between these three regimes is
rather fuzzy.
The hybrid model, which has a deterministic component
for the bright signals and a Gaussian-stochastic 
component for the remaining confusion background,
is the best model for the majority of cases.
%

\subsubsection{Fourth-order correlation approach for non-Gaussian backgrounds}
\label{s:fourth-order}

In this section, we briefly describe a fourth-order
correlation approach for detecting non-Gaussian stochastic 
signals, originally proposed in~\cite{Seto:2009}.
The key idea is that by forming a particular combination
of data from 4 detectors (the {\em excess kurtosis}), one can 
separate the non-Gaussian contribution to the background 
from any Gaussian-distributed component.
This approach requires that the noise in the four 
detectors be uncorrelated with one another, but it
does not require that the noise be Gaussian.
Here we sketch out the calculation for 4 colocated and 
coaligned detectors, which we will assume have 
isotropic antenna patterns, so that the contribution 
from the gravitational-wave signal is the same in each 
detector, and is independent of direction on the sky.
These simplifying assumptions are not essential for 
this approach;
the calculation for separated and misalinged detectors 
with non-isotropic response functions can also be 
done~\cite{Seto:2009}.

Let's begin then by denoting the output of the four
detectors $I=1,2,3,4$ in the Fourier domain by
\be
\tilde d_I = \tilde n_I + \tilde h\,,
\qquad
\tilde h = \tilde g + \sum_{i=1}^n \tilde b_i\,,
\ee
where $\tilde n_I$ denotes the noise in detector $I$ and
$\tilde h$ denotes the total gravitational-wave contribution,
which has a Gaussian-stochastic component $\tilde g$, and
a non-Gaussian component formed from the superposition
of short-duration burst signals $\tilde b_i$, $i=1,2,\cdots, n$.
We assume that the noise in the detectors are uncorrelated
with one another and with the gravitational-wave signals,
and that the individual gravitational-wave signals are
also uncorrelated amongst themselves.
The (random) number of bursts present in a particular 
segment of data is determined by a Poisson distribution
\be
P(n) = \frac{\lambda^n e^{-\lambda}}{n!}\,,
\ee
where
\be
\lambda = \langle n\rangle 
= \sum_{n=0}^\infty nP(n)\,,
\ee
is the expected number of bursts in segment duration $T_{\rm seg}$.
The 4th-order combination of data that we consider is 
\be
{\cal K}
\equiv
\langle \tilde d_1 \tilde d_2 \tilde d_3^* \tilde d_4^*\rangle
-\langle \tilde d_1 \tilde d_2\rangle\langle \tilde d_3^* \tilde d_4^*\rangle
-\langle \tilde d_1 \tilde d_3^*\rangle\langle \tilde d_2 \tilde d_4^*\rangle
-\langle \tilde d_1 \tilde d_4^*\rangle\langle \tilde d_2 \tilde d_3^*\rangle\,,
\label{e:kurtosis}
\ee
where angle brackets $\langle\ \rangle$ can be thought of
as either expectation value (i.e., ensemble average) or 
as an average over the Fourier components of the data, 
i.e., as an {\em estimator} of the expected correlations.
Since the noise in the detectors are uncorrelated with everything, 
the only contributions to ${\cal K}$ will come from 
expectation values of 
products of $\tilde h = \tilde g +\sum_i \tilde b_i$ with itself.
Calculating the quadratic terms that enter (\ref{e:kurtosis}),
we find:
\be
\begin{aligned}
\langle\tilde d_I\tilde d_J\rangle 
&= \langle \tilde g\tilde g\rangle
+ \lambda\langle \tilde b\tilde b\rangle\,,
\\
\langle\tilde d_I\tilde d_J^*\rangle 
&= \langle \tilde g\tilde g^*\rangle
+ \lambda\langle \tilde b\tilde b^*\rangle\,,
\end{aligned}
\ee
where we used
\be
\left\langle \sum_i\sum_j \tilde b_i \tilde b_j\right\rangle
=\left\langle \sum_i \tilde b_i \tilde b_i\right\rangle
= \lambda \langle\tilde b\tilde b\rangle\,,
\ee
which assumes that all the bursts have the same mean-square value, 
$\langle \tilde b_i\tilde b_i\rangle \equiv \langle \tilde b\tilde b\rangle$.
For the 4th-order term, we find:
\begin{multline}
\langle\tilde d_1\tilde d_2\tilde d_3^*\tilde d_4^*\rangle 
=\langle \tilde g\tilde g\tilde g^*\tilde g^*\rangle
+\lambda^2\left[
|\langle \tilde b\tilde b\rangle|^2
+2\langle \tilde b\tilde b^*\rangle^2
\right]
\\
+\lambda\left[
\langle \tilde b\tilde b\tilde b^*\tilde b^*\rangle
+\langle \tilde g\tilde g\rangle \langle\tilde b\tilde b\rangle^*
+\langle \tilde g\tilde g\rangle^*\langle\tilde b\tilde b\rangle
+4\langle \tilde g\tilde g^*\rangle\langle\tilde b\tilde b^*\rangle
\right]\,.
\end{multline}
Substituting these results back into expression (\ref{e:kurtosis})
yields:
\be
{\cal K} = \lambda\langle \tilde b \tilde b\tilde b^*\tilde b^*\rangle\,,
\ee
where we used
\be
\langle \tilde g\tilde g\tilde g^*\tilde g^*\rangle
-|\langle \tilde g\tilde g\rangle|^2
-2\langle \tilde g\tilde g^*\rangle^2 
=0\,,
\ee
for the Gaussian-stochastic signal component $\tilde g$.
Thus, both the detector noise and the Gaussian-stochastic
component of the signal have dropped out of the expression 
for ${\cal K}$, leaving only the contribution from the 
non-Gaussian component of the background.

As mentioned already, the above calculation can be extended to
the case of separated and misaligned detectors~\cite{Seto:2009}.
In so doing, one obtains expressions for 
{\em generalized} (4th-order) overlap functions, 
which are sky-averages of the 
product of the response functions for four different detectors.
The expected value of the 4th-order detection statistic for 
this more general analysis involves generalized overlap 
functions for both the (squared) overall intensity and 
circular polarization components of the non-Gaussian background.
We will discuss circular polarization in the following section, 
but in the simpler context of Gaussian-stationary isotropic backgrounds.
Readers should see~\cite{Seto:2009} for more details regarding 
circular polarization in the context of non-Gaussian stochastic 
signals discussed above.

\subsection{Circular polarization}
\label{s:circular}

Up until now, we have only considered {\em unpolarized}
stochastic backgrounds.
That is, we 
have assumed that the gravitational-wave power
in the $+$ and $\times$ polarization modes are equal 
(on average) and are statistically independent of 
one another (i.e., there are no correlations between 
the $+$ and $\times$ polarization modes).
It is possible, however, for some processes in the 
early Universe to give rise to {\em parity violations} 
\cite{Alexander-et-al:2006},
which would manifest themselves as an asymmetry in 
the amount of right and left {\em circularly} polarized
gravitational waves.
Following \cite{Seto-Taruya:2007, Seto-Taruya:2008}, 
we now describe how to generalize our cross-correlation
methods to look for evidence of circular polarization 
in a stochastic background.

\subsubsection{Polarization correlation matrix}
\label{s:pol_corr}

Let us start by writing down the quadratic 
expectation values for the Fourier components 
$h_{ab}(f,\hat n)$ of the metric perturbations 
$h_{ab}(t,\vec x)$ for a {\em polarized anisotropic} 
Gaussian-stationary background.
(We will restrict attention to isotropic backgrounds
later on.)
It turns out that these expectation values can be 
written in terms of the {\em Stokes' parameters} 
$I$, $Q$, $U$, and $V$, which are
defined for a monochromatic plane gravitational
wave in Appendix~\ref{s:polarization_tensors}.
If we expand $h_{ab}(f,\hat n)$ 
in terms of the {\em linear} polarization basis
tensors $e^A_{ab}(\hat n)$, where $A=\{+,\times\}$,
we have
\be
\langle h_A(f,\hat n) h_{A'}^*(f',\hat n')\rangle
=\frac{1}{2}
S_h^{AA'}(f,\hat n)
\delta(f-f')
\delta^2(\hat n,\hat n')\,,
\label{e:aniso_pol_AA'}
\ee
where
\be
S_h^{AA'}(f,\hat n)
=\frac{1}{2}\left[
\begin{array}{cc}
I(f,\hat n) + Q(f,\hat n) & U(f,\hat h)-iV(f,\hat n)
\\
U(f,\hat n)+iV(f,\hat n) & I(f,\hat n) - Q(f,\hat n)
\\ 
\end{array}
\right]\,.
\ee
If instead we expand $h_{ab}(f,\hat n)$ in terms 
of the {\em circular} polarization basis tensors
$e^C_{ab}(\hat n)$, where $C=\{R,L\}$, then
\be
\langle h_C(f,\hat n) h_{C'}^*(f',\hat n')\rangle
=\frac{1}{2}
S_h^{CC'}(f,\hat n)
\delta(f-f')
\delta^2(\hat n,\hat n')\,,
\label{e:aniso_pol_CC'}
\ee
where
\be
S_h^{CC'}(f, \hat n)
=\frac{1}{2}\left[
\begin{array}{cc}
I(f,\hat n) + V(f,\hat n) & Q(f,\hat h)-iU(f,\hat n)
\\
Q(f,\hat n)+iU(f,\hat n) & I(f,\hat n) - V(f,\hat n)
\\ 
\end{array}
\right]\,.
\ee
This second representation of the polarization correlation 
matrix is sometimes more convenient when one is searching 
for evidence of circular polarization in the background, 
as $V$ is a measure of a possible asymmetry 
between the right and left circular polarization components:
\be
\langle h_R(f,\hat n) h_{R'}^*(f',\hat n')\rangle
-\langle h_L(f,\hat n) h_{L'}^*(f',\hat n')\rangle
=\frac{1}{2}
V(f,\hat n)
\delta(f-f')
\delta^2(\hat n,\hat n')\,.
\ee
The factor of 1/2 on the right-hand side of the above
equation, as compared to (\ref{e:Stokes_RL}), is for
one-sided power spectra.

As discussed in Appendix~\ref{s:polarization_tensors}, the 
Stokes' parameters $I$ and $V$ are ordinary scalar 
(spin 0) fields on the sphere, while $Q$ and $U$ 
transform like spin 4 fields under a rotation of 
the unit vectors $\{\hat l,\hat m\}$ tangent to the
sphere.
Thus, $I$ and $V$ can be written as linear combinations
of the ordinary spherical harmonics $Y_{lm}(\hat n)$:
\be
\begin{aligned}
I(f,\hat n) &= \sum_{l=0}^\infty\sum_{m=-l}^l I_{lm}(f) Y_{lm}(\hat n)\,,
\\
V(f,\hat n) &= \sum_{l=0}^\infty\sum_{m=-l}^l V_{lm}(f) Y_{lm}(\hat n)\,,
\\
\end{aligned}
\ee
while $Q\pm iU$ can be written as linear
combination of the spin-weighted $\pm 4$ spherical
harmonics ${}_{\pm 4}Y_{lm}(\hat n)$:
\be
Q(f,\hat n) \pm i U(f,\hat n) 
= \sum_{l=4}^\infty \sum_{m=-l}^l C^{\pm}_{lm}(f)\, 
{}_{\pm 4}Y_{lm}(\hat n)\,.
\ee
Note that the expansions for $Q\pm i U$ start at $l=4$, 
which means that the $Q$, $U$ components of the 
polarization correlation matrix vanish if the background
is isotropic (i.e., has only a contribution from the
monopole $l=0$, $m=0$).
So for simplicity, we will restrict our attention to
polarized {\em isotropic backgrounds}, for which the 
circular polarization correlation matrix becomes diagonal
and the quadratic exprectation values reduce to:
\be
\langle h_C(f,\hat n) h_{C'}^*(f',\hat n')\rangle
=\frac{1}{8\pi}
S_h^{C}(f)
\delta_{CC'}
\delta(f-f')
\delta^2(\hat n,\hat n')\,,
\label{e:iso_pol}
\ee
where
\be
\begin{aligned}
S_h^{R}(f)&\equiv \frac{1}{2}(I(f) + V(f))\,,
\\
S_h^{L}(f)&\equiv \frac{1}{2}(I(f) - V(f))\,.
\end{aligned}
\ee
Note that 
\be
S_h^R(f) + S_h^L(f) = I(f) 
\equiv S_h(f)\,,
\ee
which is just the total strain power spectral 
density for the gravitational-wave background.

\subsubsection{Overlap functions}
\label{s:circular_overlap}

Given (\ref{e:iso_pol}), we are now in a position to 
calculate the expected value of the product of the 
Fourier transforms of the response of two detectors
$I$ and $J$ to such a background.
Similar to (\ref{e:responseRA}), we can write the 
response of detector $I$ as
\be
\tilde h_I(f) = \int d^2\Omega_{\hat n}\>
\left(R^R(f,\hat n) h_R(f,\hat n)+R^L(f,\hat n) h_L(f,\hat n)\right)\,,
\ee
where $R$, $L$ label the right and left circular
polarization states for both the Fourier components
and the detector response functions.
Writing down a similar expression for the response
of detector $J$, and using (\ref{e:iso_pol}) to 
evaluate the expected value of the product of the
responses, we find
\be
\langle \tilde h_I(f)\tilde h_J^*(f')\rangle
= \frac{1}{2}\delta(f-f')
\left[\Gamma^{(I)}_{IJ}(f)I(f) + \Gamma^{(V)}_{IJ}(f)V(f)\right]\,,
\label{e:IGammaI+VGammaV}
\ee
where 
\be
\begin{aligned}
\Gamma^{(I)}_{IJ}(f) 
&\equiv\frac{1}{8\pi}\int d^2\Omega_{\hat n}\>
\left[
R^R_I(f,\hat n) R^{R*}_J(f,\hat n) +
R^L_I(f,\hat n) R^{L*}_J(f,\hat n)
\right]\,,
\\
\Gamma^{(V)}_{IJ}(f) 
&\equiv\frac{1}{8\pi}\int d^2\Omega_{\hat n}\>
\left[
R^R_I(f,\hat n) R^{R*}_J(f,\hat n) -
R^L_I(f,\hat n) R^{L*}_J(f,\hat n)
\right]\,,
\end{aligned}
\ee
are the overlap functions for the $I$ and $V$ 
Stokes parameters for 
a polarized isotropic stochastic background.
Using
\be
\begin{aligned}
R^R &= \frac{1}{\sqrt{2}}\left(R^++iR^\times\right)\,,
\\
R^L &= \frac{1}{\sqrt{2}}\left(R^+-iR^\times\right)\,,
\label{e:RR,RL}
\end{aligned}
\ee
we can also write the above overlap functions as
\be
\begin{aligned}
\Gamma^{(I)}_{IJ}(f) 
&\equiv\frac{1}{8\pi}\int d^2\Omega_{\hat n}\>
\left[
R^+_I(f,\hat n) R^{+*}_J(f,\hat n) +
R^\times_I(f,\hat n) R^{\times*}_J(f,\hat n)
\right]\,,
\\
\Gamma^{(V)}_{IJ}(f) 
&\equiv\frac{i}{8\pi}\int d^2\Omega_{\hat n}\>
\left[
R^+_I(f,\hat n) R^{+*}_J(f,\hat n) -
R^\times_I(f,\hat n) R^{\times*}_J(f,\hat n)
\right]\,.
\end{aligned}
\ee
Note that $\Gamma^{(I)}_{IJ}(f)$ is identical to the 
ordinary overlap function $\Gamma_{IJ}(f)$ for an
isotropic background (\ref{e:GammaIJ}).

\begin{figure}[h!tbp]
\begin{center}
\includegraphics[angle=0,width=0.5\columnwidth]{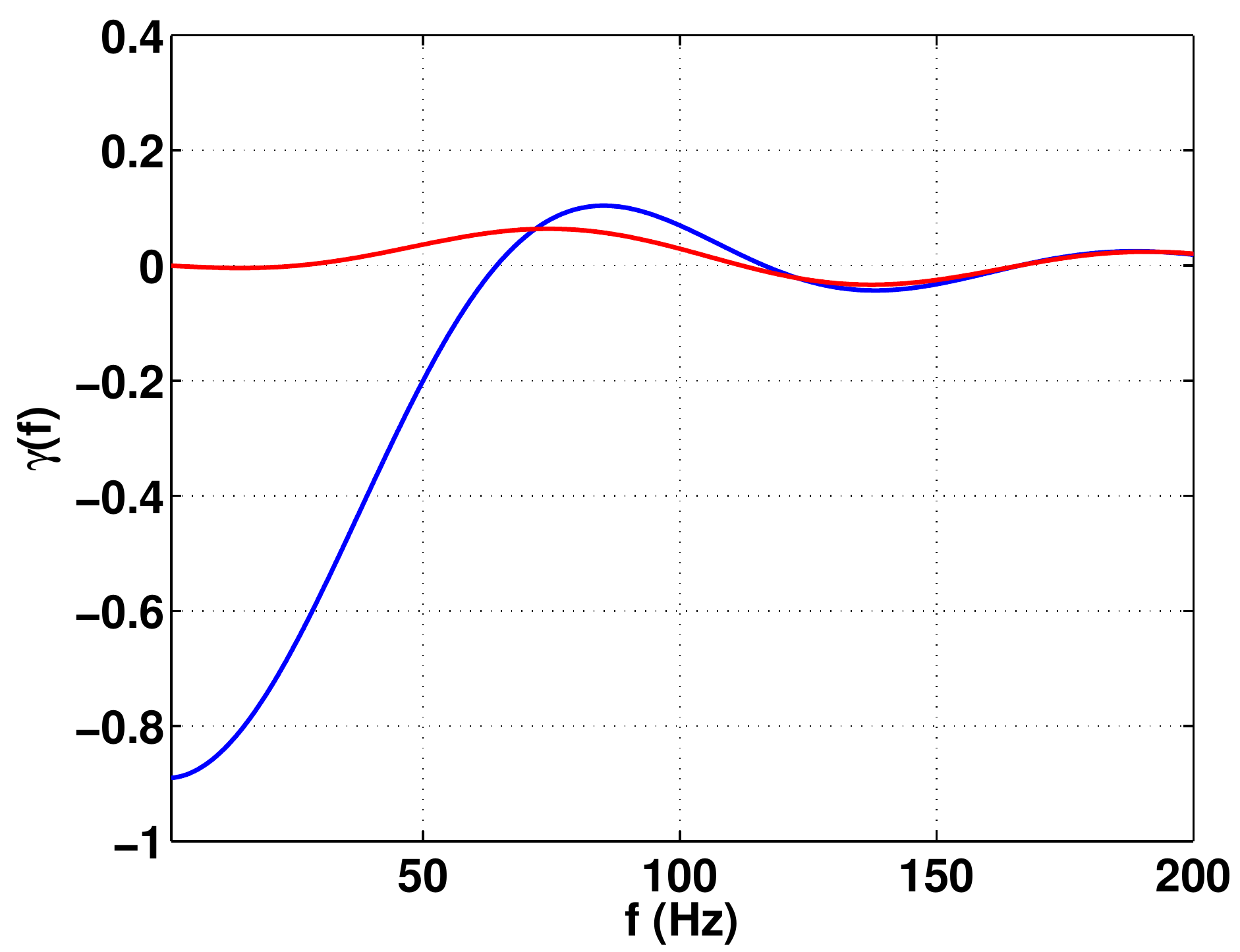}\\
\includegraphics[angle=0,width=0.5\columnwidth]{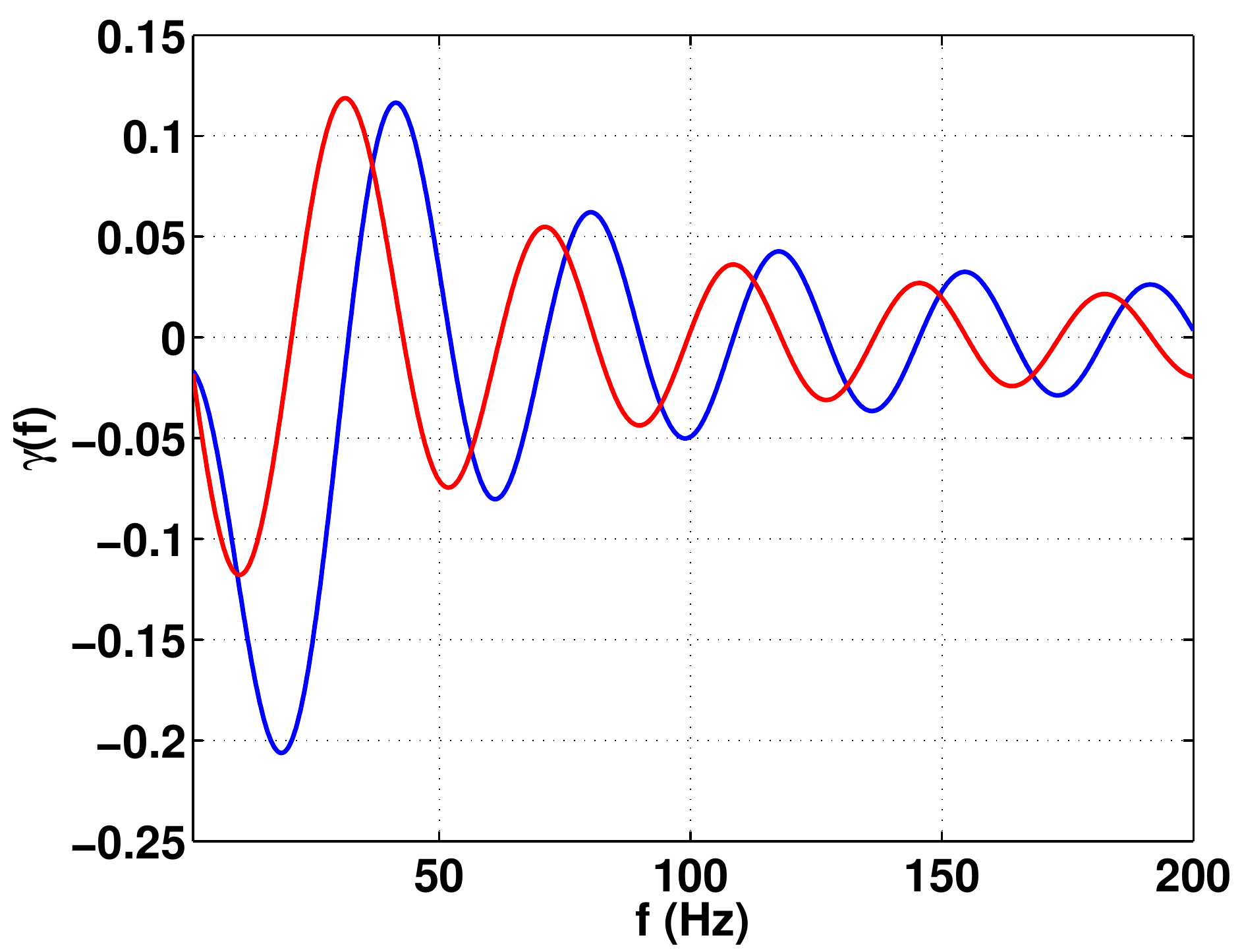}\\
\includegraphics[angle=0,width=0.5\columnwidth]{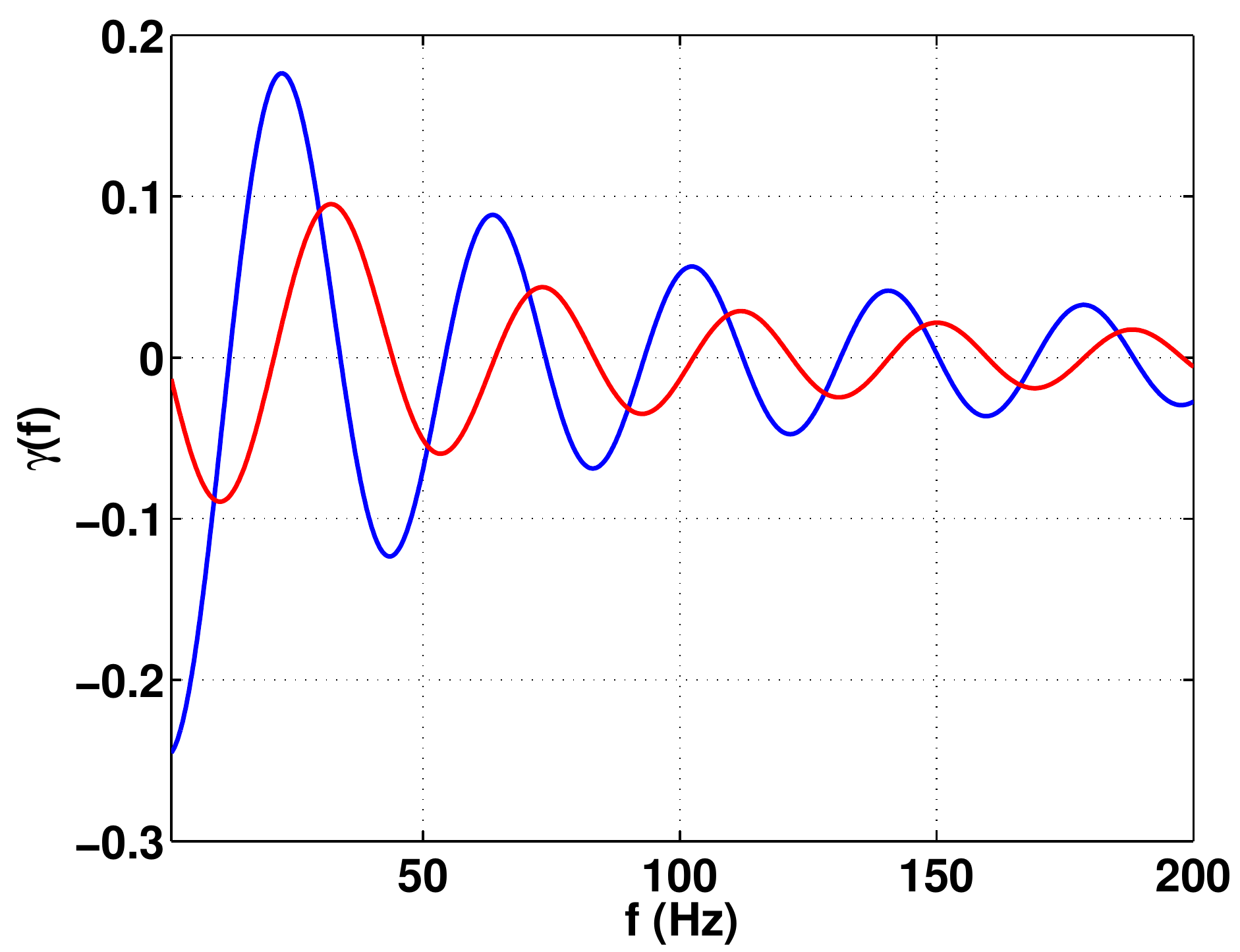}
\caption{Normalized overlap functions for the $I$ and $V$
Stokes' parameters for the 
LIGO Hanford-LIGO Livingston detector pair (top panel);
for the LIGO Hanford-Virgo detector pair (middle panel);
for the LIGO Livingston-Virgo detector pair (bottom panel).
The $I$ overlap functions are shown in blue; 
the $V$ overlap functions are shown in red.
Note the change in scale of the vertical axes.}
\label{f:overlapIV}
\end{center}
\end{figure}

Figure~\ref{f:overlapIV} show plots of the $I$ 
and $V$ overlap functions for the LIGO-Virgo
detector pairs,
using the small-antenna limit for the strain response functions.
The overlap functions have been normalized 
(\ref{e:gamma_normalized}) so that $\gamma^{(I)}_{IJ}(f)=1$
for colocated and coaligned detectors.
Similar plots can be made for other interferometer pairs, 
by simply using the appropriate response functions for 
those detectors.

NOTE: For pulsar timing, one can show that 
$\Gamma^{(V)}_{IJ}(f) = 0$ for any pair of pulsars.
This means that one cannot detect the presence of a 
circularly polarized stochastic background using a
pulsar timing array if one restricts attention to just
the isotropic component of the background.
One must include higher-order multipoles in the 
analysis---i.e., do an {\em anisotropic} search as
discussed in Section~\ref{s:anisotropic}.
Such an analysis for anisotropic polarized backgrounds
using pulsar timing arrays is given in \cite{Kato-Soda:2016}.
In that paper, they extend the analysis of 
\cite{Mingarelli-et-al:2013} to include circular polarization.
See \cite{Kato-Soda:2016} for additional details.

\subsubsection{Component separation: ML estimates of $I$ and $V$}
\label{s:comp_sep_circular}

As shown in \cite{Seto-Taruya:2007, Seto-Taruya:2008},
in order to separate the $I(f)$ and $V(f)$ contributions 
to a polarized isotropic background at each frequency
$f$, we will need to analyze data from at least two 
independent baselines (so three or more detectors).
In what follows, we will use the notation $\alpha=1,2,\cdots, N_b$
to denote the individual baselines (detector pairs)
and $\alpha_1$, $\alpha_2$ to denote the two detectors
that constitute that baseline.
The formalism we adopt here is similar to that for constructing
maximum-likelihood estimators of gravitational-wave
power for unpolarized anisotropic backgrounds (Section~\ref{s:ML}).
For a general discussion of component separation for isotropic
backgrounds, see \cite{Parida-et-al:2016}.

As usual, we begin by cross-correlating the data from
pairs of detectors for the independent baselines:
\be
\hat C_\alpha(f) \equiv 
\frac{2}{T}\,
\tilde d_{\alpha_1}(f) \tilde d^*_{\alpha_2}(f)\,,
\ee
where 
\be
\tilde d_{\alpha_I}(f) 
= \tilde h_{\alpha_I}(\tilde f) + \tilde n_{\alpha_I}(f)\,,
\quad I=1,2\,,
\ee
are the Fourier transforms of the time-domain data 
$d_{\alpha_I}(t)$, and $T$ is the duration of the data.
Assuming that the noise in the individual detectors are 
uncorrelated with one another, we can easily 
calculate the expected value of $\hat C_\alpha(f)$ 
using our previous result (\ref{e:IGammaI+VGammaV}).
The result is
\be
\langle
\hat C_\alpha(f)
\rangle 
=\Gamma^{(I)}_\alpha(f)I(f)+ \Gamma^{(V)}_\alpha(f)V(f)\,.
\ee
We will write this equation abstractly as a matrix
equation
\be
\langle \hat C\rangle = M {\cal S}\,,
\label{e:<Chat>_stokes}
\ee
where 
\be
\hat C = \left[
\begin{array}{c}
\hat C_1 \\
\hat C_2 \\
\vdots \\
\hat C_{N_b} \\
\end{array}
\right]\,,
\quad
M \equiv
\left[
\begin{array}{cc}
\Gamma_1^{(I)} &
\Gamma_1^{(V)} 
\\
\Gamma_2^{(I)} &
\Gamma_2^{(V)} 
\\
\vdots & \vdots
\\
\Gamma_{N_b}^{(I)} &
\Gamma_{N_b}^{(V)} 
\end{array}
\right]\,,
\quad
{\cal S}\equiv
\left[
\begin{array}{c}
I \\
V \\
\end{array}
\right]\,.
\ee
In this notation,
$\hat C$ is an $N_f N_b\times 1$ data vector,
$M$ is an $N_f N_b\times 2 N_f$ 
detector network response matrix,
and ${\cal S}$ is an $2N_f\times 1$ vector containing the
unknown Stokes' parameters, which we want to estimate 
from the data.%
\footnote{At times it will be convenient to think of $M$
as an $N_f\times N_f$ block diagonal matrix
with $N_b\times 2$ blocks, one for each frequency.
At other times, it will be convenient to think
of $M$ as an $N_b\times 2$ block matrix with 
diagonal $N_f\times N_f$ blocks.
The calculations we need to do usually determine
which representation is most appropriate.
(Similar statements can be made for the vectors
$\hat C$ and ${\cal S}$.)}

We also need an expression for the noise covariance 
matrix ${\cal N}$ for the cross-correlated data $\hat C$.
In the weak-signal limit, the covariance matrix is
approximately diagonal with matrix elements
\be
\begin{aligned}
{\cal N}_{\alpha\alpha'}(f,f') 
&\equiv \langle \hat C_\alpha(f) \hat C^*_{\alpha'}(f')\rangle
-\langle \hat C_\alpha(f)\rangle\langle \hat C^*_{\alpha'}(f')\rangle
\\
&\approx 
\delta_{\alpha\alpha'}
\delta_{ff'}
P_{n_{\alpha_1}}(f) P_{n_{\alpha_2}}(f)\,,
\label{e:calN_alpha,alpha'}
\end{aligned}
\ee
where $P_{n_{\alpha_I}}(f)$ are the one-sided power spectral densities
of the noise in the detectors. 
If we treat the noise power spectra as known quantities 
(or if we estimate the noise power spectra 
from the auto-correlated output of each detector), 
we can write down a likelihood function for the 
cross-correlated data given the signal model (\ref{e:<Chat>_stokes}).
Assuming a Gaussian-stationary distribution for the noise, 
we have
\be
p(\hat C|{\cal S}) \propto
\exp\left[-\frac{1}{2}(\hat C - M{\cal S})^\dagger 
{\cal N}^{-1} (\hat C - M{\cal S})\right]\,.
\label{e:likehood_S}
\ee
This likelihood has exactly the same form as that 
in (\ref{e:likelihood_P(k)}), so the maximum-likehood 
estimators for the Stokes' parameters 
${\cal S} = [I, V]^T$ also have the same form:
\be
\hat {\cal S}
=F^{-1} X\,,
\label{e:Shat}
\ee
where
\be
F\equiv M^\dagger {\cal N}^{-1} M\,,
\qquad
X\equiv M^\dagger {\cal N}^{-1} \hat C\,,
\label{e:F,X_S}
\ee
with $M$ and ${\cal N}$ given above.
As before, inverting $F$ may require some sort of
regularization, e.g., using
singular-value decomposition (Section~\ref{s:svd_power}).
If that's the case then $F^{-1}$ should be replaced
in the above formula by its pseudo-inverse $F^+$.
The uncertainty in the maximum likelihood recoved 
values is given by the covariance matrix
\be
\langle \hat{\cal S}\hat{\cal S}^\dagger\rangle-
\langle \hat{\cal S}\rangle\langle \hat{\cal S}^\dagger\rangle
\approx F^{-1}\,,
\ee
where we are again assuming the weak-signal limit.

\subsubsection{Example: Component separation for two baselines}
\label{s:example-comp-sep}

As an explicit example, we now write down the 
maximum-likelihood estimators for the Stokes'
parameters ${\cal S} = [I, V]^T$ for a detector
network consisting of 
two baselines $\alpha$ and $\beta$.
For this case, the detector 
network response matrix $M$ is a 
square $2 N_f\times 2 N_f$ matrix, which we 
assume has non-zero determinant.
Then it follows simply from the definitions 
(\ref{e:F,X_S}) of $F$ and $X$ that
\be 
\hat{\cal S} = F^{-1} X  = M^{-1}\hat C\,,
\ee
for which 
\be
\begin{aligned}
\hat I(f)
&= 
\left(\Gamma^{(I)}_\alpha\Gamma^{(V)}_\beta - \Gamma^{(I)}_\beta\Gamma^{(V)}_\alpha\right)^{-1}
\left[\Gamma^{(V)}_\beta \hat C_\alpha - \Gamma^{(V)}_\alpha \hat C_\beta\right]\,,
\\
\hat V(f)
&=
\left(\Gamma^{(I)}_\alpha\Gamma^{(V)}_\beta - \Gamma^{(I)}_\beta\Gamma^{(V)}_\alpha\right)^{-1}
\left[-\Gamma^{(I)}_\beta \hat C_\alpha + \Gamma^{(I)}_\alpha \hat C_\beta\right]\,.
\end{aligned}
\ee
The marginalized uncertainties in these estimates
are obtained by taking the diagonal elements of the 
inverse of the Fisher matrix:
\be
\begin{aligned}
\sigma_{\hat I}^2 
&= (F^{-1})_{II}
= \frac{N_\alpha\,(\Gamma^{(V)}_\beta)^2+
N_\beta\,(\Gamma^{(V)}_\alpha)^2}
{\left(\Gamma^{(I)}_\alpha \Gamma^{(V)}_\beta -
\Gamma^{(I)}_\beta \Gamma^{(V)}_\alpha\right)^2}\,,
\\
\sigma_{\hat V}^2 
&= (F^{-1})_{VV}
= \frac{N_\alpha\,(\Gamma^{(I)}_\beta)^2+
N_\beta\,(\Gamma^{(I)}_\alpha)^2}
{\left(\Gamma^{(I)}_\alpha \Gamma^{(V)}_\beta -
\Gamma^{(I)}_\beta \Gamma^{(V)}_\alpha\right)^2}\,,
\end{aligned}
\label{e:sigma2I,V}
\ee
where $N_\alpha$, $N_\beta$, defined by 
$N_\alpha(f)\equiv P_{n_{\alpha_1}}(f) P_{n_{\alpha_2}}(f)$
(and similarly for $N_\beta$), is a diagonal element of the 
noise covariance matrix ${\cal N}$ (\ref{e:calN_alpha,alpha'}).

\subsubsection{Effective overlap functions for $I$ and $V$ 
for multiple baselines}
\label{s:gamma_eff_circ}

\begin{figure}[h!tbp]
\begin{center}
\includegraphics[angle=0,width=0.6\columnwidth]{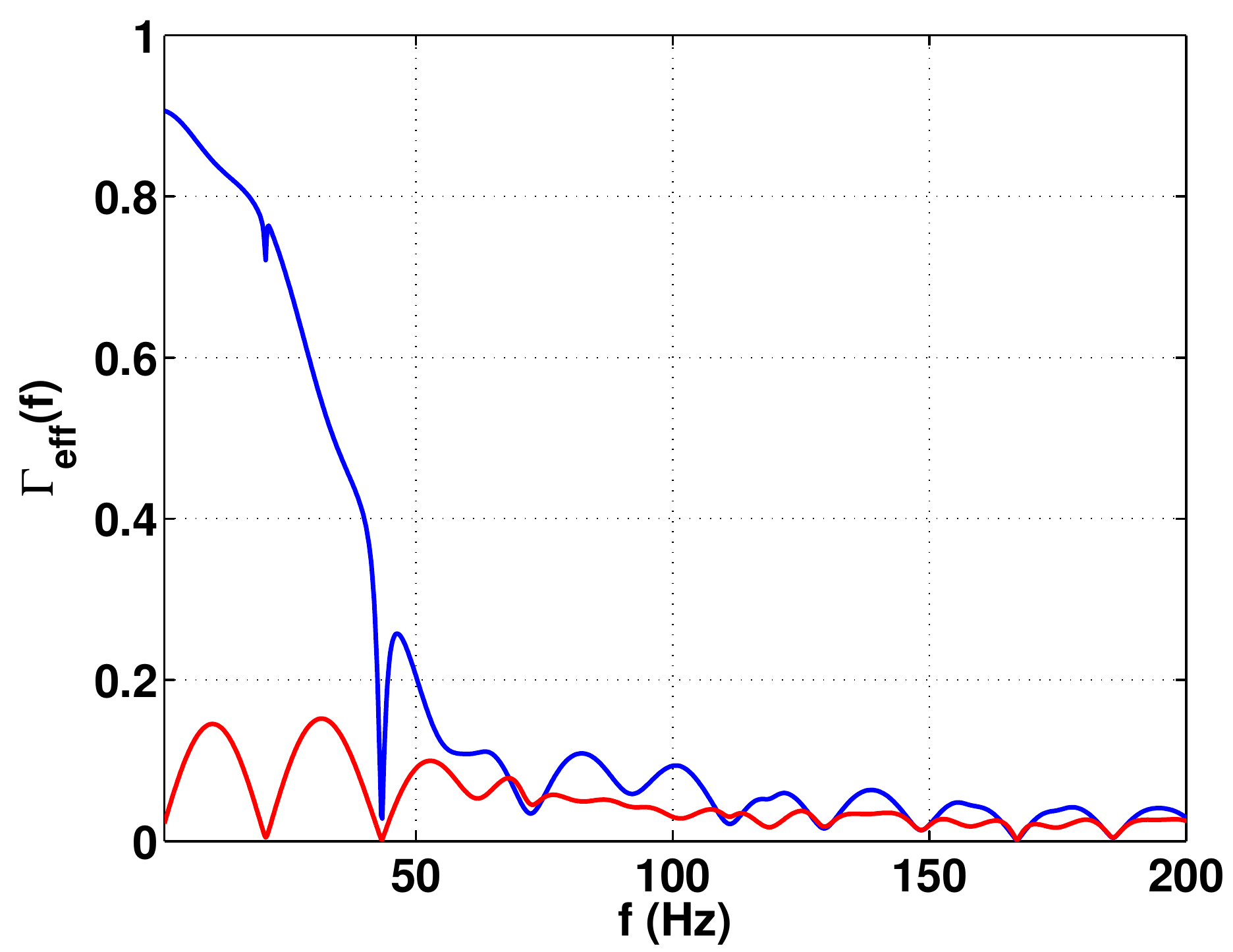}
\caption{Effective overlap functions for $I$ and $V$
for the multibaseline network formed from the LIGO Hanford,
LIGO Livingston, and Virgo detectors.
$\Gamma_{\rm eff}^{(I)}(f)$ is shown in blue;
$\Gamma_{\rm eff}^{(V)}(f)$ is shown in red.}
\label{f:overlap_eff}
\end{center}
\end{figure}

The above expressions for the uncertainties in 
the estimates of $I$ and $V$ can easily be extended
to the case of an arbitray number of baselines 
$\alpha = 1,2,\cdots, N_b$.
For multiple baselines with noise spectra
$N_\alpha(f)\equiv P_{n_{\alpha_1}}(f) P_{n_{\alpha_2}}(f)$,
one can show that 
\be
F =
\left[
\begin{array}{cc}
\sum_\alpha N_\alpha^{-1}(\Gamma^{(I)}_\alpha)^2 &
\sum_\alpha N_\alpha^{-1}\Gamma^{(I)}_\alpha \Gamma^{(V)}_\alpha \\
\sum_\alpha N_\alpha^{-1}\Gamma^{(V)}_\alpha \Gamma^{(I)}_\alpha &
\sum_\alpha N_\alpha^{-1}(\Gamma^{(V)}_\alpha)^2 \\
\end{array}
\right]\,.
\ee
Let us assume that the determinant of the 
$2\times 2$ matrices for each frequency (which we will denote
by $\bar F$) are not equal to zero.
Then
\be
\begin{aligned}
\sigma_{\hat I}^2 
&= (\bar F^{-1})_{II}
= \frac{1}{\det(\bar F)}\,
\sum_\alpha N_\alpha^{-1}(\Gamma^{(V)}_\alpha)^2\,,
\\
\sigma_{\hat V}^2 
&= (\bar F^{-1})_{VV}
= \frac{1}{\det(\bar F)}\,
\sum_\alpha N_\alpha^{-1}(\Gamma^{(I)}_\alpha)^2\,.
\label{e:sigma2I,V_network}
\end{aligned}
\ee
Following \cite{Seto-Taruya:2008}, we can now define 
{\em effective} overlap functions for $I$ and $V$ 
associated with a multibaseline
detector network by basically inverting the above 
uncertainties.
For simplicity, we will assume that the noise power 
spectra for the detectors are equal to one another 
so that $N_\alpha\equiv N$ can be factored out of the 
above expressions.
We then define 
\be
\begin{aligned}
\Gamma_{\rm eff}^{(I)}(f) \equiv  \sqrt{N}\sigma_{\hat I}^{-1}
&=\left(
\frac{N^2\det(\bar F)}
{\sum_\alpha(\Gamma^{(V)}_\alpha)^2}\right)^{1/2}\,,
\\
\Gamma_{\rm eff}^{(V)}(f) \equiv  \sqrt{N}\sigma_{\hat V}^{-1}
&=\left(
\frac{N^2\det(\bar F)}
{\sum_\alpha(\Gamma^{(I)}_\alpha)^2}\right)^{1/2}\,.
\end{aligned}
\ee
These quantities give us an indication of how 
sensitive the multibaseline network is to 
extracting the $I$ and $V$ components of the background.
Plots of $\Gamma_{\rm eff}^{(I)}(f)$ and $\Gamma_{\rm eff}^{(V)}(f)$ 
are shown in Figure~\ref{f:overlap_eff} for the 
multibaseline network formed from the 
LIGO Hanford, LIGO Livingston, and Virgo detectors.
Recall that the overlap functions for the individual 
detectors pairs are shown in Figure~\ref{f:overlapIV}.
Dips in sensitivity correspond to frequencies 
where the determinant of $\bar F$ is zero (or close to zero).
%

\subsection{non-GR polarization modes: Preliminaries}
\label{s:altpol}

\begin{figure}[h!tbp]
\begin{center}
\includegraphics[angle=0,width=0.85\columnwidth]{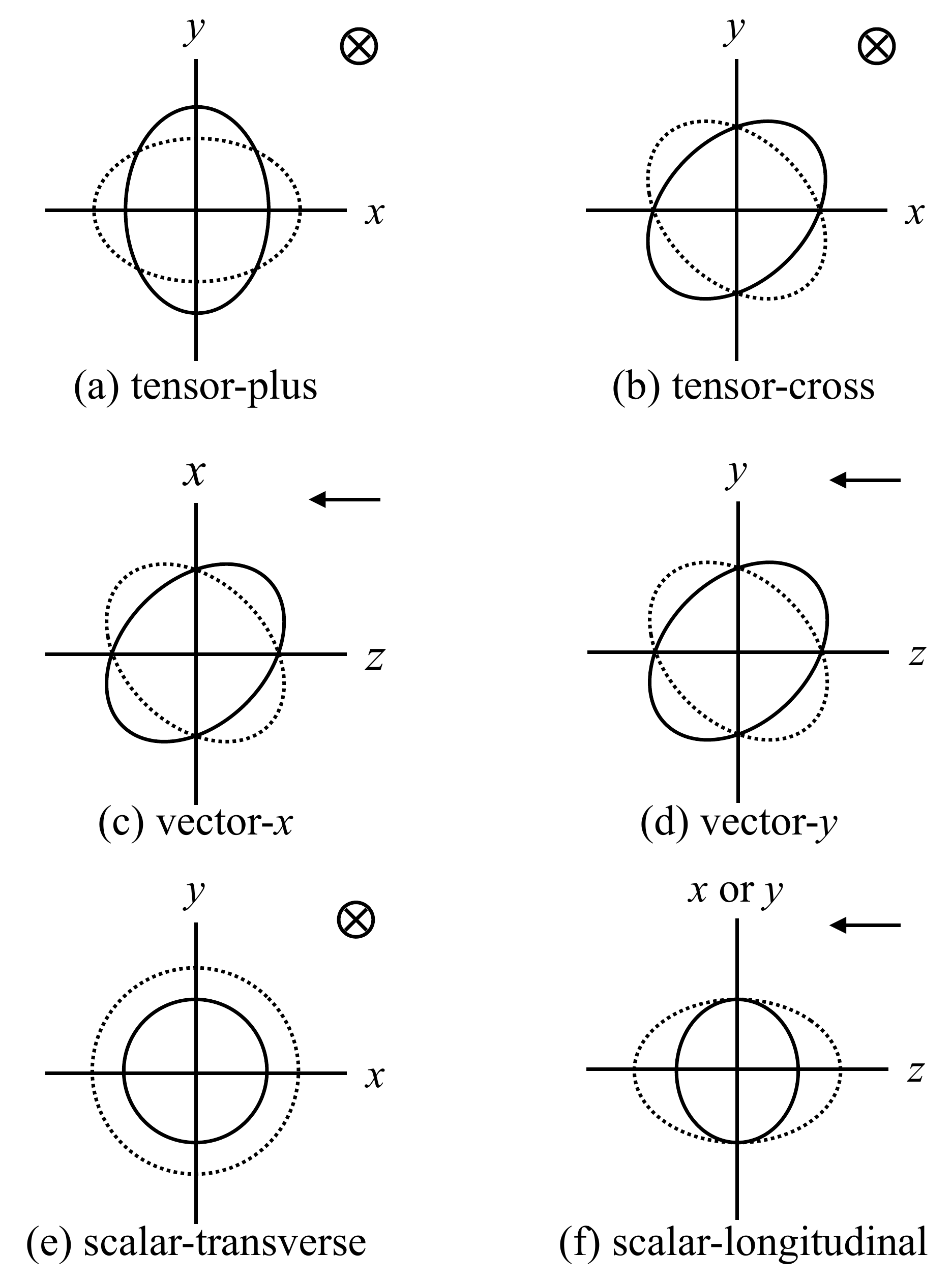}
\caption{Graphical representation of the six different polarization modes.
The circle with a cross or arrow represents the direction of propagation of 
the gravitational wave.
The solid and dotted circles and ellipses 
denote deformations to a ring of particles
$180^\circ$ out of phase with one another.
Adapted from Figure~1 in \cite{Nishizawa-et-al:2009}.}
\label{f:eab_graphical}
\end{center}
\end{figure}

In a general metric theory of gravity, there
are six possible polarization modes:
The standard $+$ and $\times$ {\em tensor} 
modes predicted by general relativity (GR);
two {\em vector} (or ``shear") modes, 
which we will denote by $X$ and $Y$; 
and two {\em scalar} modes:
a ``breathing" mode $B$
and a pure longitudinal mode $L$
(see, e.g., \cite{Nishizawa-et-al:2009}).
The tensor and breathing modes are {\em transverse} 
to the direction of propagation, while the two 
vector modes and the scalar longitudinal mode have 
{\em longitudinal} components (parallel to the
direction of propagation).
See Figure~\ref{f:eab_graphical}.

In terms of the orthonormal vectors 
$\{\hat n,\hat l,\hat m\}$
defined by (\ref{e:nlm_def}), 
the polarization basis tensors for the  
six different polarization modes are:
\be
\begin{aligned}
&e_{ab}^{+}(\hat n)
=\hat l_a\hat l_b-\hat m_a\hat m_b\,,
\quad
&e_{ab}^{\times}(\hat n)
=\hat l_a\hat m_b + \hat m_a\hat l_b\,.
\\
&e_{ab}^{X}(\hat n)
=\hat l_a\hat n_b+\hat n_a\hat l_b\,,
\quad
&e_{ab}^{Y}(\hat n)
=\hat m_a\hat n_b + \hat n_a\hat m_b\,,
\\
&e_{ab}^{B}(\hat n)
=\hat l_a\hat l_b+\hat m_a\hat m_b\,,
\quad
&e_{ab}^{L}(\hat n)
=\sqrt{2}\,\hat n_a\hat n_b\,.
\label{e:eab_altpol_def}
\end{aligned}
\ee
We will denote these collectively as 
$e^A_{ab}(\hat n)$, where $A=\{+,\times,X,Y,B,L\}$.
In a coordinate system where $\hat n$ points along the 
$z$-axis, and $\hat l$ and $\hat m$ point along the 
$x$ and $y$ axes, the polarization tensors can be
represented by the following $3\times 3$ matrices:
\be
\begin{aligned}
&e_{ab}^+=
\left[
\begin{array}{ccc}
1 & 0 & 0\\
0 & -1 & 0\\
0 & 0 & 0\\
\end{array}
\right]\,,
\quad
&e_{ab}^\times=
\left[
\begin{array}{ccc}
0 & 1 & 0\\
1 & 0 & 0\\
0 & 0 & 0\\
\end{array}
\right]\,,
\\
&e_{ab}^X=
\left[
\begin{array}{ccc}
0 & 0 & 1\\
0 & 0 & 0\\
1 & 0 & 0\\
\end{array}
\right]\,,
\quad
&e_{ab}^Y=
\left[
\begin{array}{ccc}
0 & 0 & 0\\
0 & 0 & 1\\
0 & 1 & 0\\
\end{array}
\right]\,,
\\
&e_{ab}^B=
\left[
\begin{array}{ccc}
1 & 0 & 0\\
0 & 1 & 0\\
0 & 0 & 0\\
\end{array}
\right]\,,
\quad
&e_{ab}^L=
\left[
\begin{array}{ccc}
0 & 0 & 0\\
0 & 0 & 0\\
0 & 0 & \sqrt{2}\\
\end{array}
\right]\,.
\label{e:eab_alppol_matrices}
\end{aligned}
\ee
%

\subsubsection{Tranformation of the polarization tensors
under a rotation about $\hat n$}

We have already seen 
(Appendix~\ref{s:polarization_tensors})
that under a rotation of the unit vectors
$\{\hat l, \hat m\}$ by an angle $\psi$ around $\hat n$, 
the polarization tensors $e^+_{ab}(\hat n)$, $e^\times_{ab}(\hat n)$ 
transform to:
\be
\begin{aligned}
\epsilon^+_{ab}(\hat n,\psi) 
&=\cos 2\psi\, e^+_{ab}(\hat n)+\sin 2\psi\, e^\times_{ab}(\hat n)\,,
\\
\epsilon^\times_{ab}(\hat n,\psi) 
&=-\sin 2\psi\, e^+_{ab}(\hat n)+\cos 2\psi\, e^\times_{ab}(\hat n)\,.
\end{aligned}
\label{e:spin2-transform}
\ee 
This reflects the spin 2 nature of the
tensor modes $+$, $\times$ in general relativity.
Similarly, under the same rotation, the polarization tensors
$e^X_{ab}(\hat n)$, $e^Y_{ab}(\hat n)$ transform to:
\be
\begin{aligned}
\epsilon^X_{ab}(\hat n,\psi) 
&=\cos \psi\, e^X_{ab}(\hat n)+\sin \psi\, e^Y_{ab}(\hat n)\,,
\\
\epsilon^Y_{ab}(\hat n,\psi) 
&=-\sin \psi\, e^X_{ab}(\hat n)+\cos \psi\, e^Y_{ab}(\hat n)\,,
\end{aligned}
\label{e:spin1-transform}
\ee 
while $e^B_{ab}(\hat n)$, $e^L_{ab}(\hat n)$ are left
unchanged:
\be
\begin{aligned}
\epsilon^B_{ab}(\hat n,\psi) & = e^B_{ab}(\hat n)\,,
\\
\epsilon^L_{ab}(\hat n,\psi) & = e^L_{ab}(\hat n)\,.
\end{aligned}
\label{e:spin0-transform}
\ee
These last two transformations correspond to the spin~1
nature of the vector modes $X$, $Y$, and the 
spin~0 nature of the scalar modes $B$, $L$.

\subsubsection{Polarization and spherical harmonic basis expansions}
\label{e:altpol_sphharm}

For the tensor modes $+$, $\times$, we found 
(Section~\ref{s:sphericalharmonicbasis})
that it was convenient to expand the Fourier 
components $h_{ab}(f,\hat k)$ of the metric perturbations
$h_{ab}(t,\vec x)$ 
in terms of either the polarization basis tensors:
\be
h_{ab}(f,\hat n) = h_+(f,\hat n) e^+_{ab}(\hat n)
+ h_\times(f,\hat n) e^\times_{ab}(\hat n)\,,
\ee
or the gradient and curl tensor 
spherical harmonics:
\be
h_{ab}(f,\hat n)
=\sum_{l=2}^\infty \sum_{m=-l}^l
\left[a^G_{(lm)}(f)Y^G_{(lm)ab}(\hat n)
+a^C_{(lm)}(f)Y^C_{(lm)ab}(\hat n)\right]\,.
\ee
Recall that $Y^G$ and $Y^C$ are related to 
spin-weight~$\pm 2$ spherical harmonics as
described in Appendices~\ref{s:grad-curl-tensor}
and \ref{s:spinweightedY}.
For the vector and scalar modes we can 
perform similar expansions, provided we 
use appropriately defined tensor spherical
harmonics, which transform properly under rotations.
For the vector modes $X$, $Y$, we need to 
use the {\em vector-gradient} and
{\em vector-curl} spherical harmonics
$Y^{V_G}$, 
$Y^{V_C}$, which are defined
in terms of spin-weight~$\pm 1$ spherical 
harmonics (Appendices~\ref{s:grad-curl-vector} and
\ref{s:spinweightedY}).
For the scalar modes, we can use
\be
\begin{aligned}
Y^B_{(lm)ab}(\hat n)\equiv
\frac{1}{\sqrt{2}} Y_{lm}(\hat n) e^B_{ab}(\hat n)\,,
\qquad
Y^L_{(lm)ab}(\hat n)\equiv
\frac{1}{\sqrt{2}} Y_{lm}(\hat n) e^L_{ab}(\hat n)\,,
\end{aligned}
\ee
which are defined in terms of ordinary (scalar)
spherical harmonics.
In terms of these definitions, we can write
the expansions in compact form
\be
h_{ab}(f,\hat n) = \sum_A h_A(f,\hat n) e^A_{ab}(\hat n)\,,
\label{e:altpol_exp_pol}
\ee
or
\be
h_{ab}(f,\hat n)
= \sum_P \sum_{(lm)}
a^P_{(lm)}(f)Y^P_{(lm)ab}(\hat n)\,,
\label{e:altpol_exp_sph}
\ee
where $A=\{+,\times, X, Y, B, L\}$ and
$P=\{G,C, V_G, V_C, B, L\}$ or some subsets 
thereof.
Note that $\sum_{(lm)}$ is shorthand for
\be
\sum_{l=2}^\infty \sum_{m=-l}^l\,,
\qquad
\sum_{l=1}^\infty \sum_{m=-l}^l\,,
\qquad
\sum_{l=0}^\infty \sum_{m=-l}^l\,,
\ee
for the tensor, vector, and scalar modes, respectively.

\subsubsection{Detector response}

The detector response functions corresponding to
the above two expansions 
(\ref{e:altpol_exp_pol}) and (\ref{e:altpol_exp_sph})
are:
\be
R^A(f,\hat n) = R^{ab}(f,\hat n) e^A_{ab}(\hat n)\,,
\ee
and
\be
R^P_{(lm)}(f) =\int d^2\Omega_{\hat n}\>
R^{ab}(f, \hat n) Y^P_{(lm)ab}(\hat n)\,.
\ee
In terms of these response functions, the detector
response (in the frequency domain) to a 
gravitational-wave background (\ref{e:planewave}) is:
\be
\tilde h(f) = 
\int d^2\Omega_{\hat n}\>
\sum_A R^A(f,\hat n) h_A(f,\hat n)\,,
\ee
or
\be
\tilde h(f) =
\sum_P \sum_{(lm)}
R^P_{(lm)}(f)a^P_{(lm)}(f)\,.
\ee
%

\subsubsection{Searches for non-GR polarizations using different detectors}
\label{s:search_w_diff_detectors}

Evidence for non-GR polarization modes can show up in searches 
for {\em either} deterministic or stochastic gravitational-wave signals.
Whether these alternative polarization modes are first discovered from 
the observation of gravitational waves from a resolvable source 
(like a binary black hole merger) or from a stochastic background 
depends in part on the type and number of detectors making the 
observations.
For example, individual binary black hole mergers (GW150914 and 
GW151226) have already been observed by advanced LIGO.
But it was not possible to extract information about the 
polarization of the waves, since the two LIGO interferometers
are effectively co-aligned (and hence see the {\em same} 
polarization).
Adding Virgo, KAGRA, and LIGO-India to the global network will 
eventually allow for the extraction of this polarization information.
Pulsar timing arrays, on the other hand, are expected to 
first detect a stochastic background from the inspirals of 
SMBHBs in the centers of distant galaxies~\cite{Rosado:2015epa}.
So if evidence of alternative polarization modes are discovered 
by pulsar timing, it will most-likely first come from stochastic 
background observations.

In the following sections, we describe stochastic background
search methods for non-GR polarization modes using both 
ground-based interferometers (Section~\ref{s:altpolIFO})
and pulsar timing arrays (Section~\ref{s:altpolPTA}).
We will calculate antenna patterns, overlap 
functions, and discuss component separation for 
the tensor, vector, and scalar polarization modes.
For ground-based interferometers, our discussion 
will be based on~\cite{Nishizawa-et-al:2009}.
For pulsar timing arrays, see~\cite{Lee-et-al:2008, 
Chamberlin-Siemens:2012, Gair-et-al:2015}.

\subsection{Searches for non-GR polarizations using ground-based detectors}
\label{s:altpolIFO}

We now describe cross-correlation searches for 
non-GR polarization modes using a network of
ground-based laser interferometers.
For additional details, see \cite{Nishizawa-et-al:2009}.

\subsubsection{Response functions}
\label{s:response_altpolIFO}

For ground-based interferometers in the small
antenna limit, the strain response functions
$R^A(f,\hat n)$ for the different polarization 
modes $A=\{+,\times, X,Y, B, L\}$ are given by
\be
R^A(f,\hat n) \simeq 
\frac{1}{2}(u^a u^b - v^a v^b)e^A_{ab}(\hat n)\,,
\ee
where $\hat u$, $\hat v$ are unit vectors pointing
in the direction of the arms of the interferometer,
and where we have chosen the origin of coordinates to be 
at the vertex of the interferometer.
Note that there is no frequency dependence of 
the response function in the small-antenna limit.
Assuming a $90^\circ$ opening angle between the
interferometer arms, and choosing a coordinate 
system such that $\hat u$ and $\hat v$ point in the
$\hat x$ and $\hat y$ direction, we find
\be
\begin{aligned}
R^+(\hat n) 
& = \frac{1}{2}(1+\cos^2\theta)\cos 2\phi\,,
\qquad
&R^\times(\hat n) 
= -\cos\theta\sin 2\phi\,,
\\
R^X(\hat n) 
& = \sin\theta\cos\theta\cos2\phi\,,
\qquad
&R^Y(\hat n) 
= -\sin\theta\sin 2\phi\,,
\\
R^B(\hat n) 
& = -\frac{1}{2}\sin^2\theta\cos2\phi\,,
\qquad
&R^L(\hat n) 
= \frac{1}{\sqrt{2}}\sin^2\theta\cos2\phi\,,
\end{aligned}
\ee 
where we used (\ref{e:nlm_def}) for our definition
of $\{\hat n,\hat l,\hat m\}$.

\begin{figure}[h!tbp]
\begin{center}
\includegraphics[angle=0,width=0.75\columnwidth]{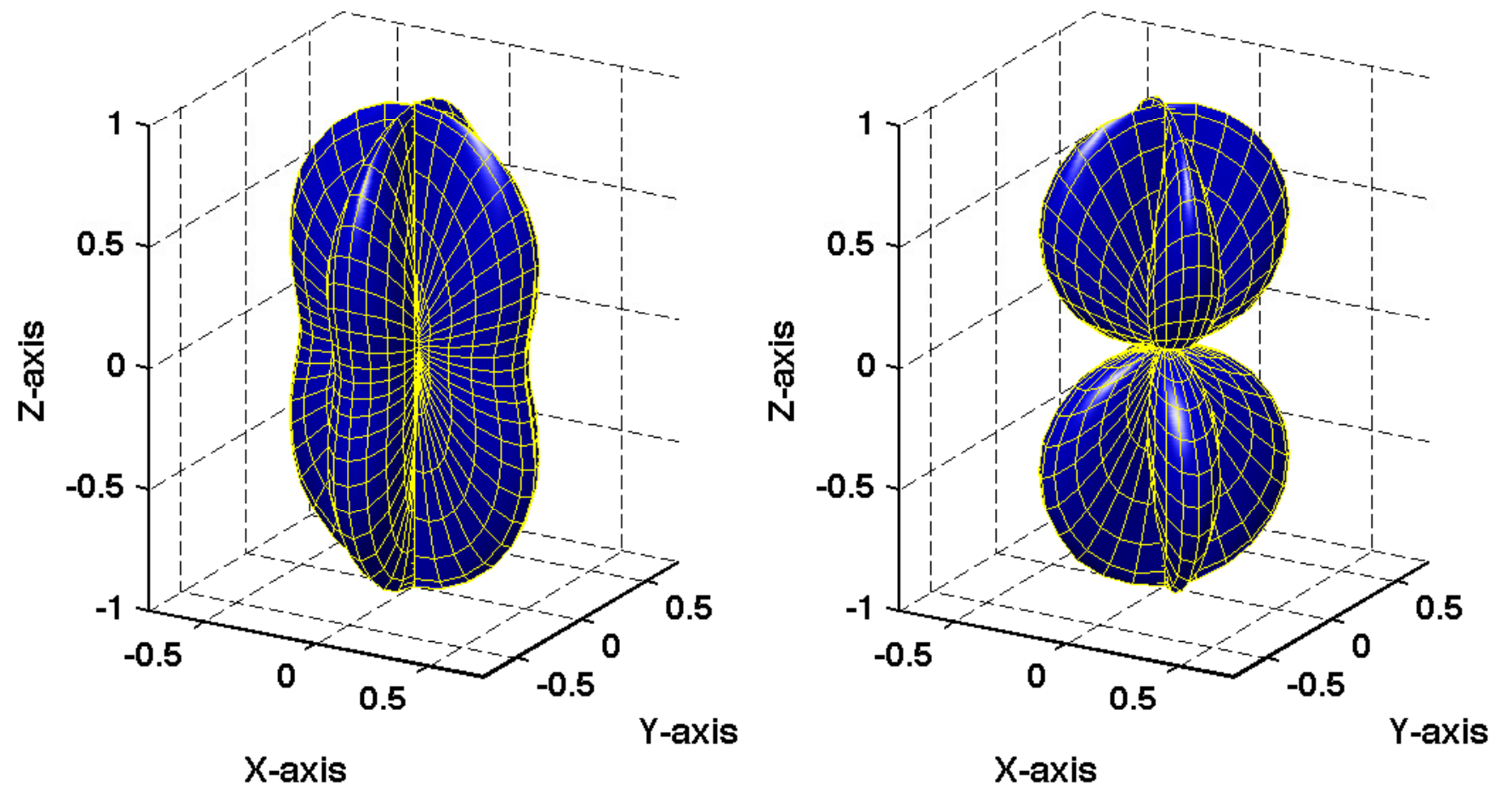}
\includegraphics[angle=0,width=0.75\columnwidth]{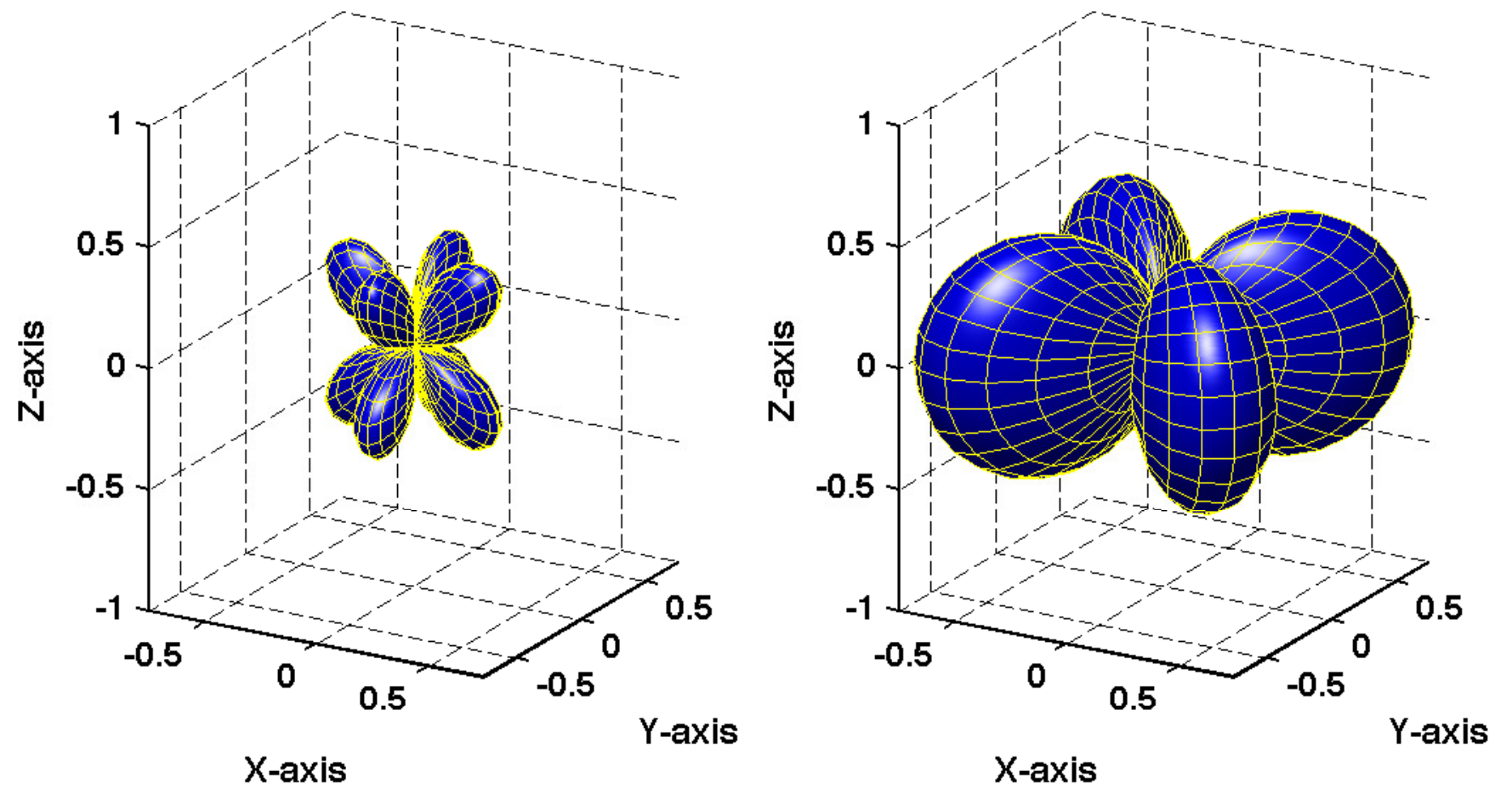}
\includegraphics[angle=0,width=0.75\columnwidth]{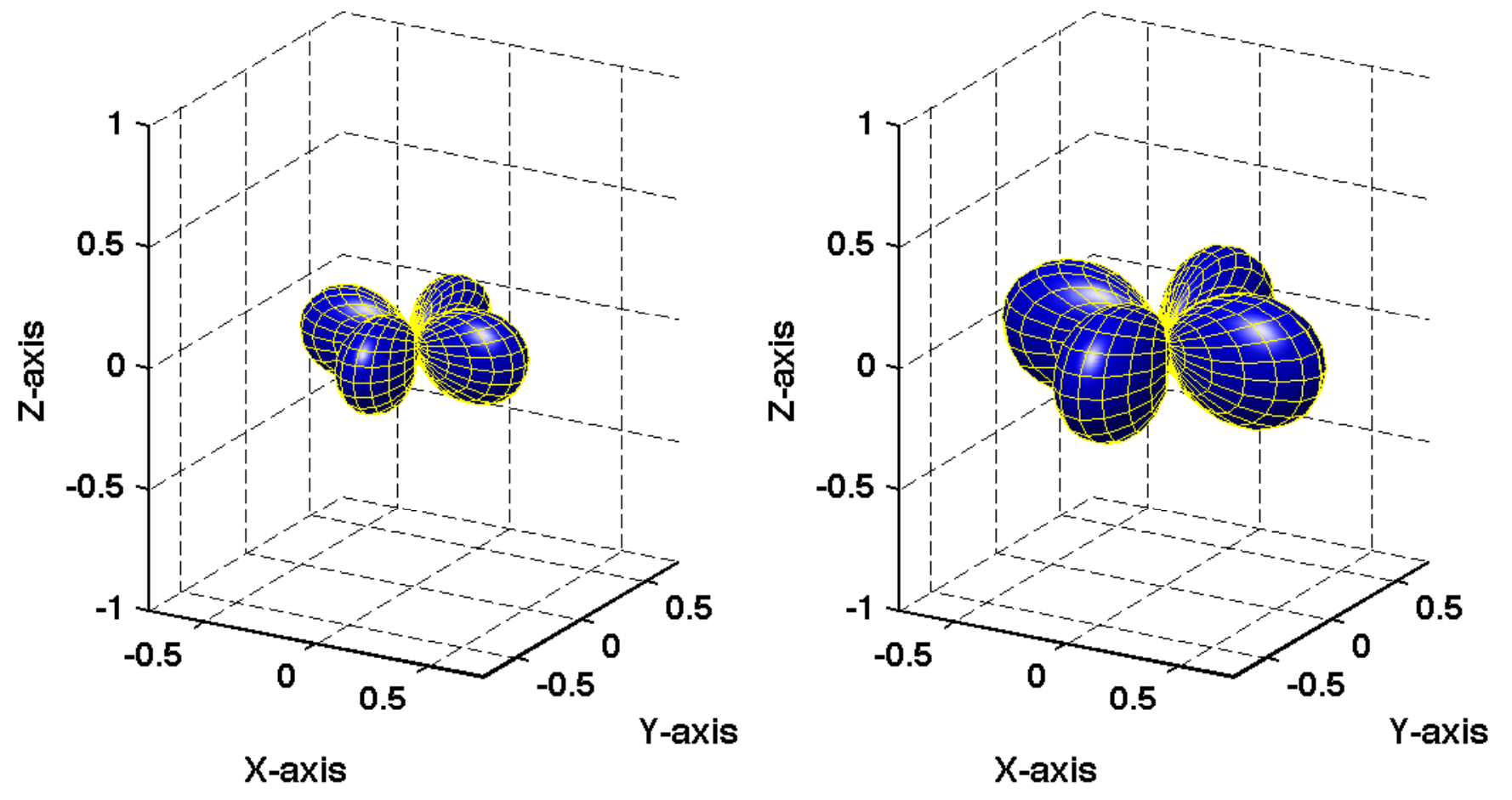}
\caption{Antenna patterns for Michelson interferometer strain
response $|R^A(\hat n)|$ evaluated in the small-antenna limit, $f=0$.
The top two plots correspond to the two tensor modes, $A=+, \times$.
The middle two plots correspond to the two vector modes, $A=X, Y$.
The bottom two plots correspond to the two scalar modes, $A=B, L$.
The interferometer arms point in the $\hat x$ and $\hat y$ directions.}
\label{f:pattern3dAltPolIFO}
\end{center}
\end{figure}

\begin{figure}[h!tbp]
\begin{center}
\includegraphics[angle=0,width=0.32\columnwidth]{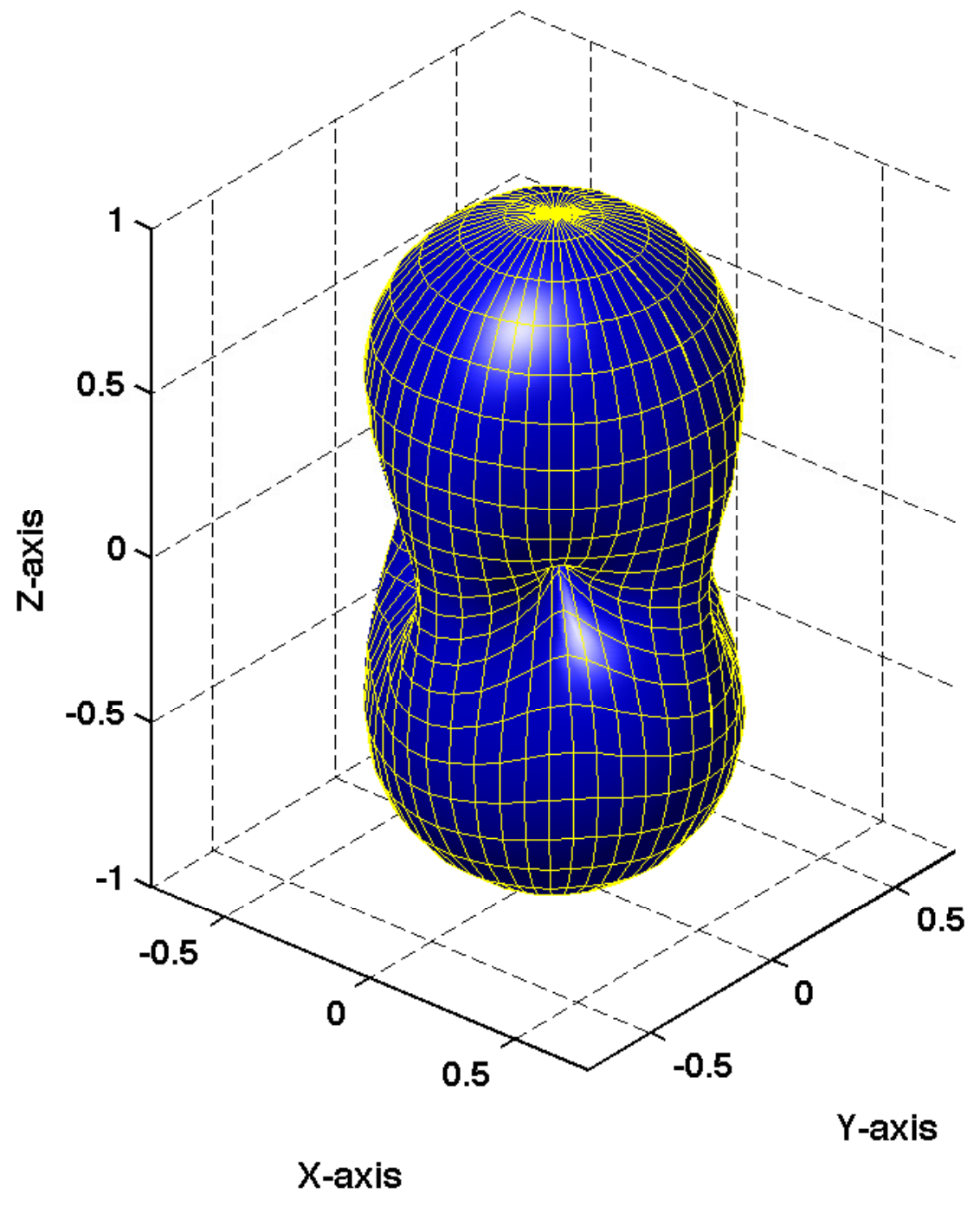}
\includegraphics[angle=0,width=0.32\columnwidth]{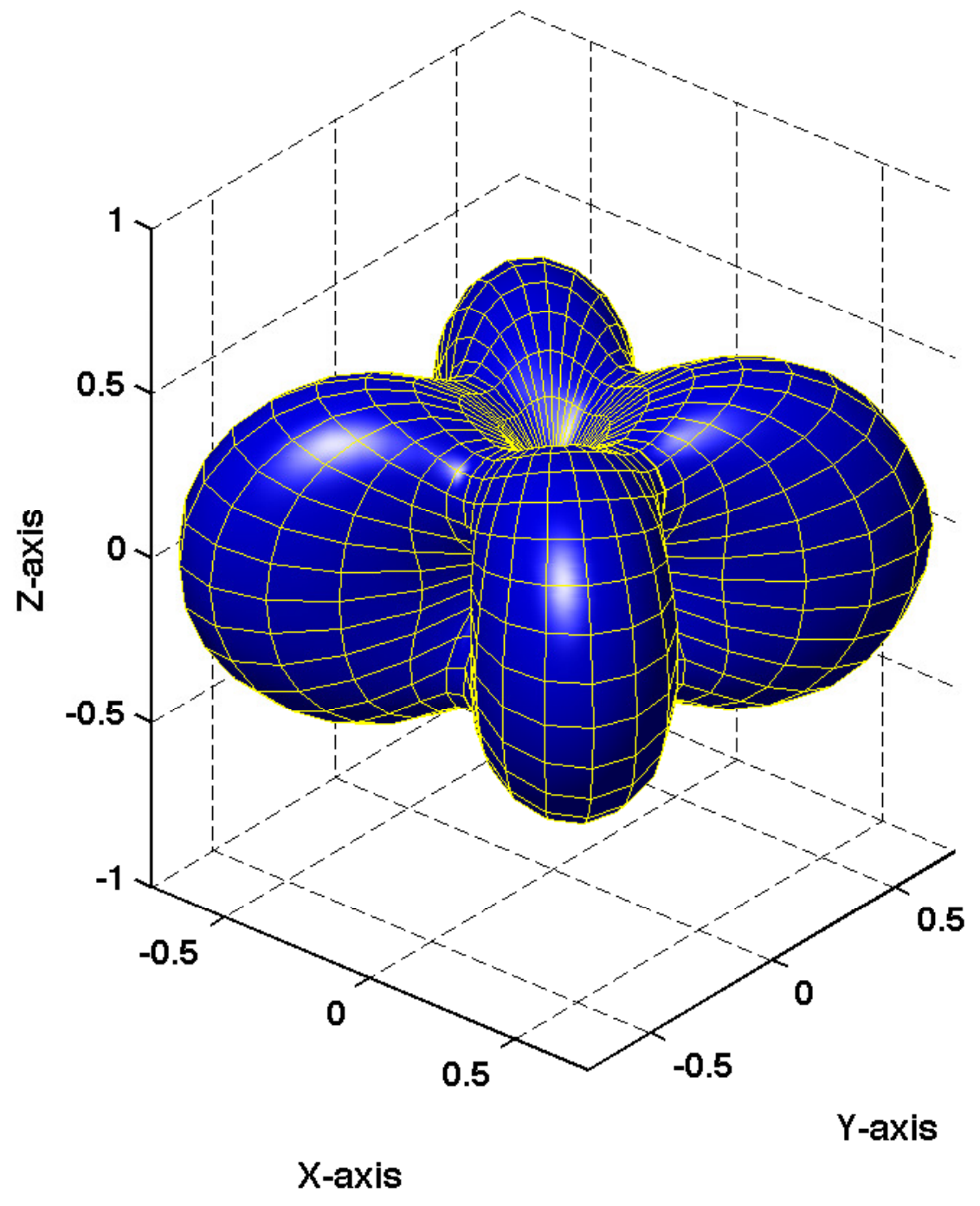}
\includegraphics[angle=0,width=0.32\columnwidth]{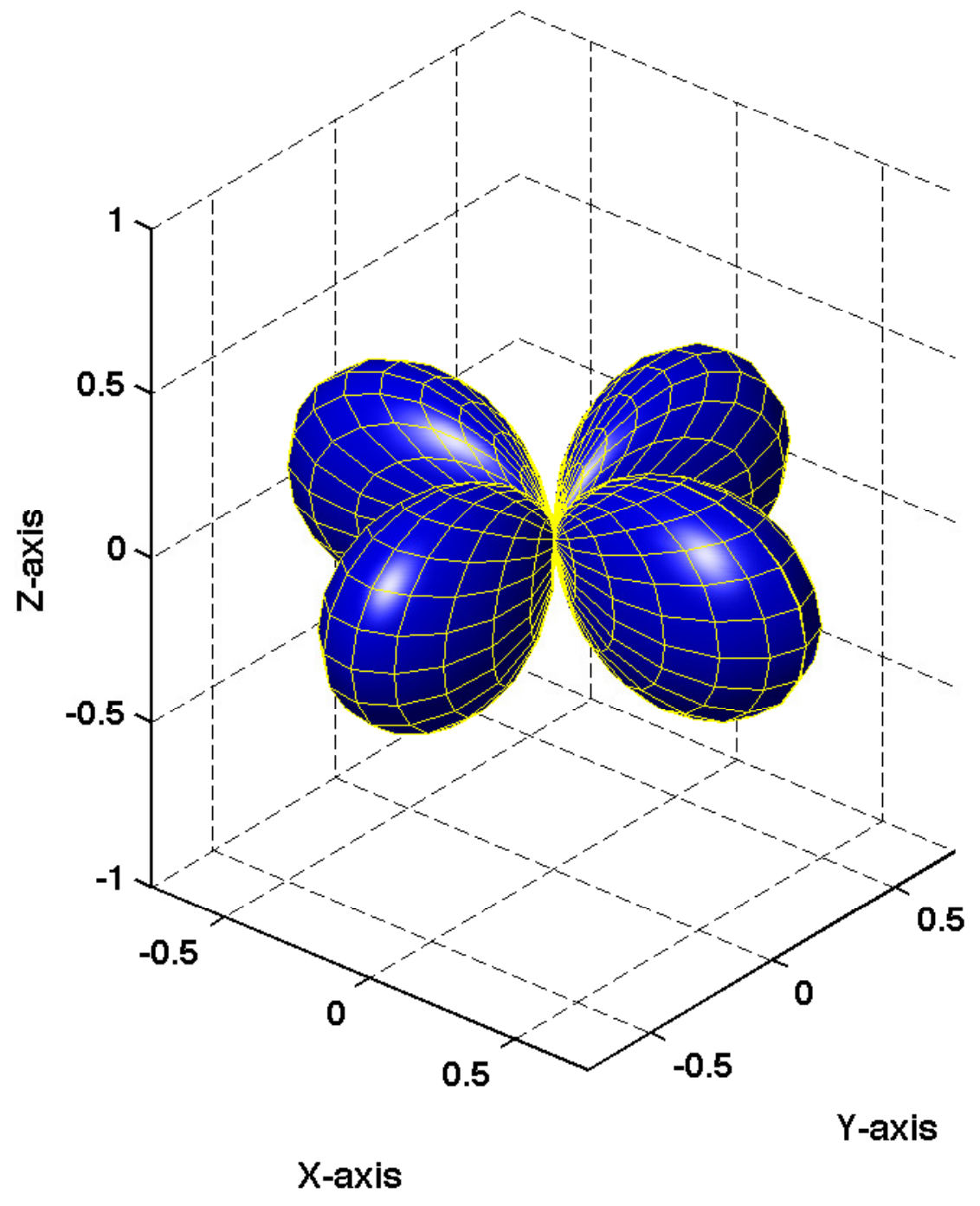}
\caption{Antenna patterns for Michelson interferometer strain
response to unpolarized gravitational waves for tensor (left plot),
vector (middle plot), ans scalar modes (right plot), 
evaluated in the small antenna limit, $f=0$.
The interferometer arms point in the $\hat x$ and $\hat y$ directions.}
\label{f:pattern3dAltPolIFO_peanut}
\end{center}
\end{figure}

From these expressions, we see that the response 
functions for the breathing and
longitudinal modes differ only by a constant multiplicative
factor of $-\sqrt{2}$.
This degeneracy means that we will not be able to 
distinguish these two polarization modes using 
ground-based interferometers.
Plots of the antenna patterns $|R^A(\hat n)|$ for the 
six different polarization modes are shown in 
Figure~\ref{f:pattern3dAltPolIFO}.
Note that the overall magnitude of the response gets smaller
as one moves from tensor, to vector, to scalar polarization
modes.
In Figure~\ref{f:pattern3dAltPolIFO_peanut}, we plot the ``peanut"
antenna patterns for the response to unpolarized gravitational
waves for the tensor, vector, and scalar modes, respectively.
By unpolarized we simply mean that the incident 
gravitational waves have equal power in the $+$ and $\times$ 
polarizations for the tensor modes;
equal power in the $X$ and $Y$ polarizations for the vector modes,
and equal power in the $B$ and $L$ polarizations for the scalar modes.
%

\subsubsection{Overlap functions}
\label{s:overlap_altpolIFO}

Similar to what we did in Section~\ref{s:circular_overlap},
let us assume that the stochastic background is 
{\em independently polarized}, but is otherwise 
Gaussian-stationary and isotropic.
This means that the quadratic expectation values of the 
Fourier components of the metric perturbations can be 
written as 
\be
\langle h_A(f,\hat n) h_{A'}^*(f',\hat n')\rangle
=\frac{1}{8\pi}
S_h^{A}(f)
\delta_{AA'}
\delta(f-f')
\delta^2(\hat n,\hat n')\,,
\label{e:iso_altpol}
\ee
where $A=\{+,\times,X,Y,B,L\}$.
The functions $S_h^A(f)$ are such that 
\be
\begin{aligned}
S^{(T)}_h(f) &= S_h^+(f) + S_h^\times(f)\,,
\\
S^{(V)}_h(f) &= S_h^X(f) + S_h^Y(f)\,,
\\
S^{(S)}_h(f) &= S_h^B(f) + S_h^L(f)\,,
\end{aligned}
\ee
are the one-sided strain spectral densities for the 
tensor, vector, and scalar modes individually.
For simplicity, we will also assume that the tensor,
vector, and scalar modes are individually unpolarized
so that $S_h^+(f)=S_h^\times(f)$, $S_h^X(f)=S_h^Y(f)$, etc.
All of these assumptions together define the 
stochastic signal  model for this example.

The above expectation values (\ref{e:iso_altpol})
can now be used to calculate the expected value 
of the correlated response of two detectors to such 
a background.
Writing the response of detector $I$ as 
\be
\tilde h_I(f) = \int d^2\Omega_{\hat n}\> \sum_A R_I^A(f,\hat n) h_A(f,\hat n)\,,
\ee
it follows (as we have done many times before) that
\be
\langle \tilde h_I(f)\tilde h_J^*(f')\rangle
= \frac{1}{2}\delta(f-f')
\left[\Gamma^{(T)}_{IJ}(f)S_h^{(T)}(f) 
+ \Gamma^{(V)}_{IJ}(f)S_h^{(V)}(f)
+ \Gamma^{(S)}_{IJ}(f)S_h^{(S)}(f)\right]\,,
\label{e:T+V+S}
\ee
where
\be
\begin{aligned}
\Gamma^{(T)}_{IJ}(f) 
&\equiv\frac{1}{8\pi}\int d^2\Omega_{\hat n}\>
\left[
R^+_I(f,\hat n) R^{+*}_J(f,\hat n) +
R^\times_I(f,\hat n) R^{\times*}_J(f,\hat n)
\right]\,,
\\
\Gamma^{(V)}_{IJ}(f) 
&\equiv\frac{1}{8\pi}\int d^2\Omega_{\hat n}\>
\left[
R^X_I(f,\hat n) R^{X*}_J(f,\hat n) +
R^Y_I(f,\hat n) R^{Y*}_J(f,\hat n)
\right]\,,
\\
\Gamma^{(S)}_{IJ}(f) 
&\equiv\frac{1}{8\pi}\int d^2\Omega_{\hat n}\>
\left[
R^B_I(f,\hat n) R^{B*}_J(f,\hat n) +
R^L_I(f,\hat n) R^{L*}_J(f,\hat n)
\right]\,,
\end{aligned}
\label{e:gammaTVS_altpolIFO}
\ee
are the corresponding overlap functions for the 
tensor, vector, and scalar modes $\{T,V,S\}$.
Note that $\Gamma^{(T)}_{IJ}(f)$ is identical to the 
ordinary overlap function $\Gamma_{IJ}(f)$ for an
isotropic background (\ref{e:GammaIJ}).

\begin{figure}[h!tbp]
\begin{center}
\includegraphics[angle=0,width=0.5\columnwidth]{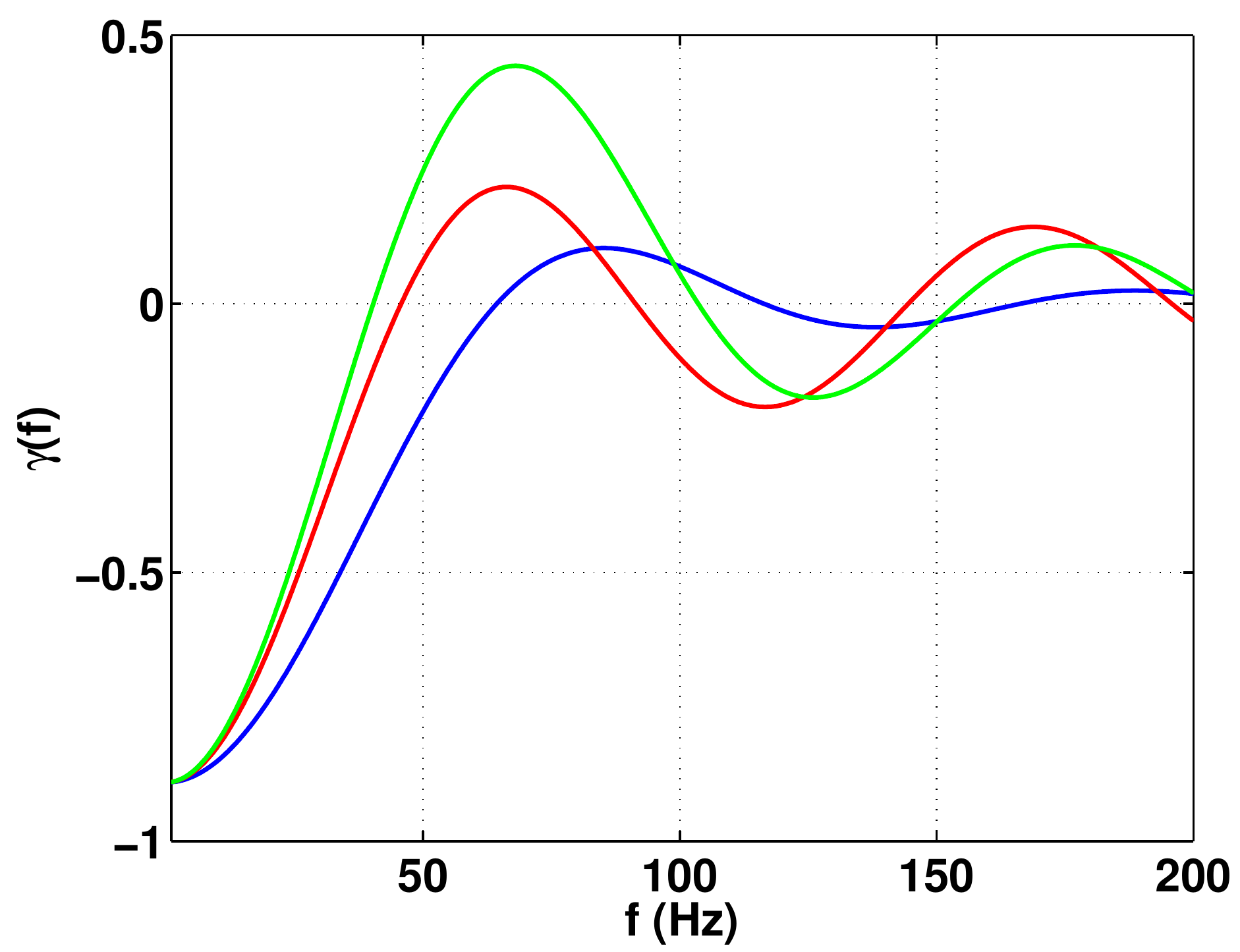}\\
\includegraphics[angle=0,width=0.5\columnwidth]{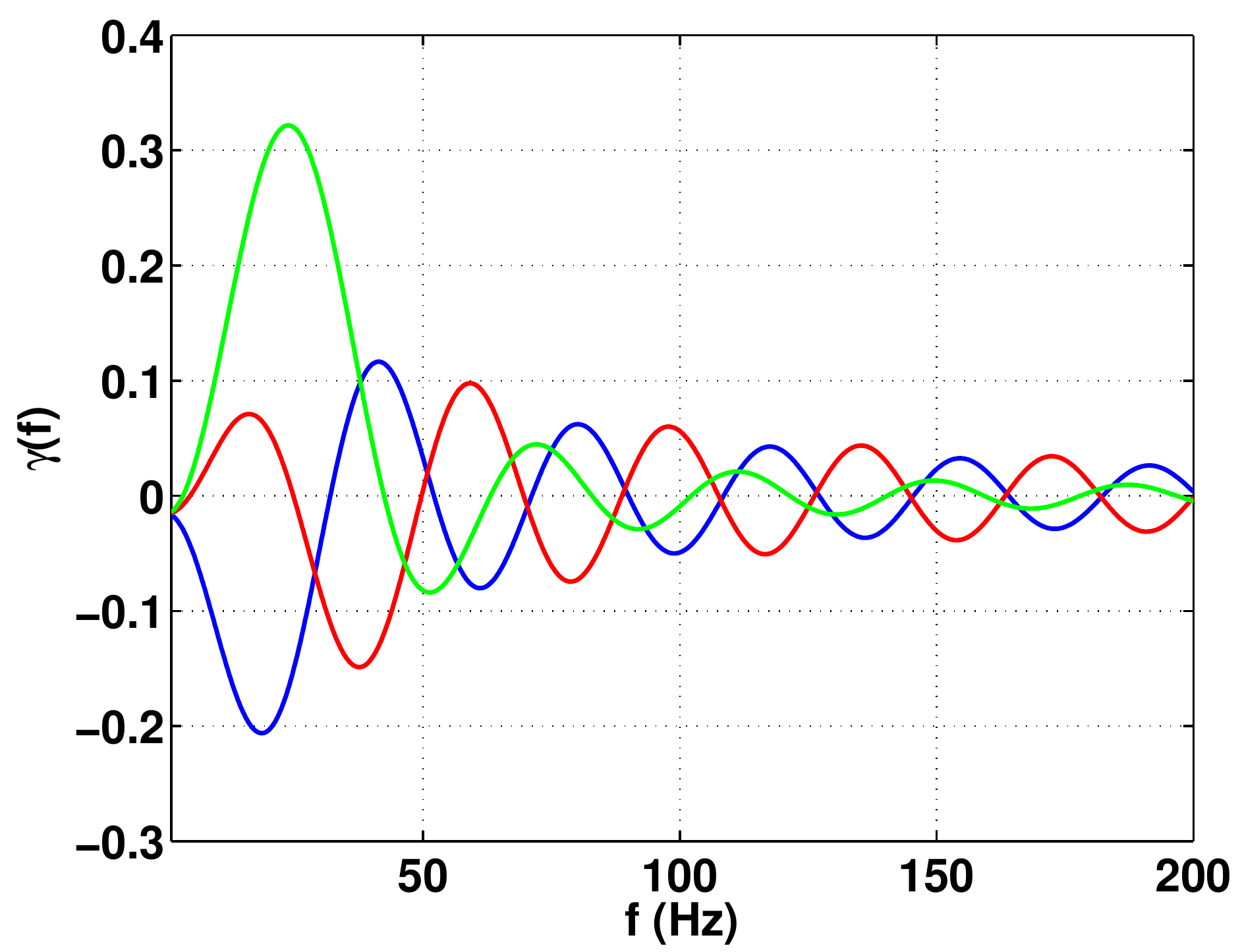}\\
\includegraphics[angle=0,width=0.5\columnwidth]{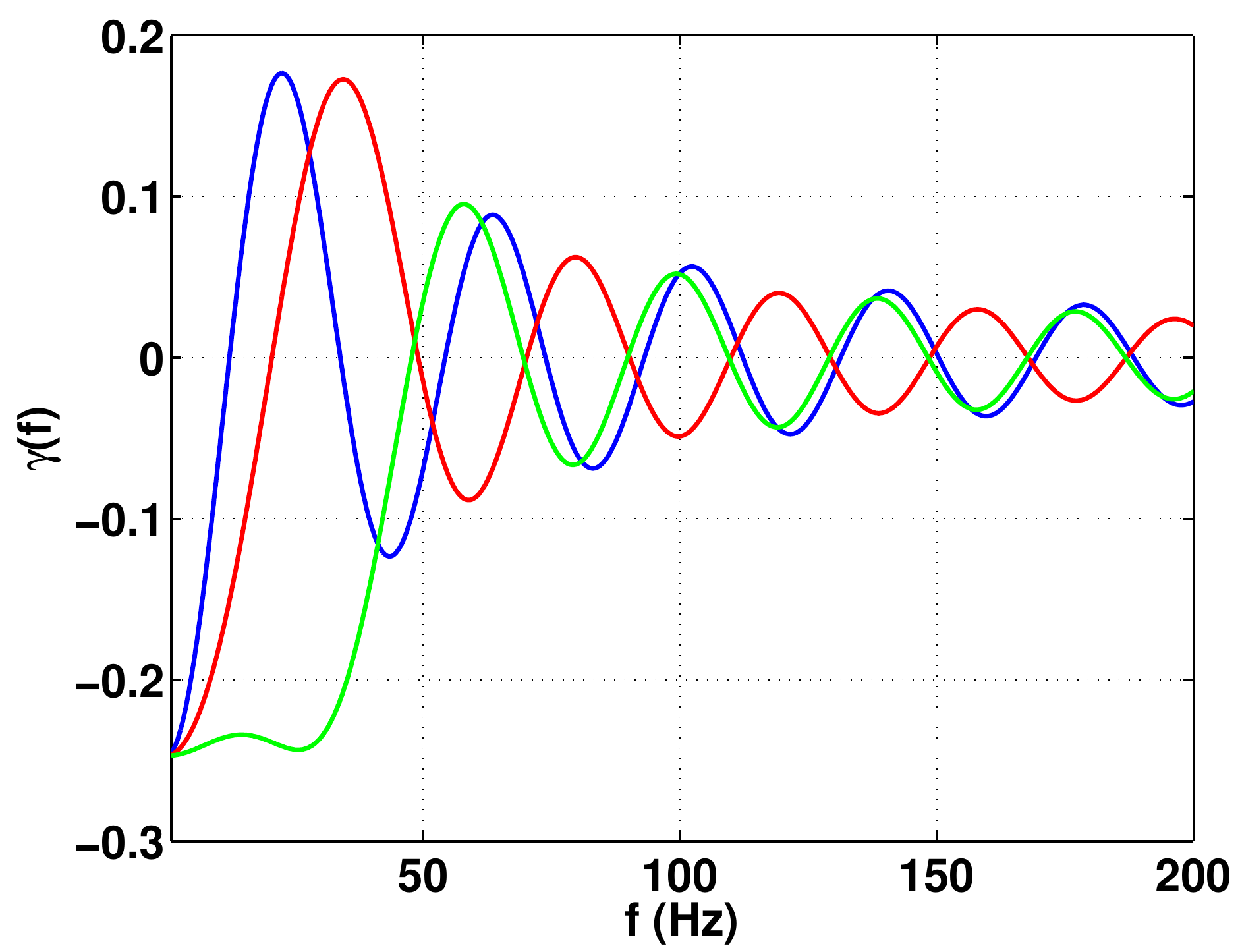}\\
\caption{Normalized overlap functions for unpolarized
tensor, vector, and scalar modes for the 
LIGO Hanford-LIGO Livingston detector pair (top panel);
for the LIGO Hanford-Virgo detector pair (middle panel); and
for the LIGO Livingston-Virgo detector pair (bottom panel).
The tensor overlap functions are shown in blue;
the vector overlap functions are shown in red;
the scalar overlap functions are shown in green.
These overlap functions were calculated in the small-antenna limit.}
\label{f:overlapTVS}
\end{center}
\end{figure}

Figure~\ref{f:overlapTVS} show plots of the 
tensor, vector, and scalar overlap functions
for the three different LIGO-Virgo detector pairs.
The overlap functions have been normalized 
so that they equal 1 for colocated and 
coaligned detectors.
This requires multiplying $\Gamma_{IJ}(f)$ by 
a factor of $5$ for the tensor and vector 
overlap functions (\ref{e:gamma_normalized}), 
but by a factor of 10 for the scalar overlap functions.

\subsubsection{Component separation: ML estimates of $S_h^{(T)}$, $S_h^{(V)}$, and $S_h^{(S)}$}
\label{s:comp_sep_altpolIFO}

Proceeding along the same lines as in 
Section~\ref{s:comp_sep_circular}, 
we now describe a method
for separating the tensor, vector, and scalar contributions
to the total strain spectral density.
As shown in \cite{Nishizawa-et-al:2009}, we will need
to analyze data from at least three independent 
baselines (so at least three detectors) to separate the
tensor, vector, and scalar contributions at each frequency $f$.
As before, we will adopt 
the notation $\alpha=1,2,\cdots, N_b$
to denote the individual baselines (detector pairs)
and $\alpha_1$, $\alpha_2$ to denote the two detectors
that constitute that baseline.

Our starting point is again the cross-correlated 
data from pairs of detectors in the network:
\be
\hat C_\alpha(f) \equiv 
\frac{2}{T}\,\tilde d_{\alpha_1}(f) \tilde d^*_{\alpha_2}(f)\,,
\ee
where 
\be
\tilde d_{\alpha_I}(f) 
= \tilde h_{\alpha_I}(\tilde f) + \tilde n_{\alpha_I}(f)\,,
\quad I=1,2\,.
\ee
Assuming that the noise in the individual detectors 
are uncorrelated with one another, it follows that
\be
\langle
\hat C_\alpha(f)
\rangle 
=
\Gamma^{(T)}_\alpha(f)S_h^{(T)}(f)+
\Gamma^{(V)}_\alpha(f)S_h^{(V)}(f)+
\Gamma^{(S)}_\alpha(f)S_h^{(S)}(f)\,.
\label{e:<C>_A}
\ee
In addition,
\be
\begin{aligned}
{\cal N}_{\alpha\alpha'}(f,f') 
&\equiv \langle \hat C_\alpha(f) \hat C^*_{\alpha'}(f')\rangle
-\langle \hat C_\alpha(f)\rangle\langle \hat C^*_{\alpha'}(f')\rangle
\\
&\approx 
\delta_{\alpha\alpha'}
\delta_{ff'}
P_{n_{\alpha_1}}(f) P_{n_{\alpha_2}}(f)\,,
\end{aligned}
\label{e:N_altpol}
\ee
where $P_{n_{\alpha_I}}(f)$ are the one-sided power 
spectral densities of the noise in the detectors, and
where we have assumed again that the gravitational-wave 
signal is weak compared to the detector noise.
As we did in Section~\ref{s:comp_sep_circular}
we can write down a likelihood function for the
cross-correlated data given the signal model
(\ref{e:<C>_A}):
\be
p(\hat C|{\cal A}) \propto
\exp\left[-\frac{1}{2}(\hat C - M{\cal A})^\dagger 
{\cal N}^{-1} (\hat C - M{\cal A})\right]\,.
\label{e:likehood_A}
\ee
Here we have adopted the matrix notation:
\be
M \equiv
\left[
\begin{array}{ccc}
\Gamma_1^{(T)} &
\Gamma_1^{(V)} & 
\Gamma_1^{(S)}
\\
\Gamma_2^{(T)} &
\Gamma_2^{(V)} & 
\Gamma_2^{(S)} 
\\
\vdots & \vdots & \vdots
\\
\Gamma_{N_b}^{(T)} &
\Gamma_{N_b}^{(V)} & 
\Gamma_{N_b}^{(S)} 
\end{array}
\right]\,,
\quad
{\cal A}\equiv
\left[
\begin{array}{c}
S_h^{(T)} \\
S_h^{(V)} \\
S_h^{(S)} 
\end{array}
\right]\,.
\label{e:M_altpol}
\ee
Since ${\cal A}$ enters quadratically in the 
exponential, we have the usual expression for the 
maximum-likehood estimators:
\be
\hat {\cal A}
=F^{-1} X\,,
\label{e:Ahat}
\ee
where
\be
F\equiv M^\dagger {\cal N}^{-1} M\,,
\qquad
X\equiv M^\dagger {\cal N}^{-1} \hat C\,,
\label{e:F,X_A}
\ee
with $M$ and ${\cal N}$ given above, and with the standard
proviso about possibly having to use 
singular-value decomposition to invert $F$.
The uncertainty in the maximum-likelihood recoved 
values is given by the covariance matrix
\be
\langle \hat{\cal A}\hat{\cal A}^\dagger\rangle-
\langle \hat{\cal A}\rangle\langle \hat{\cal A}^\dagger\rangle
\approx F^{-1}\,,
\ee
which we will use below to define {\em effective}
overlap functions for the tensor, vector, and scalar 
modes for a multibaseline network of detectors.

\subsubsection{Effective overlap functions for multiple baselines}
\label{s:gamma_eff}

\begin{figure}[h!tbp]
\begin{center}
\includegraphics[angle=0,width=0.6\columnwidth]{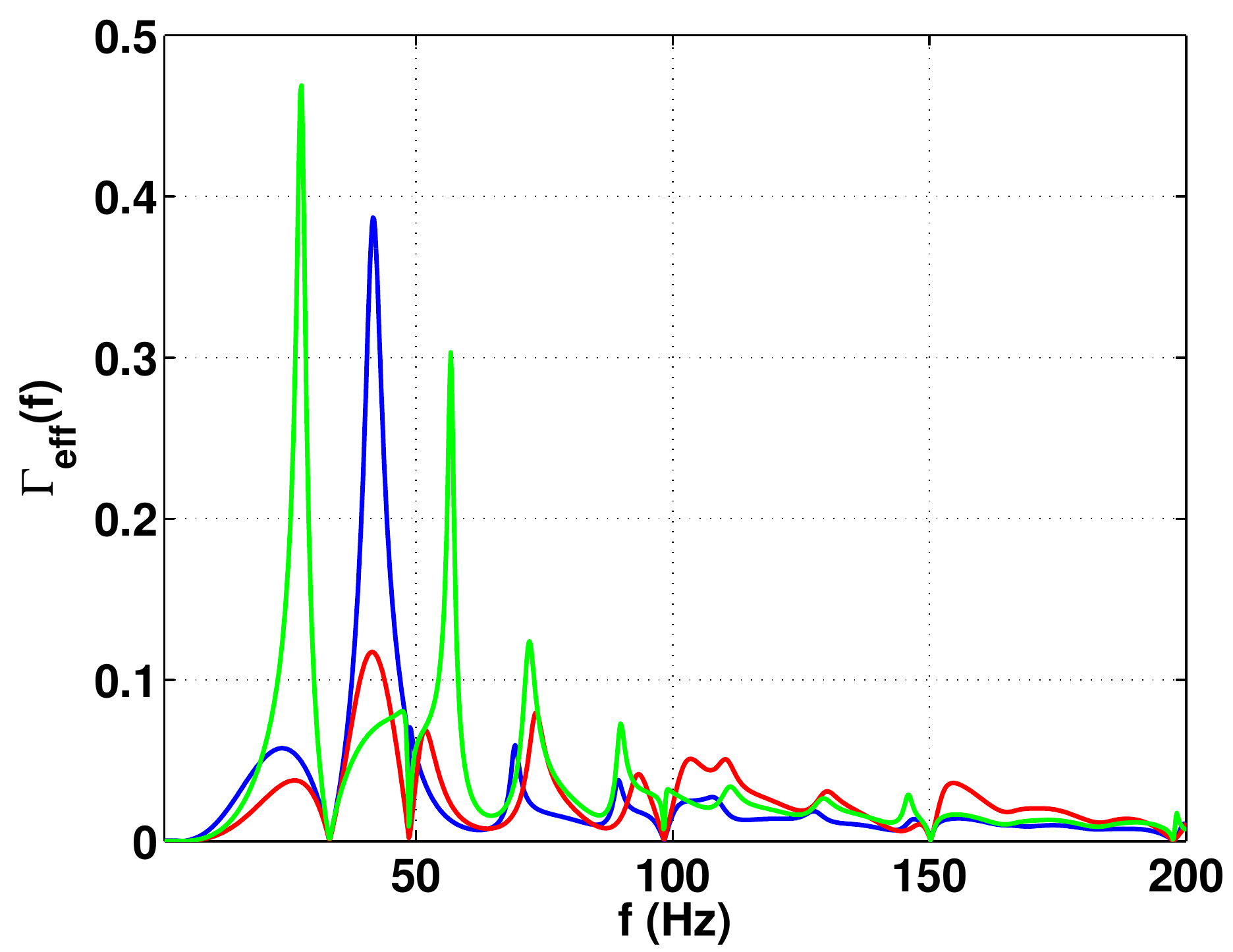}
\caption{Effective overlap functions for 
$S_h^{(T)}$, $S_h^{(V)}$, $S_h^{(S)}$,
for the multibaseline network formed from the LIGO Hanford,
LIGO Livingston, and Virgo detectors.
$\Gamma_{\rm eff}^{(T)}(f)$ is shown in blue;
$\Gamma_{\rm eff}^{(V)}(f)$ is shown in red;
$\Gamma_{\rm eff}^{(S)}(f)$ is shown in green.}
\label{f:overlap_eff_altpol}
\end{center}
\end{figure}

For a multibaseline network of detectors, one has
\be
F=\left[
\begin{array}{ccc}
\sum_\alpha N_\alpha^{-1}(\Gamma^{(T)}_\alpha)^2 &
\sum_\alpha N_\alpha^{-1}\Gamma^{(T)}_\alpha \Gamma^{(V)}_\alpha &
\sum_\alpha N_\alpha^{-1}\Gamma^{(T)}_\alpha \Gamma^{(S)}_\alpha \\
\sum_\alpha N_\alpha^{-1}\Gamma^{(V)}_\alpha \Gamma^{(T)}_\alpha &
\sum_\alpha N_\alpha^{-1}(\Gamma^{(V)}_\alpha)^2 &
\sum_\alpha N_\alpha^{-1}\Gamma^{(V)}_\alpha \Gamma^{(S)}_\alpha \\
\sum_\alpha N_\alpha^{-1}\Gamma^{(S)}_\alpha \Gamma^{(T)}_\alpha &
\sum_\alpha N_\alpha^{-1}\Gamma^{(S)}_\alpha \Gamma^{(V)}_\alpha &
\sum_\alpha N_\alpha^{-1}(\Gamma^{(S)}_\alpha)^2 \\
\end{array}
\right]\,,
\ee
where $N_\alpha(f)\equiv P_{n_{\alpha_1}}(f) P_{n_{\alpha_2}}(f)$.
Let us assume that the determinant of the 
$3\times 3$ matrices for each frequency 
(which we will denote by $\bar F$) 
are not equal to zero.
Then the uncertainties in the estimators of 
$S_h^{(T)}$, $S_h^{(V)}$, and $S_h^{(S)}$
can be written as

\be
\begin{aligned}
\sigma_{\hat T}^2 
&= (\bar F^{-1})_{TT}
= \frac{1}{\det(\bar F)}
\left(
{\sum_\alpha N_\alpha^{-1}(\Gamma^{(V)}_\alpha)^2}
\sum_{\alpha'} N_{\alpha'}^{-1}(\Gamma^{(S)}_{\alpha'})^2-
\left(\sum_\alpha N_\alpha^{-1}\Gamma^{(S)}_\alpha \Gamma^{(V)}_\alpha\right)^2
\right)\,,
\\
\sigma_{\hat V}^2 
&= (\bar F^{-1})_{VV}
= \frac{1}{\det(\bar F)}
\left(
{\sum_\alpha N_\alpha^{-1}(\Gamma^{(T)}_\alpha)^2}
\sum_{\alpha'} N_{\alpha'}^{-1}(\Gamma^{(S)}_{\alpha'})^2-
\left(\sum_\alpha N_\alpha^{-1}\Gamma^{(S)}_\alpha \Gamma^{(T)}_\alpha\right)^2
\right)\,,
\\
\sigma_{\hat S}^2 
&= (\bar F^{-1})_{SS}
= \frac{1}{\det(\bar F)}
\left(
{\sum_\alpha N_\alpha^{-1}(\Gamma^{(T)}_\alpha)^2}
\sum_{\alpha'} N_{\alpha'}^{-1}(\Gamma^{(V)}_{\alpha'})^2-
\left(\sum_\alpha N_\alpha^{-1}\Gamma^{(V)}_\alpha \Gamma^{(T)}_\alpha\right)^2
\right)\,.
\\
\end{aligned}
\ee
Following \cite{Nishizawa-et-al:2009}, we can now 
define the {\em effective} overlap functions for 
the the tensor, vector, and scalar modes, associated 
with a multibaseline detector network.
As we did in Section~\ref{s:gamma_eff_circ}, we 
will assume for simplicity that the noise power 
spectra for the detectors are equal to one another
so that $N_\alpha\equiv N$  can be factored out of
the above expressions.
We then define
\be
\begin{aligned}
\Gamma_{\rm eff}^{(T)}(f) \equiv  \sigma_{\hat T}^{-1}\sqrt{N}
&=
\left(
\frac{N^3 \det(\bar F)}
{\sum_\alpha(\Gamma^{(V)}_\alpha)^2
\sum_{\alpha'}(\Gamma^{(S)}_{\alpha'})^2-
\left(\sum_\alpha \Gamma^{(S)}_\alpha \Gamma^{(V)}_\alpha\right)^2}
\right)^{1/2}\,,
\\
\Gamma_{\rm eff}^{(V)}(f) \equiv  \sigma_{\hat V}^{-1}\sqrt{N}
&=
\left(
\frac{N^3 \det(\bar F)}
{\sum_\alpha(\Gamma^{(T)}_\alpha)^2
\sum_{\alpha'}(\Gamma^{(S)}_{\alpha'})^2-
\left(\sum_\alpha \Gamma^{(S)}_\alpha \Gamma^{(T)}_\alpha\right)^2}
\right)^{1/2}\,,
\\
\Gamma_{\rm eff}^{(S)}(f) \equiv  \sigma_{\hat S}^{-1}\sqrt{N}
&=
\left(
\frac{N^3 \det(\bar F)}
{\sum_\alpha(\Gamma^{(T)}_\alpha)^2
\sum_{\alpha'}(\Gamma^{(V)}_{\alpha'})^2-
\left(\sum_\alpha \Gamma^{(V)}_\alpha \Gamma^{(T)}_\alpha\right)^2}
\right)^{1/2}\,.
\end{aligned}
\ee
%
Plots of $\Gamma_{\rm eff}^{(T)}(f)$,
$\Gamma_{\rm eff}^{(V)}(f)$, and
$\Gamma_{\rm eff}^{(S)}(f)$
are shown in Figure~\ref{f:overlap_eff_altpol} for the 
multibaseline network formed from the 
LIGO Hanford, LIGO Livingston, and Virgo detectors.
Dips in sensitivity correspond to frequencies 
where the determinant of $\bar F$ is zero (or close to zero).
%

\subsection{Searches for non-GR polarizations using pulsar timing arrays}
\label{s:altpolPTA}

As discussed in Section~\ref{s:search_w_diff_detectors}
it is also possible to search for non-GR polarizations
using a pulsar timing array.
Although the general concepts are the same as those for
ground-based interferometers, there are some important 
differences, as the vector and scalar
longitudinal polarization modes require keeping the pulsar 
term in the response functions to avoid possible singularities.
We shall see below that the sensitivity to the vector and 
scalar longitudinal modes increases dramatically when 
cross-correlating data from pairs of pulsars with 
small angular separations.
For additional details, see
\cite{Lee-et-al:2008, Chamberlin-Siemens:2012, Gair-et-al:2015}.

\subsubsection{Polarization basis response functions}
\label{s:response_pol_altpolPTA}

For pulsar timing, the response functions for 
Doppler frequency measurements for the different 
polarization modes $A=\{+,\times, X,Y,B,L\}$
are given by
\be
R^A(f,\hat n) = \frac{1}{2}
\frac{\hat p^a\hat p^b}{1-\hat n\cdot \hat p}
e^A_{ab}(\hat n)
\left[1-e^{-\frac{i2\pi fL}{c}(1-\hat n\cdot\hat p)}
\right]\,,
\label{e:RA_altpolPTA}
\ee
where $\hat p$ points in the direction to the 
pulsar and $L$ is its distance from Earth
(see Section~\ref{s:one-way-tracking} with 
$\hat p=-\hat u$).
Without loss of generality, we have assumed 
that the location of the measurement is at the 
origin of coordinates.
Note that we have kept the {\em pulsar term}
(the second term in the square brackets) since,
as we shall see below, it is
needed to get finite expressions for the 
response and overlap functions for the vector
and scalar longitudinal modes.

Choosing our coordinate system so that $\hat z$ 
points along $\hat p$,
we find:
\be
\begin{aligned}
R^+(f,\hat n) 
&=\frac{1}{2}(1+\cos\theta)
\left[1-e^{-\frac{i2\pi fL}{c}(1-\cos\theta)}\right]\,,
\\
R^\times(f,\hat n) 
&=0\,,
\\
R^X(f,\hat n) 
&=-\frac{\sin\theta\cos\theta}{1-\cos\theta}
\left[1-e^{-\frac{i2\pi fL}{c}(1-\cos\theta)}\right]\,,
\\
R^Y(f,\hat n)
&=0\,,
\\
R^B(f,\hat n) 
&=\frac{1}{2}(1+\cos\theta)
\left[1-e^{-\frac{i2\pi fL}{c}(1-\cos\theta)}\right]\,,
\\
R^L(f,\hat n) 
&=\frac{1}{\sqrt{2}}\frac{\cos^2\theta}{1-\cos\theta}
\left[1-e^{-\frac{i2\pi fL}{c}(1-\cos\theta)}\right]\,,
\end{aligned}
\label{e:RAlist_altpolPTA}
\ee
where we used (\ref{e:nlm_def}) for our definitions
of $\{\hat n,\hat l,\hat m\}$.
Note that the response functions for the 
breathing mode $B$ and the tensor $+$ mode 
have the same form for our particular 
choice of $\{\hat l,\hat m\}$.
This is not a problem, however, as we can 
still distinguish these modes due to their
different behavior under rotations.
The difference between the breathing
and tensor modes becomes more apparent 
in terms of the spherical harmonic basis
response functions 
$R^B_{(lm)}(f)$ and $R^G_{(lm)}(f)$, which
are given in (\ref{e:RPlist_altpolPTA}).

If we did not include the pulsar terms
in the above expressions, 
then the response functions for both the vector 
and scalar longitudinal modes would become
singular at $\theta = 0$ (i.e., $\cos\theta=1$).%
\footnote{This corresponds to the direction 
to the pulsar and the direction to the source 
of the gravitational wave being the same.  
For this case, the radio pulse from the pulsar 
and the gravitational wave travel in phase with 
one another from the pulsar to Earth.
It is as if the radio pulse ``surfs" the 
gravitational wave~\cite{Chamberlin-Siemens:2012}.}
The factor of $\sin\theta$ in 
the numerator for $R^X(f,\hat n)$ ``softens" the 
$(1-\cos\theta)^{-1}$ singularity to 
$(1-\cos\theta)^{-1/2}$, so 
that it becomes integrable when 
calculating the vector longitudinal 
overlap functions~\cite{Lee-et-al:2008, Chamberlin-Siemens:2012, 
Gair-et-al:2015}.
(We will discuss this in more detail in
Section~\ref{s:overlap_altpolPTA}.)
By keeping the pulsar term we remove these 
singularities as can be seen by expanding the
full expressions in (\ref{e:RAlist_altpolPTA})
for $\theta\ll 1$:
\be
\begin{aligned}
R^+(f,\hat n) 
&\approx i y \theta^2/2\,,
\\
R^\times(f,\hat n) 
&=0\,,
\\
R^X(f,\hat n) 
&\approx -i y \theta\,,
\\
R^Y(f,\hat n)
&=0\,,
\\
R^B(f,\hat n) 
&\approx i y \theta^2/2\,,
\\
R^L(f,\hat n) 
&\approx i y/\sqrt{2}\,,
\end{aligned}
\ee
where $y\equiv 2\pi fL/c$, and we have assumed 
that $y\theta^2$ is also sufficiently small that 
we could Taylor expand the exponential.
Since the typical distance to a pulsar is 
a few kiloparsecs and $f= 3 \times 10^{-9}~{\rm Hz}$ 
for 10~yr of observation, we have 
$y~\sim 10^4$, which means 
$\theta \lesssim 10^{-2}$ for the above expansions to be valid.
Thus, for small angular separations between the 
direction to the pulsar and the direction
to the gravitational wave, the reponse to the 
scalar-longitudinal modes will be more than an 
order-of-magnitude larger than that for the 
vector modes, and several orders-of-magnitude 
larger than that for both the tensor and breathing modes.
This increased sensitivity of the scalar longitudinal
and vector longitudinal modes  will also become 
apparent when we calculate the overlap functions 
for a pair of pulsars (see Section~\ref{s:overlap_altpolPTA} 
and Figure~\ref{f:overlap_altpolPTA}).

\subsubsection{Spherical harmonic basis response functions}
\label{s:response_sph_altpolPTA}

It is also interesting to calculate the Doppler-frequency 
response functions for the
tensor spherical harmonic components 
$P=\{G, C, V_G, V_C, B, L\}$.
The general expression is given by:
\be
R^P_{(lm)}(f) = \int d^2\Omega_{\hat n}\>
\frac{1}{2}
\frac{\hat p^a\hat p^b}{1-\hat n\cdot \hat p}
Y^P_{(lm)ab}(\hat n)
\left[1-e^{-\frac{i2\pi fL}{c}(1-\hat n\cdot\hat p)}
\right]\,.
\label{e:RP_altpol}
\ee
As shown in \cite{Gair-et-al:2015}, the above 
integral can be evaluated and then simplified by 
taking the limit $y \gg 1$, which as we mentioned 
above is valid for typical pulsars.
The final results (taken from that paper) are:
\be
\begin{aligned}
R^G_{(lm)}(f) 
&\approx 2\pi\,{}^{(2)}\!N_l
Y_{lm}(\hat p)\,,
&\qquad l=2,3, \cdots\,,
\\
R^C_{(lm)}(f) 
&\approx 0\,,
&\qquad l=2,3, \cdots\,,
\\
R^{V_G}_{(lm)}(f) 
&\approx 2\pi\,{}^{(1)}\!N_l 
\left[1-\frac{2}{3}\delta_{l1}\right]
Y_{lm}(\hat p)\,,
&\qquad l=1,2, \cdots\,,
\\
R^{V_C}_{(lm)}(f) 
&\approx 0\,,
&\qquad l=1,2, \cdots\,,
\\
R^B_{(lm)}(f) 
&\approx 2\pi\, \frac{1}{\sqrt{2}}
\left[\delta_{l0} + \frac{1}{3}\delta_{l1}\right]
Y_{lm}(\hat p)\,,
&\qquad l=0,1, \cdots\,,
\\
R^L_{(lm)}(f) 
&\approx 
2\pi \left[-\delta_{l0}-\frac{1}{3} \delta_{l1} + \frac{1}{2}\bar H_l(y)\right]
Y_{lm}(\hat p)\,,
&\qquad l=0,1, \cdots\,,
\end{aligned}
\label{e:RPlist_altpolPTA}
\ee
where ${}^{(1)}\!N_l$ and ${}^{(2)}\!N_l$ are constants defined by
(\ref{e:1N}) and (\ref{e:2N}), and 
\be
\bar H_l(y) \equiv \int_{-1}^1 dx\>
\frac{1}{(1-x)} P_l(x) \left[1-e^{-iy(1-x)}\right]\,.
\ee
There are several important features to highlight
about these expressions:
(i) All of the response functions depend in the same
way on the angular position of the pulsar, which is 
simply $Y_{lm}(\hat p)$.
(ii) Just as we saw earlier (\ref{e:RP_PTA}) 
that the response to the 
tensor curl mode is zero, so too is the 
response to the vector curl mode.
{\em Thus, pulsar timing arrays are also insensitive to the 
curl component of the the vector-longitudinal modes.}
(iii) In the limit $y\gg 1$, only the response 
to the scalar-longitudinal mode has 
frequency dependence (via $y$).
(iv) The response to the breathing mode has 
non-zero contributions only from $l=0$ and $l=1$.
In terms of power (which is effectively the square 
of the response), this means that  pulsar timing 
observations will be insensitive to anisotropies 
in power in the breathing mode beyond quadrupole 
(i.e., $l=2$).
 
\subsubsection{Overlap functions}
\label{s:overlap_altpolPTA}

To calculate the overlap functions for non-GR 
polarization modes for pulsar timing arrays, 
we will proceed as we did in 
Section~\ref{s:overlap_altpolIFO}, assuming 
that the stochastic background is independently 
polarized, but is otherwise Gaussian-stationary
and isotropic.
(Extensions to {\em anisotropic} backgrounds
will be briefly mentioned in Section~\ref{s:comp_sep_altpolPTA}.
Details can be found in \cite{Gair-et-al:2015}.)
Making these assumptions, the quadratic expectation
values of the Fourier coefficients $h_A(f,\hat n)$
take the form
\be
\langle h_A(f,\hat n) h_{A'}^*(f',\hat n')\rangle
=\frac{1}{8\pi}
S_h^{A}(f)
\delta_{AA'}
\delta(f-f')
\delta^2(\hat n,\hat n')\,,
\ee
where $S^A_h(f)$ are the one-sided strain spectral
densities for the individual polarization modes.
The overlap functions can then be calculated in
the usual way, leading to 
\be
\langle \tilde h_I(f)\tilde h_J^*(f')\rangle
=\frac{1}{2}\delta(f-f')\sum_A \Gamma^A_{IJ}(f) S_h^A(f)\,,
\ee
where
\be
\Gamma^{A}_{IJ}(f) \equiv \frac{1}{4\pi}
\int d^2\Omega_n R^A_I(f,\hat n)R_J^{A*}(f,\hat n)\,.
\ee
Note the factor of $1/4\pi$ as compared to 
$1/8\pi$ in (\ref{e:gammaTVS_altpolIFO}), 
and that there is no summation over $A$ on the 
right-hand side of this expression.

For simplicity we will also assume as before 
that the tensor modes $\{+,\times\}$ and the 
vector-longitudinal modes $\{X, Y\}$ are 
unpolarized, so that 
\be
\begin{aligned}
S_h^+(f)=S_h^\times(f)\equiv \frac{1}{2} S_h^{(T)}(f)\,,
\\
S_h^X(f)=S_h^Y(f)\equiv \frac{1}{2} S_h^{(V)}(f)\,.
\end{aligned}
\ee
Then we can define:
\be
\begin{aligned}
\Gamma^{(T)}_{IJ}(f) 
&\equiv\frac{1}{8\pi}\int d^2\Omega_{\hat n}\>
\left[
R^+_I(f,\hat n) R^{+*}_J(f,\hat n) +
R^\times_I(f,\hat n) R^{\times*}_J(f,\hat n)
\right]\,,
\\
\Gamma^{(V)}_{IJ}(f) 
&\equiv\frac{1}{8\pi}\int d^2\Omega_{\hat n}\>
\left[
R^X_I(f,\hat n) R^{X*}_J(f,\hat n) +
R^Y_I(f,\hat n) R^{Y*}_J(f,\hat n)
\right]\,,
\end{aligned}
\ee
for the unpolarized tensor and vector-longitudinal components.
But we will keep the breathing and scalar-longitudinal 
overlap functions separate:
\be
\begin{aligned}
\Gamma^{B}_{IJ}(f) 
&\equiv\frac{1}{4\pi}\int d^2\Omega_{\hat n}\>
R^B_I(f,\hat n) R^{B*}_J(f,\hat n)\,,
\\
\Gamma^{L}_{IJ}(f) 
&\equiv\frac{1}{4\pi}\int d^2\Omega_{\hat n}\>
R^L_I(f,\hat n) R^{L*}_J(f,\hat n)\,,
\end{aligned}
\label{e:gammaB,L}
\ee
given the complications that arise when trying
to explicitly calculate $\Gamma^L_{IJ}(f)$.

As noted in Section~\ref{s:example-HD}, the 
overlap function for the tensor modes can be
calculated analytically~\cite{Hellings-Downs:1983},
without needing to include the pulsar term in the 
response functions:
\be
\Gamma^{(T)}_{IJ}
= \frac{1}{3}
\left[\frac{3}{2}\left(\frac{1-\cos\zeta_{IJ}}{2}\right)
\ln\left(\frac{1-\cos\zeta_{IJ}}{2}\right)
-\frac{1}{4}\left(\frac{1-\cos\zeta_{IJ}}{2}\right)
+\frac{1}{2}\right]\,,
\ee
where $\zeta_{IJ}$ is the angle between two Earth-pulsar 
baselines, i.e.,
$\cos\zeta_{IJ} = \hat p_I\cdot\hat p_J$.
The above expression differs from (\ref{e:HD_normalized}) 
by an overall normalization.
The overlap functions for the breathing mode and
for the vector longitudinal modes can be 
also be calculated analytically, again without
needing to include the pulsar term in the reponse.
For the breathing mode we have
\be
\Gamma^B_{IJ}
= \frac{2}{3}
\left[\frac{3}{8} + \frac{1}{8}\cos\zeta_{IJ}\right]\,.
\ee
For the vector-longitudinal modes we 
have~\cite{Lee-et-al:2008, Gair-et-al:2015}
\be
\Gamma^{(V)}_{IJ}
= \frac{1}{3}
\left[
\frac{3}{2}\ln\left(\frac{2}
{1-\cos\zeta_{IJ}}\right) - 2\cos\zeta_{IJ} - \frac{3}{2}
\right]\,,
\ee
where we have assumed here that the angular separation 
$\zeta_{IJ}$ is not too small.
In the limit $\zeta_{IJ}\rightarrow 0$, the above 
expression for $\Gamma^{(V)}_{IJ}$ diverges, which
means that we need to include the pulsar terms in 
the response functions to handle that case.
Doing so results in an expression that is finite, 
but depends on the frequency $f$
via the distances to the pulsars, 
$2\pi fL_I/c$ and $2\pi fL_J/c$.
(See Appendix J of \cite{Gair-et-al:2015} for an
analytic expression for $\Gamma^{(V)}_{IJ}(f)$ in
the limit $\zeta_{IJ}\rightarrow 0$.)

Finally, for the scalar longitudinal overlap 
function $\Gamma^L_{IJ}(f)$, there is no known 
analytic expression for the integral in (\ref{e:gammaB,L}),
except in the limit of codirectional ($\zeta_{IJ}=0$)
and anti-directional ($\zeta_{IJ}=\pi$) 
pulsars~\cite{Lee-et-al:2008, Chamberlin-Siemens:2012, 
Gair-et-al:2015}.
The pulsar terms need to be included in the 
scalar-longitudinal response functions for all cases
to obtain a finite result, which again depend 
on the frequency $f$ via the distances to the pulsars.
A semi-analytic expression for 
$\Gamma^L_{IJ}(f)$ is derived in \cite{Gair-et-al:2015},
which is valid in the $2\pi fL/c \gg 1$ limit.
The semi-analytic expression effectively replaces
the double integral over directions on the sky 
$\hat n=(\theta,\phi)$ with just a single numerical 
integration over $\theta$.
See \cite{Gair-et-al:2015} for additional details
regarding that calculation.

\begin{figure}[h!tbp]
\begin{center}
\includegraphics[angle=0,width=0.49\columnwidth]{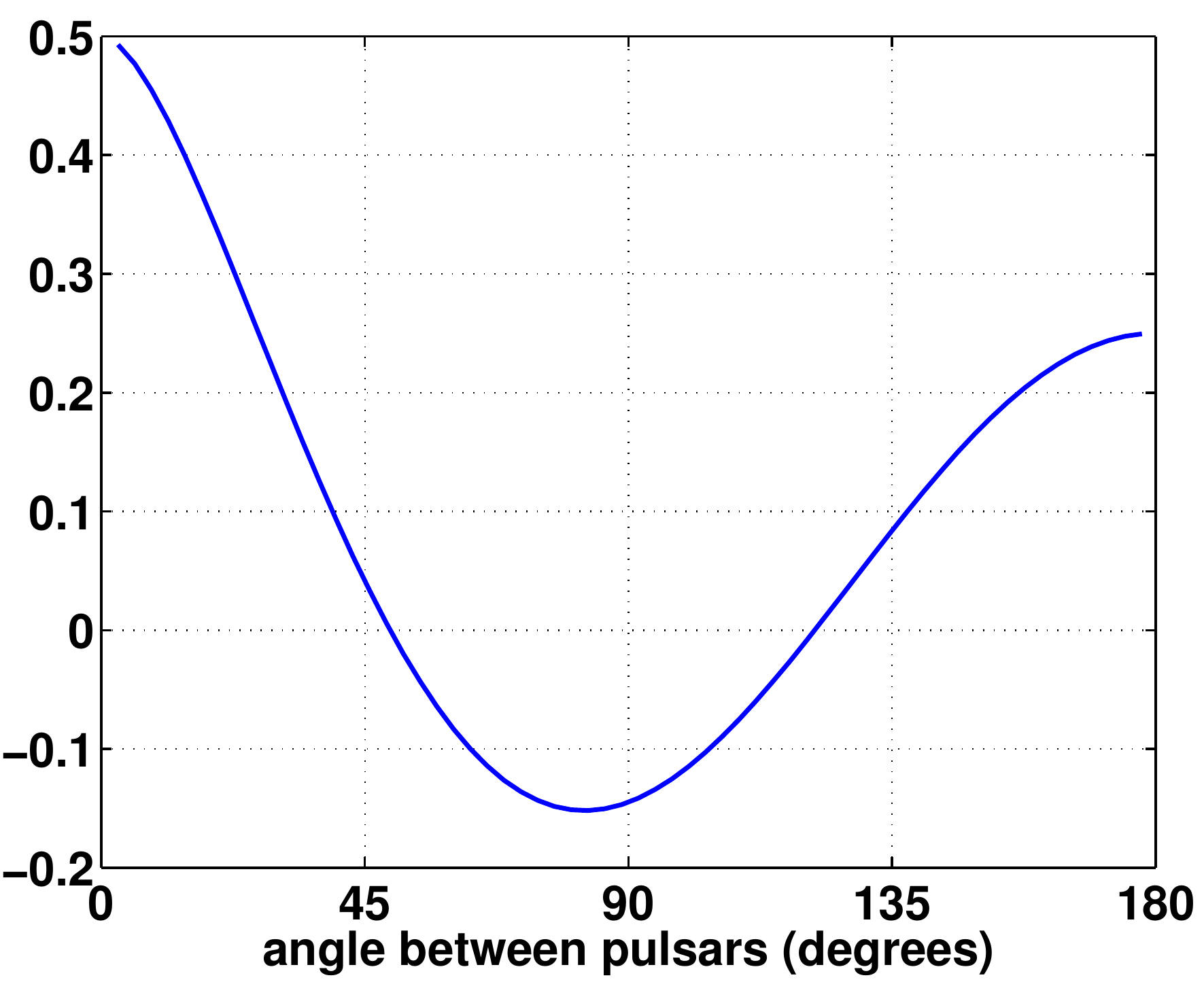}
\includegraphics[angle=0,width=0.49\columnwidth]{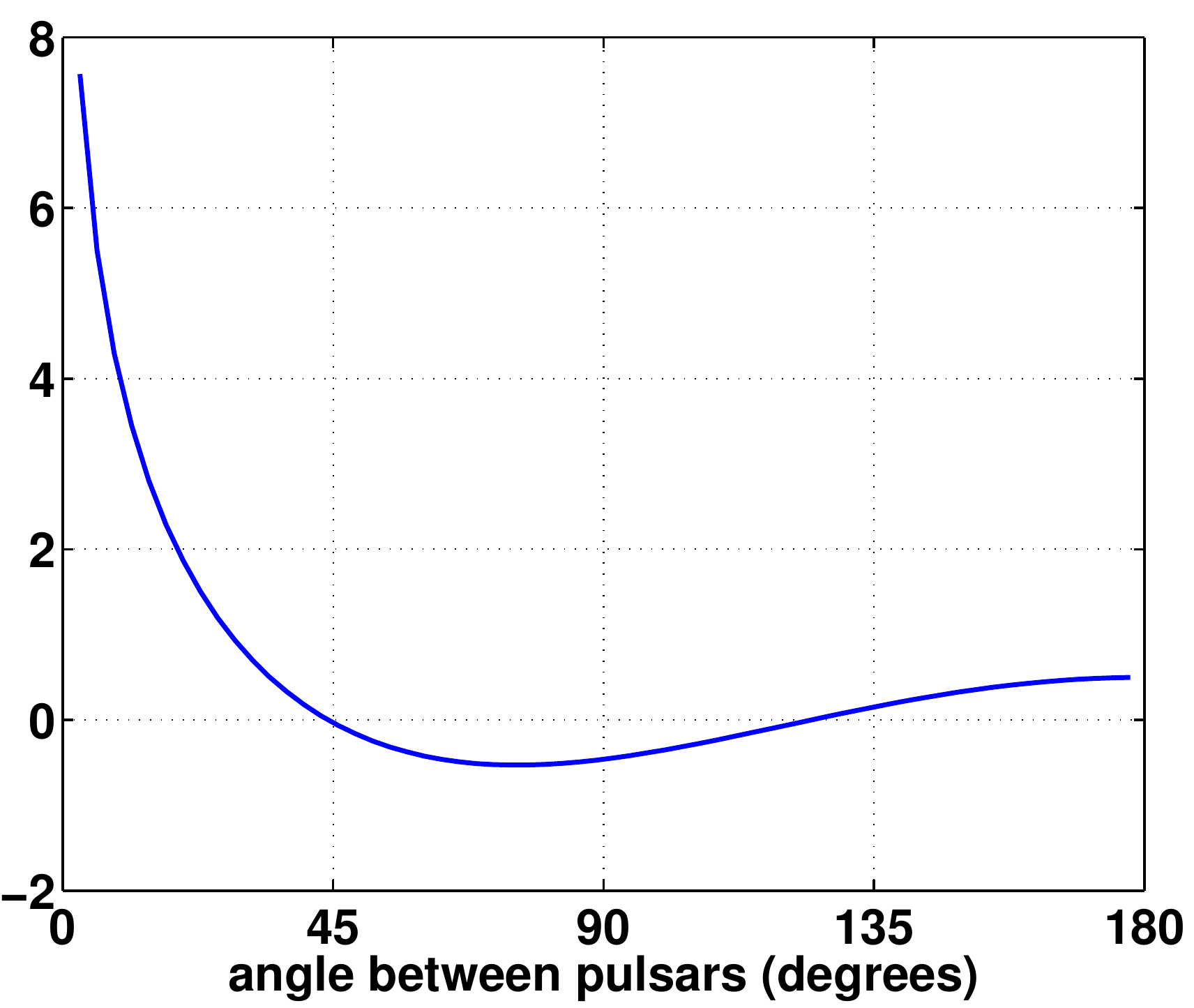}
\includegraphics[angle=0,width=0.49\columnwidth]{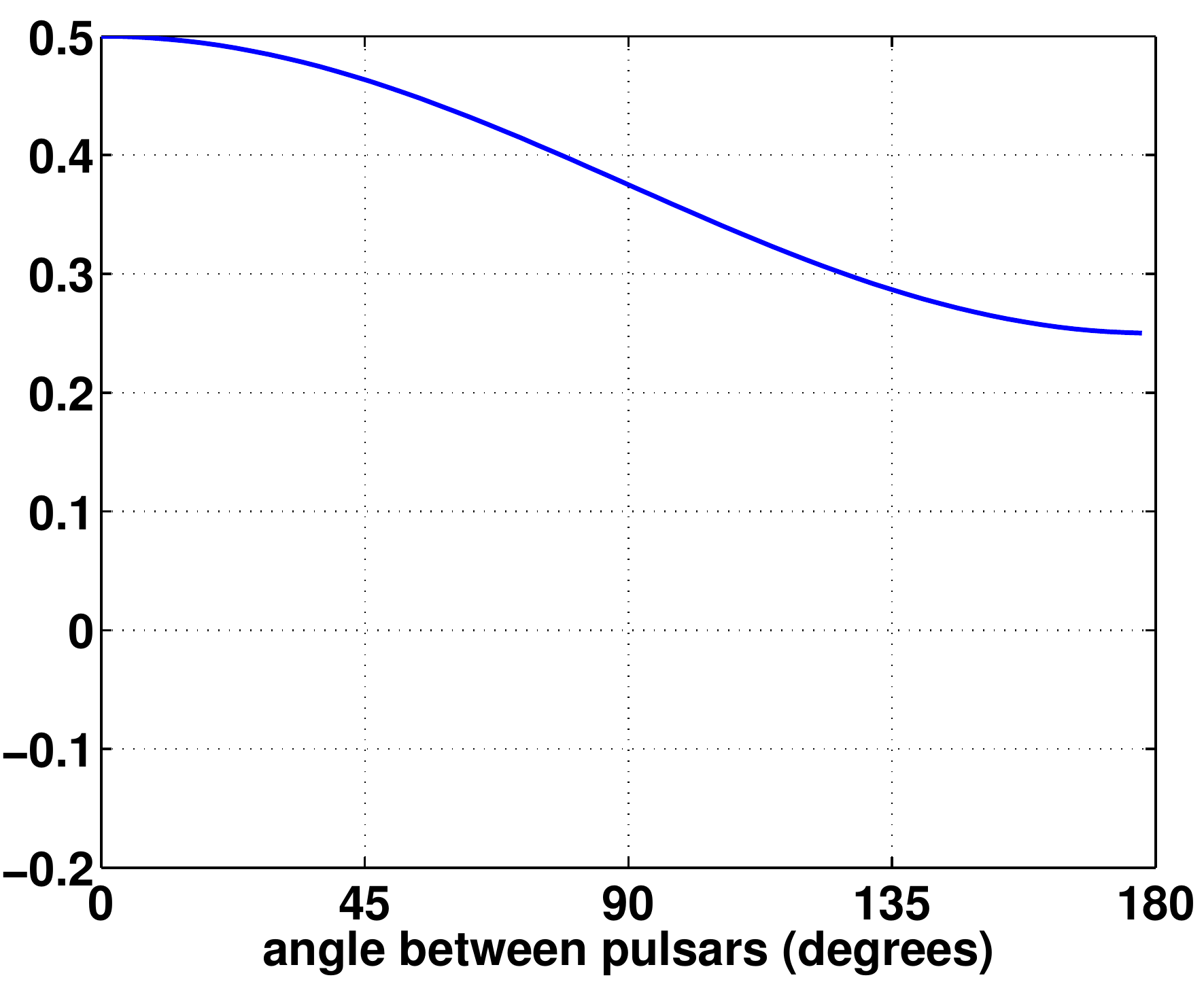}
\includegraphics[angle=0,width=0.49\columnwidth]{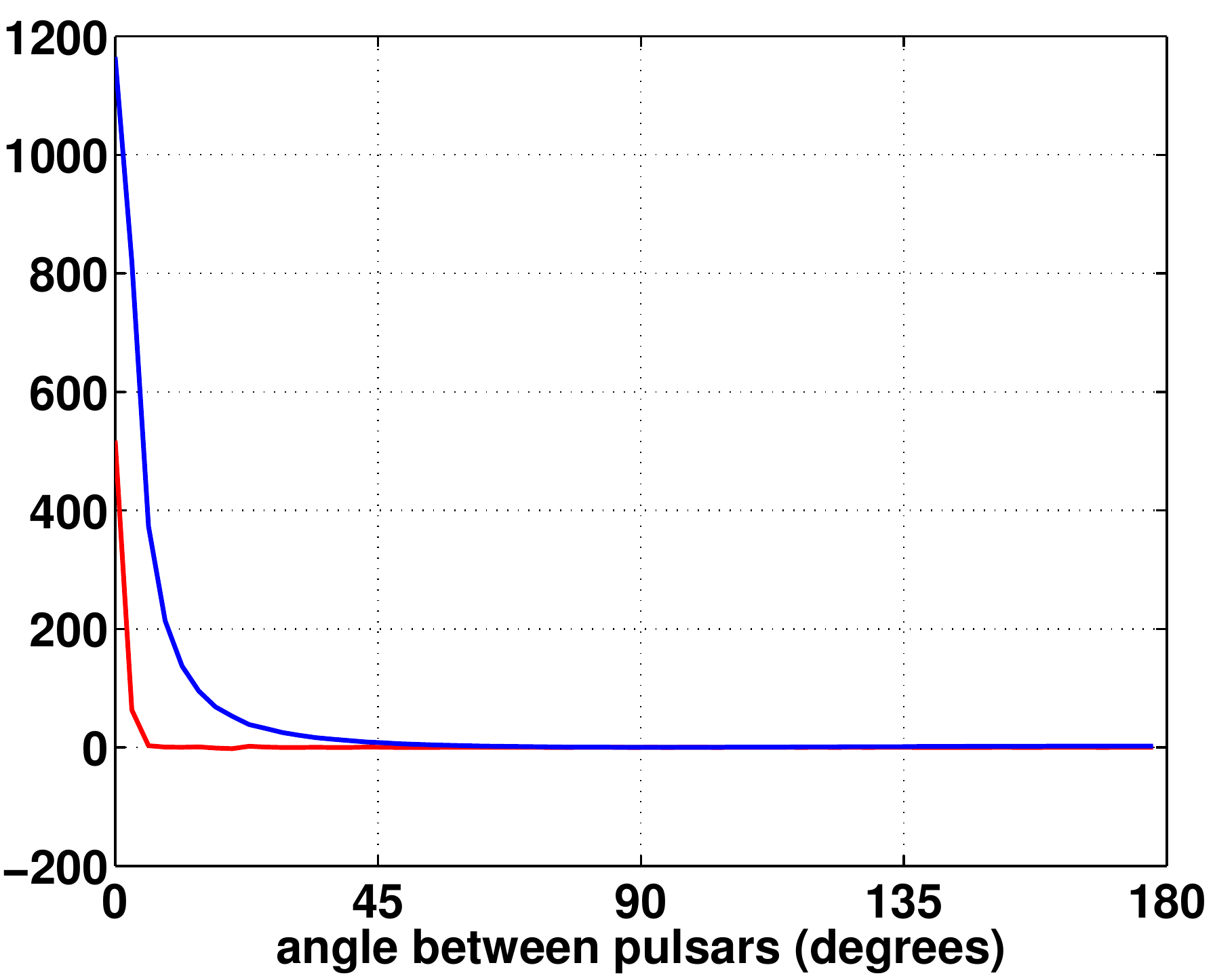}
\caption{Normalized overlap functions for the 
tensor (upper left),
vector-longitudinal (upper right), 
breathing (lower left), 
and scalar-longitudinal (lower right) polarization
modes, plotted as functions of the angular
separation between pairs of pulsars.
The blue and red curves in the lower right-hand
plot correspond to the real and imaginary parts of
the scalar-longitudinal overlap function.
Note the different vertical scales for the 
vector-longitudinal and scalar-longitudinal 
overlap functions, compared to those for the 
tensor and breathing modes.}
\label{f:overlap_altpolPTA}
\end{center}
\end{figure}

Plots of the normalized overlap functions for the 
tensor, vector-longitudinal, breathing and 
scalar-longitudinal modes 
are shown in Figure~\ref{f:overlap_altpolPTA},
plotted as functions of the angular separation $\zeta$
between pairs of pulsars.
The normalization is the same for each overlap 
function, chosen so that the tensor overlap function
agrees with the normalized Hellings and Downs curve 
(\ref{e:HD_normalized}).
The plots for the tensor, vector-longitudinal, 
and breathing modes are all real and do not 
depend on frequency;
the plot for the scalar-longitudinal modes has
both real and imaginary components (imaginary shown in red), 
and depends on frequency via the distances to 
the pulsars. 
For the scalar-longitudinal overlap function, we 
chose $y_1=1000$ and $y_2=2000$ for all pulsar pairs,
where $y\equiv 2\pi fL/c$, and we did the 
integration numerically over both $\theta$ and $\phi$.
Note the different vertical scales for the 
vector-longitudinal and scalar-longitudinal overlap functions,
compared to those for the tensor and breathing modes.
For small angular separations, the sensitivity to
vector-longitudinal modes is roughly an order of 
magnitude larger than that for the tensor and 
breathing modes, while the sensitivity to the scalar-longitudinal 
mode is several orders-of-magnitude larger.
This is consistent with what we found for the 
response functions,
as discussed at the end of Section~\ref{s:response_pol_altpolPTA}.

\subsubsection{Component separation and anisotropic backgrounds}
\label{s:comp_sep_altpolPTA}

As shown in \cite{Gair-et-al:2015}, the above 
calculations for non-GR polarization modes
can be extended to {\em anisotropic} stochastic
backgrounds.
The spherical harmonic components of the overlap 
functions
\be
\Gamma^A_{lm}(f)
=\frac{1}{4\pi}\int d^2\Omega_{\hat n}\> 
Y_{lm}(\hat n)R^A_I(f,\hat n)R^{A*}_J(f,\hat n)
\ee
can be calculated {\em analytically} for 
the tensor, vector, and breathing polarization
modes for all values of $l$ and $m$,
while the components of the scalar longitudinal 
overlap function admit only
semi-analytic expressions.
(This is similar to what we described in the 
previous section in the context of 
an isotropic background.)
Plots of the first few spherical harmonics 
components, as a function of the angular
separation $\zeta_{IJ}$ between a pair of 
pulsars, are 
given in Figures~1, 5, 2, and 3 of \cite{Gair-et-al:2015}.

The ability to separate the contributions to 
the background from the different 
polarization modes depends crucially on the form 
of the spherical harmonic basis response 
functions $R^P_{(lm)}(f)$, 
where $P=\{G,C, V_G, V_C, B,L\}$.
These were defined in (\ref{e:RP_altpol})
and have the $y\equiv 2\pi fL/c \gg 1$ 
limiting expressions 
given in (\ref{e:RPlist_altpolPTA}).
Recall that the $(lm)$ indices here 
correspond to an expansion of the Fourier components 
of the metric perturbations in terms of tensor 
(spin 2), vector (spin 1), and scalar (spin 0) 
spherical harmonics:
\be
h_{ab}(f,\hat n) 
= \sum_P \sum_{(lm)} a_{(lm)}^P(f) Y^P_{(lm)ab}(\hat n)\,,
\ee
for which
\be
\tilde h_I(f) =
\sum_P \sum_{(lm)}
R^P_{I(lm)}(f)a^P_{(lm)}(f)
\ee
is the response of pulsar $I$ to the background.
The expansion coefficients 
$a_{(lm)}^P(f)$ give the contributions
of the different 
polarization modes to the background, and
$R^P_{I(lm)}(f)$ are the response functions 
for those particular coefficients.
For an angular resolution of order 
$180^\circ/l_{\rm max}$, the total number of 
modes that are (in principle) accessible to a
pulsar timing array with a sufficient 
number of pulsars is
\be
N_m= 3(l_{\rm max}+1)^2 - 1\,.
\ee
This expression uses the result that 
the response to the curl modes for both the 
tensor and vector components are identically
zero, as is the response to the breathing 
modes for $l\ge 2$.
Since a pulsar timing array having $N_p$
pulsars can measure at most $2N_p$ real 
components of the background 
(as discussed in Section~\ref{s:basis_skies}), 
we see that
at least $N_p = N_m$ pulsars are required
to measure the $N_m$ (complex) components.
%
%

But as noted in Section~\ref{s:response_sph_altpolPTA},
all of the response functions $R^P_{(lm)}(f)$ 
depend on the direction $\hat p$ to the 
pulsar in exactly the same way, 
being proportional to $Y_{lm}(\hat p)$.
This degeneracy complicates the extraction 
of the different polarization modes.
For the tensor and breathing modes, the
degeneracy is broken since 
pulsar timing arrays typically operate 
in a regime where $y\gg 1$, 
for which the pulsar term can be ignored in 
the response functions for these modes.
In that limit, a pulsar timing array is only
sensitive to breathing modes with $l=0, 1$,
while the tensor modes are non-zero only for
$l\ge 2$.
On the other hand, the scalar-longitudinal
and vector-longitudinal modes can only be 
distinguished from the tensor and breathing
modes if there are multiple pulsars along 
the same line of sight, or if there is a 
known correlation between the expansion 
coefficients $a^P_{(lm)}(f)$ at 
different frequencies, e.g.,
a power-law spectrum.
For either of these two cases, we can exploit
the frequency dependence of the pulsar term,
which is more significant for the
longitudinal modes of the background.
Keeping all of the frequency-dependent terms~\cite{Gair-et-al:2015}:
\begin{multline}
R^{L}_{(lm)}(f) = 
2\pi(-1)^l
\bigg\{
-\delta_{l0} + \frac{1}{3} \delta_{l1}
\\
+(-i)^l {\rm e}^{-iy} \left[ \left(1-i\frac{l}{y}\right)j_l(y)
+ i j_{l+1}(y)\right]+ \frac{1}{2}H_l(y)\bigg\}
Y_{lm}(\hat p)\,,
\label{e:RSL-full}
\end{multline}
and
\begin{multline}
R^{V_G}_{(lm)}(f) 
=\pi(-1)^l\,{}^{(1)}\!N_l
\bigg\{\frac{4}{3} \delta_{l1} + 2 (-i)^l {\rm e}^{-iy}
\bigg[\left(1-\frac{il}{y}\right) (l+1)j_l(y) 
\\
- (y-i(2l+3))j_{l+1}(y) - iy j_{l+2}(y) \bigg] \bigg\}
Y_{lm}(\hat p)\,,
\label{e:RVL-full}
\end{multline}
for the scalar-longitudinal and vector-longitudinal
response functions, where $j_l(y)$ denotes a 
spherical Bessel functions of order $l$ 
and $y\equiv 2\pi fL/c$.
If we take the $y\gg 1$ limit of these equations,
we recover the approximate expressions given in
(\ref{e:RPlist_altpolPTA}).
But to separate the various components of the 
background, we need to use these more complicated 
expressions to break the angular-direction degeneracy.

A quantitative analysis of the sensitivity of a
phase-coherent mapping search 
(Section~\ref{s:phase-coherent})
to the different components $a^P_{(lm)}(f)$ 
of a stochastic background is given in \cite{Gair-et-al:2015}.
The results of that analysis are 
summarized in Table~\ref{t:comp_sep_altpol}, which
is taken from that paper.
The entries in the table show how the uncertainties in 
our measurements change as we search for:
(i) only the tensor modes,
(ii) both tensor and breathing modes,
(iii) tensor, breathing, and scalar-longitudinal modes, 
and (iv) all possible modes. 
The uncertainties were obtained by taking
the square root of the diagonal elements 
of the inverse of the Fisher matrix, following
the general prescription described in 
Section~\ref{s:likelihood_phase}.
For this calculation, 30~pulsars were
distributed randomly on the sky, with distances
chosen at random, uniformly between 1 and 10~kpc. 
There was only a single frequency component,
$f_0=3\times 10^{-9}~{\rm Hz}$,
and the measurement uncertainty (associated with 
pulse time of arrivals)
was assumed to be the same for all the pulsars in the array.
The background was also assumed to contain modes
with equal intrinsic amplitudes up to 
$l_{\rm max}=2$, so that the total number
of modes $N_m=26$ was less than the number of 
pulsars in the array.
This gave a fully-determined system of equations
that needed to be solved.
\begin{table}
\caption{Relative uncertainties for the tensor, breathing, scalar-longitudinal, 
and vector-longitudinal polarization modes searched for separately or in various 
combinations for $l_{\rm max}=2$ and $N_p=30$ pulsars.
This table is adapted from Table~II in \cite{Gair-et-al:2015}.}
\label{t:comp_sep_altpol}
\centering
\begin{tabular}{l | c c c c c c c c c}
\toprule
& \multicolumn{9}{c} {$(l,m)$ mode} \\ 
& $(0,0)$ & $(1,-1)$ & $(1,0)$ & $(1,1)$ & $(2,-2)$ & $(2,-1)$ & $(2,0)$ & $(2,1)$ & $(2,2)$ \\
\midrule
tensor & $-$ & $-$ & $-$ & $-$ & 0.44 & 0.38 & 0.32 & 0.38 & 0.44 \\
& & & & & & & & & \\
tensor & $-$ & $-$ & $-$ & $-$ & 0.49 & 0.39 & 0.37 & 0.39 & 0.49 \\
breathing & 0.16 & 0.53 & 0.46 & 0.53 & $-$ & $-$ & $-$ & $-$ & $-$ \\
& & & & & & & & & \\
tensor & $-$ & $-$ & $-$ & $-$ & 16.2 & 10.5 & 11.4 & 10.5 & 16.2 \\
breathing & 4.36 & 16.1 & 14.1 & 16.1 & $-$ & $-$ & $-$ & $-$ & $-$ \\
longitudinal & 0.71 & 0.96 & 0.84 & 0.96 & 1.21 & 0.78 & 0.86 & 0.78 & 1.21 \\
& & & & & & & & & \\
tensor & $-$ & $-$ & $-$ & $-$ & 1.4e5 & 5.4e4 & 8.0e4 & 5.4e4 & 1.4e5 \\
breathing & 18.4 & 9.4e4 & 6.2e4 & 9.4e4 & $-$ & $-$ & $-$ & $-$ & $-$ \\
longitudinal & 3.08 & 11.5 & 8.68 & 11.5 & 20.9 & 7.51 & 11.9 & 7.52 & 20.9 \\
vector & $-$ & 6.6e4 & 4.4e4 & 6.6e4 & 7.0e4 & 2.7e4 & 4.0e4 & 2.7e4 & 7.0e4 \\
\bottomrule
\end{tabular}
\end{table}

The entries in the table reflect our expectations 
for recovering the different modes of the background.
Namely, there is little change in our ability to 
recover the tensor modes when the breathing modes are
also included in the analysis.
This is because the tensor modes are non-zero only for
$l\ge 2$, while the response to the breathing modes
is non-zero only for $l=0,1$.
Adding the scalar-longitudinal modes to the analysis 
worsens the recovery of the tensor and breathing 
modes by about an order of magnitude, 
as the scalar-longitudinal modes can also have
non-zero values for all values of $l$.
(There are simply more parameters to recover.)
But one is still able to break the degeneracy as the
response to the scalar-longitudinal modes depends 
{\em strongly} on the distances to the pulsars.
The uncertainity in the recovery of the 
scalar-longitudinal modes is about an order of magnitude
less than that for the tensor and breathing modes, since the
analysis assumes equal intrinsic amplitudes for all the
modes, while the correlated response to the 
scalar-longitudinal modes is much larger for small
angular separations between the pulsars 
(Section~\ref{s:overlap_altpolPTA} and Figure~\ref{f:overlap_altpolPTA}).
Finally, adding the vector-longitudinal modes to 
the analysis weakens
the recovery of the scalar-longitudinal
modes by about an order of magnitude, 
again because more parameters need to be recovered.
However, it {\em severely worsens} the recovery of 
all the other modes, because of the degeneracy in the
response on the angular direction to the pulsars.
There is some dependence on frequency for the vector-longitudinal
response, as indicated in (\ref{e:RVL-full}), but it is much
weaker than the frequency dependence of the scalar-longitudinal
modes.
So the degeneracy is not broken nearly as strongly for these modes.
See \cite{Gair-et-al:2015} for more details.

\subsection{Other searches}
\label{s:others}

It is also possible to use the general 
cross-correlation techniques 
described in Section~\ref{s:corr} to search 
for signals that don't really constitute a stochastic 
gravitational-wave background.
Using a stochastic-based cross-correlation method to 
search for such signals is not optimal, but it
still gives valid results for detection statistics 
or estimators of signal parameters, with error bars 
that properly reflect the uncertainty in these 
quantities.
It is just that these error bars are {\em larger} 
than those for an optimal (minimum variance) search,
which is better ``tuned" for the signal.
Below we briefly describe how the general cross-correlation 
method can be used to search for
(i) long-duration unmodelled transients and 
(ii) persistent (or continuous) gravitational
waves from targeted sources.

\subsubsection{Searches for long-duration unmodelled transients}
\label{s:STAMP}

The Stochastic Transient Analysis Multi-detector 
Pipeline~\cite{Thrane-et-al:2011}
(STAMP for short) is a cross-correlation search for
unmodelled long-duration transient signals (``bursts") 
that last on order a few seconds to several hours 
or longer.
The duration of these transients are long compared 
to the typical merger signal from inspiralling binaries 
(tens of milliseconds to a few seconds), 
but short compared to the persistent 
quasi-monochromatic signals that one expects from
e.g., rotating (non-axisymmetric) neutron stars.
STAMP was developed in the context of ground-based
interferometers, but the general method, which we 
briefly describe below, is also valid for other types 
of gravitational-wave detectors.

STAMP is effectively an adapted gravitational-wave
radiometer search (Section~\ref{s:radiometer-SHD}),
which cross-correlates data from pairs of detectors 
(\ref{e:CIJ}), weighted by the {\em inverse} of the 
integrand of the overlap 
function $\gamma_{IJ}(t;f,\hat n)$ for a particular 
direction $\hat n$ on the sky:
\be
\tilde s_{IJ}(t;f,\hat n)\equiv
\frac{2}{\tau}
\frac{\tilde d_I(t;f)\tilde d_J^*(t;f)}
{\gamma_{IJ}(t;f,\hat n)}\,,
\ee
where $\tau$ is the duration of the segments defining
the short-term Fourier transforms.
The weighting by the inverse of 
$\gamma_{IJ}(t;f,\hat n)$
is used so that the expected value of 
$\tilde s_{IJ}(t;f,\hat n)$
is just the gravitational-wave power in pixel $(t;f)$
for a point source in direction $\hat n$, which follows 
from (\ref{e:<CIJ>}).
The data $\tilde s_{IJ}(t;f,\hat n)$ for 
a single direction $\hat n$ define a 
{\em time-frequency map}.
For a typical analysis using the LIGO Hanford
and LIGO Livingston interferometer, a single
map has a frequency range from about 50 to
$\sim\!1000~{\rm Hz}$, and a time duration of 
a couple hundred seconds (or whatever the 
expected duration of the transient might be).
A strong burst signal shows up as {\em cluster} 
or {\em track} of bright pixels in the 
time-frequency map, which stands out above the noise.
The data analysis problem thus becomes a
{\em pattern recognition} problem.

The procedure for deciding whether or not a 
signal is present in the data can be broken 
down into three steps:
(i) determine if a statistically significant 
clump or track of bright pixels is present in
a time-frequency map, which requires using 
some form of pattern-recognition or 
clustering algorithm 
(see~\cite{Thrane-et-al:2011} and relevant
references cited therein);
(ii) calculate the value of the detection 
statistic $\Lambda$, obtained from a 
weighted sum of the power in the pixels 
for each cluster determined by the previous step;
(iii) compare the observed value of the 
detection statistic to a threshold value 
$\Lambda_*$, which depends on the desired false 
alarm rate.
This threshold is typically calculated by 
time-shifting the data to empirically determine 
the sampling distribution of $\Lambda$ in the 
absence of a signal.
If $\Lambda_{\rm obs}>\Lambda_*$, then 
reject the null hypothesis and claim detection 
as discussed in \ref{s:freq-hypothesis-testing}.
(Actually, in practice, this last step is a bit 
more complicated, as one typically does 
follow-up investigations using auxiliary 
instrumental and environmental channels, 
and data quality indicators.
This provides additional confidence that the 
gravitational-wave candidate is not some 
spurious instrumental or environmental artefact.)

\begin{figure}[h!tbp]
\begin{center}
\includegraphics[angle=0,width=0.49\columnwidth]{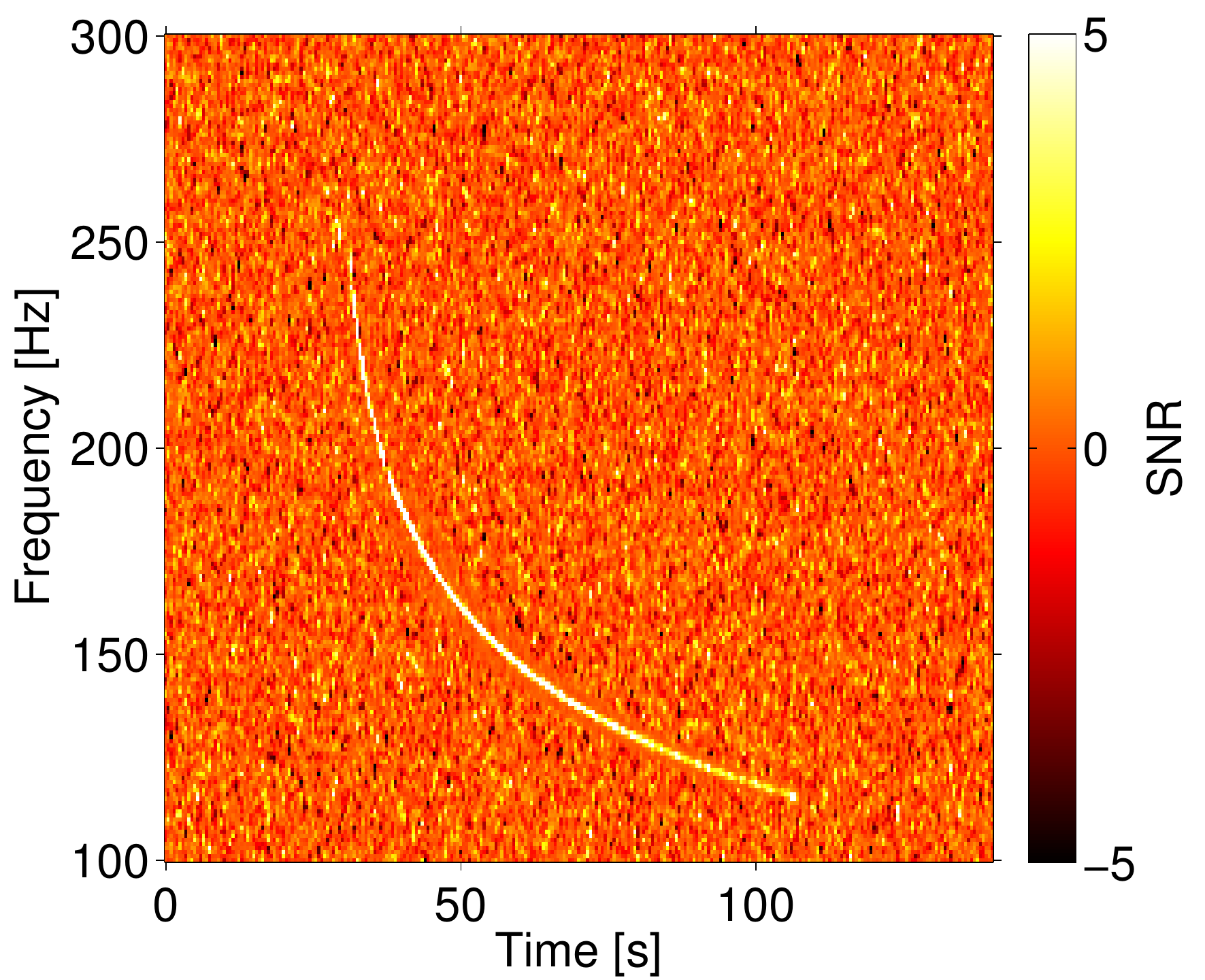}
\includegraphics[angle=0,width=0.49\columnwidth]{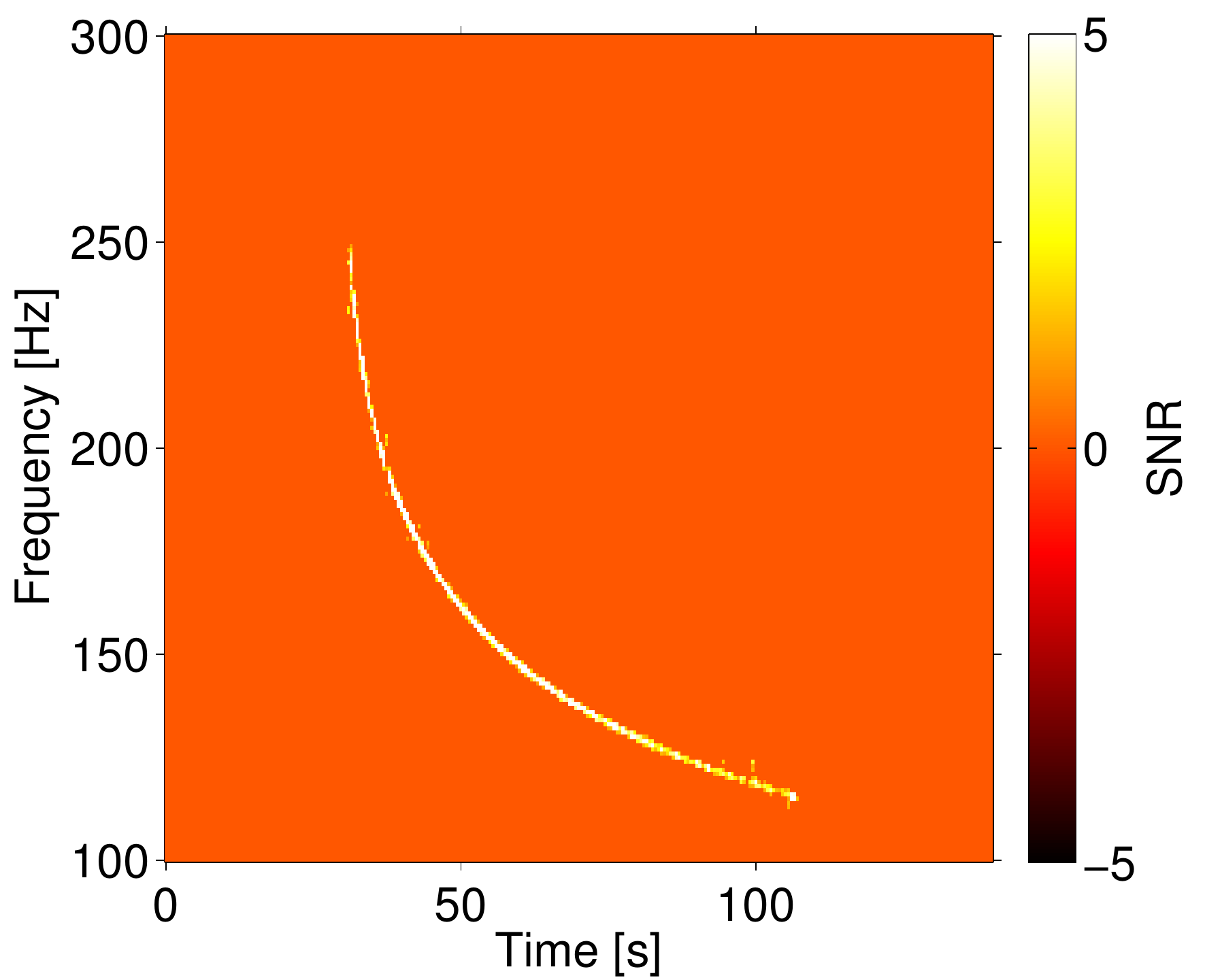}
\caption{Time-frequency maps for an injected
long-duration transient gravitational-wave signal
in noise.
Left panel: signal-to-noise ratio map before processing.
Right panel: signal-to-noise ratio map after applying a 
clustering algorithm.
Note that the noise fluctuations have effectively
been eliminated in the second plot.
Images provided by Tanner Prestegard.}
\label{f:STAMP}
\end{center}
\end{figure}

Figure~\ref{f:STAMP} is an example of a 
time-frequency map with a simulated long-duration 
gravitational-wave signal injected into simulated 
initial LIGO detector noise. 
This particular signal is an {\em accretion disk 
instability waveform}, based on a model by 
van~Putten~\cite{vanPutten:2001, vanPutten:2003, vanPutten:2008}.
The signal is a (inverse) ``chirp" in gravitational radiation 
having an exponentially decaying frequency.
(The magnitude of the signal increases with time as 
the frequency {\em decreases}.)
The injected signal is strong enough to be
seen by eye in the raw time-frequency map 
(left panel).
After applying a clustering algorithm, the
fluctuations in the detector noise have been
noticeably reduced (right panel).

Readers should see~\cite{Thrane-et-al:2011} 
for many more details regarding STAMP,
and~\cite{Abbott-et-al:STAMP-S5S6, Aasi-et-al:STAMP-GRB}
for results from analyses of LIGO data taken during 
their 5th and 6th science runs---the first paper 
describes an all-sky
search for long-duration gravitational-wave transients;
the second, a triggered-search for long-duration
gravitational-transients coincident with long duration
gamma-ray bursts.

\subsubsection{Searches for targeted-sources of continuous gravitational waves}
\label{s:corrCW}

The gravitational-wave radiometer method
(Section~\ref{s:radiometer-SHD}) can also be used to 
look for gravitational waves from persistent 
(continuous) sources at known locations on the sky, 
e.g., the galactic center, the location of SN 1987A, or from
low-mass X-ray binaries like Sco~X-1 
\cite{Abadie-et-al:S5-anisotropic, Messenger-et-al:2015, LVC:O1-anisotropic}.
For example, Sco~X-1 is expected to emit gravitational 
waves from the (suspected) 
rotating neutron star at its core, 
having non-axisymmetric distortions produced by 
the accretion of matter from the low-mass 
companion.
The parameters of this system that determine
the phase evolution of the gravitational 
radiation are not well-constrained:
(i) Since the neutron star at the
core has not been observed to emit pulsations 
in the radio or any electromagnetic band, the 
orbital parameters of the binary are estimated 
instead from optical observations of the low-mass
companion~\cite{Steeghs-Casares:2002, Galloway-et-al:2014}.
These observations do not constrain the orbital 
parameters as tightly as being able to directly 
monitor the spin frequency of the neutron star.
(ii) The intrinsic spin evolution of the
neutron star also has large uncertainties
due to the high rate of accretion from the 
low-mass companion star.
Both of these features translate into a 
{\em large} parameter space volume over 
which to search, making fully-coherent 
matched-filter searches for the 
gravitational-wave signal 
computationally challenging~\cite{Messenger-et-al:2015}.

Nonetheless, for such sources, one can perform a 
{\em narrow-band, targeted} radiometer search, 
cross-correlating data from a pair of detectors
with a filter function proportional 
to the integrand $\gamma_{IJ}(t;f,\hat n_0)$ of
the overlap function evaluated at the direction
$\hat n_0$ to the source on the sky:
\be
\hat C_{IJ}(f) =\frac{2}{\tau} \sum_t 
\tilde d_I(t;f) \tilde d_J^*(t; f)
\frac{{\cal N}_{IJ}(f)\gamma_{IJ}(t; f,\hat n_0)}{P_{n_I}(t;f) P_{n_J}(t; f)}\,,
\ee
where 
\be
{\cal N}_{IJ}(f) \equiv \left[\sum_t 
\frac{\gamma^2_{IJ}(t; f,\hat n_0)}{P_{n_I}(t;f) P_{n_J}(t; f)}\right]^{-1}\,.
\ee
The search is narrow-band in the sense that one
doesn't integrate over the whole frequency band 
of the detectors, but looks instead for evidence
of a gravitational wave in narrow frequency 
bins that span the sensitive band of the detector.
The weighted cross-correlations are summed over time, 
to build up signal-to-noise ratio, since the 
source is assumed to be persistent.
The frequentist detection statistic is the 
squared signal-to-noise ratio of the 
cross-correlated power contained in each narrow 
frequency band:
\be
\Lambda(d) = \frac{|\hat C_{IJ}(f)|^2}{{\rm Var}[\hat C_{IJ}(f)]}
\approx \frac{|\hat C_{IJ}(f)|^2}{{\cal N}_{IJ}(f)}\,,
\ee
where we used the result that the variance of
the cross-correlation estimator $\hat C_{IJ}(f)$ 
equals the normalization factor ${\cal N}_{IJ}(f)$ 
in the weak-signal limit.
This modified radiometer search is {\em robust}
in the sense that it makes minimal assumptions about the 
source.
The detection efficiency of the search could be 
improved if one had additional information about the 
signal (e.g., if one knew that the radiation was 
circularly polarized), which could then be included 
in the stochastic signal model.

\section{Real-world complications}
\label{s:complications}

\begin{quotation}
Experience with real-world data, however, soon convinces one that 
both stationarity and Gaussianity are fairy tales invented for the 
amusement of undergraduates.
{\em D.J.~Thompson}~\cite{389899}
\end{quotation}

\noindent
The analyses described in the previous sections assumed that the
instrument noise is stationary, Gaussian distributed, and uncorrelated
between detectors. The analyses also implicitly assumed that the data
were regularly sampled and devoid of gaps, facilitating an easy
transition between the recorded time series and the frequency domain
where many of the analyses are performed. In practice, all of these
assumptions are violated to varying degrees, and the analyses of real
data require additional care. Analyses that assume stationary,
Gaussian noise can produced biased results when applied to more
complicated real-world data sets.

\subsection{Observatory-specific challenges}

To begin the discussion, 
we highlight some of the challenges associated with 
real-world data, which are specific to the different 
observational domains---e.g., ground-based detectors,
space-based detectors, and pulsar timing.
Then, in the following subsections, we discuss the
complications in more detail, and suggest ways to 
deal with or mitigate these problems.

\subsubsection{Ground-based interferometers}

Analysis of data from the first and second generation ground-based
interferometers have shown that the data are neither perfectly
stationary nor Gaussian. The non-stationarity can be broadly
categorized as having two components: slow, adiabatic drifts in the
noise spectrum with time; and short-duration noise transients,
referred to as {\em glitches}~\cite{Blackburn-et-al:2008}, 
which have compact support in
time-frequency. These glitches are also the dominant cause of
non-Gaussianity in the noise distributions, giving rise to long 
``tails" (large amplitude events with non-negligible probability),
which extend past a core distribution that is well described as
Gaussian. The data are evenly sampled by design, though there are
often large gaps between data segments due to ``loss of lock"
(the interferometer being knocked out of data-taking mode due 
to an environmental disturbance or instrumental malfunction), 
scheduled maintenance, etc.

An analysis of LIGO-Virgo data that assumes the noise spectrum is
constant over days or weeks would produce biased results.  In
practice, the data is analyzed using $\sim 1$ minute-long segments.
Glitches, on the other hand, do not pose a significant problem for
stochastic searches as they are rarely coherent between detectors.
Glitches are a more serious problem for searches that target short
duration, deterministic signals.

\subsubsection{Pulsar timing arrays}

Pulsar timing data are, in many ways, far more challenging to
analyze~\cite{vanHaasteren-Levin:2010}.
The lack of dedicated telescope facilities, and the practical
constraints associated with making the observations, result in data
that are irregularly sampled. Moreover, the very long observation
timelines (years to decades) and the mixture of facilities yield data
sets that have been collected using a variety of receivers, data
recorders, and pulse folding schemes. The heterogeneity of the 
observations causes the data to be non-stationary.
In addition, the characteristic period of the gravitational waves 
searched for is of order the duration of the observations.
Thus, Fourier domain methods for pulsar timing analyses have, at best, 
limited formal utility.

An additional complication for pulsar timing analyses 
is that a complicated deterministic 
timing model that predicts the time of arrival of each pulse has 
to be subtracted from the data to produce the timing residuals used in the
gravitational-wave analyses. The timing model includes a pulsar
spindown model and a detailed pulse propagation model that accounts
for the relative motion of the Earth and pulsar.  Many of the pulsars
are in binary systems, so the timing model has to include relativistic
orbital motion, and propagation effects such as the Shapiro time delay.
Since errors in the timing model are strongly correlated with the 
gravitational-wave signal, subtracting the timing model unfortunately
removes part of the signal as well.
Subtraction of the timing model also introduces non-stationarity into 
the data~\cite{vanHaasteren-Levin:2013}, again making time-domain 
analyses the only possibility~\cite{vanHaasteren:2008yh}.

\subsubsection{Space-based detectors}

For future space detectors we can only guess at the nature of the
noise. Results from the LISA Pathfinder mission provide some
insight~\cite{Pathfinder:2016}, 
but only for a subset of the detector components, and for
somewhat different flight hardware. The data will be regularly
sampled, but data gaps are expected due to re-pointing of the
communication antennae and orbit adjustments. Possible sources of
non-stationarity include variations in the solar wind, thermal
variations, and tidal perturbations from the Earth and other solar
system bodies. The plans for the first space interferometers envision
a single array of 3 spacecraft with 6 laser links. From these links
three noise-orthogonal signal channels can be synthesized, but these
combinations are also signal orthogonal, and so cross-correlation 
cannot be used to detect a signal.

\subsection{Non-stationary noise}
\label{s:nonstationarynoise}

Data from existing gravitational-wave detectors, including bars,
interferometers, and pulsar timing, exhibit various degrees of
non-stationarity. Here we give examples relevant to ground-based
interferometers, but the situation is similar for the other detection
techniques.

Non-stationary behavior can manifest itself in many forms, and there are no
doubt many factors that contribute to the non-stationarity seen in
interferometer data. Nonetheless, a simple two-part model does a good
job of capturing the bulk of the non-stationary features. The two-part
model consists of a slowly-varying noise spectral density $S_n(t;f)$,
and localized noise transients or ``glitches''. The slow drift in the
spectrum can be modeled as a locally-stationary noise
process~\cite{2011arXiv1109.4174D}, which has the nice feature that
for small enough time segments, the data in each segment can be
treated as stationary. The glitch contribution to the non-stationarity
poses more of a challenge, as the non-stationarity persists even for
short data segments.

\subsubsection{Local stationarity}
\label{s:localstationarity}

\begin{figure}[h!tbp]
\begin{center}
\subfigure[]{\includegraphics[width=.49\textwidth]{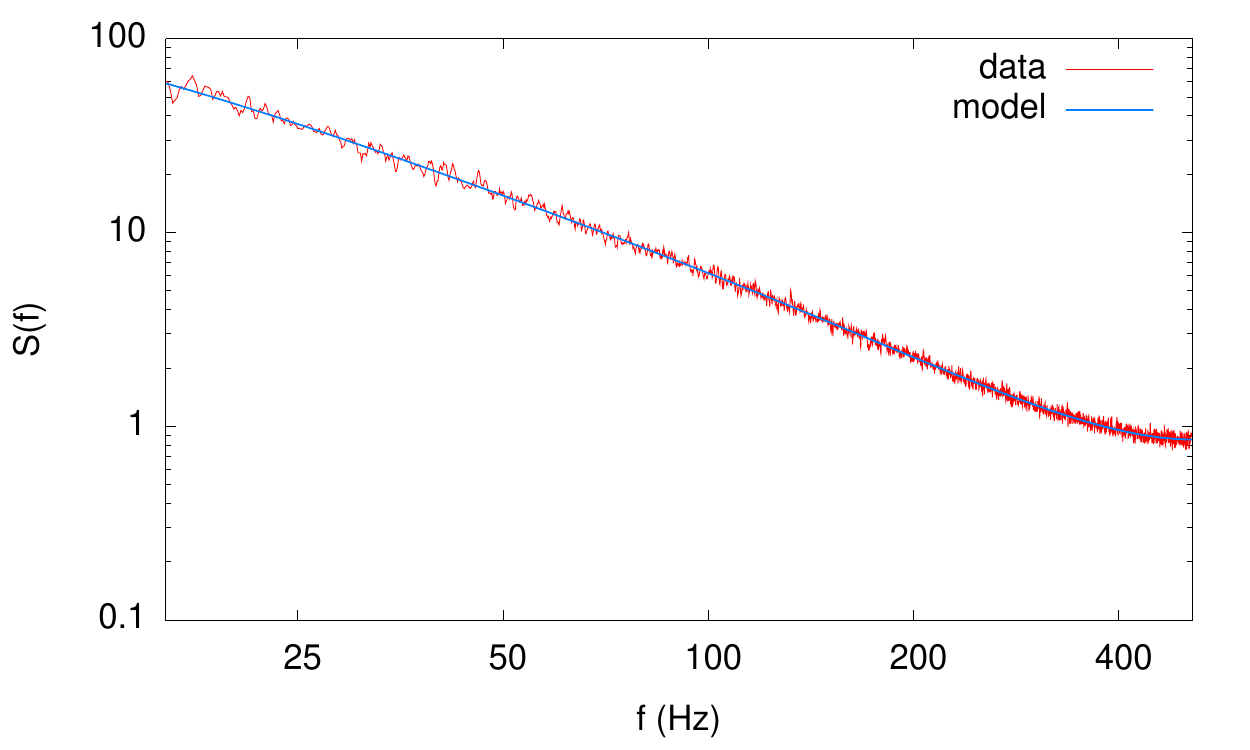}}
\subfigure[]{\includegraphics[width=.49\textwidth]{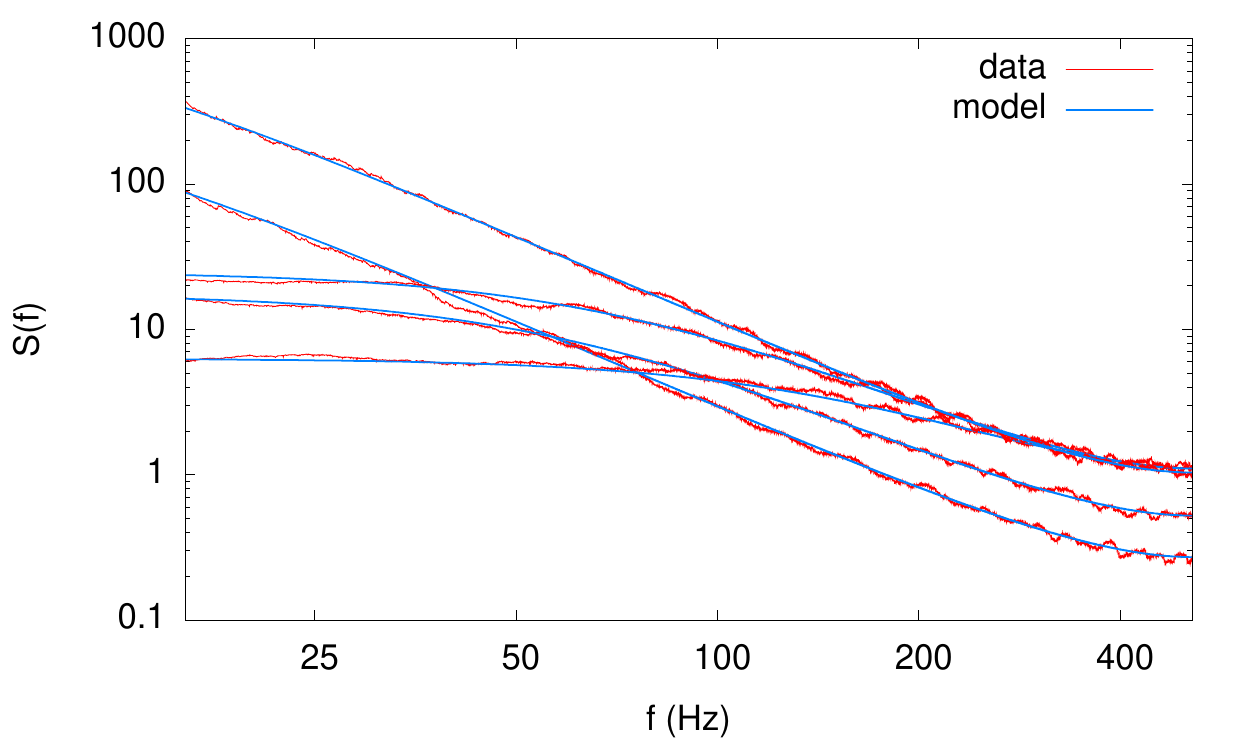}}
\caption{Spectra for the locally-stationary AR(1) model. Panel
  (a) shows the smoothed spectrum computed using the full data set
  compared to the time average of the theoretical spectrum.  Panel (b)
  shows smoothed spectra from the $1^{\rm st}$, $8^{\rm th}$, $16^{\rm
    th}$, $24^{\rm th}$ and $32^{\rm nd}$ time segments compared to
  the theoretical $S(t;f)$ computed at the central time for each
  segment.}
\label{comp:spec}
\end{center}
\end{figure}

To illustrate the two-component description of non-stationary data, we
begin with a toy model of a locally-stationary red noise
process. Later we will add a model for the impulsive, glitch
component (Section~\ref{s:glitches}).  
Consider an auto-regressive AR(1) process of the form:
\begin{equation}\label{ar}
x_t = q(t) x_{t-1} + \epsilon(t) \delta\,,
\end{equation}
where $\delta \sim N(0,1)$ is a unit-variance 
Gaussian deviate, and $q(t)$ and
$\epsilon(t)$ are slowly-varying functions of time. The local power
spectrum $S(t;f)$ for this process has the form
\begin{equation}\label{spec}
S(t;f)  = \frac{\epsilon^2(t)}{1+ q^2(t) - 2 q(t) \cos(\pi f/f_{\rm N})} \,,
\end{equation}
where $f_{\rm N}$ is the Nyquist frequency. For a data segment of duration
$T$ with $N$ samples, $f_{\rm N} = N/(2 T)=1/(2\Delta t)$. Figure~\ref{comp:spec} shows
the average and local spectra for $T=1024$ seconds of data sampled at
$1024$ Hz with 
\be
\begin{aligned}
q(t) = q_0 \frac{(1+ \alpha \cos(2 \pi t/T))}{(1+\alpha)}\,,
\qquad
\epsilon(t) = \epsilon_0 (1+t/T)\,,
\end{aligned}
\ee
and $q_0=0.95$, $\alpha=0.4$, and $\epsilon_0 = 1$.
The local spectra are computed using 32-second segments of data that
are smoothed and compared to the predicted spectra (\ref{spec}).  The
smoothed average spectrum is computed using the full data set and
compared to the theoretical average spectrum
\begin{equation}
S(f)  = \frac{1}{T} \int_0^T S(t;f) \, dt \,.
\end{equation}
The high degree of non-stationarity is clearly apparent from the
several orders of magnitude variation in the spectra across different
segments of data. In LIGO stochastic background analyses, a ``delta
sigma'' cut is used to discard segments of data that exhibit
significant non-stationarity.  The square-root of the variance
(\ref{e:varOmegabetahat}) of the cross-correlation statistic is compared
between three consecutive short segments of data (each typically 
60~seconds long), and if the levels differ by more than 20\%--30\% those
segments are not used in the analysis~\cite{S3HLiso, S4HLiso}.

\begin{figure}[h!tbp]
\begin{center}
\subfigure[]{\includegraphics[width=.49\textwidth]{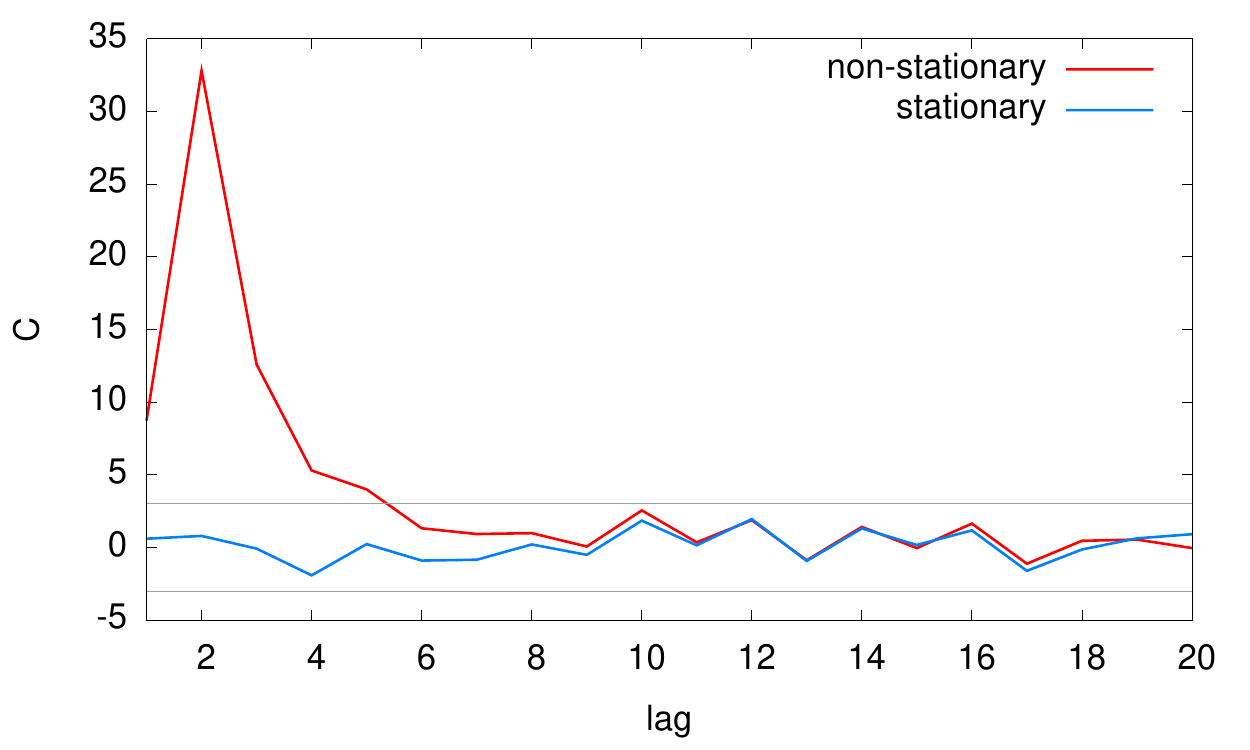}}
\subfigure[]{\includegraphics[width=.49\textwidth]{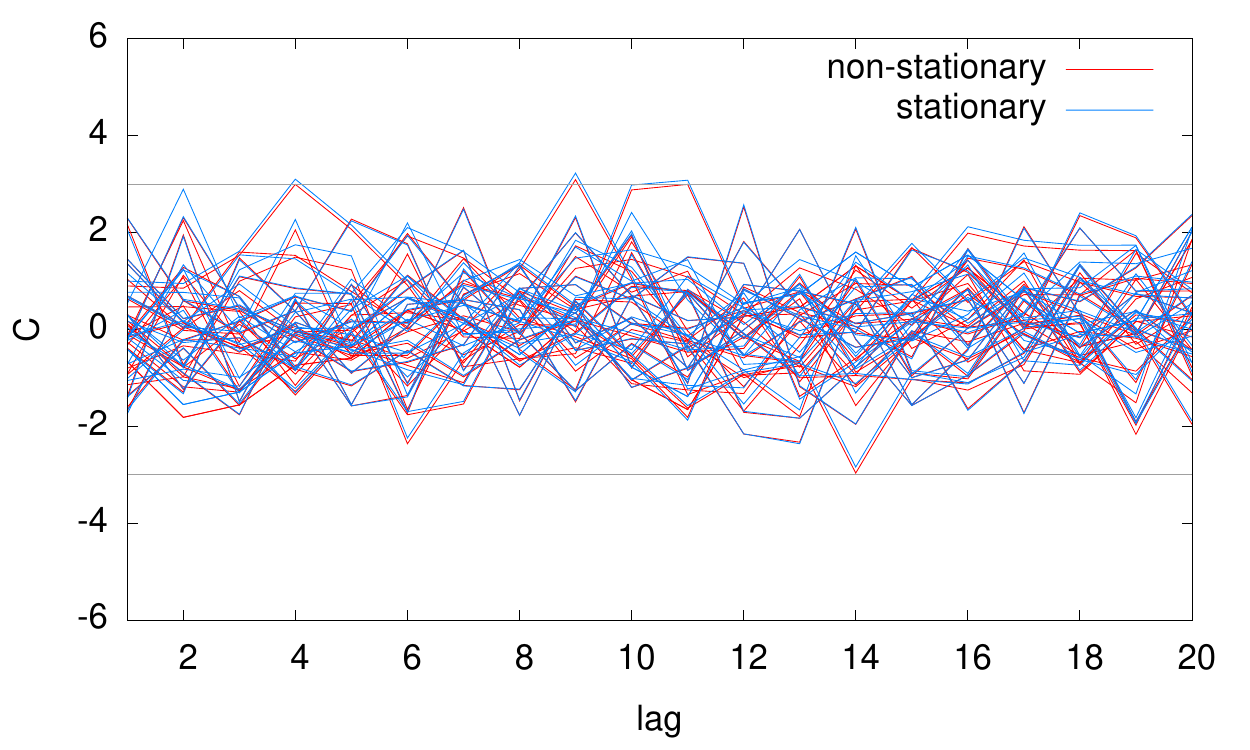}}
\caption{Autocorrelation of the whitened Fourier coefficients as a
  function of lag for stationary and locally-stationary AR(1)
  models. Panel (a) is a comparison for the full data sets, and panel
  (b) is for each of the 32 sub-segments. The locally-stationary data
  show clear departures from stationarity in the full data set, but
  are consistent with stationarity in the shorter sub-segments of
  data.}
\label{comp:autoc}
\end{center}
\end{figure}

The degree of non-stationarity can be measured from the
auto-correlation of the whitened Fourier coefficients $\bar{x}_f =
\tilde{x}_f/\sqrt{\hat{S}(f)}$, where $\hat{S}(f)$ is estimated from
the smoothed power spectra.  The auto-correlation at lag $k$ is
defined by
\be 
c(k) \equiv \frac{1}{2N} \sum_{i=1}^N (\bar{x}_i
\bar{x}^*_{i+k} + \bar{x}^*_i \bar{x}_{i+k}) \,.  
\ee
For stationary, Gaussian noise in the large-$N$ limit, $c(k)$ for $k >
0$ is Gaussian distributed with zero mean and variance
$\sigma^2=1/N$~\cite{2009arXiv0911.4744D}. It is convenient to use the
scaled auto-covariance $C(k) \equiv\sqrt{N} c(k)$, which has unit
variance for stationary, Gaussian noise. Figure~\ref{comp:autoc}
compares $C(k)$ computed for the locally-stationary AR(1) model
shown in Figure~\ref{comp:spec}, and a stationary AR(1) model with
$q(t)=q_0$ and $\epsilon(t) = \epsilon_0$.
The locally-stationary model shows clear departures from stationarity
when the auto-correlation is computed using the full data set (as
evidenced by the large autocorrelations for small lags), while the
data in each of the 32 sub-segments shows no signs of
non-stationarity.

\begin{figure}[h!tbp]
\begin{center}
\subfigure[]{\includegraphics[width=.49\textwidth]{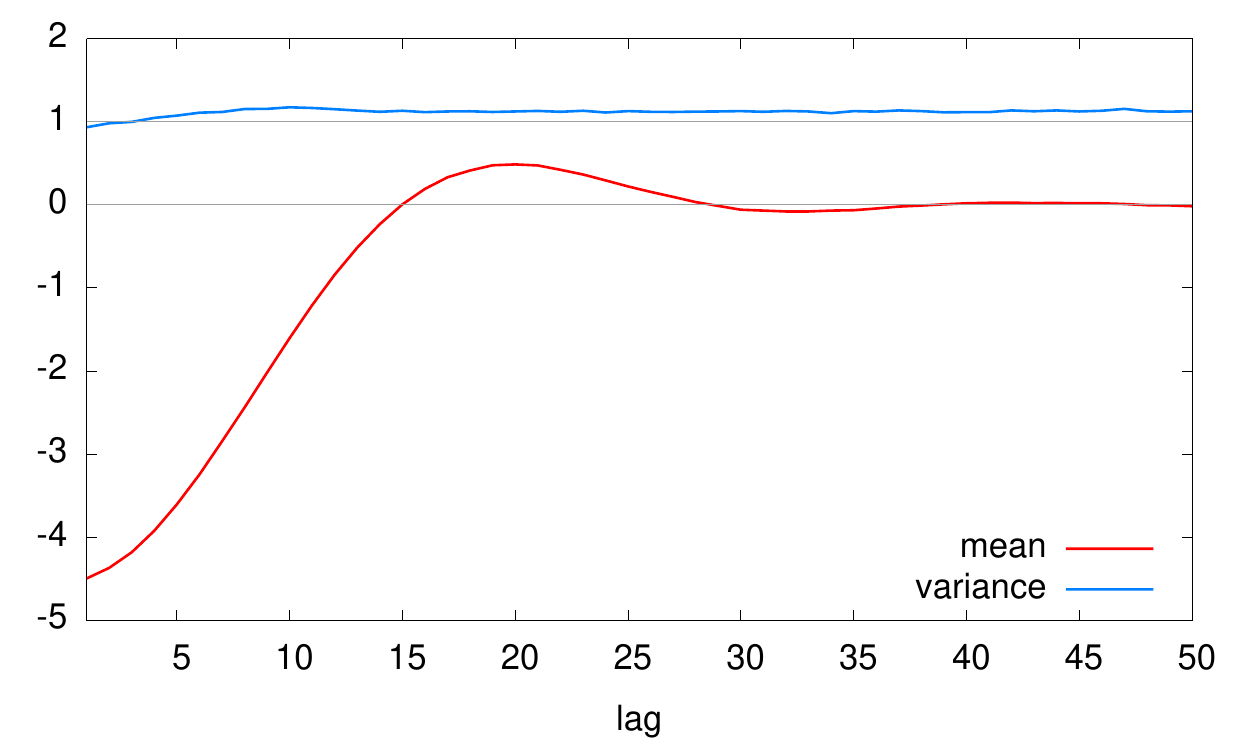}}
\subfigure[]{\includegraphics[width=.49\textwidth]{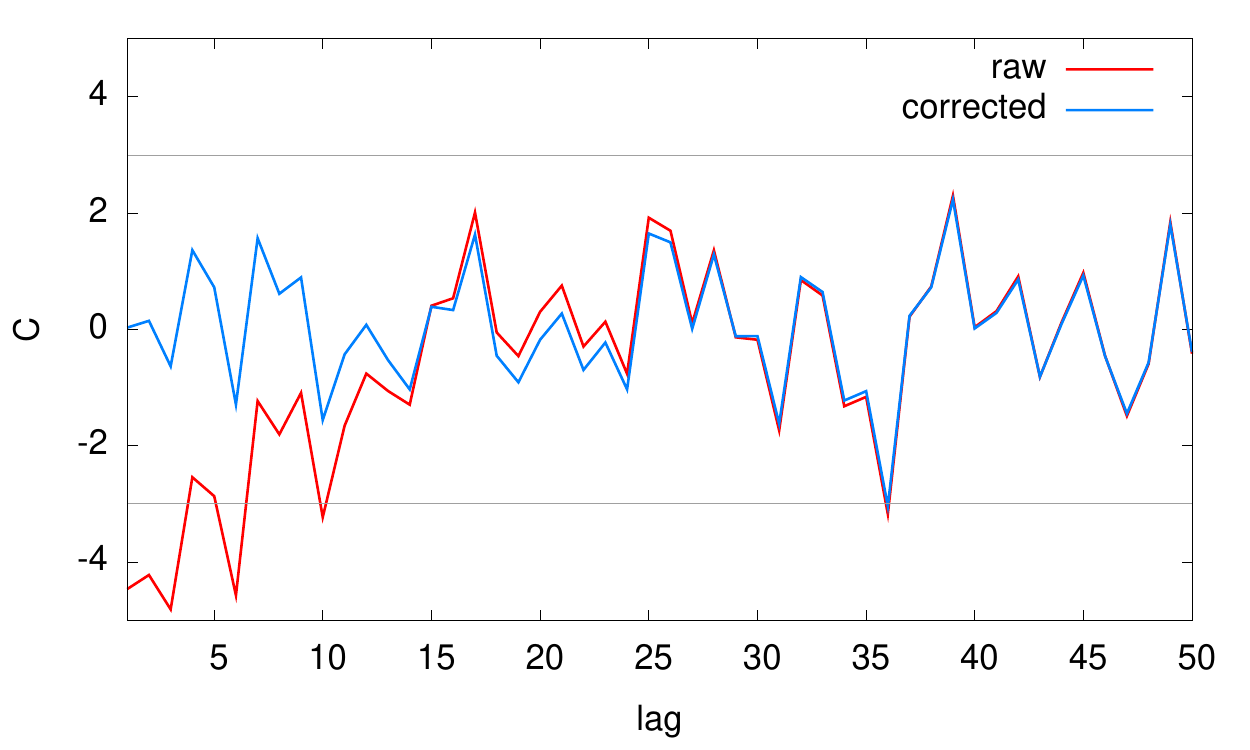}}
\caption{Panel (a) shows the mean and variance of the autocorrelation
  for stationary, Gaussian noise caused by a Tukey window. Panel (b)
  shows the raw and corrected autocorrelation for a stationary,
  Gaussian noise process.}
\label{comp:tukey}
\end{center}
\end{figure}

One note of caution in using the Fourier autocorrelation $C(k)$ as an
indicator of non-stationarity is that any window that is applied to
the time-domain data to lessen spectral leakage in the Fourier
transform necessarily makes the data non-stationary. Choosing a 
window function (Appendix~\ref{s:windowing}) that is unity across 
most of the samples, such as a Tukey window (\ref{e:tukeywindow}),
lessens the taper-induced
non-stationarity, but does not eliminate the effect. The solution is
to apply a correction to the autocorrelation that accounts for the
window. Figure~\ref{comp:tukey} shows the impact that a Tukey window
has on the mean and variance of the Fourier autocorrelation $C(k)$.
In this simulation $N= 32768$ samples were used with a Tukey window
that is constant across the central 90\% of the samples. By
subtracting the mean and scaling by the square-root of the variance
caused by the Tukey window, the non-stationarity caused by the filter
can be corrected for.
%

\subsubsection{Glitches}
\label{s:glitches}

\begin{figure}[h!tbp]
\begin{center}
\subfigure[]{\includegraphics[width=0.9\textwidth]{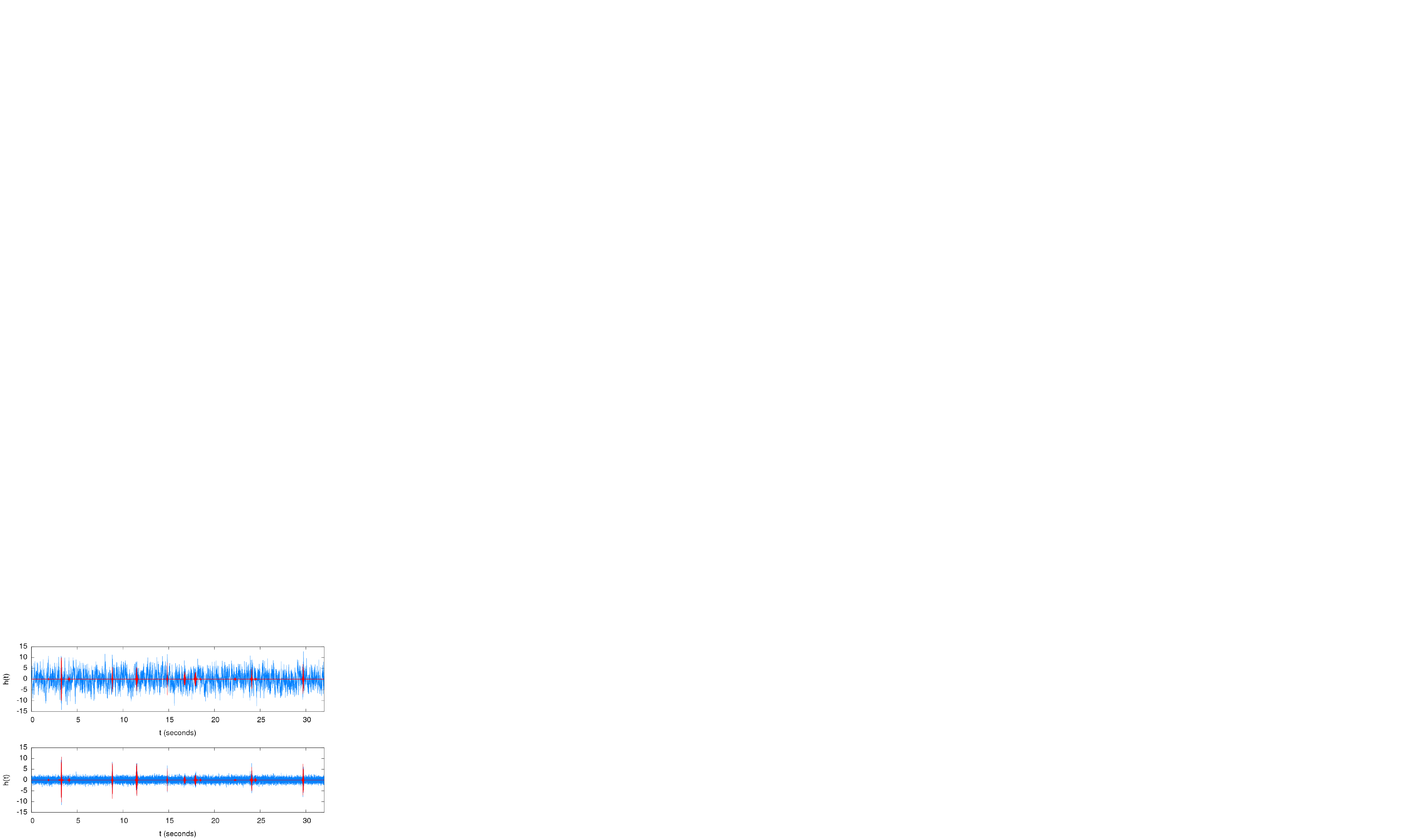}}
\subfigure[]{\includegraphics[width=0.9\textwidth]{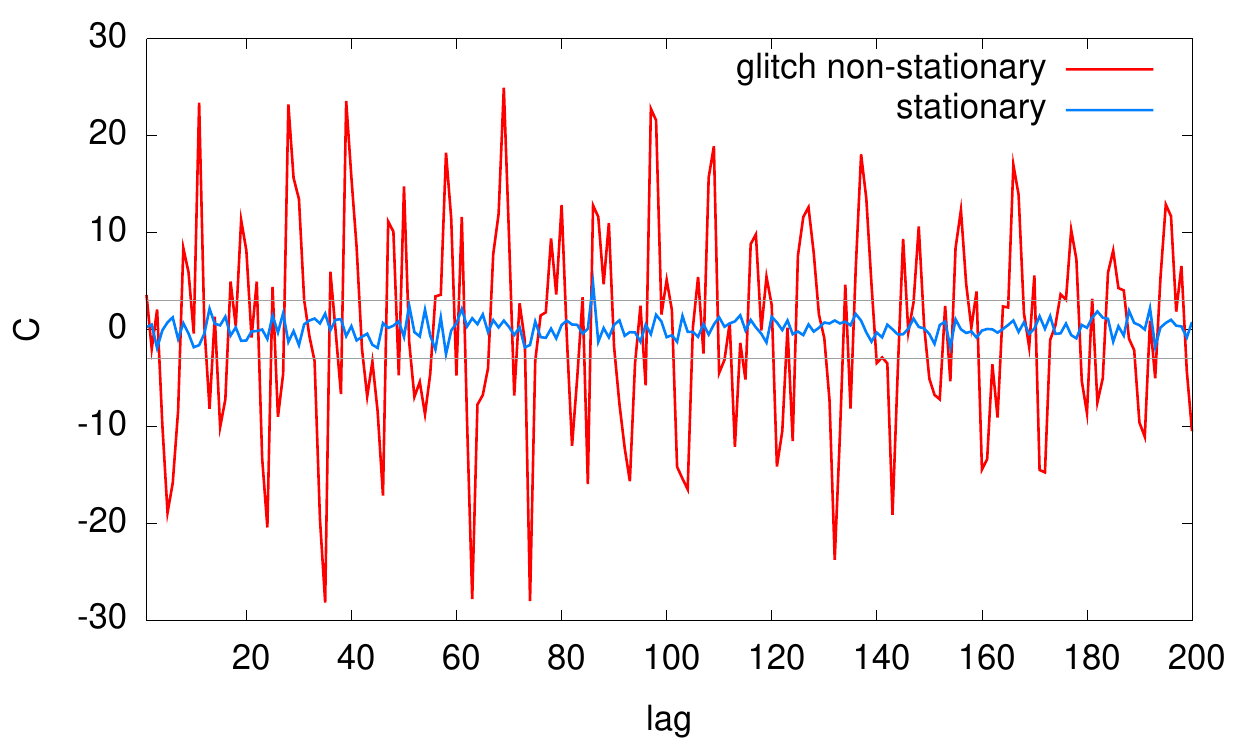}}
\caption{Panel (a)
shows simulated stationary AR(1) data with non-stationary noise 
transients, or glitches, highlighted in red. 
The upper panel is the raw data, while the lower panel has been whitened by the
estimated amplitude spectral density. 
Panel (b) shows the autocorrelation for the stationary AR(1) data 
without glitches in blue, and with glitches in red.}
\label{comp:glitch}
\end{center}
\end{figure}

To model the second form of non-stationarity caused by short-duration
noise transients, we add Gaussian-enveloped noise bursts to stationary
AR(1) data. The bursts are simulated by generating white noise in
the time domain, that is then multiplied by a Gaussian window centered
at time $t_0$ with width $\sigma_t$. The data is then Fourier
transformed, and the Fourier coefficients are multiplied by a Gaussian
window centered at $f_0$ with width $\sigma_f$.  In the simulation,
the central times were drawn from a Poisson process with a rate of
$0.5\, {\rm Hz}$, and the central frequencies were drawn from a uniform
distribution $U[0,f_{\rm N}]$.  
The duration and bandwidth were also drawn from uniform
distributions: $\sigma_t \sim U[0.01\,{\rm s},0.05\,{\rm s}]$,
$\sigma_f \sim U[2\,{\rm Hz}, 50\,{\rm Hz}]$. The signal-to-noise
ratio of the bursts was drawn from the distribution
\be
p({\rm SNR}) = \frac{ {\rm SNR}}{ 2\, {\rm SNR}_*^2 \left(1+ \frac{{\rm SNR}}{2\, {\rm SNR}_*}\right)^3} \, .
\ee
This form for the ${\rm SNR}$ distribution is used by the BayesWave
algorithm~\cite{bayeswave} as a prior on the amplitude of
glitches. The truncated power-law form for $p({\rm SNR})$ is motivated
by the distribution of glitches seen in real
data. Figure~\ref{comp:glitch} shows a 32-second segment of simulated
data, and the dramatic effect that the glitches have on the
autocorrelation of the Fourier transform. Unlike the 
locally-stationary noise process, 
which only introduced correlations for small
lags, the glitches produce a much larger deviation from stationarity
that extends to large lags.

\subsection{Non-Gaussian noise}
\label{s:nongaussiannoise}

Gaussian noise processes are ubiquitous in nature, and provide a
remarkably good model for the data seen in gravitational-wave
detectors. Properly whitened gravitational-wave data typically have 
a Gaussian core that accounts for the bulk of the samples, along with a
small number of outliers in the tails of the distribution. Even these
small departures can severely impact analyses that assume perfectly
Gaussian distributions.

Gauss developed the {\em least-squares} (maximum-likelihood) data analysis
technique in an effort to determine the orbit of the newly discovered
dwarf planet Ceres. Gauss showed that if measurement errors are: 
(i) more likely small than large, 
(ii) symmetric, and 
(iii) have zero mean, 
then they follow a normal distribution (first described by de Moivre in
1733). Gauss' proof relied on the law of large numbers: he assumed
that under repeated measurements the most-likely value of a quantity
is given by the mean of the measured values. The assumptions used in
Gauss' derivation were placed on a firmer footing by Laplace, who
derived the central limit theorem, which states that the arithmetic
mean of a sufficiently large number of independent random deviates
will be approximately normally distributed, regardless of the
underlying distributions the deviates are drawn from, so long as the
distributions have finite first and second moments. The central limit
theorem is often invoked to explain the ubiquity of Gaussian
measurement errors. While the classic central limit theorem applies to
noise contributions that are fundamentally stochastic (such as those
with a quantum origin), a variant of the central limit theorem also
applies to the sum of a large number of {\em deterministic} effects, 
so long as the deterministic processes obey certain
conditions~\cite{ImkellervonStorch2005}.

\begin{figure}[h!tbp]
\begin{center}
\subfigure[]{\includegraphics[width=.49\textwidth]{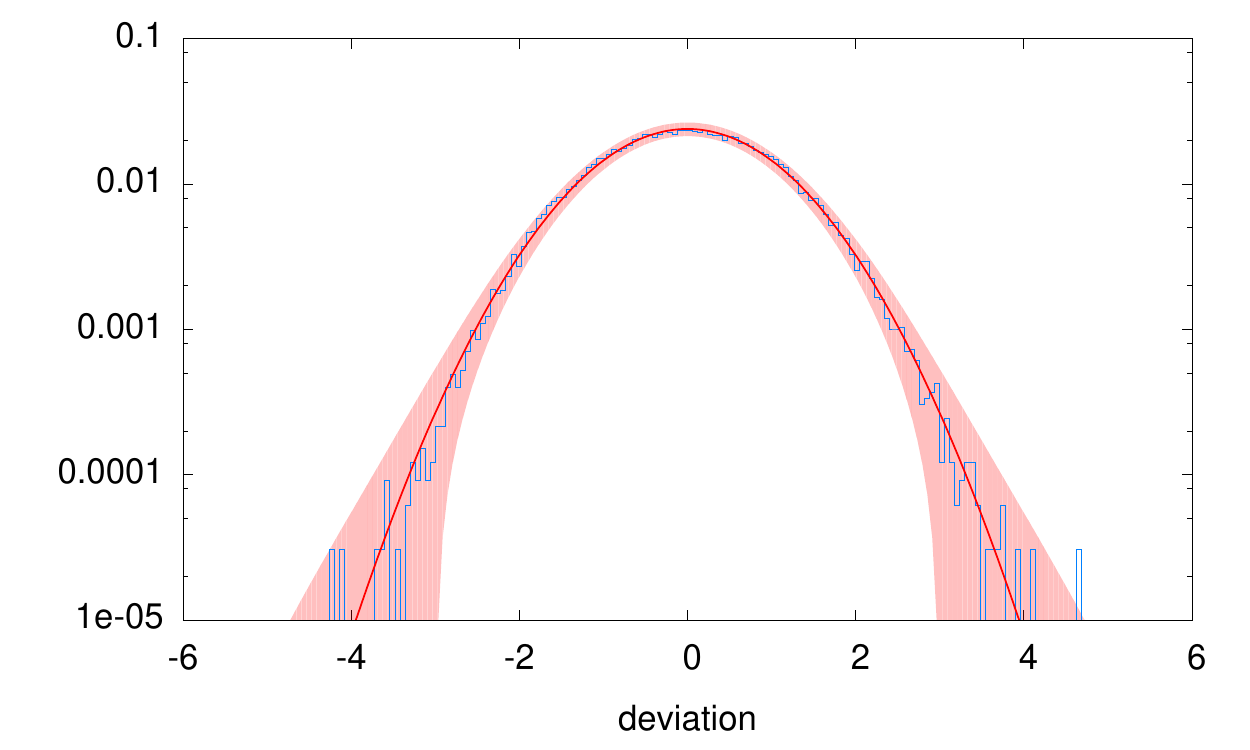}}
\subfigure[]{\includegraphics[width=.49\textwidth]{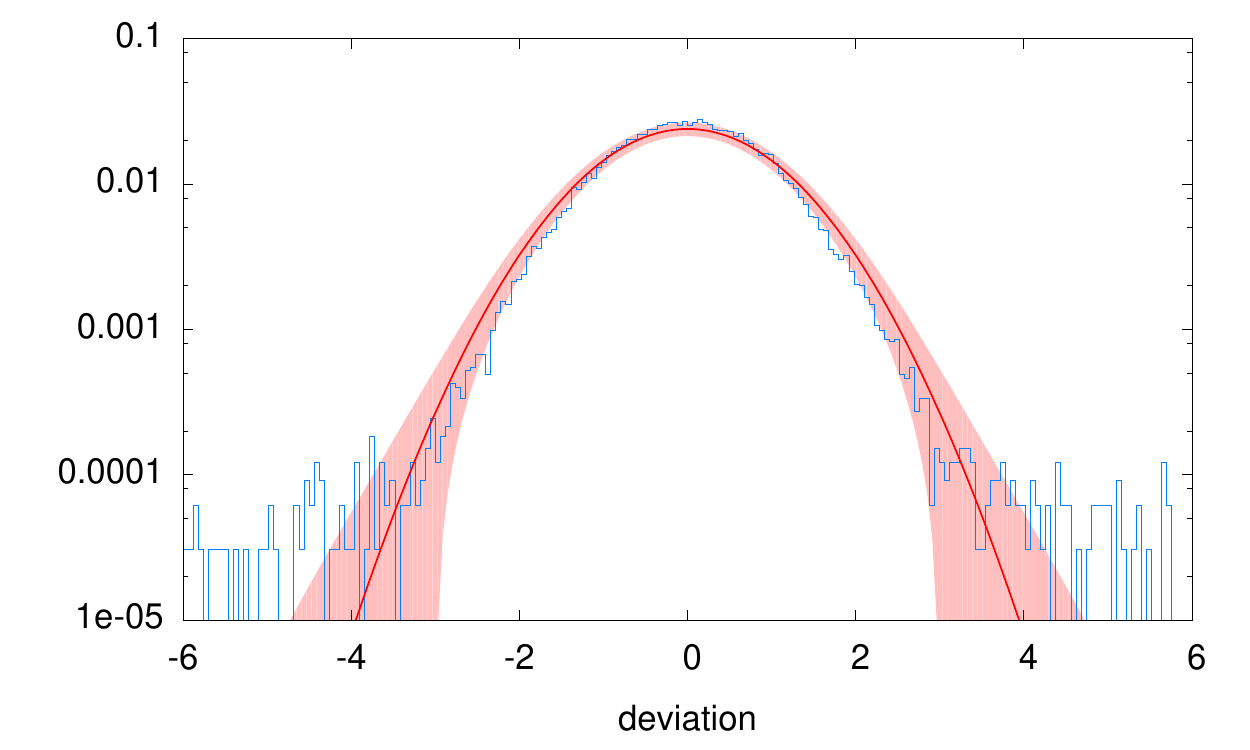}}
\caption{Histograms of the whitened data samples for the simulated
  data shown in Figure~\ref{comp:glitch}. A reference $N(0,1)$
  Gaussian distribution is shown as a red line. The light red band
  denotes the 3-sigma confidence interval for the finite number of
  samples used to produce the histograms. Panel (a) uses the whitened
  Fourier coefficients, while panel (b) uses the whitened time-domain
  samples. While the non-Gaussianity is most apparent in the time
  domain, both distributions fail the Anderson-Darling test for
  Gaussianity.}
\label{comp:gauss}
\end{center}
\end{figure}

Since gravitational-wave data typically have highly-colored spectra,
one cannot simply compare the distribution of samples in time or
frequency to a Gaussian distribution. The data first have to be 
whitened. This can be done by dividing the Fourier coefficients by 
the square-root of an estimate of the power spectra, 
and inverse Fourier transforming the
result to arrive at a whitened time series. Figure~\ref{comp:gauss}
shows histograms of the whitened Fourier-domain and time-domain
samples for the simulated data shown in Figure~\ref{comp:glitch}. 
By eye, the frequency-domain samples appear fairly Gaussian, while the
time-domain samples show clear departures from Gaussianity. Applying
the Anderson-Darling test~\cite{doi:10.1080/01621459.1954.10501232} 
to both sets of samples indicates that the
Gaussian hypothesis is rejected in both cases, with a $p$-value of
$p=2.6\times 10^{-5}$ for the Fourier-domain samples and $p<10^{-20}$ 
for the time-domain samples. Applying the same analysis to
the locally-stationary AR(1) model generated using 32 seconds of
data (i.e., setting $T=32$ s in the model for $q(t)$ and
$\epsilon(t)$), we find that the whitened Fourier coefficients
generally pass the Anderson-Darling test, while the whitened time-domain 
samples do not. Overall, glitches cause much larger departures
from Gaussianity than adiabatic variation in the noise levels.

To-date, there have been no detailed studies of the effects of
non-stationary and non-Gaussian noise on stochastic background
analyses beyond the theoretical investigations 
in~\cite{Allen-et-al:2002, Allen:2002jw, Himemoto:2006hw}.
However, a variety of checks have been applied to the
LIGO-Virgo analyses using time-shifted data and hardware and software
signal injections, and the results were found to be consistent with
the performance expected for stationary, Gaussian noise~\cite{S3HLiso,
  S4HLiso}. In particular, the distribution of the residuals of the
cross-correlation detection statistic, formed by subtracting the mean
and scaling by the square root of the variance, have been shown to be
Gaussian distributed~\cite{S4HLiso}.

\subsection{Gaps and irregular sampling}
\label{s:irregularsampling}

Data gaps and irregular sampling do not significantly impact the
analyses of interferometer data, but pose a major challenge to pulsar
timing analyses.

\subsubsection{Interferometer data}

Interferometer data are regularly sampled, and gaps in the data pose
no great challenge since the non-stationarity already demands that the
analysis be performed on short segments of coincident data. The main
difficulty working with short segments of data is accounting for the
filters that need to be applied to suppress spectral
leakage~\cite{S3HLiso, Lazz:2004}.

\subsubsection{Pulsar timing data}

The collection of pulsar timing data is constrained by telescope,
funding, and personnel availability. A large number of pulsars are now
observed fairly regularly, with observations occurring every 2--3
weeks.  Older data sets are less regularly sampled, and often have
gaps of months or even years~\cite{2015ApJ...813...65T}. Moreover, the
sensitivity of the instruments varies significantly over time, making
the data highly non-stationary, thus obviating the benefit of
performing the analyses in the frequency domain. For these reasons,
modern pulsar timing analyses are conducted directly in the time 
domain~\cite{vanHaasteren:2008yh}.

Noise modeling for pulsar timing has become increasingly more
sophisticated~\cite{2014MNRAS.437.3004L, 2015ApJ...813...65T}, but in
broad strokes, the two main terms in the noise model are: (i) 
measurement errors $\sigma_i$ in each time-of-arrival measurement, 
which are assumed to be uncorrelated between time samples $i$ and $j$, and 
(ii) a stationary red noise component $S_{ij}$ that depends on the lag
$\vert i - j\vert$~\cite{vanHaasteren:2008yh}. These contribute to the
time-domain noise correlation matrix $C_n$, which appears in the Gaussian
likelihood~(\ref{gauss_likelihood}):
\be
(C_{n})_{ij} = \sigma_i^2 \delta_{ij} + S_{ij} \, .
\ee
The data gaps and irregular sampling imply that the time lags $\vert i
- j\vert$ take on a wide range of values, and do not come in multiples
of a fixed sample rate $\Delta t$. Inverting the large noise matrix
$C_n$ to compute the likelihood can be very expensive unless clever
tricks are 
used~\cite{vanHaasteren-Vallisneri:2014,vanHaasteren-Vallisneri:2015}.

\subsection{Advanced noise modeling}
\label{s:unifiednoise}

The traditional approach to noise modeling has been to assume a simple
model, such as the noise being stationary and Gaussian, and then measure
the consequences this has on the analyses using Monte Carlo studies of
time-shifted data and simulated signals. An alternative approach is to
develop more flexible noise models that can account for various types
of non-stationarity and non-Gaussianity.

\begin{figure}[h!tbp]
\begin{center}
\subfigure[]{\includegraphics[width=.49\textwidth]{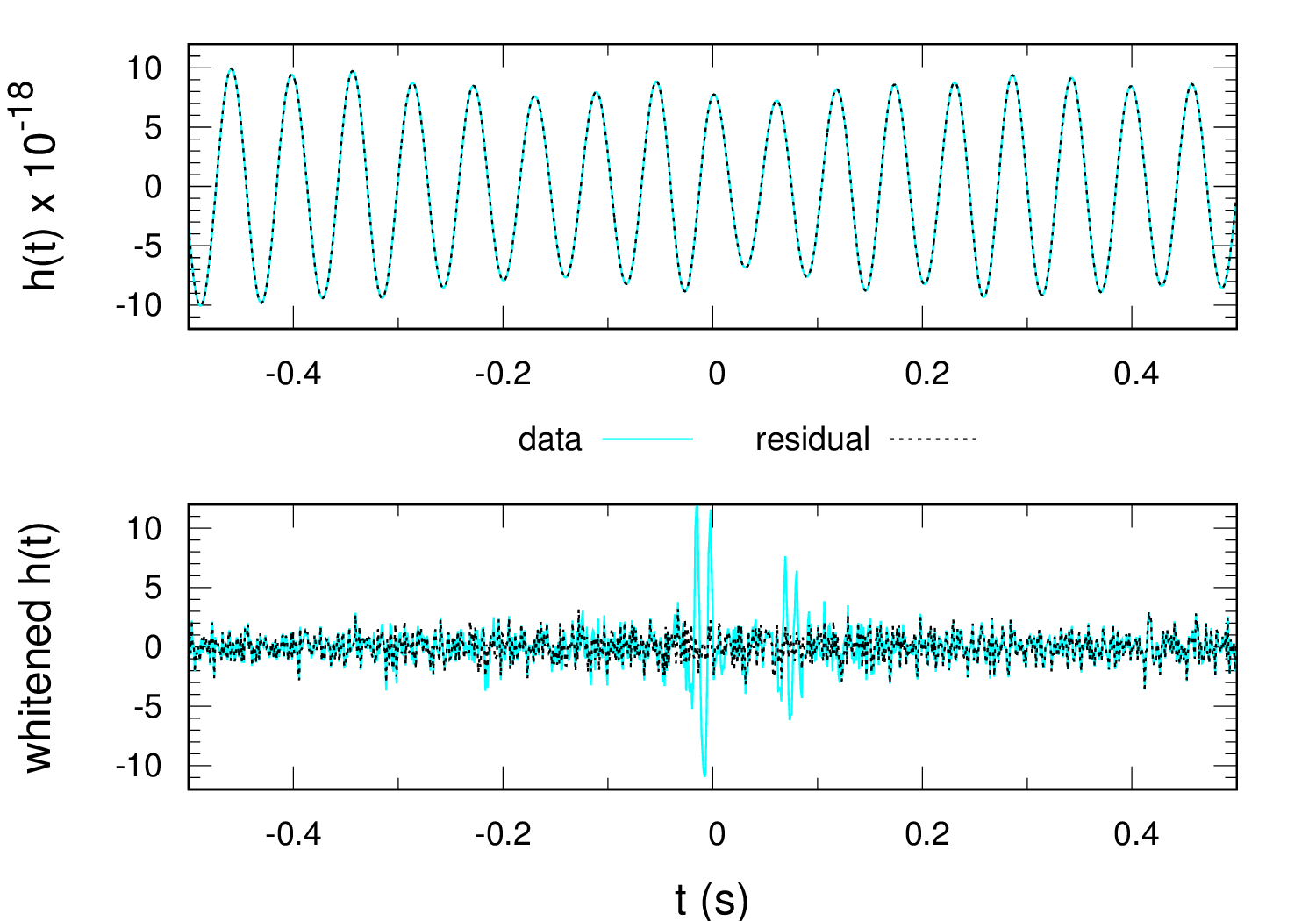}}
\subfigure[]{\includegraphics[width=.49\textwidth]{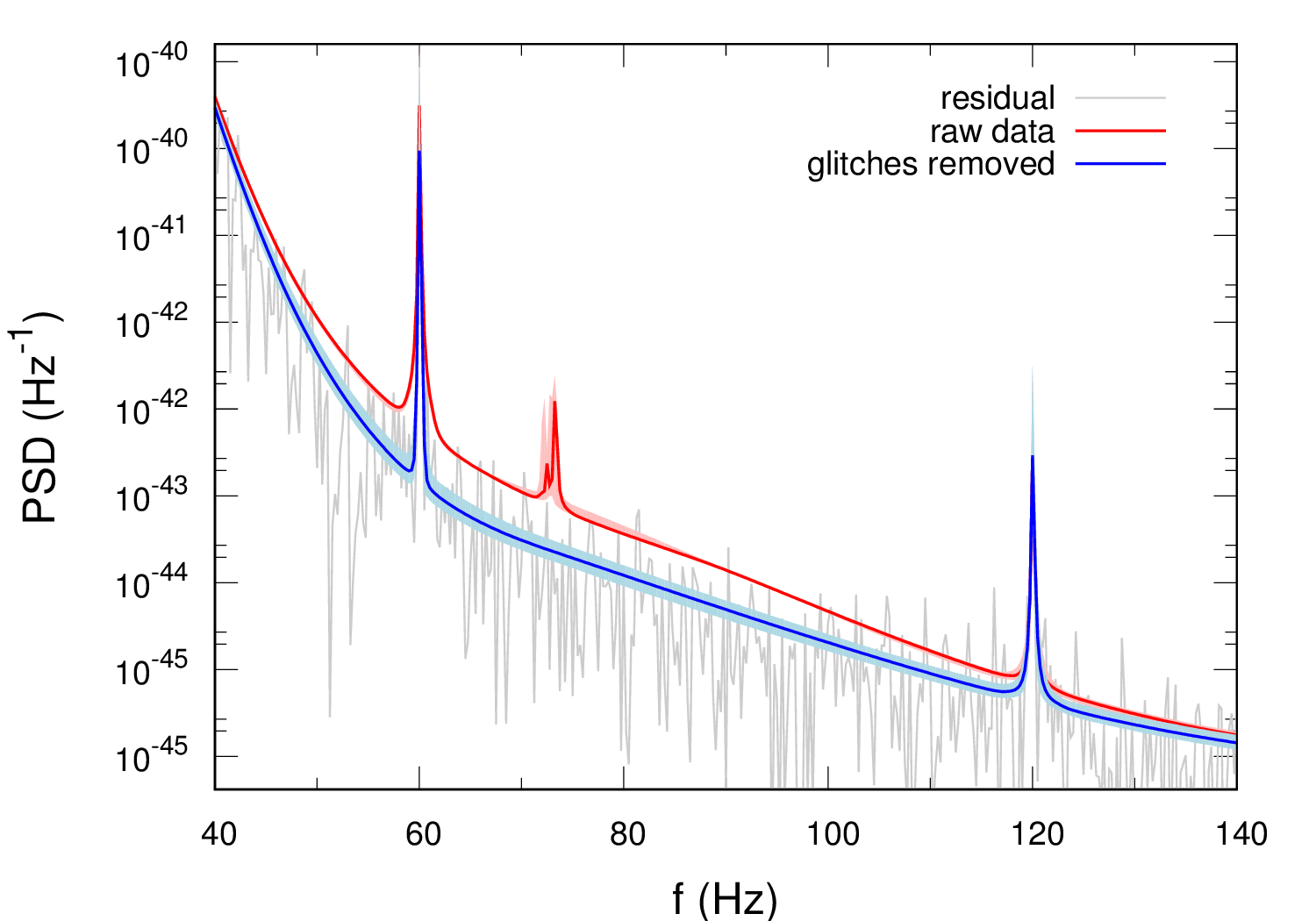}}
\caption{Panel (a) shows a 1-second sample of LIGO S6 data. The upper
  plot shows the raw data and the lower plot shows the data whitened by
  the median {\em BayesLine} spectra with glitch subtraction by the {\em BayesWave} algorithm.
  The solid aqua line is the data before glitch removal, and the dotted black line is after glitch removal. Panel (b) shows the median and
  90\% credible bands for the spectral model with (blue) and without (red) glitch subtraction. The grey line shows the power spectra of the data
  after glitch removal. Images provided by Tyson Littenberg.}
\label{comp:bw1}
\end{center}
\end{figure}

\begin{figure}[h!tbp]
\begin{center}
\subfigure[]{\includegraphics[width=.49\textwidth]{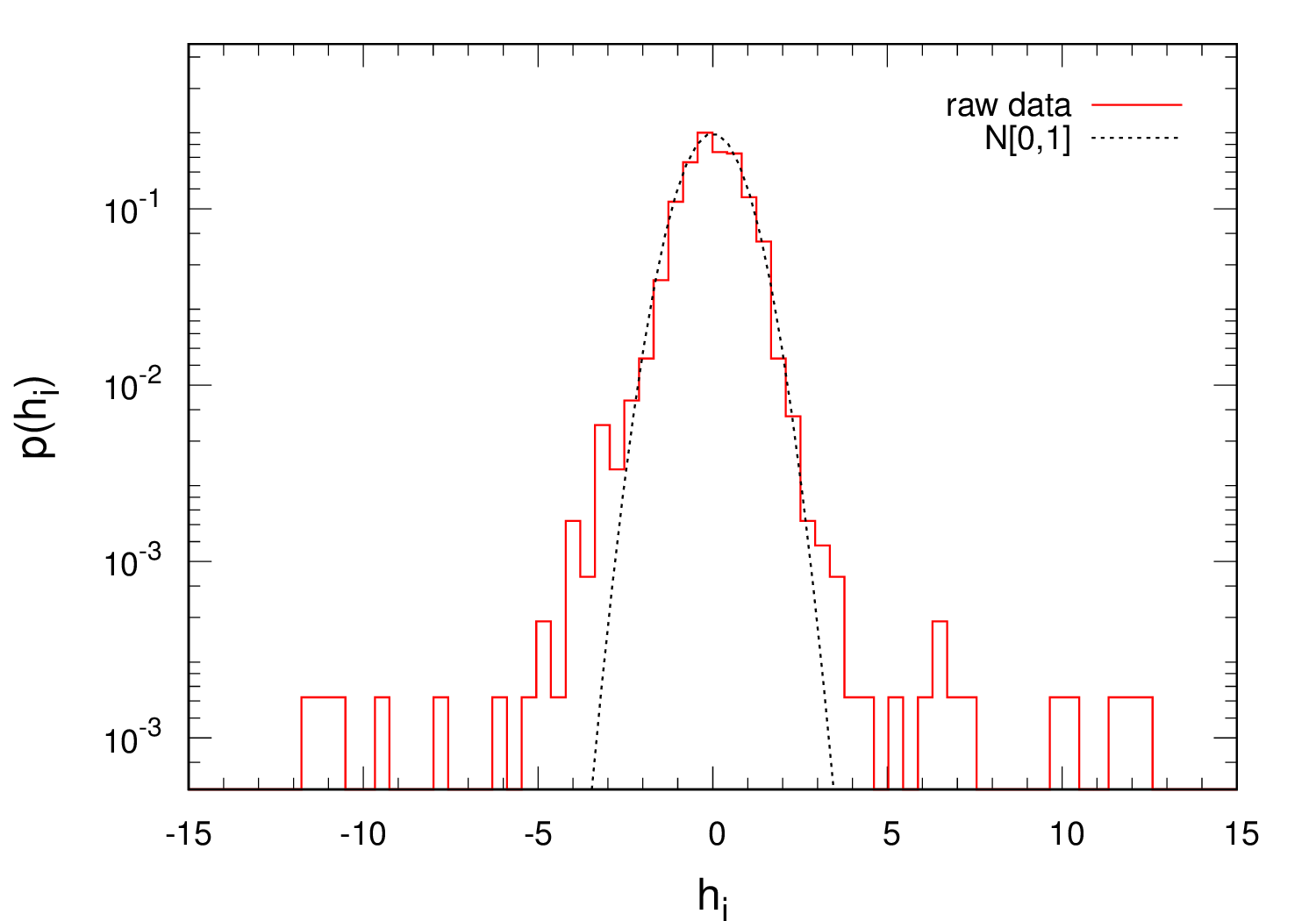}}
\subfigure[]{\includegraphics[width=.49\textwidth]{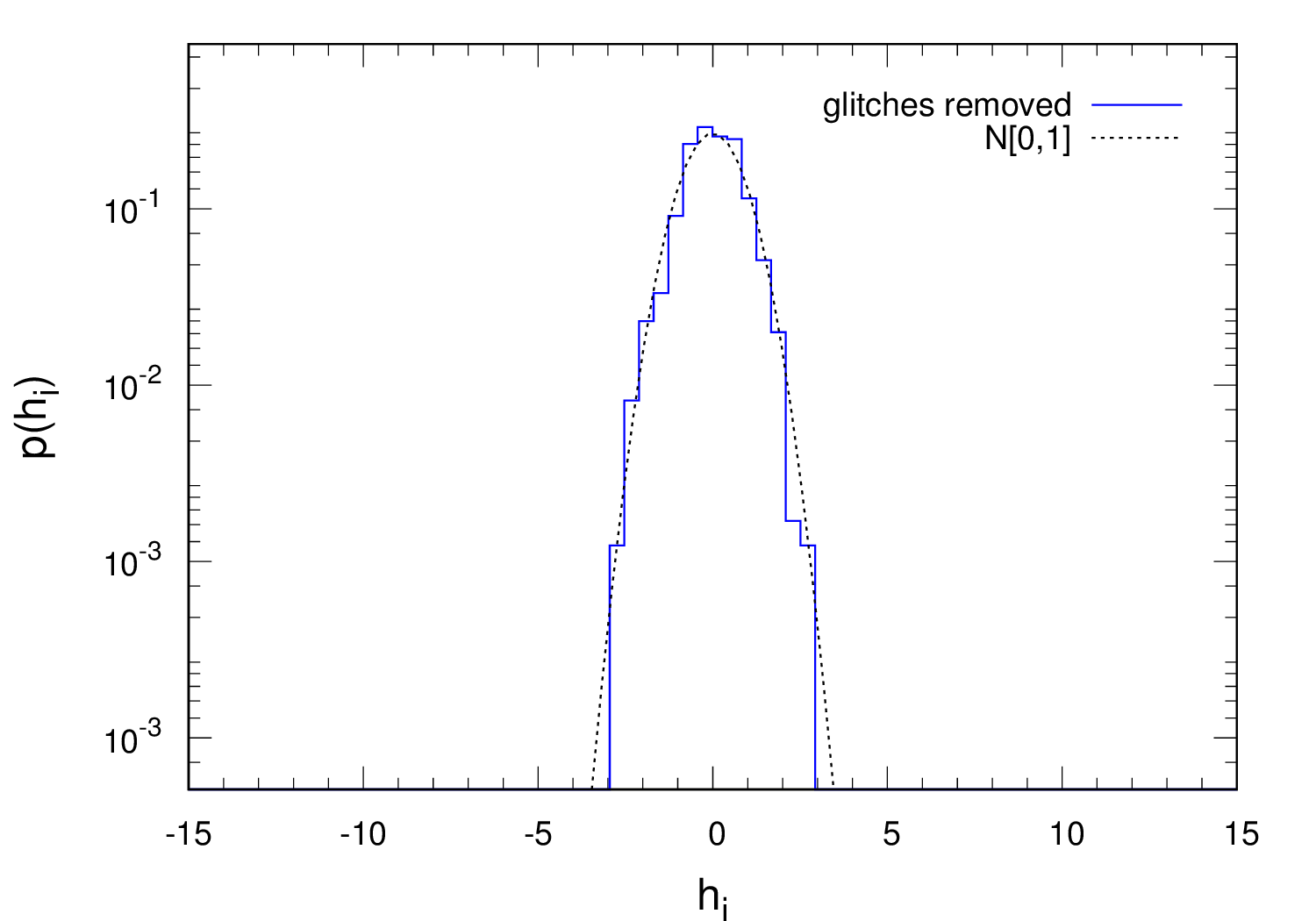}}
\caption{Histograms of the whitened time-domain data
  shown in Figure~\ref{comp:bw1}. Panel (a) is without glitch
  subtraction, while panel (b) is with glitch subtraction. Images provided by Tyson Littenberg.}
\label{comp:bw2}
\end{center}
\end{figure}

One such approach is the {\em BayesWave/BayesLine} algorithm, which uses a
two-part noise model composed of a stationary, Gaussian component
$S(f)$, and short duration glitches, $g(t)$, modeled as Gaussian-enveloped 
sinusoids~\cite{bayeswave, bayesline}. The spectral model
$S(f)$ is based on a cubic-spline with a variable number of control
points to model the smoothly-varying part of the spectrum, and a
collection of truncated Lorentzians to model sharp line features. The
optimal number and placement of the control points and Lorentizians is
determined from the data using a trans-dimensional Markov Chain Monte
Carlo technique. The same technique is used to determine the number of
sine-Gaussian glitches and their parameters (central time and
frequency, duration, etc.). This approach has been applied to both LIGO
data~\cite{bayeswave, bayesline} and pulsar timing
data~\cite{Ellis:2016mtg}.  Figure~\ref{comp:bw1} demonstrates the
application of the {\em BayesWave} and {\em BayesLine} algorithms to data from the LIGO
Hanford detector during the S6 science run of the initial LIGO
detectors. Removing the glitches has a significant impact on the
inferred power spectra. Figure~\ref{comp:bw2} displays histograms of
the whitened Fourier coefficients for the data shown in
Figure~\ref{comp:bw1} with and without glitch removal.

Additional models for non-stationary and non-Gaussian noise have been
considered by several authors. The detection of deterministic and
stochastic signals was considered 
in~\cite{Allen-et-al:2002, Allen:2002jw, Himemoto:2006hw} for a
variety of non-Gaussian noise models, including exponential and
two-component Gaussian models.  
The two-component Gaussian model combined with a
non-stationary glitch model was studied in~\cite{Littenberg:2010gf}. 
Student's $t$-distribution was considered in~\cite{Rover:2011qd}.  
A non-stationary and non-Gaussian noise
model was derived in~\cite{Principe:2008zz} based on a Poisson
distribution of sine-Gaussian glitches.

\subsection{Correlated noise }
\label{s:correlatednoise}

The standard cross-correlation statistic for detecting stochastic
backgrounds relies on the noise in each detector being
uncorrelated. If we return to the simple model for 
colocated and coaligned detectors, with white Gaussian noise 
and a white Gaussian signal (Section~\ref{s:white}), 
but now introduce a correlated noise component
$S_{n_{12}}$, then the correlation matrix for the signal-plus-noise 
model becomes 
\be C = \left[
\begin{array}{cc}
(S_{n_1} +S_h)\,\unit_{N\times N} & (S_h + S_{n_{12}})\,\unit_{N\times N} 
\\
(S_h+S_{n_{12}}) \,\unit_{N\times N} & (S_{n_2} +S_h)\,\unit_{N\times N}
\\
\end{array}
\right]\, ,
\ee
yielding the maximum likelihood solution
\be
\begin{aligned}
&\hat S_h  \equiv \frac{1}{N}\sum_{i=1}^N d_{1i} d_{2i} -  S_{n_{12}} \,,
\\
&\hat S_{n_1} \equiv \frac{1}{N}\sum_{i=1}^N d_{1i}^2 - \hat S_h\,,
\\
&\hat S_{n_2} \equiv \frac{1}{N}\sum_{i=1}^N d_{2i}^2 - \hat S_h\,.
\end{aligned}
\ee
We see that there is a degeneracy between the estimate for the signal
$\hat S_h$ and the correlated noise $S_{n_{12}}$, with no way to
separate the two components.  Correlated noise with the same spectrum
as the signal presents a {\em fundamental} limit to the detection of
stochastic signals.

If the spectral shape of either, or preferably both, the signal and
the correlated noise are known, then it is possible to separate the
contributions using techniques similar to those that are used to
separate the primordial cosmic-microwave-background signal from
foreground contamination~\cite{Bennett:2003bz}. When the cause of the
correlated noise is not fully understood, or when searching for
signals with arbitrary spectral shapes, spectrum-based component
separation will not be possible.

Several sources of correlated noise have been hypothesized, and in
some cases observed, for both interferometer and pulsar timing
analyses. Some of the correlations are due to the
electronics~\cite{S3HLiso}, such as correlations between harmonics of
the $60 \, {\rm Hz}$ AC power lines between the LIGO Hanford and LIGO
Livingston detectors, and correlations at multiples of 
$16\, {\rm Hz}$ from the data sampling referenced to clocks on the 
Global Positioning System satellites. These narrow-band correlations are easily
removed using notch filters. Correlations in the global time standard
can also impact pulsar timing observations, as can errors in the
ephemeris used in the timing model.

\subsubsection{Schumann resonances}
\label{s:schumann}

\begin{figure}[h!tbp]
\begin{center}
\subfigure[]{\includegraphics[width=.49\textwidth]{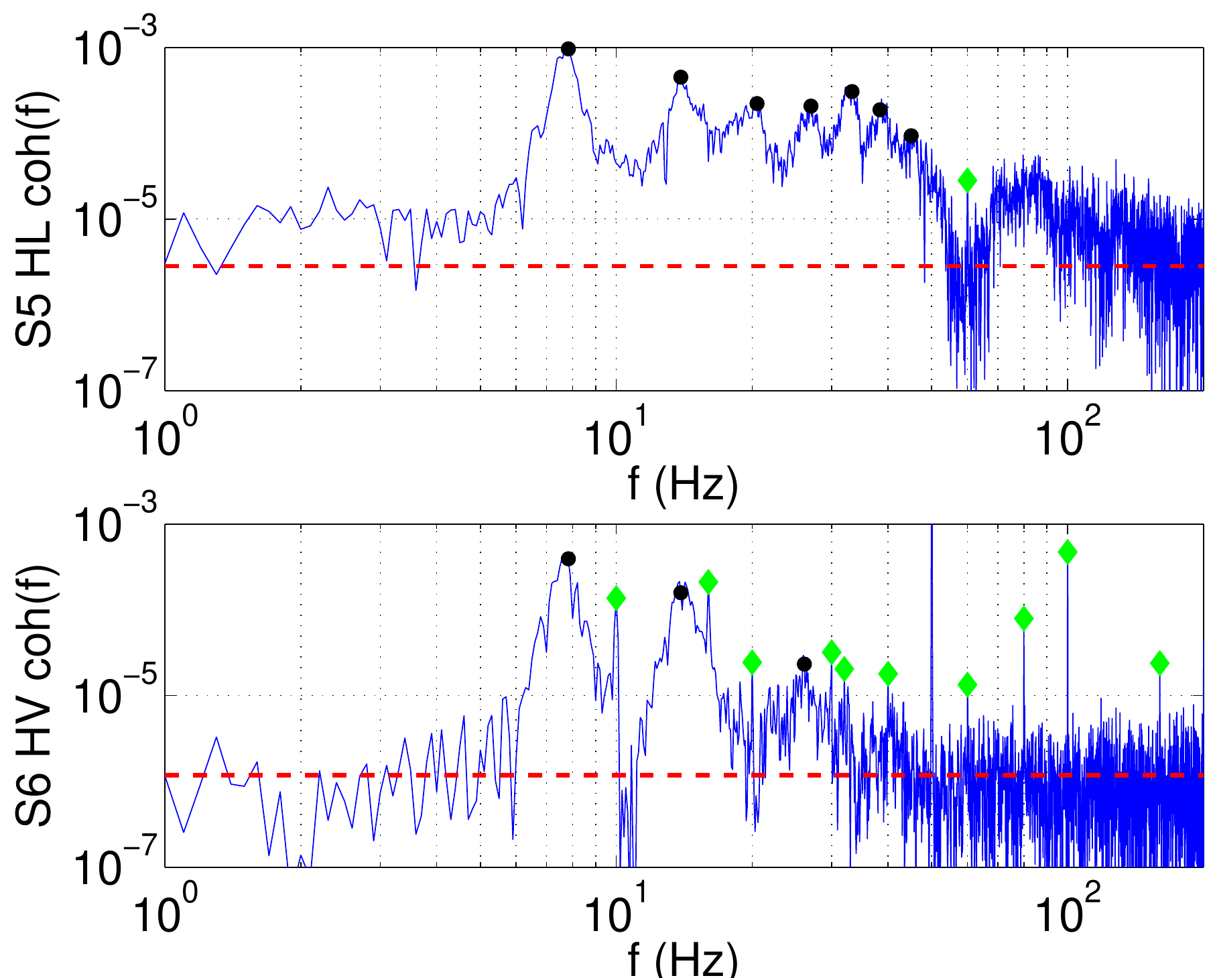}}
\subfigure[]{\includegraphics[width=.49\textwidth]{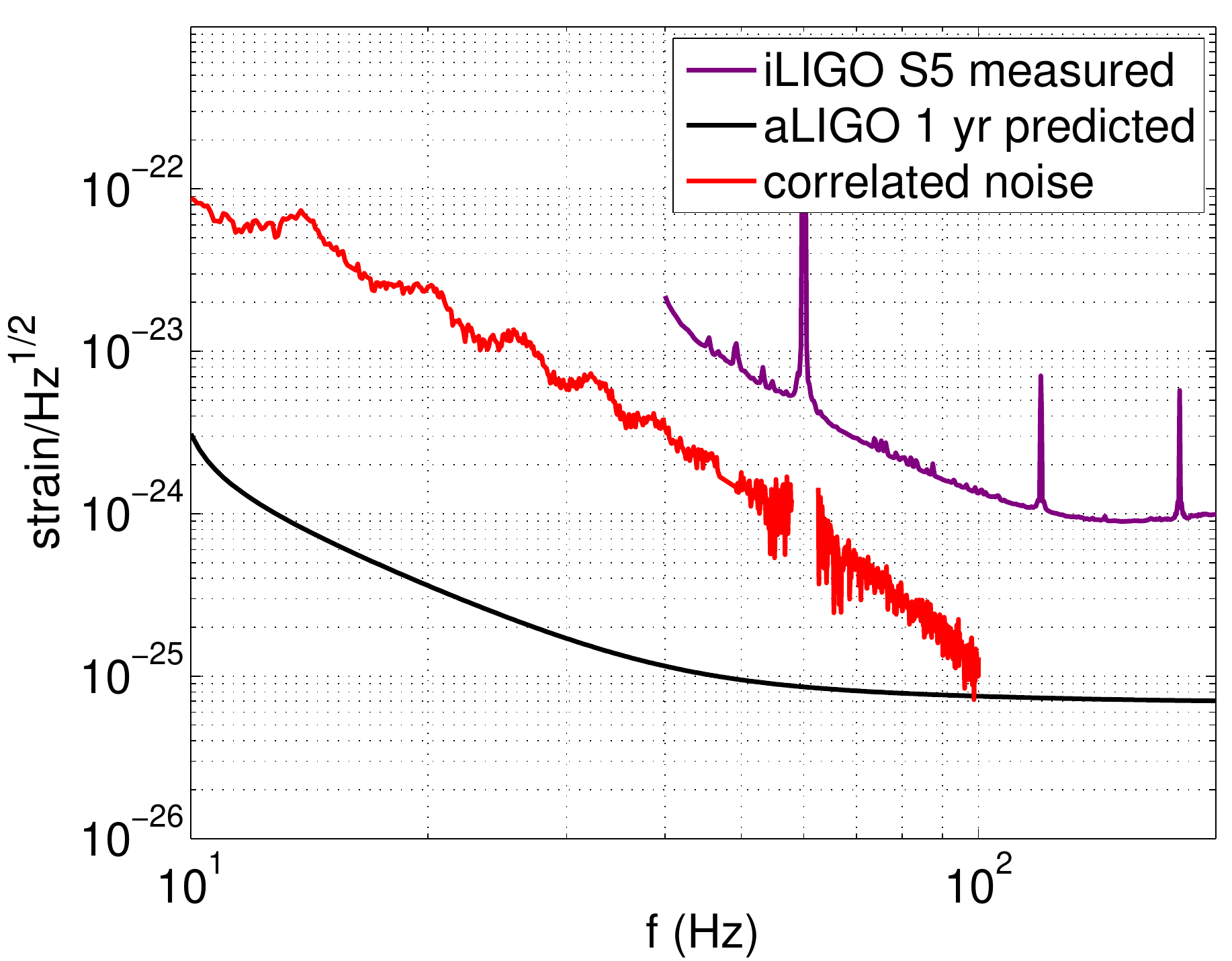}}
\caption{Panel (a) shows the cross-correlation of magnetometer
  readings between the LIGO-Hanford and LIGO-Livingston sites (HL),
  and also the LIGO-Hanford and Virgo sites (HV). The
  peaks indicated by black dots are due to Schumann resonances. The
  green dots mark peaks that are due to correlations caused by the
  electronics. Panel (b) shows the amplitude spectra of the initial
  and advanced LIGO detectors compared to the estimated level of the
  correlated noise due to Schumann resonances. The correlated noise
  level in advanced LIGO should be lower due to differences in the
  design (notably the lack of magnets attached to the mirrors). Images
  reproduced with permission from \cite{schumann1}, copyright by APS.}
\label{comp:sch}
\end{center}
\end{figure}

One possible broad-band source of correlated noise for ground-based 
interferometers that has received considerable
attention~\cite{schumann1, schumann2, schumann3} are Schumann
resonances in the Earth's magnetic field caused by lightning
strikes. These resonances can produce coherent oscillations over
thousands of kilometers, and have been observed to produced
correlations in magnetometer readings at the LIGO and Virgo
sites~\cite{schumann1}, as shown in panel (a) of Figure~\ref{comp:sch}. 
The spectrum of the correlations induced in the detector output depend on
both the spectrum of the time-varying magnetic field, and the
couplings to the instrument. The estimated spectrum of correlated
noise in the initial LIGO detectors from Schumann resonances is shown
in panel (b) of Figure~\ref{comp:sch}.  
The estimated spectrum lies below the
initial LIGO noise curve, but above the design noise curve for the
advanced instruments. The situation is not as dire as it looks, however, 
since the advanced LIGO detectors have a different design that should have
weaker coupling to magnetic fields. Nonetheless, Schumann resonances
may end up being a limiting factor for advanced LIGO stochastic
searches, and efforts are underway to model and subtract their
effects~\cite{schumann3}. Correlated noise is a much larger problem
for colocated detectors, such as the $2\ {\rm km}$ and $4\ {\rm km}$ 
initial LIGO detectors that shared the Hanford site. There it
was found that correlated noise prevented the data at frequencies
below $460\, {\rm Hz}$ from being used for stochastic background
searches~\cite{Aasi-et-al:H1H2}.

Perhaps the greatest challenge comes from correlated noise sources of
unknown origin. Such noise sources may be well below the auto-correlated
noise level in each detector, and thus very hard to detect outside of
the cross-correlation analysis. One way of separating these noise
sources from a stochastic signal is to build a large number of
interferometers at many locations around the world.  Each pair of
detectors will then have a unique overlap function for
gravitational-wave signals that will differ from the spatial
correlation pattern of the noise (unless we are incredibly unlucky!).
In principle, the difference in the frequency-dependent spatial
correlation patterns of the signal and the noise will allow the two
components to be separated.

\subsection{What can one do with a single detector (e.g., LISA)?}
\label{s:singledetector}

The discovery of the cosmic microwave background was described in a
paper with the unassuming title ``A Measurement of Excess Antenna
Temperature at 4080 Mc/s''~\cite{1965ApJ...142..419P}.  Penzias and
Wilson used a single microwave horn, and announced the result
after convincing themselves that no instrumental noise sources,
including pigeon droppings, could be responsible for the excess noise
seen in the data. In principle, the same approach could be used to
detect a stochastic gravitational-wave signal using a single
instrument. 

Single-detector detection techniques will be put to the test when the
first space-based gravitational-wave interferometer is launched, since
(unless the funding landscape changes dramatically) the instrument
will be a single array of 3 spacecraft.  Assuming that pairs of laser
links operate between each pair of spacecraft, it will be possible to
synthesize multiple interferometry signals from the phase
readouts~\cite{Estabrook:2000ef}. One particular combination of the
phase readouts, called the $T$ channel, corresponds to a Sagnac
interferometer, and is relatively insensitive to low-frequency
gravitational waves, forming an approximate null channel (see
Section~\ref{s:null} for a discussion of null channels). Other
combinations, such as the so-called $A$ and $E$
channels~\cite{Prince:2002hp}, are much more sensitive to
gravitational-wave signals. Using the Sagnac $T$ to measure the
instrument noise, the relative power levels in the $\{A,E,T\}$
channels can be used to separate a stochastic signal from 
instrument noise~\cite{Tinto:2001ii}.

LISA-type observatories operate as synthetic interferometers by
forming gravitational-wave observables in post-processing using
different combinations of the phasemeter readouts from each
inter-spacecraft laser link. The combinations synthesize effective
equal-path-length interferometers to cancel the otherwise overwhelming
laser frequency noise.  These combinations have to account for the
unequal and time-varying distances between the spacecraft.

\begin{figure}[h!tbp]
\begin{center}
\subfigure[]{\includegraphics[width=.44\textwidth]{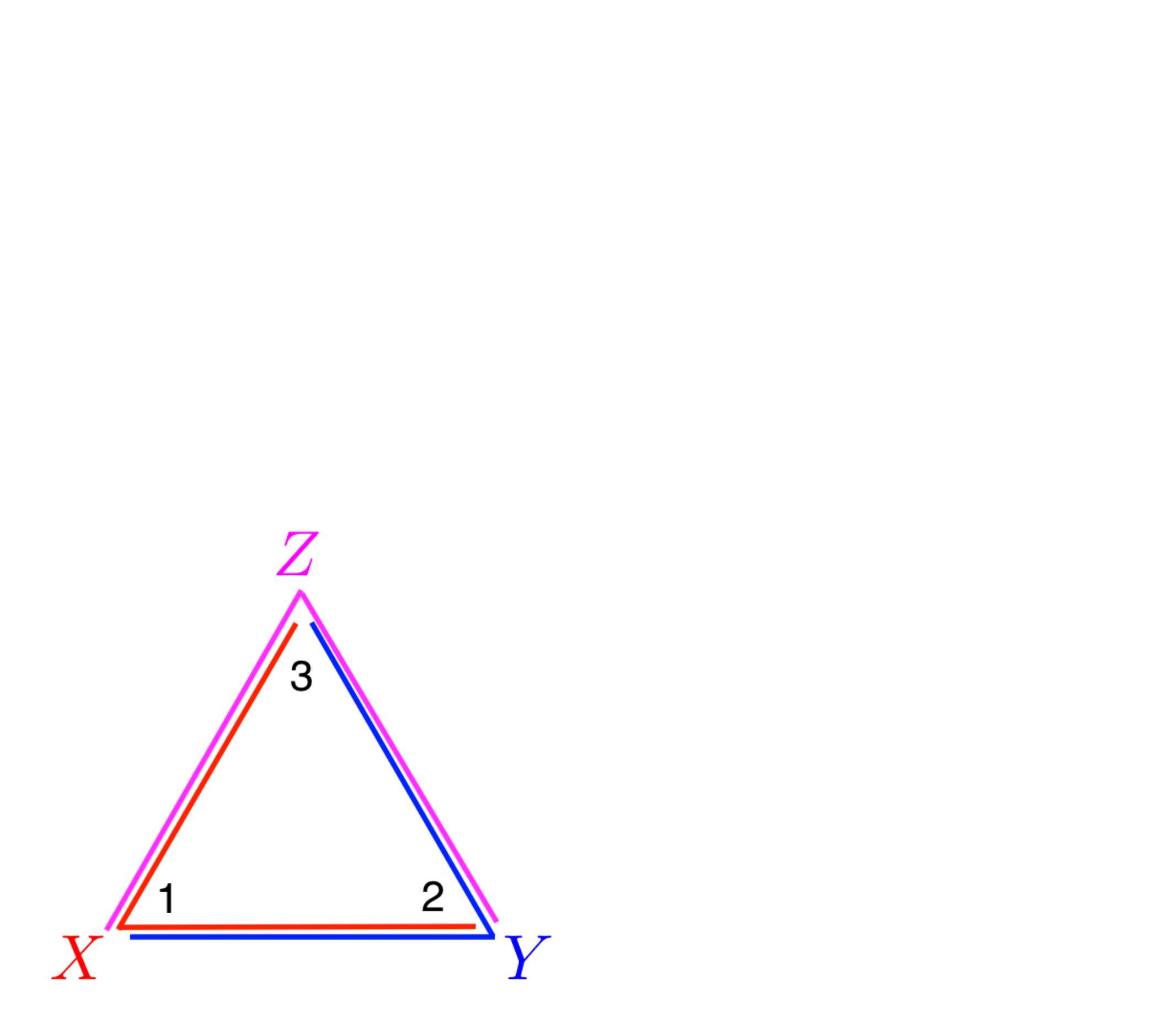}}
\subfigure[]{\includegraphics[width=.54\textwidth]{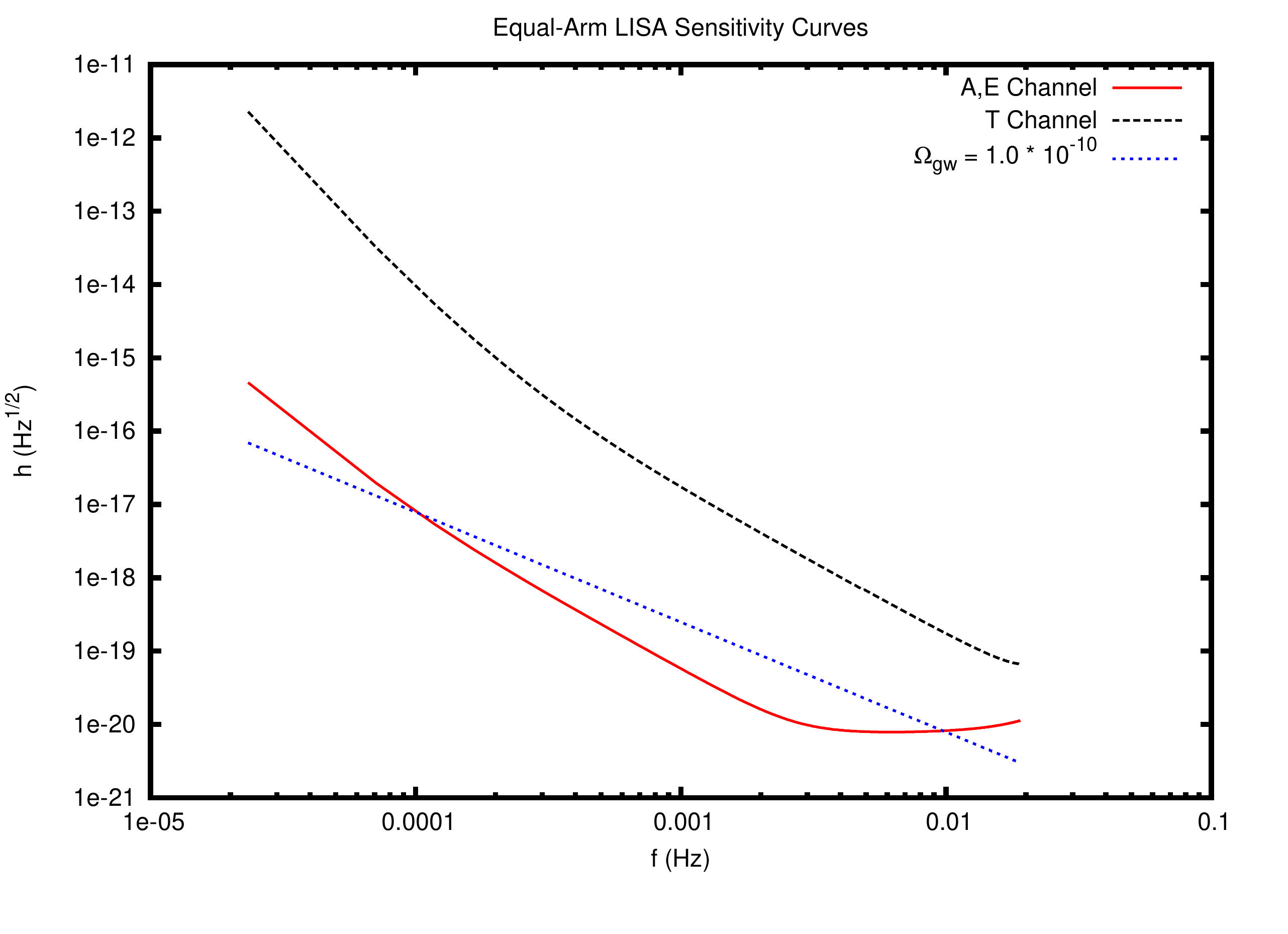}}
\caption{Panel (a) shows the geometry of a LISA-like space
  interferometer and the laser paths for the synthetic Michelson
  interferometers $X,Y,Z$. Panel (b) shows sensitivity curves for the
  $A,E,T$ interferometry variables compared to a scale-invariant
  background, $\Omega_{\rm gw}(f)=\Omega_0={\rm const}$,  
  with $\Omega_0 = 10^{-10}$. The Sagnac-like $T$
  channel is far less sensitive than the Michelson-like $A,E$
  channels, and can be used to measure the instrumental noise levels.
  (Panel (b) is adapted from~\cite{Adams-Cornish:2010}).}
\label{comp:lisa2}
\end{center}
\end{figure}

In the conceptually simpler equal-arm-length limit, the Michelson-type
signal extracted from vertex 1 (see panel (a) of Figure~\ref{comp:lisa2}) 
is given by
\be
X(t) = M_1(t) - M_1(t-2L)\,,
\ee
where
\be
M_1(t) = \Phi_{12}(t-L) +  \Phi_{21}(t) - \Phi_{13}(t-L) -  \Phi_{31}(t)\,,
\ee
and $\Phi_{ij}(t)$ is the readout from the phasemeter on spacecraft
$j$ that receives light from spacecraft $i$.  Permuting the spacecraft
labels $\{1,2,3\}$ yields equivalent expressions for the Michelson
observables $Y$ and $Z$, as shown in panel (a) of
Figure~\ref{comp:lisa2}. The phasemeter readouts $\Phi_{ij}(t)$ are
impacted by acceleration noise $S^a_{ij}$ and position noise
$S^p_{ij}$.  When the noise levels in each spacecraft are equal, there
exist noise-orthogonal combinations~\cite{Prince:2002hp, Adams-Cornish:2010}:
\be
\begin{aligned}
\label{AET}
A &\equiv \frac{1}{3} (2X-Y-Z)\,,
\\
E &\equiv \frac{1}{\sqrt{3}} (Z-Y)\,,
\\
T &\equiv \frac{1}{3} (X+Y+Z)\,.
\end{aligned}
\ee
Note that these variables are only noise-orthogonal in the symmetric
noise limit. For example, the position noise contribution to the
cross-spectra $\langle AE \rangle$ is given by
\be
\langle AE \rangle = 
-\frac{4}{3\sqrt{3}}\sin^2\left(\frac{f}{f_*}\right) 
\bigg(2\cos\left(\frac{f}{f_*}\right)+1\bigg)\bigg(S^p_{13}-S^p_{12}+S^p_{31}-S^p_{21}\bigg)\, ,
\ee
which vanishes when $\{S^p_{13},S^p_{12},S^p_{31},S^p_{21}\}$ are
equal, but not otherwise~\cite{Adams-Cornish:2010}. The synthetic
interferometers $A,E$ are rotated by 45 degrees with respect to each
other, and provide instantaneous measurements of the $+$ and $\times$
polarization states. The Sagnac-like $T$ channel is relatively
insensitive to gravitational waves for frequencies below the transfer
frequency $f_*\equiv c/(2\pi L)$. 
The $T$ channel can be used to infer the instrument
noise level, so that any excess in the $A,E$ channels can then be
confidently attributed to gravitational waves~\cite{Tinto:2001ii}.
For frequencies $f \ll f_*$ the $\{A,E,T\}$ channels have uncorrelated
responses to unpolarized, isotropic stochastic gravitational-wave 
signals.

There are some subtleties associated with using the $T$ channel as a
noise reference as the noise combinations in $T$ differ from those in
$A,E$.  For example, the acceleration noise appears in $T$
as~\cite{Adams-Cornish:2010}:
\be 
\label{tta}
\langle TT \rangle = 
\frac{16}{9}\sin^2\left(\frac{f}{f_*}\right)
\bigg(1-\cos\left(\frac{f}{f_*}\right)\bigg)^2 
\bigg(S^a_{12}+S^a_{13}+S^a_{31}+S^a_{32}+S^a_{23}+S^a_{21}\bigg)\,,
\ee
while the acceleration noise appears in $A$ and $E$ as
\begin{eqnarray}\label{aaa}
\langle AA \rangle 
&=& \frac{16}{9}\sin^2\left(\frac{f}{f_*}\right)
\Bigg\{\cos\left(\frac{f}{f_*}\right) \bigg[4\big(S^a_{12}+S^a_{13}+S^a_{31}+S^a_{21}\big)-2\big(S^a_{23}+S^a_{32}\big)\bigg] 
\nonumber \\       
& & +\cos\left(\frac{2f}{f_*}\right)
\bigg[\frac{1}{2}(S^a_{12}+S^a_{13}+S^a_{23}+S^a_{32})+2(S^a_{31}+S^a_{21})\bigg] 
\nonumber \\
& & +\frac{9}{2}(S^a_{12}+S^a_{13})+3(S^a_{31}+S^a_{21})+\frac{3}{2}(S^a_{23}+S^a_{32})\Bigg\}\,,
\end{eqnarray}
and
\begin{multline}
\label{eea}
\langle EE \rangle =   
\frac{16}{3}\sin^2\left(\frac{f}{f_*}\right)\bigg\{ S^a_{23}+S^a_{32}+S^a_{21}+S^a_{31} 
+2\cos\left(\frac{f}{f_*}\right)\bigg(S^a_{23}+S^a_{32}\bigg) 
\\			 
+\cos^2\left(\frac{f}{f_*}\right)\bigg(S^a_{23}+S^a_{32}+S^a_{12}+S^a_{13}\bigg)\bigg\}\,.
\end{multline}
In the ideal case where the noise levels are the same in each link,
$T$ provides a measurement of the average noise, which can then be
used as an estimator for the noise in $A,E$.  An analysis that assumes
common noise levels will overstate the sensitivity to a signal.  A
more conservative approach is allow for unequal noise levels and to
infer the individual contributions from the data. For example, if one
link is particularly noisy, it will dominate the noise in $T$, and
enter unequally in $A$ and $E$, making it possible to identify the bad
link and account for it in the analysis.

Bayesian inference is ideally suited to the task of jointly inferring
the signal and noise levels using models that fold in prior knowledge
of the signals and instrument
components~\cite{Adams-Cornish:2010}. The separation is aided by the
difference in the transfer functions for the signal and the
noise. Analytic expressions for the signal transfer or auto-correlated 
response functions (which are proportional to $\Gamma_{II}$ from 
Section~\ref{s:autocorrelatedresponse}) can be derived in the 
low-frequency limit $f \ll f_*$:
\be
{\cal R}_{TT} = 4\sin^2\left(\frac{f}{f_*}\right)\bigg[\frac{1}{12096}
\left(\frac{f}{f_*}\right)^6  -\frac{61}{4354560}\left(\frac{f}{f_*}\right)^8 + ...\bigg]\, ,
\ee
and
\begin{multline}
{\cal R}_{AA} = {\cal R}_{EE} 
= 4\sin^2\left(\frac{f}{f_*}\right)\bigg[\frac{3}{10}-\frac{169}{1680}\left(\frac{f}{f_*}\right)^2 
+\frac{85}{6048}\left(\frac{f}{f_*}\right)^4   
\\
- \frac{178273}{159667200}\left(\frac{f}{f_*}\right)^6     
+ \frac{19121}{24766560000}\left(\frac{f}{f_*}\right)^8 + ...\bigg] \,.
\end{multline}
Note that these signal transfer functions are very different from the
acceleration noise transfer functions given in~(\ref{tta}), (\ref{aaa}), 
(\ref{eea}). The difference in the transfer functions,
combined with priors on the functional form of the power spectral
density of the noise and signal, allows for the detection of signals
that are significantly below the noise level, even when there are not
enough links to form the $T$ channel~\cite{Adams-Cornish:2010}. The
sensitivity decreases for less informative priors. In the limit that
the priors allow for arbitrarily complicated functional forms for the
noise and signal spectra---forms so {\em contrived} that they can compensate
for the differences in the transfer functions---it becomes impossible
to separate signal from noise. In practice, a combination of
pre-flight and on-board testing, combined with physical modeling, will
hopefully constrain the noise model sufficiently to inform the
analysis and allow for component separation.

An additional complication for space interferometers operating in the
mHz frequency range are the millions of astrophysical signals that can
drown-out a cosmologically-generated 
stochastic background. While the brightest signals from
massive black hole mergers, stellar captures, and galactic binaries can
be identified and subtracted, a large number of weaker overlapping
signals will remain, creating a residual {\em confusion noise}. The
largest source of confusion noise is expected to come from millions of
compact white-dwarf binaries in our galaxy. The annual modulation of the
white-dwarf confusion noise due to the motion of the LISA spacecraft
(see Figure~\ref{f:cyclostationary}) 
will allow for this component to be separated from an isotropic
stochastic background, though at the cost of reduced sensitivity to the
background~\cite{Adams:2013qma}.

\section{Prospects for detection}
\label{s:obs}

\begin{quotation}
It's tough to make predictions, especially about the future.
{\em Yogi Berra}
\end{quotation}

\noindent
The detection of the binary black hole merger signals GW150914 and GW151226
give us confidence that stochastic gravitational waves will be
detected in the not-to-distant future.  Not only do they show that our
basic measurement principles are sound, they also point to the
existence of a much larger population of weaker signals from more
distant sources that will combine to form a stochastic background that
may be detected by 2020~\cite{TheLIGOScientific:2016-stochastic}. 
Indeed, a
confusion background from the superposition of weaker signals
eventually becomes the limiting noise source for detecting individual
systems~\cite{Barack:2004wc}.  As a general rule of thumb, individual
bright systems will be detected before the background for transient
signals (those that are in-band for a fraction of the observation
time), while the reverse is true for long-lived signals, such as the
slowly evolving supermassive black-hole binaries targeted by pulsar timing
arrays~\cite{Rosado:2015epa}. The prospects for detecting more exotic
stochastic signals, such as those from phase transitions in the early
Universe or inflation, are much less certain, but are worth pursing
for their high scientific value. In this section we begin with a brief
review of detection sensitivities curves across the gravitational-wave
spectrum, followed by a review of the current limits and prospects for
detection in each observational window.

\subsection{Detection sensitivity curves}
\label{s:detection-sensitivity-curves}

Detector sensitivity curves provide a useful visual indicator of the
sensitivity of an instrument to potential gravitational-wave sources. 
A good pedagogical description of the
various types of sensitivity curve in common use can be found 
in~\cite{Moore:2014lga}.  
Here we provide a more condensed summary.

\begin{figure}[h!tbp]
\begin{center}
\subfigure[]{\includegraphics[width=.45\textwidth]{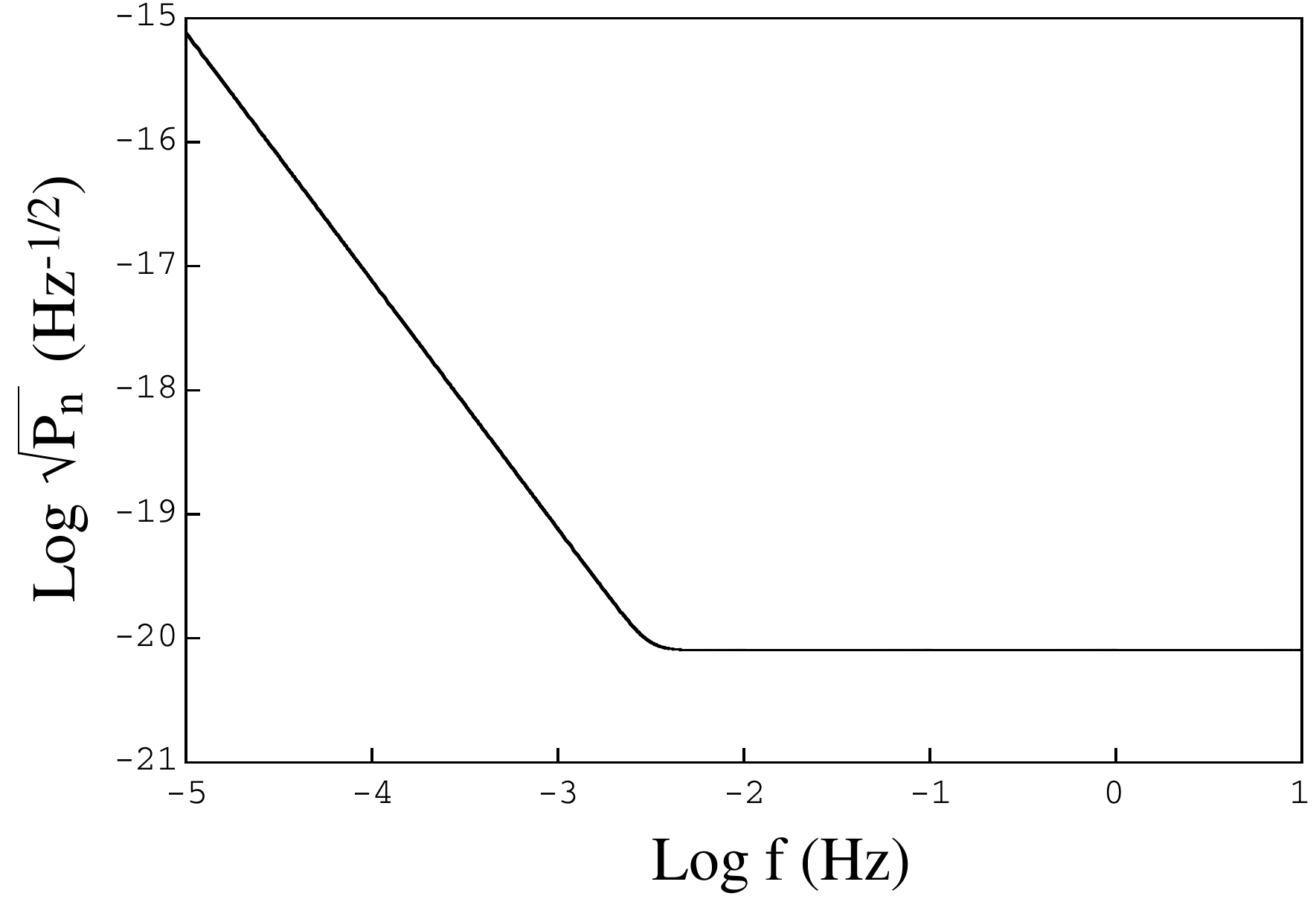}}
\subfigure[]{\includegraphics[width=.45\textwidth]{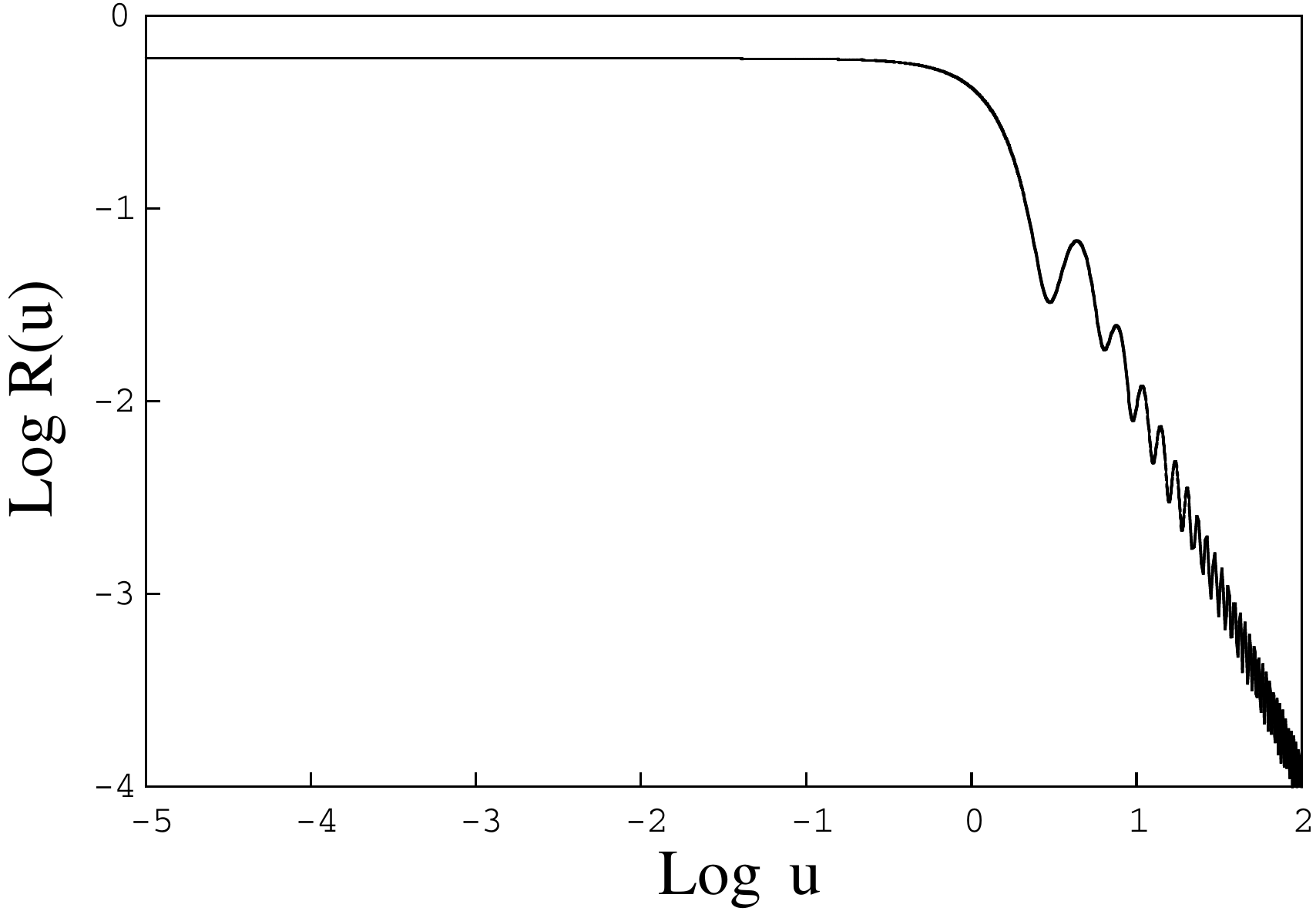}}
\subfigure[]{\includegraphics[width=.45\textwidth]{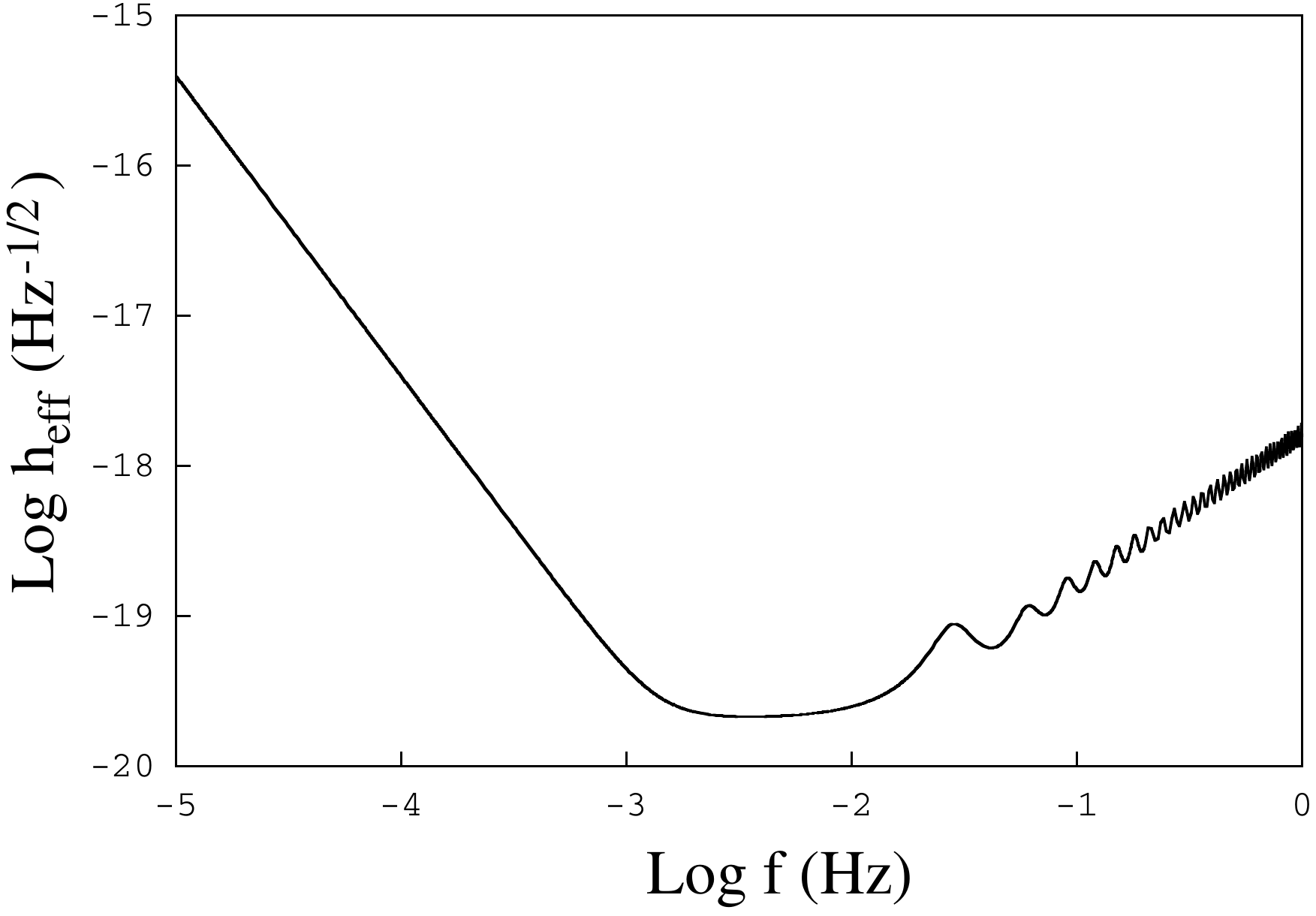}}
\subfigure[]{\includegraphics[width=.46\textwidth]{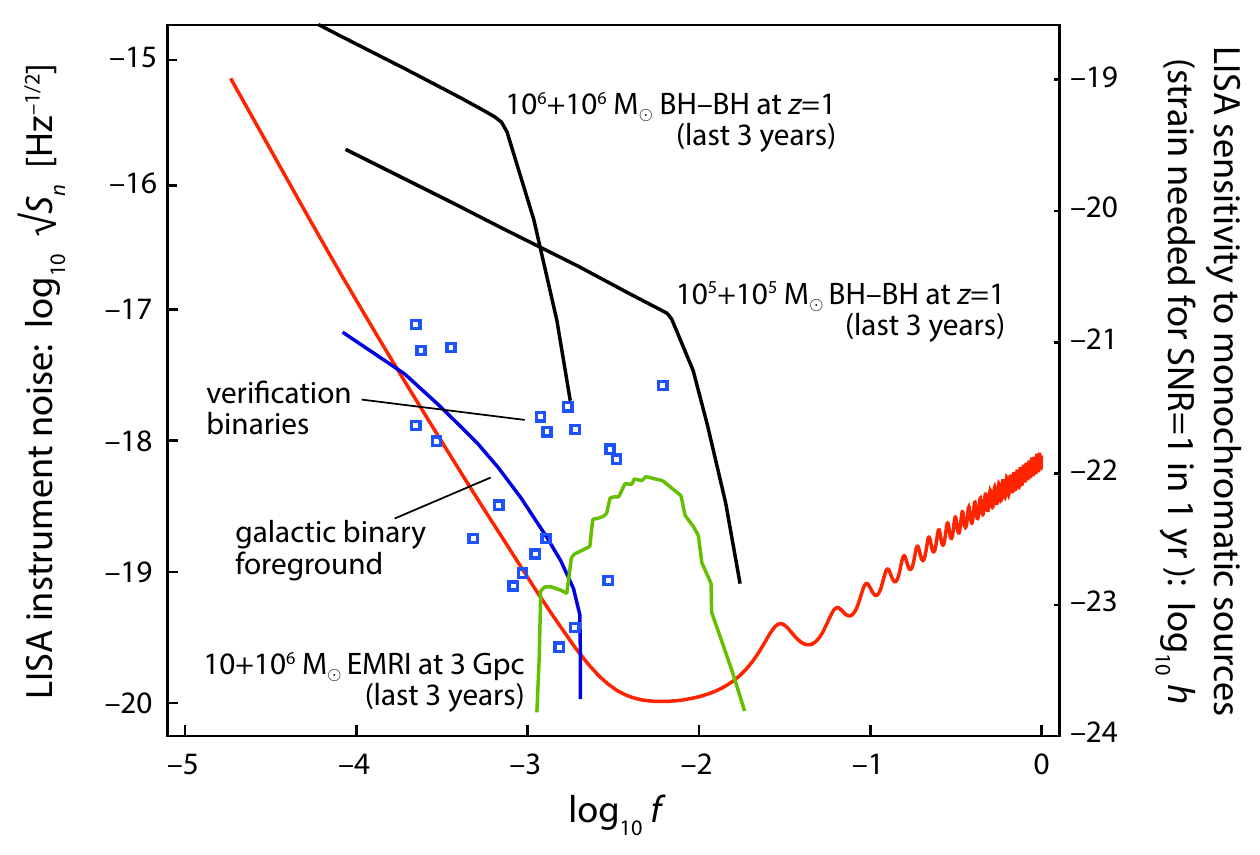}}
\caption{Constructing a sensitivity curve for the LISA detector. Panel
  (a) shows the amplitude spectral density of the noise. Panel (b)
  shows the sky and polarization-averaged response function.  Panel
  (c) shows the sensitivity curve found by dividing the noise spectral
  density by the response function. Panel (d) compares the
  filtered effective signal strength $\sqrt{2fTS_h(f)}$ 
  for various signals to the LISA sensitivity curve $\sqrt{S_n(f)}$.
  Panels (a)--(c): Image reproduced with permission from
  \cite{Larson:1999we}, copyright by APS. Panel (d): Image provided by
  M.~Vallisneri.}
\label{obs:lisa}
\end{center}
\end{figure}

The simplest type of sensitivity curve is a plot of the power spectral
density of the detector noise $P_n(f)$, or its amplitude spectral density 
$\sqrt{P_n(f)}$. 
(Recall that the mean-squared noise in the band $[f_1,f_2]$
is just the integral of $P_n(f)$ over that band.)
But plots of $P_n(f)$ or $\sqrt{P_n(f)}$
can be misleading since they do not
take into account the frequency-dependent response to gravitational
waves seen in Figure~\ref{f:gammaII_extended}. A better quantity to
plot is the sky and polarization-averaged amplitude 
spectral density
\be
h_{\rm eff}(f) \equiv \sqrt{S_n(f)} = \sqrt{P_n(f)/{\cal R}(f)}\, ,
\ee
which has units of ${\rm strain}/\sqrt{{\rm Hz}}$, or the 
corresponding (dimensionless) characteristic strain noise
\be
h_n(f) \equiv \sqrt{f S_n(f)}\,,
\label{e:hn}
\ee
where ${\cal R}(f) \equiv \Gamma_{II}(f)$ is the
transfer function defined in~(\ref{e:Ph}) and (\ref{e:GammaII}). 
Figure~\ref{obs:lisa}
shows the construction of a LISA sensitivity curve from $P_n(f)$ and
${\cal R}(u)$, where $u=f/f_*$ and $f_*= c/(2 \pi L)$. 
Note that for LIGO the factor ${\cal R}(f)$ is usually not included in
sensitivity plots since $f_* \simeq 12 \, {\rm kHz}$, and ${\cal R}(f)$ 
is effectively constant across the LIGO band.

\begin{figure}[h!tbp]
\begin{center}
\includegraphics[width=.7\textwidth]{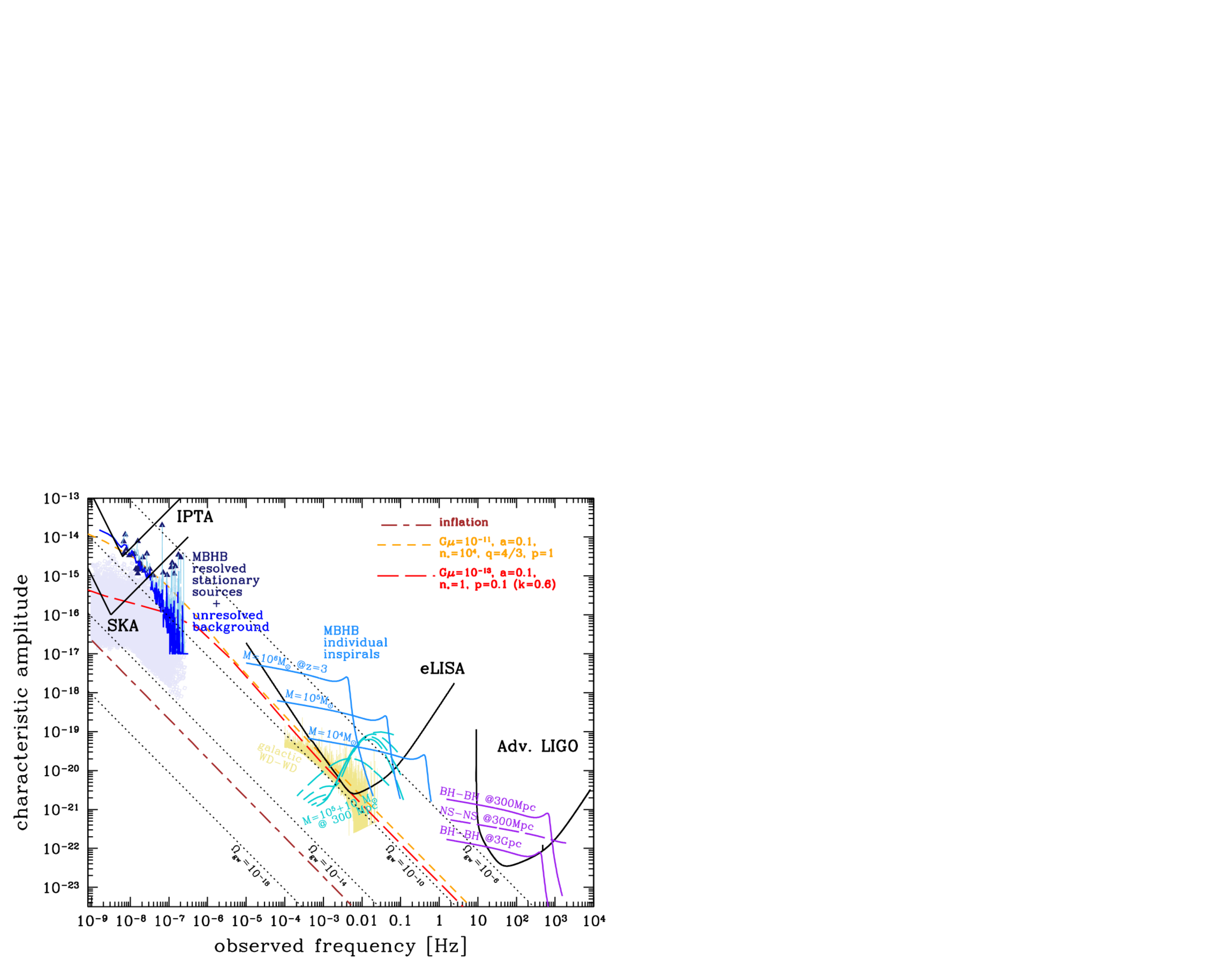}
\caption{Examples of detector sensitivity curves compared to potential
  gravitational-wave signals, comparing the characteristic strain 
  signal $h_c(f)$ to the characteristic strain noise $h_n(f)$.
  Image reproduced with permission from \cite{Janssen:2014dka},
  copyright by the authors.}
  \label{obs:examples}
  \end{center}
\end{figure}

The amplitude spectral density sensitivity curve $h_{\rm eff}(f)$ has to be
interpreted with some care, as simply comparing this curve to the
amplitude spectral density of a signal does not immediately convey how
detectable the signal is, as the likelihood function and detection statistics
derived from the likelihood function involve integrals over frequency. The
problem is compounded by the necessity to plot the sensitivity curves
on a log-log scale, where ``integration-by-eye'' misses the increase
in the number of frequency bins per logarithmic frequency
interval. Rather than plot the raw signals, it is more informative to
show quantities that account for the detection techniques being
used. For example, the amplitude signal-to-noise ratio $\rho$ for a 
deterministic signal $\tilde{h}(f)$ is given by
\be
\rho^2 = \int_{f=0}^\infty 
\frac{4 \vert \tilde{h}(f) \vert^2}{P_n(f)} \, df 
=  \int_{f=0}^\infty 
\frac{4 f  \vert  \tilde{h}(f) \vert^2}{P_n(f)} \, d(\ln f) \, .
\ee
Averaging over sky location and polarization we have
\be
\overline{\rho^2} 
=  \int_{f=0}^\infty 
\frac{4 f \tilde h_{\rm rss}^2(f) {\cal R}(f)} {P_n(f)} \,d(\ln f)
=   \int_{f=0}^\infty 
\frac{  (2 f  T) S_h(f)}{S_n(f)} \, d(\ln f) \, , 
\label{obs:snrd}
\ee
where 
$\tilde h_{\rm rss}^2(f) \equiv \vert \tilde h_+(f) \vert^2 + \vert \tilde h_\times(f) \vert^2$,
and $S_h(f)$ is the power spectral density of the gravitational-wave signal,
\be
S_h(f) \equiv \frac{2 \tilde h_{\rm rss}^2(f)}{T} \, .
\ee
The quantity $(2 f  T) S_h(f)/S_n(f)$ is the contribution to the 
square of the signal-to-noise ratio per logarithmic frequency interval.
The factor of $2 f T$ describes the boost that we get by coherently 
integrating the signal over many cycles. For deterministic signals
the amplitude signal-to-noise ratio grows as $T^{1/2}$. 
Since sensitivity curves are 
usually plotted in terms of the amplitude spectral density $h_{\rm eff}(f) = \sqrt{S_n(f)}$,
it is natural to plot signals in terms of the square-root of the numerator 
of~(\ref{obs:snrd}). Representative LISA sources are represented
in this way in panel (d) of Figure~\ref{obs:lisa}. 
An alternative choice is to plot both of these quantities multiplied 
by the square-root of the frequency, which yield the characteristic strain
for the signal, $h_c(f)$, as well as for the noise, $h_n(f)$.
Examples of characteristic strain sensitivity curves are
shown in Figure~\ref{obs:examples}. 

\begin{figure}[h!tbp]
\begin{center}
\subfigure[]{\includegraphics[width=.48\textwidth]{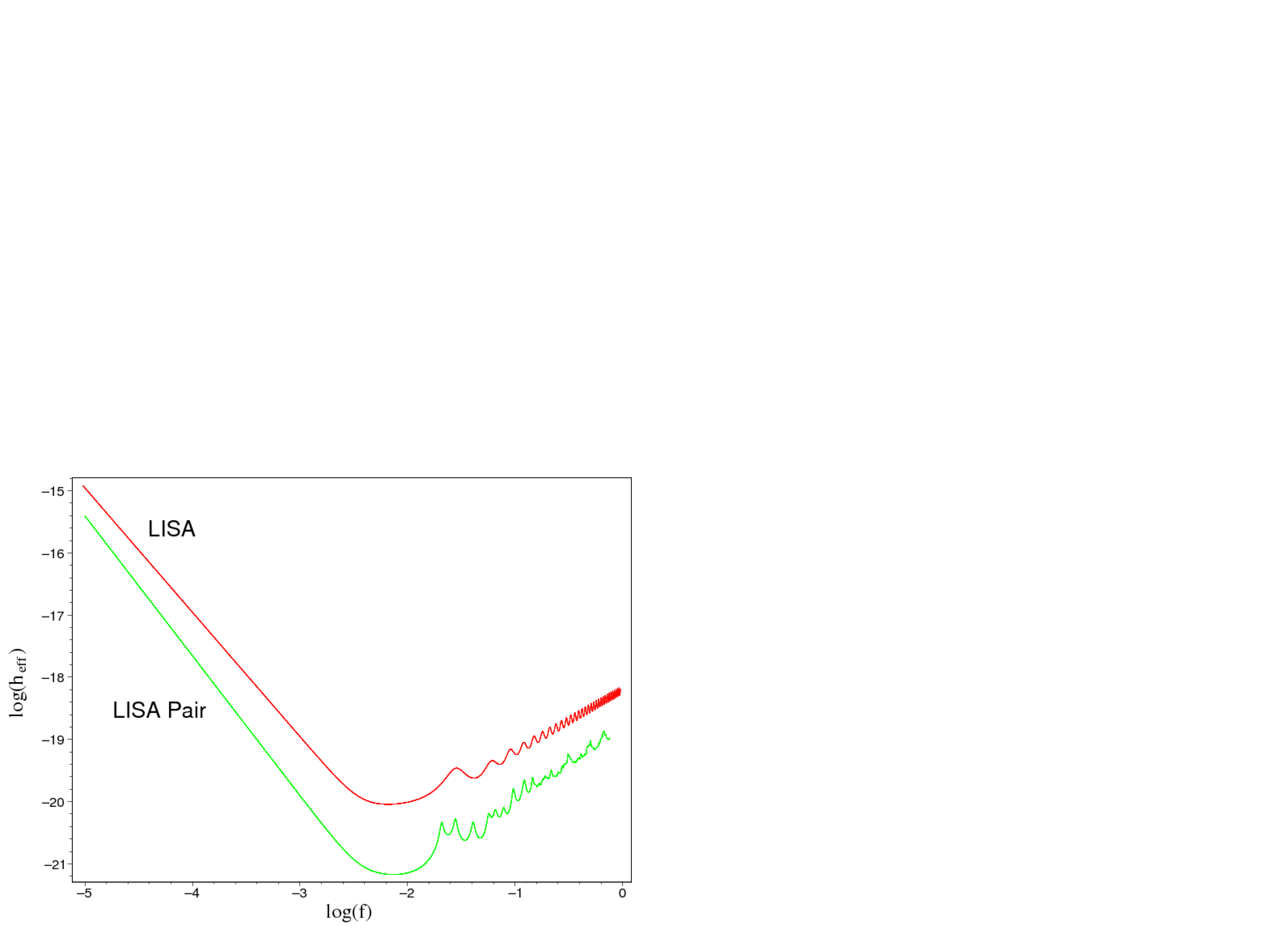}}
\subfigure[]{\includegraphics[width=.48\textwidth]{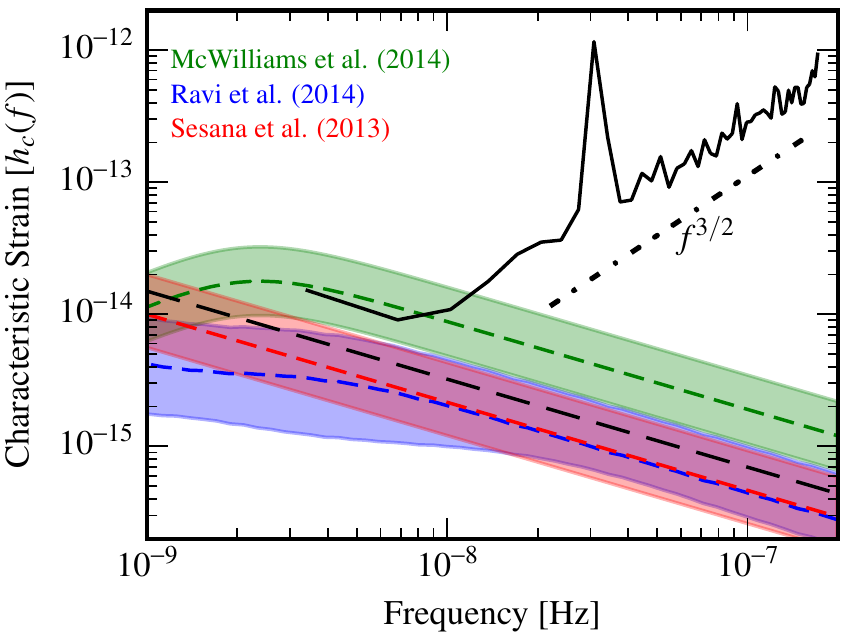}}
\caption{Panel (a) compares $h_{\rm eff}(f)$ for an isotropic stochastic
  background for a single LISA detector to
  that for a pair of LISA detectors arranged in a crossed-star configuration 
  using an observation time of one year. 
  Panel (b) compares the
  per-frequency-bin $(\Delta f = 1/T)$ upper limits on an isotropic
  stochastic background derived from the NANOGrav 9-year data set
  (solid black line) to three astrophysical models for the signal from
  supermassive black hole binaries. The upturn in the bound at low
  frequencies and the spike at $f=1/{\rm year}$ are due to the timing
  model acting as a filter on the signal. 
  Images reproduced with permission from \cite{Cornish:2001bb},
  copyright by APS (Panel (a)); and from \cite{Arzoumanian:2015liz},
  copyright by AAS (Panel (b)).}
\label{obs:heff}
\end{center}
\end{figure}

For isotropic stochastic signals, the sky location and 
polarization-averaged signal-to-noise ratio $\rho$ is
\be
\rho^2 =
2T \int_{f=0}^\infty df\>\sum_{I=1}^M\sum_{J>I}^M 
\frac{\Gamma_{IJ}^2(f) S_h^2(f)}{P_{n_I}(f)P_{n_J}(f)} 
= \int_{f=0}^\infty  
\frac{ (2f T) S_h^2(f)}{S^2_{\rm net}(f)}\, d(\ln f) \,,
\ee
where
\be
S_{\rm net}(f)\equiv
\left[
\sum_{I=1}^M\sum_{J>I}^M 
\frac{\Gamma_{IJ}^2(f)}{P_{n_I}(f)P_{n_J}(f)}
\right]^{-1/2}\,.
\ee
Note that for stochastic signals, $\rho$ is a {\em power} 
signal-to-noise ratio.
Similar to the amplitude signal-to-noise ratio for 
deterministic signals,
the power signal-to-noise ratio for stochastic signals 
grows as $T^{1/2}$.
(This assumes we are in the weak-signal limit, and that the
effective low-frequency cutoff does not change with time. 
See~\cite{Siemens-et-al:2013} for a more complicated scaling that
occurs for pulsar timing arrays.) Following the same logic as was
applied to deterministic signals, it would be natural to plot $(2f
T)^{1/4} \sqrt{S_h(f)}$ against sensitivity curves defined by
$\sqrt{S_{\rm net}(f)}$. Unfortunately, such conventions are not
uniformly applied, and the factor of $(2f T)^{1/4}$ is often applied
to $\sqrt{S_{\rm net}(f)}$ instead:
\be
h_{\rm eff}(f) \equiv \frac{1}{(2 T f)^{1/4}} \sqrt{S_{\rm net}(f)}\,.
\ee
A plot of $h_{\rm eff}(f)$, averaged over a logarithmic frequency
interval $\Delta f = f/10$, for a crossed pair of LISA-like detectors
is shown in Figure~\ref{obs:heff}.  Also shown in this figure are the
related per-frequency-bin upper bounds that are quoted by pulsar
timing groups using fixed frequency intervals $\Delta f = 1/T$.

\begin{figure}[h!tbp]
\begin{center}
\subfigure[]{\includegraphics[width=.48\textwidth]{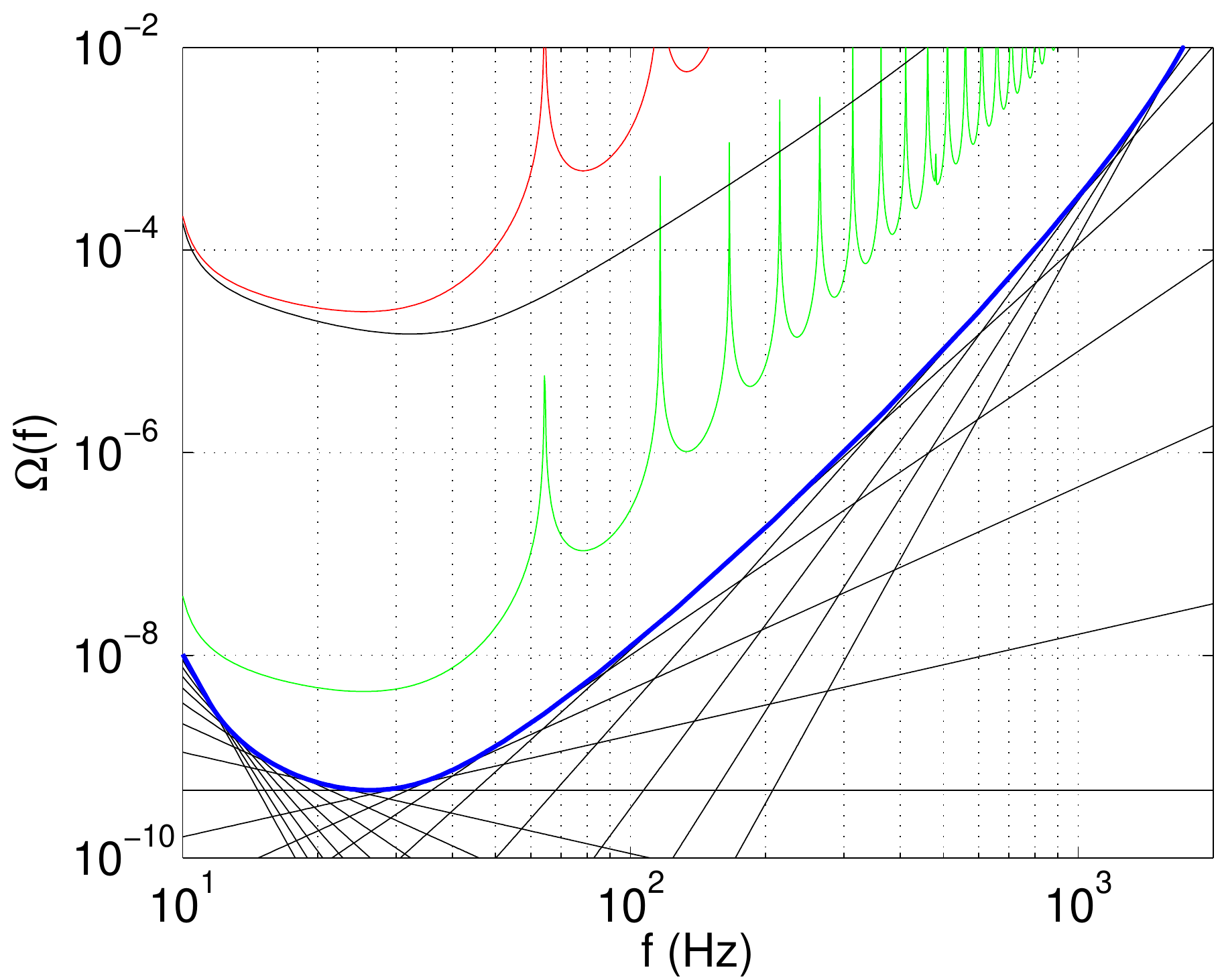}}
\subfigure[]{\includegraphics[width=.48\textwidth, height=.37\textwidth]{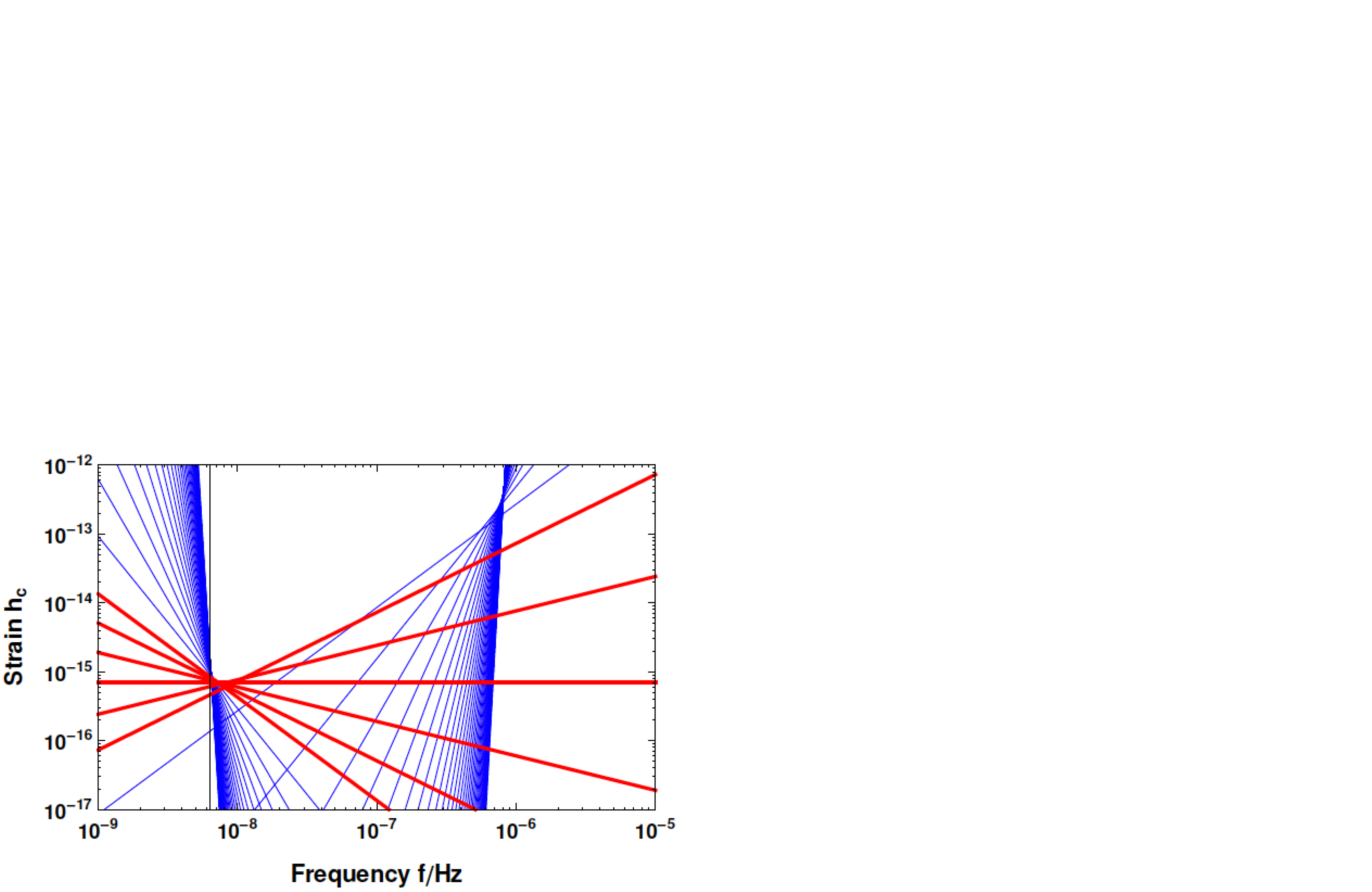}}
\caption{Panel (a) shows the sensitivity of the advanced LIGO
  Hanford--Livingston detector pair 
  in terms of gravitational-wave energy density
  $\Omega_{\rm gw}(f)$ using a variety of methods. The blue line is
  the sensitivity to isotropic stochastic signals with power-law spectra, 
  formed from the
  envelope of backgrounds with a wide range of spectral slopes (shown
  as straight black lines). Also shown as a black curve is the noise
  spectral density of a single LIGO detector 
  converted to units of $\Omega(f)$. 
  The red and green lines are variants of $h_{\rm eff}(f)$, again
  converted to units of $\Omega(f)$. The lower green
  curve is for an observation time of one year and $\Delta f =
  0.25\,{\rm Hz}$. 
  Panel (b) shows the characteristic
  strain sensitivity for a hypothetical pulsar timing array formed
  from the envelope of a large number of power law models. The red
  lines show a subset of the power law models used. The upper and
  lower frequency limits to the sensitivity are set by the observation cadence
  and the observation time, respectively. 
  Images reproduced with permission from \cite{Thrane-Romano:2013},
  copyright by APS (Panel a); and from \cite{Moore:2014eua}, copyright by
  IOP (Panel b).}
\label{obs:pow}
\end{center}
\end{figure}

The most common form of sensitivity curve for stochastic backgrounds
compares predictions of the gravitational-wave energy density
$\Omega_{\rm gw}(f)$ to the equivalent noise energy density
$\Omega_n(f) \equiv 2 \pi^2 f^3 S_n(f) /(3 H_0^2)$. These plots have the
advantage of being easy to produce and explain, but they do not fully
capture the boost that comes from integrating over frequencies.  An
alternative form of sensitivity curve that better represents the
analysis procedure uses the envelope of limits that can be placed on
power-law stochastic backgrounds~\cite{Thrane-Romano:2013, Moore:2014eua}. 
This method has the advantage of incorporating the
integrated nature of the detection statistic. Examples for advanced
LIGO and PTAs are shown in Figure~\ref{obs:pow}.

\subsection{Current observational results}
\label{s:observational-results}

\subsubsection{CMB isotropy}
\label{s:cmb-isotropy}

The cosmic microwave background (CMB) provides a snapshot of the
Universe $\approx\!400,000$ years after the big bang. During this epoch, 
the dense, hot plasma that filled the early Universe dilutes and cools to
the point where electrons and ions combine to form a neutral gas that
is transparent to photons.  Maps of the CMB contain a record of the
conditions when the CMB photons were last scattered.

Gravitational waves propagating through the early Universe, referred
to as tensor perturbations in the CMB literature, can leave an imprint
in the temperature and polarization pattern when CMB photons scatter
off the tidally-squeezed plasma. The challenge is to separate out the
contributions from primordial scalar, vector, and tensor perturbations,
and to separate these primordial contributions from subsequent
scattering by dust grains and hot gas.

Observations by the {\em COBE, WMAP} and {\em Planck} missions, along
with a host of ground-based and ballon-borne experiments, have
provided strong evidence in support for the inflation paradigm, where
the Universe undergoes a short period of extremely rapid expansion
driven by some, as yet unknown, {\em inflaton} field. To keep the
discussion brief, we focus our review on the standard single-field
``slow-roll'' inflation model, and direct the reader to more
extensive CMB-focused reviews, e.g., \cite{Kamionkowski-Kovetz:2015},
that cover more exotic models.

The rapid expansion of some small patch of the very early Universe
will erase any initial anisotropy and inhomogeneity, allowing the
patch to be modeled by a flat Friedmann-Lemaitre-Robertson-Walker
(FLRW) metric with scale factor $a(t)$. 
The Einstein equations for a FLRW Universe containing an
inflaton field $\phi$ with potential $V(\phi)$ are given by%
\footnote{For our discussion of inflation, we will work in 
{\em particle physics units} where both $c=1$ and $\hbar=1$.  
In place of using Newton's gravitational constant $G$, we will 
use the {\em reduced} Planck mass 
$M_{\rm Pl} \equiv (\hbar c/8\pi G)^{1/2} = 2.435\times 10^{18}~{\rm GeV}/c^2$.
In these units $M_{\rm Pl}^{2} = 1/8\pi G$, which simplifies several
of the formulae.
If you want to reinstate all of the relevant factors of $\hbar$ and $c$, 
note that the inflaton field $\phi$ has dimensions of energy and the inflaton 
potential $V(\phi)$ has dimensions of energy density.}
\be
H^2 
= \left(\frac{\dot a}{a}\right)^2 
= \frac{1}{3 M_{\rm Pl}^2} \left( \frac{1}{2} \dot\phi^2 + V(\phi) \right) \, ,
\ee
and
\be
\ddot \phi + 3 H \dot \phi + V_{,\phi} = 0\,.
\ee
In the slow-roll regime, the kinetic energy of the inflaton field
$\frac{1}{2} \dot\phi^2$ is assumed to be much smaller than the
potential energy $V(\phi)$, with $\phi$ having reached ``terminal
velocity'', such that $\ddot \phi \ll H \dot \phi$. Thus,
\be 3 H \dot \phi \simeq -V_{,\phi}
\quad {\rm and} \quad 
H^2 \simeq \frac{V}{3 M_{\rm Pl}^2} \, .  
\ee 
Necessary conditions for these approximations to
hold can be expressed in terms of a Taylor-series expansion of the
inflaton potential, leading to conditions on the first and second
derivatives of the potential: 
\be 
\epsilon_V \equiv \frac{M_{\rm Pl}^2 V_{,\phi}^2 }{2 V^2} \ll 1 \,,
\qquad
\eta_V \equiv \frac{M_{\rm Pl}^2 V_{,\phi\phi} }{V} \ll 1 \, .  
\ee 
The solution of the Einstein
equations for slow-roll inflation is well-approximated by an
exponentially de~Sitter Universe. Quantum fluctuations in the
otherwise smooth inflaton field and gravitational field give rise to
scalar and tensor perturbations, which leave their imprint in the
CMB. On large scales the power spectra for the scalar and tensor
fluctuations can be written as 
\be P_s(k) = A_s \left( \frac{k}{k_*}
\right)^{n_s(k)-1} \quad {\rm and} \quad P_t(k) = A_t \left(
\frac{k}{k_*} \right)^{n_t(k)}\,,
\ee 
where the reference wavenumber
$k_*=2\pi/\lambda_*$ is typically chosen to correspond to 
wavelengths $\lambda_* \sim
100 \, {\rm Mpc}$. The spectral indices $n_s(k)$ and $n_t(k)$ are
usually written in terms of a power-series expansion in $\ln k$: 
\be
n_s(k) = n_s + \frac{1}{2} \frac{ d n_s}{d \ln k}
\ln\left(\frac{k}{k_*} \right) + \frac{1}{6} \frac{ d^2n_s}{d \ln k^2}
\ln\left(\frac{k}{k_*} \right)^2 + \cdots \,.
\ee
The amplitude and
spectral indices are related to the energy scale for inflation, $V$,
and the slow-roll parameters $\epsilon_V$ and $\eta_V$: 
\be A_s \simeq\frac{V}{24 \pi^2 M_{\rm Pl}^4\epsilon_V} 
\quad {\rm and} \quad 
A_t \simeq \frac{2V}{3 \pi^2 M_{\rm Pl}^4} \,, 
\ee
and 
\be n_s \simeq 1 + 2 \eta_V - 6\epsilon_V 
\quad {\rm and} \quad 
n_t \simeq -2 \epsilon_V \,.  
\ee
Measuring $A_s$, $A_t$, and $n_s$ fixes the energy scale of inflation,
$V$, and the two leading terms in the Taylor-series expansion of the
inflaton potential, $V_{,\phi}$ and $V_{,\phi\phi}$. Additionally
measuring $n_t$ would provide a consistency check for the slow-roll
model.

One challenge in measuring $P_s(k)$ and $P_t(k)$ is that the scalar
and tensor perturbations both source temperature and polarization
anisotropies in the CMB radiation.  Another challenge is that
foreground gas and dust can also contribute to the temperature and
polarization anisotropies. The various components can be teased apart
by observing a wide range of CMB energies across a wide range of
angular scales.

The primordial contribution to the CMB follows a black-body spectrum,
while the dominant foreground contribution from gas and dust have very
different spectra. By observing at multiple CMB wavelengths the
primordial and foreground contributions can be separated. Separating
the scalar and tensor contributions to the primordial component of 
the temperature anisotropies can be achieved by making maps that cover
a wide range of angular scales, while separating their contributions
to the polarization anisotropies can be achieved by decomposing the
signal into curl-free $E$-modes and divergence-free $B$-modes, and
using measurements made on a wide range of angular scales. For a more
in-depth description, see Chapter 27 of the Review of Particle
Physics~\cite{Agashe:2014kda}.

\begin{figure}[h!tbp]
\begin{center}
\subfigure[]{\includegraphics[width=.49\textwidth]{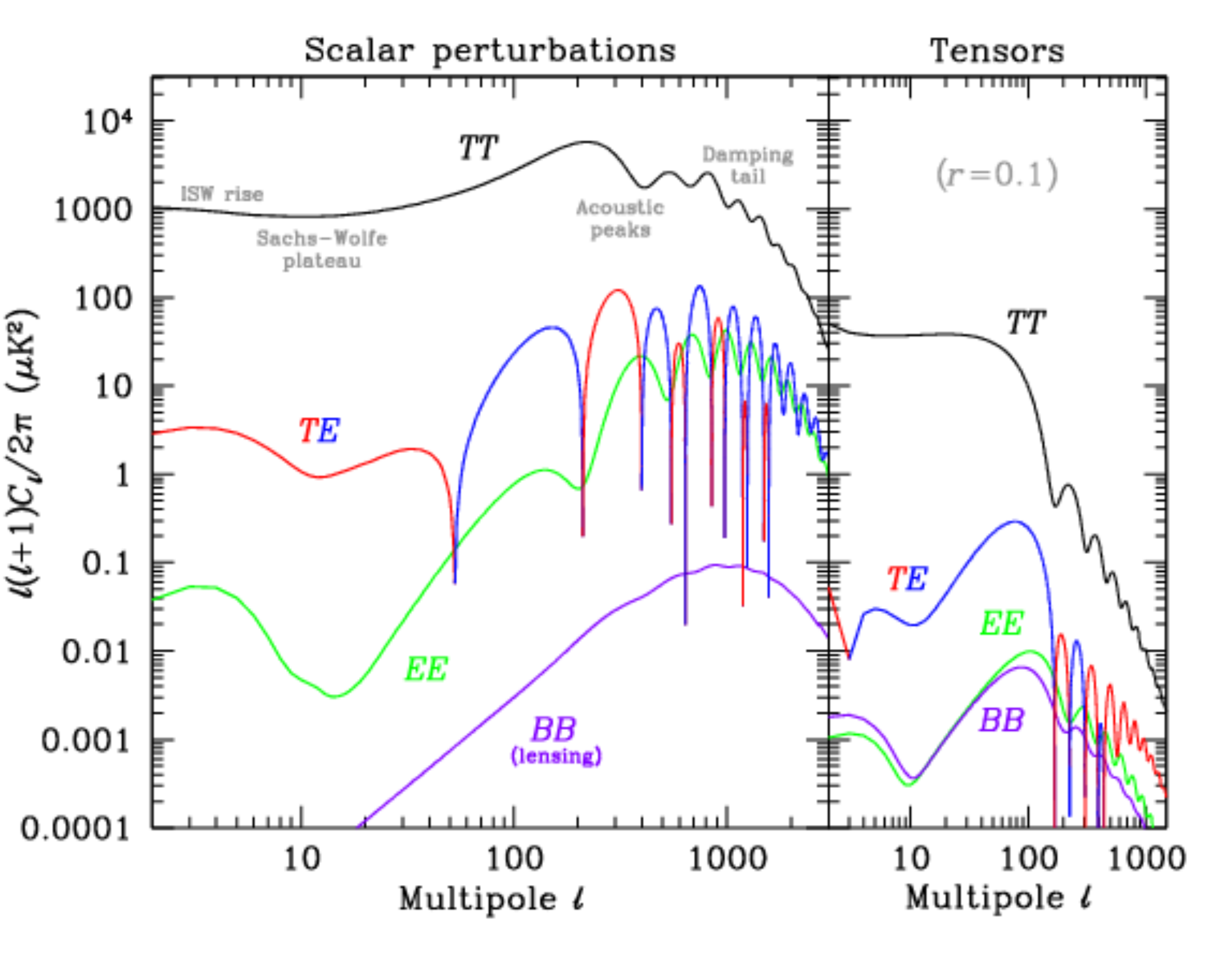}}
\subfigure[]{\includegraphics[width=.49\textwidth]{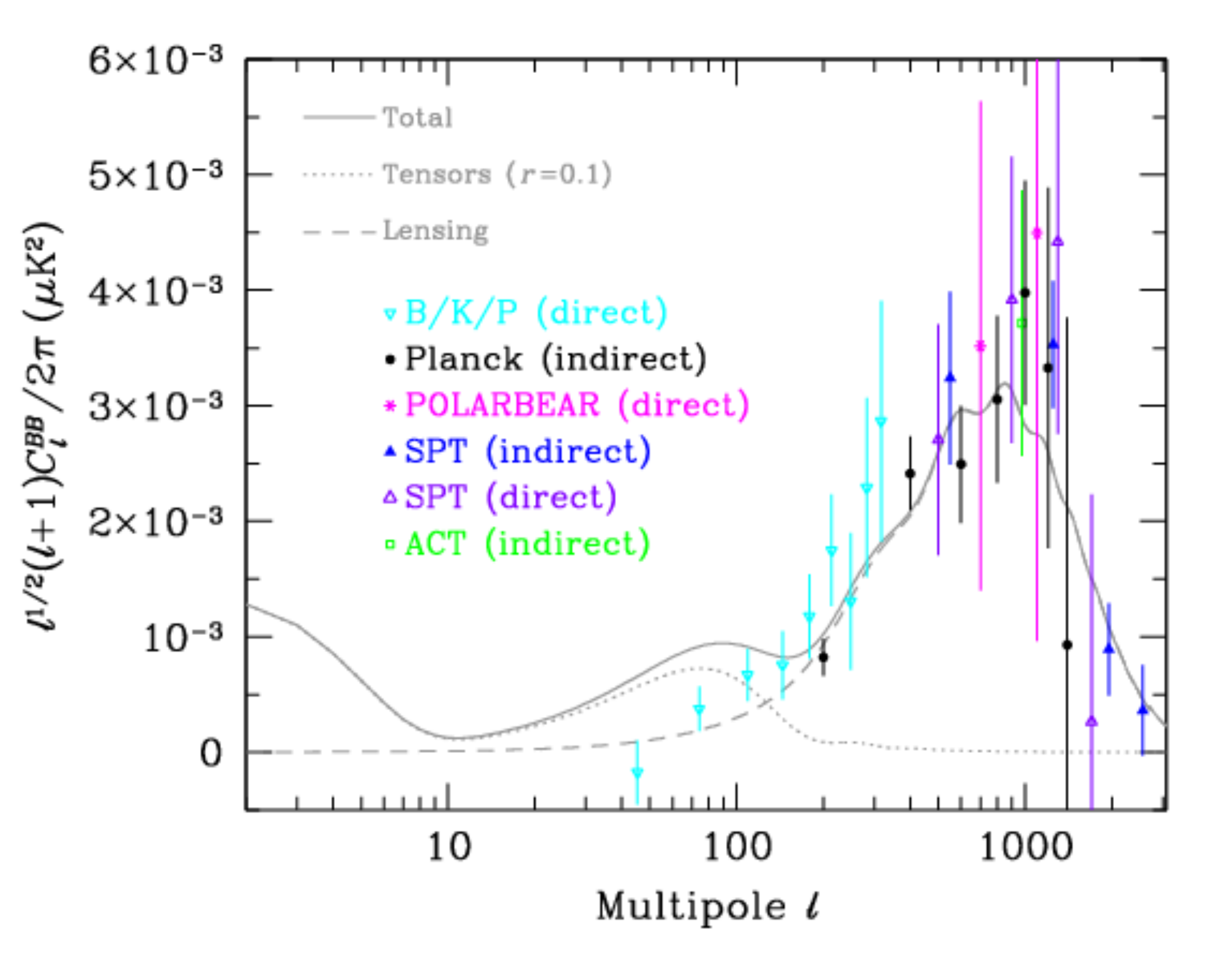}}
\caption{Panel (a) shows the theoretical predictions for the
  temperature and polarization cross-spectra from scalar and tensor
  perturbations for the best fit $\Lambda$CDM model from {\em Planck},
  assuming a tensor-to-scalar ratio of $r=0.1$. The curves are labeled by
  type: $TT$ labels the temperature power spectrum, while $TE$ labels the
  temperature-$E$-mode cross spectrum and so on. Panel (b) compares
  recent measurements of the $BB$ spectrum to the theoretical
  prediction. Images reproduced with permission from \cite{Agashe:2014kda},
    copyright by UC Regents.}
\label{obs:cmbspec}
\end{center}
\end{figure}

The scalar and tensor contributions to the large-scale temperature
anisotropy can be computed using linear perturbation theory. The
anisotropy due to tensor fluctuations arises solely from the
gravitational potential differences on the last-scattering surface,
while the anisotropy due to scalar fluctuations is more complicated,
and include contributions from the excitation of sound waves in
addition to variations in the gravitational potential. As the
co-moving horizon grows, tensor modes that have wavelengths shorter
than the horizon size redshift and lose energy. Consequently, the
tensor contribution to the CMB anisotropy drops by roughly two orders
of magnitude between angular scales $\ell=2$ and $\ell = 200$, while
the scalar contribution, after an initial dip, grows until reaching
the first acoustic peak at $\ell \simeq 220$.  Plots of the predicted
scalar and tensor contributions to the tenperature ($TT$) power spectra
using the best fit $\Lambda$CDM model from {\em Planck} are shown in
panel (a) of Figure~\ref{obs:cmbspec}. By comparing the CMB anisotropy
at very large scales ($\ell \sim 2$--$10$) and degree scales 
($\ell \sim 200$), 
it is possible to constrain the {\em tensor-to-scalar ratio}~\cite{Knox:1994qj}:
\be
r \equiv \frac{A_t}{A_s}\,.
\ee
In practice, a more sophisticated joint
analysis is performed using all available CMB data (often combined
with other data sets, such as maps of large-scale structure, weak
lensing, and measurements of the expansion history), simultaneously
fitting for a large number of cosmological parameters. The 
{\em Planck} temperature map, combined with weak lensing data, provide a 
precise measurement for the amplitude and spectral index of the scalar
perturbations: 
\be
\ln A_s = -19.928 \pm 0.057\,,
\qquad
n_s = 0.9603 \pm 0.0073\,,
\ee
and a bound on the tensor-to-scalar ratio:
\be
r < 0.12 \quad (95\%\ {\rm confidence})\,, 
\ee
using a pivot scale of $k_* = 0.002\, {\rm Mpc}^{-1}$. The
{\em Planck} bound on $r$ is the most stringent possible using CMB
temperature data~\cite{Knox:1994qj}.
(In fact, it beats the theoretical limit slightly since 
the analysis also used weak lensing and {\em WMAP} polarization data.) 
In order to improve on this bound, or to detect the tensor 
contribution, CMB polarization data must also be used.

\begin{figure}[h!tbp]
\includegraphics[width=1.0\textwidth]{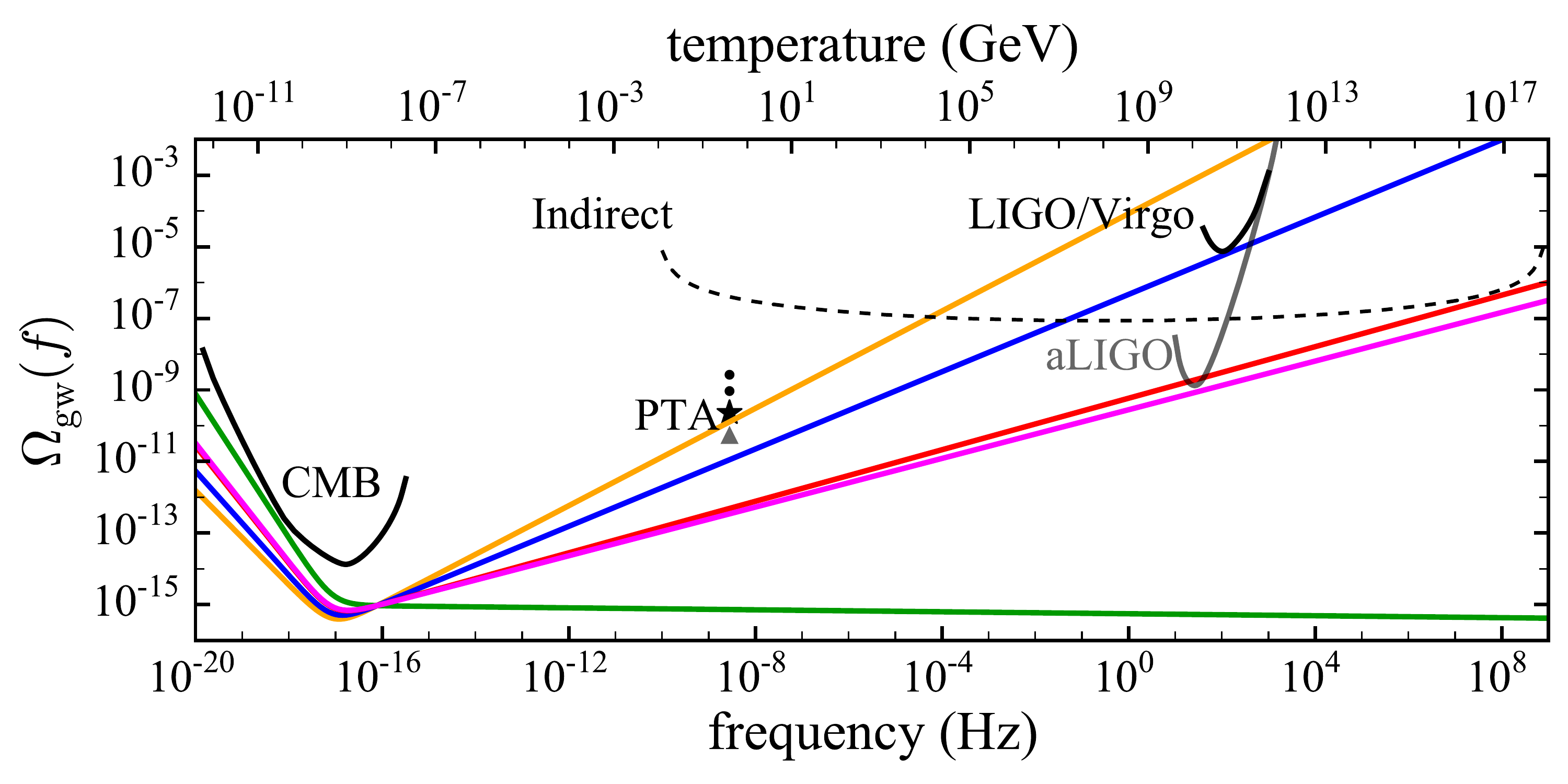}
\caption{Current and projected bounds on $\Omega_{\rm gw}(f)$ from CMB
  measurements, pulsar timing observations, and ground based
  interferometers. The curve marked ``CMB'' shows the projected
  sensitivity of the {\em Planck} satellite to primordial $B$-mode
  polarization anisotropies. The black star marked ``PTA'' is the
  current 95\% upper limit from the Parkes pulsar timing array.  The
  LIGO and aLIGO sensitivity curves were produced using the power-law
  envelope method~\cite{Thrane-Romano:2013}. The
  curve labeled ``indirect bounds'' was produced by converting bounds
  on the total gravitational-wave energy density from CMB temperature
  and polarization power spectra, weak lensing, baryon acoustic
  oscillations, and Big Bang nucleosynthesis to bounds on the energy
  density per logarithmic frequency interval using power-law
  models. The colored lines are theoretical predictions for the
  primordial background assuming $r=0.11$ and for spectral slopes $n_t
  = 0.68$ (orange curve), $n_t= 0.54$ (blue), $n_t= 0.36$ (red), and
  $n_t = 0.34$ (magenta). The prediction for the simple slow-roll
  inflation model discussed in this section, $n_t= -r/8$, is shown in
  green.
  Image reproduced with permission from \cite{Lasky:2015lej},
    copyright by the authors.}
\label{obs:bounds}
\end{figure}

The {\em Planck} bound on $r$ can be mapped into constraints on the
gravitational-wave energy density via~\cite{Turner:1993vb, Lasky:2015lej}:
\be
\Omega_{\rm gw}(f) = \frac{3 r A_s \Omega_r}{128} 
\left( \frac{f}{f_*}\right)^{n_t} 
\left[ \frac{1}{2} \left(\frac{f_{\rm eq}}{f} \right)^2 + \frac{16}{9} \right] \, ,
\ee
where $f = c k/(2\pi)$, $f_{\rm eq} \equiv \sqrt{2} H_0\Omega_m /(2 \pi
\sqrt{\Omega_r})$ is the frequency of a horizon-scale mode when matter and
radiation have the same density, and $\Omega_m$ and $\Omega_r$ are the
matter and radiation density today, in units of the critical
density. The projected {\em Planck} bound from the $B$-mode power
spectrum, along with existing and projected bounds from pulsar timing
and aLIGO are shown in Figure~\ref{obs:bounds}, which is taken from~\cite{Lasky:2015lej}. 
Also shown are curves
for theoretical models with a large tensor-to-scalar ratio ($r=0.11$) 
and a range of spectral tilts $n_t$.

Coherent motion in the primordial plasma can polarize the CMB photons
through Thomson scattering. Scalar perturbations source curl-free
$E$-mode polarization anisotropies, while the tensor perturbations 
source divergence-free $B$-mode polarization anisotropies, in addition to
$E$-modes. In principle, by decomposing the polarization into $E$ and
$B$ components, and using observations across a range of angular
scales, it should be possible to separate the scalar and tensor
contributions. In practice, the measurements are extremely challenging
due to the weakness of the signals (nano-Kelvin or smaller
polarization fluctuations as compared to micro-Kelvin temperature
fluctuations) and foreground noise. The main noise contributions come
from gravitational lensing, which converts a fraction of the much
larger $E$-mode anisotropy into $B$-modes, and scattering by dust
grains, which can convert unpolarized CMB radiation into $E$ and $B$
modes. Both of these potential noise sources have recently been
detected~\cite{Hanson:2013hsb, Ade:2015tva}. The detection of $B$-mode
polarization on large angular scales by {\em BICEP2} was originally
interpreted as having a primordial origin~\cite{Ade:2014xna}, but a
joint analysis using {\em Planck} dust maps~\cite{Ade:2015tva} showed
the signal to be consistent with foreground noise.

While detecting the primordial $B$-mode contribution is very
challenging, the pay-off is very large, as measuring the amplitude of
the tensor perturbations, $A_t$, fixes the energy scale of inflation,
and can be used to strongly constrain models of inflation.

\subsubsection{Pulsar timing}
\label{s:pulsar-timing}

Pulsar timing observations have made tremendous progress in the past
ten years and are now producing limits that seriously constrain
astrophysical models for supermassive black hole mergers. The current
observations are most sensitive at $f\sim 10^{-8}\ \mr{Hz}$, so we
choose a reference frequency of $f_{\rm ref} = 10^{-8}\ \mr{Hz}$, and
quote the latest bounds on $\Omega_{\rm gw}(f) = \Omega_\beta
(f/f_{\rm ref})^\beta$ in terms of bounds on $\Omega_\beta$ for a
Hubble constant value of 
$H_0 = 70~{\rm km}\,{\rm s}^{-1}\,{\rm Mpc}^{-1}$.

For a scale invariant ($n_t=0$) cosmological background, $\beta = 0$.
The most recent 95\% confidence limits on such a background 
are~\cite{Lentati:2015qwp, Arzoumanian:2015liz, 2015Sci...349.1522S, Lasky:2015lej}:
\be
\begin{array}{ll}
\Omega_0 < 1.2 \times 10^{-9}
&({\rm EPTA})\,,
\\
\Omega_0 < 8.5 \times 10^{-10}
&({\rm NANOGrav})\,,
\\
\Omega_0 < 2.1 \times 10^{-10} 
&({\rm PPTA})\,.
\end{array}
\ee
For a stochastic background from a population of black hole binaries
on quasi-circular orbits driven by gravitational-wave emission,
$\beta=2/3$.
The most recent 95\% confidence limits on such a background 
are~\cite{Lentati:2015qwp, Arzoumanian:2015liz, 2015Sci...349.1522S}:
\be
\begin{array}{ll}
\Omega_{2/3} < 5.4 \times 10^{-9}
&({\rm EPTA})\,,
\\
\Omega_{2/3} < 1.3 \times 10^{-9} 
&({\rm NANOGrav})\,,
\\
\Omega_{2/3} < 6.0 \times 10^{-10}
&({\rm PPTA})\,.
\end{array}
\ee
%

\subsubsection{Spacecraft Doppler tracking}
\label{s:doppler_tracking}

Spacecraft Doppler tracking~\cite{lrr-2006-1} operates on the same
principles as pulsar timing, with a precision on-board clock and radio
telemetry replacing the regular lighthouse-like radio emission of a
pulsar. The $\sim\!1$--$10$ AU Earth-spacecraft separation places
spacecraft Doppler tracking between pulsar timing and future LISA-like
missions in terms of baseline and gravitational-wave frequency
coverage. In principle, a fleet of spacecraft each equipped with
accurate clocks and high-power radio transmitters could be used to
perform the same type of cross-correlation analysis used in pulsar
timing, but to-date the analyses have been limited to single
spacecraft studies.

The most stringent bounds come from using the Cassini spacecraft, and
place a bound on the strength of a stochastic gravitational-wave
background at frequencies of order one over the transit time to the
spacecraft~\cite{Abbate:2003stu}:
\begin{equation}
\Omega_{\mr{gw}}(f)<0.027
\quad\mr{for}\quad
10^{-6}< f< 10^{-3}\ \mr{Hz} \, .
\end{equation}
%

\subsubsection{Interferometer bounds}
\label{s:ifo}

Data from the initial LIGO and Virgo observation runs, and
more recently, from advanced LIGO's first observing run (O1),
have been used to place constraints on the fractional energy 
density of isotropic stochastic backgrounds across multiple frequency 
bands between $20-1726$ Hz. The bounds are quoted in terms of 
$\Omega_{\rm gw}(f) =\Omega_\beta (f/f_{\rm ref})^\beta$ 
for $\beta=0$ (flat in energy density), 
$\beta=3$ (flat in strain spectral density), and 
$\beta=2/3$ (appropriate for a stochastic signal from a 
population of inspiralling binaries).
The $\beta=0$ bounds are quoted for the lower frequency bands, 
where the
sensitivity is greatest for signals with this slope, while the
$\beta=3$ bounds are quoted for the higher frequency bands. 
The $\beta=2/3$ bound is motivated by the detection of multiple
binary black hole mergers during O1,
which implies that stellar-remnant black holes may produce a 
detectable stochastic signal from the superposition of many individually
undetected sources~\cite{TheLIGOScientific:2016-stochastic}.  
The bounds assume a Hubble constant value of 
$H_0 = 68~{\rm km}\,{\rm s}^{-1}\,{\rm Mpc}^{-1}$.

\medskip
\noindent{\bf Initial LIGO and Virgo data}
\medskip

\noindent
Combining the initial LIGO and Virgo data, 
the most stringent 95\%-confidence upper limits for 
$\beta=0$ are~\cite{Aasi:2014zwg}:
\begin{equation}
\begin{array}{lll}
\Omega_{\mr{gw}}(f)< 5.6 \times 10^{-6}
&{\rm for}&41.5 < f< 169.25\ \mr{Hz}\,,
\\
\Omega_{\mr{gw}}(f)< 1.8 \times 10^{-4}
&{\rm for}&170 < f< 600\ \mr{Hz}\,.
\end{array}
\end{equation}
The bounds for $\beta=3$ are~\cite{Aasi:2014zwg, Aasi-et-al:H1H2}:
\begin{equation}
\begin{array}{lll}
\Omega_{\mr{gw}}(f)< 7.7 \times 10^{-4}\left(\frac{f}{900 \, \mr{Hz}}\right)^3
&{\rm for}&460 < f< 1000\ \mr{Hz}\,,
\\
\Omega_{\mr{gw}}(f)< 1.0 \times 10^{0}\left(\frac{f}{1300 \, \mr{Hz}}\right)^3
&{\rm for}&1000 < f< 1726\ \mr{Hz}\,.
\end{array}
\end{equation}
We note that the $\beta=3$ bound for the $460 < f< 1000\ \mr{Hz}$
frequency band comes from a correlation analysis using the colocated
2 km and 4 km Hanford detectors~\cite{Aasi-et-al:H1H2}.

\medskip
\noindent{\bf Advanced LIGO's first observing run O1}
\medskip

\noindent
The analysis of data from LIGO's first observing run O1
improves on the above limits for $\beta=0$ and $\beta=3$
at lower frequencies~\cite{LVC:O1-isotropic}:
\begin{equation}
\begin{array}{lll}
\Omega_{\mr{gw}}(f)< 1.7 \times 10^{-7}
&{\rm for}&20 < f< 85.8\ \mr{Hz}\,,
\\
\Omega_{\mr{gw}}(f)< 1.7 \times 10^{-8}\left(\frac{f}{25 \, \mr{Hz}}\right)^3
&{\rm for}&20 < f< 305\ \mr{Hz}\,.
\end{array}
\end{equation}
The data was also used to place a 
limit on stochastic signals with spectral slope
$\beta=2/3$, appropriate for stochastic signals from inspiralling
binaries~\cite{LVC:O1-isotropic}:
\begin{equation}
\begin{array}{lll}
\Omega_{\mr{gw}}(f)< 1.3 \times 10^{-7}\left(\frac{f}{25 \, \mr{Hz}}\right)^{2/3}
&{\rm for}&20 < f< 98.2\ \mr{Hz}\,.
\end{array}
\end{equation}

\subsubsection{Bounds on anisotropic backgrounds}
\label{s:aniso-bounds}

\begin{figure}[h!tbp]
\begin{center}
\includegraphics[width=.49\textwidth]{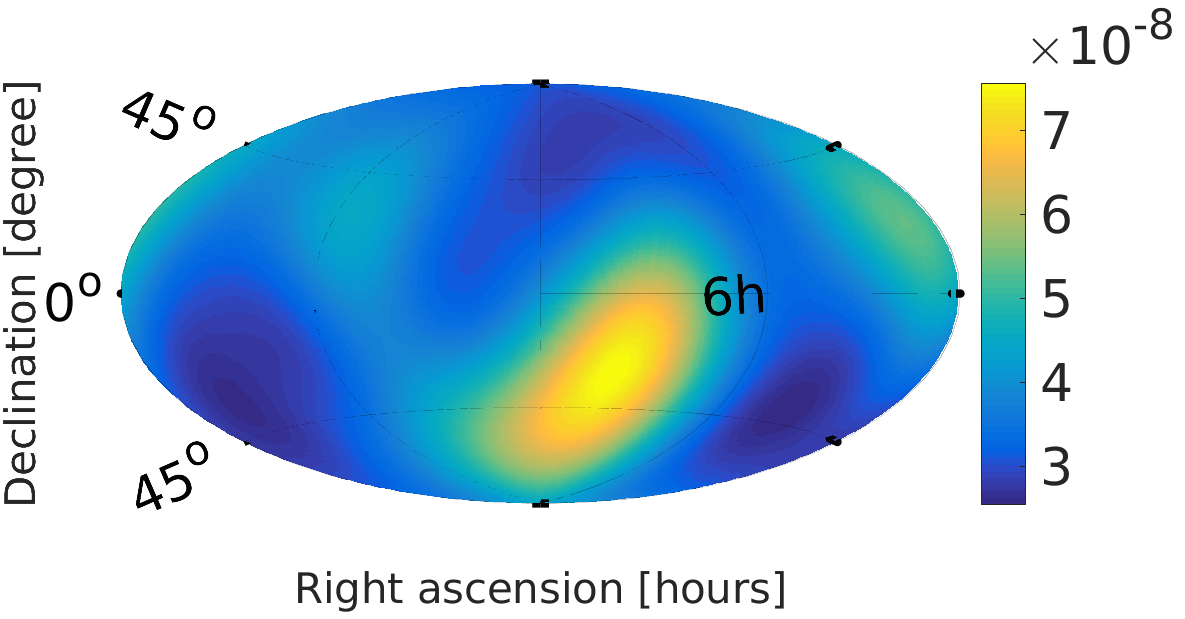}
\includegraphics[width=.49\textwidth]{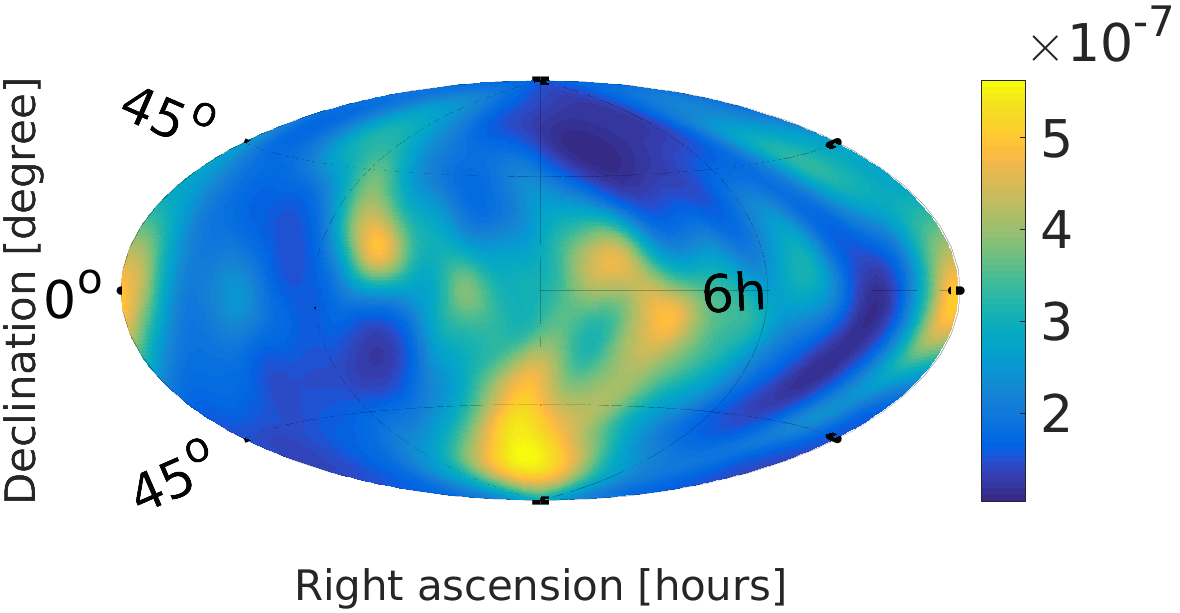}
\includegraphics[width=.49\textwidth]{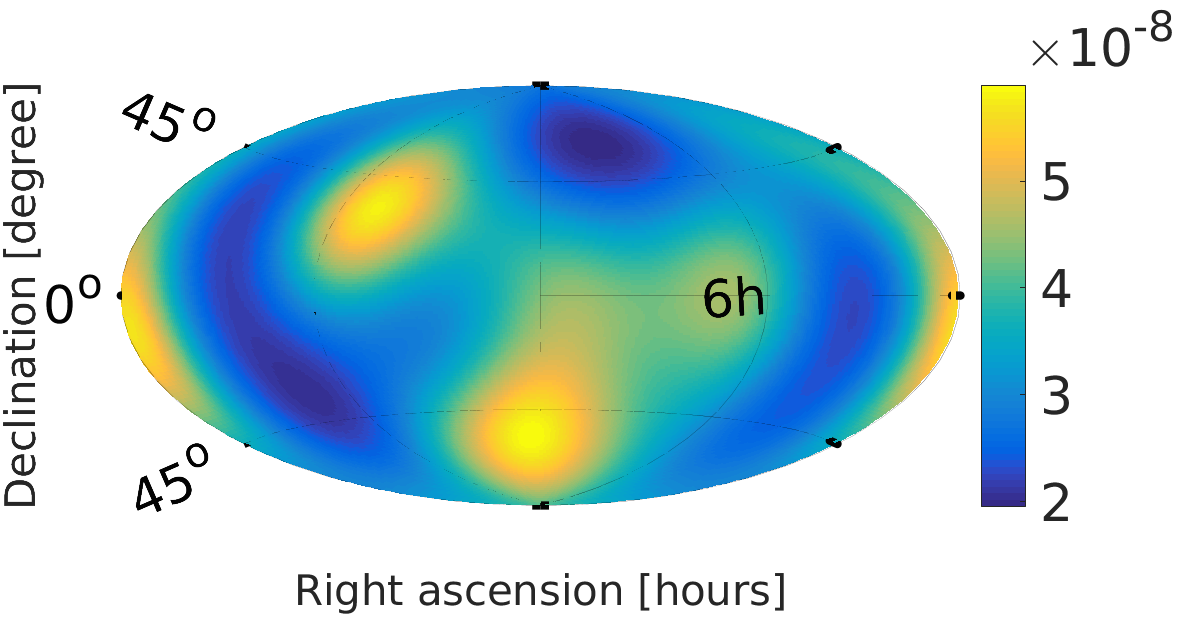}
\includegraphics[width=.49\textwidth]{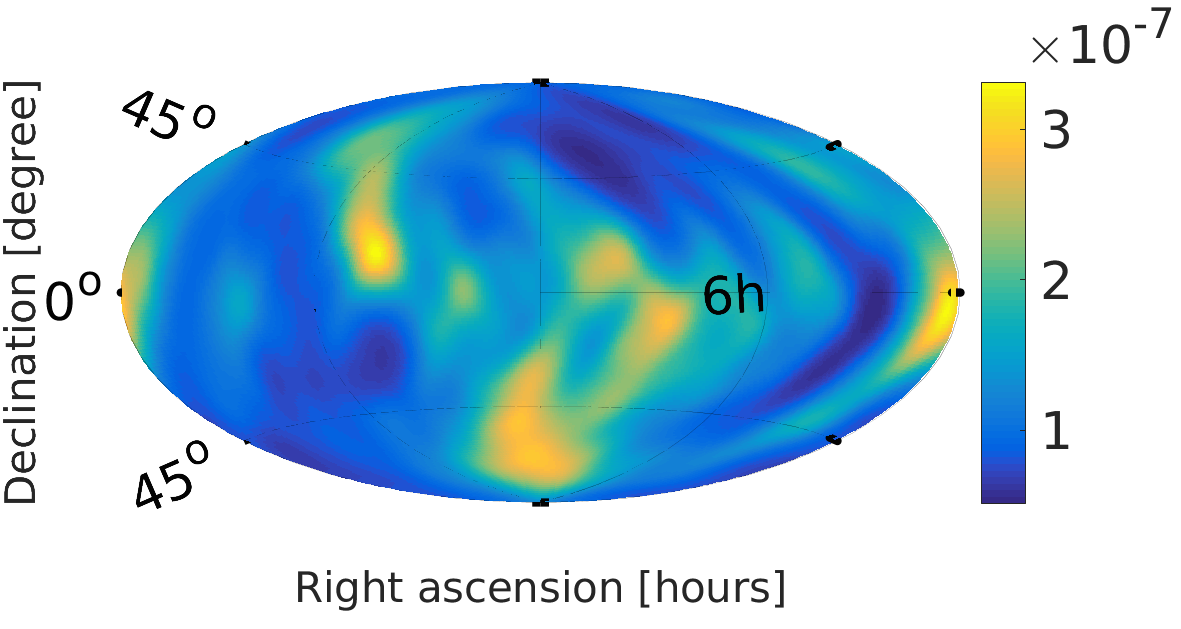}
\includegraphics[width=.49\textwidth]{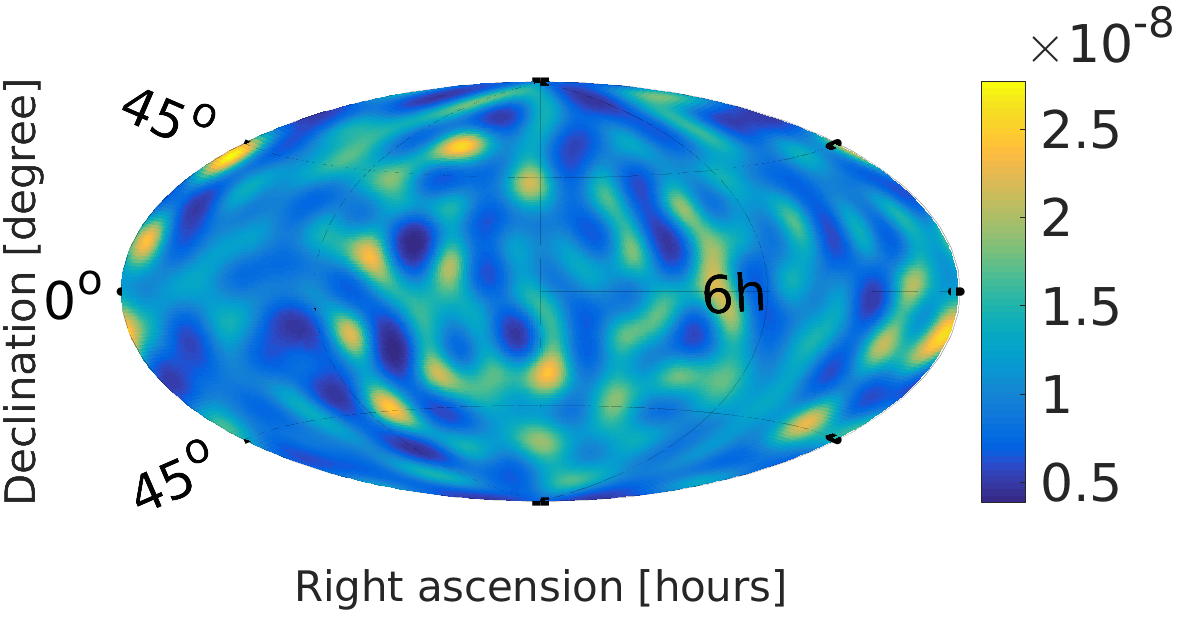}
\includegraphics[width=.49\textwidth]{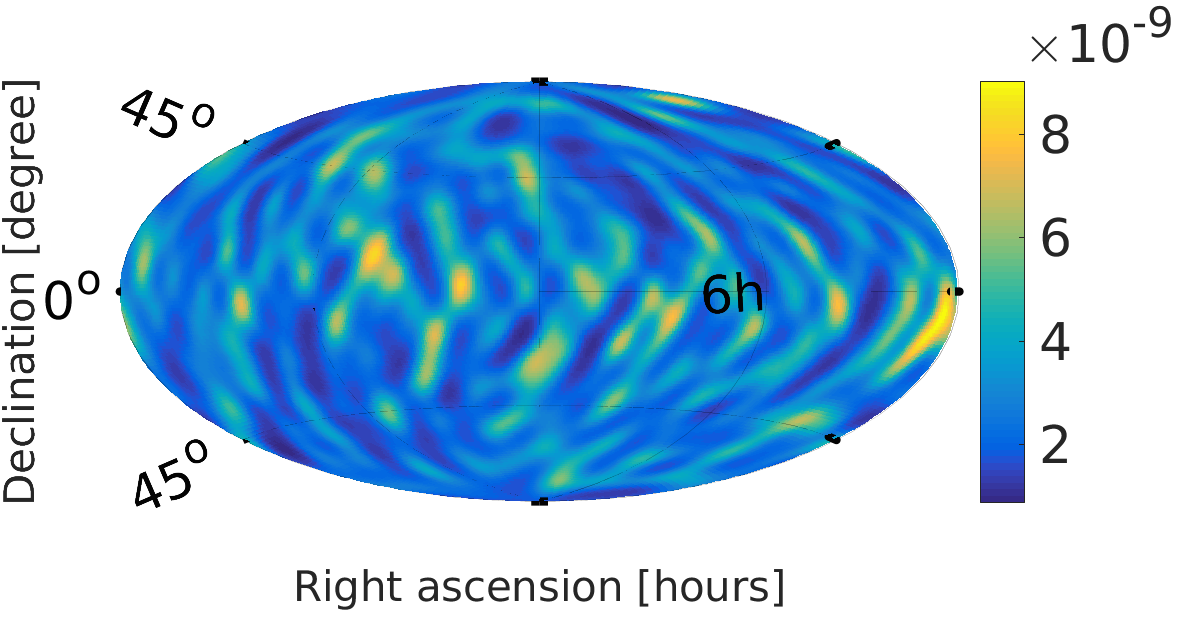}
\caption{90\% confidence-level upper-limit maps on 
  gravitational-wave power for anisotropic backgrounds having
  spectral indices $\beta = 0$, 2/3, and 3 
  (first, second, and third row, respectively).
  The data analyzed were from advanced LIGO's first 
  observational run O1~\cite{LVC:O1-anisotropic}.
  Left column: UL maps on the fractional energy density 
  $\Omega_\beta(\hat n)$ expressed in units of 
  $\mr{sr}^{-1}$, constructed using the spherical 
  harmonic decomposition method out to $l_{\rm max}=3$, 4, and 16.
  Right column: UL maps on the energy flux 
  $F_{\beta, \hat n_0}$ expressed in units of 
  ${\rm erg}\,{\rm cm}^{-2}\,{\rm s}^{-1}\,{\rm Hz}^{-1}$,
  constructed using the radiometer method, which assumes a
  point-source signal model.
  Figure adapted from \cite{LVC:O1-anisotropic}.}
\label{f:aniso-bounds-LIGO}
\end{center}
\end{figure}

\begin{figure}[h!tbp]
\begin{center}
\includegraphics[width=.6\textwidth]{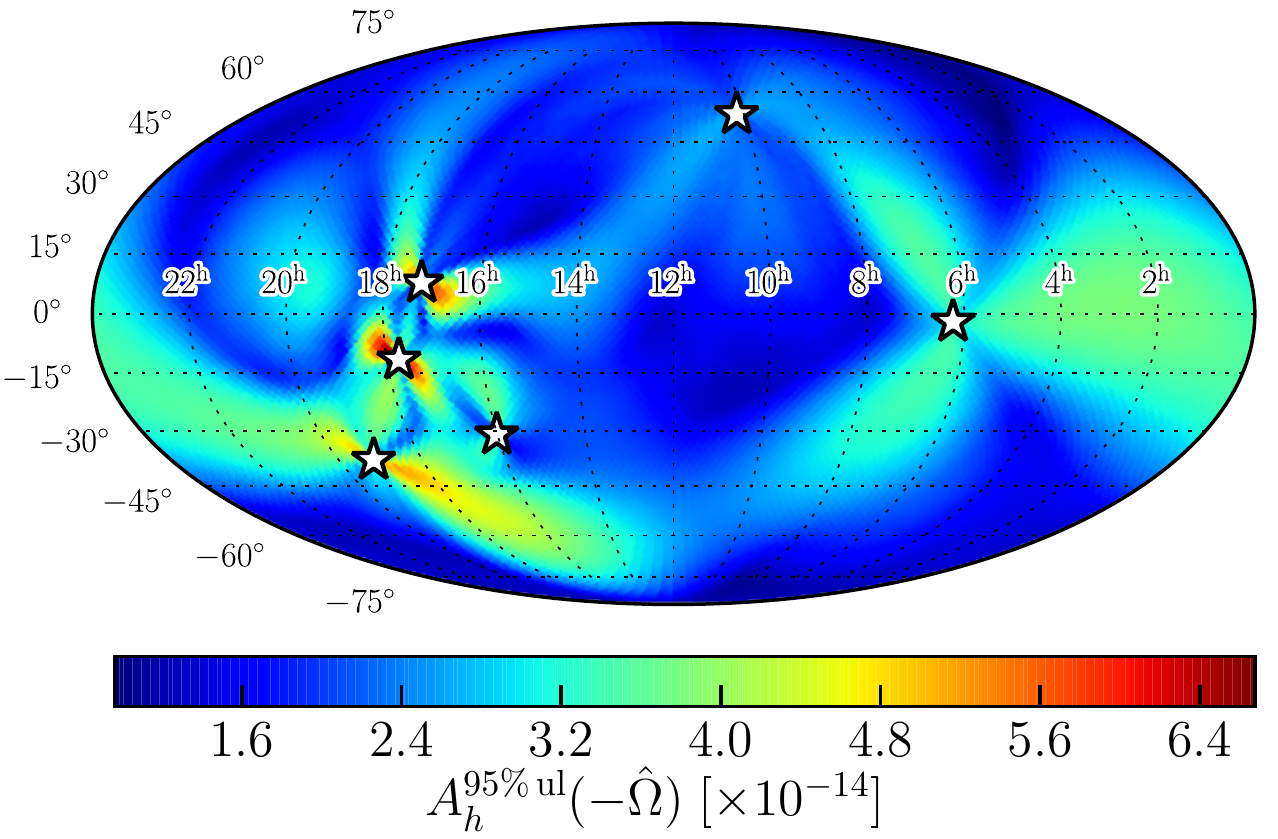}
\caption{95\% confidence-level upper-limit map on the 
  characteristic strain amplitude for an 
  anisotropic background having spectral index $\alpha=-2/3$.  
  The white stars show the location of the EPTA pulsars used 
  for the analysis.
  Image reproduced with permission from \cite{Taylor-et-al:EPTA-anisotropic},
    copyright by APS.}
\label{f:aniso-bounds-EPTA}
\end{center}
\end{figure}

Constraints on anisotropic backgrounds have also been set using 
data from both initial and advanced 
LIGO~\cite{Abadie-et-al:S5-anisotropic, LVC:O1-anisotropic} and
from the European Pulsar Timing Array~\cite{Taylor-et-al:EPTA-anisotropic}.
The corresponding upper-limit maps for advanced LIGO's first
observing run (O1) and from the EPTA data are shown in
Figures~\ref{f:aniso-bounds-LIGO} and~\ref{f:aniso-bounds-EPTA}, respectively.

The upper-limit maps shown in Figure~\ref{f:aniso-bounds-LIGO}
are for advanced LIGO's first observational run~\cite{LVC:O1-anisotropic}.
The maps were constructed using both the spherical harmonic 
decomposition method (left column) and the radiometer method (right column).
(These methods are described in Section~\ref{s:radiometer-SHD}.)
The three rows correspond to anisotropic backgrounds having 
spectral indices $\beta=0$, 2/3, and 3, respectively.
The spherical harmonic decomposition maps have 
$l_{\rm max}=3$, 4, and 16, respectively, and the upper limits 
are on
\be
\Omega_\beta(\hat n) 
\equiv \frac{2\pi^2}{3 H_0^2} f_{\rm ref}^3{\cal P}(\hat n)\,,
\qquad
{\cal P}(\hat n) = \sum_{l=0}^{l_{\rm max}} \sum_{m=-l}^l
{\cal P}_{lm} Y_{lm}(\hat n)\,,
\ee
expressed in units of fractional energy density per sterardian,
${\rm sr}^{-1}$.
These limits can be used, for example, to put a constraint on 
the integrated fractional energy density:
\be
\Omega_{\rm gw}(f) = \int d^2\Omega_{\hat n}\>
\Omega_{\beta}(\hat n) \left(\frac{f}{f_{\rm ref}}\right)^\beta\,.
\ee
The radiometer maps give upper limits on the 
energy flux
\be
F_{\beta, \hat n_0}
\equiv \frac{c^3 \pi}{4 G}f_{\rm ref}^2 {\cal P}_{\hat n_0}\,,
\qquad
{\cal P}(\hat n) = 
{\cal P}_{\hat n_0}\,\delta^2(\hat n,\hat n_0)\,,
\ee
expressed in units of 
${\rm erg}\,{\rm cm}^{-2}\,{\rm s}^{-1}\,{\rm Hz}^{-1}$.
Here, $G$ is Newton's gravitational constant, 
and ${\cal P}_{\hat n_0}$ is the signal power of a
single point source in direction $\hat n_0$
(which is the radiometer signal model).%
\footnote{One should think of a radiometer upper-limit
map as a convenient way of representing upper limits 
for a {\em collection} of individual point-source 
signal models, one for each point on the sky.
As described in Section~\ref{s:radiometer-SHD}, the 
radiometer analysis ignores correlations between neighboring
pixels on the sky, completely side-stepping the 
deconvolution problem associated with a non-trivial 
point spread function for the search.
In other words, each pixel of a radiometer map 
corresponds to a separate analysis.}
The reference frequency for all the maps is $f_{\rm ref}=25~{\rm Hz}$,
corresponding to the most sensitive part of the frequency band
for a stochastic search at advanced LIGO design sensitivity.
All the searches include frequencies $20<f<500~{\rm Hz}$, which more
than cover the regions of 99\% sensitivity for each spectral index.

The upper-limit map shown in Figure~\ref{f:aniso-bounds-EPTA} is 
for the 2015 European Pulsar Timing 
Array data~\cite{Taylor-et-al:EPTA-anisotropic}.
The map shows the 95\% confidence-level upper limits on the 
(dimensionless) 
amplitude $A_h$ of the characteristic strain (\ref{e:hc-A}):
\be
h_c(f) = A_h \left(\frac{f}{{\rm yr}^{-1}}\right)^{-2/3}\,,
\ee
for $\sim\! 2< f< 90~{\rm nHz}$.
The spectral index $\alpha = -2/3$ is
appropriate for a stochastic background formed from the 
superposition of gravitational-wave-driven, circular,
inspiraling supermassive black-hole binaries, which is 
an expected source at the nano-Hz frequencies probed by 
pulsar timing arrays.
The corresponding spectral index for the fractional 
energy density in gravitational waves, 
$\Omega_{\rm gw}(f)$, 
is $\beta = 2/3$ (Section~\ref{s:characteristic-strain}).

\section*{Acknowledgements}
\label{s:acknowledgements}

JDR acknowledges support from National Science Foundation Awards
PHY-1205585, CREST HRD-1242090, PHY-1505861. NJC acknowledges support
from National Science Foundation Awards PHY-1306702 and PHY-1607343, and NASA award
NNX16AB98G. JDR and NJC acknowledge support from the National Science Foundation NANOGrav
Physics Frontier Center, NSF PFC-1430284.
We also thank members of the LIGO-Virgo stochastic working group 
and members of NANOGrav for countless discussions related to all 
things stochastic.
Special thanks go out to Bruce Allen, 
Matt Benacquista, Nelson Christensen, Gwynne Crowder, Yuri Levin,
Tyson Littenberg, Chris Messenger, 
Soumya Mohanty, Tanner Prestegard, Eric Thrane, and 
Michele Vallisneri, who either commented on parts of the text or 
provided figures for us to use.
Special thanks also go out to an anonymous referee for many comments 
and useful suggestions for improving parts of the text.
This research made use of Python and its standard libraries: 
numpy and matplotlib.  
We also made use of MEALPix (a Matlab implementation of 
HEALPix~\cite{HEALPix}), developed by the GWAstro Research Group 
and available from {\tt http://gwastro.psu.edu}.
Finally, we thank the editors of \textit{Living Reviews in Relativity} 
(especially Bala Iyer and Frank Schulz) for their incredible 
patience while this article was being written.
This document has been assigned LIGO Document Control Center number 
LIGO-P1600242.

\appendix

\section{Freedom in the choice of polarization basis tensors}
\label{s:polarization_tensors}

\subsection{Linear polarization}
\label{s:linear-pol-basis}

\begin{figure}[h!tbp]
\begin{center}
\includegraphics[angle=0,width=0.7\textwidth]{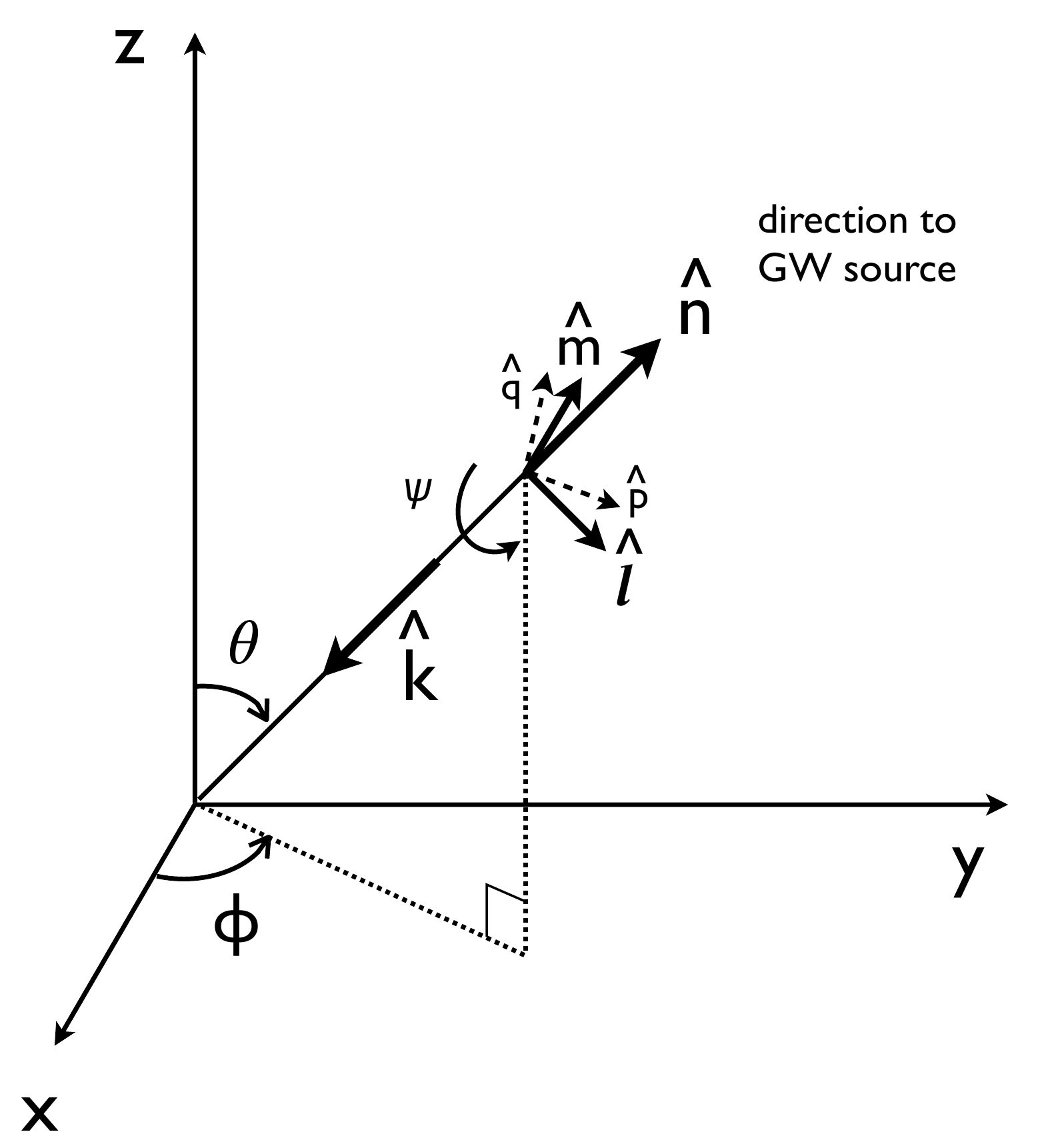}
\caption{Different choices for the unit vectors
perpendicular to $\hat n$.
By rotating the unit vectors $\hat l$, $\hat m$ 
by the angle $\psi$ in the plane orthogonal to 
$\hat n$, one obtains new unit vectors, 
$\hat p$, $\hat q$, in terms of which new 
polarization basis tensors, $\epsilon^+_{ab}(\hat n,\psi)$,
$\epsilon^\times_{ab}(\hat n,\psi)$, are defined.}
\label{f:nlm_convention_app}
\end{center}
\end{figure}

In the main text, we chose the $A=+,\times$ polarization
basis tensors to be
\be
\begin{aligned}
e_{ab}^+(\hat n) 
&=\hat l_a\hat l_b - \hat m_a\hat m_b\,,
\\
e_{ab}^\times(\hat n) 
&=\hat l_a\hat m_b + \hat m_a\hat l_b\,,
\end{aligned}
\ee
where $\hat n$ is the direction to the gravitational-wave
source, and $\hat l$, $\hat m$ are unit vectors 
tangent to the sphere:
\be
\begin{aligned}
\hat n
&=\sin\theta\cos\phi\,\hat x
+\sin\theta\sin\phi\,\hat y
+\cos\theta\,\hat z
\equiv \hat r\,,
\\
\hat l
&=\cos\theta\cos\phi\,\hat x
+\cos\theta\sin\phi\,\hat y
-\sin\theta\,\hat z 
\equiv \hat\theta\,,
\\
\hat m
&=-\sin\phi\,\hat x
+\cos\phi\,\hat y
\equiv \hat\phi\,.
\end{aligned}
\ee
This particular choice for the vectors $\hat l$, $\hat m$,
perpendicular 
to $\hat n$ is somewhat arbitrary, as
one can rotate these vectors by an angle $\psi$ in the plane
orthogonal to ${\hat n}$, preserving the triple
as a right-handed orthonormal triad.
(For a gravitational-wave source with a symmetry axis,
such as a binary system or rotating neutron star,
the angle $\psi$ can be interpreted as the
{\em polarization angle} of the source.)
See Figure~\ref{f:nlm_convention_app}.
Under such a rotation, $\hat l$ and $\hat m$
transform to new unit vectors
\be
\begin{aligned}
&\hat p
\equiv\cos\psi\,\hat l + \sin\psi\,\hat m\,,
\\
&\hat q
\equiv-\sin\psi\,\hat l + \cos\psi\,\hat m\,,
\end{aligned}
\ee
leading to new polarization tensors
\be
\begin{aligned}
\epsilon_{ab}^+(\hat n,\psi)
&\equiv
\hat p_a\hat p_b - \hat q_a\hat q_b\,,
\\
\epsilon_{ab}^\times(\hat n,\psi)
&\equiv
\hat p_a\hat q_b + \hat q_a\hat p_b\,.
\end{aligned}
\ee
The new polarization tensors are related to the original
ones via
\be
\begin{aligned}
\epsilon_{ab}^+({\hat n},\psi) 
&=\cos 2\psi\,e_{ab}^+({\hat n}) +
\sin 2\psi\,e_{ab}^\times({\hat n})\,,
\\
\epsilon_{ab}^\times({\hat n},\psi) 
&=-\sin 2\psi\,e_{ab}^+({\hat n}) +
\cos 2\psi\,e_{ab}^\times({\hat n})\,.
\end{aligned}
\ee
%

\subsection{Circular polarization}
\label{s:circular-pol-basis}

The form of the above transformation suggests a more
convenient basis of polarization tensors.
Namely, if we define the complex combinations
\be
\begin{aligned}
{e}_{ab}^R
&\equiv\frac{1}{\sqrt{2}}
\left(
e_{ab}^+ + i\,e_{ab}^\times
\right)\,,
\\
{e}_{ab}^L 
&\equiv
\frac{1}{\sqrt{2}}
\left(
e_{ab}^+ - i\,e_{ab}^\times
\right)\,,
\label{e:eReL}
\end{aligned}
\ee
or, equivalently,
\be
\begin{aligned}
{e}_{ab}^R
&\equiv\frac{1}{\sqrt{2}}(\hat l_a+i\hat m_a)(\hat l_b+i\hat m_b)
\,,
\\
{e}_{ab}^L 
&\equiv\frac{1}{\sqrt{2}}(\hat l_a-i\hat m_a)(\hat l_b-i\hat m_b)\,,
\end{aligned}
\ee
then under the above rotation by $\psi$,
\be
\begin{aligned}
\epsilon_{ab}^R({\hat n},\psi)
&=e^{-i 2\psi}\, {e}_{ab}^R({\hat n})\,,
\\
\epsilon_{ab}^L({\hat n},\psi)
&=e^{i 2\psi}\, {e}_{ab}^L({\hat n})\,.
\label{e:eReL_transformation}
\end{aligned}
\ee
The tensors $e^R_{ab}$, $e^L_{ab}$ correspond to
{\em right} and {\em left} circularly polarized waves when
looking down on the $\{\hat l,\hat m\}$ plane in the 
$-\hat n$-direction.
(The deformation ellipse for $e^R_{ab}$ would rotate 
to the right, i.e., clockwise, when viewed in this direction.)
The fact that the right and left circularly polarized
waves transform by a simple phase factor involving $2\psi$
is a manifestation of the spin-two nature of the graviton
\cite{Weinberg:1972}.
Indeed, one can show that the scalar field
$e^R_{ab}(\hat n)h^{ab}(f,\hat n)$ can be written as a
linear combination of spin-weight $+2$ spherical 
harmonics ${}_2Y_{lm}(\hat n)$, while
$e^L_{ab}(\hat n)h^{ab}(f,\hat n)$ can be written as
a linear combination of spin-weight $-2$ spherical 
harmonics ${}_{-2}Y_{lm}(\hat n)$.
(See Appendices~\ref{s:spinweightedY}, \ref{s:grad-curl-vector},
\ref{s:grad-curl-tensor} for more details regarding 
spin-weighted and vector and tensor spherical harmonics.)

The Fourier components $h_{ab}(f,\hat n)$ of the 
metric perturbations $h_{ab}(t,\vec x)$ can be expanded
in terms of either the linear polarization basis tensors:
\be
h_{ab}(f,\hat n) = 
h_+(f,\hat n)e^+_{ab}(\hat n)+
h_\times (f,\hat n)e^\times_{ab}(\hat n)\,,
\label{e:hab_linpol_expansion}
\ee
or the circular polarization basis tensors:
\be
h_{ab}(f,\hat n) = 
h_R(f,\hat n)e^R_{ab}(\hat n)+
h_L (f,\hat n)e^L_{ab}(\hat n)\,.
\label{e:hab_circpol_expansion}
\ee
The expansion coefficients $h_R$, $h_L$ are related to 
$h_+$, $h_\times$ via:
\be
\begin{aligned}
h_R
&= \frac{1}{\sqrt{2}}
\left( h_+ - i h_\times\right)\,,
\\
h_L
&= \frac{1}{\sqrt{2}}
\left( h_+ + i h_\times\right)\,.
\label{e:hRhL}
\end{aligned}
\ee
Note the sign change on the right-hand side of (\ref{e:hRhL})
compared to (\ref{e:eReL}).

\subsection{Polarization matrix and Stokes' parameters}
\label{s:stokes}

For a single monochromatic plane wave, the expansion
coefficients $h_+$, $h_\times$ or $h_R$, $h_L$ are (complex-valued) constants.
The polarization content of the plane wave is 
encoded in terms of the $2\times 2$ (Hermitian) polarization matrix 
\be
J_{BB'} \equiv h_B h_{B'}^*\,,
\ee
where $B$ labels either the linear polarization 
components $A\equiv\{+,\times\}$ or 
circular polarization components $C\equiv\{R,L\}$.
For linear polarization, the matrix elements have the form
\be
J_{AA'} = 
\frac{1}{2}\left[
\begin{array}{cc}
I+Q & U-iV\\
U+iV & I-Q\\
\end{array}
\right]\,,
\ee
where
$I$, $Q$, $U$, $V$ are the 
{\em Stokes' parameters} \cite{Jackson:1998}:
\be
\begin{aligned}
I&=|h_+|^2 + |h_\times|^2\,,
\\
Q&=|h_+|^2 - |h_\times|^2\,,
\\
U&=h_+h_\times^* + h_\times h_+^*\,,
\\
V&=i(h_+ h_\times^* - h_\times h_+^*)\,.
\end{aligned}
\label{e:Stokes_+x}
\ee
For circular polarization, we have 
\be
J_{CC'} = 
\frac{1}{2}\left[
\begin{array}{cc}
I+V & Q-iU\\
Q+iU & I-V\\
\end{array}
\right]\,,
\ee
where 
\be
\begin{aligned}
I&=|h_R|^2 + |h_L|^2\,,
\\
Q&=h_Rh_L^* + h_L h_R^*\,,
\\
U&=i(h_R h_L^* - h_L h_R^*)\,,
\\
V&=|h_R|^2 - |h_L|^2\,.
\end{aligned}
\label{e:Stokes_RL}
\ee
Note that $I$ is the total intensity of the wave,
$Q$ is a measure of linear polarization,
$|h_+|^2-|h_\times|^2$, and
$V$ is a measure of circular polarization,
$|h_R|^2 - |h_L|^2$.
Since a stochastic gravitational-wave background is a 
linear {\em superposition} of plane waves having 
different frequencies and coming from different 
directions on the sky, the matrix elements of $J$ 
will be replaced by quadractic 
{\em expectation values}, e.g., 
$\langle h_+(f,\hat n) h_\times^*(f',\hat n')\rangle$,
which will also depend on whether the background is 
stationary or anisotropic, etc.

Given the transformation properties 
(\ref{e:eReL_transformation}) of $e^R_{ab}$, $e^L_{ab}$, 
and the definition (\ref{e:hRhL}) of $h_R$, $h_L$, 
it follows that $h_R$, $h_L$ transform to 
\be
\begin{aligned}
\bar h_R &= e^{i2\psi} h_R\,,
\\
\bar h_L &= e^{-i2\psi} h_L\,,
\end{aligned}
\label{e:hRhL_transformation}
\ee
under a rotation of the basis vectors 
$\{\hat l,\hat m\}$ by $\psi$.
From these equations and expressions (\ref{e:Stokes_RL})
for the Stokes parameters, we can further show that
$I$, $Q$, $U$, $V$ transform to 
\be
\begin{aligned}
\bar I &= I\,,
\\
\bar V &= V\,,
\\
\bar Q+i\bar U &= e^{-i4\psi}(Q+iU)\,,
\\
\bar Q-i\bar U &= e^{i4\psi}(Q-iU)\,,
\end{aligned}
\label{e:Stokes_transformation}
\ee
under a rotation by $\psi$.
Thus, $I$ and $V$ are ordinary scalar 
(spin~0) functions on the sphere, while
$Q\pm iU$ are spin~4 fields, and 
can be written as linear combinations of 
spin-weight~$\pm 4$ spherical harmonics 
${}_{\pm 4}Y_{lm}(\hat n)$.
This has relevance for searches for circularly or 
linearly polarized stochastic backgrounds, as 
circular polarization, $V$, is present in the isotropic
component of the background, while 
linear polarization, $Q$, is not~\cite{Seto:2008}.

\section{Some standard results for Gaussian random variables}
\label{s:basics}

The statistical properties of a random variable $X$ are completely 
determined by its probability distribution $p_X(x)$.
The {\em moments} of the distribution 
$\langle X\rangle$, $\langle X^2\rangle$, $\langle X^3\rangle\,, \cdots$,
are defined by
\begin{equation}
\langle X^n\rangle \equiv \int_{-\infty}^\infty dx\, x^n p_X(x)\,.
\end{equation}
The first moment $\langle X\rangle$ is the expected 
(or mean) value of $X$, and is often denoted by $\mu$; 
the second moment is related to the variance $\sigma^2$
via the formula
$\langle X^2\rangle = \sigma^2 + \langle X\rangle^2$.
The {\em characteristic function} of the probability 
distribution is defined by the Fourier transform:
\begin{equation}
\varphi_X(t)\equiv\int_{-\infty}^\infty dx\,e^{itx}\,p_X(x)\,.
\label{e:characteristic}
\end{equation}
Note that by expanding the exponential
\begin{equation}
\varphi_X(t) = 
1 +it\langle X\rangle +\frac{i^2t^2}{2!}\langle X^2\rangle
+\cdots
\,.
\label{e:phi_expansion}
\end{equation}
This means that the moments $\langle X^n\rangle$ can 
be obtained by simply differentiating $\varphi_X(t)$:
\begin{equation}
\langle X^n\rangle = 
i^{-n}\left[\frac{d^n}{dt^n}\varphi_X(t)\right]\bigg|_{t=0}\,.
\end{equation}
If the moments are all finite and the expansion (\ref{e:phi_expansion})
is absolutely convergent near the origin, then
the probability distribution $p_X(x)$ is simply the inverse 
Fourier transform of $\varphi_X(t)$:
\begin{equation}
p_X(x) = \frac{1}{2\pi}\int_{-\infty}^\infty dt\,e^{-itx}\,\varphi_X(t)\,.
\end{equation}
A similar result can be obtained for a {\em one-sided} 
probability distribution $p_X(x)$ (e.g., defined only for $x\ge 0$)
by working with Laplace transformations instead.

If $X$ is a {\em Gaussian} random variable, then 
\begin{equation}
p_X(x) = \frac{1}{\sqrt{2\pi}\sigma}\,
e^{-\frac{1}{2}\frac{(x-\mu)^2}{\sigma^2}}\,.
\label{e:gaussian}
\end{equation}
The parameters $\mu$ and $\sigma^2$ are just the mean and variance
of $X$:
\begin{equation}
\mu=\langle X\rangle\,,
\qquad
\sigma^2 =\langle X^2\rangle -\langle X\rangle^2\,.
\end{equation}
A nice property of Gaussian distributions is that all
third and higher-order moments can be expressed as a sum
of products of the first two moments.
For example, for a single Gaussian random variable $X$,
\be
\begin{aligned}
\langle X^3\rangle 
&=
3 \langle X\rangle \langle X^2\rangle 
-2 \langle X\rangle^3\,,
\\
\langle X^4\rangle 
&=
4\langle X\rangle \langle X^3\rangle
+3\langle X^2\rangle^2 
-12\langle X\rangle^2 \langle X^2\rangle
+6\langle X\rangle^4\,,
\\
&\vdots 
\end{aligned}
\ee
More generally, for $n\ge 3$ 
these relations can be obtained by solving the
equations
\begin{equation}
\left[\frac{d^n}{dt^n}\ln\varphi_X(t)\right]\bigg|_{t=0} = 0
\label{e:exp-general}
\end{equation}
for $\langle X^n\rangle$, 
where $\varphi_X(t)$ is given by the right-hand side of (\ref{e:phi_expansion}).
The fact that the derivatives are actually equal to zero 
follows from the specific
form for the characteristic function for a Gaussian distribution:
\begin{equation}
\varphi_X(t)=\exp\left[i\mu t-\frac{\sigma^2 t^2}{2}\right]\,.
\end{equation}
Since $\ln\varphi_X(t)$ is quadratic in $t$, all third and
higher-order derivatives vanish.

A {\em multivariate Gaussian} distribution is a 
generalization of (\ref{e:gaussian}) to a set of random variable 
$\mb{X}\equiv\{X_1, X_2,\cdots, X_N\}$.
The joint probability density function is given by
\begin{equation}
p_{\mb{X}}(x_1,x_2,\cdots, x_N) = 
\frac{1}{\sqrt{\det(2\pi C)}}\,
e^{-\frac{1}{2}\sum_{i,j}
(x_i-\mu_i) \left(C^{-1}\right)_{ij} (x_j-\mu_j)}\,,
\end{equation}
where $\mu_i=\langle X_i\rangle$ are the mean values, and 
\begin{equation}
C_{ij}=\langle X_i X_j\rangle - \langle X_i\rangle \langle X_j\rangle
\end{equation}
are the elements of the {\em covariance matrix} $C$.
For a zero-mean multivariate Gaussian distribution, all
of the odd-ordered moments are identically zero.
In addition,
\begin{equation}
\langle X_1 X_2 X_3 X_4\rangle =
\langle X_1 X_2\rangle\langle X_3 X_4\rangle+
\langle X_1 X_3\rangle\langle X_2 X_4\rangle+
\langle X_1 X_4\rangle\langle X_2 X_3\rangle\,.
\label{e:<abcd>}
\end{equation}
We will use several of the above results repeatedly 
throughout the main text, as most of the probability 
distributions that we work with are multivariate-Gaussian.

\section{Definitions and tests for stationarity and Gaussianity}
\label{s:real}

Here we provide definitions of what it means for data to be stationary
and Gaussian, and highlight some tests for these properties.
Ascertaining whether or not data are stationary and Gaussian can be
challenging as the tests rely on comparison with alternative models,
and some models are better at picking up certain forms of
non-stationarity and non-Gaussianity than others.

\subsection{Definition of stationarity}

A stationary stochastic process has statistical properties that do not
depend on time: the joint statistical properties of the sample
$\{x_{t_1},\dots,x_{t_k}\}$ are identical to the joint statistical
properties of the sample $\{x_{t_1+\tau},\dots,x_{t_k+\tau}\}$ for all
$\tau$ and $k$. In particular, the joint distribution of $(x_t,x_s)$
depends only on the lag $|t-s|$, and not on $t$ or $s$, and all
higher-order moments are strictly independent of time. A less
restrictive, and more practical notion, is that of {\em weak or
second-order stationarity}, which asserts that the mean and variance
are constant, and that the auto-covariance ${\rm cov}(x_t,x_{t+\tau})$
depends only on the lag $\tau$.

\subsection{Definition of Gaussianity}

A continuous random variable $X$ is said to be a Gaussian, or {\em
normal}, random variable $X\sim N(\mu,\sigma^2)$ if its probability density
function is given by
\begin{equation}
p_X(x) = \frac{1}{\sqrt{2\pi} \sigma} e^{-\frac{1}{2}\frac{(x-\mu)^2}{\sigma^2}} \, .
\end{equation}
The multivariate generalization to a collection of continuous random
variables ${\mb X}\equiv \{X_1,X_2,\cdots,X_N\}$
is given in terms of a Gaussian probability density
function with covariance matrix $C$:
\begin{equation}
p_{\mb X}(x_1,x_2,\cdots,x_N) 
= \frac{1}{\sqrt{{\rm det}(2\pi C)} }e^{-\frac{1}{2} 
\sum_{i,j} (x_i -\mu_i)\left(C^{-1}\right)_{ij} (x_j-\mu_j)} \, .
\end{equation}
See Appendix~\ref{s:basics} for additional statistical properties of
Gaussian random variables.

\subsection{Tests for stationarity}

There exists a vast literature on tests of non-stationarity of 
time-series data. The simplest tests for non-stationarity are qualitative
in nature and involve looking at plots of the mean, variance, and
auto-correlation as a function of time (for example, by using a sliding
window of some duration to select the samples used to compute these
quantities).  The difficulty with this approach is deciding on what
constitutes acceptable levels of variation. The concept of
time-varying correlations and time-varying spectral densities are well
defined and useful concepts for {\em locally-stationary}
processes~\cite{2011arXiv1109.4174D}, but less so for other forms
of non-stationarity (Section~\ref{s:localstationarity}).

It is unclear whether many of the more powerful quantitative tests for
non-stationarity are useful for gravitational-wave data analysis.  For
example, commonly used tests, such as the augmented Dickey-Fuller test
and the Phillips-Perron test, which test to see if the data follow a
``unit root'' auto-regressive process, do not appear to be
particularly applicable since the noise encountered in
gravitational-wave experiments usually exhibits high auto-correlation,
and thus has roots that are naturally close to unity, which poses a
challenge for these tests~\cite{Muller2005195}.

The most useful tests, at least for evenly-sampled gravitational-wave
data, are those based on evolutionary spectral estimates, or
correlations in the Fourier coefficients.  The Priestley-Subba~Rao
test~\cite{10.2307/2984336}, and modern variants based on
wavelets~\cite{RSSB:RSSB231}, use window functions to compute spectral
estimates as a function of time. A statistical test is then used to
assess if the spectral estimates are consistent with stationarity. The
second type of test is based on the fact that second-order stationary
time series produce uncorrelated Fourier series (which is why most
gravitational-wave analyses are performed in the Fourier domain).
Statistical tests can be used to decide whether the the level of
correlation between Fourier coefficients indicates that the data are
non-stationary~\cite{2009arXiv0911.4744D}.

\subsection{Tests for Gaussianity}

There are a large number of tests for Gaussianity described in the
literature that are in regular use. These tests are based on different
properties of the Gaussian distribution, and the power of the tests
differ depending on the nature of the non-Gaussianity.

Three of the most widely used frequentist tests are the Shapiro--Wilk
test, the Anderson--Darling test, and the Lilliefors test (a modified
Kolmogorov--Smirnov test).  The Shapiro--Wilk test is a regression
test that out-performs other tests on small data sets, but is
challenging to apply to the large data sets encountered in gravitational-wave data
analysis. Both the Anderson--Darling and the Lilliefors test are based
on the distance between the hypothesized cumulative distribution
function (in this case, that of a Gaussian distribution) and the
cumulative distribution function of the data. The Anderson--Darling
test performs almost as well, and sometimes better, than the
Shapiro--Wilk test~\cite{Seier2011}, and can be used on large data sets.

Bayesian tests for Gaussianity can be performed by computing the Bayes
factors between competing models for the data, in this case the
Gaussian distribution and some more general alternative such as 
Student's $t$-distribution~\cite{SPIEGELHALTER01011980, kruschke2013bayesian}. 
This approach has been applied to gravitational-wave data
analysis~\cite{Littenberg:2010gf, Cornish-Romano:2015}.

\section{Discretely-sampled data}
\label{s:discretely-sampled-data}

In this appendix, we describe the relationship between 
continuous functions of time and frequency (used 
throughout most of the article) and their 
discretely-sampled counterparts.
This is needed to cast a theoretical analysis into one 
that can be run on a digital computer, which naturally 
works with a finite number of discrete samples.
Although we will focus attention on topics that are 
most-relevant to searches for stochastic 
gravitational-wave backgrounds, much of what we say 
here is general and
relevant to many other signal processing applications.
We refer interested readers to 
e.g., \cite{Oppenheim-Schafer:1999, Press:1992, Gregory:2005} 
for more thorough discussions of these topics.

\subsection{Discretely-sampled time-series}
\label{s:discrete-time-series}

In the majority of the text, we represented the output of a 
detector by a time-series, e.g., $x(t)$, which was a function 
of a {\em continuous} time parameter $t$.
Usually, the range of $t$ was infinite 
(from $-\infty$ to $\infty$), 
although sometimes we would restrict attention to a finite
duration, $t\in[t_0,t_0+T]$, where $t_0$ was some
initial time (usually $t_0=0$), and $T$ was the length 
of an analysis segment or the total duration of an observation.
The Fourier transform of $x(t)$ (assumed to be defined for all 
$t$) was defined as
\begin{equation}
\tilde x(f)\equiv \int_{-\infty}^\infty dt\> 
x(t)\,e^{-i 2\pi f t}\,,
\label{e:FT}
\end{equation}
with inverse Fourier transform
\begin{equation}
x(t)=\int_{-\infty}^\infty df\> 
\tilde x(f)\,e^{i 2\pi f t}\,.
\end{equation}
The signal power in the frequency band $f$ to $f+df$ is 
proportional to $|\tilde x(f)|^2\,df$.

In practice, any real time-series will be 
{\em discretely-sampled}.
This means that a continuous function of time $x(t)$ 
will be represented by a set of discrete values
\begin{equation}
x_k\equiv x(t_k)\,,
\end{equation}
where
\begin{equation}
t_k=t_0+k\Delta t\,,
\quad\quad k=0,\pm 1,\pm 2,\cdots\,.
\end{equation}
For now we allow $k$ to take on an infinite set
of values, corresponding to time-series having
an infinite duration; 
shortly, we will restrict attention to a 
discretely-sampled time-series having a {\em finite} duration.
Here we have assumed regularly-sampled data 
(i.e., the time interval between adjacent samples $x_k$ and $x_{k+1}$ 
is a constant $\Delta t$), although for some cases
(e.g., pulsar timing)
the data samples $x_k$ will correspond to 
irregularly-spaced times.
Although it is more difficult to compute power spectra for 
irregularly-spaced time series, there do exist algorithms---like 
the Lomb-Scargle algorithm~\cite{Lomb:1976, Scargle:1982}---which 
can be used for this purpose.
Also, to simplify the analysis slightly in what follows, we will 
set $t_0=0$.

A convenient way of representing a discretely-sampled 
time-series is to multiply the continuous 
function $x(t)$ by an infinite sum of Dirac delta functions,
called the {\em Dirac comb}:
\begin{equation}
\Delta_{\Delta t}(t)\equiv
\sum_{k=-\infty}^\infty \delta(t-k\Delta t)\,.
\end{equation}
The function
\begin{equation}
x_d(t)
\equiv\Delta t\,\sum_{k=-\infty}^\infty x_k\,\delta(t-k\Delta t)
\end{equation}
is a continuous time-series representation of the 
discretely-sampled data $x_k=x(k\Delta t)$.
(The multiplicative factor $\Delta t$ is included
so that $x_d(t)$ has the same dimensions as $x(t)$.)
Using the above expression, it immediately follows that
the Fourier transform of $x_d(t)$ is
\begin{equation}
\tilde x_d(f)
=\Delta t\sum_{k=-\infty}^\infty 
x_k\, e^{-i2\pi fk\Delta t}\,.
\label{e:tilde_x_d}
\end{equation}
Alternatively, since $x_d(t)$ is a product of two 
functions in the time domain, its Fourier transform
is the {\em convolution} of the Fourier transforms
$\tilde x(f)$ and $\widetilde\Delta_{\Delta t}(f)$ in the
frequency domain:
\begin{equation}
\tilde x_d(f)=\Delta t\int_{-\infty}^\infty
df'\>
\tilde x(f-f')\widetilde\Delta_{\Delta t}(f')\,.
\end{equation}
But since the Fourier transform of the Dirac comb is another
Dirac comb,
\begin{equation}
\widetilde\Delta_{\Delta t}(f')=\frac{1}{\Delta t}
\sum_{k=-\infty}^\infty \delta\left(f'-k/\Delta t\right)\,,
\end{equation}
it follows that 
\begin{equation}
\tilde x_d(f) =\sum_{k=-\infty}^\infty \tilde x(f-k/\Delta t)\,.
\label{e:aliasing}
\end{equation}
This is the relation between the Fourier transforms of 
the continuous and discretely-sampled time-series.
Note that $\tilde x_d(f)$ is periodic in $f$ with period 
$1/\Delta t$.

One can interpret (\ref{e:aliasing}) as follows:
If $x(t)$ and $\Delta t$ are 
such that $\tilde x(f)=0$
outside $[-f_{\rm N}, f_{\rm N}]$,
where $f_{\rm N}\equiv1/(2\Delta t)$ is the {\em Nyquist}
critical frequency, then the Fourier transform of the 
discretely-sampled data is {\em identical} to that of 
the original continuous time-series for 
$f\in[-f_{\rm N}, f_{\rm N}]$.
Otherwise, there is {\em aliasing} of power from outside 
the Nyquist band
making $|\tilde x_d(f)|>|\tilde x(f)|$ for 
$f\in[-f_{\rm N}, f_{\rm N}]$.
In other words, $\tilde x_d(f)=\tilde x(f)$ for 
$f\in[-f_{\rm N}, f_{\rm N}]$ if and only if
the following two conditions hold:
\begin{enumerate}[(i)]
\item $x(t)$ is {\em band-limited}---i.e., 
$\tilde x(f)=0$ for $|f|\ge f_{\rm max}$, 
where $f_{\rm max}$ is some finite frequency,

\item 
the sampling rate $1/\Delta t$ is sufficiently
large that $f_{\rm max}<f_{\rm N}$, or, 
equivalently, $\Delta t < 1/(2 f_{\rm max})$.

\end{enumerate}
For the special case where $x(t)$ happens to be periodic, 
the condition for no aliasing is to sample at least twice 
per period.

For a band-limited signal sampled so that $f_{\rm max}<f_{\rm N}$,
we can recover the continuous time-series $x(t)$ from the
discrete samples $x_k$.
The explicit reconstruction formula is obtained by taking
the inverse Fourier transform of $\tilde x(f)$, replacing 
$\tilde x(f)$ by $\tilde x_d(f)$ for $f\in[-f_{\rm N},f_{\rm N}]$,
and then using (\ref{e:tilde_x_d}) to get an expression 
involving $x_k$.
The final result is
\begin{equation}
x(t) = \sum_{k=-\infty}^\infty
x_k\, \mr{sinc}\left[\pi(t-k\Delta t)/\Delta t\right]\,,
\end{equation}
where $\mr{sinc}\,(x)\equiv \sin(x)/x$.
Note that the reconstruction formula involves a sum over
an {\em infinite} set of $x_k$.  
This is a consequence of $x(t)$ being band-limited, since a 
function of compact support in the frequency domain must 
have infinite support in the time domain.
 
\subsection{Windowing}
\label{s:windowing}

\begin{figure}[htbp!]
\begin{center}
\subfigure[]{\includegraphics[angle=0,width=0.49\textwidth]{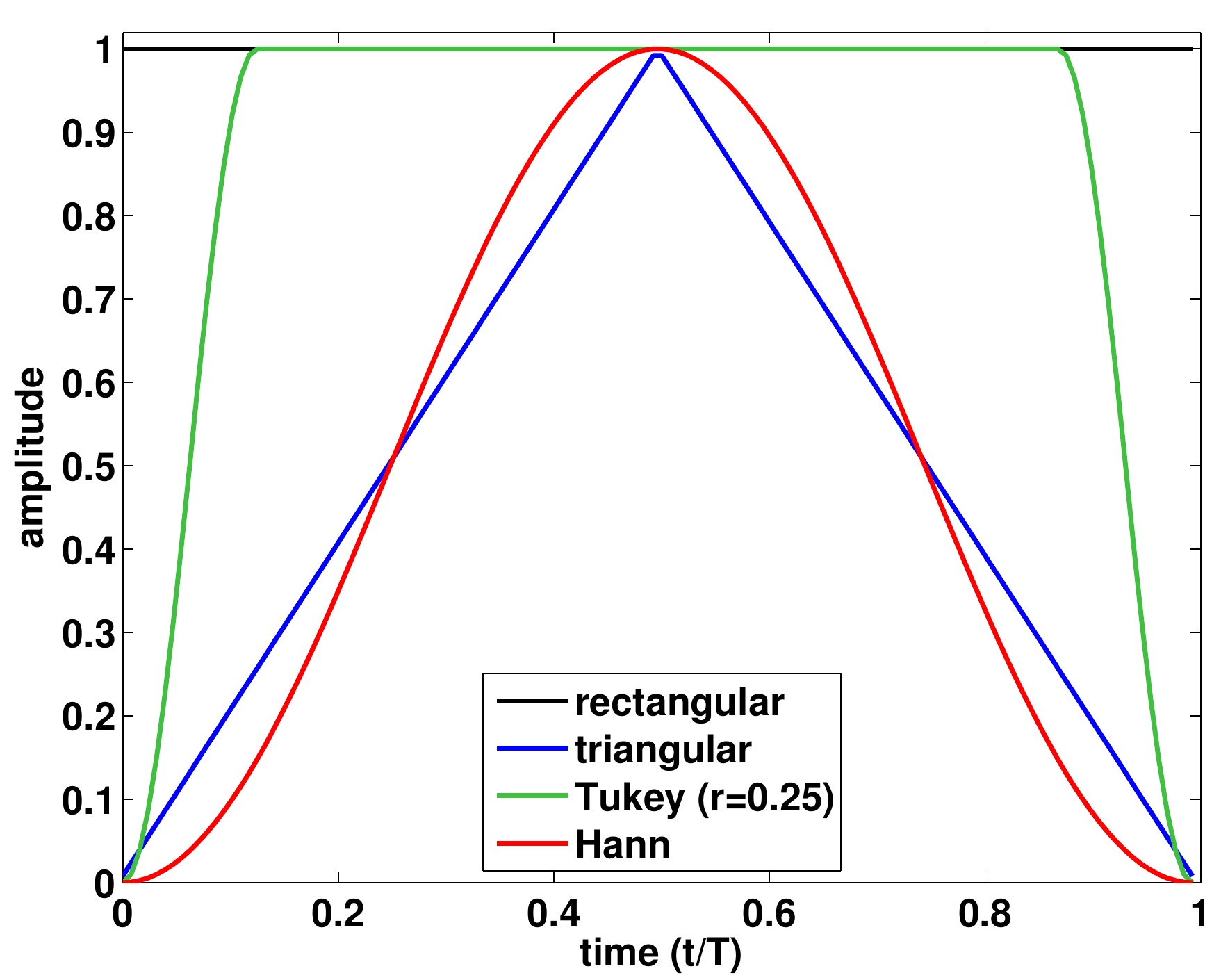}}
\subfigure[]{\includegraphics[angle=0,width=0.49\textwidth]{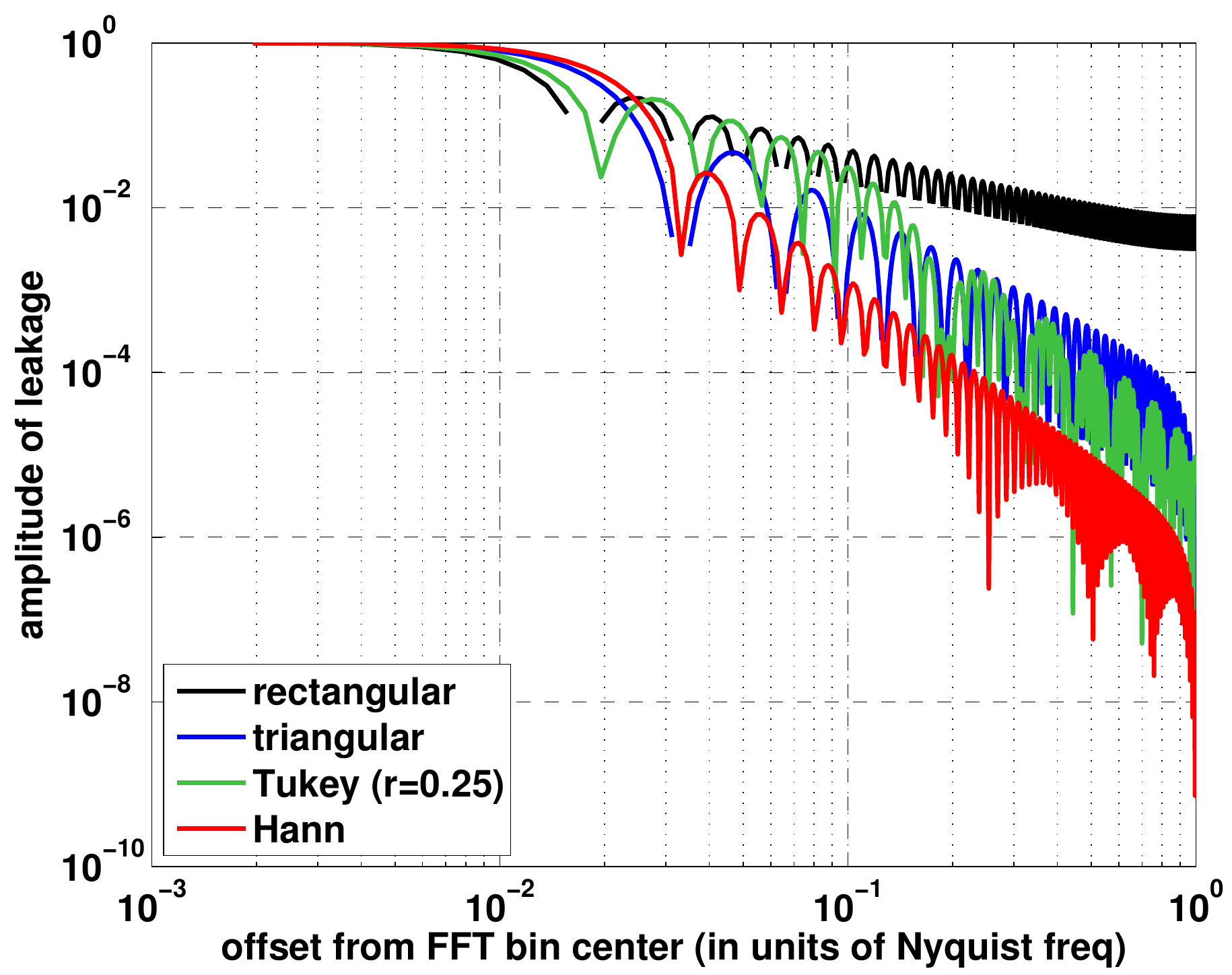}}
\caption{
Panel (a): Plots of window functions defined in the main text.
Panel (b): Amplitude of spectral leakage for different windows
as a function of the offset from the bin center in units of the
Nyquist frequency $f_{\rm N}\equiv 1/(2\Delta t)$.
The dips arise because we are using a finite number of samples 
($N=128$) to represent the windows.}
\label{f:windows-leakage}
\end{center}
\end{figure}

In addition to being discretely-sampled, real-world signals
are non-zero for only a finite duration $T$.
Mathematically, the simplest way to do this is to multiply 
an infinite-duration time-series $x(t)$ by a {\em rectangular} 
(or {\em top-hat}) function
\begin{equation}
w(t)\equiv\bigg\{
\begin{array}{cc}
1 & 0\le t\le T\\
0 & \mr{otherwise}
\end{array}\,,
\label{e:rectangular}
\end{equation}
which simply sets $x(t)$ to zero outside the interval $[0,T]$.
The rectangular function is a special case of a more
general class of so-called {\em window} functions 
(or {\em tapers}), all of which set the signal to zero outside 
the interval $[0,T]$.
Other examples include:
\medskip

\noindent
\underline{Triangular window}:
\be
w(t)\equiv 1-|2t/T-1|\,,
\label{e:triangularwindow}
\ee
\underline{Tukey window}:
\be
w(t)\equiv
\left\{
\begin{array}{ll}
\frac{1}{2}
\left[1-\cos\left(\frac{2\pi}{r}\frac{t}{T}\right)\right]\,,
& 0\le \frac{t}{T} \le \frac{r}{2}
\\
1\,,
&
\frac{r}{2} \le \frac{t}{T} \le 1 - \frac{r}{2}
\\
\frac{1}{2}
\left[1+\cos\left(\frac{2\pi}{r}\left[\frac{t}{T}-\left(1-\frac{r}{2}\right)\right]\right)\right]\,,
& 1-\frac{r}{2} \le \frac{t}{T} \le 1
\\
\end{array}
\right.\,,
\label{e:tukeywindow}
\ee
\underline{Hann window}:
\be\
w(t)\equiv
\frac{1}{2}(1-\cos(2\pi t/T))\,.
\label{e:hannwindow}
\ee
All of these windows taper the signal so that it ``ramps-up" and 
``ramps-down" at the start and end of the interval.
(See Figure~\ref{f:windows-leakage}, panel (a).)
For example, the Tukey window (\ref{e:tukeywindow}) is defined
by a parameter $r$, which specifies the fraction of time that 
the window ramps-up to unity and then back down to zero, with 
a cosine-like taper.
For $r=1$, the Tukey window becomes a Hann window (\ref{e:hannwindow}).
Several other common window functions are also used in
signal processing applications;
see e.g., \cite{Oppenheim-Schafer:1999, Press:1992}
for more details. 

Given a time-series $x(t)$ and a choice of window function 
$w(t)$, we define the {\em windowed time-series} by
\begin{equation}
x_w(t)\equiv w(t)x(t)\,.
\end{equation}
Since $x_w(t)$ is just a product of two functions in the 
time domain, its Fourier transform $\tilde x(f)$ 
is the convolution of the Fourier transforms 
$\tilde x(f)$ and $\tilde w(f)$ in the frequency domain:
\begin{equation}
\tilde x_w(f)=\int_{-\infty}^\infty df'\>
\tilde x(f-f')\tilde w(f')\,.
\end{equation}
Since $w(t)$ has compact support in the time domain,
$\tilde w(f)$ has infinite support in the frequency 
domain, meaning
that the power in the 
windowed time-series $\tilde x_w(f)$
will contain power in $\tilde x(f')$ 
from frequencies $f'~(\ne f)$ as well.
This {\em smearing} or {\em leakage} of power 
exists for {\em any} type of window, 
although the extent of the leakage depends on the 
shape of the window as shown in 
Figure~\ref{f:windows-leakage}, panel (b).%
\footnote{The normalized leakage of a window $w(t)$ 
is defined as $|\tilde w(f)|/|\tilde w(0)|$.}
The rectangular window has the largest spectral leakage 
of all the windows, while the Hann window has the smallest 
leakage (several orders of magnitude suppression) for 
large frequency offsets, due to its smooth turn-on and 
turn-off.
In general, there is a trade-off between spectral leakage
and the loss of time-domain data due to the windowing.  
The Tukey window provides a nice balance in that 
spectral leakage can been strongly suppressed while only 
affecting a small fraction of the time domain samples. 
If one needs greater suppression but cannot
afford to lose more data, one can use Hann windows that 
overlap by 50\% (see e.g., \cite{Lazz:2004} in the context
of stochastic background searches using LIGO data).

Windowing can also be applied to a discretely-sampled
time-series, leading to a time-series which
is represented by a {\em finite} number of discrete
samples $x_k$, where $k=0,1,\cdots N-1$ and $T=N\Delta t$.
Similar to what we saw in the previous subsection, 
this finite, 
discretely-sampled time-series can be conveniently 
represented by a continuous time-series by multiplying
by the Dirac comb.
Explicitly,
\begin{equation}
x_{dw}(t) 
=\Delta t\sum_{k=0}^{N-1}w_k x_k \,\delta(t-k\Delta t)\,,
\end{equation}
where $x_k=x(k\Delta t)$ and $w_k=w(k\Delta t)$.
Note that this function can also be written as
\begin{equation}
x_{dw}(t) = w(t)x_d(t) = w_d(t)x(t)\,,
\end{equation}
from which it immediately follows that
\begin{equation}
\tilde x_{dw}(f)
=\int_{-\infty}^\infty df'\>
\tilde x_d(f-f')\tilde w(f')
=\int_{-\infty}^\infty df'\>
\tilde x(f-f')\tilde w_d(f')
\label{e:tilde_x_dw}\,.
\end{equation}

As a specific example, let $w(t)$ be the rectangular
window defined by (\ref{e:rectangular}).
Then it is easy to show that
\begin{equation}
\tilde w(f) = e^{-i\pi fT}\,T\mr{sinc}(\pi fT)\,.
\end{equation}
In addition, one can show that the discretized-version 
of the rectangular window has Fourier transform%
\footnote{If the rectangular window is defined to be 
non-zero for $t\in[-T/2,T/2]$ instead of $[0,T]$, then
$\tilde w_d(f)=T\mc{D}_N(f\Delta t)$, which does
not include the phase factor on the right-hand side of 
(\ref{e:tilde_w_d}).}
\begin{equation}
\tilde w_d(f) 
=\Delta t\sum_{k=0}^{N-1} e^{-i2\pi f k\Delta t}
=e^{-i\pi(N-1)f\Delta t}\,T\mc{D}_N(f\Delta t)\,,
\label{e:tilde_w_d}
\end{equation}
where
\begin{equation}
\mc{D}_N(x) 
\equiv\frac{1}{N}\frac{\sin(N\pi x)}{\sin(\pi x)}
=\sum_{k=-\infty}^\infty
\mr{sinc}[\pi(x-k)N]
\label{e:dirichlet}
\end{equation}
is the {\em Dirichlet kernel}~\cite{Percival-Walden:1993}.
Hence, for a rectangular window, (\ref{e:tilde_x_dw})
has the explicit form
\begin{equation}
\tilde x_{dw}(f)=\int_{-\infty}^\infty df'\ 
\tilde{x}(f-f')
e^{-i\pi(N-1) f'\Delta t}\,
T\mc{D}_N(f'\Delta t)\,,
\label{e:tilde_x_dw_explicit}
\end{equation}
which relates the Fourier transform of the infinite-duration, 
continuous time-series $x(t)$ to the Fourier transform of the 
finite-duration, discretely-sampled time-series $x_{dw}(t)$.

\subsection{Discrete Fourier transform}
\label{s:DFT}

Just as any real-world signal processing algorithm must
deal with finite-duration, discretely-sampled time-series data
$x_k$, where $k=0,1,\cdots, N-1$,
so too must frequency-series (like $\tilde x_{dw}(f)$)
be represented by a finite set of discrete values.
From our earlier discussion (Section~\ref{s:discrete-time-series})
about aliasing, we know that the Nyquist frequency, 
$f_{\rm N}=1/(2\Delta t)$, is the maximum frequency of 
a band-limited signal that can be faithfully represented 
with discrete samples $x_k$ taken with sampling period 
$\Delta t$.
In addition, the frequency resolution $\Delta f$ of 
the Fourier transform of a finite-duration signal
is limited to $\Delta f\equiv 1/T$, where $T$ is the 
total duration of the signal,
since it is meaningless to talk about the Fourier 
components corresponding to periods
greater than the total observation time.
Thus, the best we can do in practice is 
to evaluate the Fourier transform 
of the finite set of discretely-sampled time-series data
$x_k$ at the discrete frequencies
\begin{equation}
f_j
\equiv j\Delta f
=\frac{j}{N\Delta t}\,,
\qquad
j=-N/2,-N/2+1,\cdots, N/2-1\,,
\end{equation}
which lie in the frequency band $[-f_{\rm N},f_{\rm N} - \Delta f]$.
If $N$ is odd, the index $j$ runs from $-(N-1)/2$ 
to $(N-1)/2$.
(In what follows, we will assume that $N$ is even.)

The {\em discrete Fourier transform} (DFT) of $x_k$, 
where $k=0,1,\cdots,N-1$, is defined to be 
\be
\mr{DFT}(x_j)
\equiv \sum_{k=0}^{N-1}x_k\,e^{-i2\pi jk/N}\,,
\qquad j=-N/2,-N/2+1,\cdots, N/2-1\,.
\ee
Note that the kernel of this transformation
\be
U_{jk} \equiv \frac{1}{\sqrt{N}}e^{-i2\pi jk/N}\,,
\ee
is a {\em unitary} matrix, and thus satisfies
\be
U^{-1} = U^\dagger\,,
\qquad |{\rm det}(U)| =1\,,
\ee
as a consequence of the identity
\begin{equation}
\frac{1}{N}\sum_{l=0}^{N-1} e^{-i 2\pi(j-k)l/N}=\delta_{jk}\,.
\end{equation}
The inverse transformation (from the $\mr{DFT}(x_j)$ back to $x_k$) 
is thus
\begin{equation}
x_k 
=\frac{1}{N}\sum_{j=-N/2}^{N/2-1} \mr{DFT}(x_j)\, e^{2\pi jk/N}\,,
\qquad k=0,1,\cdots,N-1\,,
\end{equation}
Using the above results, one can also show that
\be
\sum_{k=0}^{N-1} |x_k|^2 = \frac{1}{N}\sum_{j=-N/2}^{N/2-1}|{\rm DFT}(x_j)|^2\,,
\label{e:parseval-discrete}
\ee
which is called {\em Parseval's theorem}.
Parseval's theorem is what tells us that the total power in a 
signal is the same when calculated in either the time domain or 
the frequency domain.
(More on this below.)

\subsection{DFTs and discretely-sampled Fourier transforms}
\label{s:DFTs-FTs}

To make the connection between the DFT of a set of discrete 
samples $x_k = x(k\Delta t)$ and the Fourier transform 
$\tilde x(f)$ of the underlying continuous time-series $x(t)$, 
we first define
\be
\tilde x_j \equiv \Delta t\, \mr{DFT}(x_j)\,.
\ee
The factor of $\Delta t$ gives $\tilde x_j$ and $\tilde x(f)$
the same units.
Using (\ref{e:tilde_x_d}), one can show that
\begin{equation}
\tilde x_j=\tilde x_{dw}(f_j)\,,
\label{e:exact_relation}
\end{equation}
where the window function $w(t)$ entering the definition of
$\tilde x_{dw}(f)$ is the (trivial)
rectangular window on $[0,T]$.
Thus, up to a factor of $\Delta t$, the DFT of a finite
set of discretely-sampled data is just the the Fourier 
transform of the discretized, rectangular-windowed data 
evaluated at the discrete frequencies $f_j$.

An explicit relation between $\tilde x_j$ and the Fourier
transform $\tilde x(f)$ of the infinite-duration, 
continuous time-series $x(t)$ is more complicated than 
(\ref{e:exact_relation}), due to the leakage of power 
from $\tilde x(f)$ into $\tilde x_{dw}(f)$, as discussed 
in the previous section.
From (\ref{e:tilde_x_dw_explicit}) it follows that 
\begin{equation}
\tilde x_j=\int_{-\infty}^\infty df'\ 
\tilde{x}(f_j-f')
e^{-i\pi(N-1) f'\Delta t}\,
T\mc{D}_N(f'\Delta t)\,.
\end{equation}
But since
$T\mc{D}_N(f'\Delta t)$ is typically 
well-approximated by the Dirac delta function $\delta(f')$, 
we have the {\em approximate} relation
\begin{equation}
\tilde x_j\simeq \tilde x(f_j)\,.
\label{e:approx_DFT_FT}
\end{equation}
Finally, note that if the $x_k$ are real, as they will be 
if they are discrete samples of a real-valued time-series $x(t)$,
then
\be
\tilde x_{-j} = \tilde x_j^*\,,
\quad{\rm for}\quad 
j=0,1,\cdots,N/2-1\,.
\ee
So no information is lost if we restrict attention to non-negative
frequencies
\be
f_j = j\Delta f =  \frac{j}{N\Delta t}\,,
\quad{\rm\  where}\quad
j=0,1,\cdots, N/2-1\,.
\ee

\subsection{Discrete power spectra}
\label{s:discretepowerspectra}

Suppose we are given $N$ samples 
$n_k$, $k=0,1,\cdots, N-1$, of a real-valued, stationary 
random process, e.g., detector noise or a stochastic signal.
Then we define its discrete power spectrum as
\be
{S_n}_j\equiv \frac{2}{T}|\tilde n_j|^2\,,
\qquad j=0,1,\cdots, N/2-1\,,
\label{e:Pj}
\ee
where $\tilde n_j\equiv \Delta t\,\mr{DFT}(n_j)$.
The factor of $2$ has been included to make it a
{\em one-sided} power spectrum, for which Parseval's
theorem (\ref{e:parseval-discrete}) takes the form:
\be
\sum_{j=0}^{N/2-1} \Delta f\, {S_n}_j 
=\frac{1}{N}\sum_{k=0}^{N-1}\, |n_k|^2\,.
\label{e:parseval-discrete-1sided}
\ee
Using the approximate relation (\ref{e:approx_DFT_FT}), it
follows that ${S_n}_j\simeq {S_n}(f_j)$, where $S_n(f)$ is the 
power spectrum of the underlying continuous time series 
$n(t)$.
With this correspondence, we see that 
(\ref{e:parseval-discrete-1sided}) is the discretized
version of
\be
\int_0^{f_{\rm N}}df\> S_n(f) 
= \frac{1}{T}\int_{0}^{T} dt\> |n(t)|^2\,,
\label{e:parseval-continuous}
\ee
which is the continuous version of Parseval's theorem.
Similarly, the expectation values 
\be
\langle \tilde n(f)\tilde n^*(f')\rangle
=\frac{1}{2}\delta(f-f')\,S_n(f)
\ee
for the continuous functions become
\be
\langle \tilde n_j \tilde n_{j'}^*\rangle
\simeq \frac{T}{2}\delta_{jj'}{S_n}_{j'}\,,
\ee
where we used
\be
\delta(f_j-f_{j'}) = \delta((j-j')\Delta f) = \frac{1}{\Delta f}\delta_{jj'}
=T\delta_{jj'}\,.
\ee
%

\subsection{Discrete and continuous probability distributions}
\label{s:discrete-continuous-probability}

Suppose further that the $N$ samples $n_k$, $k=0,1,\cdots,N-1$,
are Gaussian distributed with zero mean and covariance matrix
\be
(C_n)_{kk'} \equiv \langle n_k n_{k'}\rangle\,.
\ee
Then the probability distribution for 
$n\equiv (n_0,n_1,\cdots, n_{N-1})^T$ is
\be
p(n) = \frac{1}{\sqrt{\det(2\pi C_n)}}
\exp\left[-\frac{1}{2} n^\dagger C_n^{-1} n\right]\,,
\ee
with volume element
\be
d^Nn = \prod_{k=0}^{N-1} dn_k\,.
\ee
Using the above results, one can show that in the limit
of large $N$, the DFT (approximately) diagonalizes the covariance 
matrix $C_n$:
\be
U C_n U^{-1} 
\simeq
\frac{1}{2\Delta t}{\rm diag}({S_n}_{k})\,.
\ee
From this, one can then show that the probability 
distribution for the discrete frequency components 
$\tilde{n}\equiv \left(\tilde n_0, \tilde n_1,\cdots, \tilde n_{N/2-1}\right)^T$ 
is given by
\be
p(\tilde{n}) \simeq 
\prod_{j=0}^{N/2-1}
\frac{2}{\pi T {S_n}_{j}}
\exp\left[-\frac{2|\tilde n_j|^2}{T {S_n}_{j}}\right]
=
\prod_{j=0}^{N/2-1}
\frac{2}{\pi T {S_n}_{j}}
\exp\left[-\frac{1}{2}\frac{(\Re\tilde n_j)^2 + (\Im\tilde n_j)^2}{T {S_n}_{j}/4}\right]\,,
\label{e:pntilde}
\ee
with volume element
\be
d^{N/2} \tilde n
\equiv
\prod_{j=0}^{N/2-1} d(\Re{\tilde n}_j)d(\Im{\tilde n}_j)\,.
\ee
In the continuum limit:
\be
-\frac{1}{2} n^\dagger C_n^{-1} n
\simeq - \sum_{j=0}^{N/2-1} \frac{2|\tilde n_j|^2}{T {S_n}_{j}}
\simeq
-\frac{1}{2}(\tilde n|\tilde n)\,,
\ee
where
\be
(\tilde g|\tilde k) \equiv
2\int_0^\infty df\>
(S_n(f))^{-1}\left[\tilde g^*(f)\tilde k(f)+ \tilde g(f)\tilde k^*(f)\right]
\ee
is the noise-weighted inner product of $\tilde g(f)$, $\tilde k(f)$.
See \cite{Cutler:1998} for more details regarding the noise-weighted inner 
product in the continuum limit.

\section{Ordinary (scalar) and spin-weighted spherical harmonics}
\label{s:spinweightedY}

This appendix, adapted from \cite{Gair-et-al:2015}, 
summarizes some useful relations involving spin-weighted
and ordinary spherical harmonics, ${}_sY_{lm}(\hat n)$ and $Y_{lm}(\hat n)$.
For more details, see e.g., 
\cite{Goldberg:1967} and \cite{delCastillo}.
Note that for our analyses, we can restrict attention to spin-weighted 
spherical harmonics having {\em integral} spin weight $s$, even though 
spin-weighted spherical harmonics with half-integral spin weight do exist.
\medskip

\noindent
Ordinary spherical harmonics:
\be
Y_{lm}(\hat n)\equiv
Y_{lm}(\theta,\phi) = N_l^m
P_l^m(\cos\theta)e^{im\phi}\,,
\quad
{\rm where}\ 
N_l^m = \sqrt{\frac{2l+1}{4\pi}\frac{(l-m)!}{(l+m)!}}\,.
\label{e:Nlm}
\ee
Relation of spin-weighted spherical harmonics 
to ordinary spherical harmonics:
\be
\begin{aligned}
{}_sY_{lm}(\theta,\phi) 
&=\sqrt{\frac{(l-s)!}{(l+s)!}}\,
\edth^s Y_{lm}(\theta,\phi)
\quad{\rm for}\quad
0\le s\le l\,,
\\
{}_sY_{lm}(\theta,\phi) 
&=\sqrt{\frac{(l+s)!}{(l-s)!}}\,
(-1)^s
\overline{\edth}{}^{-s} Y_{lm}(\theta,\phi)
\quad{\rm for}\quad
-l\le s\le 0\,,
\end{aligned}
\ee
where
\be
\begin{aligned}
\edth \eta
&=-(\sin\theta)^s
\left[\frac{\partial}{\partial\theta}
+i\csc\theta\frac{\partial}{\partial\phi}\right]
(\sin\theta)^{-s}\eta\,,
\\
\overline{\edth} \eta
&=-(\sin\theta)^{-s}
\left[\frac{\partial}{\partial\theta}
-i\csc\theta\frac{\partial}{\partial\phi}\right]
(\sin\theta)^{s}\eta\,,
\end{aligned}
\ee
and $\eta=\eta(\theta,\phi)$ is a spin-$s$ scalar field.\\

\noindent
%
%
%
Complex conjugate:
\be
{}_sY_{lm}^*(\theta,\phi) = (-1)^{m+s}\,{}_{-s}Y_{l,-m}(\theta, \phi)\,.
\ee
Relation to Wigner rotation matrices:
\be
D^l{}_{m'm}(\phi,\theta,\psi)=
(-1)^{m'} \sqrt{\frac{4\pi}{2l+1}}\,{}_mY_{l,-m'}(\theta,\phi)e^{-im\psi}\,,
\label{e:WignerD}
\ee
or
\be
\left[D^l{}_{m'm}(\phi,\theta,\psi)\right]^*=
(-1)^{m} \sqrt{\frac{4\pi}{2l+1}}\,{}_{-m}Y_{l,m'}(\theta,\phi)e^{im\psi}\,.
\label{e:WignerD_CC}
\ee
Parity transformation:
\be
{}_sY_{lm}(\pi-\theta,\phi+\pi) = (-1)^l\,{}_{-s}Y_{lm}(\theta, \phi)\,.
\ee
\noindent
Orthonormality (for fixed $s$):
\be
\int d^2\Omega_{\hat n}\>
{}_sY_{lm}(\hat n) \,{}_{s}Y_{l'm'}^*(\hat n)
\equiv\int_0^{2\pi} d\phi\int_0^\pi \sin\theta\, d\theta\>
{}_sY_{lm}(\theta,\phi) \,{}_sY_{l'm'}^*(\theta,\phi)
= \delta_{ll'}\delta_{mm'}\,.
\ee
Addition theorem for spin-weighted spherical harmonics:
\be
\sum_{m=-l}^l {}_sY_{lm}(\theta_1,\phi_1)\,{}_{s'}Y_{lm}^*(\theta_2,\phi_2)
=(-1)^{-s'}\sqrt{\frac{2l+1}{4\pi}}\,{}_{-s'}Y_{ls}(\theta_3,\phi_3)
e^{is'\chi_3}\,,
\ee
where
\be
\cos\theta_3 = \cos\theta_1\cos\theta_2 + \sin\theta_1\sin\theta_2\cos(\phi_2-\phi_1)\,,
\ee
and
\be
\begin{aligned}
e^{-i(\phi_3+\chi_3)/2}
&=\frac{\cos\frac{1}{2}(\phi_2-\phi_1)\cos\frac{1}{2}(\theta_2-\theta_1)
-i\sin\frac{1}{2}(\phi_2-\phi_1)\cos\frac{1}{2}(\theta_1+\theta_2)}
{\sqrt{\cos^2\frac{1}{2}(\phi_2-\phi_1)\cos^2\frac{1}{2}(\theta_2-\theta_1)
+\sin^2\frac{1}{2}(\phi_2-\phi_1)\cos^2\frac{1}{2}(\theta_1+\theta_2)}}\,,
\\
e^{i(\phi_3-\chi_3)/2}
&=\frac{\cos\frac{1}{2}(\phi_2-\phi_1)\sin\frac{1}{2}(\theta_2-\theta_1)
+i\sin\frac{1}{2}(\phi_2-\phi_1)\sin\frac{1}{2}(\theta_1+\theta_2)}
{\sqrt{\cos^2\frac{1}{2}(\phi_2-\phi_1)\sin^2\frac{1}{2}(\theta_2-\theta_1)
+\sin^2\frac{1}{2}(\phi_2-\phi_1)\sin^2\frac{1}{2}(\theta_1+\theta_2)}}\,.
\end{aligned}
\ee
Addition theorem for ordinary spherical harmonics:
\be
\sum_{m=-l}^l Y_{lm}(\hat n_1) Y_{lm}^*(\hat n_2)
=\frac{2l+1}{4\pi}\,P_l(\hat n_1\cdot\hat n_2)\,.
\label{e:addition_theorem_0Y}
\ee
Integral of a product of spin-weighted spherical harmonics:
\begin{multline}
\int d^2\Omega_{\hat n}\>\>
{}_{s_1}Y_{l_1m_1}(\hat n) \,
{}_{s_2}Y_{l_2m_3}(\hat n) \,
{}_{s_3}Y_{l_3m_3}(\hat n) \,
\\
= \sqrt{\frac{(2l_1+1)(2l_2+1)(2l_3+1)}{4\pi}}
\left( \begin{array}{ccc}l_1&l_2&l_3\\m_1&m_2&m_3\end{array} \right)
\left( \begin{array}{ccc}l_1&l_2&l_3\\-s_1&-s_2&-s_3\end{array} \right)\,,
\end{multline}
where 
$\left( \begin{array}{ccc}l_1&l_2&l_3\\m_1&m_2&m_3 \end{array} \right)$
is a Wigner 3-$j$ symbol~\cite{Wigner:1959, Messiah:1962}.
%
%

\section{Gradient and curl rank-1 (vector) spherical harmonics}
\label{s:grad-curl-vector}

The gradient and curl rank-1 (vector) spherical harmonics are defined 
for $l\ge 1$ by:
\be
\begin{aligned}
Y^{G}_{(lm)a} &\equiv \frac{1}{2} {}^{(1)}\!N_l\partial_a Y_{lm}
=\frac{1}{2} {}^{(1)}\!N_l\left(\frac{\partial Y_{lm}}{\partial\theta}\,\hat\theta_a +
\frac{1}{\sin\theta}\frac{\partial Y_{lm}}{\partial\phi}\,\hat\phi_a\right),
\\
Y^{C}_{(lm)a} &\equiv \frac{1}{2} {}^{(1)}\!N_l(\partial_b Y_{lm})\epsilon^b{}_a
=\frac{1}{2} {}^{(1)}\!N_l\left(-\frac{1}{\sin\theta}\frac{\partial Y_{lm}}{\partial\phi}\,\hat\theta_a
+\frac{\partial Y_{lm}}{\partial\theta}\,\hat\phi_a\right),
\label{e:YGClma}
\end{aligned}
\ee
where $\hat\theta$ and $\hat\phi$ are the standard unit vectors tangent
to the 2-sphere
\be
\begin{aligned}
\hat\theta
&=\cos\theta\cos\phi\,\hat x+
\cos\theta\sin\phi\,\hat y-
\sin\theta\,\hat z\,,
\\
\hat\phi
&=-\sin\phi\,\hat x+
\cos\phi\,\hat y\,,
\label{e:thetahat_phihat}
\end{aligned}
\ee
${}^{(1)}\!N_l$ is a normalization constant
\be
{}^{(1)}\!N_l = \sqrt{\frac{2(l-1)!}{(l+1)!}}\,,
\label{e:1N}
\ee
and $\epsilon_{ab}$ is the Levi-Civita anti-symmetric tensor
\be
\epsilon_{ab} 
= \sqrt{g} \left( \begin{array}{cc}0&1\\-1&0\end{array}\right)\,,
\qquad
g\equiv {\rm det}(g_{ab})\,.
\label{e:levi-civita}
\ee
Following standard practice, we use the metric tensor 
$g_{ab}$ on the 2-sphere 
and its inverse $g^{ab}$ to ``lower" and ``raise" tensor
indices---e.g., $\epsilon^{c}{}_b \equiv g^{ca}\epsilon_{ab}$. 
In standard spherical coordinates $(\theta,\phi)$,
\be
g_{ab}=\left(
\begin{array}{cc}
1&0\\
0&\sin^2\theta\\
\end{array}
\right)\,,
\qquad
\sqrt{g}=\sin\theta\,.
\label{e:gab-2sphere}
\ee
The gradient and curl spherical harmonics are related to the 
spin-weight $\pm1$ spherical harmonics 
\be
{}_{\pm1}Y_{lm}(\theta, \phi) 
=\sqrt{\frac{(l-1)!}{(l+1)!}}
\frac{N_l^m}{\sqrt{1-x^2}} 
\left( \pm(1-x^2) \frac{dP_l^m}{dx} + m P_l^m(x)\right) {\rm e}^{im\phi}\,,
\ee
where $x=\cos\theta$, via
\be
Y^G_{(lm)a} \pm iY^C_{(lm)a}
=\pm\frac{1}{\sqrt{2}}(\hat\theta_a \pm i \hat\phi_a)\,{}_{\mp 1}Y_{lm}\,,
\ee
or, equivalently, 
\be
\begin{aligned}
Y^{G}_{(lm)a}
&=\frac{1}{2\sqrt{2}}\left[
\left({}_{-1}Y_{lm} - {}_{1}Y_{lm}\right) \hat\theta_a
+i  \left({}_{-1}Y_{lm} + {}_{1}Y_{lm}\right) \hat\phi_a\right],
\\
Y^{C}_{(lm)a}
&=\frac{1}{2\sqrt{2}}\left[
\left({}_{-1}Y_{lm} - {}_{1}Y_{lm}\right) \hat\phi_a
-i  \left({}_{-1}Y_{lm} + {}_{1}Y_{lm}\right) \hat\theta_a\right].
\end{aligned}
\ee

For decompositions of vector-longitudinal backgrounds, as discussed in the main text, 
it will be convenient to construct rank-2 tensor fields
\be
\begin{aligned}
Y^{V_G}_{(lm)ab}
&= Y^G_{(lm)a}\hat n_b+ Y^G_{(lm)b}\hat n_a\,,
\\
Y^{V_C}_{(lm)ab}
&= Y^C_{(lm)a}\hat n_b+ Y^C_{(lm)b}\hat n_a\,,
\label{e:YVGVClmab}
\end{aligned}
\ee
where $\hat n$ is the unit radial vector
orthogonal to the surface of the 2-sphere:
\be
\hat n = \sin\theta\cos\phi\,\hat x+\sin\theta\sin\phi\,\hat y
+\cos\theta\,\hat z\,.
\label{e:nhat}
\ee
These fields satisfy the following orthonormality relations
\be
\begin{aligned}
\int d^2\Omega_{\hat n}\> 
Y^{V_G}_{(lm)ab} (\hat n) Y^{V_G}_{(l'm')}{}^{ab\,*}(\hat n) 
&=\delta_{ll'}\delta_{mm'}\,, 
\\
\int d^2\Omega_{\hat n}\> 
Y^{V_C}_{(lm)ab} (\hat n) Y^{V_C}_{(l'm')}{}^{ab\,*}(\hat n) 
&=\delta_{ll'}\delta_{mm'}\,,
\\
\int d^2\Omega_{\hat n}\> 
Y^{V_G}_{(lm)ab} (\hat n) Y^{V_C}_{(l'm')}{}^{ab\,*}(\hat n) 
&=0\,.
\label{e:YVGVCOrthog}
\end{aligned}
\ee
%

\section{Gradient and curl rank-2 (tensor) spherical harmonics}
\label{s:grad-curl-tensor}

The gradient and curl rank-2 (tensor) spherical harmonics are defined for 
$l\ge 2$ by:
\be
\begin{aligned}
Y^G_{(lm)ab} &= {}^{(2)}\!N_l 
\left(Y_{(lm);ab} - \frac{1}{2} g_{ab}  Y_{(lm);c}{}^{c} \right)\,,
\\
Y^C_{(lm)ab} &= \frac{{}^{(2)}\!N_l}{2} 
\left(Y_{(lm);ac}\epsilon^c{}_b +  Y_{(lm);bc} \epsilon^c{}_a \right)\,,
\label{e:YGClmab}
\end{aligned}
\ee
where a semicolon denotes covariant derivative on the 2-sphere, 
$\epsilon_{ab}$ is the Levi-Civita anti-symmetric tensor~(\ref{e:levi-civita}),
$g_{ab}$ is the metric tensor on the 2-sphere~(\ref{e:gab-2sphere}), 
and ${}^{(2)}\!N_l$ is a normalization constant
\be
{}^{(2)}\!N_l = \sqrt{\frac{2 (l-2)!}{(l+2)!}}\,.
\label{e:2N}
\ee
Using the standard polarization tensors on the 2-sphere:
\be
\begin{aligned}
e^+_{ab}(\hat n) 
&= \hat\theta_a\hat\theta_b -\hat\phi_a\hat\phi_b\,,
\\
e^\times_{ab}(\hat n) 
&= \hat\theta_a\hat\phi_b +\hat\phi_a\hat\theta_b\,,
\end{aligned}
\ee
where $\hat\theta$, $\hat\phi$ are given by 
(\ref{e:thetahat_phihat}) and $\hat n$ by (\ref{e:nhat}),
we have~\cite{Hu-White:1997}:
\be
\begin{aligned}
Y^G_{(lm)ab}(\hat n) 
&= \frac{{}^{(2)}\!N_l}{2} \left[ W_{(lm)}(\hat n) e_{ab}^+(\hat n) 
+ X_{(lm)}(\hat n) e_{ab}^\times(\hat n) \right]\,, 
\\
Y^C_{(lm)ab}(\hat n) 
&= \frac{{}^{(2)}\!N_l}{2} \left[ W_{(lm)}(\hat n)e_{ab}^\times(\hat n) 
- X_{(lm)}(\hat n) e_{ab}^+(\hat n) \right]\,,
\label{e:YGCdef}
\end{aligned}
\ee
where
\be
\begin{aligned}
W_{(lm)}(\hat n) 
&= \left( \frac{\partial^2}{\partial \theta^2} 
- \cot\theta\frac{\partial}{\partial\theta} 
+\frac{m^2}{\sin^2\theta} \right)Y_{lm}(\hat n) 
= \left( 2 \frac{\partial^2}{\partial \theta^2} 
+ l(l+1) \right) Y_{lm}(\hat n)\,,
\\
X_{(lm)}(\hat n) 
&= \frac{2 i m}{\sin\theta} \left( \frac{\partial}{\partial\theta} 
- \cot\theta\right) Y_{lm}(\hat n)\,.
\label{e:WX}
\end{aligned}
\ee
%
%
%
%
%
These functions enter the expression for the 
spin-weight~$\pm 2$ spherical harmonics 
\cite{NewmanPenrose:1966, Goldberg:1967}:
\begin{equation}
{}_{\pm2}Y_{lm}(\hat n)
=\frac{{}^{(2)}\!N_l}{\sqrt{2}} \left[
W_{(lm)}(\hat n)\pm i X_{(lm)}(\hat n)\right]\,,
\label{e:Ypm2}
\end{equation}
which are related to the gradient and curl spherical harmonics via
\begin{align}
Y^G_{(lm)ab}(\hat n) \pm i Y^C_{(lm)ab}(\hat n)
&=\frac{1}{\sqrt{2}}
\left(e_{ab}^+(\hat n) \pm i e_{ab}^\times(\hat n)\right)
\,{}_{\mp 2}Y_{lm}(\hat n)\,.
\label{e:YG+iYC}
\end{align}
Note that the gradient and curl spherical harmonics satisfy
the orthonormality relations
\be
\begin{aligned}
\int_{S^2} d^2\Omega_{\hat n}\> 
Y^{G}_{(lm)ab} (\hat n) Y^{G}_{(l'm')}{}^{ab\,*}(\hat n) 
&=\delta_{ll'}\delta_{mm'}\,, 
\\
\int_{S^2} d^2\Omega_{\hat n}\> 
Y^{C}_{(lm)ab} (\hat n) Y^{C}_{(l'm')}{}^{ab\,*}(\hat n) 
&=\delta_{ll'}\delta_{mm'}\,,
\\
\int_{S^2} d^2\Omega_{\hat n}\> 
Y^{G}_{(lm)ab} (\hat n) Y^{C}_{(l'm')}{}^{ab\,*}(\hat n) 
&=0\,.
\label{e:YGCOrthog}
\end{aligned}
\ee
%

\section{Translation between $\hat n$ and $\hat k$ conventions}
\label{s:translation}

Numerous papers on detecting stochastic gravitational-wave backgrounds
have adopted the convention where the polarization tensors and detector 
response functions are functions of the {\em direction of propagation} 
of the gravitational wave, $\hat k$, where $\hat k$ points radially
{\em outward}.
In this article, we have adopted instead the convention where 
plane wave expansions, polarization tensors, and response functions
are written in terms of the {\em direction to the source} of the 
gravitational wave, $\hat n$, where again $\hat n$ points
radially outward.
In both approaches, the unit vectors $\hat l$ and $\hat m$, which are
perpendicular to $\hat k$ (or $\hat n$) and are used to define 
the polarization tensors, are typically chosen to be the standard 
spherical polar coordinate unit vectors $\hat\theta$ and $\hat\phi$.
Thus, the polarization tensors $e^{+,\times}_{ab}(\hat n)$ and 
$e^{+,\times}_{ab}(\hat k)$ are the same for both conventions.
What is different is the expression for an individual 
plane wave---either $e^{i2\pi f(t-\hat k\cdot\vec x/c)}$ or 
$e^{i2\pi f(t+\hat n\cdot\vec x/c)}$---as the direction of propagation
of the wave is opposite the direction to the source.

In this appendix, we summarize how the expressions for the response 
functions $R^{ab}(f,\hat n)$, $R^A(f, \hat n)$, and $R^P_{(lm)}(f)$, 
given in previous sections are related to similar quantities 
calculated in other papers that use the $\hat k$-convention.
For completeness, we will write down expressions for the vector 
and scalar polarization modes (Section~\ref{s:altpol})
in addition to the standard tensor ($+$, $\times$ or grad and curl)
modes in general relativity.
We will denote quantities calculated using the $\hat k$-convention
with an overbar, e.g., $\bar R^A(f,\hat k)$.

\subsection{General relationship between the response functions}

Plane wave expansion:
\be
h_{ab}(t,\vec x) = \int_{-\infty}^\infty \int d^2\Omega_{\hat n}\>
h_{ab}(f,\hat n) e^{i2\pi f(t+\hat n\cdot\vec x/c)}\,.
\ee
Detector response:
\be
\begin{aligned}
h(t) 
&= \int_{-\infty}^\infty d\tau\int d^3 y\>
R^{ab}(\tau, \vec y) h_{ab}(t-\tau, \vec x-\vec y)
\\
&=\int_{-\infty}^\infty df\int d^2\Omega_{\hat n}\>
R^{ab}(f,\hat n) h_{ab}(f,\hat n)e^{i2\pi ft}\,,
\end{aligned}
\ee
where
\be
R^{ab}(f,\hat n) = e^{i2\pi f\hat n\cdot\vec x/c}
\int_{-\infty}^\infty d\tau\int d^3 y\>
R^{ab}(\tau, \vec y) e^{-i2\pi f(\tau +\hat n\cdot \vec y/c)}\,.
\ee
Note that compared to an expansion in terms of the 
direction of propagation $\hat k$, we have:
\be
R^{ab}(f,\hat n) = \bar R^{ab}(f,\hat k)\big|_{\hat k=-\hat n}\,.
\label{e:Rab_reln}
\ee
This is the general relationship between the response
functions for the two approaches.

\subsection{Polarization basis response functions}

The response functions in the polarization basis
are given by:
\be
R^A(f,\hat n) = R^{ab}(f,\hat n) e^A_{ab}(\hat n)\,,
\ee
where $A=\{+,\times, X, Y, B, L\}$ label 
the tensor, vector, and scalar polarization modes
(two for each).
Since the polarization basis tensors 
$e^A_{ab}(\hat n)$ are the same for the two approaches,
it follows from (\ref{e:Rab_reln}) that
\be
R^A(f,\hat n) = \bar R^{ab}(f,\hat k) 
e^A_{ab}(\hat n)\big|_{\hat k=-\hat n}\,.
\ee
If we further use the transformation properties of
the polarization basis tensors $e^A_{ab}(\hat n)$ under a 
parity transformation
(i.e., $\hat n\rightarrow -\hat n$) we have:
\be
\begin{aligned}
R^+(f,\hat n) 
&=\bar R^+(f,\hat k)\big|_{\hat k=-\hat n}\,,
\\
R^\times(f,\hat n) 
&=-\bar R^\times(f,\hat k)\big|_{\hat k=-\hat n}\,,
\\
R^X(f,\hat n) 
&=-\bar R^X(f,\hat k)\big|_{\hat k=-\hat n}\,,
\\
R^Y(f,\hat n) 
&=\bar R^Y(f,\hat k)\big|_{\hat k=-\hat n}\,,
\\
R^B(f,\hat n) 
&=\bar R^B(f,\hat k)\big|_{\hat k=-\hat n}\,,
\\
R^L(f,\hat n) 
&=\bar R^L(f,\hat k)\big|_{\hat k=-\hat n}\,.
\\
\end{aligned}
\ee
Note that in terms of standard angular coordinates
$(\theta,\phi)$ on the sphere, the substitution
$\hat k= -\hat n$ corresponds to 
\be
\theta\rightarrow \pi -\theta\,,
\qquad
\phi\rightarrow \phi+\pi\,,
\ee
for which
\be
\begin{aligned}
&\sin\theta \rightarrow \sin\theta\,,
\\
&\cos\theta \rightarrow -\cos\theta\,,
\\
&\sin\phi \rightarrow -\sin\phi\,,
\\
&\cos\phi \rightarrow -\cos\phi\,.
\end{aligned}
\ee
%

\subsection{Spherical harmonic basis response functions}

The response functions in the spherical harmonic 
basis are given by:
\be
R^P_{(lm)}(f) = \int d^2\Omega_{\hat n}\>
R^{ab}(f,\hat n) Y^P_{(lm)ab}(\hat n)\,,
\ee
where
$P=\{G,C,V_G,V_C, B,L\}$ label the tensor, 
vector, and scalar spherical harmonic modes.
If we use the transformation properties of 
the spherical harmonics
$Y^P_{(lm)ab}(\hat n)$ under a parity transformation,
it follows that:
\be
\begin{aligned}
R^G_{(lm)}(f)
&=(-1)^l\bar R^G_{(lm)}(f)\,,
\\
R^C_{(lm)}(f)
&=(-1)^{l+1}\bar R^C_{(lm)}(f)\,,
\\
R^{V_G}_{(lm)}(f)
&=(-1)^l\bar R^{V_G}_{(lm)}(f)\,,
\\
R^{V_C}_{(lm)}(f)
&=(-1)^{l+1}\bar R^{V_C}_{(lm)}(f)\,,
\\
R^B_{(lm)}(f)
&=(-1)^l\bar R^B_{(lm)}(f)\,,
\\
R^L_{(lm)}(f)
&=(-1)^l\bar R^L_{(lm)}(f)\,.
\end{aligned}
\ee
Thus, the curl modes (both tensor and vector) involve a factor of 
$(-1)^{l+1}$, while all the other modes involve a factor of $(-1)^l$.
 


\end{document}